\documentclass[a4paper,11pt]{article}

\usepackage{verbatim}
\usepackage{jheppub}
\usepackage[T1]{fontenc}
\usepackage{slashed}
\usepackage[utf8]{inputenc}
\usepackage{array}
\usepackage{wrapfig}
\usepackage{multirow}
\usepackage{tabu}
\usepackage{amsmath,amssymb,amsthm,amsfonts}
\usepackage{enumitem}
\usepackage{graphicx}
\usepackage{placeins}
\usepackage{longtable}
\usepackage{float}
\usepackage{caption}
\usepackage{subcaption}
\usepackage[dvipsnames]{xcolor}
\usepackage{tikz,tikz-3dplot}
\usepackage[compat=1.0.0]{tikz-feynman}
\usetikzlibrary{intersections,backgrounds,calc,arrows.meta,shapes.geometric,positioning,arrows.meta,decorations.pathmorphing,decorations.markings}
\usepackage{soul}
\usepackage[usestackEOL]{stackengine}
\usepackage{cancel}
\usepackage{mathrsfs}
\usepackage{lipsum}

\tikzset{
quark/.style={postaction={decorate},
  decoration={markings,mark=at position .5 with {\Huge \arrow[#1]{latex}}}},
scalar/.style={dashed,postaction={decorate},
  decoration={markings,mark=at position .5 with {\arrow[#1]{latex}}}},
gluon/.style={decorate,
 decoration={coil,amplitude=5pt, segment length=10pt, pre length=0cm, post length=0cm}},
boson/.style={-latex,decorate, decoration={snake, segment length=4pt, amplitude=1.8pt, pre length=.1cm, post length=.25cm}},
photon/.style={decorate, decoration={snake, segment length=10pt, amplitude=5pt, pre length=.1cm, post length=.1cm}},
dphoton/.style={decorate, decoration={snake, segment length=4pt, amplitude=1.8pt, pre length=.1cm, post length=.25cm},-latex}
}

\def\R{\mathcal{R}}
\def\I{\mathcal{I}}
\def\T{\mathcal{T}}
\def\r{\boldsymbol{r}}
\def\x{\boldsymbol{x}}
\def\v{\boldsymbol{v}}
\def\w{\boldsymbol{w}}

\def\s{\boldsymbol{s}}
\def\t{\boldsymbol{t}}

\usetikzlibrary{3d}           
\usetikzlibrary{perspective}  

\newcommand*\circled[1]{\tikz[baseline=(char.base)]{\node[shape=circle,draw,inner sep=2pt] (char) {#1};}}
\newcommand{\leftrarrows}{\mathrel{\raise.75ex\hbox{\oalign{%
  $\scriptstyle\leftarrow$\cr
  \vrule width0pt height.5ex$\hfil\scriptstyle\relbar$\cr}}}}
\newcommand{\lrightarrows}{\mathrel{\raise.75ex\hbox{\oalign{%
  $\scriptstyle\relbar$\hfil\cr
  $\scriptstyle\vrule width0pt height.5ex\smash\rightarrow$\cr}}}}
\newcommand{\Rrelbar}{\mathrel{\raise.75ex\hbox{\oalign{%
  $\scriptstyle\relbar$\cr
  \vrule width0pt height.5ex$\scriptstyle\relbar$}}}}

\newcommand\greencheckmark[1][]{%
  \tikz[scale=0.4,#1]{\fill(0,.35) -- (.25,0) -- (1,.7) -- (.25,.15) -- cycle;}%
}
\newcommand\crossmark[1][]{%
  \tikz[scale=0.4,#1]{
    \fill(0,0)--(0.1,0) .. controls (0.5,0.4) .. (1,0.7)--(0.9,0.7) ..  controls (0.5,0.5) ..(0,0.1) --cycle;
    \fill(1,0.1)--(0.9,0.1) .. controls (0.5,0.3) .. (0,0.7)--(0.1,0.7) .. controls (0.5,0.4) ..(1,0.2) --cycle;
  }%
}

\makeatletter
\def\leftrightarrowsfill@{\arrowfill@\leftrarrows\Rrelbar\lrightarrows}
\newcommand{\xleftrightarrows}[2][]{\ext@arrow 3399\leftrightarrowsfill@{#1}{#2}}
\makeatother

\newtheorem{theorem}{Theorem}[section]
\newtheorem{corollary}{Corollary}[theorem]
\newtheorem{lemma}[theorem]{Lemma}
\newtheorem{proposition}{Proposition}[section]

\theoremstyle{definition}

\theoremstyle{remark}
\newtheorem*{remark}{Remark}

\title{\textbf Identifying regions in wide-angle scattering via graph-theoretical approaches}

\author[a,b]{Yao Ma}

\affiliation[a]{Institute for Theoretical Physics, ETH Zürich, 8093 Zürich, Switzerland\footnote{Current Affiliation}}
\affiliation[b]{Higgs Centre for Theoretical Physics, School of Physics and Astronomy,\\
The University of Edinburgh, Edinburgh EH9 3FD, Scotland, U.K.}

\emailAdd{yaomay@phys.ethz.ch}

\abstract{The method of regions, which provides a systematic approach for computing Feynman integrals involving multiple kinematic scales, proposes that a Feynman integral can be approximated and even reproduced by summing over integrals expanded in certain regions. A modern perspective of the method of regions considers any given Feynman integral as a specific Newton polytope, defined as the convex hull of the points associated with Symanzik polynomials. The regions then correspond one-to-one with the lower facets of this polytope.

As Symanzik polynomials correspond to the spanning trees and spanning 2-trees of the Feynman graph, a graph-theoretical study of these polynomials may allow us to identify the complete set of regions for a given expansion. In this work, our primary focus is on three specific expansions: the on-shell expansion of generic wide-angle scattering, the soft expansion of generic wide-angle scattering, and the mass expansion of heavy-to-light decay. For each of these expansions, we employ graph-theoretical approaches to derive the generic forms of the regions involved in the method of regions. The results, applicable to all orders, offer insights that can be leveraged to investigate various aspects of scattering amplitudes.}

\begin{document}
\maketitle
\flushbottom

\section{Introduction}
\label{section-introduction}

Evaluating a given Feynman integral $\I$ with multiple kinematic scales can be rather challenging at high orders, and it is sometimes natural to approximate such an integral by expanding it in the ratios of the small and large scales. Among the various techniques of obtaining the asymptotic expansion, the method of regions (MoR), also referred to as the expansion by regions or the strategy of regions in other literature, allows one to rewrite the original Feynman integral $\I$ as the sum over integrals that are expanded in certain regions~\cite{BnkSmn97}. Namely,
\begin{eqnarray}
\mathcal{I}= \mathcal{I}^{(R_1)}+\mathcal{I}^{(R_2)}+\dots+\mathcal{I}^{(R_n)},
\label{MoR_definition}
\end{eqnarray}
where $R_1,\dots,R_n$ are the regions, and $\I^{(R_1)},\dots,\I^{(R_n)}$ are the corresponding expanded integrals. The MoR states that in each $\I^{(R_i)}$, the integrand is expanded into a Taylor series with respect to the small parameters in $R_i$, while the integration measure is unchanged. Once the regions are known, the problem of evaluating $\I$ is reduced to that of evaluating $\I^{(R_1)},\dots,\I^{(R_n)}$ to certain orders, which can be much simpler.

For a given expansion of a Feynman integral, one fundamental aspect of applying the MoR is to identify the set of relevant regions. In the context of asymptotic expansions in Euclidean space, such as the large mass and large momentum expansions, it has been demonstrated by Smirnov that each of the regions in the expansion corresponds to a specific assignment of large loop momenta within an \emph{asymptotically irreducible subgraph}~\cite{Smn90,Smn94,Smn23}. This correspondence is built upon the earlier work from the 1980s~\cite{CtkGrshnTch82,Ctk83,GrshnLrnTch83,Grshn86,GrshnLrn87,Ctk88I,Ctk88II,SmthDVr88,Grshn89}, which explain that an asymptotic expansion can be achieved by identifying subgraphs whose loop momenta are of the same order of magnitude as the large masses or momenta. Furthermore, a mathematical proof has been established that summing over expansions, with respect to these subgraphs, reproduces the original Feynman integral.

However, in Minkowski space, the presence of lightlike momenta introduces new regions where loop momenta can be collinear, soft, and so on. This complexity makes the identification of relevant regions in an expansion a subtle task. While the MoR is applicable for many processes in Minkowski space, such as the threshold limit $q^2 \rightarrow 4m^2$ for two-loop self-energy and vertex graphs~\cite{BnkSmn97}, the Sudakov limits for two-loop vertex graphs~\cite{SmnRkmt99}, and so on, the identification of these regions has traditionally been performed on a case-by-case basis, often relying on heuristic methods based on examples and experience. It is worth noticing that in ref.~\cite{Jtz11}, where general arguments explaining why the MoR works are provided, the regions are considered as certain disjoint subsets forming a partition of the entire integration measure. This provides insights into identifying the set of regions for a generic Feynman integral, although it remains a challenging endeavor.

For certain types of Feynman integrals, a systematic approach to the regions can be achieved through parameter representation~\cite{Plp08,PakSmn11,JtzSmnSmn12}, in which regions are characterized by specific scalings of the parameters. In this manuscript, we will focus on the Lee-Pomeransky parameterization~\cite{LeePmrsk13}, where the integral information is encapsulated in the Lee-Pomeransky polynomial denoted as $\mathcal{P}\equiv \mathcal{U}+\mathcal{F}$, with $\mathcal{U}$ and $\mathcal{F}$ being the first and second Symanzik polynomials. The core element of this approach is to view any Feynman integral as a Newton polytope $\Delta(\mathcal{P})$, defined as the convex hull of a certain set of points. Each point in this set represents the powers of $x_i$ in a given monomial of $\mathcal{P}$. Each region is then related to a \emph{lower facet} of this Newton polytope, such that its corresponding parameter scaling can be read off from the \emph{region vector} which is normal to this lower facet~\cite{SmnvSmnSmn19,AnthnrySkrRmn19,Smn21book,AnthnryPalRmnSkr22book,HrchJnsSlk22}. In essence, this geometric approach to the MoR translates the task of identifying relevant regions of a Feynman integral, into the task of identifying the lower facets of the corresponding Newton polytope.

Note that this approach effectively captures all the regions in cases where the monomials of $\mathcal{P}$ are all of the same sign, or in cases where there is an analytic continuation from such a Euclidean domain. This is because cancellations within $\mathcal{P}$ can be excluded, and we are allowed to focus solely on the facets of the Newton polytope. However, in situations where the monomials of $\mathcal{P}$ must have distinct signs, regions that are due to cancellations within $\mathcal{P}$ may be relevant as well. In such cases, a suitable change of integration variables is required before constructing the Newton polytope~\cite{JtzSmnSmn12,SmnvSmnSmn19,AnthnrySkrRmn19,Smn21book,AnthnryPalRmnSkr22book}. We note that, although indefinite $\mathcal{P}$ are quite natural in practice, for most cases in the context of wide-angle scattering, all the regions are still from the lower facets of the Newton polytope. As a result, the aforementioned Newton polytope approach will guarantee to identify the complete list of regions. We will explain this in more detail later (in section~\ref{section-validity_Newton_polytope_approach}).

It is then natural to raise the following question: \emph{for any expansion of interest, particularly in the context of wide-angle scattering, can we establish a general rule, which governs all the regions and specifies all the relevant modes}? Notably, computer codes based on the Newton polytope approach, such as Asy2~\cite{JtzSmnSmn12} (part of the FIESTA program~\cite{SmnTtyk09FIESTA,SmnSmnTtyk11FIESTA2,Smn14FIESTA3,Smn16FIESTA4,Smn22FIESTA5}), ASPIRE~\cite{AnthnrySkrRmn19}, and pySecDec~\cite{BrwkHrchJahnJnsKrnSlk18,HrchJnsSlk22}, have already been developed to identify regions. They have proven quite useful when the edges are not too numerous. Therefore, addressing this question requires a deep understanding of the output of these computer codes. This understanding becomes particularly valuable when a large number of edges are involved, thus being associated with a high-dimensional Newton polytope, for which computer codes may struggle to produce timely results. This question, therefore, delves into the Newton polytope approach, and its resolution surpasses the capabilities of current computer codes.

This paper aims to address this question by providing an all-order analysis, predicting the complete list of region vectors for any graph in the context of wide-angle scattering. Furthermore, each region can be interpreted in both parameter and momentum representation. The underlying idea is based on the observation that the terms of $\mathcal{P}$ are characterized by the \emph{spanning (2-)trees} of the graph~$G$, with the leading terms specifically related to the \emph{minimum spanning (2-)trees} of~$G$. Consequently, the problem of identifying the lower facets of $\Delta(\mathcal{P})$ can be reformulated as follows: \emph{we aim to find all the possible scalings of the Lee-Pomeransky parameters $x_1,\dots,x_N$, such that the points, which are associated with the leading terms--equivalently the minimum spanning (2-)trees of $G$--span a lower facet of the Newton polytope $\Delta(\mathcal{P})$.}

\begin{figure}[t]
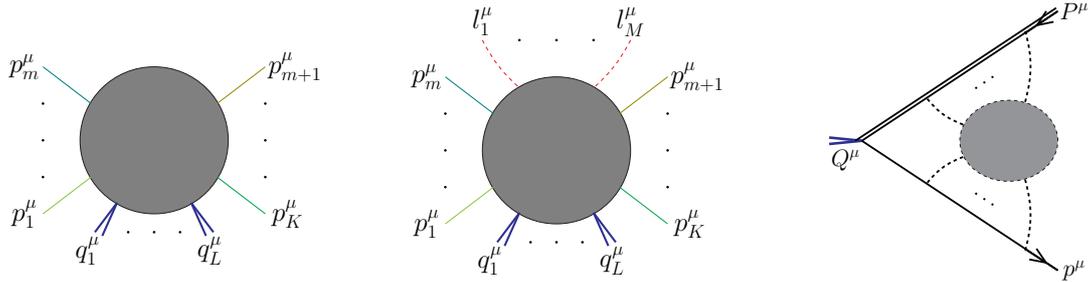

\centering
\begin{subfigure}[b]{0.32\textwidth}
\centering
\include{figs/problem_to_study_onshell}
\vspace{-1em}
\caption{Wide-angle scattering with external momenta $\{p_i^\mu\}_{i=1,\dots,K}$ and $\{q_j^\mu\}_{i=1,\dots,L}$.}
\label{figure-problem_to_study_onshell}
\end{subfigure}
\hfill
\begin{subfigure}[b]{0.32\textwidth}
\centering
\include{figs/problem_to_study_threshold}
\vspace{-1.35em}
\caption{Wide-angle scattering with external momenta~$\{p_i^\mu\}_{i=1,\dots,K}$, $\{q_j^\mu\}_{i=1,\dots,L}$, and $\{l_k\}_{k=1,\dots,M}$.}
\label{figure-problem_to_study_threshold}
\end{subfigure}
\hfill
\begin{subfigure}[b]{0.3\textwidth}
\centering
\include{figs/problem_to_study_mass}
\vspace{-1.5em}
\caption{The heavy-to-light decay processes with external momenta $Q^\mu$, $P^\mu$, and $p^\mu$.}
\label{figure-problem_to_study_mass}
\end{subfigure}
\caption{The graphs which we will focus on in the study of (a) the on-shell expansion, (b) the soft expansion, and (c) the mass expansions, respectively. The gray blob in each figure contains an arbitrary number of internal edges and vertices.}
\label{figure-problems_to_study}
\end{figure}

Throughout this paper, we will focus on three specific expansions. The first is the \emph{on-shell expansion} of generic wide-angle scattering (see figure~\ref{figure-problem_to_study_onshell}). We will consider Feynman graphs with $(K+L)$ non-exceptional external momenta, $p_1^\mu, \dots, p_K^\mu, q_1^\mu, \dots, q_L^\mu$, contributing to off-shell Green’s functions with \emph{massless} fields in Minkowski spacetime. The on-shell expansion is defined by considering the limit $p_i^2\rightarrow 0$ meanwhile keeping all the other invariants to be $\mathcal{O}(Q^2)$. More precisely, by introducing $\lambda\ll 1$ as a scaling parameter,
\begin{eqnarray}
    &&\hspace{-2em}\textbf{on-shell expansion:}\nonumber\\
    &&p_i^2\sim \lambda Q^2\ \ (i=1,\dots,K),\quad q_j^2\sim Q^2\ \ (j=1,\dots,L),\quad p_{i_1}\cdot p_{i_2}\sim Q^2\ \ (i_1\neq i_2).
\label{eq:wideangle_onshell_kinematics}
\end{eqnarray}
Note that the wide-angle condition is incorporated in $p_{i_1}\cdot p_{i_2}\sim Q^2$ above, which implies that the angle between the two three-momenta $\boldsymbol{p}_i$ and $\boldsymbol{p}_j$ is $\mathcal{O}(1)$.

The second expansion we will focus on is the \emph{soft expansion} of generic wide-angle scattering (see figure~\ref{figure-problem_to_study_threshold}). The Feynman graphs have $K+L+M$ external momenta denoted as $p_1^\mu,\dots,p_K^\mu, q_1^\mu,\dots,q_L^\mu, l_1^\mu,\dots,l_M^\mu$, and all its propagators are \emph{massless}. Notably, each $q_j^\mu$ is off shell, and each $p_i^\mu$ and $l_k^\mu$ is strictly on the lightcone, namely, $p_i^2 = l_k^2 = 0$. The key distinction between $p_i^\mu$ and $l_k^\mu$ lies in the fact that each $p_i^\mu$ is collinear to one specific direction, whereas each $l_k^\mu$ is soft. The soft expansion is then defined by the limit $l_k^\mu\rightarrow 0$. More precisely, by introducing the small parameter $\lambda\ll 1$,
\begin{subequations}
\label{eq:wideangle_soft_kinematics}
\begin{align}
    & \hspace{-1em}\textbf{soft expansion:}\nonumber\\
    & p_i^2=0\ \ (i=1,\dots,K), \quad q_j^2\sim Q^2\ \ (j=1,\dots,L), \quad l_k^2=0\ \ (k=1,\dots,M),\\
    & p_{i_1}\cdot p_{i_2}\sim Q^2\ \ (i_1\neq i_2), \quad p_i\cdot l_k\sim q_j\cdot l_k\sim \lambda Q^2, \quad l_{k_1}\cdot l_{k_2}\sim \lambda^2 Q^2\ \ (k_1\neq k_2).
\end{align}
\end{subequations}

The third expansion is a specific \emph{mass expansion} of the heavy-to-light decay processes (see figure~\ref{figure-problem_to_study_mass}, which is a special type of wide-angle scattering). Here the Feynman graphs feature three external momenta: one off-shell momentum $Q^\mu$, one on-shell momentum $P^\mu$ with a large invariant mass $M$, and another on-shell momentum $p^\mu$ with a small mass~$m$. Recent progress in this frontier includes studies such as refs.~\cite{BigiGbn16,SzfrCznck16,Chen18,Gambino:2020jvv,GaoLiZhu13,BrchsfClMnk13,ChenWang18,DattaRanaRvdrSkr23,ChenLiLiWangWangWu23}. In this context, there is a path connecting $Q^\mu$ and $P^\mu$, represented by the double solid line and composed exclusively of mass-$M$ edges; there is a path connecting $Q^\mu$ and $p^\mu$, represented by the single solid line and composed exclusively of mass-$m$ edges; meanwhile, all the other edges of $G$, represented by the dashed lines, are massless. By introducing the small parameter $\lambda\ll 1$, we have
\begin{eqnarray}
    \textbf{mass expansion:} \qquad P^2=M^2\sim Q^2,\quad p^2=m^2\sim \lambda Q^2,\quad P\cdot p\sim Q^2.
\label{eq:decay_mass_kinematics}
\end{eqnarray}

Essentially, regions of a given expansion are intricately linked to the infrared structure of the Feynman integral in the corresponding kinematic limit, which is described by the \emph{Landau equations}~\cite{Lnd59}. Solutions to the Landau equations identify specific submanifolds within the integration space, commonly referred to as \emph{pinch surfaces}~\cite{Stm78I,LbyStm78,Stm95book,Stm96lectures,ClsSprStm04,Cls11book}. These surfaces are named so because certain integration contours are ``pinched'' at each of these manifolds, resulting in singularities of the Feynman integrand. Recent research has shown significant interest in the Landau equations~\cite{Brn09,BlchKrm10,AbrBrtDuhrGrd17cuts,Cls20,BghfKrm23,Mbr20,HndtMLdSwtzVrg22,MzrTln22,BjlVrgvHpl23,FlgTBbdl22,FvlMzrTln23revisited,FvlMzrTln23determinants}, e.g., ref.~\cite{GrdHzgJnsMaSchlk22} which investigates the relation between the regions in the on-shell expansion and the pinch surfaces. A key proposition of that work states that each region vector must adhere to specific forms that correspond to the pinch surfaces. While there is much evidence supporting this proposition based on examples, it remained unclear whether it holds at all orders. One of our main results here is to provide rigorous justification for this proposition.

This paper is organized as follows. In section~\ref{section-general_setup}, we introduce the general setup of the whole analysis in this work. This includes brief introductions to the Newton polytope approach and graph theory, along with some general conclusions that will be applied later in the text. Using this knowledge, in section~\ref{section-regions_onshell}, we study the on-shell expansion of generic wide-angle scattering. We will present a rigorous, all-order proof, demonstrating that each region follows the same configuration as a pinch surface, and only the hard, collinear, and soft modes are relevant.\footnote{Using the lightcone coordinate, each momentum $k^\mu = (k^+,k^-,\boldsymbol{k}_\perp)$ scales as $Q(1,1,1)$ if it is in the hard mode, scales as $Q(1,\lambda,\lambda^{1/2})$ if it is in the collinear-to-$+$ mode, and scales as $Q(\lambda,\lambda,\lambda)$ if it is in the soft mode. In some other literature, this soft mode is referred to as the ``infrared mode''~\cite{Kcmsk89} or the ``ultrasoft mode''~\cite{Smn02book,SmnRkmt99,Jtz11,Smn99,KuhnPeninSmn00}.} By combining this with the results in ref.~\cite{GrdHzgJnsMaSchlk22}, which discusses further constraints on the subgraphs, we derive the \emph{on-shell-expansion region theorem}. Then, in section~\ref{section-regions_soft_expansion}, we include soft momenta in the external kinematics and study the soft expansion of generic wide-angle scattering. Based on various examples, we propose that the regions still follow the same configurations as the pinch surfaces, while only the hard, jet, and soft modes are relevant. Building on this proposition, we further show that each such region must satisfy certain additional requirements. These results are summarized as the \emph{soft-expansion region proposition}. In section~\ref{section-regions_mass_expansion}, we focus on the mass expansion of heavy-to-light decay processes. From some typical examples, we observe that the regions involve not only the hard, collinear and soft modes, but possibly the semihard, soft-collinear, soft$^2$ modes, etc. The emergence of these modes must follow certain rules, from which we classify the regions into three types: I, IIA, and IIB. For planar graphs, the general prescription of the regions is equivalent to the \emph{terrace formalism}, which we will propose and verify through examples. We conclude by summarizing these findings and discussing potential avenues for future research in section~\ref{section-conclusions_outlook}.

Some details are presented in the appendices. In appendix~\ref{appendix-Prim_algorithm}, we discuss the relation between lemma~\ref{lemma-weight_hierarchical_partition_tree_structure} and Prim's algorithm, which finds a minimum spanning tree for a weighted graph. In appendix~\ref{appendix-details_in_proof}, we explain some details in the proof of the on-shell-expansion region theorem. Finally, in appendix~\ref{appendix-mass_expansion_examples} we present some examples in the mass expansion, and manifest all their associated region vectors.

\section{General setup}
\label{section-general_setup}

In this section, we lay the foundation for our analysis by introducing key concepts from both the Newton polytope approach and graph theory. Additionally, we present some fundamental lemmas that will be useful in the subsequent sections, setting the stage for a comprehensive understanding of the regions and formalisms developed throughout this paper.

To achieve this, we begin in section~\ref{section-Newton_polytope_approach} by reviewing the Newton polytope approach to the MoR. In section~\ref{section-some_basics_graph_theory}, we introduce some basic graph-theoretical concepts. We then clarify the basic idea of our later analysis by developing two fundamental criteria in section~\ref{section-two_fundamental_criteria}, and a key concept of ``weight-ordered partition'' of a given graph in section~\ref{section-weight_ordered_partition}. Finally, we summarize some essential concepts and notations in section~\ref{section-notations}.

\subsection{The Newton polytope approach}
\label{section-Newton_polytope_approach}

We now provide a concise overview of the Newton polytope approach to the MoR. Below, we will start by employing the Lee-Pomeransky representation of a Feynman integral, defining the expansions and Newton polytopes. Subsequently, we will clarify why regions correspond to the lower facets of the associated Newton polytope. Following that, we will present a few one-loop examples for illustration. At the end of this subsection, we discuss the validity of the Newton polytope approach for expansions in the wide-angle kinematics.

\subsubsection{Lee-Pomeransky representation}
\label{section-Lee_Pomearnsky_representation}

Let us consider any scalar Feynman integral corresponding to the graph $G$, dimensionally regularized in $D=4-2\epsilon$ dimensions:
\begin{eqnarray}
    \I\equiv \mathcal{C}\cdot \int \prod_{i=1}^L \frac{d^Dk_i}{i\pi^{D/2}} \prod_{e\in G} \frac{1}{D_e^{\nu_e}},
\end{eqnarray}
where $\mathcal{C}$ is an overall factor, $L$ is the number of independent loop momenta, $e$ represents the edges of $G$, and $D_e = l_e^2-m_2^2$, with $l_e^\mu$ the line momentum of $e$. Following the conventions of ref.~\cite{GrdHzgJnsMaSchlk22}, we use $\s$ to denote the set of scalar products formed among the external momenta of $G$. The integral $\I$ is then a function of $\s$, and the Lee-Pomeransky representation of $\I(\s)$ is~\cite{LeePmrsk13}
\begin{equation}
\I(\s) = \mathcal{C}'\cdot \int_0^\infty \left( \prod_{e\in G} \frac{dx_e}{x_e}\right)\cdot \left(\prod_{e\in G} x_e^{\nu_e} \right) \cdot \Big( \mathcal{P}(\x;\s) \Big)^{-D/2},
\label{lee_pomeransky_definition}
\end{equation}
where the overall factor $\mathcal{C}'\equiv \frac{\Gamma(D/2)}{\Gamma((L+1)D/2-\sum_{e\in G}\nu_e) \prod_{e\in G}\Gamma(\nu_e)} \cdot \mathcal{C}$. In the integrand above, we have used $\x$ to denote the set of Lee-Pomeransky parameters $x_1,\dots,x_N$, with $N$ the number of edges of $G$. $\mathcal{P}(\x;\s)$ is called the Lee-Pomeransky polynomial, defined as
\begin{eqnarray}
\label{lee_pomeransky_integrand_definition}
    \mathcal{P}(\x;\s)\equiv \mathcal{U}(\x)+\mathcal{F}(\x;\s),
\end{eqnarray}
where $\mathcal{U}(\x)$ and $\mathcal{F}(\x;\s)$ are the first and second Symanzik polynomials, given by
\begin{eqnarray}
\mathcal{U}(\x)=\sum_{T^1}^{}\prod_{e\notin T^1}^{}x_e,\qquad \mathcal{F}(\x;\s)=\sum_{T^2}^{} (-s_{T^2}^{}) \prod_{e\notin T^2}^{}x_e +\mathcal{U}(\x)\sum_{e}^{}m_e^2 x_e\ . \label{UFterm_general_expression}
\end{eqnarray}
The notations $T^1$ and $T^2$ respectively denote a \emph{spanning tree} and a \emph{spanning 2-tree} of the graph $G$. The symbol $s_{T^2}^{}$ is the square of the total momentum flowing between the components of the spanning 2-tree $T^2$. For each $\mathcal{F}$ term, the associated factor $(-s_{T^2})$ or $m_e^2$ is referred to as its \emph{kinematic factor}. In a particular expansion, eq.~(\ref{UFterm_general_expression}) may be simplified. For example, in the on-shell expansion we will investigate in detail in section~\ref{section-regions_onshell}, the terms in $\mathcal{U}(\x)\sum_{e}^{}m_e^2 x_e$ vanish because all the propagators are massless, and the $s_{T^2}^{}$ can only be $p_i^2$ or $q_j^2$ for some $i,j$, in line with eq.~(\ref{eq:wideangle_onshell_kinematics}).

As we will see later in section~\ref{section-Newton_polyope_lower_facets}, the Lee-Pomeransky polynomial $\mathcal{P}(\x;\s)$ contains all the information needed for constructing the Newton polytope, which further leads to the identification of regions. This allows us to focus solely on scalar Feynman integrals without loss of generality, because any tensor Feynman integral can be reduced to a sum over scalar integrals according to ref.~\cite{Trsv96}, each of which has the same Lee-Pomeransky polynomial $\mathcal{P}(\x;\s)$ and possibly different dimensions.

A region in momentum representation is characterized by a particular scaling of the loop momenta $k$, which is translated into the scaling of $\x$ in Lee-Pomeransky representation. As is demonstrated in ref.~\cite{GrdHzgJnsMaSchlk22}, the scaling limit of each Lee-Pomeransky parameters $x_e$ is related to the scaling of the off-shellness of the corresponding line momentum $l_e^\mu$:
\begin{equation}
\label{eq:LP_offshellness_relation}
x_e \sim D_e^{-1} = \left(l_e^2-m_e^2\right)^{-1}.
\end{equation}
In other words, given the modes in momentum space, one can establish the scaling rule of the Lee-Pomeransky parameters, and vice versa.\footnote{From the scaling of $x_e$, one can immediately obtain the scaling of $l_e^2-m_e^2$ using eq.~(\ref{eq:LP_offshellness_relation}). To further determine the scaling of the components of $l_e^\mu$, there may be ambiguities, and one may need the momentum conservation, the Landau equation, etc. We will see some examples and discuss in more detail in section~\ref{section-regions_mass_expansion}.}

\subsubsection{Expansions}
We next define the expansions of $\I(\s)$ around generic kinematic limits, for example, those in eqs.~(\ref{eq:wideangle_onshell_kinematics})-(\ref{eq:decay_mass_kinematics}). To this end, we partition the set of the Mandelstam variables $\s$ as follows:
\begin{eqnarray}
\label{eq:Mandelstam_variable_partition}
    \s \equiv \s^{(0)}\sqcup \t^{(1)}\sqcup \t^{(2)}\sqcup \dots \t^{(n)}.
\end{eqnarray}
Note that here we use the disjoint union symbol ``$\sqcup$'' to emphasize that any two sets on the right-hand side of eq.~(\ref{eq:Mandelstam_variable_partition}) do not intersect; in principle, one can also use ``$\cup$''. Moreover, for each $s'\in \s^{(0)}$ and $t'\in \t^{(i)}$ (for some $i\in \{1,\dots,n\}$), we have $t'/ s'\sim \lambda^i$, which means that all the variables in the set $\t^{(i)}$ approach zero at the same speed as $\lambda\to 0$. By introducing an $\left | \s \right |$-dimensional scaling vector $\w$:
\begin{equation}
\label{kinematic_variables_scaling}
    \w\equiv \left(w_1,w_2,w_3,\dots\right),\qquad w_j=\left\{\begin{matrix}
    i\quad s_j\in \t^{(i)}\\ 
    0\quad s_j\in \s^{(0)}
    \end{matrix}\right.,
\end{equation}
any kinematic limit can be well described by $s_j\sim \lambda^{w_j} Q^2$ under a suitable partition in (\ref{eq:Mandelstam_variable_partition}), where $Q^2$ is some hard scale from $\s^{(0)}$. For example, for the kinematic limit of the on-shell expansion, eq.~(\ref{eq:wideangle_onshell_kinematics}), we have $n=1$: all the $p_i^2$ are in $\t^{(1)}$ while the rest are in $\s^{(0)}$. For the soft expansion (\ref{eq:wideangle_soft_kinematics}), we have $n=2$: the variables $l_{k_1}\cdot l_{k_2}$ are in $\t^{(2)}$, the variables $p_i\cdot l_k$ and $q_j\cdot l_k$ are in $\t^{(1)}$, and the rest (except $p_i^2=l_k^2=0$, which are not present in $\mathcal{P}(\x;\s)$) are in~$\s^{(0)}$. For the mass expansion (\ref{eq:decay_mass_kinematics}), we have $n=1$: $p^2$($=m^2$) is in $\t^{(1)}$ while $P^2$($=M^2$), $Q^2$, and $P\cdot p$ are all in~$\s^{(0)}$.

As has been explained above, each region $R$ is characterized by a specific scaling of the parameters $\x$. To formulate, $R$ is associated with an $N$-dimensional vector
\begin{eqnarray}
    \boldsymbol{u}_R \equiv \left( u_{R,1},\dots,u_{R,N} \right),
\end{eqnarray}
such that $x_e\sim \lambda^{u_{R,e}}$ for each $e\in G$. Given the scaling of $\s$ and $\x$, each term of $\mathcal{P}(\x;\s)$ is then of a certain scale in $\lambda$. Those terms with the lowest power of $\lambda$ are called the \emph{leading terms}, because they are dominant as $\lambda\to 0$. We denote the sum of all the leading terms as $\mathcal{P}^{(R)}(\x;\s)$, which is equal to the sum over $\mathcal{U}^{(R)}(\x)$ and $\mathcal{F}^{(R)}(\x;\s)$, the leading contribution of the first and second Symanzik polynomials respectively. In line with eq.~(\ref{lee_pomeransky_definition}), we further define:
\begin{eqnarray}
\label{leading_lee_pomeransky_integrand_definition}
    I(\x;\s) \equiv \left( \prod_{e\in G} x_e^{\nu_e} \right) \Big( \mathcal{P}(\x;\s) \Big)^{-D/2},\quad I_0^{(R)}(\x;\s) \equiv \left( \prod_{e\in G} x_e^{\nu_e} \right) \left( \mathcal{P}^{(R)}(\x;\s) \right)^{-D/2}.
\end{eqnarray}
The function $I_0^{(R)}(\x;\s)$, by definition, describes the leading contribution of the asymptotic behavior of $I(\x;\s)$ in the region $R$. To extract the scale of $\lambda$ corresponding to each term of $\mathcal{P}(\x;\s)$, we can rescale $\x\to \lambda^{\boldsymbol{u}_R}\x$ and $\s\to \lambda^{\boldsymbol{w}}\s$, where
\begin{eqnarray}
\lambda^{\boldsymbol{b}} \boldsymbol{a} \equiv (\lambda^{b_1} a_1,\lambda^{b_2} a_2,\lambda^{b_3} a_3,\dots).
\label{lambda_power_b_times_a}
\end{eqnarray}
Then the function $I(\lambda^{\boldsymbol{u}_R}\x; \lambda^{\boldsymbol{w}}\s)$ can be expanded as follows as $\lambda\to 0$
\begin{eqnarray}
    I(\lambda^{\boldsymbol{u}_R}\x; \lambda^{\boldsymbol{w}}\s) \ \overset{\lambda\to 0}{\sim}\ \lambda^{p_R(\epsilon)}  I^{(R)}(\x;\s) + \text{non-leading},
\end{eqnarray}
where the exponent $p_R^{}(\epsilon)$ is a linear function of the dimensional regularization parameter~$\epsilon$. The expansion of $\I(\s)$ in the region $R$, denoted as $\T^{(R)} \I(\s)$, can then be defined by the following Taylor expansion in $\lambda$:
\begin{eqnarray}
\T^{(R)} \I(\s) \equiv \left. \mathcal{C}'\cdot \int_0^\infty \left( \prod_{e\in G} \frac{dx_e}{x_e} \right) T_\lambda \left( \lambda^{-p_R^{}(\epsilon)} I(\lambda^{\boldsymbol{u}_R}\x;\lambda^{\w}\s) \right) \right|_{\lambda=1},
\label{integral_expansion_operator_definition}
\end{eqnarray}
where $T_\lambda\equiv \sum_{n=0}^\infty \lambda^n T_{\lambda,n}$ and $T_{\lambda,n} (\cdots)= \frac{1}{n!} \frac{d^n}{d\lambda^n} (\cdots)|_{\lambda=0}$ are regular Taylor expansion operators. In other words, for each region $R$, the relevant contribution arises from the Taylor expansion in $\lambda$ after we rescale the Lee-Pomeransky parameters $x_i\to \lambda^{u_i} x_i$ and the Mandelstam variables $s_i\to \lambda^{w_i} s_i$. Note that $\lambda$ serves as a bookkeeping parameter, which is finally set to $1$. Let us denote the set of relevant regions by $\mathcal{R}[\I(\s)]$, then the statement of the MoR, eq.~(\ref{MoR_definition}), can be described as:
\begin{eqnarray}
\label{eq:MoR_description_expansions}
    \mathcal{I}(\s) &&= \sum_{R\in\mathcal{R}[\I(\s)]} \T^{(R)} \I(\s) \nonumber\\
    &&= \sum_{R \in \R[\I(\s)]}\mathcal{C}'\cdot \int_0^\infty \left( \prod_{e\in G} \frac{dx_e}{x_e} \right) \left. T_\lambda \left( \lambda^{-p_R^{}(\epsilon)} I(\lambda^{\boldsymbol{u}_R}\x;\lambda^{\w}\s) \right) \right|_{\lambda=1}.
\end{eqnarray}
The vectors $\boldsymbol{w}$ and $\s$ can be directly taken from eqs.~(\ref{eq:Mandelstam_variable_partition}) and~(\ref{kinematic_variables_scaling}). It then remains to identify the list of regions $\mathcal{R}[\I(\s)]$ and their associated vectors $\boldsymbol{u}_R$. Below we will demonstrate how the Newton polytope approach achieves this.

\subsubsection{Newton polytopes and their lower facets}
\label{section-Newton_polyope_lower_facets}

The Newton polytope corresponding to the integral $\I(\s)$ can be obtained as follows. First, let us consider the Lee-Pomeransky polynomial with only the Mandelstam variables rescaled as $\s\to \lambda^{\boldsymbol{w}}\s$:
\begin{align}
&\mathcal{P}(\boldsymbol{x},\lambda^{\w} \s) = \sum_{i} c_i(\s)\, x_1^{r_{i,1}} \dots x_N^{r_{i,N}}\, \lambda^{r_{i,N+1}},
\label{LP_polynomial_rescaling_momenta}
\end{align}
where the powers $r_{i,j}$ take the value of natural numbers. For each $i$, we define the $(N+1)$-dimensional exponent vector
\begin{eqnarray}
    \boldsymbol{r}_i\equiv (\hat{\r}_i; r_{i,N+1})\equiv (r_{i,1}, \ldots, r_{i,N}; r_{i,N+1}).
\end{eqnarray}
We then define the Newton polytope as the \emph{convex hull} of the vertices given by all the exponent vectors $\r_i$, and denote it as $\Delta(\mathcal{P})$. Namely, suppose there are $m$ terms in $\mathcal{P}(\x;\s)$,
\begin{eqnarray}
\label{eq:convex_hull}
\Delta(\mathcal{P}) \overset{\text{def}}{=} \mathrm{convHull}(\boldsymbol{r}_1,\ldots,\boldsymbol{r}_m) = \left\{ \sum_i^m \alpha_i \boldsymbol{r}_i\ \Big|\ \alpha_i \geqslant 0 \textup{ and } \sum_i^m \alpha_i = 1 \right\}.
\end{eqnarray}
Accordingly, the Newton polytope $\Delta(\mathcal{P})$ is an $(N+1)$-dimensional polytope enclosing all points defined by the exponent vectors $\boldsymbol{r}_i$. The first $N$ dimensions correspond to the Lee-Pomeransky parameters $x_1, \ldots, x_N$, which are integrated over, while the $(N+1)$th dimension corresponds to the expansion parameter, $\lambda$.

A key observation is that each leading polynomial $\mathcal{P}^{(R)}$, where $R$ is a region contributing to $\I(\s)$ in eq.~(\ref{eq:MoR_description_expansions}), corresponds to a \emph{lower facet} of $\Delta(\mathcal{P})$. The lower facets of $\Delta(\mathcal{P})$ are defined as those $N$-dimensional faces of $\Delta(\mathcal{P})$, such that for each of them, the $(N+1)$th component of its inward-pointing normal vector $\boldsymbol{v}$ is positive.
To see the correspondence between regions and lower facets, we consider the polynomial
\begin{eqnarray}
    \mathcal{P}(\lambda^{\boldsymbol{u}_R} \boldsymbol{x}, \lambda^{\w} \s) =&& \sum_i c_i(\s)\, \lambda^{\boldsymbol{u}_R \cdot \hat{\r}_i}\, x_1^{r_{i,1}} \dots x_N^{r_{i,N}}\, \lambda^{r_{i,N+1}}\nonumber\\
    =&& \sum_i c_i(\s)\, \lambda^{\boldsymbol{v}_R \cdot \r_i} x_1^{r_{i,1}} \dots x_N^{r_{i,N}},
\label{eq:LPscalingsubstitution}
\end{eqnarray}
where the vector $\v_R$, which we call the \emph{region vector}, is defined as
\begin{eqnarray}
    \v_R\equiv (\boldsymbol{u}_R; 1) = (u_{R,1},\dots,u_{R,N}; 1).
\end{eqnarray}
From above, the scaling of each term of $\mathcal{P}(\x;\s)$, in a region $R$, is manifested as $\lambda^{\v_{R}\cdot \r_i}$.

Since $\mathcal{P}(\lambda^{\boldsymbol{u}_R} \x; \lambda^{\boldsymbol{w}}\s) \overset{\lambda\to 0}{\sim} \mathcal{P}^{(R)}(\lambda^{\boldsymbol{u}_R} \x; \lambda^{\boldsymbol{w}}\s)$ by definition, if we further assume that no cancellation takes place in $\mathcal{P}(\lambda^{\boldsymbol{u}_R} \x; \lambda^{\boldsymbol{w}}\s)$, it follows that the $\mathcal{P}^{(R)}(\x;\s)$ terms are all of the maximum $\lambda^{\v_{R}\cdot \r_i}$. Equivalently, they correspond to exactly those points $\r_i$ that \emph{minimize} the product $\boldsymbol{v}_R \cdot \boldsymbol{r}_i$ in eq.~(\ref{eq:LPscalingsubstitution}). Those points must be on a boundary of the Newton polytope $\Delta(\mathcal{P})$, which, more precisely, can only be an $N$-dimensional facet of $\Delta(\mathcal{P})$.
The reason why ($N-k$)-dimensional faces of $\Delta(\mathcal{P})$ ($k=1,2,\dots$) do not contribute is that, their corresponding dimensionally-regularized Feynman integral is scaleless, and hence set to zero in dimensional regularization:
\begin{eqnarray}
I_0^{(R)}(c^{\boldsymbol{u}} \x;\s) = c^q I_0^{(R)}(\x;\s)\quad \Rightarrow \quad \int [d\x] I_0^{(R)}(\x;\s)=0
\label{eq:uf_scaleless}
\end{eqnarray}
for some $c,q\in\mathbb{R}$ ($c\neq 1$) and $\boldsymbol{u} \in \mathbb{R}^N$.

From now on, we will use $f_R$ to denote the ($N$-dimensional) lower facet associated to~$R$. Then for any point $\r_1\in f_R$, hence corresponding to a leading term, and $\r_2\in \Delta(\mathcal{P}) \setminus f_R$, hence corresponding to a non-leading term, by definition we have:
\begin{eqnarray}
\label{eq:leading_nonleading_points}
    \boldsymbol{v}_R \cdot \boldsymbol{r}_1 < \boldsymbol{v}_R \cdot \boldsymbol{r}_2\quad \Leftrightarrow \quad (\r_2 - \r_1)\cdot \v_R >0.
\end{eqnarray}
In order that the inequality above holds for any $\r_1\in f_R$ and $\r_2\in \Delta(\mathcal{P}) \setminus f_R$, the vector $\v_R$ must be normal to $f_R$ and points inward $\Delta(\mathcal{P})$. Since the $(N+1)$th entry of $\v_R$ is positive (more precisely, equal to $1$), it follows that $f_R$ must be a lower facet of $\Delta(\mathcal{P})$.

In identifying the region vectors above, we have been using the Lee-Pomeransky representation, where the Newton polytope is constructed from the polynomial $\mathcal{P}(\x;\s) = \mathcal{U}(\x)+\mathcal{F}(\x;\s)$. Alternatively, one may use the Feynman representation, and consider the Newton polytope constructed from $\mathcal{U}(\x)\mathcal{F}(\x;\s)$, for example, in refs.~\cite{PakSmn11,JtzSmnSmn12}. It has been shown in ref.~\cite{SmnvSmnSmn19} that the results of these two approaches lead to identical regions.

\subsubsection{One-loop examples}
\label{section-one_loop_examples}

\begin{figure}[t]
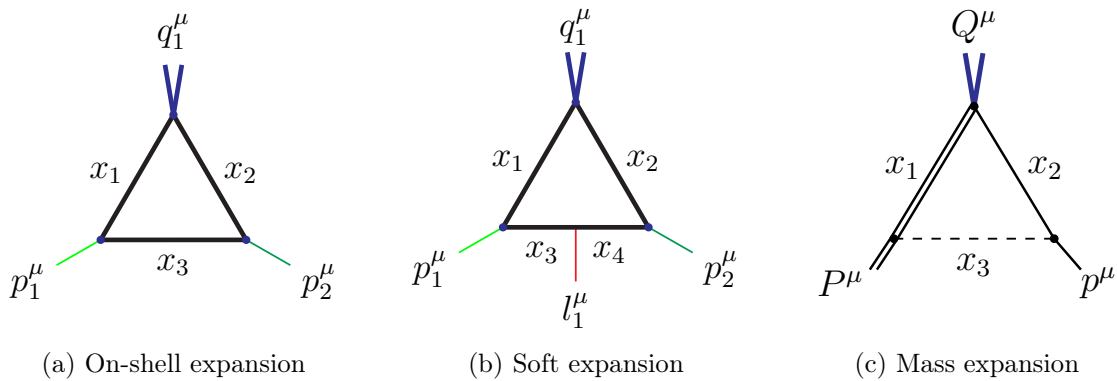

\centering
\begin{subfigure}[b]{0.32\textwidth}
\centering
\include{figs/coleman_norton_restriction_g}
\vspace{-2em}
\caption{On-shell expansion}
\label{one_loop_graph_example_onshell}
\end{subfigure}
\hfill
\begin{subfigure}[b]{0.32\textwidth}
\centering
\include{figs/coleman_norton_restriction_1}
\vspace{-3em}
\caption{Soft expansion}
\label{one_loop_graph_example_threshold}
\end{subfigure}
\hfill
\begin{subfigure}[b]{0.3\textwidth}
\centering
\include{figs/coleman_norton_restriction_2}
\vspace{-2em}
\caption{Mass expansion}
\label{one_loop_graph_example_mass}
\end{subfigure}
\caption{Three examples of the one-loop Sudakov form factor}
\label{figure-one_loop_graph_examples}
\end{figure}

Let us now demonstrate the concepts introduced in sections~\ref{section-Lee_Pomearnsky_representation}-\ref{section-Newton_polyope_lower_facets} using some one-loop examples. We shall focus on the three graphs in figure~\ref{figure-one_loop_graph_examples}, which are the simplest examples whose Feynman integrals have multiple regions.

We start by focusing on the triangle graph (figure~\ref{one_loop_graph_example_onshell}) in the on-shell limit $p_1^2\sim p_2^2\sim \lambda q_1^2$ as $\lambda\to 0$. Thus $\s^{(0)} = \{ q_1^2 \}$ and $\t^{(1)} = \{ p_1^2, p_2^2 \}$, and the Lee-Pomeransky polynomial is
\begin{eqnarray}
    \mathcal{P}(\x;\s) = x_1+x_2+x_3 + (-p_1^2)x_1x_3+ (-p_2^2)x_2x_3+ (-q_1^2)x_1x_2.
\end{eqnarray}
The 4-dimensional Newton polytope $\Delta (\mathcal{P})$ is defined as the convex hull of the six points $\r_i = (r_{i,1}, r_{i,2}, r_{i,3}; r_{i,4})$, whose entries are the exponents of $\{ x_1,x_2,x_3; \lambda\}$ taken from the monomials of $\mathcal{P}(\x;\s)$:
\begin{align}
\Delta (\mathcal{P}) = \mathrm{convHull}(\r_1,\r_2,\r_3,\r_4,\r_5,\r_6),
\end{align}
with $\r_1 = (1,0,0,0),\, \r_2 = (0,1,0,0),\, \r_3 = (0,0,1,0),\, \r_4=(1,0,1,1),\, \r_5=(0,1,1,1)$, and $\r_6=(1,1,0,0)$. The vectors, which are normal to the lower facets of $\Delta(\mathcal{P})$ and pointing inward $\Delta(\mathcal{P})$, can be obtained from the computer codes (for example, we use pySecDec here) directly, which are:
\begin{center}
\begin{tabular}{ll}
 $\boldsymbol{v}_H = (0,0,0;1),$\quad &\quad $\boldsymbol{v}_S = (-1,-1,-2;1),$  \\ 
 $\boldsymbol{v}_{C_1} = (-1,0,-1;1),$\quad &\quad $\boldsymbol{v}_{C_2} = (0,-1,-1;1),$
\end{tabular}
\end{center}
Note that the last entry of each $\v_R$ is positive by definition, and we have fixed it as $1$. Above we have denoted $R=H,S,C_1,C_2$, representing the hard region, the soft region, the collinear-to-$p_1$ region, and the collinear-to-$p_2$ region, respectively. The correspondence between the region vectors above and the regions in momentum space can be seen by employing eq.~(\ref{eq:LP_offshellness_relation}).

Using the same strategy, we can derive the regions for the soft expansion of figure~\ref{one_loop_graph_example_threshold} and the mass expansion of figure~\ref{one_loop_graph_example_mass}. The corresponding Mandelstam variables, Symanzik polynomials and region vectors are summarized in table~\ref{table-oneloop_Sudakov_form_factor_result}.
\begin{table}[t]
\begin{center}
\begin{tabular}{ |c||c|c|c|c| } 
\hline
 & \multirow{2}{8em}{on-shell expansion} & \multirow{2}{7em}{\ soft expansion} & \multirow{2}{7em}{mass expansion} \\
 & & & \\
\hline
$\t^{(1)}$ & $p_1^2,\quad p_2^2$ & $p_1\cdot l_1,\quad p_2\cdot l_1$ & $p^2=m^2$ \\
\hline
$\s^{(0)}$ & $q_1^2,\quad p_1\cdot p_2$ & $q_1^2 = 2p_1\cdot p_2$ & $P^2=M^2,\quad Q^2$ \\
\hline
$\mathcal{U}(\x)$ & $x_1+x_2+x_3$ & $x_1+x_2+x_3+x_4$  & $x_1+x_2+x_3$ \\
\hline
\multirow{3}{3.5em}{$\mathcal{F}(\x;\s)$} & \multirow{3}{10.5em}{$\begin{matrix}
(-p_1^2)x_1x_3 +(-p_2^2)x_2x_3\\ 
+(-q_1^2) x_1x_2
\end{matrix}$} & $(-(p_1+l_1)^2)x_1x_4$ & \multirow{3}{10em}{$\begin{matrix}
M^2x_1^2 +m^2x_2^2\\ 
+(M^2+m^2-Q^2)x_1x_2
\end{matrix}$}\\ 
 &  & $+(-(p_2+l_1)^2)x_2x_3$ & \\
 &  & $+(-q_1^2) x_1x_2$ & \\
\hline
\multirow{4}{1.5em}{$\boldsymbol{v}_R$} & $(0,0,0;1)$ & $(0,0,0;1)$ &\\
 & $(-1,-1,-2;1)$ & $(-1,-1,-2;1)$ & $(0,0,0;1)$ \\
 & $(-1,0,-1;1)$ & $(-1,0,-1;1)$ & $(0,-1,-1;1)$ \\
 & $(0,-1,-1;1)$ & $(0,-1,-1;1)$ & \\
\hline
\end{tabular}
\end{center}
\vspace{-1em}\caption{The Mandelstam invariants, the Symanzik polynomials, and the region vectors associated with figures~\ref{one_loop_graph_example_onshell}, \ref{one_loop_graph_example_threshold}, and \ref{one_loop_graph_example_mass}, respectively.}
\label{table-oneloop_Sudakov_form_factor_result}
\end{table}

From the results, it is natural to postulate that all the regions can be categorized into two types: a single \emph{hard region}, where every Lee-Pomeransky parameter scales as $\lambda^0$, and a set of \emph{infrared regions}, for which some parameters scale as $\lambda^a$ with $a$ being negative. Each of the infrared regions above corresponds to a pinch surface, which are characterized by the hard, jet and soft subgraphs of $G$. Later, from the all-order analyses in section~\ref{section-regions_onshell}-\ref{section-regions_mass_expansion}, we will see that for the on-shell expansion of figure~\ref{figure-problem_to_study_onshell} and the soft expansion of figure~\ref{figure-problem_to_study_threshold}, all the regions only involve the hard, collinear and soft modes. In contrast, for the mass expansion of figure~\ref{figure-problem_to_study_mass}, other modes such as the semihard mode, the semicollinear mode, the soft-collinear mode, and so on, are possibly involved in a region.

\subsubsection{The validity of the Newton polytope approach}
\label{section-validity_Newton_polytope_approach}

In establishing the correspondence between regions and the lower facets of the Newton polytope, we have assumed that for each region $R$, the leading terms and the leading polynomial $\mathcal{P}^{(R)}$ have the same (maximum) scale in $\lambda$, implying no cancellations among these leading terms. Such regions correspond to endpoints in parameter space, and we refer to them as \emph{endpoint-type regions}. This assumption, that all regions are of the endpoint type, holds automatically when all terms of $\mathcal{P}(\x;\s)$ have the same sign or in scenarios where there exists an analytic continuation from such a Euclidean domain. Expansions with solely Euclidean kinematics clearly belong to this type, such as the large mass expansion, the large momentum expansion, etc. Another type of graphs is the massless $2\to 2$ planar wide-angle scattering, where the independent kinematic factors $(-p_i^2)$ ($i=1,2,3,4$), $(-s_{12})$, and $(-s_{13})$ can be simultaneously chosen positive. In such cases, every term in the Lee-Pomeransky polynomial is positive.

If the terms of $\mathcal{P}(\x;\s)$ have indefinite signs, however, it is possible that certain regions arise from the cancellation among these terms, which are hidden inside the Newton polytope $\Delta(\mathcal{P})$ rather than corresponding to its lower facets. In fact, such regions correspond to pinches in the parameter space, and we refer to them as the \emph{pinch-type regions}. Examples of pinch-type regions include the potential region in self-energy graphs, the Glauber region in the Drell-Yan process, etc., as have been discussed in ref.~\cite{JtzSmnSmn12,AnthnrySkrRmn19}. To address these regions using the Newton polytope approach, one must employ a suitable change of integration variables and consider the Newton polytope in the space of these new parameters. While this method has been successful in identifying the potential and Glauber regions at the one-loop level, a systematic approach to the relevant change(s) of variables, especially for multiloop graphs, remains an open challenge.

It is worth noting that for most cases, distinct signs in $\mathcal{P}(\x;\s)$ do not imply pinch-type regions, because no cancellation can be compatible with a solution of the Landau equations for generic kinematics (namely, arbitrary values of independent Mandelstam variables). One example is provided in appendix A of ref.~\cite{GrdHzgJnsMaSchlk22}, where the four-point nonplanar double-box graph is analyzed in detail. In the on-shell kinematic limit, the coefficients are necessarily of different signs, but potential cancellations within $\mathcal{P}(\x;\s)$ do not correspond to any solutions of the Landau equations. As a result, no pinch-type regions are relevant in that example.

One may wonder whether pinch-type regions are needed for wide-angle scattering in general. This question is highly technical and will be investigated in detail in the forthcoming paper, ref.~\cite{GrdHzgJnsMa24}. The conclusion is that pinch-type regions can emerge, but only for very limited types of graphs, all of which feature the configuration of the so-called ``Landshoff scattering'', i.e., the hard scattering takes place at distinct points. These graphs are nonplanar and occur at the three-loop level. Meanwhile, although these graphs have pinch-type regions, most of the regions are still of the endpoint type. We emphasize that in this paper, we will focus exclusively on endpoint-type regions, which are the majority of regions for generic wide-angle scattering.

\subsection{Some basic concepts of graph theory}
\label{section-some_basics_graph_theory}

In this subsection, we introduce some basic graph theory concepts and notations, which will be repeatedly used in the later sections.

A graph $\Gamma$ consists of a set of \emph{vertices} $\mathcal{V}(\Gamma)$, a set of \emph{edges} $\mathcal{E}(\Gamma)$, and a set of external momenta that attach to $\Gamma$. For any other graph $\gamma$, we call that $\gamma$ is a \emph{subgraph} of $G$ and write $\gamma\subset \Gamma$, if each vertex, edge, and external momentum of $\gamma$ also belongs to $\Gamma$.

For any two graphs $\Gamma_1$ and $\Gamma_2$, we define their \emph{union} $\Gamma_1\cup \Gamma_2$ as the graph consisting of the vertices in $\mathcal{V}(\Gamma_1)\cup \mathcal{V}(\Gamma_2)$, the edges in $\mathcal{E}(\Gamma_1)\cup \mathcal{E}(\Gamma_2)$, and the external momenta attaching to either $\Gamma_1$ and $\Gamma_2$. We define their \emph{intersection} $\Gamma_1\cap \Gamma_2$ as the graph consisting of the vertices in $\mathcal{V}(\Gamma_1)\cap \mathcal{V}(\Gamma_2)$, the edges in $\mathcal{E}(\Gamma_1)\cap \mathcal{E}(\Gamma_2)$, and the external momenta attaching to both $\Gamma_1$ and $\Gamma_2$. Similarly, we denote the graph as $\Gamma_1\setminus \Gamma_2$, which consists of the vertices in $\mathcal{V}(\Gamma_1)\setminus \mathcal{V}(\Gamma_2)$, the edges in $\mathcal{E}(\Gamma_1)\setminus \mathcal{E}(\Gamma_2)$, and those external momenta attaching to $\Gamma_1$ but not to $\Gamma_2$. In contrast, $\Gamma_1/\Gamma_2$ is the graph obtained from $\Gamma_2$ by contracting $\Gamma_1$ to a vertex.

We say that a graph $\Gamma_1$ is \emph{adjacent} to another graph $\Gamma_2$ (or, $\Gamma_1$ and $\Gamma_2$ are adjacent to each other), if there is a vertex $v\in \Gamma_1\cup \Gamma_2$ and two edges $e_1\in \Gamma_1$ and $e_2\in \Gamma_2$, such that $e_1$ and $e_2$ are both incident with $v$.\footnote{In graph theory, an edge is incident with a vertex (or a vertex is incident with an edge) if the vertex is one of the two vertices the edge connects.} Note that from this definition, it is possible that $\Gamma_1$ and $\Gamma_2$ are adjacent meanwhile $\Gamma_1\cap \Gamma_2 = \varnothing$.

Any connected graph containing no loops is called a \emph{tree}. Any graph with two connected components, each of which is a tree, is called a \emph{2-tree}. If a tree $T$ contains all the vertices of a graph $\Gamma$, then we call $T$ a spanning tree of $\Gamma$. In the later context, we will use $G$ to denote the entire Feynman graph, and use $T^1$ and $T^2$ to denote spanning trees and spanning 2-trees of $G$, respectively. In particular, the spanning tree $T^1$ satisfy the following properties.
\begin{enumerate}
    \item \emph{For any two vertices $v_1,v_2\in G$, there is a unique path $P\subset T^1$ connecting $v_1$ and $v_2$.}
    \item \emph{For any edge $e\in G\setminus T^1$, there is a unique loop in the graph $T^1\cup e$.}
    \item \emph{For any subgraph $\gamma\subset G$, each $T^1$ and $T^2$ in eq.~(\ref{UFterm_general_expression}) satisfies}
    \begin{eqnarray}
        n_{\gamma} \geqslant L(\gamma),
    \label{eq:removed_edge_number_constraint1}
    \end{eqnarray}
    \emph{with $L(\gamma)$ the number of loops in $\gamma$, and $n_\gamma$ the number of edges in $\gamma\setminus T^1$ or $\gamma\setminus T^2$.}
\end{enumerate}

Let us associate each edge $e\in G$ with a real number $w(e)$, which we refer to as the \emph{weight} of~$e$. We then focus on Feynman graphs with only massless propagators, where $\mathcal{F}(\x;\s)$ can be simplified from eq.~(\ref{UFterm_general_expression}):
\begin{eqnarray}
\label{UFterm_massless_general_expression}
    \text{Massless:}\qquad \mathcal{U}(\x)=\sum_{T^1}^{}\prod_{e\notin T^1}^{}x_e,\qquad \mathcal{F}(\x;\s) = \sum_{T^2}^{} (-s_{T^2}^{}) \prod_{e\notin T^2}^{}x_e.
\end{eqnarray}
We define the weight of a spanning tree $T^1$ or a spanning 2-tree $T^2$ above as follows: $w(T^1)$ is the sum over $w(e)$ where $e$ is \emph{not} in $T^1$; in contrast, $w(T^2)$ is the sum over $w(e)$ where $e$ is \emph{not} in $T^2$, plus a kinematic contribution $n$ if the kinematic factor $s_{T^2}\sim \lambda^n Q^2$. Namely,
\begin{subequations}
    \begin{align}
        & w(T^1)\equiv \sum_{e\notin T^1} w(e);
        \label{eq:definition_spanning_tree_weight}\\
        & w(T^2)\equiv \sum_{e\notin T^2} w(e) + n\quad \text{ if }\ \  s_{T^2}\sim \lambda^n Q^2.
        \label{eq:definition_spanning_2_tree_weight}
    \end{align}
\end{subequations}
Among all these $T^1$ and $T^2$, we select those with the minimum weight, and refer to them as the \emph{minimum spanning }(\emph{2-})\emph{trees} of $G$.

Based on the conceptual setups in section~\ref{section-Newton_polytope_approach}, the spanning (2-)trees of $G$ are closely related to the points in the Newton polytope $\Delta(\mathcal{P})$. For convenience, from now on we use $\x^{\r}$ to denote a generic term of $\mathcal{P}(\x;\s)$, whose corresponding point in the Newton polytope is~$\r$, and corresponding spanning (2-)tree is $T(\r)$. In particular, if $\x^{\r}$ is in the leading Lee-Pomeransky polynomial $\mathcal{P}^{(R)}(\x;\s)$, then $\r$ in the $N$-dimensional lower facet $f_R$, and $T(\r)$ is a minimum spanning (2-)trees of $G$.

With these notations, we can rewrite eq.~(\ref{eq:definition_spanning_2_tree_weight}) as
\begin{eqnarray}
\label{eq:definition_spanning_2_tree_weight_rewritten}
    w(T^2(\r))\equiv \sum_{e\notin T^2(\r)} w(e) + r_{N+1},
\end{eqnarray}
because the ($N+1$)th entry of $\r$ is exactly the power of $\lambda$ after the kinematic factors are rescaled as $\s\to \lambda^{\w} \s$ (see eq.~(\ref{LP_polynomial_rescaling_momenta})). It is also clear from above that $r_{N+1}\in \mathbb{N}$.

\subsection{Two fundamental criteria}
\label{section-two_fundamental_criteria}

In this subsection, we introduce two fundamental criteria that describe the defining properties of the lower facets. These criteria are central in the upcoming text, as they form the basis for every lemma we will prove. The first criterion concerns the dimensionality of the facets. Since the Newton polytope is $N+1$ dimensional and each its facet is $N$ dimensional, the vector normal to $f_R$ is unique up to rescaling. We refer to this as the \emph{facet criterion}.
\begin{center}
\framebox{\Longstack[l]{\textbf{Facet criterion:}\\\qquad up to rescaling, the region vector $\v_R$ is the unique vector normal to $f_R$.}}
\end{center}

One direct result from this criterion is the inhomogeneity of the leading Lee-Pomeransky polynomial $\mathcal{P}^{(R)}(\x;\s)$: for any subset of the parameters, $\x' \subseteq \x$, the polynomial $\mathcal{P}^{(R)} (\x;\s)$ cannot be a homogeneous function of $\x'$. Otherwise, in addition to $\v_R$, the following vector will also be normal to $f_R$:
\begin{eqnarray}
\label{eq:facet_criterion_violation_additional_vector}
\v'= (v'_1,v'_2,\dots,v'_N;0),
\end{eqnarray}
where for each $i$, the entry $v'_i=1$ if $x_i\in \x'$ and $v'_i=0$ if $x_i\notin \x'$. The vectors $\v_R$ and $\v'$ are certainly not proportional to each other, which violates the facet criterion. In terms of graphs, the inhomogeneity of $\mathcal{P}^{(R)}$ indicates that for any $\gamma\subseteq G$, there must be two $\mathcal{P}^{(R)}$ terms, such that $n_{\gamma}$ is not of the same value for them.

Further properties regarding the leading Lee-Pomeransky polynomial $\mathcal{P}^{(R)}(\x;\s)$ can be derived, particularly for massless Feynman graphs, and summarize the results in the following lemma.
\begin{lemma}
For a massless Feynman graph $G$, given any region $R$ we have
\begin{enumerate}
    \item [1,] for any $\gamma\subseteq G$, there must exist a $\mathcal{P}^{(R)}$ term such that $n_{\gamma}>L(\gamma)$;
    \item [2,] $\mathcal{P}^{(R)}(\x;\s)$ must include both $\mathcal{U}^{(R)}$ terms and $\mathcal{F}^{(R)}$ terms;
    \item [3,] let us define $\v_H\equiv (0,\dots,0;1)$, then $\mathcal{P}^{(R)}(\x;\s)$ must include both $\mathcal{U}^{(R)}$ terms and some $\mathcal{F}^{(R)}$ terms with vanishing kinematic factors as $\lambda\to 0$, unless $\v_R=\v_H$.
\end{enumerate}
\label{lemma-existence_certain_terms_P}
\end{lemma}
\begin{proof}
To justify the first statement, let us recall from the relation (\ref{eq:removed_edge_number_constraint1}) that $n_{\gamma}\geqslant L(\gamma)$ holds for any $\gamma\subseteq G$. If all the $\mathcal{P}^{(R)}$ terms satisfy the condition $n_{\gamma}= L(\gamma)$, the facet criterion would be violated from above. Therefore, there must exist at least one $\mathcal{P}^{(R)}(\x;\s)$ term, for which $n_{\gamma}>L(\gamma)$ holds.

To justify the second statement, we set $\gamma=G$. Note that the degree of $\mathcal{U}^{(R)}(\x)$ is $L(G)$ and the degree of $\mathcal{F}^{(R)}(\x;\s)$ is $L(G)+1$. The inhomogeneity of $\mathcal{P}^{(R)}$ then immediately implies that both $\mathcal{U}^{(R)}$ and $\mathcal{F}^{(R)}$ terms exist.

Finally, if each $\mathcal{P}^{(R)}$ term $\x^{\r}$ is either a $\mathcal{U}^{(R)}$ term or an $\mathcal{F}^{(R)}$ terms with a kinematic factor not vanishing as $\lambda\to 0$, then such kinematic factors is $\mathcal{O}(\lambda^0 Q^2)$, and it then follows that the last entry of $\r$ is exactly $0$. Therefore, apart from $\v_R$, another vector $\v_H=(0,\dots,0;1)$ would also be normal to the corresponding lower facet $f_R$. This would violate the facet criterion unless $\v_R=\v_H$. Therefore, we have justified the third statement.
\end{proof}

After examining the implications arising from the dimensionality of $f_R$, our attention shifts to another characteristic of $f_R$: it is a ``lower'' facet rather than an upper facet. This brings us to the second criterion, which we term the \emph{minimum-weight criterion}.
\begin{center}
\framebox{\Longstack[l]{\textbf{Minimum-weight criterion:}\\\qquad$w(T(\r))\leqslant w(T(\r'))$ for any $\x^{\r}\in \mathcal{P}^{(R)}(\x;\s)$ and $\x^{\r'}\in \mathcal{P}(\x;\s)$.}}
\end{center}

This criterion plays a crucial role in our subsequent analysis, particularly in the process of eliminating terms of certain forms from $\mathcal{P}^{(R)}$. The basic idea is following: for each $\mathcal{P}^{(R)}$ term represented as $\x^{\r}$, if we can modify the spanning (2-)tree $T(\r)$ by adding and/or deleting some edges in such a way that the resulting graph $T(\r')$ corresponds to another $\mathcal{P}$ term $\x^{\r'}$ and $w(T(\r')) < w(T(\r))$, then the minimum-weight criterion is violated, which indicates that $\x^{\r}$ cannot be a $\mathcal{P}^{(R)}$ term.

As a first example, let us now apply the minimum-weight criterion to some massless Feynman graphs. In this scenario, a connected subgraph $\gamma$ has internal edges with ``large weights'' yet is only adjacent to edges with ``small weights''.
\begin{lemma}
For any massless Feynman graph $G$ and region $R$, if there is a connected subgraph $\gamma \subset G$ satisfying:
\begin{eqnarray}
\textup{min}\left \{ w(e)\big| e\in\gamma \right \} > \textup{max}\left \{ w(e)\big| e\textup{ is adjacent to } \gamma \right \},
\label{lemma3_assumption}
\end{eqnarray}
then there must exist an $\mathcal{F}^{(R)}$ term $\x^{\r}$, such that both components of $T^2(\r)$ contain some vertices of $\gamma$. Moreover, if $\textup{min}\left \{ w(e)\big| e\in\gamma \right \} > -1$, then $\x^{\r}$ is in $\mathcal{F}^{(q^2,R)}(\x;\s)$.
\label{lemma-heavy_subgraph_inside_light_graph}
\end{lemma}

\begin{proof}
From lemma~\ref{lemma-existence_certain_terms_P}, the relation $n_{\gamma}>L(\gamma)$ must hold for some leading term $\x^{\r}$.

If $\x^{\r}$ is a $\mathcal{U}^{(R)}$ term, then all the vertices of $\gamma$ lie in the spanning tree $T^1(\r)$. Moreover, $n_{\gamma}>L(\gamma)$ implies that $\gamma\cap T^1(\r)$ is disconnected, and since $\gamma$ is connected, there is an edge $e\in \gamma\setminus T^1(\r)$, whose endpoints $v_1$ and $v_2$ are in two distinct components of $T^1(\r)$ respectively. Similarly, since $T^1(\r)$ is connected, there is a path $P\subset T^1(\r)$ that joins $v_1$ and $v_2$, containing some edge $e'$ that is adjacent to $\gamma$. Now we consider the graph obtained from $T^1(\r)$ by adding $e$ and deleting $e'$, namely, $T^1(\r)\cup e\setminus e'$. Note that it is still a spanning tree of $G$ because the graph $T^1(\r)\cup e$ contains a unique loop including $e'$, which is ``broken'' after $e'$ is deleted. Let us denote its corresponding $\mathcal{U}^{(R)}$ term by $\x^{\r'}$, i.e. $T^1(\r') = T^1(\r)\cup e\setminus e'$, whose weight is
\begin{eqnarray}
w(T^1(\r')) = w(T^1(\r)) -w(e) +w(e')< w(T^1(\r)),
\end{eqnarray}
where we have used $w(e)>w(e')$ from the relation (\ref{lemma3_assumption}). The result $w(T^1(\r'))< w(T^1(\r))$ violates the minimum-weight criterion, so $\x^{\r}$ cannot be a $\mathcal{U}^{(R)}$ term.

In other words, the term $\x^{\r}$ belongs to $\mathcal{F}^{(R)}(\x;\s)$. Using precisely the same logic as above, the vertices of $\gamma$ cannot simultaneously lie in the same component of $T^2(\r)$, otherwise another spanning 2-tree $T^2(\r')$ can be obtained, with $w(T^2(\r')) < w(T^2(\r))$. This again violates the minimum-weight criterion. Therefore, the vertices of $\gamma$ must lie in the two distinct components of $T^2(\r)$.

Finally, if $\textup{min}\left\{ w(e)\big| e\in \gamma \right\}>-1$, let us consider the spanning tree $T^1(\r'')\equiv T^2(\r)\cup e'$, where $e'$ is an edge of $\gamma$ whose endpoints are in distinct components of $T^2(\r)$. According to the minimum-weight criterion, we have
\begin{eqnarray}
w(T^2(\r))\leqslant w(T^1(\r'')) = w(T^2(\r)) -w(e')-r_{N+1},
\end{eqnarray}
where $r_{N+1}$ is the kinematic contribution we need to subtract from $T^2(\r)$ (see eq.~(\ref{eq:definition_spanning_2_tree_weight_rewritten})). The inequality above implies that $r_{N+1}\leqslant -w(e') <1$. Since $r_{N+1}\in \mathbb{N}$, we have $r_{N+1}=0$, and the term $\x^{\r}$ can only be in $\mathcal{F}^{(q^2,R)}(\x;\s)$.
\end{proof}

Lemma~\ref{lemma-heavy_subgraph_inside_light_graph} has a strong implication: for regions of massless graphs, the weight of each edge is either zero or negative. We formalize this statement into the following lemma.

\begin{lemma}[Absence of positive weights]
For any massless Feynman graph $G$ and region~$R$, all the first $N$ entries of the region vector $\v_R$ are either zero or negative.
\label{lemma-no_positive_weights}
\end{lemma}

\begin{proof}
Let us consider a particular subgraph $\Gamma\subseteq G$, which consists of edges with the maximum weight $w_\text{max}$ and the endpoints of these edges. Then we take $\gamma$ as any connected component of $\Gamma$. By definition, the inequality (\ref{lemma3_assumption}) holds since $w_\text{max}$ is the maximum weight. Then according to lemma~\ref{lemma-heavy_subgraph_inside_light_graph}, there is an $\mathcal{F}^{(R)}$ term $\x^{\r}$, such that in the corresponding spanning 2-tree $T^2(\r)$, the vertices of $\gamma$ lie in distinct components of $T^2(\r)$. Since $\gamma$ is connected, there must be an edge $e'\in \gamma\setminus T^2(\r)$, whose endpoints are in the two components of $T^2(\r)$ respectively.

If $G$ contains some positive-weight edges, then $w_\text{max}>0$. Let us consider the spanning tree $T^2(\r)\cup e'$, whose weight is
\begin{eqnarray}
w(T^2(\r))- w_\text{max} -\r'_{N+1} < w(T^2(\r)),
\end{eqnarray}
where we have used $r_{N+1}\in \mathbb{N}$. As each spanning tree of $G$ corresponds to a $\mathcal{U}$ term, the inequality above implies that there is another term in the Lee-Pomeransky polynomial whose weight is smaller than $w(T^2(\r))$, which violates the minimum-weight criterion since $\x^{\r}$ is a leading term. As a result, for any $e\in G$, $w(e)\leqslant 0$.
\end{proof}

\subsection{Weight-ordered partition and contracted graphs}
\label{section-weight_ordered_partition}

When exploring the properties of a weighted graph, a natural approach is to partition its vertices and edges into subgraphs based on their weights. To implement this, let us assume that the edges of a graph $\Gamma$ have weights $w_1, w_2, \ldots, w_M$, where $w_1 < w_2 < \ldots < w_M$. Then, we define the \emph{weight-ordered partition} of $\Gamma$ as follows (please note that the symbol $\sqcup$ below emphasizes that the graphs in the union are non-intersecting):
\begin{eqnarray}
\Gamma = \bigsqcup_{i=1}^M \Gamma_i,\qquad \Gamma_i= \bigsqcup_{j=1}^{m_i} \gamma_{i,j},
\label{eq:weight_hierarchical_partition_relation}
\end{eqnarray}
such that
\begin{enumerate}
    \item [(1)] an edge $e\in \Gamma_i$ if and only if $w(e)=w_i$;
    \item [(2)] a vertex $v\in \Gamma_i$ if and only if $v$ is incident with at least one edge of $\Gamma_i$ but no edges of $\Gamma_{i'}$ where $i'>i$;
    \item [(3)] for each given $i$, $\gamma_{i,1}, \gamma_{i,2},\dots, \gamma_{i,m_i}$ are the connected components of $\Gamma_i$.
\end{enumerate}
We refer to the subgraphs $\{\Gamma_i\}$ or $\{\gamma_{i,j}\}$ as the \emph{weight-ordered subgraphs} of $\Gamma$. From the construction above, $\gamma_{i_1,j_1} \cap \gamma_{i_2,j_2} =\varnothing$ unless $(i_1,j_1)= (i_2,j_2)$.

Given any weight-ordered partition of $\Gamma$ such that $\Gamma = \bigsqcup_{i=1}^M \bigsqcup_{j=1}^{m_i} \gamma_{i,j}$ as in eq.~(\ref{eq:weight_hierarchical_partition_relation}), we now define the \emph{contracted graphs} $\widetilde{\gamma}_{i,j}$ that corresponds to each weight-ordered subgraph $\gamma_{i,j}$: $\widetilde{\gamma}_{i,j}$ is obtained from $\gamma_{i,j}$ by introducing an \emph{auxiliary vertex} and identifying all the vertices of $\cup_{i'>i} \Gamma_{i'}$, which are adjacent to $\gamma_{i,j}$, with this auxiliary vertex.
\begin{figure}[t]
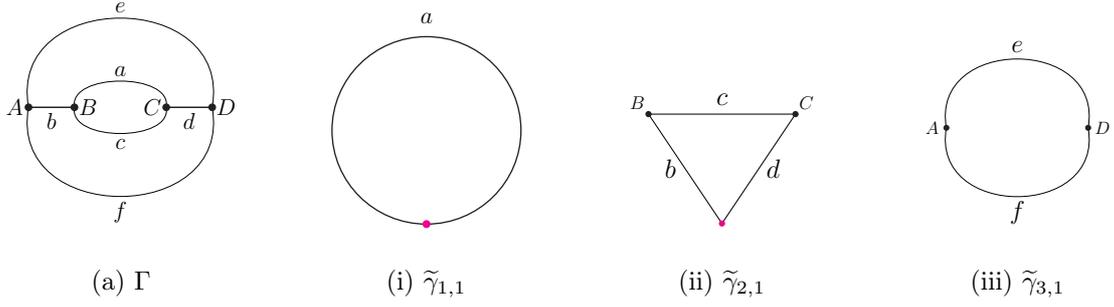

\centering
\begin{subfigure}[b]{0.22\textwidth}
\centering
\include{figs/weighted_hierarchical_example}
\vspace{-2em}
\caption{$\Gamma$}
\label{weighted_hierarchical_example}
\end{subfigure}
\hfill
\setcounter{subfigure}{0}
\renewcommand\thesubfigure{\roman{subfigure}}
\begin{subfigure}[b]{0.22\textwidth}
\centering
\include{figs/weighted_hierarchical_contracted_first}
\vspace{-2em}
\caption{$\widetilde{\gamma}_{1,1}$}
\label{weighted_hierarchical_contracted_first}
\end{subfigure}
\hfill
\begin{subfigure}[b]{0.22\textwidth}
\centering
\include{figs/weighted_hierarchical_contracted_second}
\vspace{-2em}
\caption{$\widetilde{\gamma}_{2,1}$}
\label{weighted_hierarchical_contracted_second}
\end{subfigure}
\hfill
\begin{subfigure}[b]{0.22\textwidth}
\centering
\include{figs/weighted_hierarchical_contracted_third}
\vspace{-2em}
\caption{$\widetilde{\gamma}_{3,1}$}
\label{weighted_hierarchical_contracted_third}
\end{subfigure}
\caption{A graph $\Gamma$ and its corresponding contracted graphs. (a): the graph $\Gamma$, where $w(a)<w(b)=w(c)=w(d)<w(e)=w(f)$. (i)-(iii): the contracted graphs $\widetilde{\gamma}_{1,1}$, $\widetilde{\gamma}_{2,1}$, and~$\widetilde{\gamma}_{3,1}$. Note that the auxiliary vertices of $\widetilde{\gamma}_{1,1}$ and $\widetilde{\gamma}_{2,1}$ are marked in pink.}
\label{figure-weighted_hierarchical_graph}
\end{figure}
An example of the weight-ordered partition and the contracted graphs of $\Gamma$ is shown  in figure~\ref{figure-weighted_hierarchical_graph}. The graph $\Gamma$, which can be seen as part of the entire Feynman graph $G$, has vertices $A$-$D$ and edges $a$-$f$ with weights $w(a)<w(b)=w(c)=w(d)<w(e)=w(f)$. By definition, the weight-ordered subgraphs are:
\begin{align}
\label{eq:weighted_hierarchical_subgraphs_example}
    \begin{split}
        &\Gamma_1 = \gamma_{1,1} = ( \varnothing, \{a\});\\
        &\Gamma_2 = \gamma_{2,1} = (\{B,C\}, \{b,c,d\});\\
        &\Gamma_3 = \gamma_{3,1} = (\{A,D\}, \{e,f\}).
    \end{split}
\end{align}
Note that each $\Gamma_i$ has precisely one connected component, so in this particular case, we have $\Gamma_i = \gamma_{i,1}$. The corresponding contracted graphs $\widetilde{\gamma}_{1,1}$, $\widetilde{\gamma}_{2,1}$ and $\widetilde{\gamma}_{3,1}$ are shown in figures~\ref{weighted_hierarchical_contracted_first}, \ref{weighted_hierarchical_contracted_second}, and \ref{weighted_hierarchical_contracted_third}, respectively. If there is an auxiliary vertex in a given contracted graph, it is marked in pink.


The weight-ordered partition of $\Gamma$ leads to the following crucial property: if the connectivity of $\Gamma_1,\dots,\Gamma_M$ satisfies certain conditions, any minimum spanning tree of $\Gamma$ would correspond to a minimum spanning tree of $\gamma_{i,j}$ for each $i,j$, and vice versa. Let us now formalize this property into the following lemma.
\begin{lemma}
\label{lemma-weight_hierarchical_partition_tree_structure}
For any weighted graph $\Gamma$ such that all the $M$ graphs
\begin{eqnarray}
\bigcup_{i=1}^M \Gamma_i\ ,\ \ \bigcup_{i=2}^M \Gamma_i\ ,\ \dots\ ,\ \ \Gamma_M
\label{eq:theorem_weight_hierarchical_tree_structure_connectivity}
\end{eqnarray}
are connected, then for any spanning tree of $\Gamma$, $T^1$,
\begin{eqnarray}
\widetilde{\gamma}_{i,j}\cap T^1\text{ is a spanning tree of }\widetilde{\gamma}_{i,j}\ (\forall i,j) \ \ \Leftrightarrow \ \ T^1\text{ is a minimum spanning tree of }\Gamma. \nonumber
\end{eqnarray}
\end{lemma}

First, we would like to mention that the proof of ($\Rightarrow$) is equivalent to Prim's algorithm of constructing minimum spanning trees~\cite{Jnk30,Prim57,Dkstr59}, a well-understood method, as we will explain in appendix~\ref{appendix-Prim_algorithm}. Below we focus on proving ($\Leftarrow$), i.e., $\widetilde{\gamma}_{i,j}\cap T^1$ is a spanning tree of $\widetilde{\gamma}_{i,j}$ for each $i,j$ provided that $T^1$ is a minimum spanning tree of $\Gamma$. To establish this, we need to demonstrate two essential properties of $\widetilde{\gamma}_{i,j}\cap T^1$: (1) it is a connected graph; (2) it contains no loops. We will prove both properties by contradiction.

\begin{proof}[Proof of the connectedness]
We first show the connectedness of $\widetilde{\gamma}_{i,j}\cap T^1$ by contradiction. Suppose that there exists a pair $(i_0,j_0)$ such that $\widetilde{\gamma}_{i_0,j_0}^{} \cap T^1$ has two or more connected components. Let us then denote any one component \emph{not} including the auxiliary vertex of $\widetilde{\gamma}_{i_0,j_0}^{}$ as $\gamma'_{i_0,j_0}$. The following two statements ensue. Firstly, $\gamma_{i_0,j_0}\cap T^1$ is also disconnected, one of whose components is $\gamma'_{i_0,j_0}$. Secondly, all the edges adjacent to $\gamma'_{i_0,j_0}$ must belong to $\cup_{i'\leqslant i_0} \Gamma_{i'}$, otherwise the auxiliary vertex would be in $\gamma'_{i_0,j_0}$. These two statements further imply that all the edges of $T^1$, which are adjacent to $\gamma'_{i_0,j_0}$, belong to $\cup_{i'< i_0} \Gamma_{i'}$.

Since $\gamma_{i_0,j_0}^{}$ is connected by definition, there exists an edge $e_0\in \gamma_{i_0,j_0}^{}$ such that the endpoints of $e_0$, denoted as $v_1$ and $v_2$, are in $\gamma'_{i_0,j_0}$ and $\gamma''_{i_0,j_0}$ respectively, where $\gamma''_{i_0,j_0}$ is any other component of $\gamma_{i_0,j_0}^{} \cap T^1$. Meanwhile, since $T^1$ is connected, there is a path $P\subset T^1$ joining $v_1$ and $v_2$, which contains some edge $e'\in \cup_{i'<i_0} \Gamma_{i'}$ from our reasoning above. It then follows that $T^1\cup e_0$ contains a loop, as there are two paths joining $v_1$ and $v_2$ in $T^1\cup e_0$: one is $P$ and the other is $e_0$. The graph ${T'}^1\equiv T^1\cup e_0\setminus e'$ is then another spanning tree of~$\Gamma$, whose weight is
\begin{eqnarray}
w({T'}^1)= w(T^1)- w(e_0) +w(e') <w(T^1).
\label{eq:theorem_weight_hierarchical_tree_structure_proof1}
\end{eqnarray}
In deriving the inequality above we have used $w(e_0)>w(e')$ because $e_0\in \gamma_{i_0,j_0}^{}$ and $e'\in \cup_{i'<i_0} \gamma_{i'}$. However, this inequality violates the minimum-weight criterion, implying that $\widetilde{\gamma}_{i_0,j_0}\cap T^1$ can only have one connected component. Therefore, we have proved that $\widetilde{\gamma}_{i,j}\cap T^1$ is connected for every $i,j$.
\end{proof}

\begin{figure}[t]
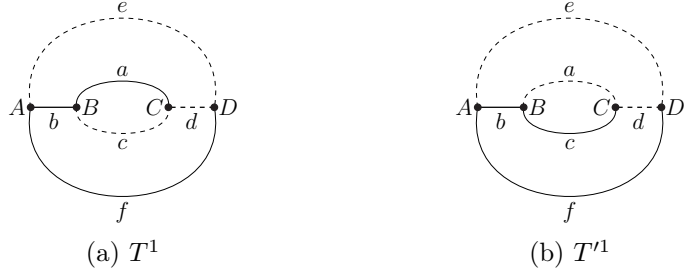

\centering
\hspace{-1em}
\begin{subfigure}[b]{0.22\textwidth}
\centering
\include{figs/weighted_hierarchical_tree1}
\vspace{-3em}
\caption{$T^1$}
\label{weighted_hierarchical_tree1}
\end{subfigure}
\hspace{6em}
\begin{subfigure}[b]{0.22\textwidth}
\centering
\include{figs/weighted_hierarchical_tree2}
\vspace{-3em}
\caption{$T'^1$}
\label{weighted_hierarchical_tree2}
\end{subfigure}
\caption{Illustration of the analysis in proving the connectedness of $\widetilde{\gamma}_{i,j}\cap T^1$, where we choose $w(a)<w(b)=w(c)=w(d)<w(e)=w(f)$ as above. (a) A spanning tree $T^1\subset \Gamma$ such that $\widetilde{\gamma}_{2,1}\cap T^1$ is disconnected: it consists of the edge $b$, the vertices $B,C$, and the auxiliary vertex. (b) Another spanning tree $T'^1\equiv T^1\cup e_0\setminus e'$ where $e_0=c$ and $e'=a$, satisfying $w(T'^1)<w(T^1)$.}
\label{figure-weighted_hierarchical_tree_theorem}
\end{figure}

An example of the analysis above is illustrated by figure~\ref{figure-weighted_hierarchical_tree_theorem}, where $T^1$ (figure~\ref{weighted_hierarchical_tree1}) consists of all the vertices and the solid edges, with the weight structure the same as that shown in figure~\ref{figure-weighted_hierarchical_graph}, i.e., $w(a)<w(b)=w(c)=w(d)<w(e)=w(f)$. Let us now focus on the graph $\gamma_{2,1}\cap T^1$, which consists of the edge $b$ and the vertices $B$ and $C$ (according to eq.~(\ref{eq:weighted_hierarchical_subgraphs_example})). By identifying $A$ (the only vertex of $\Gamma_3$ adjacent to $\gamma_{2,1}\cap T^1$) with the auxiliary vertex, $\gamma_{2,1}$ becomes $\widetilde{\gamma}_{2,1}$. The graph $\widetilde{\gamma}_{2,1}\cap T^1$ is then disconnected: one of its components consists of the edge $b$, the vertex $B$ and the auxiliary vertex, while the other consists solely of the vertex $C$. Using the analysis in the proof above, we have $v_1=C$, $v_2=B$, $e_0=c$, and $e'=a$. The graph $T'^1\equiv T^1\cup e_0\setminus e'$ (figure~\ref{weighted_hierarchical_tree2}) is then another spanning tree of $\Gamma$ whose weight is smaller than $w(T^1)$.

We emphasize that, from now on, in a figure illustrating a spanning (2-)tree $T$, a solid edge $e$ indicates $e \in T$, while a dashed edge $e$ indicates $e \notin T$. Figure~\ref{figure-weighted_hierarchical_tree_theorem} is an example.

Below we show that $\widetilde{\gamma}_{i,j}\cap T^1$ contains no loops for any $i,j$.
\begin{proof}[Proof of the absence of loops]
Suppose there is some $(i_0,j_0)$ such that there is a loop in $\widetilde{\gamma}_{i_0,j_0}\cap T^1$. In this case, either of the two possibilities occurs: (1) there is a loop in $\gamma_{i_0,j_0}\cap T^1$; (2) there is a path in $P\subset \gamma_{i_0,j_0}\cap T^1$ that connects two vertices $v_1,v_2\in \cup_{i'>i_0}\gamma_{i'}$. The first possibility contradicts the definition of a spanning tree and is thus ruled out.

For the second possibility, the tree structure of $T^1$ dictates that $\cup_{i'>i_0}\gamma_{i'} \cap T^1$ must be disconnected. Furthermore, $v_1$ and $v_2$ are in its distinct components; otherwise, two paths in $T^1$ would connect $v_1$ and $v_2$, forming a loop in $T^1$. For simplicity, here we assume that there are exactly two connected components of $\cup_{i'>i_0}\gamma_{i'} \cap T^1$. Since $\cup_{i'>i_0}\gamma_{i'}$ is connected, there exists an edge $e_*\in \cup_{i'>i_0}\gamma_{i'}$ such that its endpoints are in the two components of $\cup_{i'>i_0}\gamma_{i'} \cap T^1$ respectively. Similar to our analysis in eq.~(\ref{eq:theorem_weight_hierarchical_tree_structure_proof1}), we select any edge $e_0\in P$ and consider another spanning tree ${T''}^1\equiv T^1\cup e_*\setminus e_0$, with a weight given by
\begin{eqnarray}
w({T''}^1)= w(T^1)+ w(e_0) -w(e_*) <w(T^1),
\end{eqnarray}
because $e_0\in\gamma_{i_0,j_0}$ and $e_*\in \cup_{i'>i_0}\gamma_{i'}$, implying $w(e_0)<w(e_*)$ by definition. Again, the minimum-weight criterion is violated, thus $\widetilde{\gamma}_{i,j}\cap T^1$ contains no loops for any $i,j$.
\end{proof}

We conclude from above that $\widetilde{\gamma}_{i,j}\cap T^1$ is a spanning tree of $\widetilde{\gamma}_{i,j}$ for each $i,j$. It is crucial to underscore the necessity of the condition that all the graphs in (\ref{eq:theorem_weight_hierarchical_tree_structure_connectivity}) are connected. In our example depicted in figure~\ref{figure-weighted_hierarchical_graph}, it is evident from eq.~(\ref{eq:weighted_hierarchical_subgraphs_example}) that $\Gamma_3$, $\Gamma_2\cup \Gamma_3$ and  $\Gamma_1\cup \Gamma_2\cup \Gamma_3$ are all connected, and this condition is automatically satisfied. In contrast, if we change the weight structure of $\Gamma$ into $w(e)=w(f)<w(a)=w(c)< w(b)=w(d)$ instead, then
\begin{align}
\label{eq:weighted_hierarchical_subgraphs_another_example}
    \begin{split}
        &\gamma_{1,1} = ( \varnothing, \{e\}),\qquad \gamma_{1,2} = ( \varnothing, \{f\});\\
        &\gamma_{2,1} = ( \varnothing, \{a\}),\qquad \gamma_{2,2} = ( \varnothing, \{c\});\\
        &\gamma_{3,1} = (\{A,B\}, \{b\}),\qquad \gamma_{3,2} = (\{C,D\}, \{d\}).
    \end{split}
\end{align}
The graph $\Gamma_3 = (\{A,B,C,D\}, \{b,d\})$, which is obviously disconnected. One minimum spanning tree $T^1$ is shown below:
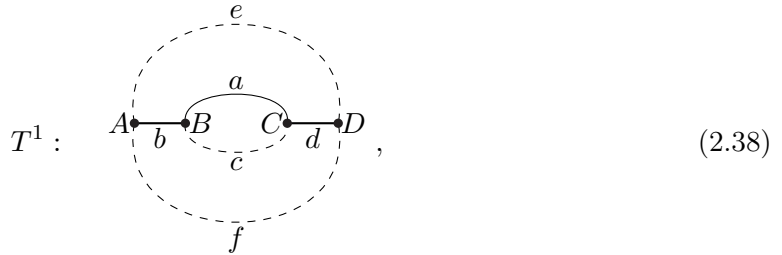
\begin{equation}
T^1:\quad
\begin{tikzpicture}[baseline=18ex, scale=0.67]
\path (3,5) edge [dashed, bend left=100, looseness=1.7] (7,5) {};
\path (3,5) edge [dashed, bend right=100, looseness=1.7] (7,5) {};
\draw (3,5) edge [thick] (4,5) node [] {};
\draw (7,5) edge [thick] (6,5) node [] {};
\path (4,5) edge [bend left=100] (6,5) {};
\path (4,5) edge [dashed, bend right=100] (6,5) {};
\node (a) at (5,5.8) {$a$};
\node (b) at (3.5,4.7) {$b$};
\node (c) at (5,4.2) {$c$};
\node (d) at (6.5,4.7) {$d$};
\node (e) at (5,7.2) {$e$};
\node (f) at (5,2.7) {$f$};
\node (A) at (2.7,5) {$A$};
\node (B) at (4.3,5) {$B$};
\node (C) at (5.7,5) {$C$};
\node (D) at (7.3,5) {$D$};
\draw[fill,thick,color=Black] (3,5) circle (2pt);
\draw[fill,thick,color=Black] (4,5) circle (2pt);
\draw[fill,thick,color=Black] (6,5) circle (2pt);
\draw[fill,thick,color=Black] (7,5) circle (2pt);
\end{tikzpicture},
\label{weighted_hierarchical_subgraphs_another_example_ST}
\end{equation}
however, here $\widetilde{\gamma}_{2,1}\cap T^1$, which is obtained from $\gamma_{2,1}$ by identifying $B$ and $C$ with the auxiliary vertex, would contain a loop, thus it is not a tree graph.

We will later employ lemma~\ref{lemma-weight_hierarchical_partition_tree_structure} to investigate the soft substructure of a region in the on-shell expansion, a key step toward proving the on-shell-expansion region theorem.

\subsection{Notations}
\label{section-notations}

In order to facilitate a comprehensive understanding of the subsequent sections, this subsection provides a concise summary of some essential concepts and notations introduced in the previous subsections. These fundamental elements will serve as building blocks for the analysis in the following sections. The summarized concepts can be found in table~\ref{table:symbols_notations} for quick reference and clarity.
\begin{table}[t]
\begin{center}
\begin{tabular}{ |c|c||c|c| } 
\hline
\multirow{2}{*}{Symbol} & \multirow{2}{*}{Description} & \multirow{2}{*}{Symbol} & \multirow{2}{*}{Description} \\
 & & & \\
\hline
$v$ & vertex & $R$ & region \\
\hline
$e$ & edge & $\x^{\r}$ & term of $\mathcal{P}(\x;\s)$ \\
\hline
$\mathcal{V}$ & set of vertices & $\mathcal{P}^{(R)}(\x;\s)$ & the leading LP polynomial \\
\hline
$\mathcal{E}$ & set of edges & $\mathcal{U}^{(R)}(\x)$ & the leading $\mathcal{U}$ polynomial \\
\hline
$G$ & Feynman graph & $\mathcal{F}^{(R)}(\x;\s)$ & the leading $\mathcal{F}$ polynomial \\
\hline
$\gamma$ & subgraph of $G$ & \multirow{2}{*}{$T^n(\r)$}  & the spanning $n$-tree corres- \\
\cline{1-2}
$V(\gamma)$ & number of vertices of $\gamma$ & & ponding to $\x^{\r}$ ($n=1,2$) \\
\hline
$N(\gamma)$ & number of edges of $\gamma$ & $w(e)$  & weight of the edge $e$ \\
\hline
$L(\gamma)$ & number of loops of $\gamma$ & $w(T^n(\r))$  & weight of $T^n(\r)$ ($n=1,2$) \\
\hline
$T^1$ & spanning tree & $P$ & path \\
\hline
$T^2$ & spanning 2-tree & $H$ & the hard subgraph of $G$ \\
\hline
$t$ & subtree & $J_i$ & the $i$th jet of $G$\\
\hline
$G(t)$ & graph generated by $t$ & $S$ & the soft subgraph of $G$ \\
\hline
\end{tabular}
\end{center}
\vspace{-1em}\caption{Some key concepts and their notations used in the later sections.}
\label{table:symbols_notations}
\end{table}

Please note we will use $t$ to denote a ``subtree'' and $G(t)$ to denote a ``graph generated by $t$'', and use $H$, $J_i$, and $S$ to denote the hard, $i$th jet, and soft subgraphs of $G$. These concepts will be introduced later in section~\ref{section-generic_form_region_rigorous_proof}.

\bigbreak
Let us end section~\ref{section-general_setup} by emphasizing that lemmas~\ref{lemma-existence_certain_terms_P}-\ref{lemma-no_positive_weights} are applicable only in the context of massless Feynman graphs, while lemma~\ref{lemma-weight_hierarchical_partition_tree_structure} also applies to graphs with massive propagators. In the upcoming text, these lemmas will be employed in the analysis of the on-shell expansion (in section~\ref{section-regions_onshell}) and soft expansion (in section~\ref{section-regions_soft_expansion}) of generic wide-angle scattering, where only massless Feynman graphs are involved. For the mass expansion of the heavy-to-light decay (in section~\ref{section-regions_mass_expansion}), we will only propose a formalism for deriving the regions rather than presenting a rigorous proof. One would need to extend lemmas~\ref{lemma-existence_certain_terms_P}-\ref{lemma-no_positive_weights} to massive cases if a rigorous analysis is required.

\section{Regions in wide-angle scattering: the on-shell expansion}
\label{section-regions_onshell}

This section centers on the on-shell expansion of wide-angle scattering, providing an all-order analysis of the generic form of the region vectors based on the graph-theoretical properties of the $\mathcal{P}^{(R)}$ terms. Specifically, we demonstrate that each region can be characterized by a pinch surface shown in figure~\ref{figure-onshell_region_H_J_S_precise}. This conclusion, in combination with the constraints on the subgraphs from ref.~\cite{GrdHzgJnsMaSchlk22}, leads to the ``on-shell-expansion region theorem'' as summarized at the end of this section. This theorem indicates how the exact list of region vectors can be read off from the graph~$G$.

This section is structured as follows. We start in section~\ref{section-motivations_onshell} by exploring the on-shell expansion of a $2\times 2$ fishnet example. Throughout this exploration, we make observations that motivate an all-order study. We next investigate the general properties of the regions appearing in the on-shell expansion of wide-angle scattering in section~\ref{section-generic_form_region_rigorous_proof}. The analysis is valid to all orders and is based on the graph-theoretical properties of the leading Lee-Pomeransky polynomials. To visualize how these theories work, we also provide some three-loop examples in section~\ref{section-examples_from_3loop_graph}. We then combine the results with those of ref.~\cite{GrdHzgJnsMaSchlk22} in section~\ref{section-further_restrictions_region_vectors}, which describes additional requirements for each region, completing the proof of the on-shell-expansion region theorem. Finally, the results are summarized in section~\ref{section-summary_onshell_regions} for clarity.

\subsection{Motivations from a fishnet graph example}
\label{section-motivations_onshell}

\begin{figure}[t]
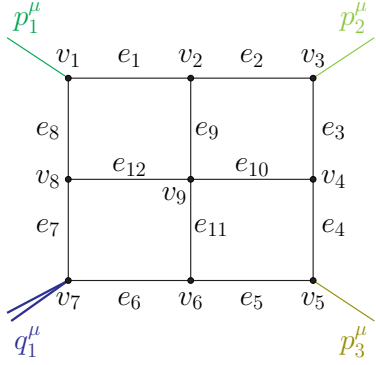

\centering
\include{figs/fishnet_2times2_onshell_3onshell_1offshell}
\vspace{-3em}\caption{The $2\times 2$ fishnet graph $G_{2\times 2}$ in the on-shell expansion, with its four external momenta $p_1^\mu,p_2^\mu,p_3^\mu$, and $q_1^\mu$ satisfying $p_1^2\sim p_2^2\sim p_3^2\sim \lambda Q^2$ and $q_1^2\sim Q^2$. The edges are labelled by $e_1,\dots,e_{12}$ and the vertices are labelled by $v_1,\dots,v_9$.}
\label{figure-fishnet_2times2_onshell_3onshell_1offshell}
\end{figure}

Let us start from a $2\times 2$ fishnet graph with three on-shell external momenta $p_1^\mu,p_2^\mu,p_3^\mu$ and an off-shell external momentum $q_1^\mu$, as shown in figure~\ref{figure-fishnet_2times2_onshell_3onshell_1offshell}. As one can check by computer codes, there are 120 regions in total for this figure. In order to investigate some generic properties of the leading polynomial $\mathcal{P}^{(R)}(\x;\s)$, let us focus on one specific region $R$, as described below.
\begin{equation}
R:\quad
,
\label{fishnet_2times2_onshell_region_example}
\end{equation}
where the red, green and blue colors indicate that the weights of the corresponding edges are $0$, $-1$ and $-2$, respectively.

The corresponding leading polynomials $\mathcal{U}^{(R)}$, $\mathcal{F}^{(p^2,R)}$ and $\mathcal{F}^{(q^2,R)}$ are then
\begin{equation}
\begin{aligned}
&\mathcal{U}^{(R)}\text{ terms:}
\\
&
%
.
\end{aligned}
\label{eq:onshell_region_good_example_Fq2term}
\end{equation}
The following properties regarding these leading terms can be verified directly by checking the graphs in (\ref{eq:onshell_region_good_example_Uterm})-(\ref{eq:onshell_region_good_example_Fq2term}):
\begin{itemize}
    \item \emph{For each $\mathcal{U}^{(R)}$ term, the corresponding spanning tree $T^1$ has a ``heavy trunk''.} Namely, for the path $P\subset T^1$ connecting any two external momenta of $G_{2\times 2}$, the weight of each edge $e\in P$ is either $0$ or $-1$. More precisely, if we order the edges of $P$ according to the distance from one external momentum, then their corresponding weights must be of the form
    \begin{eqnarray}
    \{ \underset{N_1}{\underbrace{-1,\dots,-1}}\ ,\ \underset{N_2}{\underbrace{0,\dots,0}}\ ,\ \underset{N_3}{\underbrace{-1\dots,-1}} \},
    \end{eqnarray}
    where some of the $N_i$ can be zero.
    
    For example, let us consider the path $P\in T^1(\r)$ where $\x^{\r}$ is the first term of eq.~(\ref{eq:onshell_region_good_example_Uterm}), such that $P$ joins $p_1^\mu$ and $p_2^\mu$. Then $P$ consists of the edges $\{ e_8, e_{12}, e_9, e_2 \}$, which are ordered according to their distance from $p_1^\mu$. The corresponding weights can be directly read: $\{ -1, 0, -1, -1\}$.
    
    \item \emph{The $\mathcal{U}^{(R)}$ and $\mathcal{F}^{(p^2,R)}$ terms can be related by adding/deleting one edge with $w=-1$.} In detail, given any $T^1$ corresponding to a $\mathcal{U}^{(R)}$ term, there is an edge $e\in T^1$ such that the spanning 2-tree $T^1\setminus e$ corresponds to a $\mathcal{F}^{(p^2,R)}$ term. Similarly, given any spanning 2-tree $T^2$ corresponding to an $\mathcal{F}^{(p^2,R)}$ term, there is an edge $e\in G_{2\times 2}\setminus T^2$ such that the spanning tree $T^2\cup e$ corresponds to a $\mathcal{U}^{(R)}$ term.
    
    For example, the first $\mathcal{U}^{(R)}$ term in eq.~(\ref{eq:onshell_region_good_example_Uterm}) and the first $\mathcal{F}^{(p^2,R)}$ term in eq.~(\ref{eq:onshell_region_good_example_Fp2term}) can be related through
    
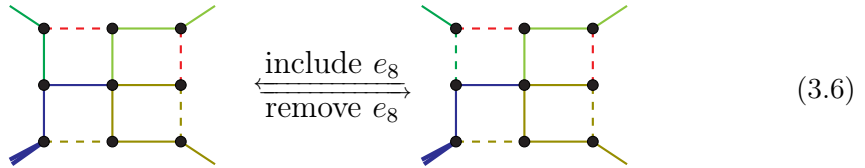
\begin{equation}
        \begin{tikzpicture}[baseline=8ex, scale=0.3]
        \draw [dashed, thick,color=Red] (2,7.5) -- (5,7.5);
        \draw [thick,color=LimeGreen] (8,7.5) -- (5,7.5);
        \draw [dashed, thick,color=Red] (8,5) -- (8,7.5);
        \draw [dashed, thick,color=olive] (8,2.5) -- (8,5);
        \draw [thick,color=olive] (5,2.5) -- (8,2.5);
        \draw [dashed, thick,color=olive] (2,2.5) -- (5,2.5);
        \draw [thick,color=Blue] (2,5) -- (2,2.5);
        \draw [thick,color=Green] (2,5) -- (2,7.5);
        \draw [thick,color=LimeGreen] (5,7.5) -- (5,5);
        \draw [thick,color=olive] (5,5) -- (8,5);
        \draw [thick,color=olive] (5,2.5) -- (5,5);
        \draw [thick,color=Blue] (5,5) -- (2,5);
        \draw [thick,color=Green] (0.5,8.5) -- (2,7.5);
        \draw [thick,color=LimeGreen] (9.5,8.5) -- (8,7.5);
        \draw [thick,color=olive] (9.5,1.5) -- (8,2.5);
        \draw [ultra thick, color=Blue] (0.5,1.7) -- (2,2.5);
        \draw [ultra thick, color=Blue] (0.6,1.5) -- (2,2.5);
        
        \node [draw,circle,minimum size=4pt,fill=Black,inner sep=0pt,outer sep=0pt] () at (2,7.5) {};
        \node [draw,circle,minimum size=4pt,fill=Black,inner sep=0pt,outer sep=0pt] () at (2,5) {};
        \node [draw,circle,minimum size=4pt,fill=Black,inner sep=0pt,outer sep=0pt] () at (2,2.5) {};
        \node [draw,circle,minimum size=4pt,fill=Black,inner sep=0pt,outer sep=0pt] () at (5,7.5) {};
        \node [draw,circle,minimum size=4pt,fill=Black,inner sep=0pt,outer sep=0pt] () at (5,5) {};
        \node [draw,circle,minimum size=4pt,fill=Black,inner sep=0pt,outer sep=0pt] () at (5,2.5) {};
        \node [draw,circle,minimum size=4pt,fill=Black,inner sep=0pt,outer sep=0pt] () at (8,7.5) {};
        \node [draw,circle,minimum size=4pt,fill=Black,inner sep=0pt,outer sep=0pt] () at (8,5) {};
        \node [draw,circle,minimum size=4pt,fill=Black,inner sep=0pt,outer sep=0pt] () at (8,2.5) {};
        \end{tikzpicture}
        \quad \xleftrightarrows[\text{\large remove $e_8$}]{\text{\large include $e_8$}} 
        \begin{tikzpicture}[baseline=8ex, scale=0.3]
        \draw [dashed, thick,color=Red] (2,7.5) -- (5,7.5);
        \draw [thick,color=LimeGreen] (8,7.5) -- (5,7.5);
        \draw [dashed, thick,color=Red] (8,5) -- (8,7.5);
        \draw [dashed, thick,color=olive] (8,2.5) -- (8,5);
        \draw [thick,color=olive] (5,2.5) -- (8,2.5);
        \draw [dashed, thick,color=olive] (2,2.5) -- (5,2.5);
        \draw [thick,color=Blue] (2,5) -- (2,2.5);
        \draw [dashed, thick,color=Green] (2,5) -- (2,7.5);
        \draw [thick,color=LimeGreen] (5,7.5) -- (5,5);
        \draw [thick,color=olive] (5,5) -- (8,5);
        \draw [thick,color=olive] (5,2.5) -- (5,5);
        \draw [thick,color=Blue] (5,5) -- (2,5);
        \draw [thick,color=Green] (0.5,8.5) -- (2,7.5);
        \draw [thick,color=LimeGreen] (9.5,8.5) -- (8,7.5);
        \draw [thick,color=olive] (9.5,1.5) -- (8,2.5);
        \draw [ultra thick, color=Blue] (0.5,1.7) -- (2,2.5);
        \draw [ultra thick, color=Blue] (0.6,1.5) -- (2,2.5);
        
        \node [draw,circle,minimum size=4pt,fill=Black,inner sep=0pt,outer sep=0pt] () at (2,7.5) {};
        \node [draw,circle,minimum size=4pt,fill=Black,inner sep=0pt,outer sep=0pt] () at (2,5) {};
        \node [draw,circle,minimum size=4pt,fill=Black,inner sep=0pt,outer sep=0pt] () at (2,2.5) {};
        \node [draw,circle,minimum size=4pt,fill=Black,inner sep=0pt,outer sep=0pt] () at (5,7.5) {};
        \node [draw,circle,minimum size=4pt,fill=Black,inner sep=0pt,outer sep=0pt] () at (5,5) {};
        \node [draw,circle,minimum size=4pt,fill=Black,inner sep=0pt,outer sep=0pt] () at (5,2.5) {};
        \node [draw,circle,minimum size=4pt,fill=Black,inner sep=0pt,outer sep=0pt] () at (8,7.5) {};
        \node [draw,circle,minimum size=4pt,fill=Black,inner sep=0pt,outer sep=0pt] () at (8,5) {};
        \node [draw,circle,minimum size=4pt,fill=Black,inner sep=0pt,outer sep=0pt] () at (8,2.5) {};
        \end{tikzpicture}
    \label{eq:UandFp2_terms_duality_example}
    \end{equation}
    
    \item \emph{The $\mathcal{U}^{(R)}$ and $\mathcal{F}^{(p^2,R)}$ terms are characterized by the numbers of the removed hard, jet and soft edges.} For the fishnet example above, each term in eq.~(\ref{eq:onshell_region_good_example_Uterm}) has two soft edges and two jet edges removed, and each term in eq.~(\ref{eq:onshell_region_good_example_Fp2term}) has two soft edges and two jet edges removed.
    
    \emph{In comparison, there can be up to two types of $\mathcal{F}^{(q^2,R)}$ terms.} For example, in each of the first 11 terms in eq.~(\ref{eq:onshell_region_good_example_Fq2term}), one hard edge, two jet edges and two soft edges are removed from $G_{2\times 2}$, while in the remaining 13 terms, four jet edges and one soft edge are removed.
\end{itemize}

These observations motivate us to establish general properties of $R$, which will be discussed later in section~\ref{section-generic_form_region_rigorous_proof}. Next, we focus on two configurations that do not appear as regions of the on-shell expansion. In other words, the parameter-space points corresponding to the terms of the leading polynomials do not suffice to span a lower facet of $\Delta(\mathcal{P})$. We explain this in detail below.

\begin{figure}[t]
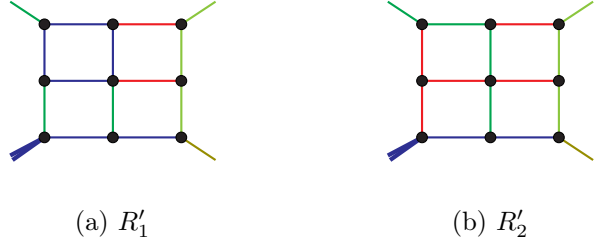

\centering
\hspace{0em}
\begin{subfigure}[b]{0.32\textwidth}
\centering
\include{figs/onshell_region_bad_example1}
\caption{$R'_1$}
\label{figure-onshell_region_bad_example1}
\end{subfigure}
\begin{subfigure}[b]{0.32\textwidth}
\centering
\hspace{10em}
\include{figs/onshell_region_bad_example2}
\caption{$R'_2$}
\label{figure-onshell_region_bad_example2}
\end{subfigure}
\caption{Two non-region examples of $G_{2\times 2}$ (figure~\ref{figure-fishnet_2times2_onshell_3onshell_1offshell}). In the first configuration~$R'_1$, there is a hard loop inside a jet (marked by green lines). In the configuration~$R'_2$, one connected components of the soft subgraph, which consists of $e_7,e_8,e_{12}$ and the vertex simultaneously incident with these edges, is attached to exactly one jet.}
\label{figure-onshell_region_bad_examples}
\end{figure}

In the first example (figure~\ref{figure-onshell_region_bad_example1}), there is a hard loop inside a jet subgraph. We denote this configuration as $R'_1$, and the leading Symanzik polynomials $\mathcal{U}^{(R'_1)}$ and $\mathcal{F}^{(R'_1)}$ are
\begin{align}
    \begin{split}
        \mathcal{U}^{(R'_1)}&= x_2 x_{10} (x_7+x_{11}) (x_1+x_8+x_9+x_{12});\\
        \mathcal{F}^{(R'_1)}&= (-p_1^2) x_2 x_7 x_{10} x_{11} (x_1+x_8+x_9+x_{12})\\
        &+ (-p_2^2) x_2 x_{10} (x_3+x_4) (x_7+x_{11}) (x_1+x_8+x_9+x_{12})\\
        &+ (-q_1^2) x_2 x_6 x_7 x_{10} (x_1+x_8+x_9+x_{12})\\
        &+ \left( -(p_2+p_3)^2 \right) x_2 x_{10} (x_1x_5 + x_5x_{11} + x_6x_{11}) (x_1+x_8+x_9+x_{12})\\
        &+ \left( -(p_1+p_2)^2 \right) \left[ (x_3x_{10} + x_4x_{10} + x_2x_3) x_7 x_{11} (x_1+x_8+x_9+x_{12}) \right].
    \end{split}
    \label{eq:figure-onshell_region_bad_example1}
\end{align}
As clear from above, the leading polynomial $\mathcal{P}^{(R'_1)}\equiv \mathcal{U}^{(R'_1)} +\mathcal{F}^{(R'_1)}$ is a homogeneous function of the parameters $x_1,x_8,x_9,x_{12}$. In other words, on defining $f_{R'_1}$ as the lower face spanned by the $\mathcal{P}^{(R'_1)}$ terms, we find least two vectors that are normal to $f_{R'_1}$:
\begin{align}
\begin{split}
    &\v_{R'_1}=(0,-2,-1,-1,0,0,-1,0,0,-2,-1,0;1)\\
    &\v'_{R'_1}=(1,0,0,0,0,0,0,1,1,0,0,1;0).
\end{split}
\end{align}
The first vector $\v_{R'_1}$ describes the defining property of $\mathcal{P}^{(R'_1)}$, i.e. all the terms of $\mathcal{P}^{(R'_1)}$ corresponds to the minimum value of $\boldsymbol{r} \cdot \v_{R'}$. The second vector $\v'_{R'_1}$ is due to the homogeneity of $\mathcal{P}^{(R'_1)}$, which is from the violation of the facet criterion (see eq.~(\ref{eq:facet_criterion_violation_additional_vector})). The existence of these two distinct vectors simultaneously implies that $\text{dim} (f_{R'_1}) \leqslant N-1$, thus $f_{R'_1}$ cannot be an $N$-dimensional facet of $\Delta(\mathcal{P})$.

In the second example (figure~\ref{figure-onshell_region_bad_example2}), one connected component of the soft subgraph (the edges $e_7,e_8,e_{12}$) is attached to only one of the jets. We denote this configuration as $R'_2$, and the leading Symanzik polynomials $\mathcal{U}^{(R'_2)}$ and $\mathcal{F}^{(R'_2)}$ are
\begin{align}
    \begin{split}
        \mathcal{U}^{(R'_2)}&= x_2 x_{10} (x_7x_8 + x_7x_{12} + x_8x_{12});\\
        \mathcal{F}^{(R'_2)}&= (-p_1^2) x_2 x_{10} (x_1 + x_9 + x_{11}) (x_7x_8 + x_7x_{12} + x_8x_{12})\\
        &+ (-p_2^2) x_2 x_{10} (x_3+x_4) (x_7x_8 + x_7x_{12} + x_8x_{12})\\
        &+ (-q_1^2) x_2 x_6 x_{10} (x_7x_8 + x_7x_{12} + x_8x_{12})\\
        &+ \left( -(p_2+p_3)^2 \right) x_2 x_5 x_{10} (x_7x_8 + x_7x_{12} + x_8x_{12})\\
        &+ \left( -(p_1+p_2)^2 \right) \left[ x_2 x_4 x_{11} (x_7x_8 + x_7x_{12} + x_8x_{12}) \right.\\
        &\qquad \qquad \qquad \quad \left. +x_{10} (x_3 + x_4) (x_9 + x_{11}) (x_7x_8 + x_7x_{12} + x_8x_{12}) \right].
    \end{split}
    \label{eq:figure-onshell_region_bad_example2}
\end{align}
Similar as the previous case, $\mathcal{P}^{(R'_2)}$ is a homogeneous function of $x_7,x_8,x_{12}$. Then at least two vectors are normal to the lower face $f_{R'_2}$:
\begin{align}
\begin{split}
    &\v_{R'_2}=(-1,-2,-1,-1,0,0,-2,-2,-1,-2,-1,-2;1)\\
    &\v'_{R'_2}=(0,0,0,0,0,0,1,1,0,0,0,1;0).
\end{split}
\end{align}
Again, the two vectors above are respectively due to the defining property and the homogeneity of $\mathcal{P}^{(R'_2)}(\x;\s)$. The lower face $f_{R'_2}$ thus cannot be a lower facet either.

\bigbreak
Let us briefly summarize the observations in this subsection. We started from a specific region $R$ of a $2\times 2$ fishnet graph $G_{2\times 2}$ in eq.~(\ref{fishnet_2times2_onshell_region_example}), and pointed out three properties of the $\mathcal{U}^{(R)}$, $\mathcal{F}^{(p^2,R)}$ and $\mathcal{F}^{(q^2,R)}$ terms. We then focused on two other configurations of $G_{2\times 2}$, namely, $R'_1$ and $R'_2$, and explained the reason why they are not the regions in the on-shell expansion. These analyses motivate us to obtain similar conclusions for a generic wide-angle scattering graph.

\subsection{The generic form of a region vector: a rigorous proof}
\label{section-generic_form_region_rigorous_proof}

Inspired by the examples above, we now aim to provide an all-order proof that each region in the on-shell expansion can be identified with a solution of the Landau equations. The proof is based on the graph-theoretical properties of the leading Symanzik polynomials and, specifically the facet criterion and the minimum-weight criterion introduced in section~\ref{section-two_fundamental_criteria}.

The entire analysis is long and technical, which we organize as follows for clarity. We start in section~\ref{section-some_properties_leading_terms} by investigating some properties of the $\mathcal{U}^{(R)}$ and $\mathcal{F}^{(p^2,R)}$ terms (lemmas~\ref{lemma-onshell_basic_weight_structure_Uterm}-\ref{lemma-onshell_heavy_in_light_constraint} and their corollaries). Next, in section~\ref{section-canonical_leading_terms} we define the canonical $\mathcal{U}^{(R)}$ and $\mathcal{F}^{(p^2,R)}$ terms, whose properties (lemma~\ref{lemma-onshell_Fp2_vertices_universal_property} and its corollaries) allow us to define the subgraphs corresponding to any given region. Then, in section~\ref{section-subgraphs_associated_each_region}, we provide the definition and some basic properties of the hard ($H$), jet ($J$) and soft ($S$) subgraphs (lemmas~\ref{lemma-onshell_hard_subgraph_connected}, \ref{lemma-onshell_subgraphs_tree_structures} and their corollaries). In particular, we delve into the weight structure of the soft subgraph $S$ in section~\ref{section-soft_substructure}, from which the $\mathcal{U}^{(R)}$ and $\mathcal{F}^{(p^2,R)}$ terms can be characterized (lemmas~\ref{lemma-onshell_sij_configuration_constraints}, \ref{lemma-onshell_Si_empty} and their corollaries). In order to characterized the $\mathcal{F}^{(q^2,R)}$ terms, the general configuration of their corresponding spanning 2-trees are studied in section~\ref{section-constraints_Fq2R_terms} (lemmas~\ref{lemma-onshell_Fq2Rterms_kgeq1_tree_structures}, \ref{lemma-onshell_Fq2Rterm_principal_path_properties} and their corollaries). These analyses culminate in the characterization of all the leading terms, further determining of the weight structure of any region $R$ (lemma~\ref{lemma-onshell_leading_terms_forms} and its corollary) in section~\ref{section-determining_hard_soft_weights}.

\begin{figure}[t]
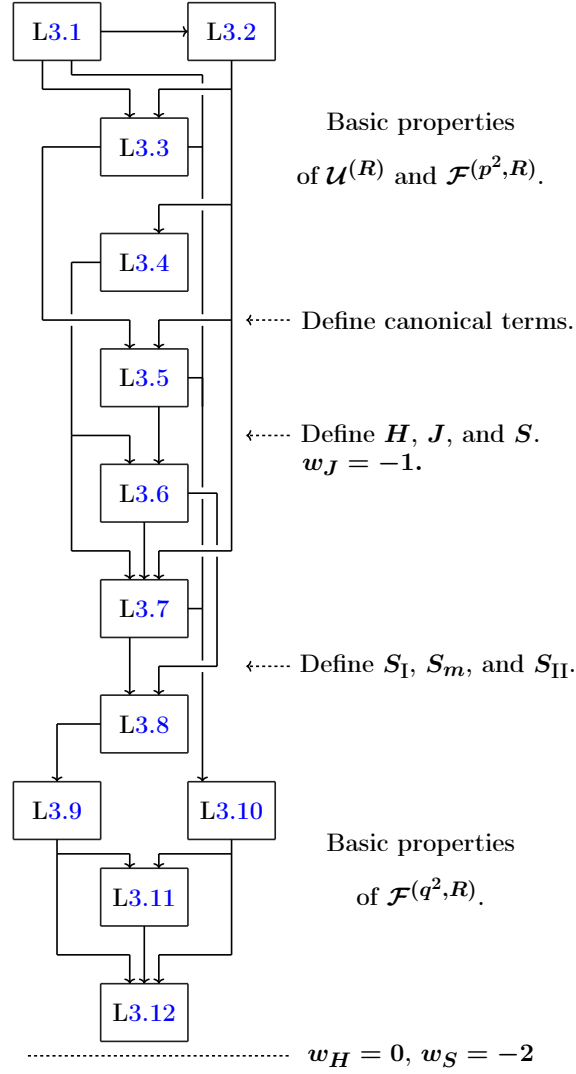

\centering
\include{figs/flowchart}
\vspace{-3em}
\caption{A flowchart outlining the proof of the on-shell-expansion region theorem, based on lemmas~\ref{lemma-onshell_basic_weight_structure_Uterm}-\ref{lemma-onshell_leading_terms_forms} and their associated corollaries. The arrows represent the dependencies between the lemmas. For instance, the derivation of lemma~\ref{lemma-onshell_Fp2_internal_less_equal_minusone} relies on lemmas~\ref{lemma-onshell_basic_weight_structure_Uterm} and~\ref{lemma-onshell_Fp2_external_less_equal_minusone}.}
\label{figure-flowchart}
\end{figure}

A flowchart illustrating the derivation of the aforementioned lemmas is presented in figure~\ref{figure-flowchart}.

\subsubsection{Some properties of the leading terms}
\label{section-some_properties_leading_terms}

For any spanning tree $T^1(\r)$ where $\x^{\r}$ is a $\mathcal{U}$ term, we define a specific momentum flow which we refer to as the \emph{$T^1(\r)$ flow} as follows: we let the external momenta $p_1^\mu,\dots,p_K^\mu$ and $q_1^\mu,\dots,q_L^\mu$ flow into (and out of) $T^1(\r)$, set to zero all the line momenta of $G\setminus T^1(\r)$, and impose momentum conservation at every vertex of $T^1(\r)$. For example, the $T^1(\r)$ flow with $\x^{\r} = x_1 x_3 x_4 x_6$, where the parameters are shown in figure~\ref{figure-fishnet_2times2_onshell_3onshell_1offshell}, is
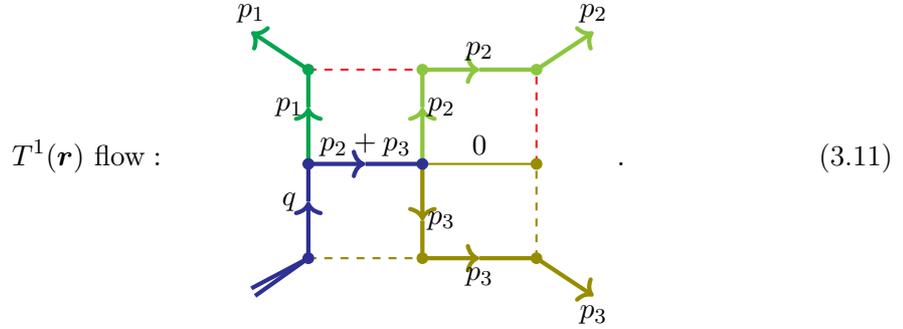
\begin{equation}
T^1(\r)\text{ flow}:\qquad
    \begin{tikzpicture}[baseline=15ex, scale=0.5]
    \draw [dashed, thick,color=Red] (2,7.5) -- (5,7.5);
    \draw [->] [ultra thick,color=LimeGreen] (5,7.5) -- (6.5,7.5); \draw [ultra thick,color=LimeGreen] (6.5,7.5) -- (8,7.5);
    \draw [dashed, thick,color=Red] (8,5) -- (8,7.5);
    \draw [dashed, thick,color=olive] (8,2.5) -- (8,5);
    \draw [->] [ultra thick,color=olive] (5,2.5) -- (6.5,2.5); \draw [ultra thick,color=olive] (6.5,2.5) -- (8,2.5);
    \draw [dashed, thick,color=olive] (2,2.5) -- (5,2.5);
    \draw [->] [ultra thick,color=Blue] (2,2.5) -- (2,4); \draw [ultra thick,color=Blue] (2,4) -- (2,5);
    \draw [->] [ultra thick,color=Green] (2,5) -- (2,6.5); \draw [ultra thick,color=Green] (2,6.5) -- (2,7.5);
    \draw [->] [ultra thick,color=LimeGreen] (5,5) -- (5,6.5); \draw [ultra thick,color=LimeGreen] (5,6.5) -- (5,7.5);
    \draw [thick,color=olive] (5,5) -- (8,5);
    \draw [->] [ultra thick,color=olive] (5,5) -- (5,3.5); \draw [ultra thick,color=olive] (5,3.5) -- (5,2.5);
    \draw [->] [ultra thick,color=Blue] (2,5) -- (3.5,5); \draw [ultra thick,color=Blue] (3.5,5) -- (5,5);
    \draw [->] [ultra thick,color=Green] (2,7.5) -- (0.5,8.5);
    \draw [->] [ultra thick,color=LimeGreen] (8,7.5) -- (9.5,8.5);
    \draw [->] [ultra thick,color=olive] (8,2.5) -- (9.5,1.5);
    \draw [ultra thick, color=Blue] (0.5,1.7) -- (2,2.5);
    \draw [ultra thick, color=Blue] (0.6,1.5) -- (2,2.5);
    
    \node [draw, circle, minimum size=4pt, color=Green, fill=Green, inner sep=0pt, outer sep=0pt] () at (2,7.5) {};
    \node [draw, circle, minimum size=4pt, color=Blue, fill=Blue, inner sep=0pt, outer sep=0pt] () at (2,5) {};
    \node [draw, circle, minimum size=4pt, color=Blue, fill=Blue, inner sep=0pt, outer sep=0pt] () at (2,2.5) {};
    \node [draw, circle, minimum size=4pt, color=LimeGreen, fill=LimeGreen, inner sep=0pt, outer sep=0pt] () at (5,7.5) {};
    \node [draw, circle, minimum size=4pt, color=Blue, fill=Blue, inner sep=0pt, outer sep=0pt] () at (5,5) {};
    \node [draw, circle, minimum size=4pt, color=olive, fill=olive, inner sep=0pt, outer sep=0pt] () at (5,2.5) {};
    \node [draw, circle, minimum size=4pt, color=LimeGreen, fill=LimeGreen, inner sep=0pt, outer sep=0pt] () at (8,7.5) {};
    \node [draw, circle, minimum size=4pt, color=olive, fill=olive, inner sep=0pt, outer sep=0pt] () at (8,5) {};
    \node [draw, circle, minimum size=4pt, color=olive, fill=olive, inner sep=0pt, outer sep=0pt] () at (8,2.5) {};
    
    \node () at (1.5,4) {$q$};
    \node () at (1.5,6.5) {$p_1$};
    \node () at (0.5,9) {$p_1$};
    \node () at (3.5,5.5) {$p_2+p_3$};
    \node () at (5.5,6.5) {$p_2$};
    \node () at (6.5,5.5) {$0$};
    \node () at (5.5,3.5) {$p_3$};
    \node () at (6.5,8) {$p_2$};
    \node () at (6.5,2) {$p_3$};
    \node () at (9.5,9) {$p_2$};
    \node () at (9.5,1) {$p_3$};
    \end{tikzpicture}.
\label{Uterm_spanning_tree_structure}
\end{equation}
Note that the edges carrying nonzero momenta in this $T^1(\r)$ flow is marked in bold.

For each given $T^1(\r)$ flow, we now partition the edges and vertices of $T^1(\r)$ into the associated \emph{trunk} and \emph{branch} subgraphs.
\begin{itemize}
    \item An edge $e\in T^1(\r)$ belongs to the \emph{$q$ trunk}, \emph{$p_i$ trunk}, or \emph{branch} of $T^1(\r)$ respectively, if its momentum in the $T^1(\r)$ flow is off shell, $p_i^\mu$ or zero.
    
    \item A vertex $v\in T^1(\r)$ belongs to the branch of $T^1(\r)$ if $v$ is incident with the edges of the branch subgraph only. It belongs to the $p_i$ trunk of $T^1(\r)$ if $v$ is incident with one or more edges of the $p_i$ trunk, and possibly edges of the branch, but no edges from the $q$ trunk or any $p_j$ trunks ($j\neq i$). Otherwise it belongs to the $q$ trunk of $T^1(\r)$.

    \item For each $i\in\{1,\dots,K\}$, the on-shell external momentum $p_i^\mu$ belongs to the $p_i$ trunk, while all the off-shell external momenta $q_1^\mu,\dots,q_L^\mu$ belong to the $q$ trunk.
    
    \item Each component of the branch is called the $p_i$ ($q$) \emph{branch} if it is adjacent to the $p_i$ ($q$) trunk of $T^1(\r)$.
\end{itemize}
For example, in the figure of (\ref{Uterm_spanning_tree_structure}), the edges $e_8$ and the vertex $v_1$ (see the labelling of vertices and edges in figure~\ref{figure-fishnet_2times2_onshell_3onshell_1offshell}) are in the $p_1$ trunk. The edges $e_2,e_9$ and the vertices $v_2,v_3$ are in the $p_2$ trunk. The edges $e_5,e_{11}$ and the vertices $v_5,v_6$ are in the $p_3$ trunk. The edges $e_7,e_{12}$ and the vertices $v_7,v_8,v_9$ are in the $q$ trunk. The edge $e_{10}$ and the vertex $v_4$ are in the $q$ branch. Meanwhile, all the $p_i$ branches are empty.

From the definitions above, it is clear that for a wide-angle scattering graph $G$ with external momenta $p_1^\mu,\dots,p_K^\mu$ and $q_1^\mu,\dots,q_L^\mu$ whose kinematics are defined in eq.~(\ref{eq:wideangle_onshell_kinematics}), each $p_i$ trunk for a given $i$ is connected and attached by the external momentum $p_i^\mu$, while the $q$ trunk is connected and attached by all the external momenta $q_1^\mu,\dots,q_L^\mu$. In contrast, the branch subgraph can be disconnected, and cannot be attached by any external momenta.

\begin{figure}[t]
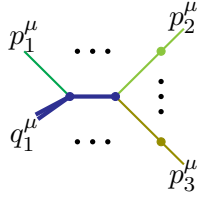
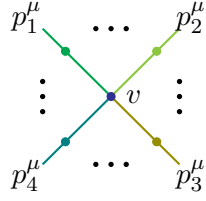

\centering
\begin{subfigure}[b]{0.3\textwidth}
\centering
\include{figs/onshell_spanning_tree_simple_p_trunk}
\caption{An example of spanning tree whose $p_1$~trunk includes the momentum $p_1^\mu$ only.}
\label{figure-onshell_spanning_tree_simple_p_trunk}
\end{subfigure}
\hspace{4em}
\begin{subfigure}[b]{0.3\textwidth}
\centering
\include{figs/onshell_spanning_tree_simple_q_trunk}
\caption{An example of spanning tree whose $q$~trunk consists of a single vertex $v$.}
\label{figure-onshell_spanning_tree_simple_q_trunk}
\end{subfigure}
\caption{Examples of some special $p_i$ and $q$ trunks.}
\label{figure-onshell_spanning_tree_simple_trunks}
\end{figure}

Let us discuss some special cases of the $p_i$ and $q$ trunks. The simplest $p_i$ trunk consists of the external momentum $p_i^\mu$ only, and contains no internal vertices or edges of $G$. This happens when $p_i^\mu$ attaches to the $q$ trunk directly (for example, $p_1^\mu$ in figure~\ref{figure-onshell_spanning_tree_simple_p_trunk}). The simplest $q$ trunk consists of a single vertex $v$ only and contains no internal edges of $G$. This happens when four or more $p$ trunks are incident with this vertex (for example, the blue vertex in the middle of figure~\ref{figure-onshell_spanning_tree_simple_q_trunk}).

We are now ready to study the first lemma in the derivation of the on-shell-expansion region theorem. This lemma imposes constraints on the weight structure of the trunk and branch subgraphs of $T^1(\r)$, where $\x^{\r}$ is any $\mathcal{U}^{(R)}$ term.
\begin{lemma}
\label{lemma-onshell_basic_weight_structure_Uterm}
For any edge $e\in T^1(\r)$ where $\x^{\r}$ is a $\mathcal{U}^{(R)}$ term, we have
\begin{itemize}
    \item $-1\leqslant w(e)\leqslant 0$ if $e$ is in the $p_i$ trunk of $T^1(\r)$;
    \item $w(e)=0$ if $e$ is in the $q$ trunk of $T^1(\r)$.
\end{itemize}
\end{lemma}

\begin{proof}
In the case that $e$ belongs to the $p_i$ trunk of $T^1(\r)$, the spanning 2-tree $T^2\equiv T^1(\r)\setminus e$ corresponds to an $\mathcal{F}^{(p_i^2)}$ term, because one of its components is attached by a single external momentum $p_i^\mu$. We denote this $\mathcal{F}^{(p_i^2)}$ term as $\x^{\r'}$, then
\begin{subequations}
\label{lemma1_proof_step1}
\begin{align}
    & w(T^1(\r))\leqslant w(T^2(\r'));
    \label{lemma1_proof_step1a}\\
    & w(T^2(\r')) = w(T^1(\r)) +w(e) -r_{N+1} +r'_{N+1};
    \label{lemma1_proof_step1b}\\
    & r'_{N+1}=1,\qquad r_{N+1}=0.
    \label{lemma1_proof_step1c}
\end{align}
\end{subequations}
The inequality (\ref{lemma1_proof_step1a}) is due to the minimum-weight criterion. The equalities in (\ref{lemma1_proof_step1b}) and (\ref{lemma1_proof_step1c}) are from the defining relation between $T^1(\r)$ and $T^2(\r')$. Since $T^2(\r')$ is obtained from $T^1(\r)$ by removing $e$ from it, the vectors $\r$ and $\r'$ differ in two entries: the one linked to $e$ is $0$ for $\r$ and $1$ for $\r'$, while the last ($N+1$-th) entry is $0$ for $\r$ and $1$ for~$\r'$. By combining (\ref{lemma1_proof_step1a})-(\ref{lemma1_proof_step1c}) together we obtain $w(e) \geqslant -1$.

If $e$ belongs to the $q$ trunk of $T^1(\r)$, the spanning 2-tree $T^1(\r)\setminus e$ corresponds to an $\mathcal{F}^{(q^2)}$ term, which we denote as $\x^{\r''}$. The following relations then hold:
\begin{subequations}
\label{lemma1_proof_step2}
\begin{align}
    & w(T^1(\r))\leqslant w(T^2(\r''));
    \label{lemma1_proof_step2a}\\
    & w(T^2(\r''))= w(T^1(\r)) +w(e) -r_{N+1} +r''_{N+1};
    \label{lemma1_proof_step2b}\\
    & r''_{N+1}=r_{N+1}=0.
    \label{lemma1_proof_step2c}
\end{align}
\end{subequations}
These relations lead to $w(e) \geqslant 0$. According to theorem~\ref{lemma-no_positive_weights} which states that $w(e)\leqslant 0$, it is clear that $w(e)=0$.
\end{proof}

As a direct application of this lemma, let us consider the (unique) path $P\subset T^1(\r)$ that joins any two external momenta of $G$, where $\x^{\r}$ is a $\mathcal{U}^{(R)}$ term. The weight structure of this path is constrained by the following corollary.
\begin{corollary}
For any $\x^{\r}\in \mathcal{U}^{(R)}(\x)$ and any edge $e\in P$, where $P\subset T^1(\r)$ is the path connecting any two external momenta of $G$, we have $-1\leqslant w(e)\leqslant 0$.
\label{lemma-onshell_basic_weight_structure_Uterm-corollary}
\end{corollary}
\begin{proof}
Since $P$ joins two external momenta of $G$, each edge of $P$ is either in the $p$ or the $q$ trunk of $T^1(\r)$ by definition. With this observation and lemma~\ref{lemma-onshell_basic_weight_structure_Uterm}, we immediately obtain $-1\leqslant w(e)\leqslant 0$.
\end{proof}

These results regarding the $\mathcal{U}^{(R)}$ terms also lead us to establish some fundamental properties of the $\mathcal{F}^{(p_i^2,R)}$ terms. From now on, for any $\mathcal{F}^{(p_i^2,R)}$ term $\x^{\r}$, we will use the notation $t(\r;p_i)$ to represent the component of $T^2(\r)$, which is attached by $p_i^\mu$. We will use $t(\r;\widehat{p}_i)$ to denote the other component of $T^2(\r)$, which is attached by all the other external momenta except $p_i^\mu$. The following lemma highlights the constraints on the weight of any edge that connects $t(\r;p_i)$ and $t(\r;\widehat{p}_i)$.
\begin{lemma}
For any spanning 2-tree $T^2(\r)$ where $\x^{\r}\in \mathcal{F}^{(p_i^2,R)}(\x;\s)$, all the edges whose endpoints are in $t(\r;p_i)$ and $t(\r;\widehat{p}_i)$ respectively must satisfy $w\leqslant -1$. Furthermore, among all these edges, at least one of them satisfies $w = -1$.
\label{lemma-onshell_Fp2_external_less_equal_minusone}
\end{lemma}

\begin{proof}
Let us use $\mathcal{E}_i$ to denote the set of edges whose endpoints are in $t(\r;p_i)$ and $t(\r;\widehat{p}_i)$ respectively. By definition, all these edges are in $G\setminus T^2(\r)$. For any $e\in \mathcal{E}_i$ we consider the spanning tree $T^1(\r')\equiv T^2(\r)\cup e$, and denote its corresponding $\mathcal{U}$ term as $\x^{\r'}$. Similar to eqs.~(\ref{lemma1_proof_step1}) and (\ref{lemma1_proof_step2}), we can derive the following relations
\begin{subequations}
\label{lemma2_proof_step1}
\begin{align}
    & w(T^2(\r))\leqslant w(T^1(\r'));
    \label{lemma2_proof_step2a}\\
    & w(T^1(\r'))-r'_{N+1} +w(e) = w(T^2(\r)) -r_{N+1};
    \label{lemma2_proof_step2b}\\
    & r'_{N+1}=0,\qquad r_{N+1}=1.
    \label{lemma2_proof_step2c}
\end{align}
\end{subequations}
These relations further yield $w(e)\leqslant -1$.

To show the existence of an edge $e_0\in \mathcal{E}_i$ such that $w(e_0) = -1$, let us consider any spanning tree $T^1(\r_0)$ where $\x^{\r_0}\in \mathcal{U}^{(R)}(\x)$ and the path $P\subset T^1(\r_0)$ that connects $p_i$ and any other external momentum of $G$. On one hand, according to corollary~\ref{lemma-onshell_basic_weight_structure_Uterm-corollary}, for any edge $e\in P$ we have $w(e) \geqslant -1$. On the other hand, since the path $P$ joins $p_i$ and another external momentum of $G$, there must be at least one edge $e_0\in P$ whose endpoints are in $t(\r;p_i)$ and $t(\r;\widehat{p}_i)$ respectively, thus $w(e_0) \leqslant -1$ as we have just derived above. In conclusion, $w(e_0) = -1$.
\end{proof}

Lemma~\ref{lemma-onshell_Fp2_external_less_equal_minusone} further implies the correspondence between $\mathcal{U}^{(R)}$ terms and $\mathcal{F}^{(p_i^2,R)}$ terms, which is summarized in the corollary below.

\begin{corollary}
For any $\mathcal{F}^{(p_i^2,R)}$ term $\x^{\r_1}$, there is an edge $e_0\in G\setminus T^2(\r_1)$, such that the spanning tree $T^2(\r_1)\cup e$ corresponds to a $\mathcal{U}^{(R)}$ term.

For any $\mathcal{U}^{(R)}$ term $\x^{\r_2}$ such that there is an edge $e_0$ in the $p_i$ trunk of $T^1(\r_2)$ with $w(e_0)=-1$, the spanning 2-tree $T^1(\r_2) \setminus e_0$ corresponds to an $\mathcal{F}^{(p_i^2,R)}$ term.
\label{lemma-onshell_Fp2_external_less_equal_minusone_corollary1}
\end{corollary}
\begin{proof}
Let us consider the first statement of the lemma. From lemma~\ref{lemma-onshell_Fp2_external_less_equal_minusone}, there exists an edge $e_0\in G\setminus T^2(\r_1)$, such that $e_0$ connects $t(\r;p_i)$ and $t(\r;\widehat{p}_i)$, and $w(e_0)=-1$. The graph $T^2(\r_1)\cup e_0$ is then a spanning tree of $G$, whose weight, by definition, is
\begin{eqnarray}
w(T^2(\r_1)) -w(e) -r_{1,N+1} +r_{2,N+1} =w(T^2(\r_1))
\end{eqnarray}
Note that we have used $w(e_0)=-1$, $r_{1,N+1}=1$, and $r_{2,N+1}=0$ above. It then follows that this spanning tree is also of the minimum weight, hence corresponding to a $\mathcal{U}^{(R)}$ term.

The second statement of the lemma can be proved similarly.
\end{proof}

Corollary~\ref{lemma-onshell_Fp2_external_less_equal_minusone_corollary1} can be considered as a generalization of our observation in eq.~(\ref{eq:UandFp2_terms_duality_example}), which implies that we can obtain an $\mathcal{F}^{(p_i^2,R)}$ term~$\x^{\r_1}$ from some certain $\mathcal{U}^{(R)}$ terms~$\x^{\r_2}$, and vice versa. Based on this $\mathcal{U}^{(R)}$-$\mathcal{F}^{(p^2,R)}$ correspondence, we can apply corollary~\ref{lemma-onshell_basic_weight_structure_Uterm-corollary} to investigate the property of the $\mathcal{F}^{(p_i,R)}$ terms, which we summarize in the following corollary.
\begin{corollary}
For any $\mathcal{F}^{(p_i^2,R)}$ term $\x^{\r_1}$ and any two external momenta of $G$ except~$p_i^\mu$, the unique path $P\subset t(\r_1;\widehat{p}_i)$ connecting them satisfies the following condition: for any $e\in P$, $-1\leqslant w(e)\leqslant 0$.
\label{lemma-onshell_Fp2_external_less_equal_minusone_corollary2}
\end{corollary}
\begin{proof}
We consider the $\mathcal{U}^{(R)}$ term $\x^{\r_2}$ such that $T^2(\r_1)\cup e_0= T^1(\r_2)$, where $e_0$ is the edge connecting the components of $T^2(\r_1)$ with $w(e_0)=-1$. By definition, the path $P\subset t(\r_1;\widehat{p}_i)$ is also contained in $T^1(\r_2)$, so according to corollary~\ref{lemma-onshell_basic_weight_structure_Uterm-corollary}, we have $-1\leqslant w(e)\leqslant 0$ for any $e\in P$. This corollary is hence proved.
\end{proof}

Before going on to investigate more properties of the leading terms, let us now introduce a few concepts as follows. We call any subgraph of $G$ a \emph{subtree} of $G$ if it is a tree graph. For example, given any $\x^{\r}\in \mathcal{F}^{(p_i^2,R)}(\x;\s)$, the two components of $T^2(\r)$, i.e. $t(\r;p_i)$ and $t(\r;\widehat{p}_i)$, are both subtrees of $G$. Suppose $t'$ is a subtree of $G$, we define the subgraph of $G$ that is \emph{generated} by $t'$, i.e. $G(t')$, as follows:
\begin{enumerate}
    \item [(1)] any vertex $v\in G(t')$\quad $\Leftrightarrow$\quad $v\in t'$;
    \item [(2)] any edge $e\in G(t')$\quad $\Leftrightarrow$\quad $e\in G$, and both endpoints of $e$ are in $t'$.
\end{enumerate}
We further define the \emph{external edges} of $G(t')$ as those edges whose endpoints are in $G(t')$ and $G\setminus G(t')$, respectively.

With this concept, we can thus view lemma~\ref{lemma-onshell_Fp2_external_less_equal_minusone} as follows: given any $\mathcal{F}^{(p_i^2,R)}$ term $\x^{\r}$, the \emph{external} edges of $G(t(\r;p_i))$ all satisfy $w\leqslant -1$, and in particular, some of them satisfy $w=-1$. It is then natural to inquire whether the same conclusion holds for the \emph{internal} edges of $G(t(\r;p_i))$. The answer is positive, as summarized in the following lemma.

\begin{lemma}
Any edge $e\in G(t(\r;p_i))$, where $\x^{\r}\in \mathcal{F}^{(p_i^2,R)}(\x;\s)$, satisfies $w(e) \leqslant -1$.
\label{lemma-onshell_Fp2_internal_less_equal_minusone}
\end{lemma}

\begin{proof}
We prove it by contradiction. Assume that there is an $\mathcal{F}^{(p_i^2,R)}$ term $\x^{\r_1}$, such that the graph $G(t(\r_1;p_i))$ contains some edges satisfying $w>-1$. Among all these edge we focus on a subset of them, denoted as $\mathcal{E}_0$, such that each element in $\mathcal{E}_0$ has the maximum weight $w_0$. We further use $\Gamma_0$ to denote the graph consisting of edges in $\mathcal{E}_0$ and their endpoints. From this definition, $\Gamma_0 \subset G(t(\r_1;p_i))$, and $w(e)=w_0>-1$ for any $e\in \Gamma_0$; in addition, $w_0> w(e')$ for any $e'\in G(t(\r_1;p_i))\setminus \Gamma_0$ by definition.

Then we choose any connected component of $\Gamma_0$ and denote it as $\gamma_0$. According to lemma~\ref{lemma-heavy_subgraph_inside_light_graph}, there is an $\mathcal{F}^{(q^2,R)}$ term $\x^{\r_2}$, such that both two components of $T^2(\r_2)$ contain some vertices of $\gamma_0$. Since $\gamma_0$ is connected, there must be an edge $e_0\in \gamma_0$ such that the endpoints of $e_0$ are respectively in the two distinct components of $T^2(\r_2)$. For convenience, we denote these two endpoints as $A$ and $B$ according to the following rule: $A$ is in the component of $T^2(\r_2)$ that is attached by the momentum $p_i^\mu$, while $B$ is in the other component of $T^2(\r_2)$. Furthermore, the two components of $T^2(\r_2)$ including $A$ and $B$ respectively, are denoted as $t(\r_2;A)$ and $t(\r_2;B)$.

For clarity, let us emphasize that $\x^{\r_1}$ is an $\mathcal{F}^{(p_i^2,R)}$ term while $\x^{\r_2}$ is an $\mathcal{F}^{(q^2,R)}$ term. Both $A$ and $B$ are in one component of $T^2(\r_1)$, which is $t(\r;p_i)$, because $e_0\in \gamma_0\subseteq \Gamma_0\subset t(\r;p_i)$. In contrast, $A$ and $B$ are in the two distinct components of $T^2(\r_2)$, which we have denoted as $t(\r_2;A)$ and $t(\r_2;B)$, respectively. These notions are sketched in figure~\ref{onshell_lemma3_proof_general_picture}, which can be thought of as embedded in a generic wide-angle scattering graph.
\begin{figure}[t]
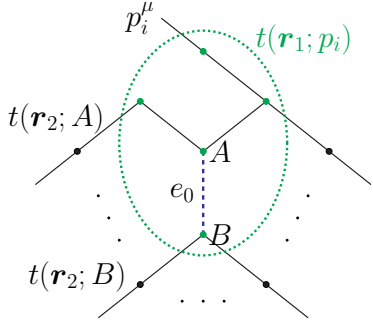

\centering
\include{figs/onshell_lemma3_proof_general_picture}
\vspace{-3em}
\caption{An example showing the configurations of~$t(\r_2;A)$ and $t(\r_2;B)$. For the two components of $T^2(\r_2)$, $t(\r_2;A)$ is the one that includes $A$ and is attached by $p_i^\mu$, and $t(\r_2;B)$ is the other one including $B$. Note that the tree graph $t(\r_1;p_i)$, encircled by dotted green curves, includes all the green vertices.}
\label{onshell_lemma3_proof_general_picture}
\end{figure}

Below, we shall discuss all the possible configurations of $t(\r_2;B)$, and explain that for each possibility, one can always find another $\mathcal{F}^{(q^2)}$ term whose corresponding spanning 2-tree, $T_*^2$, satisfies $w(T_*^2)<w(T^2(\r_2))$. Since $\x^{\r_2}$ is leading by definition, such an inequality clearly violates the minimum-weight criterion, completing the proof by contradiction.

\begin{itemize}
    \item [\textbf{I.}] If an off-shell external momentum $q_j^\mu$ attaches to $t(\r_2;B)$ (see figure~\ref{onshell_lemma3_proof_config_I}), then there is a unique path $P_B\subset t(\r_2;B)$ connecting $B$ and $q_j^\mu$. Since $B\in t(\r_1;p_i)$ while $q_j^\mu$ is not attached to $t(\r_1;p_i)$, there must be an edge $e'\subset P_B$ connecting the two components of $T^2(\r_1)$. From lemma~\ref{lemma-onshell_Fp2_external_less_equal_minusone}, $w(e') \leqslant -1$, and we then consider the spanning 2-tree
    \begin{eqnarray}
    T_*^2\equiv T^2(\r_2)\cup e_0 \setminus e'.
    \end{eqnarray}
    By definition, one of its components is attached by $q_j^\mu$ and possibly other external momenta, meanwhile, the other component is attached by all the external momenta of $t(\r_2;A)$, whose sum is off shell, and possibly other external momenta. Therefore, the momentum flowing between the components of $T_*^2$ must be off shell, indicating that $T_*^2$ corresponds to an $\mathcal{F}^{(q^2)}$ term (see figure~\ref{onshell_lemma3_proof_config_I_comparison}). The weight of $T_*^2$ is then
    \begin{eqnarray}
    w(T_*^2) = w(T^2(\r_2)) +w(e') -w(e_0)< w(T^2(\r_2)).
    \end{eqnarray}
    In the inequality above, we have used $w(e') \leqslant -1< w(e_0)$. Hence a contradiction to the minimum-weight criterion once more, leading to the conclusion that $t(\r_2;B)$ cannot be attached by any off-shell external momentum.
\end{itemize}
\begin{figure}[t]
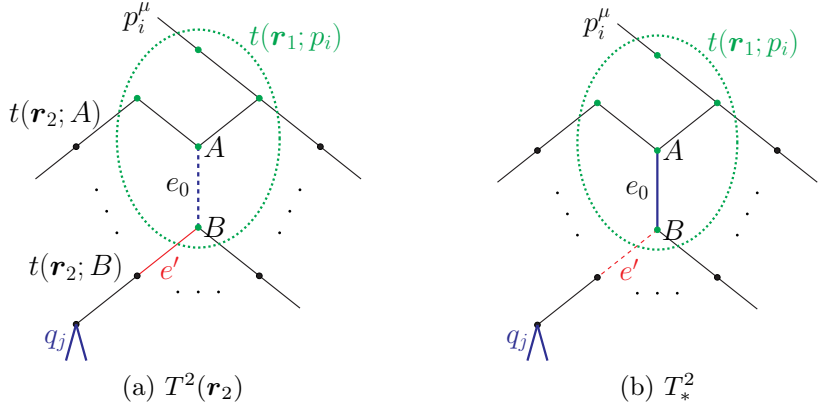

\centering
\hspace{-2em}
\begin{subfigure}[b]{0.32\textwidth}
\centering
\include{figs/onshell_lemma3_proof_config_I}
\vspace{-3em}
\caption{$T^2(\r_2)$}
\label{onshell_lemma3_proof_config_I}
\end{subfigure}
\hspace{3em}
\begin{subfigure}[b]{0.32\textwidth}
\centering
\include{figs/onshell_lemma3_proof_config_I_comparison}
\vspace{-3em}
\caption{$T_*^2$}
\label{onshell_lemma3_proof_config_I_comparison}
\end{subfigure}
\caption{The comparison of $T^2(\r_2)$ and $T_*^2$ in case~\textbf{I}.}
\label{figure-onshell_lemma3_proof_I}
\end{figure}

We then only need to consider the possibility that $t(\r_2;B)$ is attached by on-shell external momenta only. To be specific, let us denote these on-shell external momenta by $p_{k_1}^\mu,\dots,p_{k_n}^\mu$ ($n\geqslant 2$ in order that $\x^{\r_2}$ is an $\mathcal{F}^{(q^2)}$ term). For each $a\in\{1,\dots,n\}$, we further use $P_a$ to denote the path in $t(\r_2;B)$, which joins the vertex $B$ and the external momenta $p_{k_a}^\mu$. It then follows from the tree structure of $t(\r_2;B)$ that, for any given $a\neq b$, $P_a\cap P_b$ must be a path in $t(\r_2;B)$, one of whose endpoints is $B$.

\begin{itemize}
    \item [\textbf{II.}] If there is a pair $a\neq b$ such that not all the vertices of $P_a\cap P_b$ are in $t(\r_1;p_i)$, then there must be an edge $e'\in P_a\cap P_b$ whose endpoints are in $t(\r_1;p_i)$ and $t(\r_1;\widehat{p}_i)$ respectively (see figure~\ref{onshell_lemma3_proof_config_II}). From lemma~\ref{lemma-onshell_Fp2_external_less_equal_minusone}, $w(e')\leqslant -1$. Similar to the analysis for case~\textbf{I}, we then focus on the spanning 2-tree $T_*^2\equiv t(\r_2;B)\cup e_0\setminus e'$ (see figure~\ref{onshell_lemma3_proof_config_II_comparison}).
    
    By construction, the vertex $B$ is located in one component of $T_*^2$, attached by all the external momenta of $t(\r_2;A)$ and possibly other external momenta, the sum of which is off shell. Meanwhile, the other component of~$T_*^2$ is attached by $p_{k_a}^\mu$, $p_{k_b}^\mu$, and possibly other external momenta. Therefore, the momentum flow between the components of $T_*^2$ must be off shell, indicating that $T_*^2$ corresponds to an $\mathcal{F}^{(q^2)}$ term. Once again, we have
    \begin{eqnarray}
    w(T_*^2) = w(T^2(\r_2)) +w(e') -w(e_0) <w(T^2(\r_2)),
    \end{eqnarray}
    and the violation of the minimum-weight criterion occurs, ruling out this possibility.
\end{itemize}
\begin{figure}[t]
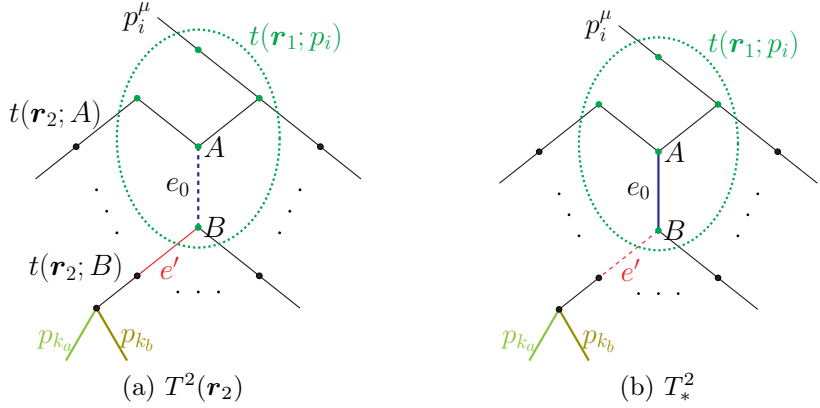

\centering
\hspace{-2em}
\begin{subfigure}[b]{0.32\textwidth}
\centering
\include{figs/onshell_lemma3_proof_config_II}
\vspace{-3em}
\caption{$T^2(\r_2)$}
\label{onshell_lemma3_proof_config_II}
\end{subfigure}
\hspace{3em}
\begin{subfigure}[b]{0.32\textwidth}
\centering
\include{figs/onshell_lemma3_proof_config_II_comparison}
\vspace{-3em}
\caption{$T_*^2$}
\label{onshell_lemma3_proof_config_II_comparison}
\end{subfigure}
\caption{The comparison of $T^2(\r_2)$ and $T_*^2$ in case~\textbf{II}.}
\label{figure-onshell_lemma3_proof_II}
\end{figure}

With cases~\textbf{I} and~\textbf{II} ruled out, the remaining possible configurations of $T^2(\r_2)$ can be described by the following conditions: the external momenta attaching to $t(\r_2;B)$ are $p_{k_1}^\mu,\dots,p_{k_n}^\mu$ ($n\geqslant 2$); furthermore, for each choice of $a,b\in \{1,\dots,n\}$, the vertices of $P_a\cap P_b$ all belong to $t(\r_1;p_i)$, where $P_i\subset t(\r_2;B)$ represents the path connecting $B$ and $p_{k_i}^\mu$. These conditions imply that, for each path $P_a$, there is an edge $e_a\in P_a$ connecting the components of $T^2(\r_1)$, and $e_1,\dots,e_n$ are all distinct from each other.

With this observation, we consider the graph $\Gamma'$ obtained by removing $e_1,\dots,e_n$ from $t(\r_2;B)$, namely, $\Gamma'\equiv t(\r_2;B)\setminus \big(e_1 \cup\cdots\cup e_n \big)$. It has $n+1$ components: one of them contains the vertex $B$, and the remaining $n$ of them are attached by the external momenta $p_{k_1}^\mu,\dots,p_{k_n}^\mu$ respectively, which we denote as $\gamma_1,\dots,\gamma_n$.

As guaranteed by corollary~\ref{lemma-onshell_Fp2_external_less_equal_minusone_corollary2}, for any two momenta $p_{k_a}^\mu,\ p_{k_b}^\mu \in \{p_{k_1}^\mu,\dots,p_{k_n}^\mu\}$, there is a path $P_{ab}\subset t(\r_1;\widehat{p}_i)$ connecting them, with $-1\leqslant w(e)\leqslant 0$ for each edge $e\in P_{ab}$. According to this definition, the endpoints of $P_{ab}$ are both in $t(\r_2;B)$. Below, we will discuss the following two possibilities regarding the positions of the other vertices of $P_{ab}$.

\begin{itemize}
    \item [\textbf{III.}] If all the vertices of $P_{ab}$ are in $t(\r_2;B)$, then there exists an edge $e_{ab}\in P_{ab}$, whose endpoints are in $\gamma_a$ and $\gamma_b$ respectively (see figure~\ref{onshell_lemma3_proof_config_III}). We then modify $T^2(\r_2)$ by adding $e_0$ and $e_{ab}$ to it and removing $e_a$ and $e_b$ from it. The result, denoted as $T_*^2$, is a spanning 2-tree. Namely,
    \begin{eqnarray}
    T_*^2\equiv T^2(\r_2) \cup(e_0\cup e_{ab}) \setminus (e_a\cup e_b).
    \end{eqnarray}
    Note that one component of $T_*^2$ is attached by $p_{k_a}^\mu$ and $p_{k_b}^\mu$, and the other is attached by $p_{k_i}^\mu$ ($i\neq a,b$) and all the external momenta of $t(\r_2;A)$. Therefore, the momentum flowing between the two connected components of $T_*^2$ must be off shell, and $T_*^2$ corresponds to an $\mathcal{F}^{(q^2,R)}$ term (see figure~\ref{onshell_lemma3_proof_config_III_comparison}). The weight of $T_*^2$ is hence
    \begin{eqnarray}
    w(T_*^2) = w(T^2(\r_2)) +w(e_c) +w(e_d) -w(e_0) -w(e_{cd}) <w(T^2(\r_2)),
    \end{eqnarray}
    where we have used $w(e_0)>-1$, $w(e_{cd})\geqslant -1$, and $w(e_c), w(e_d)\leqslant -1$. Again we see the violation of the minimum-weight criterion, and this case is ruled out.
\begin{figure}[t]
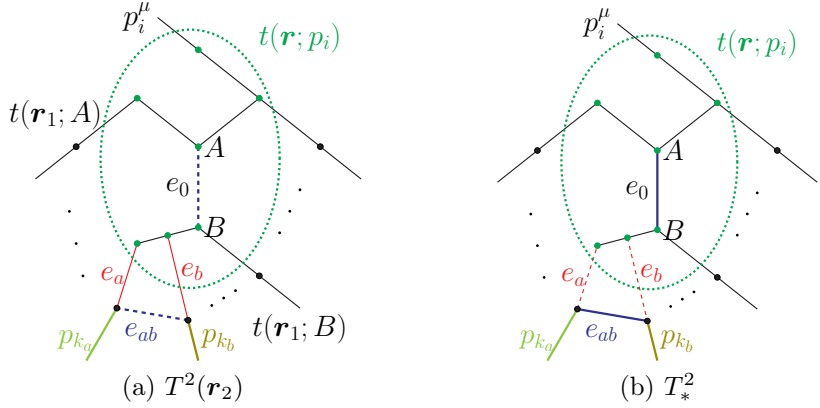

\centering
\hspace{-2em}
\begin{subfigure}[b]{0.32\textwidth}
\centering
\include{figs/onshell_lemma3_proof_config_III}
\vspace{-3em}
\caption{$T^2(\r_2)$}
\label{onshell_lemma3_proof_config_III}
\end{subfigure}
\hspace{3em}
\begin{subfigure}[b]{0.32\textwidth}
\centering
\include{figs/onshell_lemma3_proof_config_III_comparison}
\vspace{-3em}
\caption{$T_*^2$}
\label{onshell_lemma3_proof_config_III_comparison}
\end{subfigure}
\caption{The comparison of $T^2(\r_2)$ and $T_*^2$ in case~\textbf{III}.}
\label{figure-onshell_lemma3_proof_III}
\end{figure}
\end{itemize}    

In constructing $T_*^2$ above, the main idea is to include $P_{ab}$ in one component of $T_*^2$, which is attached by $p_{k_a}^\mu$ and $p_{k_b}^\mu$. As we will see below, a similar analysis is applicable to the case where some vertices of $P_{ab}$ are included in $t(\r_2;A)$.

\begin{itemize}
    \item [\textbf{IV.}] If some vertices of $P_{ab}$ are in $t(\r_2;A)$, then these vertices cannot be the endpoints of $P_{ab}$, which are in $t(\r_2;B)$ by definition. Without loss of generality, we assume that $P_{ab}$ starts from $p_{k_a}^\mu$, leaves $\gamma_a$ at vertex $B'$, enters $t(\r_2;A)$ at vertex $A'$, then leaves $t(\r_2;A)$ at vertex $A''$, and finally enters $\gamma_b$ at vertex $B''$ (see figure~\ref{onshell_lemma3_proof_config_IV}). For later convenience, let us denote the edge of $P_{ab}$ which connects $A'$ and $B'$ by $e'_*$, and the edge connecting $A''$ and $B''$ as $e''_*$.
    
    The first observation is that $w(e'_*), w(e''_*) \geqslant -1$, because $e'_*,e''_*\in P_{ab}$, and every edge of $P_{ab}$ satisfies $-1\leqslant w\leqslant 0$. Then we recall that $A\in t(\r_1;p_i)$ and $A',A''\in t(\r_1;\widehat{p}_i)$, so the path $P'\subset t(\r_2;A)$ connecting $A$ and $A'$ must contain some edge $e'_a$, such that the endpoints of $e'_a$ are in $t(\r_1;p_i)$ and $t(\r_1;\widehat{p}_i)$ respectively. Furthermore, $w(e'_a)\leqslant -1$ according to lemma~\ref{lemma-onshell_Fp2_external_less_equal_minusone}.
    
    From the observations above, let us consider the spanning 2-tree $T_*^2$ obtained from $T^2(\r_2)$ by adding $e_0,e'_*,e''_*$ to it and removing $e'_a,e_a,e_b$ from it. Namely,
    \begin{eqnarray}
    T_*^2\equiv T^2(\r_2)\cup (e_0\cup e'_*\cup e''_*) \setminus (e'_a\cup e_a\cup e_b).
    \end{eqnarray}
    It then follows that the momentum flowing between the components of $T_*^2$ is exactly $(p_{k_a}+p_{k_b})^\mu$, which is off shell (see figure~\ref{onshell_lemma3_proof_config_IV_comparison}). Then $T_*^2$ corresponds to an $\mathcal{F}^{(q^2)}$ term, whose weight is
    \begin{align}
        w(T_*^2) = w(T^2(\r_2)) +w(e'_a) +w(e_a) +w(e_b) -w(e_0) -w(e'_*) -w(e''_*) <w(T^2(\r_2)).
    \label{onshel_lemma3_proof_caseIV}
    \end{align}
    In deriving the inequality above we have used the defining properties $w(e_0)>-1$, $w(e'_*), w(e''_*)\geqslant -1$, and $w(e'_a), w(e_a), w(e_b)\leqslant -1$. Again, the minimum-weight criterion is violated, and the possibility of this case is also ruled out.
\end{itemize}
\begin{figure}[t]
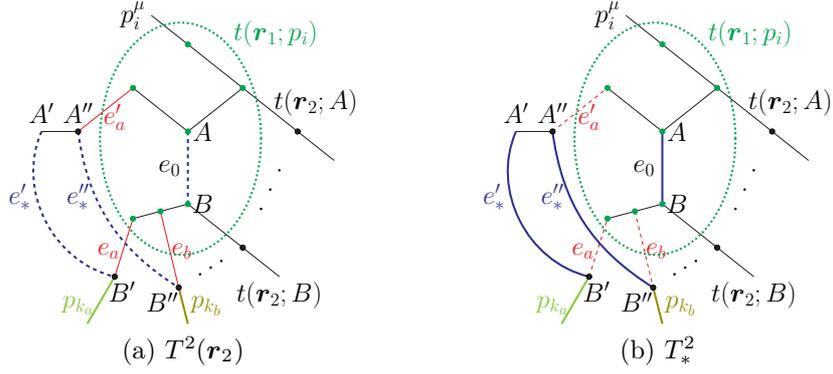

\centering
\begin{subfigure}[b]{0.32\textwidth}
\centering
\include{figs/onshell_lemma3_proof_config_IV}
\vspace{-3em}
\caption{$T^2(\r_2)$}
\label{onshell_lemma3_proof_config_IV}
\end{subfigure}
\hspace{3em}
\begin{subfigure}[b]{0.32\textwidth}
\centering
\include{figs/onshell_lemma3_proof_config_IV_comparison}
\vspace{-3em}
\caption{$T_*^2$}
\label{onshell_lemma3_proof_config_IV_comparison}
\end{subfigure}
\caption{The comparison of $T^2(\r_2)$ and $T_*^2$ in case~\textbf{IV}.}
\label{figure-onshell_lemma3_proof_IV}
\end{figure}

To summarize, in the cases~\textbf{I}-\textbf{IV} above we have thoroughly examined all potential configurations in which $e_0$, an edge of $G(t(\r_1;p_i))$ with $w(e_0)>-1$, connects the two distinct components of $T^2(\r_2)$. In each case, we identify a violation of the minimum-weight criterion. Consequently, we establish that $w(e)\leqslant -1$ for every edge $e\in G(t(\r_1;p_i))$ where $\x^{\r_1}$ is an $\mathcal{F}^{(p_i^2,R)}$ term. Thus, we have proved this lemma.
\end{proof}

As a direct application of lemma~\ref{lemma-onshell_Fp2_internal_less_equal_minusone}, we can clearly identify that the example illustrated in figure~\ref{figure-onshell_region_bad_example1}, where a hard loop is present inside a jet, does not qualify as a valid region. Specifically, the configuration in that example falls within case~\textbf{II}, as described in the preceding proof.

Furthermore, as another practical application, we can infer that for any $\mathcal{F}^{(p_i^2,R)}$ term~$\x^{\r}$, there exists a path ``passing through'' $t(\r;p_i)$, whose edges all satisfy $w=-1$.

\begin{corollary}
\label{lemma-onshell_Fp2_internal_less_equal_minusone_corollary1}
For any $\mathcal{F}^{(p_i^2,R)}$ term $\x^{\r}$, let $e_0$ be an edge whose endpoints are in the two distinct components of $T^2(\r)$, and $w(e_0)=-1$. Then all the edges in the path $P\subset t(\r;p_i)$, which connects the external momentum $p_i^\mu$ and the endpoint of $e_0$ within $t(\r;p_i)$, satisfy $w=-1$.
\end{corollary}
\begin{proof}
Let us first recall the content of lemma~\ref{lemma-onshell_Fp2_external_less_equal_minusone}, which guarantees the existence of an edge $e_0$ that connects the two components of $T^2(\r)$ with $w(e_0)=-1$. For convenience, we use $v_0$ to denote the endpoint of $e_0$ which is in $t(\r;p_i)$. From corollary~\ref{lemma-onshell_Fp2_external_less_equal_minusone_corollary1}, the spanning tree $T^1\equiv T^2(\r)\cup e_0$ corresponds to a $\mathcal{U}^{(R)}$ term, and the path $P\subset t(\r;p_i)$ joining $p_i^\mu$ and~$v_0$ is part of the $p_i$ trunk of $T^1$. Now, on one hand, lemma~\ref{lemma-onshell_basic_weight_structure_Uterm} indicates that, for any edge $e\in P$, we have $e\geqslant -1$. On the other hand, given that $P\subset t(\r;p_i)$ by definition, lemma~\ref{lemma-onshell_Fp2_internal_less_equal_minusone} further implies $w(e)\leqslant -1$. It then follows that $w(e)=-1$.
\end{proof}

This corollary can be applied to analyzing the $\mathcal{U}^{(R)}$-term weight structures. As previously shown in lemma~\ref{lemma-onshell_basic_weight_structure_Uterm}, there is a constraint on the weight structure of a generic $\mathcal{U}^{(R)}$ term: every $p_i$ trunk of $T^1(\r)$, where $\x^{\r}\in \mathcal{U}^{(R)}(\x)$ consists of edges satisfying $-1\leqslant w\leqslant 0$. This constraint can be further refined as follows.

\begin{corollary}
\label{lemma-onshell_Fp2_internal_less_equal_minusone_corollary2}
In every $p_i$ trunk of $T^1(\r)$ where $\x^{\r}\in \mathcal{U}^{(R)}(\x)$, there exists a vertex $v_i$ satisfying the following two conditions:
\begin{enumerate}
    \item $w(e)=-1$ for any $e\in P_{i,\textup{out}}$, where $P_{i,\textup{out}}\subset T^1(\r)$ is the path connecting $p_i^\mu$ and $v_i$;
    \item $-1< w(e)\leqslant 0$ for any $e\in P_{i,\textup{in}}$, where $P_{i,\textup{in}}\subset T^1(\r)$ is the path connecting $v_i$ and the $q$~trunk of $T^1(\r)$.
\end{enumerate}
\end{corollary}
\begin{proof}
For any edge $e'$ in the $p_i$ trunk of $T^1(\r)$ such that $w(e')=-1$, we note that $T^1(\r)\setminus e'$ is a spanning 2-tree corresponding to an $\mathcal{F}^{(p_i^2,R)}$ term. We can directly apply corollary~\ref{lemma-onshell_Fp2_internal_less_equal_minusone_corollary1} to demonstrate that the path $P_{i,\textup{in}}\subset T^1(\r)$, which connects $p_i^\mu$ and either endpoint of $e'$, consists of edges satisfying $w=-1$. This further implies that, the union of those $w=-1$ edges within the $p_i$ trunk of $T^1(\r)$ forms a connected graph. In other words, there exists a vertex $v_i$ in the $p_i$ trunk of $T^1(\r)$, such that for each edge $e$ in the $p_i$ trunk, $w(e)=-1$ if $e$ is positioned between $p_i^\mu$ and $v_i$, and $-1<w(e)\leqslant 0$ otherwise. The two statements in the corollary are then justified.
\end{proof}

Leveraging lemma~\ref{lemma-onshell_Fp2_internal_less_equal_minusone} and its associated corollaries, we establish that for each $\mathcal{F}^{(p_i^2,R)}$ term $\x^{\r}$, there is a connected subgraph of $G(t(\r;p_i))$ including the external momentum $p_i^\mu$, and all its edges satisfy $w=-1$. Our next objective extends further: we seek to show that a specific subgraph of $G(t(\r;p_i))$, which includes $p_i^\mu$, \emph{all} the edges of $G(t(\r;p_i))$ with $w=-1$, and their endpoints, is indeed connected. To accomplish this, we need the following lemma.

\begin{lemma}
\label{lemma-onshell_heavy_in_light_constraint}
For any subgraph $\gamma\subset G$ that meets two conditions: (1) $\gamma$ contains at least one edge, and (2) $\gamma$ is attached by no external momenta, the following relation is \textbf{impossible}:
\begin{eqnarray}
\textup{min}\left \{ w(e):\ e\in\gamma \right \} \geqslant -1> \textup{max}\left \{ w(e):\ e\textup{ is adjacent to } \gamma \right \}.
\label{eq:heavy_in_light_forbidden_relation}
\end{eqnarray}
\end{lemma}

Before presenting a proof of this lemma, let us compare (\ref{eq:heavy_in_light_forbidden_relation}) with the inequality (\ref{lemma3_assumption}). The latter is expressed as $\textup{min}\left \{ w(e)\big| e\in\gamma \right \} > \textup{max}\left \{ w(e)\big| e\textup{ is adjacent to } \gamma \right \}$ and serves as a condition of lemma~\ref{lemma-heavy_subgraph_inside_light_graph}. Lemma~\ref{lemma-onshell_heavy_in_light_constraint} can be regarded as an additional constraint on this inequality, effectively eliminating the possibility of (\ref{eq:heavy_in_light_forbidden_relation}).

\begin{proof}
We shall prove this lemma by contradiction, assuming that (\ref{eq:heavy_in_light_forbidden_relation}) holds for some region $R$ and seeking contradictions arising from this assumption.

First of all, lemma~\ref{lemma-heavy_subgraph_inside_light_graph} indicates that there must be an $\mathcal{F}^{(R)}$ term $\x^{\r_1}$ such that both components of $T^2(\r_1)$ contain some vertices of $\gamma$. Moreover, $\x^{\r_1}$ cannot be an $\mathcal{F}^{(p^2,R)}$ term.
To see this, we consider the contrary, i.e. $\x^{\r_1}\in \mathcal{F}^{(p_i^2,R)}$ for some $i\in \{1,\dots,K\}$. Then for any edge $e_0\in \gamma$ that joins the components of $T^2(\r_1)$, we have $w(e_0)\leqslant -1$ as a consequence of lemma~\ref{lemma-onshell_Fp2_external_less_equal_minusone}. Meanwhile, we also have $w(e_0)\geqslant-1$ according to the first inequality of~(\ref{eq:heavy_in_light_forbidden_relation}), and as a result, $w(e_0)=-1$. On one hand, from the result of corollary~\ref{lemma-onshell_Fp2_internal_less_equal_minusone_corollary1}, there is a path in $t(\r_1;p_i)$ from $e_0$ to $p_i^\mu$, which consists of edges satisfying $w=-1$. On the other hand, since this path joins $\gamma$ and $p_i^\mu$, the second inequality of (\ref{eq:heavy_in_light_forbidden_relation}) implies that there must be some edges on this path with $w<-1$. These two statements contradict each other.

As a result, $\x^{\r_1}$ can only be an $\mathcal{F}^{(q^2,R)}$ term. For later convenience, let us denote the endpoints of $e_0$ as $A$ and $B$, and the components of $T^2(\r_1)$ containing $A$ and $B$ as $t(\r_1;A)$ and $t(\r_1;B)$, respectively.

Below we discuss the possible configurations of $T^2(\r_1)$. A first observation is that (\ref{eq:heavy_in_light_forbidden_relation}) forbids any off-shell external momenta of $G$. Otherwise, suppose $G$ has an off-shell external momentum $q^\mu$ attaching to $t(\r_1;A)$. The path $P\subset t(\r_1;A)$ joining $A$ and $q^\mu$, then must contain some edges $e'\in G\setminus \gamma$, which satisfy $w(e')<-1$ according to (\ref{eq:heavy_in_light_forbidden_relation}). The spanning 2-tree $T_*^2\equiv T^2(\r_1)\cup e_0\setminus e'$ then corresponds to another $\mathcal{F}^{(q^2)}$ term with a smaller weight, because
\begin{eqnarray}
w(T_*^2)= w(T^2(\r_1)) +w(e_j) -w(e_0)< w(T^2(\r_1)).
\label{proof_heavy_in_light_constraint_case1}
\end{eqnarray}
This clearly violates the minimum-weight criterion. It then follows that all the external momenta of $G$ must be on shell. We point out that the analysis given above resembles that of case~\textbf{I} in the proof of lemma~\ref{lemma-onshell_Fp2_internal_less_equal_minusone}.

Let us now suppose the on-shell external momenta $p_{k_a},p_{k_b},\dots,p_{k_n}$ ($n\geqslant 2$) attach to $t(\r_1;A)$, and denote $P_a\subset t(\r_1;A)$ as the path joining $A$ to $p_{k_a}$ for any $a=1,\dots,n$. If there exist $a$ and $b$ such that $P_a\cap P_b \not \subset \gamma$, then according to (\ref{eq:heavy_in_light_forbidden_relation}), $P_a\cap P_b$ must contain some edge $e'\in G\setminus \gamma$ satisfying $w(e')<-1$. The spanning 2-tree $T^2(\r_1)\cup e_0 \setminus e'$ then corresponds to another $\mathcal{F}^{(q^2)}$ term with a smaller weight. We point out that the analysis of this case resembles that of case~\textbf{II} in the proof of lemma~\ref{lemma-onshell_Fp2_internal_less_equal_minusone}.

Now the only possibility where (\ref{eq:heavy_in_light_forbidden_relation}) may hold is that all the external momenta of $G$ are on shell, and moreover, $P_a\cap P_b\subset \gamma$ for any $a$ and $b$. That is to say, starting from $A$ and ending at $p_{k_a}^\mu$ and $p_{k_b}^\mu$, the paths $P_a$ and $P_b$ get separated at a $\gamma$ vertex. Moreover, as a result of (\ref{eq:heavy_in_light_forbidden_relation}), for each $a\in \{1,\dots,n\}$, there is some edge $e'_a\in P_a\cap (G\setminus \gamma)$, which satisfies $w<-1$ due to (\ref{eq:heavy_in_light_forbidden_relation}). The configuration of $T^2(\r_1)$ can then be described by figure~\ref{figure-onshell_lemma4_heavy_in_light_Fq2}.
\begin{figure}[t]
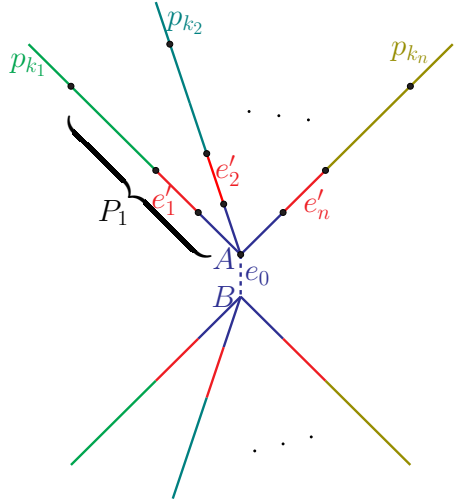

\centering
\include{figs/onshell_lemma4_heavy_in_light_Fq2}
\vspace{-3em}\caption{The potentially possible configuration of $T^2(\r_1)$ with (\ref{eq:heavy_in_light_forbidden_relation}) satisfied.}
\label{figure-onshell_lemma4_heavy_in_light_Fq2}
\end{figure}

To see that such a configuration also leads to contradictions, we would use the same method as in cases~\textbf{III} and ~\textbf{IV} in the proof of lemma~\ref{lemma-onshell_Fp2_internal_less_equal_minusone}. Now recall corollary~\ref{lemma-onshell_Fp2_external_less_equal_minusone_corollary2}, which states that for any two external momenta $p_i^\mu$ and $p_j^\mu$, there exists a path joining them, whose edges all satisfy $w\geqslant-1$. Let us further assume that all the vertices of $P_*$ are in $t(\r_1;A)$. This further implies the existence of an edge $e_*\in P_*$, such that the endpoints of $e_*$ are in $p_i^\mu$ and $p_j^\mu$ respectively (see figure~\ref{onshell_lemma4_heavy_in_light}, where $(i,j)=(1,2)$). With this observation, we modify $T^2(\r_1)$ by adding $e_0,e_*$ and removing $e'_1,e'_2$. The obtained spanning 2-tree corresponds to an $\mathcal{F}^{(q^2)}$ term $\x^{\r_2}$, because the momentum flowing between its components is off shell, more precisely, $(p_1+p_2)^\mu$ (see figure~\ref{onshell_lemma4_heavy_in_light_comparison}).
\begin{figure}[t]
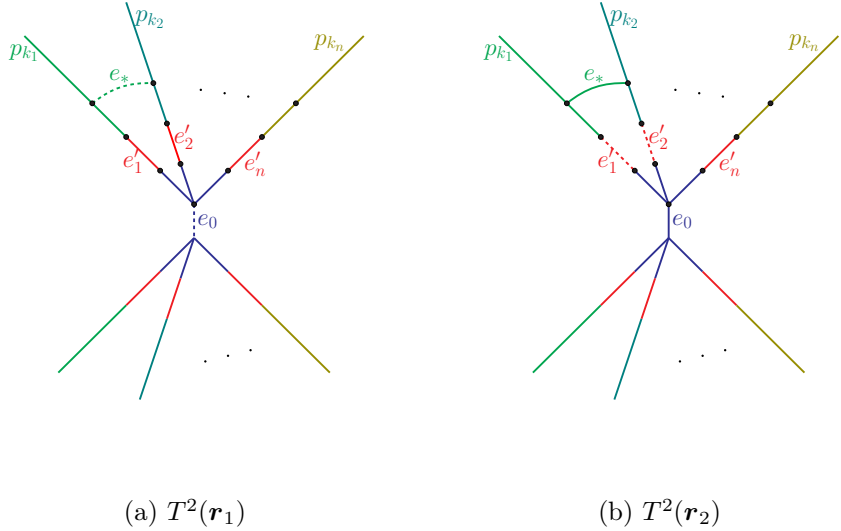

\centering
\begin{subfigure}[b]{0.32\textwidth}
\centering
\include{figs/onshell_lemma4_heavy_in_light}
\caption{$T^2(\r_1)$}
\label{onshell_lemma4_heavy_in_light}
\end{subfigure}
\hspace{3em}
\begin{subfigure}[b]{0.32\textwidth}
\centering
\include{figs/onshell_lemma4_heavy_in_light_comparison}
\caption{$T^2(\r_2)$}
\label{onshell_lemma4_heavy_in_light_comparison}
\end{subfigure}
\caption{The comparison of $T^2(\r_1)$ and $T^2(\r_2)$ when $P_a\cap P_b\subset \gamma$ for any $a$ and $b$. Note that here we have assumed that the path $P_*$, which joins $p_{k_1}(\ref{eq:heavy_in_light_forbidden_relation})$ and $p_{k_2}(\ref{eq:heavy_in_light_forbidden_relation})$ through $e_*$, is contained in $t(\r_1;A)$.}
\label{figure-onshell_lemma4_heavy_in_light}
\end{figure}
The weight of $T^2(\r_2)$ is then
\begin{eqnarray}
w(T^2(\r_2))= w(T^2(\r_1)) +w(e'_1) +w(e'_2) -w(e_0) -w(e_*)< w(T^2(\r_1)),
\label{proof_heavy_in_light_constraint_case3}
\end{eqnarray}
which violates the minimum-weight criterion. In deriving above, we have used the defining properties $w(e_0),w(e_*)\geqslant -1$ and $w(e'_1),w(e'_2)<-1$.

Let us remind ourselves that the proof here is based on the assumption that all the vertices of $P_*$ are within $t(\r_1;A)$, which is similar to cases~\textbf{III} in the proof of lemma~\ref{lemma-onshell_Fp2_internal_less_equal_minusone}. If this assumption does not hold, an analysis similar to cases~\textbf{IV} can then be applied. We shall not repeat the whole analysis here for simplicity.

In conclusion, the relation (\ref{eq:heavy_in_light_forbidden_relation}) leads to contradictions for all the possible configurations, whenever $\gamma$ has at least one edge and is attached by no external momenta. We have thus proved the lemma.
\end{proof}

An example illustrating the proof above will be given later in (\ref{eq:3loop_nonplanar_example_lemma4and6}). As a direct consequence of lemma~\ref{lemma-onshell_heavy_in_light_constraint}, the connectedness of a specific subgraph $\gamma_{J_i}\subseteq G(t(\r;p_i))$ can be shown, as described in the following corollary.

\begin{corollary}
\label{lemma-onshell_heavy_in_light_constraint_corollary1}
For any $\mathcal{F}^{(p_i^2,R)}$ term $\x^{\r}$, we define $\gamma_{J_i}^{}(\r)$ as the subgraph consisting~of:
\begin{enumerate}
    \item [(1)] all the edges $e\in G(t(\r;p_i))$ satisfying $w(e)=-1$, 
    \item [(2)] all the endpoints of these edges,
    \item [(3)] the external momentum $p_i^\mu$,
\end{enumerate}
then $\gamma_{J_i}^{}(\r)$ is connected, and $\gamma_{J_i}^{}(\r)\cap t(\r;p_i)$ is a spanning tree of $\gamma_{J_i}^{}(\r)$.
\end{corollary}

\begin{proof}
To prove the connectedness of $\gamma_{J_i}^{}(\r)$ and $\gamma_{J_i}^{}(\r)\cap t(\r;p_i)$, it suffices to show that the path $P\subset t(\r;p_i)$, which joins any two vertices of $\gamma_{J_i}^{}(\r)$, consists of edges satisfying $w=-1$. Consider the contrary, i.e. there is pair of vertices $A,B\in t(\r;p_i)$, such that the path $P_1 \subset t(\r;p_i)$ joining $A$ and $B$ contains an edge $e_1\in P_1$ with $w(e_1)\neq -1$. We further know from lemma~\ref{lemma-onshell_Fp2_internal_less_equal_minusone} that $w(e_1) <-1$.

In this case, $A$ and $B$ are in distinct connected components of $\gamma_{J_i}^{}(\r) \cap t(\r;p_i)$. One can further deduce that $A$ and $B$ must be in distinct connected components of $\gamma_{J_i}^{}(\r)$ as well. To see this, let us consider the contrary, i.e. $A$ and $B$ are in the same component of $\gamma_{J_i}^{}(\r)$. It then follows that there is a path $P_2$ in this component of $\gamma_{J_i}^{}(\r)$, which joins $A$ and $B$. Its edges, from the definition of $\gamma_{J_i}^{}(\r)$, all satisfy $w=-1$. Then there must be an edge $e_2\in P_2$, whose endpoints are respectively in the distinct components as $\gamma_{J_i}^{}(\r) \cap t(\r;p_i)$. We can then modify $T^2(\r)$ by removing $e_1$ and adding $e_2$ to obtain another spanning 2-tree $T^2(\r')$. The weight of $T^2(\r')$ is $w(T^2(\r))+w(e_1)-w(e_2)$, which is smaller than $w(T^2(\r))$. This violation of the minimum-weight criterion implies that $A$ and $B$ are in distinct connected components of~$\gamma_{J_i}^{}(\r)$.

Without loss of generality, we assume that $A$ is in the component which is not attached by $p_i^\mu$, and we denote it as $\gamma'_{J_i}$. From lemmas~\ref{lemma-onshell_Fp2_external_less_equal_minusone} and \ref{lemma-onshell_Fp2_internal_less_equal_minusone}, all the external edges of $\gamma'_{J_i}$ satisfy $w<-1$. However, this configuration is forbidden by lemma~\ref{lemma-onshell_heavy_in_light_constraint}.

Therefore, such a pair of vertices $A$ and $B$, as described at the beginning of the proof, cannot exist. We have thus proved that $\gamma_{J_i}^{}$ and $\gamma_{J_i}^{}(\r)\cap t(\r;p_i)$ must be connected. Since $t(\r;p_i)$ is a tree graph, it follows that $\gamma_{J_i}^{}(\r)\cap t(\r;p_i)$ is a spanning tree of $\gamma_{J_i}^{}(\r)$.
\end{proof}

This corollary has an essential implication. Recall that the third statement of lemma~\ref{lemma-existence_certain_terms_P} states that for any region $R$ that is not the hard region, the polynomial $\mathcal{P}^{(R)}(\x;\s)$ must include some $\mathcal{F}^{(p^2)}$ terms. For a given $\mathcal{F}^{(p_i^2)}$ term then, corollary~\ref{lemma-onshell_heavy_in_light_constraint_corollary1} states the existence of a connected subgraph of $G$ to which the external momentum $p_i^\mu$ is attached, and that the weight of every edge of this subgraph is exactly $-1$. As we will see later, this subgraph is contained in the $i$th jet subgraph $J_i$. In other words, for any region $R\neq R_h$, there should be some nontrivial jet subgraphs $J_i\subset G$.

To establish a precise definition of the jet subgraphs, our attention turns to certain special terms within $\mathcal{P}^{(R)}(\x;\s)$. These terms will be referred to as the \emph{canonical leading terms}.

\subsubsection{Canonical leading terms}
\label{section-canonical_leading_terms}

Within the various $\mathcal{U}^{(R)}$ and $\mathcal{F}^{(p_i^2,R)}$ terms for a given region $R$, we now delve into the concept of \emph{canonical $\mathcal{U}^{(R)}$ and $\mathcal{F}^{(p_i^2,R)}$ terms}, which are defined as follows.
\begin{itemize}
    \item any $\mathcal{F}^{(p_i^2,R)}$ term $\x^{\r_*}$ is a \emph{canonical $\mathcal{F}^{(p_i^2,R)}$ term}, if and only if, for any $e\in G$ connecting the two components of $T^2(\r_*)$ and satisfying $w(e)=-1$, the endpoint of $e$, situated within $t(\r_*;\widehat{p}_i)$, either belongs to the $q$ trunk of the spanning tree $T^1\equiv T^2(\r_*)\cup e$, or is incident with another edge with $w>-1$;
    
    \item any $\mathcal{U}^{(R)}$ term $\x^{\r_*}$ is called a \emph{canonical $\mathcal{U}^{(R)}$ term}, if and only if, for each $i=1,\dots,K$, there exists an edge $e_i\subset T^1(\r_*)$ such that the spanning 2-tree $T^1(\r_*) \setminus e_i$ corresponds to a canonical $\mathcal{F}^{(p_i^2,R)}$ term.
\end{itemize}
These concepts, as we will see later, help us partition $G$ into the corresponding hard, jet and soft subgraphs according to the region $R$. To illustrate, let us see some examples of canonical and non-canonical $\mathcal{F}^{(p_3^2,R)}$ terms. Among the following three spanning 2-trees, taken from the list of $\mathcal{F}^{(p_3^2,R)}$ terms in eq.~(\ref{eq:onshell_region_good_example_Fp2term}),
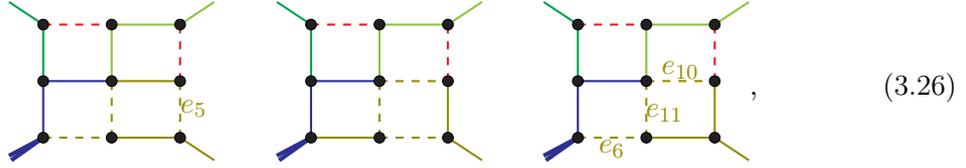
\begin{equation}
\label{two_noncanonical_one_canonical_Fp2R}
\begin{tikzpicture}[baseline=8ex, scale=0.3]

\draw [dashed, thick,color=Red] (2,7.5) -- (5,7.5);
\draw [thick,color=LimeGreen] (8,7.5) -- (5,7.5);
\draw [dashed, thick,color=Red] (8,5) -- (8,7.5);
\draw [dashed, thick,color=olive] (8,2.5) -- (8,5);
\draw [thick,color=olive] (5,2.5) -- (8,2.5);
\draw [dashed, thick,color=olive] (2,2.5) -- (5,2.5);
\draw [thick,color=Blue] (2,5) -- (2,2.5);
\draw [thick,color=Green] (2,5) -- (2,7.5);
\draw [thick,color=LimeGreen] (5,7.5) -- (5,5);
\draw [thick,color=olive] (5,5) -- (8,5);
\draw [dashed, thick,color=olive] (5,2.5) -- (5,5);
\draw [thick,color=Blue] (5,5) -- (2,5);

\draw [thick,color=Green] (0.5,8.5) -- (2,7.5);
\draw [thick,color=LimeGreen] (9.5,8.5) -- (8,7.5);
\draw [thick,color=olive] (9.5,1.5) -- (8,2.5);
\draw [ultra thick, color=Blue] (0.5,1.7) -- (2,2.5);
\draw [ultra thick, color=Blue] (0.6,1.5) -- (2,2.5);

\node [draw,circle,minimum size=4pt,fill=Black,inner sep=0pt,outer sep=0pt] () at (2,7.5) {};
\node [draw,circle,minimum size=4pt,fill=Black,inner sep=0pt,outer sep=0pt] () at (2,5) {};
\node [draw,circle,minimum size=4pt,fill=Black,inner sep=0pt,outer sep=0pt] () at (2,2.5) {};
\node [draw,circle,minimum size=4pt,fill=Black,inner sep=0pt,outer sep=0pt] () at (5,7.5) {};
\node [draw,circle,minimum size=4pt,fill=Black,inner sep=0pt,outer sep=0pt] () at (5,5) {};
\node [draw,circle,minimum size=4pt,fill=Black,inner sep=0pt,outer sep=0pt] () at (5,2.5) {};
\node [draw,circle,minimum size=4pt,fill=Black,inner sep=0pt,outer sep=0pt] () at (8,7.5) {};
\node [draw,circle,minimum size=4pt,fill=Black,inner sep=0pt,outer sep=0pt] () at (8,5) {};
\node [draw,circle,minimum size=4pt,fill=Black,inner sep=0pt,outer sep=0pt] () at (8,2.5) {};

\node () at (8.6,3.8) {\color{olive} $e_5$};
\end{tikzpicture}
\qquad 
\begin{tikzpicture}[baseline=8ex, scale=0.3]

\draw [dashed, thick,color=Red] (2,7.5) -- (5,7.5);
\draw [thick,color=LimeGreen] (8,7.5) -- (5,7.5);
\draw [dashed, thick,color=Red] (8,5) -- (8,7.5);
\draw [thick,color=olive] (8,2.5) -- (8,5);
\draw [dashed, thick,color=olive] (5,2.5) -- (8,2.5);
\draw [thick,color=olive] (2,2.5) -- (5,2.5);
\draw [thick,color=Blue] (2,5) -- (2,2.5);
\draw [thick,color=Green] (2,5) -- (2,7.5);
\draw [thick,color=LimeGreen] (5,7.5) -- (5,5);
\draw [dashed, thick,color=olive] (5,5) -- (8,5);
\draw [dashed, thick,color=olive] (5,2.5) -- (5,5);
\draw [thick,color=Blue] (5,5) -- (2,5);

\draw [thick,color=Green] (0.5,8.5) -- (2,7.5);
\draw [thick,color=LimeGreen] (9.5,8.5) -- (8,7.5);
\draw [thick,color=olive] (9.5,1.5) -- (8,2.5);
\draw [ultra thick, color=Blue] (0.5,1.7) -- (2,2.5);
\draw [ultra thick, color=Blue] (0.6,1.5) -- (2,2.5);

\node [draw,circle,minimum size=4pt,fill=Black,inner sep=0pt,outer sep=0pt] () at (2,7.5) {};
\node [draw,circle,minimum size=4pt,fill=Black,inner sep=0pt,outer sep=0pt] () at (2,5) {};
\node [draw,circle,minimum size=4pt,fill=Black,inner sep=0pt,outer sep=0pt] () at (2,2.5) {};
\node [draw,circle,minimum size=4pt,fill=Black,inner sep=0pt,outer sep=0pt] () at (5,7.5) {};
\node [draw,circle,minimum size=4pt,fill=Black,inner sep=0pt,outer sep=0pt] () at (5,5) {};
\node [draw,circle,minimum size=4pt,fill=Black,inner sep=0pt,outer sep=0pt] () at (5,2.5) {};
\node [draw,circle,minimum size=4pt,fill=Black,inner sep=0pt,outer sep=0pt] () at (8,7.5) {};
\node [draw,circle,minimum size=4pt,fill=Black,inner sep=0pt,outer sep=0pt] () at (8,5) {};
\node [draw,circle,minimum size=4pt,fill=Black,inner sep=0pt,outer sep=0pt] () at (8,2.5) {};
\end{tikzpicture}
\qquad 
\begin{tikzpicture}[baseline=8ex, scale=0.3]

\draw [dashed, thick,color=Red] (2,7.5) -- (5,7.5);
\draw [thick,color=LimeGreen] (8,7.5) -- (5,7.5);
\draw [dashed, thick,color=Red] (8,5) -- (8,7.5);
\draw [thick,color=olive] (8,2.5) -- (8,5);
\draw [thick,color=olive] (5,2.5) -- (8,2.5);
\draw [dashed, thick,color=olive] (2,2.5) -- (5,2.5);
\draw [thick,color=Blue] (2,5) -- (2,2.5);
\draw [thick,color=Green] (2,5) -- (2,7.5);
\draw [thick,color=LimeGreen] (5,7.5) -- (5,5);
\draw [dashed, thick,color=olive] (5,5) -- (8,5);
\draw [dashed, thick,color=olive] (5,2.5) -- (5,5);
\draw [thick,color=Blue] (5,5) -- (2,5);

\draw [thick,color=Green] (0.5,8.5) -- (2,7.5);
\draw [thick,color=LimeGreen] (9.5,8.5) -- (8,7.5);
\draw [thick,color=olive] (9.5,1.5) -- (8,2.5);
\draw [ultra thick, color=Blue] (0.5,1.7) -- (2,2.5);
\draw [ultra thick, color=Blue] (0.6,1.5) -- (2,2.5);

\node [draw,circle,minimum size=4pt,fill=Black,inner sep=0pt,outer sep=0pt] () at (2,7.5) {};
\node [draw,circle,minimum size=4pt,fill=Black,inner sep=0pt,outer sep=0pt] () at (2,5) {};
\node [draw,circle,minimum size=4pt,fill=Black,inner sep=0pt,outer sep=0pt] () at (2,2.5) {};
\node [draw,circle,minimum size=4pt,fill=Black,inner sep=0pt,outer sep=0pt] () at (5,7.5) {};
\node [draw,circle,minimum size=4pt,fill=Black,inner sep=0pt,outer sep=0pt] () at (5,5) {};
\node [draw,circle,minimum size=4pt,fill=Black,inner sep=0pt,outer sep=0pt] () at (5,2.5) {};
\node [draw,circle,minimum size=4pt,fill=Black,inner sep=0pt,outer sep=0pt] () at (8,7.5) {};
\node [draw,circle,minimum size=4pt,fill=Black,inner sep=0pt,outer sep=0pt] () at (8,5) {};
\node [draw,circle,minimum size=4pt,fill=Black,inner sep=0pt,outer sep=0pt] () at (8,2.5) {};

\node () at (3.5,2) {\color{olive} $e_6$};
\node () at (5.8,3.6) {\color{olive} $e_{11}$};
\node () at (6.5,5.5) {\color{olive} $e_{10}$};

\end{tikzpicture},
\end{equation}
only the third graph corresponds to a canonical $\mathcal{F}^{(p_3^2,R)}$ term, because those edges connecting the two components of the spanning 2-tree $T^2$ and satisfying $w=-1$ are exactly the dashed olive edges $e_6,e_{10},e_{11}$ (the notation here is in accordance with figure~\ref{figure-fishnet_2times2_onshell_3onshell_1offshell}). One can check that each of these edges $e_i$ ($i\in\{6,10,11\}$) is incident with a vertex in the $q$ trunk of $T^2\cup e_i$. The first and second spanning 2-trees, in comparison, do not have this property. For example, the edge $e_4$ labelled in the first graph connects the two components of the spanning 2-tree, but neither of its endpoints is in the $q$ trunk of $T^2\cup e_4$, and neither of them is incident with an edge satisfying $w>-1$.

Essentially, starting from any $\mathcal{F}^{(p_i^2,R)}$ term $\x^{\r_0}$, one can obtain a canonical $\mathcal{F}^{(p_i^2,R)}$ term $\x^{\r_*}$ through the following iterative process.
\begin{itemize}
    \item \emph{Step $0$}: If $\x^{\r_0}$ is already a canonical $\mathcal{F}^{(p_i^2,R)}$ term, the construction is completed. Otherwise we go to Step 1, and note that by definition, there is an edge $e_1$ connecting the two components of $T^2(\r_0)$ and satisfying $w(e_1)=-1$, such that both its endpoints are in the $p_i$ trunk of $T^1(\r'_0)\equiv T^2(\r_0)\cup e_1$, and incident with edges satisfying $w\leqslant -1$ only.
    
    \item \emph{Step $k$} ($k=1,2,\dots$): We then consider the spanning tree $T^1(\r'_{k-1})\equiv T^2(\r_{k-1})\cup e_k$, and focus on a special set of edges $\mathcal{E}_k$, which consists of all the edges in its $p_i$ trunk with $w=-1$. Among the elements of $\mathcal{E}_k$, we further denote one particular edge, which has the largest distance from the external momentum $p_i^\mu$, by $e'_k$. The spanning 2-tree $T^2(\r_k)\equiv T^1(\r'_{k-1})\setminus e'_k$ then corresponds to another $\mathcal{F}^{(p_i^2,R)}$ term~$\x^{\r_k}$. Note that $t(\r_{k-1};p_i)\subset t(\r_k;p_i)$.

    If $\x^{\r_k}$ is already a canonical $\mathcal{F}^{(p_i^2,R)}$ term, the construction is completed, otherwise there is an edge $e_{k+1}$ connecting the two components of $T^2(\r_k)$ and satisfying $w(e_{k+1})=-1$, such that both its endpoints are in the $p_i$ trunk of $T^1(\r'_k)\equiv T^2(\r_k)\cup e_{k+1}$, and incident with edges satisfying $w\leqslant -1$ only. We then go to the next step (\emph{Step $k+1$}).
\end{itemize}

This recursive procedure guarantees that a canonical $\mathcal{F}^{(p_i^2,R)}$ term $\x^{\r_*}$ would be obtained after a finite number of steps. Let us revisit the second graph in (\ref{two_noncanonical_one_canonical_Fp2R}) as an example:
\begin{eqnarray}
\label{deriving_canonical_Fp2Rterm_example}
&&\begin{tikzpicture}[baseline=8ex, scale=0.3]

\draw [dashed, thick,color=Red] (2,7.5) -- (5,7.5);
\draw [thick,color=LimeGreen] (8,7.5) -- (5,7.5);
\draw [dashed, thick,color=Red] (8,5) -- (8,7.5);
\draw [thick,color=olive] (8,2.5) -- (8,5);
\draw [dashed, thick,color=olive] (5,2.5) -- (8,2.5);
\draw [thick,color=olive] (2,2.5) -- (5,2.5);
\draw [thick,color=Blue] (2,5) -- (2,2.5);
\draw [thick,color=Green] (2,5) -- (2,7.5);
\draw [thick,color=LimeGreen] (5,7.5) -- (5,5);
\draw [dashed, thick,color=olive] (5,5) -- (8,5);
\draw [dashed, thick,color=olive] (5,2.5) -- (5,5);
\draw [thick,color=Blue] (5,5) -- (2,5);

\draw [thick,color=Green] (0.5,8.5) -- (2,7.5);
\draw [thick,color=LimeGreen] (9.5,8.5) -- (8,7.5);
\draw [thick,color=olive] (9.5,1.5) -- (8,2.5);
\draw [ultra thick, color=Blue] (0.5,1.7) -- (2,2.5);
\draw [ultra thick, color=Blue] (0.6,1.5) -- (2,2.5);

\node [draw,circle,minimum size=4pt,fill=Black,inner sep=0pt,outer sep=0pt] () at (2,7.5) {};
\node [draw,circle,minimum size=4pt,fill=Black,inner sep=0pt,outer sep=0pt] () at (2,5) {};
\node [draw,circle,minimum size=4pt,fill=Black,inner sep=0pt,outer sep=0pt] () at (2,2.5) {};
\node [draw,circle,minimum size=4pt,fill=Black,inner sep=0pt,outer sep=0pt] () at (5,7.5) {};
\node [draw,circle,minimum size=4pt,fill=Black,inner sep=0pt,outer sep=0pt] () at (5,5) {};
\node [draw,circle,minimum size=4pt,fill=Black,inner sep=0pt,outer sep=0pt] () at (5,2.5) {};
\node [draw,circle,minimum size=4pt,fill=Black,inner sep=0pt,outer sep=0pt] () at (8,7.5) {};
\node [draw,circle,minimum size=4pt,fill=Black,inner sep=0pt,outer sep=0pt] () at (8,5) {};
\node [draw,circle,minimum size=4pt,fill=Black,inner sep=0pt,outer sep=0pt] () at (8,2.5) {};
\end{tikzpicture}
\quad \rightarrow \quad
\begin{tikzpicture}[baseline=8ex, scale=0.3]

\draw [dashed, thick,color=Red] (2,7.5) -- (5,7.5);
\draw [thick,color=LimeGreen] (8,7.5) -- (5,7.5);
\draw [dashed, thick,color=Red] (8,5) -- (8,7.5);
\draw [thick,color=olive] (8,2.5) -- (8,5);
\draw [thick,color=olive] (5,2.5) -- (8,2.5);
\draw [thick,color=olive] (2,2.5) -- (5,2.5);
\draw [thick,color=Blue] (2,5) -- (2,2.5);
\draw [thick,color=Green] (2,5) -- (2,7.5);
\draw [thick,color=LimeGreen] (5,7.5) -- (5,5);
\draw [dashed, thick,color=olive] (5,5) -- (8,5);
\draw [dashed, thick,color=olive] (5,2.5) -- (5,5);
\draw [thick,color=Blue] (5,5) -- (2,5);

\draw [thick,color=Green] (0.5,8.5) -- (2,7.5);
\draw [thick,color=LimeGreen] (9.5,8.5) -- (8,7.5);
\draw [thick,color=olive] (9.5,1.5) -- (8,2.5);
\draw [ultra thick, color=Blue] (0.5,1.7) -- (2,2.5);
\draw [ultra thick, color=Blue] (0.6,1.5) -- (2,2.5);

\node [draw,circle,minimum size=4pt,fill=Black,inner sep=0pt,outer sep=0pt] () at (2,7.5) {};
\node [draw,circle,minimum size=4pt,fill=Black,inner sep=0pt,outer sep=0pt] () at (2,5) {};
\node [draw,circle,minimum size=4pt,fill=Black,inner sep=0pt,outer sep=0pt] () at (2,2.5) {};
\node [draw,circle,minimum size=4pt,fill=Black,inner sep=0pt,outer sep=0pt] () at (5,7.5) {};
\node [draw,circle,minimum size=4pt,fill=Black,inner sep=0pt,outer sep=0pt] () at (5,5) {};
\node [draw,circle,minimum size=4pt,fill=Black,inner sep=0pt,outer sep=0pt] () at (5,2.5) {};
\node [draw,circle,minimum size=4pt,fill=Black,inner sep=0pt,outer sep=0pt] () at (8,7.5) {};
\node [draw,circle,minimum size=4pt,fill=Black,inner sep=0pt,outer sep=0pt] () at (8,5) {};
\node [draw,circle,minimum size=4pt,fill=Black,inner sep=0pt,outer sep=0pt] () at (8,2.5) {};
\end{tikzpicture}
\quad \rightarrow \quad
\begin{tikzpicture}[baseline=8ex, scale=0.3]

\draw [dashed, thick,color=Red] (2,7.5) -- (5,7.5);
\draw [thick,color=LimeGreen] (8,7.5) -- (5,7.5);
\draw [dashed, thick,color=Red] (8,5) -- (8,7.5);
\draw [thick,color=olive] (8,2.5) -- (8,5);
\draw [thick,color=olive] (5,2.5) -- (8,2.5);
\draw [dashed, thick,color=olive] (2,2.5) -- (5,2.5);
\draw [thick,color=Blue] (2,5) -- (2,2.5);
\draw [thick,color=Green] (2,5) -- (2,7.5);
\draw [thick,color=LimeGreen] (5,7.5) -- (5,5);
\draw [dashed, thick,color=olive] (5,5) -- (8,5);
\draw [dashed, thick,color=olive] (5,2.5) -- (5,5);
\draw [thick,color=Blue] (5,5) -- (2,5);

\draw [thick,color=Green] (0.5,8.5) -- (2,7.5);
\draw [thick,color=LimeGreen] (9.5,8.5) -- (8,7.5);
\draw [thick,color=olive] (9.5,1.5) -- (8,2.5);
\draw [ultra thick, color=Blue] (0.5,1.7) -- (2,2.5);
\draw [ultra thick, color=Blue] (0.6,1.5) -- (2,2.5);

\node [draw,circle,minimum size=4pt,fill=Black,inner sep=0pt,outer sep=0pt] () at (2,7.5) {};
\node [draw,circle,minimum size=4pt,fill=Black,inner sep=0pt,outer sep=0pt] () at (2,5) {};
\node [draw,circle,minimum size=4pt,fill=Black,inner sep=0pt,outer sep=0pt] () at (2,2.5) {};
\node [draw,circle,minimum size=4pt,fill=Black,inner sep=0pt,outer sep=0pt] () at (5,7.5) {};
\node [draw,circle,minimum size=4pt,fill=Black,inner sep=0pt,outer sep=0pt] () at (5,5) {};
\node [draw,circle,minimum size=4pt,fill=Black,inner sep=0pt,outer sep=0pt] () at (5,2.5) {};
\node [draw,circle,minimum size=4pt,fill=Black,inner sep=0pt,outer sep=0pt] () at (8,7.5) {};
\node [draw,circle,minimum size=4pt,fill=Black,inner sep=0pt,outer sep=0pt] () at (8,5) {};
\node [draw,circle,minimum size=4pt,fill=Black,inner sep=0pt,outer sep=0pt] () at (8,2.5) {};
\end{tikzpicture}.\\
&&\qquad T^2(\r_0) \qquad\qquad\qquad\qquad T^1(\r'_0) \qquad\qquad\qquad\qquad T^2(\r_1)\nonumber \\
&&\ \ \underset{\text{Step 0}}{\underbrace{\qquad\qquad\qquad}}\qquad\qquad\qquad \underset{\text{Step 1}}{\underbrace{\qquad\qquad\qquad\qquad\qquad\qquad\qquad}} \nonumber
\end{eqnarray}
It is clear from above that only Step 0 and Step 1 are needed, which end with a canonical $\mathcal{F}^{(p_3^2,R)}$ term $\x^{\r_1}$.

Based on the derivation of a canonical $\mathcal{F}^{(p_i^2,R)}$ term as explained above, a canonical $\mathcal{U}^{(R)}$ term $\x^{\r_*}$ can be obtained from any $\mathcal{U}^{(R)}$ term $\x^{\r_0}$ through the following operations. For each given $i\in \{1,\dots,K\}$, if the $p_i$ trunk only contains the external momentum $p_i$, no further operation would be needed. Otherwise, we denote the set of edges in the $p_i$ trunk with $w=-1$ as $\mathcal{E}_0$, and denote the specific edge $e'_0\in \mathcal{E}_0$ which is the farthest from the external momentum $p_i^\mu$. The spanning 2-tree $T^2(\r'_0)\equiv T^1(\r_0)\setminus e'_0$ then corresponds to an $\mathcal{F}^{(p_i^2,R)}$ term $\x^{\r'_0}$.
Then one can apply the procedure of obtaining canonical $\mathcal{F}^{(p_i^2,R)}$ terms above, which ends with a canonical $\mathcal{F}^{(p_i^2,R)}$ term $\x^{\r'_*}$. From lemma~\ref{lemma-onshell_Fp2_external_less_equal_minusone}, there is an edge $e'_*$ connecting the components of $T^2(\r'_*)$ with $w(e'_*)=-1$. The spanning tree $T^1(\r_*)\equiv T^2(\r'_*)\cup e'_*$ then satisfies the following property: it corresponds to a $\mathcal{U}^{(R)}$ term, and $T^1(\r_*)\setminus e'_*$ is a canonical $\mathcal{F}^{(p_i^2,R)}$ term. After doing the same for each $i\in \{1,\dots,K\}$, a canonical $\mathcal{U}^{(R)}$ term is then obtained.

As a consequence, there exist one or more canonical $\mathcal{F}^{(p_i^2,R)}$ terms unless $\mathcal{F}^{(p_i^2,R)}=0$, and there exist one or more canonical $\mathcal{U}^{(R)}$ terms unless $\mathcal{F}^{(p_i^2,R)}=0$ for all the $i=1,\dots,K$.

\bigbreak
Note that for a given $i$, the canonical $\mathcal{F}^{(p_i^2,R)}$ terms are not uniquely defined. For example, the vector $\boldsymbol{v}_R = (-2,-2,-1,-1,0,0,-1,-1,-2,-2,-2,-2;1)$ for the $2\times 2$ fishnet graph describes the region below.
\begin{equation}
R:
\begin{tikzpicture}[baseline=8ex, scale=0.3]
\draw [thick,color=Red] (2,7.5) -- (5,7.5);
\draw [thick,color=Red] (5,7.5) -- (5,5);
\draw [thick,color=Green] (5,5) -- (2,5);
\draw [thick,color=Green] (2,5) -- (2,7.5);
\draw [thick,color=Red] (8,7.5) -- (5,7.5);
\draw [thick,color=LimeGreen] (8,5) -- (8,7.5);
\draw [thick,color=Red] (5,5) -- (8,5);
\draw [thick,color=Green] (2,5) -- (2,2.5);
\draw [thick,color=Blue] (2,2.5) -- (5,2.5);
\draw [thick,color=Green] (5,2.5) -- (5,5);
\draw [thick,color=Blue] (5,2.5) -- (8,2.5);
\draw [thick,color=LimeGreen] (8,2.5) -- (8,5);

\draw [thick,color=Green] (0.5,8.5) -- (2,7.5);
\draw [thick,color=LimeGreen] (9.5,8.5) -- (8,7.5);
\draw [thick,color=olive] (9.5,1.5) -- (8,2.5);
\draw [ultra thick, color=Blue] (0.5,1.7) -- (2,2.5);
\draw [ultra thick, color=Blue] (0.6,1.5) -- (2,2.5);

\node [draw,circle,minimum size=4pt,fill=Black,inner sep=0pt,outer sep=0pt] () at (2,7.5) {};
\node [draw,circle,minimum size=4pt,fill=Black,inner sep=0pt,outer sep=0pt] () at (2,5) {};
\node [draw,circle,minimum size=4pt,fill=Black,inner sep=0pt,outer sep=0pt] () at (2,2.5) {};
\node [draw,circle,minimum size=4pt,fill=Black,inner sep=0pt,outer sep=0pt] () at (5,7.5) {};
\node [draw,circle,minimum size=4pt,fill=Black,inner sep=0pt,outer sep=0pt] () at (5,5) {};
\node [draw,circle,minimum size=4pt,fill=Black,inner sep=0pt,outer sep=0pt] () at (5,2.5) {};
\node [draw,circle,minimum size=4pt,fill=Black,inner sep=0pt,outer sep=0pt] () at (8,7.5) {};
\node [draw,circle,minimum size=4pt,fill=Black,inner sep=0pt,outer sep=0pt] () at (8,5) {};
\node [draw,circle,minimum size=4pt,fill=Black,inner sep=0pt,outer sep=0pt] () at (8,2.5) {};
\end{tikzpicture},
\label{eq:fishnet_2times2_onshell_region_example_canonical}
\end{equation}
One can check that the two terms, $\x^{\r_1}=x_2 x_7 x_9 x_{10} x_{11}$ and $\x^{\r_2}=x_2 x_7 x_9 x_{11} x_{12}$ are both canonical $\mathcal{F}^{(p_1^2,R)}$ terms, whose spanning 2-trees are:
\begin{equation}
T^2(\r_1)=
\begin{tikzpicture}[baseline=8ex, scale=0.3]
\draw [thick,color=Red] (2,7.5) -- (5,7.5);
\draw [dashed, thick,color=Red] (5,7.5) -- (5,5);
\draw [thick,color=Green] (5,5) -- (2,5);
\draw [thick,color=Green] (2,5) -- (2,7.5);
\draw [dashed, thick,color=Red] (8,7.5) -- (5,7.5);
\draw [thick,color=LimeGreen] (8,5) -- (8,7.5);
\draw [dashed, thick,color=Red] (5,5) -- (8,5);
\draw [dashed, thick,color=Green] (2,5) -- (2,2.5);
\draw [thick,color=Blue] (2,2.5) -- (5,2.5);
\draw [dashed, thick,color=Green] (5,2.5) -- (5,5);
\draw [thick,color=Blue] (5,2.5) -- (8,2.5);
\draw [thick,color=LimeGreen] (8,2.5) -- (8,5);

\draw [thick,color=Green] (0.5,8.5) -- (2,7.5);
\draw [thick,color=LimeGreen] (9.5,8.5) -- (8,7.5);
\draw [thick,color=olive] (9.5,1.5) -- (8,2.5);
\draw [ultra thick, color=Blue] (0.5,1.7) -- (2,2.5);
\draw [ultra thick, color=Blue] (0.6,1.5) -- (2,2.5);

\node [draw, circle, minimum size=4pt, color=Green, fill=Green, inner sep=0pt, outer sep=0pt] () at (2,7.5) {};
\node [draw, circle, minimum size=4pt, color=Green, fill=Green, inner sep=0pt, outer sep=0pt] () at (2,5) {};
\node [draw,circle,minimum size=4pt,fill=Black,inner sep=0pt,outer sep=0pt] () at (2,2.5) {};
\node [draw,circle,minimum size=4pt,fill=Black,inner sep=0pt,outer sep=0pt] () at (5,7.5) {};
\node [draw, circle, minimum size=4pt, color=Green, fill=Green, inner sep=0pt, outer sep=0pt] () at (5,5) {};
\node [draw,circle,minimum size=4pt,fill=Black,inner sep=0pt,outer sep=0pt] () at (5,2.5) {};
\node [draw,circle,minimum size=4pt,fill=Black,inner sep=0pt,outer sep=0pt] () at (8,7.5) {};
\node [draw,circle,minimum size=4pt,fill=Black,inner sep=0pt,outer sep=0pt] () at (8,5) {};
\node [draw,circle,minimum size=4pt,fill=Black,inner sep=0pt,outer sep=0pt] () at (8,2.5) {};
\end{tikzpicture},
\qquad 
T^2(\r_2)=
\begin{tikzpicture}[baseline=8ex, scale=0.3]
\draw [dashed, thick,color=Red] (2,7.5) -- (5,7.5);
\draw [dashed, thick,color=Red] (5,7.5) -- (5,5);
\draw [thick,color=Green] (5,5) -- (2,5);
\draw [thick,color=Green] (2,5) -- (2,7.5);
\draw [thick,color=Red] (8,7.5) -- (5,7.5);
\draw [thick,color=LimeGreen] (8,5) -- (8,7.5);
\draw [dashed, thick,color=Red] (5,5) -- (8,5);
\draw [dashed, thick,color=Green] (2,5) -- (2,2.5);
\draw [thick,color=Blue] (2,2.5) -- (5,2.5);
\draw [dashed, thick,color=Green] (5,2.5) -- (5,5);
\draw [thick,color=Blue] (5,2.5) -- (8,2.5);
\draw [thick,color=LimeGreen] (8,2.5) -- (8,5);

\draw [thick,color=Green] (0.5,8.5) -- (2,7.5);
\draw [thick,color=LimeGreen] (9.5,8.5) -- (8,7.5);
\draw [thick,color=olive] (9.5,1.5) -- (8,2.5);
\draw [ultra thick, color=Blue] (0.5,1.7) -- (2,2.5);
\draw [ultra thick, color=Blue] (0.6,1.5) -- (2,2.5);

\node [draw, circle, minimum size=4pt, color=Green, fill=Green, inner sep=0pt, outer sep=0pt] () at (2,7.5) {};
\node [draw, circle, minimum size=4pt, color=Green, fill=Green, inner sep=0pt, outer sep=0pt] () at (2,5) {};
\node [draw,circle,minimum size=4pt,fill=Black,inner sep=0pt,outer sep=0pt] () at (2,2.5) {};
\node [draw,circle,minimum size=4pt,fill=Black,inner sep=0pt,outer sep=0pt] () at (5,7.5) {};
\node [draw, circle, minimum size=4pt, color=Green, fill=Green, inner sep=0pt, outer sep=0pt] () at (5,5) {};
\node [draw,circle,minimum size=4pt,fill=Black,inner sep=0pt,outer sep=0pt] () at (5,2.5) {};
\node [draw,circle,minimum size=4pt,fill=Black,inner sep=0pt,outer sep=0pt] () at (8,7.5) {};
\node [draw,circle,minimum size=4pt,fill=Black,inner sep=0pt,outer sep=0pt] () at (8,5) {};
\node [draw,circle,minimum size=4pt,fill=Black,inner sep=0pt,outer sep=0pt] () at (8,2.5) {};
\end{tikzpicture}.
\label{eq:canonical_Fp2_terms_examples}
\end{equation}

The example above shows the non-uniqueness of canonical $\mathcal{F}^{(p_i^2,R)}$ terms. Nevertheless, any two distinct canonical $\mathcal{F}^{(p_1^2,R)}$ terms share certain properties, as we shall explore below. To this end, we first define $\mathcal{V}_F(\r)$ as a certain set of vertices for each $\mathcal{F}^{(p_i^2,R)}$ term $\x^{\r}$: $\mathcal{V}_F(\r)$ consists of any vertex $v\in t(\r;p_i)$, such that $v$ is incident with an edge $e\in t(\r;p_i)$ which satisfies $w(e)=-1$.

It is straightforward to check that, in eq.~(\ref{eq:canonical_Fp2_terms_examples}) the sets $\mathcal{V}_F(\r_1)$ and $\mathcal{V}_F(\r_2)$ are identical, including exactly the green vertices. This motivates us to conjecture that $\mathcal{V}_F(\r_1) = \mathcal{V}_F(\r_2)$ holds for any two canonical $\mathcal{F}^{(p_i^2,R)}$ terms $\x^{\r_1}$ and $\x^{\r_2}$, as described by the following lemma.

\begin{lemma}
For each given $\mathcal{F}^{(p_i^2,R)}$ term $\x^{\r}$, we define
\begin{eqnarray}
    \mathcal{V}_F(\r)\equiv \{\ v\in t(\r;p_i)\ |\ \exists\ e\text{ such that }v\ \text{is incident with }e\text{, and }w(e)=-1\ \},\nonumber
\end{eqnarray}
then $\mathcal{V}_F({\r}_1) = \mathcal{V}_F({\r}_2)$ for any two canonical $\mathcal{F}^{(p_i^2,R)}$ terms $\x^{\r_1}$ and $\x^{\r_2}$.
\label{lemma-onshell_Fp2_vertices_universal_property}
\end{lemma}

The key of proving this lemma is to relate $\mathcal{V}_F(\r)$ to a set of vertices which does not rely on the choice of $\x^{\r}$. To this end we shall define the sets $\widehat{\mathcal{V}}_0,\widehat{\mathcal{V}}_1,\dots,\widehat{\mathcal{V}}_K$. First, $\widehat{\mathcal{V}}_0$ is defined to include the vertices $v$ satisfying either \emph{(1) there is a $\mathcal{U}^{(R)}$ term $\x^{\r}$ such that $v$ is in the $q$ trunk of $T^1(\r)$, or (2) $v$ is incident with an edge $e\in G$ such that $w(e)>-1$}. With this definition, a key observation is that any path $P\subset G\setminus \widehat{\mathcal{V}}_0$ that joins two on-shell external momentum $p_i^\mu$ and $p_j^\mu$ must contain some edges satisfying $w<-1$. This statement is shown in appendix~\ref{appendix-GminusV0_disconnected}, based on which each $\widehat{\mathcal{V}}_i$ ($i\in\{1,\dots,K\}$) is defined to include the vertices~$v$ satisfying: \emph{there exists a path $P\subset G\setminus \widehat{\mathcal{V}}_0$ connecting $v$ and the external momentum~$p_i^\mu$, such that $w(e)=-1$ for any $e\in P$}.

With these concepts, the definition of a canonical $\mathcal{F}^{(p_i^2,R)}$ term $\x^{\r_*}$ (given at the beginning of section~\ref{section-canonical_leading_terms}) can be simplified as: for any $e\in G$ connecting the two components of $T^2(\r_*)$ and satisfying $w(e)=-1$, the vertex of $e$ which is in $t(\r_*;\widehat{p}_i)$ belongs to $\widehat{\mathcal{V}}_0$. We also point out that given the graph $G$ and the region $R$, the sets $\widehat{\mathcal{V}}_0,\widehat{\mathcal{V}}_1,\dots,\widehat{\mathcal{V}}_K$ are uniquely defined, independent of the choice of leading terms.
Below we prove lemma~\ref{lemma-onshell_Fp2_vertices_universal_property}, to do which it suffices to show that $\mathcal{V}_F(\r)=\widehat{\mathcal{V}}_i$ for each given canonical $\mathcal{F}^{(p_i^2,R)}$ term~$\x^{\r}$.

\begin{proof} [Proof of $\mathcal{V}_F(\r) \supseteq \widehat{\mathcal{V}}_i$]
Let us consider a canonical $\mathcal{F}^{(p_i^2,R)}$ term $\x^{\r}$, and a vertex $v\in G$ such that $v\in \widehat{\mathcal{V}}_i$ and $v\notin \mathcal{V}_F(\r)$. From the definition of~$\widehat{\mathcal{V}}_i$, there is a path $P\subset G$ joining $v$ and $p_i^\mu$ such that $w(e)=-1$ for any $e\in P$, and $P\cap \widehat{\mathcal{V}}_0 =\varnothing$. Meanwhile, since $v\notin \mathcal{V}_F(\r)$, we have $v\in t(\r;\widehat{p}_i)$, so there is an edge $e'\in P$ whose endpoints are respectively in $t(\r;p_i)$ and $t(\r;\widehat{p}_i)$.
However, as we have pointed out above, the endpoint of $e'$, which is in $t(\r_*;\widehat{p}_i)$, must belong to $\widehat{\mathcal{V}}_0$, which contradicts $P\cap \widehat{\mathcal{V}}_0 =\varnothing$. As a result, such a vertex $v$ does not exist, and $\mathcal{V}_F(\r) \supseteq \widehat{\mathcal{V}}_i$.
\end{proof}

\begin{proof} [Proof of $\mathcal{V}_F(\r) \subseteq \widehat{\mathcal{V}}_i$]
For any canonical $\mathcal{F}^{(p_i^2,R)}$ term $\x^{\r}$ and any vertex $v\in \mathcal{V}_F(\r)$, from corollary~\ref{lemma-onshell_heavy_in_light_constraint_corollary1} there is a path $P\subset G(t(\r;p_i))$ joining $v$ and $p_i^\mu$, and each edge of $P$ satisfies $w=-1$. It then suffices to show that $P\cap \widehat{\mathcal{V}}_0 = \varnothing$, from which one can deduce that all the vertices of $P$ are in $\widehat{\mathcal{V}}_i$, thus $\mathcal{V}_F(\r) \subseteq \widehat{\mathcal{V}}_i$.

Let us consider any vertex $v_0\in P\cap \widehat{\mathcal{V}}_0$. A first observation is that $v_0$ cannot be incident with an edge $e\in G$ with $w(e)>-1$. This is because $G(t(\r;p_i))$ does not include edges with $w>-1$ (lemma~\ref{lemma-onshell_Fp2_internal_less_equal_minusone}). So the only possibility is that, there exists a $\mathcal{U}^{(R)}$ term $\x^{\r_1}$ such that four or more $p$ trunks of $T^1(\r_1)$ are incident with $v_0$, and furthermore, each of these $p$ trunks consists of edges satisfying $w=-1$. The general configuration is described in figure~\ref{figure-onshell_lemma5_proof_case3}. Below we explain why contradictions can be led from this configuration.
\begin{figure}[t]
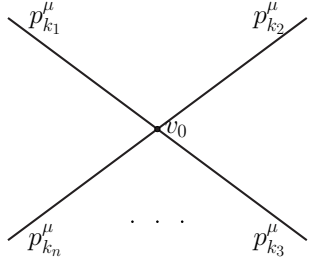

\centering
\include{figs/onshell_lemma5_proof_case3}
\vspace{-2em}
\caption{The configuration of $T^1(\r_1)$ where the $q$ trunk consists of the vertex $v_0$, and each of its $p_{k_j}$ trunks ($j=1,\dots,n\geqslant 4$) is adjacent to $v_0$. Note that potentially there are $p$ and $q$ branches in $T^1(\r)$, which we have omitted in the figure.}
\label{figure-onshell_lemma5_proof_case3}
\end{figure}

For any $e\in G\setminus T^1(\r_1)$, such that the loop in $T^1(\r_1)\cup e$ contains two of the $p_2,\dots,p_n$ trunks of $T^1(\r_1)$, say $e_i$ and $e_j$, we first show that $w(e)\leqslant -2$. This can be seen by considering the spanning 2-tree $T^2$ obtained from $T^1(\r)$ and by adding $e$ to it and remove $e_i$ and $e_j$ from it, namely, $T^1(\r)\cup e \setminus (e_i\cup e_j)$. From this construction, one component of $T^2$ has the external momenta $p_i^\mu$ and $p_j^\mu$ only. Due to momentum conservation, the other component of $T^2$ must have at least two on-shell momenta, or one off-shell momentum. In either case, the graph $T^2$ corresponds to an $\mathcal{F}^{(q^2)}$ term. We then have
\begin{eqnarray}
w(T^1(\r))\leqslant w(T^2)= w(T^1(\r)) +w(e_i) +w(e_j) -w(e),
\end{eqnarray}
where the inequality is due to the minimum-weight criterion and the equality is from the definition of $T^2$. From this, we can deduce $w(e)\leqslant w(e_j)+w(e_k) = -2$. However, corollary~\ref{lemma-onshell_Fp2_external_less_equal_minusone_corollary2} states that in the $\widehat{p}_1$ component of $T^2(\r')$, which is denoted as $T(\r';\widehat{p}_1)$, the path joining any two of the external momenta $p_2,\dots,p_n$ must consist of edges satisfying $w\geqslant -1$. These two observations ($w\leqslant-2$ and $w\geqslant -1$) contradict each other, and as a consequence, the possibility of case (3) is also ruled out.

To summarize, we have shown that the path $P_i$, or any branch component of~$T^1(\r)$ attached to $P_i$, does not include elements of $\widehat{\mathcal{V}}_0$. This completes the proof of $\mathcal{V}_F(\r) \subseteq \widehat{\mathcal{V}}_i$.
\end{proof}

Therefore, $\mathcal{V}_F(\r)= \widehat{\mathcal{V}}_i$, and the proof of lemma~\ref{lemma-onshell_Fp2_vertices_universal_property} is completed.

One can draw similar conclusions for generic canonical $\mathcal{U}^{(R)}$ terms. Namely, for any given canonical $\mathcal{U}^{(R)}$ term $\x^{\r}$, we define $\mathcal{V}_{U,i}(\r)$ \emph{to consist of any vertex $v\in G$ satisfying the following condition: there exists a path $P_i$ joining $v$ and the external momentum $p_i^\mu$, such that $w(e)=-1$ for each $e\in P_i$ and $P_i\cap \widehat{\mathcal{V}}_0 = \varnothing$}. One can apply the same technique as above and derive that $\mathcal{V}_{U,i}(\r_1)= \mathcal{V}_{U,i}(\r_2)$ for any canonical $\mathcal{U}^{(R)}$ terms $\x^{\r_1}$ and $\x^{\r_2}$. We conclude it as a corollary of lemma~\ref{lemma-onshell_Fp2_vertices_universal_property}.
\begin{corollary}
For any $i\in \{1,\dots,K\}$, $\mathcal{V}_{U,i}({\r}_1) = \mathcal{V}_{U,i}({\r}_2)$ holds if $\x^{\r_1}$ and $\x^{\r_2}$ are two canonical $\mathcal{U}^{(R)}$ terms.
\label{lemma-onshell_Fp2_vertices_universal_property_corollary1}
\end{corollary}

One can easily check the statement of this corollary by examining all the canonical $\mathcal{U}^{(R)}$ terms of eq.~(\ref{eq:fishnet_2times2_onshell_region_example_canonical}), which are
\begin{equation}
\begin{tikzpicture}[baseline=6ex, scale=0.2]
\draw [thick,color=Red] (2,7.5) -- (5,7.5);
\draw [dashed, thick,color=Red] (5,7.5) -- (5,5);
\draw [thick,color=Green] (5,5) -- (2,5);
\draw [thick,color=Green] (2,5) -- (2,7.5);
\draw [dashed, thick,color=Red] (8,7.5) -- (5,7.5);
\draw [thick,color=LimeGreen] (8,5) -- (8,7.5);
\draw [dashed, thick,color=Red] (5,5) -- (8,5);
\draw [thick,color=Green] (2,5) -- (2,2.5);
\draw [thick,color=Blue] (2,2.5) -- (5,2.5);
\draw [dashed, thick,color=Green] (5,2.5) -- (5,5);
\draw [thick,color=Blue] (5,2.5) -- (8,2.5);
\draw [thick,color=LimeGreen] (8,2.5) -- (8,5);

\draw [thick,color=Green] (0.5,8.5) -- (2,7.5);
\draw [thick,color=LimeGreen] (9.5,8.5) -- (8,7.5);
\draw [thick,color=olive] (9.5,1.5) -- (8,2.5);
\draw [ultra thick, color=Blue] (0.5,1.7) -- (2,2.5);
\draw [ultra thick, color=Blue] (0.6,1.5) -- (2,2.5);

\node [draw, circle, minimum size=4pt, color=Green, fill=Green, inner sep=0pt, outer sep=0pt] () at (2,7.5) {};
\node [draw, circle, minimum size=4pt, color=Green, fill=Green, inner sep=0pt, outer sep=0pt] () at (2,5) {};
\node [draw,circle,minimum size=4pt,fill=Black,inner sep=0pt,outer sep=0pt] () at (2,2.5) {};
\node [draw,circle,minimum size=4pt,fill=Black,inner sep=0pt,outer sep=0pt] () at (5,7.5) {};
\node [draw, circle, minimum size=4pt, color=Green, fill=Green, inner sep=0pt, outer sep=0pt] () at (5,5) {};
\node [draw,circle,minimum size=4pt,fill=Black,inner sep=0pt,outer sep=0pt] () at (5,2.5) {};
\node [draw, circle, minimum size=4pt, color=LimeGreen, fill=LimeGreen, inner sep=0pt, outer sep=0pt] () at (8,7.5) {};
\node [draw, circle, minimum size=4pt, color=LimeGreen, fill=LimeGreen, inner sep=0pt, outer sep=0pt] () at (8,5) {};
\node [draw,circle,minimum size=4pt,fill=Black,inner sep=0pt,outer sep=0pt] () at (8,2.5) {};
\end{tikzpicture}
\quad
\begin{tikzpicture}[baseline=6ex, scale=0.2]
\draw [dashed, thick,color=Red] (2,7.5) -- (5,7.5);
\draw [dashed, thick,color=Red] (5,7.5) -- (5,5);
\draw [thick,color=Green] (5,5) -- (2,5);
\draw [thick,color=Green] (2,5) -- (2,7.5);
\draw [thick,color=Red] (8,7.5) -- (5,7.5);
\draw [thick,color=LimeGreen] (8,5) -- (8,7.5);
\draw [dashed, thick,color=Red] (5,5) -- (8,5);
\draw [thick,color=Green] (2,5) -- (2,2.5);
\draw [thick,color=Blue] (2,2.5) -- (5,2.5);
\draw [dashed, thick,color=Green] (5,2.5) -- (5,5);
\draw [thick,color=Blue] (5,2.5) -- (8,2.5);
\draw [thick,color=LimeGreen] (8,2.5) -- (8,5);

\draw [thick,color=Green] (0.5,8.5) -- (2,7.5);
\draw [thick,color=LimeGreen] (9.5,8.5) -- (8,7.5);
\draw [thick,color=olive] (9.5,1.5) -- (8,2.5);
\draw [ultra thick, color=Blue] (0.5,1.7) -- (2,2.5);
\draw [ultra thick, color=Blue] (0.6,1.5) -- (2,2.5);

\node [draw, circle, minimum size=4pt, color=Green, fill=Green, inner sep=0pt, outer sep=0pt] () at (2,7.5) {};
\node [draw, circle, minimum size=4pt, color=Green, fill=Green, inner sep=0pt, outer sep=0pt] () at (2,5) {};
\node [draw,circle,minimum size=4pt,fill=Black,inner sep=0pt,outer sep=0pt] () at (2,2.5) {};
\node [draw,circle,minimum size=4pt,fill=Black,inner sep=0pt,outer sep=0pt] () at (5,7.5) {};
\node [draw, circle, minimum size=4pt, color=Green, fill=Green, inner sep=0pt, outer sep=0pt] () at (5,5) {};
\node [draw,circle,minimum size=4pt,fill=Black,inner sep=0pt,outer sep=0pt] () at (5,2.5) {};
\node [draw, circle, minimum size=4pt, color=LimeGreen, fill=LimeGreen, inner sep=0pt, outer sep=0pt] () at (8,7.5) {};
\node [draw, circle, minimum size=4pt, color=LimeGreen, fill=LimeGreen, inner sep=0pt, outer sep=0pt] () at (8,5) {};
\node [draw,circle,minimum size=4pt,fill=Black,inner sep=0pt,outer sep=0pt] () at (8,2.5) {};
\end{tikzpicture}
\quad
\begin{tikzpicture}[baseline=6ex, scale=0.2]
\draw [dashed, thick,color=Red] (2,7.5) -- (5,7.5);
\draw [thick,color=Red] (5,7.5) -- (5,5);
\draw [thick,color=Green] (5,5) -- (2,5);
\draw [thick,color=Green] (2,5) -- (2,7.5);
\draw [dashed, thick,color=Red] (8,7.5) -- (5,7.5);
\draw [thick,color=LimeGreen] (8,5) -- (8,7.5);
\draw [dashed, thick,color=Red] (5,5) -- (8,5);
\draw [thick,color=Green] (2,5) -- (2,2.5);
\draw [thick,color=Blue] (2,2.5) -- (5,2.5);
\draw [dashed, thick,color=Green] (5,2.5) -- (5,5);
\draw [thick,color=Blue] (5,2.5) -- (8,2.5);
\draw [thick,color=LimeGreen] (8,2.5) -- (8,5);

\draw [thick,color=Green] (0.5,8.5) -- (2,7.5);
\draw [thick,color=LimeGreen] (9.5,8.5) -- (8,7.5);
\draw [thick,color=olive] (9.5,1.5) -- (8,2.5);
\draw [ultra thick, color=Blue] (0.5,1.7) -- (2,2.5);
\draw [ultra thick, color=Blue] (0.6,1.5) -- (2,2.5);

\node [draw, circle, minimum size=4pt, color=Green, fill=Green, inner sep=0pt, outer sep=0pt] () at (2,7.5) {};
\node [draw, circle, minimum size=4pt, color=Green, fill=Green, inner sep=0pt, outer sep=0pt] () at (2,5) {};
\node [draw,circle,minimum size=4pt,fill=Black,inner sep=0pt,outer sep=0pt] () at (2,2.5) {};
\node [draw,circle,minimum size=4pt,fill=Black,inner sep=0pt,outer sep=0pt] () at (5,7.5) {};
\node [draw, circle, minimum size=4pt, color=Green, fill=Green, inner sep=0pt, outer sep=0pt] () at (5,5) {};
\node [draw,circle,minimum size=4pt,fill=Black,inner sep=0pt,outer sep=0pt] () at (5,2.5) {};
\node [draw, circle, minimum size=4pt, color=LimeGreen, fill=LimeGreen, inner sep=0pt, outer sep=0pt] () at (8,7.5) {};
\node [draw, circle, minimum size=4pt, color=LimeGreen, fill=LimeGreen, inner sep=0pt, outer sep=0pt] () at (8,5) {};
\node [draw,circle,minimum size=4pt,fill=Black,inner sep=0pt,outer sep=0pt] () at (8,2.5) {};
\end{tikzpicture}
\quad
\begin{tikzpicture}[baseline=6ex, scale=0.2]
\draw [thick,color=Red] (2,7.5) -- (5,7.5);
\draw [dashed, thick,color=Red] (5,7.5) -- (5,5);
\draw [thick,color=Green] (5,5) -- (2,5);
\draw [thick,color=Green] (2,5) -- (2,7.5);
\draw [dashed, thick,color=Red] (8,7.5) -- (5,7.5);
\draw [thick,color=LimeGreen] (8,5) -- (8,7.5);
\draw [dashed, thick,color=Red] (5,5) -- (8,5);
\draw [dashed, thick,color=Green] (2,5) -- (2,2.5);
\draw [thick,color=Blue] (2,2.5) -- (5,2.5);
\draw [thick,color=Green] (5,2.5) -- (5,5);
\draw [thick,color=Blue] (5,2.5) -- (8,2.5);
\draw [thick,color=LimeGreen] (8,2.5) -- (8,5);

\draw [thick,color=Green] (0.5,8.5) -- (2,7.5);
\draw [thick,color=LimeGreen] (9.5,8.5) -- (8,7.5);
\draw [thick,color=olive] (9.5,1.5) -- (8,2.5);
\draw [ultra thick, color=Blue] (0.5,1.7) -- (2,2.5);
\draw [ultra thick, color=Blue] (0.6,1.5) -- (2,2.5);

\node [draw, circle, minimum size=4pt, color=Green, fill=Green, inner sep=0pt, outer sep=0pt] () at (2,7.5) {};
\node [draw, circle, minimum size=4pt, color=Green, fill=Green, inner sep=0pt, outer sep=0pt] () at (2,5) {};
\node [draw,circle,minimum size=4pt,fill=Black,inner sep=0pt,outer sep=0pt] () at (2,2.5) {};
\node [draw,circle,minimum size=4pt,fill=Black,inner sep=0pt,outer sep=0pt] () at (5,7.5) {};
\node [draw, circle, minimum size=4pt, color=Green, fill=Green, inner sep=0pt, outer sep=0pt] () at (5,5) {};
\node [draw,circle,minimum size=4pt,fill=Black,inner sep=0pt,outer sep=0pt] () at (5,2.5) {};
\node [draw, circle, minimum size=4pt, color=LimeGreen, fill=LimeGreen, inner sep=0pt, outer sep=0pt] () at (8,7.5) {};
\node [draw, circle, minimum size=4pt, color=LimeGreen, fill=LimeGreen, inner sep=0pt, outer sep=0pt] () at (8,5) {};
\node [draw,circle,minimum size=4pt,fill=Black,inner sep=0pt,outer sep=0pt] () at (8,2.5) {};
\end{tikzpicture}
\quad
\begin{tikzpicture}[baseline=6ex, scale=0.2]
\draw [dashed, thick,color=Red] (2,7.5) -- (5,7.5);
\draw [dashed, thick,color=Red] (5,7.5) -- (5,5);
\draw [thick,color=Green] (5,5) -- (2,5);
\draw [thick,color=Green] (2,5) -- (2,7.5);
\draw [thick,color=Red] (8,7.5) -- (5,7.5);
\draw [thick,color=LimeGreen] (8,5) -- (8,7.5);
\draw [dashed, thick,color=Red] (5,5) -- (8,5);
\draw [dashed, thick,color=Green] (2,5) -- (2,2.5);
\draw [thick,color=Blue] (2,2.5) -- (5,2.5);
\draw [thick,color=Green] (5,2.5) -- (5,5);
\draw [thick,color=Blue] (5,2.5) -- (8,2.5);
\draw [thick,color=LimeGreen] (8,2.5) -- (8,5);

\draw [thick,color=Green] (0.5,8.5) -- (2,7.5);
\draw [thick,color=LimeGreen] (9.5,8.5) -- (8,7.5);
\draw [thick,color=olive] (9.5,1.5) -- (8,2.5);
\draw [ultra thick, color=Blue] (0.5,1.7) -- (2,2.5);
\draw [ultra thick, color=Blue] (0.6,1.5) -- (2,2.5);

\node [draw, circle, minimum size=4pt, color=Green, fill=Green, inner sep=0pt, outer sep=0pt] () at (2,7.5) {};
\node [draw, circle, minimum size=4pt, color=Green, fill=Green, inner sep=0pt, outer sep=0pt] () at (2,5) {};
\node [draw,circle,minimum size=4pt,fill=Black,inner sep=0pt,outer sep=0pt] () at (2,2.5) {};
\node [draw,circle,minimum size=4pt,fill=Black,inner sep=0pt,outer sep=0pt] () at (5,7.5) {};
\node [draw, circle, minimum size=4pt, color=Green, fill=Green, inner sep=0pt, outer sep=0pt] () at (5,5) {};
\node [draw,circle,minimum size=4pt,fill=Black,inner sep=0pt,outer sep=0pt] () at (5,2.5) {};
\node [draw, circle, minimum size=4pt, color=LimeGreen, fill=LimeGreen, inner sep=0pt, outer sep=0pt] () at (8,7.5) {};
\node [draw, circle, minimum size=4pt, color=LimeGreen, fill=LimeGreen, inner sep=0pt, outer sep=0pt] () at (8,5) {};
\node [draw,circle,minimum size=4pt,fill=Black,inner sep=0pt,outer sep=0pt] () at (8,2.5) {};
\end{tikzpicture}
\quad
\begin{tikzpicture}[baseline=6ex, scale=0.2]
\draw [dashed, thick,color=Red] (2,7.5) -- (5,7.5);
\draw [thick,color=Red] (5,7.5) -- (5,5);
\draw [thick,color=Green] (5,5) -- (2,5);
\draw [thick,color=Green] (2,5) -- (2,7.5);
\draw [dashed, thick,color=Red] (8,7.5) -- (5,7.5);
\draw [thick,color=LimeGreen] (8,5) -- (8,7.5);
\draw [dashed, thick,color=Red] (5,5) -- (8,5);
\draw [dashed, thick,color=Green] (2,5) -- (2,2.5);
\draw [thick,color=Blue] (2,2.5) -- (5,2.5);
\draw [thick,color=Green] (5,2.5) -- (5,5);
\draw [thick,color=Blue] (5,2.5) -- (8,2.5);
\draw [thick,color=LimeGreen] (8,2.5) -- (8,5);

\draw [thick,color=Green] (0.5,8.5) -- (2,7.5);
\draw [thick,color=LimeGreen] (9.5,8.5) -- (8,7.5);
\draw [thick,color=olive] (9.5,1.5) -- (8,2.5);
\draw [ultra thick, color=Blue] (0.5,1.7) -- (2,2.5);
\draw [ultra thick, color=Blue] (0.6,1.5) -- (2,2.5);

\node [draw, circle, minimum size=4pt, color=Green, fill=Green, inner sep=0pt, outer sep=0pt] () at (2,7.5) {};
\node [draw, circle, minimum size=4pt, color=Green, fill=Green, inner sep=0pt, outer sep=0pt] () at (2,5) {};
\node [draw,circle,minimum size=4pt,fill=Black,inner sep=0pt,outer sep=0pt] () at (2,2.5) {};
\node [draw,circle,minimum size=4pt,fill=Black,inner sep=0pt,outer sep=0pt] () at (5,7.5) {};
\node [draw, circle, minimum size=4pt, color=Green, fill=Green, inner sep=0pt, outer sep=0pt] () at (5,5) {};
\node [draw,circle,minimum size=4pt,fill=Black,inner sep=0pt,outer sep=0pt] () at (5,2.5) {};
\node [draw, circle, minimum size=4pt, color=LimeGreen, fill=LimeGreen, inner sep=0pt, outer sep=0pt] () at (8,7.5) {};
\node [draw, circle, minimum size=4pt, color=LimeGreen, fill=LimeGreen, inner sep=0pt, outer sep=0pt] () at (8,5) {};
\node [draw,circle,minimum size=4pt,fill=Black,inner sep=0pt,outer sep=0pt] () at (8,2.5) {};
\end{tikzpicture}.
\label{eq:onshell_region_canonical_Uterms_example}
\end{equation}
Corollary~\ref{lemma-onshell_Fp2_vertices_universal_property_corollary1} can then be directly verified here: for each graph, the nontrivial sets $\mathcal{V}_{U,i}(\r)$ are $\mathcal{V}_{U,1}(\r)$ (consisting of the {\color{Green}green} vertices) and $\mathcal{V}_{U,2}(\r)$ ({\color{LimeGreen}lime green}). They are the same for all the seven graphs above.

\bigbreak
The advantage of introducing the notion of canonical leading terms is follows. In order to determine the weight structure of an arbitrary region $R$ and relate it to the solution to the Landau equations, the key is to partition the entire graph $G$ into its hard, jet and soft subgraphs. The canonical leading terms provide a natural and rigorous way of defining these subgraphs. Moreover, one can use lemmas~\ref{lemma-onshell_basic_weight_structure_Uterm}-\ref{lemma-onshell_heavy_in_light_constraint} and their corresponding corollaries, which describe properties of generic $\mathcal{U}^{(R)}$ and $\mathcal{F}^{(p^2,R)}$ terms, to study the structure of the hard, jet and soft subgraphs.

\subsubsection{Subgraphs associated with each region}
\label{section-subgraphs_associated_each_region}

Given any region $R$, we now aim at assigning the vertices and edges of $G$ to the corresponding hard ($H$), jet ($J$) and soft ($S$) subgraphs.

To this end, let us first recall the notions of $\widehat{\mathcal{V}}_0, \widehat{\mathcal{V}}_1, \dots, \widehat{\mathcal{V}}_K$ defined below the statement of lemma~\ref{lemma-onshell_Fp2_vertices_universal_property}. For each $v_0\in \widehat{\mathcal{V}}_0$, either $v_0$ is incident with an edge $e\in G$ such that $w(e)>-1$, or there exists some $\mathcal{U}^{(R)}$ term $\x^{\r}$ such that $v_0$ is in the $q$ trunk of $T^1(\r)$. We then explained that the graph $G\setminus \widehat{\mathcal{V}}_0$ is disconnected, from which we further define the sets $\widehat{\mathcal{V}}_1, \dots, \widehat{\mathcal{V}}_K$: for each vertex $v_i\in \widehat{\mathcal{V}}_i$, there exists a path $P\subset G\setminus \widehat{\mathcal{V}}_0$ connecting $v_i$ and the external momentum $p_i^\mu$, such that $w(e)=-1$ for any $e\in P$.

The vertices and edges of $G$ can then be partitioned as follows.
\begin{itemize}
    \item For any vertex $v\in G$, we assign it to the \emph{$i$th jet subgraph} $J_i$ if $v\in \widehat{\mathcal{V}}_i$; we assign it to the \emph{soft subgraph} $S$ if all the edges, which $v$ is incident with, satisfy $w<-1$; otherwise we assign it to the \emph{hard subgraph} $H$.    
    \item For each edge $e\in G$, we assign it to $H$ if $w(e)\geqslant -1$ and both its endpoints are in $H$; we assign it to $J_i$ if $w(e)\geqslant -1$, and its endpoints are either both in $J_i$, or respectively in $J_i$ and $H$; otherwise we assign it to $S$.
\end{itemize}
We further denote $J=\cup_{i=1}^K J_i$. It follows clearly from above that
\begin{eqnarray}
    G= H\sqcup J\sqcup S.
\end{eqnarray}
Based on these definitions, some basic properties of $H$, $J$ and $S$ follow from lemma~\ref{lemma-onshell_Fp2_vertices_universal_property}.
\begin{corollary}
\label{lemma-onshell_Fp2_vertices_universal_property_corollary2}
For any given region $R$, the associated hard, jet and soft subgraphs satisfy the following properties.
\begin{itemize}
    \item [1.] Each jet $J_i$ is connected, and $J_i\cup J_j= \varnothing$ for $i\neq j$.
    \item [2.] Weight structure properties:
    \begin{enumerate}
        \item [2-1,] all the hard edges satisfy $w\geqslant -1$;
        \item [2-2,] all the jet edges satisfy $w=-1$;
        \item [2-3,] all the soft edges satisfy $w<-1$.
    \end{enumerate}
    \item [3.] For any spanning tree $T^1$ corresponding to a $\mathcal{U}^{(R)}$ term and any spanning 2-tree $T^2$ corresponding to a $\mathcal{F}^{(p_i^2,R)}$ term,
    \begin{enumerate}
        \item [3-1,] all the vertices in the $q$ trunk of $T^1$ must belong to $H$;
        \item [3-2,] no hard vertices belong to the $p_i$ component of $T^2$.
    \end{enumerate}
\end{itemize}
\end{corollary}
\begin{proof}
First, from the definition of $\widehat{\mathcal{V}}_i$, $J_i$ is precisely the graph $\gamma_{J_i}^{}(\r)$ defined in corollary~\ref{lemma-onshell_heavy_in_light_constraint_corollary1}, where $\r$ corresponds to a canonical $\mathcal{F}^{(p_i^2,R)}$ term. Corollary~\ref{lemma-onshell_heavy_in_light_constraint_corollary1} then guarantees $J_i$ to be connected. To see $J_i\cap J_j= \varnothing$ for any $J_i$ and $J_j$, we note that they share no vertices because $\widehat{\mathcal{V}}_i\cap \widehat{\mathcal{V}}_j = \varnothing$; they share no edges because each edge of $J_i$ is incident with at least one $J_i$ vertex but no $J_j$ vertices. Thus $J_i\cap J_j=\varnothing$.

We next justify the weight structure properties above. Statement \emph{2-1} is exactly from the definition of $H$.
Statement \emph{2-2} excludes the possibility of $w>-1$ for any jet edge $e^J$. This is because at least one endpoint of $e^J$ belongs to $\widehat{\mathcal{V}}_i$, which is exclusively incident with edges satisfying $w\leqslant -1$. (Any vertex incident with an edge with $w>-1$ would be in $\widehat{\mathcal{V}}_0$ by definition.)
To justify statement \emph{2-3}, note that any soft edge $e^S$ is defined to satisfy one of these three possibilities: (1) $e^S$ is incident with a soft vertex; (2) the endpoints of $e^S$ are both in and $H\cup J_i$ for a given $i$, and $w(e^S)<-1$; (3) the endpoints of $e$ are in $J_i$ and $J_j$ ($i\neq j$) respectively. For (1) and (2) we immediately have $w(e^S)<-1$.
For (3), let us consider any canonical $\mathcal{U}^{(R)}$ term $\x^{\r}$. According to corollary~\ref{lemma-onshell_Fp2_vertices_universal_property_corollary1}, all the elements of $\widehat{\mathcal{V}}_i$ ($\widehat{\mathcal{V}}_j$) are in the $p_i$ ($p_j$) trunk and branch of $T^1(\r)$ (figure~\ref{onshell_canonical_Uterm_nontrivial_i_j_jet}). Since $e^S$ joins $J_i$ and $J_j$, there is a loop in the graph $T^1(\r) \cup e^S$, which contains some edges $e_i\in J_i$ and $e_j\in J_j$. The graph $T^1(\r) \cup e^S \setminus \left( e_i\cup e_j \right)$ then corresponds to an $\mathcal{F}^{(q^2)}$ term $\x^{\r'}$ (figure~\ref{onshell_canonical_Uterm_nontrivial_i_j_jet_modified}). It then follows from the minimum-weight criterion that
\begin{figure}[t]
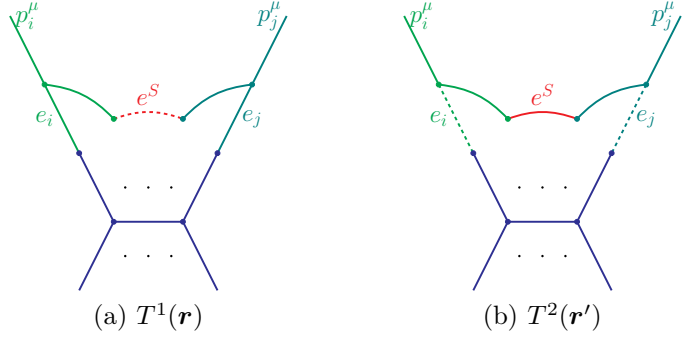

\centering
\begin{subfigure}[b]{0.25\textwidth}
\centering
\include{figs/onshell_canonical_Uterm_nontrivial_i_j_jet}
\vspace{-3em}
\caption{$T^1(\r)$}
\label{onshell_canonical_Uterm_nontrivial_i_j_jet}
\end{subfigure}
\hspace{3em}
\begin{subfigure}[b]{0.25\textwidth}
\centering
\include{figs/onshell_canonical_Uterm_nontrivial_i_j_jet_modified}
\vspace{-3em}
\caption{$T^2(\r')$}
\label{onshell_canonical_Uterm_nontrivial_i_j_jet_modified}
\end{subfigure}
\caption{The spanning tree $T^1(\r)$ and the spanning 2-tree $T^2(\r')$ in the analysis of weight structure properties, as part of the proof of corollary~\ref{lemma-onshell_Fp2_vertices_universal_property_corollary2}.}
\label{figure-onshell_lemma5_proof}
\end{figure}
\begin{eqnarray}
    w(T^1(\r))\leqslant w(T^2(\r')) = w(T^1(\r)) + w(e_i)+w(e_j) -w(e^S) \quad\Rightarrow\quad w(e^S)\leqslant -2,
\end{eqnarray}
where we have used $w(e_i)=w(e_j)=-1$. Thus every soft edge must satisfy $w<-1$.

Finally, we justify the remaining two statements in the corollary. Any vertex $v$ in the $q$ trunk of $T^1$ is defined to be in $\widehat{\mathcal{V}}_0$, which implies that $v\notin J$ and $v\notin S$, and thus $v\in H$. This justifies statement \emph{3-1}.
For any vertex $v$ in the $p_i$~component of $T^2$, lemma~\ref{lemma-onshell_Fp2_internal_less_equal_minusone} indicates that $v$ is incident with edges either (1) all satisfy $w<-1$, or (2) some of them satisfy $w=-1$ while the others satisfy $w<-1$. Then (1) implies $v\in S$, and (2) implies $v\in J_i$. As a result, the $p_i$ component of $T^2$ does not have any vertices in $H$, which justifies statement \emph{3-2}.
\end{proof}

Based on the observations in corollary~\ref{lemma-onshell_Fp2_vertices_universal_property_corollary2}, one can further show the connectedness of the hard subgraph $H$.
\begin{lemma}
The hard subgraph $H$ is connected.
\label{lemma-onshell_hard_subgraph_connected}
\end{lemma}
\begin{proof}
Let us consider any canonical $\mathcal{U}^{(R)}$ term $\x^{\r}$ and choose any vertex $v_0$ in the $q$ trunk.\footnote{As explained in section~\ref{section-some_properties_leading_terms}, the $q$ trunk of $T^1(\r)$ where $\x^{\r}\in \mathcal{U}^{(R)}(\x)$ is always nonempty.} Below we will see that for any other vertex $v_1\in H$, the path $P\subset T^1(\r)$, which joins $v_0$ and~$v_1$, must be contained in $H$.
\begin{itemize}
    \item [\textbf{I.}] If $v_1$ is in the $q$ trunk of $T^1(\r)$, then $P$ is also in the $q$ trunk. All the vertices of $P$ are then in $H$ due to \emph{3-1} in corollary~\ref{lemma-onshell_Fp2_vertices_universal_property_corollary2}. Since each edge in the $q$ trunk satisfies $w=0$, it follows from the definition that all the edges in $P$ also belong to $H$.
    
    \item [\textbf{II.}] If $v_1$ is in the $p_i$ trunk, then each edge $e\in P$ is either in the $q$ or the $p_i$ trunk, which furthermore must satisfy $w(e) >-1$. (To see this, note that any edge in the $q$ trunk satisfies $w=0$, and any edge in the $p_i$ trunk satisfies $w\geqslant-1$. If one of them, say $e_i$, satisfies $w(e_i)=-1$, then $v_1$ would be in the $p_i$ component of the spanning 2-tree $T^1(\r)\setminus e_i$.
    Then statement \emph{3-2} of corollary~\ref{lemma-onshell_Fp2_vertices_universal_property_corollary2} would contradict the assumption that $v_1\in H$.) By comparing statements \emph{2-1}, \emph{2-2} and \emph{2-3} in corollary~\ref{lemma-onshell_Fp2_vertices_universal_property_corollary2}, it is then clear that $P\subset H$.
    
    \item [\textbf{III.}] Let us now consider the possibilities that $v_1$ is in the $q$ or $p_i$ branch, and show that $P\subset H$ still holds. To this end we shall first prove by contradiction that all the edges of $P$ satisfy $w\geqslant -1$. Consider the contrary, i.e. $w<-1$ for some edges $e'\in P$. Then $T^1(\r) \setminus e'$ is a spanning 2-tree, such that $v_1$ is in one component while $v_0$ (with the $q$ trunk) is in the other.
    Let us denote the component containing $v_1$ as $t(\r;v_1)$, then a key observation is that any edge $e''\neq e'$, whose endpoints are in $t(\r;v_1)$ and $G\setminus t(\r;v_1)$ respectively, must satisfy $w(e'')<-1$. To see it, suppose that $w(e'')\geqslant -1$, then the weight of the spanning tree ${T'}^1\equiv T^1(\r)\cup e''\setminus e'$ (figure~\ref{figure-onshell_lemma6_proof}) is smaller than $w(T^1(\r))$, violating the minimum-weight criterion. In other words, all the edges adjacent to the graph $G(t(\r;v_1))$ satisfy $w<-1$.
    \begin{figure}[t]
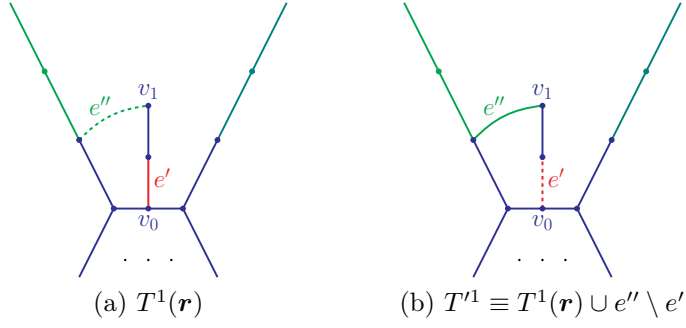

    \centering
    \begin{subfigure}[b]{0.25\textwidth}
    \centering
    \include{figs/onshell_lemma6_spanning_tree}
    \vspace{-3em}
    \caption{$T^1(\r)$}
    \label{onshell_lemma6_spanning_tree}
    \end{subfigure}
    \hspace{3em}
    \begin{subfigure}[b]{0.25\textwidth}
    \centering
    \include{figs/onshell_lemma6_spanning_tree_prime}
    \vspace{-3em}
    \caption{$T'^1\equiv T^1(\r)\cup e''\setminus e'$}
    \label{onshell_lemma6_spanning_tree_prime}
    \end{subfigure}
    \caption{The spanning trees $T^1(\r)$ and $T'^1$ in the discussion of the cases where $v_1$ is in the $q$ or $p_i$ branch, as part of the proof of lemma~\ref{lemma-onshell_hard_subgraph_connected}.}
    \label{figure-onshell_lemma6_proof}
    \end{figure}

    Since $v_1\in H$, it must be incident with some edges satisfying $w\geqslant -1$. Then the only possible configurations would be: there is a connected subgraph $\gamma_1\subset G(t(\r;v_1))$ such that $v\in\gamma_1$, and all the edges in $\gamma_1$ satisfy $w\geqslant -1$ while all the edges adjacent to $\gamma_1$ satisfy $w<-1$. However, such configurations are ruled out by lemma~\ref{lemma-onshell_heavy_in_light_constraint}. We have thus proved by contradiction, that all the edges in the path $P$ satisfy $w\geqslant -1$.
    
    A further observation is that none of these edges would be in $J$. Otherwise we take one special edge $e^{J_j}\in P\cap J_j$ (note that $j\in \{1,\dots,K\}$, which can either be identical to $i$ or not), such that any other $e'\in P\cap J_j$ is between $e^{J_j}$ and $v_1$. Let us further suppose that $e^{J_j}$ is in the $p_i$ trunk of $T^1(\r)$ (for other possibilities the following analysis can be applied similarly). Since $w(e^{J_j})=-1$ by definition, the graph $T^1(\r)\setminus e^{J_j}$ is then a spanning 2-tree corresponding to a canonical $\mathcal{F}^{(p_i^2,R)}$ term, whose $p_i$ component contains the vertex $v_1$. For this configuration, however, statement \emph{3-2} of corollary~\ref{lemma-onshell_Fp2_vertices_universal_property_corollary2} would contradict the assumption that $v_1\in H$.
    
    As a result, all the edges of $P$ are in $H$, thus $P\subset H$ must hold.
\end{itemize}

To conclude from above, we have seen that for a given $v_0$ in the $q$ trunk of $T^1(\r)$ and any other vertex $v_1\in H$, the path $P\subset T^1(\r)$ joining $v_0$ and $v_1$ is contained in $H$. This suffices to show that $H$ is connected.
\end{proof}

\begin{remark}
    We comment that the analysis above is more than sufficient to prove the connectedness of $H$: given any $v_1,v_2\in H$, we have shown that there exists a path $P\subset H$ that joins them, and furthermore, $P\subset T^1(\r)$ for a canonical $\mathcal{U}^{(R)}$ term $\x^{\r}$. We will apply this statement to the proof of lemma~\ref{lemma-onshell_subgraphs_tree_structures} later.
\end{remark}

\begin{figure}[t]
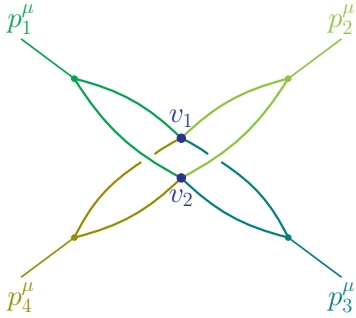

\centering
\include{figs/onshell_lemma6_configuration_unallowed}
\vspace{-2em}
\caption{A configuration that is \emph{not} an endpoint-type region in the on-shell expansion of wide-angle scattering because the hard subgraph is disconnected, which is forbidden by lemma~\ref{lemma-onshell_hard_subgraph_connected}.}
\label{figure-onshell_lemma6_configuration_unallowed}
\end{figure}

An example illustrating the proof above will be provided later in (\ref{eq:3loop_nonplanar_example_lemma4and6}). As an immediate result from lemma~\ref{lemma-onshell_hard_subgraph_connected}, the configuration in figure~\ref{figure-onshell_lemma6_configuration_unallowed} does \emph{not} correspond to a lower facet of $\Delta(\mathcal{P})$, because the hard subgraph $H$ exhibits two connected components, each being a single vertex ($v_1$ and $v_2$ respectively).

Another result from lemma~\ref{lemma-onshell_hard_subgraph_connected} is that the graph $H\cup (\cup'_i J_i)$ is also connected, where $\cup'_i$ denotes the union of any jets chosen among $\{J_1,\dots,J_K\}$. This is simply because each jet is connected and adjacent to a connected graph $H$. In particular, $H\cup J$ is connected, which, for our later convenience, we summarize into the following corollary.
\begin{corollary}
\label{lemma-onshell_hard_subgraph_connected_corollary1}
    The graph $H\cup J$ is connected.
\end{corollary}

\begin{figure}[t]
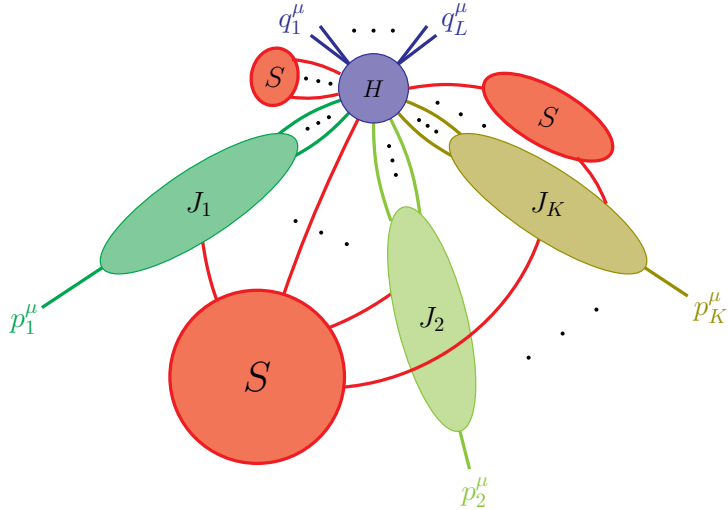

\centering
\include{figs/onshell_region_H_J_S}
\vspace{-3em}
\caption{The picture of a generic region $R$, where $H,J,S$ are the hard, jet and soft subgraphs of $G$ that are associated with $R$. The dots represent that multiple lines can connect the blobs. The hard subgraph $H$ is connected, and attached by all the off-shell external momenta $q_1^\mu,\dots,q_L^\mu$. Each jet subgraph $J_i$ is connected, and attached by the on-shell external momentum $p_i^\mu$. The soft subgraph $S$, which is adjacent to $H\cup J$, can be disconnected, and is attached by no external momenta.}
\label{figure-onshell_region_H_J_S}
\end{figure}

Using the results we have obtained so far (lemma~\ref{lemma-onshell_basic_weight_structure_Uterm}-\ref{lemma-onshell_hard_subgraph_connected} and their corollaries), we claim that the regions in the on-shell expansion for wide-angle scattering can always be described by figure~\ref{figure-onshell_region_H_J_S}. In this figure, the on-shell external momenta $p_1^\mu,\dots,p_K^\mu$ attach to the jets $J_1,\dots,J_K$. All the jets are adjacent to the hard subgraph $H$, to which all the off-shell external momenta $q_1^\mu,\dots,q_L^\mu$ attach. The hard and jet subgraphs can also interact through the soft subgraph $S$, which may have several distinct connected components.
Some properties of $H$, $J$ and $S$, which we have obtained above, are shown in the following table.
\begin{center}
\begin{tabular}{ |c||c|c|c|c|c| } 
\hline
 & external & \multirow{2}{5.3em}{connectivity} & weight & other \\
 & momenta & & of edges & properties \\
\hline
$H$ & $q_1^\mu,\dots,q_L^\mu$ & connected & $w\geqslant -1$ & \\
\hline
$J_i$ & $p_i^\mu$ & connected & $w=-1$ & $J_i\cap J_j=\varnothing$ \\
\hline
 & \multirow{3}{1em}{$\varnothing$} & & \multirow{3}{3.5em}{$w<-1$} & Each component \\ 
$S$ &  & & & attaches to \\
 &  & & & at least one $J_i$.\\
\hline
\end{tabular}
\end{center}

Two particularly useful concepts are the contracted jet and soft subgraphs. For each $i\in\{1,\dots,K\}$, the \emph{$i$th contracted jet graph} $\widetilde{J}_i$ is obtained from $J_i$ by introducing a new jet vertex and identifying it with all the vertices of $H$ that are incident with a momentum of the $i$th jet. The \emph{contracted soft graph} $\widetilde{S}$ is obtained from $S$ by introducing a new vertex and identifying it with all the vertices of $H$ and $J$ that are incident with a soft momentum. Note that $\widetilde{J}$ and $\widetilde{S}$ are closely related to the concept of ``contracted subgraphs'', as defined in section~\ref{section-weight_ordered_partition}.\footnote{As follows from the definition above, $\widetilde{S}$ is a special contracted subgraph, because the weight of any edge in $S$ is smaller than that of any edge in $H\cup J$; namely, $w(e_S)<w(e_{HJ})$ for any $e_S\in S$ and $e_{HJ}\in H\cup J$. In comparison, at this point we cannot claim that every $\widetilde{J}_i$ is a contracted subgraph, because we have not shown that $w(e_J)< w(e_H)$ for any $e_H\in H$ and $e_J\in J$, according to the table above. Nevertheless, later in section~\ref{section-determining_hard_soft_weights} we will show that $w(e_H)=0$, which automatically leads to $w(e_J)< w(e_H)$. Thus every $\widetilde{J}_i$ is indeed a contracted subgraph.}

An immediate advantage of introducing $\widetilde{J}$ and $\widetilde{S}$ is that the loop number of~$G$ can be expressed as the sum over the loop numbers of $H$, $\widetilde{J}$ and $\widetilde{S}$.
\begin{corollary}
\label{lemma-onshell_hard_subgraph_connected_corollary2}
The loop numbers of $H$, $\widetilde{J}$ and $\widetilde{S}$ satisfy the following relations:
\begin{eqnarray}
\label{eq:lemma_onshell_subgraphs_loop_relation1}
&& L(H) + L(\widetilde{J}) = L(H\cup J), \\
\label{eq:lemma_onshell_subgraphs_loop_relation2}
&& L(H) + L(\widetilde{J}) + L(\widetilde{S}) = L(G).
\end{eqnarray}
\end{corollary}
The proof of this corollary relies on the connectedness of $H$ and $H\cup J$ as we showed above, and involves direct applications of Euler's formula.
\begin{proof}
    Recall that a given graph~$\gamma$, we have used $V(\gamma)$, $N(\gamma)$, and $L(\gamma)$ to denote the number of vertices, edges, and loops of $\gamma$, respectively. To derive eq.~(\ref{eq:lemma_onshell_subgraphs_loop_relation1}), we notice that both subgraphs $H$ and $H \cup J$ are connected. Then we apply the Euler's formula for a connected graph, $L+V-N=1$, and obtain:
    \begin{subequations}
        \begin{align}
            &V(H \cup J)-N(H \cup J)+L(H \cup J)=1, \\
            &V(\widetilde{J})-N(\widetilde{J})+L(\widetilde{J})=1, \\
            &V(H)-N(H)+L(H)=1.
        \end{align}
    \end{subequations}
    Moreover, the number of vertices and edges of the $G$, $G\setminus S$ and $\widetilde{S}$ are related through:
    \begin{subequations}
        \begin{align}
            &V(H\cup J) =V(\widetilde{J})+V(H)-1, \\
            &N(H\cup J) =N(\widetilde{J})+N(H).
        \end{align}
    \end{subequations}
    From these relations we immediately arrive at $L(H \cup J) = L(H) + L(\widetilde{J})$. The proof of eq.~(\ref{eq:lemma_onshell_subgraphs_loop_relation2}) is similar.
\end{proof}

Nevertheless, the following lemma states that for any $\mathcal{U}^{(R)}$ term, the graphs $H\cap T(\r)$, $\widetilde{J}\cap T(\r)$, and $\widetilde{S}\cap T(\r)$ are the spanning trees of $H$, $\widetilde{J}$, and $\widetilde{S}$, respectively. This lemma can be seen as an analogue of lemma~\ref{lemma-weight_hierarchical_partition_tree_structure}.
\begin{lemma}
\label{lemma-onshell_subgraphs_tree_structures}
For any $\x^{\r}$ being a $\mathcal{U}^{(R)}$ or an $\mathcal{F}^{(p_i^2,R)}$ term, we have:
\begin{enumerate}
    \item [\emph{1},] the graphs $H\cap T(\r)$ and $\widetilde{S}\cap T(\r)$ are spanning trees of $H$ and $\widetilde{S}$ respectively;
    
    \item [\emph{2},] if $\x^{\r}$ is a $\mathcal{U}^{(R)}$ term, then for each $i$, $\widetilde{J}_i\cap T^1(\r)$ is a spanning tree of $\widetilde{J}_i$;
    
    \item [\emph{3},] if $\x^{\r}$ is an $\mathcal{F}^{(p_i^2,R)}$ term, then $\widetilde{J}_i\cap T^1(\r)$ is a spanning 2-tree of $\widetilde{J}_i$, and $\widetilde{J}_j\cap T^1(\r)$ (for each $j\neq i$) is a spanning tree of $\widetilde{J}_i$.
\end{enumerate}
\end{lemma}

\begin{proof}
We start by showing that $H\cap T(\r)$ is a spanning tree of $H$. If $\x^{\r}$ is a $\mathcal{U}^{(R)}$ term, then from the remark below the proof of lemma~\ref{lemma-onshell_hard_subgraph_connected}, the path $P\subset T^1(\r)$ joining any two vertices $v_0,v_1\in H$ must be contained in $H$. In other words, $H\cap T^1(\r)$ is a connected tree graph, which is hence a spanning tree of $H$. If $\x^{\r}$ is an $\mathcal{F}^{(p_i^2,R)}$ term, then from corollary~\ref{lemma-onshell_Fp2_external_less_equal_minusone_corollary1} there exists a $\mathcal{U}^{(R)}$ term $\x^{\r'}$, such that $T^1(\r') = T^2(\r) \cup e$ where $e$ joins the two components of $T^2(\r)$ and satisfies $w(e)=-1$. It then follows from the definition of jet edges that $e\in J_i$. So $H\cap T^2(\r) = H\cap T^1(\r')$. In other words, $H\cap T(\r)$ is a spanning tree of~$H$ regardless of whether $\x^{\r}$ is a $\mathcal{U}^{(R)}$ term or an $\mathcal{F}^{(p_i^2,R)}$ term.

We next show that $\widetilde{S}\cap T(\r)$ is a spanning tree of $\widetilde{S}$. Again, let us start by assuming that $\x^{\r}$ is a $\mathcal{U}^{(R)}$ term. A key observation is that \emph{no} paths in $S\cap T^1(\r)$ can connect two vertices of $H\cup J$. If such a path $P$ exists, because of the connectedness of $H\cup J$ (corollary~\ref{lemma-onshell_hard_subgraph_connected_corollary1}), there must be an edge $e^{HJ}\in H\cup J$ whose endpoints are in the two components of $T^1(\r)\setminus P$ respectively. The spanning tree $T^1(\r)\cup e^{HJ}\setminus e^S$, where $e^S$ is any edge in $P$, has a weight smaller than $w(T^1(\r))$, which violates the minimum-weight criterion. As a result, $S\cap T^1(\r)$ contains no paths joining two vertices of $H\cup J$, which implies that there exists no loops in $\widetilde{S}\cap T(\r)$. Meanwhile, $\widetilde{S}\cap T(\r)$ is connected, because it is obtained from $T^1(\r)$ by contracting its subgraph $(H\cup J)\cap T^1(\r)$ and identifying it with the auxiliary vertex. Therefore, $\widetilde{S}\cap T(\r)$ is a spanning tree of $\widetilde{S}$ for $\x^{\r}$ being a $\mathcal{U}^{(R)}$ term, which, due to the correspondence between $\mathcal{U}^{(R)}$ and $\mathcal{F}^{(p_i^2,R)}$ terms as pointed out above, also holds for $\x^{\r}$ being an $\mathcal{F}^{(p_i^2,R)}$ term.

So far we have justified statement \emph{1}. Now we consider a canonical $\mathcal{F}^{(p_i^2,R)}$ term $\x^{\r_*}$ and examine the structure of the graph $\widetilde{J}_i \cap T^2(\r_*)$. There are $V(J_i)+1$ vertices in $\widetilde{J}_i$: $N(J_i)$ of them are the $J_i$ vertices, which are in the $p_i$ component of $T^2(\r_*)$; the remaining one is the auxiliary vertex, which is in the $\widehat{p}_i$ component of $T^2(\r_*)$. Moreover, from corollary~\ref{lemma-onshell_heavy_in_light_constraint_corollary1}, the graph $J_i\cap t(\r_*;p_i)$ is connected, from which we can deduce that $\widetilde{J}_i \cap T^2(\r_*)$ is a spanning 2-tree of $\widetilde{J}_i$. Then recall that in section~\ref{section-canonical_leading_terms}, we have shown that a canonical $\mathcal{F}^{(p_i^2,R)}$ term from any $\mathcal{F}^{(p_i^2,R)}$ term $\x^{\r}$, by adding certain $n$ edges of $J_i$ and deleting another $n$ edges of $J_i$, i.e.
\begin{equation}
    T^2(\x^{\r}) \xrightarrow[]{\text{section~\ref{section-canonical_leading_terms}}} T^2(\x^{\r_*})
\label{eq:canonical_Fp2Rterm_to_FP2Rterm}
\end{equation}
A key observation is that after adding one $J_i$ edge and deleting another $J_i$ edge, a spanning 2-tree of $\widetilde{J}_i$ becomes another spanning 2-tree of $\widetilde{J}_i$. Therefore, the operations in (\ref{eq:canonical_Fp2Rterm_to_FP2Rterm}) ensures that $\widetilde{J}_i \cap T^2(\r)$ is a spanning 2-tree of $\widetilde{J}_i$ if $\x^{\r}$ is an $\mathcal{F}^{(p_i^2,R)}$ term. By using corollary~\ref{lemma-onshell_Fp2_external_less_equal_minusone_corollary1}, which maps any $\mathcal{U}^{(R)}$ term to a corresponding $\mathcal{F}^{(p_i^2,R)}$ term, it then follows that $\widetilde{J}_i \cap T^2(\r)$ is a spanning tree of $\widetilde{J}_i$ for any $\mathcal{U}^{(R)}$ term $\x^{\r}$.

We have thus proved all the statements of this lemma.
\end{proof}

Let's summarize the results into the following table:
\begin{center}
\begin{tabular}{ |c||c|c|c| } 
\hline
 & \multirow{2}{3em}{$H\cap T$} & \multirow{2}{3.3em}{$\widetilde{J}_i \cap T$} & \multirow{2}{3em}{$\widetilde{S} \cap T$} \\
 & & & \\ \hline
$\mathcal{U}^{(R)}$ terms & spanning tree & spanning tree & spanning tree \\ \hline
$\mathcal{F}^{(p_i^2,R)}$ terms & spanning tree & spanning 2-tree & spanning tree \\ \hline
\end{tabular}
\end{center}
In section~\ref{section-some_basics_graph_theory} we have defined the notion $n_\gamma$: for a leading term~$\x^{\r}$, $n_\gamma$ is the number of edges of $\gamma$ that are removed to obtain $T(\r)$. Equivalently, $n_\gamma$ is the edge number of $\gamma\setminus T(\r)$. We can then deduce $n_H$, $n_{J_i}$, and $n_S$ for each leading term from lemma~\ref{lemma-onshell_subgraphs_tree_structures}.
\begin{corollary}
\label{lemma-onshell_subgraphs_tree_structures_corollary1}
For any given region $R$, the leading terms are characterized by the following equations:
\begin{subequations}
\begin{align}
    \label{eq:lemma_onshell_leading_U_generic_form}
    \mathcal{U}^{(R)}:&\quad n_H=L(H),\quad n_{J_i} = L(\widetilde{J}_i)\ \ \forall i,\quad n_S=L(\widetilde{S});\\
    \label{eq:lemma_onshell_leading_Fp2_generic_form}
    \mathcal{F}^{(p_i^2,R)}:&\quad n_H=L(H),\quad n_{J_i} = L(\widetilde{J}_i)+1,\quad n_{J_j} = L(\widetilde{J}_j)\ \ \forall j\neq i,\quad n_S=L(\widetilde{S});\\
    \label{eq:lemma_onshell_leading_Fq2_generic_form}    
    \mathcal{F}^{(q^2,R)}:&\quad n_H +n_J = L(H\cup J) + k+1,\quad n_S=L(\widetilde{S})-k\quad (k=0,1,\dots).
\end{align}
\end{subequations}
\end{corollary}
\begin{proof}
The equations (\ref{eq:lemma_onshell_leading_U_generic_form}) and (\ref{eq:lemma_onshell_leading_Fp2_generic_form}) are direct results from lemma~\ref{lemma-onshell_subgraphs_tree_structures}. In detail, to obtain a spanning tree (spanning 2-tree) of $\gamma$ where the subgraph $\gamma= H,\widetilde{J}_i,\widetilde{S}$, one need to remove exactly $L(\gamma)$ ($L(\gamma)+1$) edges from $\gamma$. To verify eq.~(\ref{eq:lemma_onshell_leading_Fq2_generic_form}), a key observation is $n_S\leqslant L(\widetilde{S})$. This is because $n_S> L(\widetilde{S})$ would imply that the graph $\widetilde{S} \cap T$ is disconnected, one of whose components consists exclusively of the edges of $S$ which do not attach to $H\cup J$. Momentum conservation of this disconnected component implies that the total momentum flowing into it vanish, and hence it would not contribute to the Symanzik polynomials. Equivalently, $n_S=L(\widetilde{S})-k$ for some natural number $k$. Moreover, since
\begin{eqnarray}
n_H+n_J+n_S = L(G) +1 = L(H\cap J) + L(\widetilde{S}) +1,
\end{eqnarray}
for any $\mathcal{F}$ term, we immediately obtain $n_H +n_J = L(H\cup J) + k+1$. Note that in the last equality we have used the result of corollary~\ref{lemma-onshell_hard_subgraph_connected_corollary2}.
\end{proof}

Eq.~(\ref{eq:lemma_onshell_leading_U_generic_form}) indicates that every spanning tree that corresponds to a $\mathcal{U}^{(R)}$ term can be divided into the spanning trees of $H,\widetilde{J}_1,\dots,\widetilde{J}_K,\widetilde{S}$. In fact, the reverse of this statement is also true. Namely, \emph{the graph $t_H\cup t_{J_1}\cup\dots\cup t_{J_K}\cup t_S$ is a minimum spanning tree of $G$ (hence corresponds to a $\mathcal{U}^{(R)}$ term), where $t_\gamma$ is any minimum spanning tree of $\widetilde{\gamma}$, for each $\gamma\in \{H,J_1,\dots,J_K,S\}$.} (Note that $\widetilde{H}\equiv H$ here.) This is because the combination of any chosen spanning trees of $H$, $\widetilde{J}_1,\dots,\widetilde{J}_K$, $\widetilde{S}$ is a spanning tree of $G$, which, in line with eq.~(\ref{eq:lemma_onshell_leading_U_generic_form}), is of the minimum weight. This observation allows us to further study the structure of the hard subgraph.
\begin{corollary}
\label{lemma-onshell_subgraphs_tree_structures_corollary2}
    For any two vertices $v_1^H,v_2^H\in H$ that are both incident with some edges of the same jet $J_i$, there exists a path $P^H\subset H$ connecting $v_1^H$ and $v_2^H$, such that every edge of $P^H$ satisfies $w>-1$.
\end{corollary}
\begin{proof}
    Since each jet is connected, there must be two paths $P_1,P_2\subset J_i$, starting at the same vertex of $J_i$ and ending at $v_1^H$ and $v_2^H$ respectively. From the observation above, there always exists a $\mathcal{U}^{(R)}$ term $\x^{\r}$, such that $P_1\subset T^1(\r)$.
    
    We now define $P^H$ to be the path in $T^1(\r)$, which connects $v_1^H$ and $v_2^H$. By definition, $v_1^H$ is in the $p_i$ trunk, while $v_2^H$ can belong to either the trunk or the branch of $T^1(\r)$. For the most general case, let us consider the following configuration: $P^H$ starts from $v_1^H$ in the $p_i$ trunk of $T^1(\r)$, next enters the $q$ trunk at $v_3^H$, then leaves the $q$ trunk at $v_4^H$, and finally ends at $v_2^H$ in the $q$ branch (see figure~\ref{onshell_lemma10_corollary1_original}). To illustrate that $w(e^H)>-1$ holds for each edge $e^H$ in this specific $P^H$, note that any $e^H$ between $v_1^H$ and $v_3^H$ satisfies $w(e^H)>-1$ from corollary~\ref{lemma-onshell_Fp2_internal_less_equal_minusone_corollary2}, and any $e^H$ between $v_3^H$ and $v_4^H$ satisfies $w(e^H)=0>-1$ from lemma~\ref{lemma-onshell_basic_weight_structure_Uterm}. Finally, if there exists an $e^H$ between $v_4^H$ and $v_2^H$ with $w(e^H)=-1$, the configuration in figure~\ref{onshell_lemma10_corollary1_comparison} would correspond to an $\mathcal{F}^{(p_i^2,R)}$ term because its weight is equal to $w(T^1(\r))$. However, its $p_i$ component includes a hard vertex $v_2^H$, which contradicts \emph{3-2} of corollary~\ref{lemma-onshell_Fp2_vertices_universal_property_corollary2}. This contradiction implies that $w(e^H)>-1$ must hold for every edge $e^H\in P^H$.
    \begin{figure}[t]
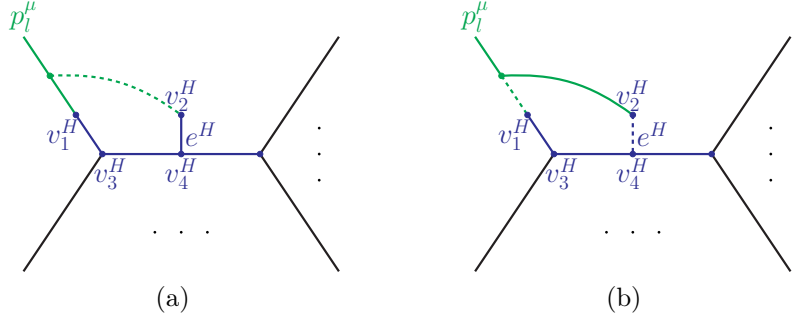

    \centering
    \begin{subfigure}[b]{0.3\textwidth}
    \centering
    \include{figs/onshell_lemma10_corollary1_original}
    \vspace{-3em}
    \caption{}
    \label{onshell_lemma10_corollary1_original}
    \end{subfigure}
    \hspace{3em}
    \begin{subfigure}[b]{0.3\textwidth}
    \centering
    \include{figs/onshell_lemma10_corollary1_comparison}
    \vspace{-3em}
    \caption{}
    \label{onshell_lemma10_corollary1_comparison}
    \end{subfigure}
    \caption{A graphic illustration of the analysis used to prove corollary~\ref{lemma-onshell_Fq2Rterms_kgeq1_tree_structures_corollary1}. (a) The spanning tree $T^1(\r')$ corresponding to a $\mathcal{U}^{(R)}$ term $\x^{\r'}$, where $P_1\subset T^1(\r)$ is the path whose endpoints are $v_1^H$ and $v_2^H$. We further suppose that there is a single edge $e^H$ between $v_4^H$ and $v_2^H$. (b) If $w(e^H)=-1$, then we can construct a spanning 2-tree corresponding to an $\mathcal{F}^{(p_l^2,R)}$ term, where $v_2^H$ is in its $p_l$ component.}
    \label{figure-onshell_lemma10_corollary1}
    \end{figure}
\end{proof}

So far, we have defined the hard, jet, and soft subgraphs associated with each given region and explored some of their fundamental properties. The weight structure of the jets is straightforward: $w(e)=-1$ for any $e\in J$. As for $H$ and $S$, at this stage, we only know that $w(e)<-1$ for $e\in S$ and $w(e)\geqslant -1$ for $e\in H$, and it remains to determine their precise weight structures. To this end, we shall consider partitioning a generic soft subgraph $S$ into subgraphs according to the weights of their edges.

\subsubsection{The soft substructure}
\label{section-soft_substructure}

Let us consider the weight-ordered partition of $S$. In accordance with eq.~(\ref{eq:weight_hierarchical_partition_relation}), we set
\begin{eqnarray}
S\equiv \bigsqcup_{i=1}^M S_i =\bigsqcup_{i=1}^M \bigsqcup_{j=1}^{m_i} s_{i,j},
\label{eq:weight_ordered_partition_S}
\end{eqnarray}
where, for each fixed $i\in \{ 1,\dots,M \}$ and each $e\in S_i$, we have $w(e)=w_i$. The weights $w_1,\dots,w_M$ satisfy $w_1<\dots<w_M<-1$. Each graph $s_{i,j}$ represents one connected component of $S_i$, meaning that $S_i= \cup_{j=1}^{m_i} s_{i,j}$.

As a first observation, the configuration of the graphs $\{s_{i,j}\}_{\tiny \begin{matrix} i=1,\dots,M\\ j=1,\dots,m_i \end{matrix}}$ is constrained by the following lemma.
\begin{lemma}
\label{lemma-onshell_sij_configuration_constraints}
For each i,j, $s_{i,j}$ is adjacent to the following graph
\begin{eqnarray}
\bigcup_{i'>i}^M S_{i'}\cup H\cup J. \nonumber
\end{eqnarray}
\end{lemma}

\begin{proof}
We shall prove this lemma by contradiction. That is, we start by assuming the existence of an $s_{a,b}$, such that all the edges adjacent to it belong to $\cup_{i'\leqslant a} S_{i'}$.

It follows from the definition of weight-ordered partitions (in section~\ref{section-weight_ordered_partition}) that, for any edge of $s_{a,b}$, its endpoints are both in $s_{a,b}$. So any $s_{a,b'}$ with $b'\neq b$ cannot be adjacent to $s_{a,b}$, otherwise $s_{a,b}$ and $s_{a,b'}$ would be contained in the same component of $S_a$. As a result, all the edges adjacent to $s_{a,b}$ belong to $\cup_{i'< a} S_{i'}$, as described by figure~\ref{figure-onshell_lemma8_proof_G}, where $s_{a,b}$ is represented by the grey blob and the edges of $\cup_{i'<a} S_{i'}$ are red.
\begin{figure}[t]
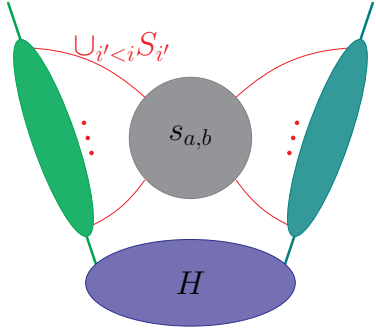

\centering
\include{figs/onshell_lemma8_proof_G}
\vspace{-2em}
\caption{The configuration of $s_{a,b}$ (grey) which is not adjacent to $H\cup J$. In this case only the edges of $\cup_{i'<i} S_{i'}$ (red) are adjacent to $s_{a,b}$.}
\label{figure-onshell_lemma8_proof_G}
\end{figure}

The statement \emph{1} of lemma~\ref{lemma-existence_certain_terms_P} indicates that $n_{s_{a,b}}> L(s_{a,b})$ holds for some leading term~$\x^{\r}$, which further implies that $s_{a,b}\cap T(\r)$ is disconnected, and there exists an edge $e_0\in s_{a,b}\setminus T(r)$ whose endpoints are respectively in distinct components of $s_{a,b}\cap T(r)$. Meanwhile, according to corollary~\ref{lemma-onshell_subgraphs_tree_structures_corollary1}, the graph $\widetilde{S}\cap T(\r)$ must be connected.\footnote{More precisely, $\widetilde{S}\cap T(\r)$ is a spanning tree of $\widetilde{S}$ if $\x^{\r}$ is a $\mathcal{U}^{(R)}$ or $\mathcal{F}^{(p^2,R)}$ term (see eqs.~(\ref{eq:lemma_onshell_leading_U_generic_form}) and (\ref{eq:lemma_onshell_leading_Fp2_generic_form})), and $\widetilde{S}\cap T(\r)$ may contain loops if $\x^{\r}$ is an $\mathcal{F}^{(q^2,R)}$ terms (see eq.~(\ref{eq:lemma_onshell_leading_Fq2_generic_form})). In either case, $\widetilde{S}\cap T(\r)$ is connected.}
Since the two connected components of $s_{a,b}\cap T(\r)$, which contain the endpoints of $e_0$, are subgraphs of $\widetilde{S}\cap T(\r)$, there must be a path $P\subset \widetilde{S}\cap T(\r)$ joining them (otherwise $\widetilde{S}\cap T(\r)$ would not be connected). As we have argued above, all the edges adjacent to $s_{a,b}$ belong to $\cup_{i'<a} S_{i'}$, then $P$ must contain some edges $e'\in \cup_{i'<a} S_{i'}$ (see figure~\ref{onshell_lemma8_proof_Fq2_term}). The graph $T(\r)\cup e_0\setminus e'$ then corresponds to another leading term, which is of the same type as~$\x^{\r}$ (see figure~\ref{onshell_lemma8_proof_Fq2_comparison}). Furthermore,
\begin{eqnarray}
w(T^2(\r')) = w(T^2(\r)) +w(e') -w(e_0) < w(T^2(\r)),
\end{eqnarray}
where we have used $w(e') < w(e_0)$ because $e'\in \cup_{i'<a} S_{i'}$ and $e_0\in s_{a,b}$. The inequality above, however, violates the minimum-weight criterion.
\begin{figure}[t]
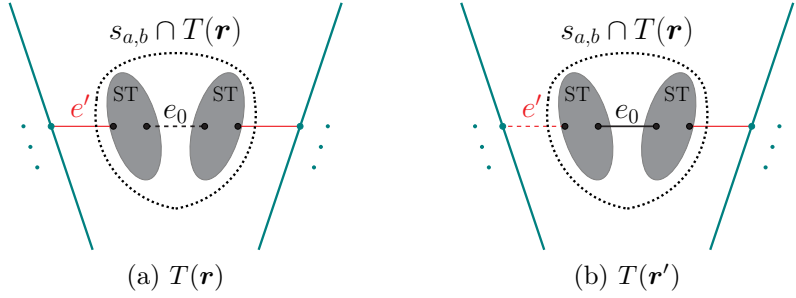

\centering
\begin{subfigure}[b]{0.3\textwidth}
\centering
\include{figs/onshell_lemma8_proof_Fq2_term}
\vspace{-3em}
\caption{$T(\r)$}
\label{onshell_lemma8_proof_Fq2_term}
\end{subfigure}
\hspace{3em}
\begin{subfigure}[b]{0.3\textwidth}
\centering
\include{figs/onshell_lemma8_proof_Fq2_comparison}
\vspace{-3em}
\caption{$T(\r')$}
\label{onshell_lemma8_proof_Fq2_comparison}
\end{subfigure}
\caption{The spanning trees $T(\r)$ and $T(\r')\equiv T(\r)\cup e_0\setminus e'$.}
\label{figure-onshell_lemma8_proof_Fq2}
\end{figure}

We have thus completed the proof by contradiction, showing that each $s_{i,j}$ must be adjacent to $\cup_{i'>i}^M S_{i'}\cup H\cup J$.
\end{proof}

Note that an example illustrating the proof above will be provided later in (\ref{eq:3loop_nonplanar_example_lemma8}).

For later convenience, let us assume that $w_{\overline{m}}=-2$ for some $\overline{m}\in \{1,\dots, M\}$ and define two subgraphs $S_\text{I},S_\text{II} \subset S$ as
\begin{align}
\label{eq:partition_S_according_to_mbar}
\begin{split}
    S_\text{I}\equiv S_1\cup\cdots \cup S_{\overline{m}-1},\qquad S_\text{II}\equiv S_{\overline{m}+1}\cup\cdots \cup S_M.
\end{split}
\end{align}
For the special case of $\overline{m}=1$, we have $S_\text{I}= \varnothing$ and $S_\text{II}= S_2\cup\dots\cup S_M$. It is clear from above that $S$ is partitioned into three subgraphs: $S= S_\text{I} \sqcup S_{\overline{m}} \sqcup S_\text{II}$. An application of lemma~\ref{lemma-onshell_sij_configuration_constraints} leads to the following constraints on the configurations of $S_{\overline{m}}$ and $S_\text{II}$.
\begin{corollary}
\label{lemma-onshell_sij_configuration_constraints_corollary1}
The soft subgraph $S$ satisfies the following conditions:
\begin{enumerate}
    \item [$\scriptstyle{\textup{\circled{1}}}$] for each $i\in \{1,\dots,M\}$, the graph $\cup_{i'>i}^M S_{i'}\cup H\cup J$ is connected;
    \item [$\scriptstyle{\textup{\circled{2}}}$] each component of the graph $S_\textup{II}$ must be adjacent to $H\cup J$, and each component of the graph $S_{\overline{m}}$ must be adjacent to $H\cup J\cup S_\textup{II}$;
    \item [$\scriptstyle{\textup{\circled{3}}}$] each component of $S_\textup{II}$ is adjacent to at most one jet.
\end{enumerate}
\end{corollary}

\begin{proof}
The proof of the first statement is straightforward, achieved by applying lemma~\ref{lemma-onshell_sij_configuration_constraints} recursively. First, for $i=M$ the graph $\cup_{i'>i}^M S_{i'}\cup H\cup J$ is exactly $H\cup J$, which is connected (corollary~\ref{lemma-onshell_hard_subgraph_connected_corollary1}). Next, for $i=M-1$ the graph $\cup_{i'>i}^M S_{i'}\cup H\cup J$ becomes $S_M\cup H\cup J$, which must also be connected since every component of $S_M$ is adjacent to $H\cup J$. The same reasoning can then be applied recursively to $i=M-2,\dots,1$ to establish the proof.

The second statement immediately follows from above. We note that, in particular, the graphs $H$, $H\cup J$, $H\cup J\cup S_\textup{II}$ and $H\cup J\cup S_\textup{II}\cup S_{\overline{m}}$ are all connected graphs.

To establish the third statement, let's consider its opposite, that one component of $S_\text{II}$, denoted as $\gamma$, is adjacent to multiple jets, including $J_i$ and $J_j$. According to lemma~\ref{lemma-weight_hierarchical_partition_tree_structure}, for any $\mathcal{U}^{(R)}$ term $\x^{\r}$, $\widetilde{\gamma}\cap T^1(\r)$ is a spanning tree of $\widetilde{\gamma}$. Let us modify $T^1(\r)$ by adding an edge $e\in \gamma$, such that there is a path $P\subset \gamma\cap T^1(\r)$ joining two jet vertices $v_i\in J_i$ and $v_j\in J_j$, and then removing two edges in $\widetilde{J}_i$ and $\widetilde{J}_j$, such that $P$ is in the same component of the obtained spanning 2-tree with $p_i^\mu$ and $p_j^\mu$. The corresponding $\mathcal{F}^{(q^2)}$ term $\x^{\r'}$ is associated with a weight of:
\begin{eqnarray}
w(T^2(\r')) = w(T^1(\r)) +2\times (-1) -w(e)< w(T^1(\r)),
\end{eqnarray}
where we have used $w(e)>-2$ (because $e\in S_\text{II}$) above. This violation of the minimum-weight criterion implies that each component of $S_\text{II}$ is adjacent to exactly one jet.
\end{proof}

Lemma~\ref{lemma-onshell_sij_configuration_constraints} and corollary~\ref{lemma-onshell_sij_configuration_constraints_corollary1} provide descriptions of the configurations of the weight-ordered subgraphs of $S$, and they lead to the following significant conclusion: the graph~$S_\text{I}$ is always empty in an on-shell expansion region.
\begin{lemma}
\label{lemma-onshell_Si_empty}
$S_\textup{I} =\varnothing$. Or equivalently, $\overline{m}=1$ in eq.~(\ref{eq:partition_S_according_to_mbar}).
\end{lemma}

\begin{proof}
We will prove this lemma by contradiction. Suppose $S_\text{I}\neq \varnothing$, which indicates that $\overline{m}>1$ in eq.~(\ref{eq:partition_S_according_to_mbar}), and $S_1\subseteq S_\text{I}$. Let us take any connected component of $S_1$ and denote it by $\gamma_1$. Below we examine the structure of $\widetilde{\gamma}_1\cap T(\r)$ for any given leading term $\x^{\r}$.

\begin{itemize}
    \item If $\x^{\r}$ is a $\mathcal{U}^{(R)}$ term, then $T^1(\r)$ is a minimum spanning tree of $G$. As shown in corollary~\ref{lemma-onshell_sij_configuration_constraints_corollary1} that $\cup_{i'>i}^M S_{i'}\cup H\cup J$ is connected for each $i$. Taking $i=1$, it follows that $G\setminus S_1$ is connected. We can then apply lemma~\ref{lemma-weight_hierarchical_partition_tree_structure}, indicating that $\widetilde{\gamma}_1\cap T^1(\r)$ is a spanning tree of $\widetilde{\gamma}_1$.

    \item If $\x^{\r}$ is an $\mathcal{F}^{(p^2,R)}$ term, then $\widetilde{\gamma}_1\cap T^2(\r)$ is also a spanning tree of $\widetilde{\gamma}_1$. This is because every $\mathcal{F}^{(p^2,R)}$ term becomes a $\mathcal{U}^{(R)}$ term after a jet edge is added to the corresponding spanning 2-tree (corollary~\ref{lemma-onshell_Fp2_external_less_equal_minusone_corollary1}), with the soft subgraph unchanged. So the result for $\mathcal{U}^{(R)}$ terms mentioned above still holds.
\end{itemize}

In other words, the polynomial $\mathcal{U}^{(R)}+\mathcal{F}^{(p^2,R)}$ satisfies a homogeneity property, namely, $n_{\gamma_1}= L(\widetilde{\gamma}_1)$. To meet the facet criterion, there must exist a $\mathcal{F}^{(q^2,R)}$ term, denoted as $\x^{\r}$, for which $n_{\gamma_1}\neq L(\widetilde{\gamma}_1)$ for this $\x^{\r_*}$, otherwise the homogeneity of $\mathcal{P}^{(R)}$ would impose an extra constraint on the facet. A consequence of $n_{\gamma_1}\neq L(\widetilde{\gamma}_1)$ is that $\gamma_1 \cap T^2(\r_*)$ is \emph{not} a spanning tree of $\gamma_1$: either it is disconnected, or it contains some loops. It is important to note that $\widetilde{\gamma}_1 \cap T^2(\r_*)$ cannot be disconnected, otherwise one of its components would consist solely of edges from $\gamma_1$ which do not attach to any other subgraphs of $G\setminus \gamma_1$. Momentum conservation of this disconnected component would lead to a vanishing total momentum, thus such a $T^2(\r_*)$ would not contribute to the Symanzik polynomials. Therefore, $\widetilde{\gamma}_1 \cap T^2(\r_*)$ must contain some loops. Each loop of $\widetilde{\gamma}_1 \cap T^2(\r_*)$ corresponds to a path in $\gamma_1\cap T^2(\r_*)$ with endpoints are in $G\setminus S_1$. On obtaining $\widetilde{\gamma}_1$ from $\gamma_1$, these endpoints are identified with the auxiliary vertex.

Let us consider the configuration where there are $k(\geqslant 1)$ loops in $\widetilde{\gamma}_1\cap T^2(\r_*)$, and the paths in $\gamma_1\cap T^2(\r_*)$, corresponding to these loops, are labelled by $A_1 B_1,\dots,A_k B_k$, where $A_i,B_i\in G\setminus S_1$ represent the endpoints of the $i$th path for each $i\in \{1,\dots,k\}$. A key observation is that, for each $i$, there exists another path $C_iD_i$, such that
\begin{align}
    \textup{(1)}\ \ A_iB_i\subseteq C_iD_i\subset S\cap T^2(\r_*);\qquad \textup{(2)}\ \ C_i,D_i \in H\cup J.\nonumber
\end{align}
A proof of this statement is given in appendix~\ref{appendix-details_lemma9_proof}. In figure~\ref{figure-onshell_lemma9_path_ABCD_examples} we show three examples of $T^2(\r_*)$ and specify the corresponding paths $A_iB_i$ and $C_iD_i$. In figure~\ref{onshell_lemma9_path_ABCD_example1} $S=S_1 = \gamma_1$, so $A_i,B_i\in H\cup J$ automatically, and $A_iB_i = C_iD_i$. In figure~\ref{onshell_lemma9_path_ABCD_example2} $B_i\in H\cup J$ while $A_i\in S\setminus \gamma_1$, so $D_i$ is identical to $B_i$ while $C_i\in H\cup J$ differs from $A_i$. In figure~\ref{onshell_lemma9_path_ABCD_example3} $A_i,B_i\in S\setminus \gamma_1$, which thus differ from $C_i,D_i\in H\cup J$.
\begin{figure}[t]
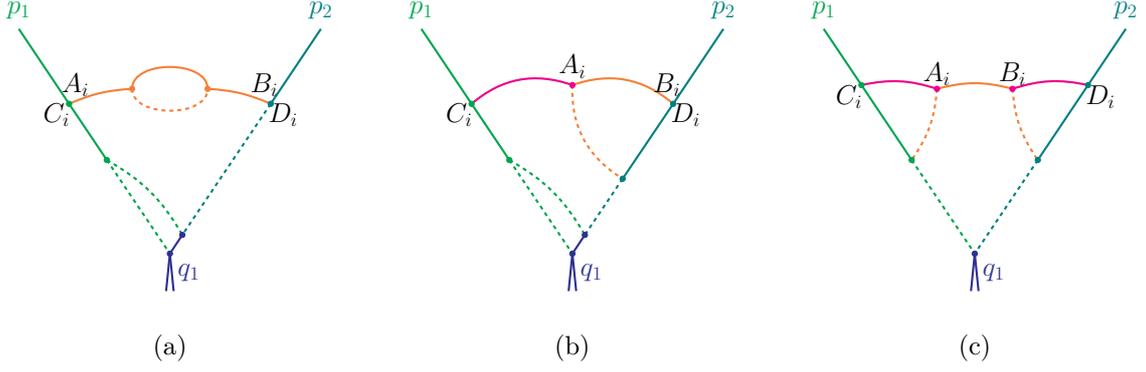

\centering
\begin{subfigure}[b]{0.3\textwidth}
\centering
\include{figs/onshell_lemma9_path_ABCD_example1}
\vspace{-2em}
\caption{}
\label{onshell_lemma9_path_ABCD_example1}
\end{subfigure}
\hfill
\begin{subfigure}[b]{0.3\textwidth}
\centering
\include{figs/onshell_lemma9_path_ABCD_example2}
\vspace{-2em}
\caption{}
\label{onshell_lemma9_path_ABCD_example2}
\end{subfigure}
\hfill
\begin{subfigure}[b]{0.3\textwidth}
\centering
\include{figs/onshell_lemma9_path_ABCD_example3}
\vspace{-2em}
\caption{}
\label{onshell_lemma9_path_ABCD_example3}
\end{subfigure}
\caption{Three examples of the spanning 2-tree $T^2(\r_*)$ ($\x^{\r_*}$ is an $\mathcal{F}^{(q^2,R)}$ term) containing a path $A_iB_i\in \gamma_1\cap T^2(\r_*)$, such that the endpoints $A_i$ and $B_i$ are both in $G\setminus S_1$. Note that the graph $S_1$ is colored in \textbf{\color{Orange} orange}, while $S\setminus S_1$ is in \textbf{\color{Magenta} magenta}.}
\label{figure-onshell_lemma9_path_ABCD_examples}
\end{figure}

This observation implies that there are $k$ paths in $T^2(\r_*)\cap S$ that join $H\cup J$. Since $T^2(\r_*)$ is a spanning 2-tree, $(H\cup J)\cap T^2(\r_*)$ has $k+2$ components in total. Equivalently, there exist $k+1$ edges $e_1^{HJ},\dots,e_{k+1}^{HJ} \in H\cup J$, such that $T^2(\r_*) \cup (e_1^{HJ}\cup\dots\cup e_{k+1}^{HJ})$ is a connected tree graph. We then take any $k$ edges $e_1^{\gamma_1},\dots,e_k^{\gamma_1}\in \gamma_1$, such that $e_i^{\gamma_1}\in A_iB_i$ for each $i$. The graph
\begin{eqnarray}
T^2(\r_*) \cup (e_1^{HJ}\cup\dots\cup e_{k+1}^{HJ}) \setminus (e_1^{\gamma_1}\cup\dots\cup e_k^{\gamma_1})\nonumber
\end{eqnarray}
is then a spanning tree of $G$, which corresponds to a $\mathcal{U}$ term $\x^{\r'}$, and $w(T^2(\r_*))$ can be related to $w(T^1(\r'))$ as:
\begin{eqnarray}
w(T^1(\r')) -w(T^2(\r_*)) &&= w(e_1^{\gamma_1}) +\dots+ w(e_k^{\gamma_1}) -w(e_1^{HJ}) -\dots-w(e_{k+1}^{HJ}) \nonumber\\
&&< -2\cdot k - (-1)\cdot (k+1) =-k+1 \leqslant 0.
\end{eqnarray}
In the first inequality, we have used that $w(e)<-2$ for $e\in \gamma_1$ and $w(e)\geqslant -1$ for $e\in H\cup J$. This direct calculation shows that $w(T^2(\r')) <w(T^2(\r_*))$, implying that $\x^{\r_*}$ would not contribute to the leading polynomial $\mathcal{P}^{(R)}(\x;\s)$.

We have thus excluded the possibility of $w<-2$ for any edge, and derived $S_\text{I}=\varnothing$.
\end{proof}

Note that an example illustrating the proof above will be provided later in (\ref{eq:3loop_nonplanar_example_lemma9}).

As a consequence of lemma~\ref{lemma-onshell_Si_empty}, each region divides the entire graph $G$ into four subgraphs: $H$, $J$, $S_\text{II}$, and $S_1$. The weight structures of $J$ and $S_1$ have already been determined: $w(e)=-1$ for any $e\in J$ and $w(e)=-2$ for any $e\in S_1$. Now, it is necessary to establish the weight structures of $H$ and $S_\text{II}$. To achieve this, we group $\{s_{i,j}\}$ into two graphs, $S_\text{II}^H$ and $S_\text{II}^J$. This involves the following three steps.
\begin{enumerate}
    \item We begin by assigning the graphs $\{s_{M,j}\}_{j=1,\dots,m_M^{}}$. From lemma~\ref{lemma-onshell_sij_configuration_constraints}, each given $s_{M,j}$ must be adjacent to $H\cup J$. Then for each $k\in \{1,\dots, K\}$, the union of those $s_{M,j}$ that are adjacent to $J_k$ is denoted as $S_M^{J_k}$. The union of the remaining $s_{M,j}$ (each of which is adjacent to $H$ but \emph{not} to $J$) is denoted as $S_M^H$. By defining $S_M^J\equiv \cup_{k=1}^K S_M^{J_k}$, we have effectively partitioned $S_M$ into $S_M=S_M^H\sqcup S_M^J$.
    
    \item Following the same procedure, we assign the graphs $\{s_{M-1,j}\},\dots,\{s_{2,j}\}$ recursively. For each $k\in \{1,\dots,K\}$, the union of those $s_{i,j}$ that are adjacent to $\cup_{i'>i}^M S_{i'}^{J_k}\cup J_k$ is denoted as $S_i^{J_k}$. The union of the remaining $s_{i,j}$ (each of which is adjacent to $\cup_{i'>i}^M S_{i'}^H\cup H$ but not to $\cup_{i'>i}^M S_{i'}^{J_k}\cup J_k$) is denoted as $S_i^H$.
    
    We further define $S_i^J \equiv \cup_{k=1}^K S_M^{J_k}$. From these constructions, we have
    \begin{eqnarray}
    \label{eq:definition_Si}
        S_i=S_i^H\sqcup S_i^J, \qquad \forall i\in \{2,\dots,M\}.
    \end{eqnarray}
    
    \item By defining
    \begin{eqnarray}
    S_\text{II}^H \equiv \bigcup_{i=2}^M S_i^H, \qquad S_\text{II}^{J_k} \equiv \bigcup_{i=2}^M S_i^{J_k}, \qquad S_\text{II}^J \equiv \bigcup_{i=2}^M S_i^J.
    \end{eqnarray}
    It follows clearly that $S_\text{II} = S_\text{II}^H\sqcup S_\text{II}^J$.
\end{enumerate}
Let's illustrate these concepts with the Sudakov form factor below, where the external momenta are $p_1^\mu$, $p_2^\mu$ (on-shell), and $q_1^\mu$ (off-shell). In this example, we have $S = S_\text{II}$, consisting of edges $e_1,\dots,e_8$ (labelled by the red numbers 1-8 respectively) and all the red vertices, as shown below:
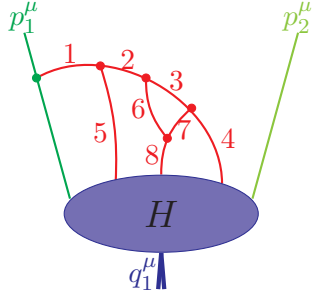
\begin{equation}
\begin{tikzpicture}[baseline=8ex, scale=0.4]

\draw [thick,color=Green] (0.5,8.5) -- (2,3);
\draw [thick,color=LimeGreen] (9.5,8.5) -- (8,3);
\draw [ultra thick, color=Blue] (4.9,0) -- (5,1.2);
\draw [ultra thick, color=Blue] (5.1,0) -- (5,1.2);

\path (0.9,7) edge [thick, Red, bend left=60] (7,3.5) {};
\path (6,6) edge [thick, Red, bend right=25] (5,3.8) {};
\path (4.5,7) edge [thick, Red, bend right=20] (5.2,5) {};
\path (3,7.4) edge [thick, Red, bend left=10] (3.5,3.5) {};

\node [draw, circle, minimum size=3pt, color=Green, fill=Green, inner sep=0pt, outer sep=0pt] () at (0.9,7) {};
\node [draw, circle, minimum size=3pt, color=Red, fill=Red, inner sep=0pt, outer sep=0pt] () at (3,7.4) {};
\node [draw, circle, minimum size=3pt, color=Red, fill=Red, inner sep=0pt, outer sep=0pt] () at (4.5,7) {};
\node [draw, circle, minimum size=3pt, color=Red, fill=Red, inner sep=0pt, outer sep=0pt] () at (6,6) {};
\node [draw, circle, minimum size=3pt, color=Red, fill=Red, inner sep=0pt, outer sep=0pt] () at (5.2,5) {};
\node [draw, ellipse, minimum height=2.67em, minimum width=6.67em, color=Blue, fill=Blue!60, inner sep=0pt,outer sep=0pt] () at (5,2.5) {};

\node () at (2,7.8) {\color{Red} $1$};
\node () at (3.9,7.6) {\color{Red} $2$};
\node () at (5.5,6.9) {\color{Red} $3$};
\node () at (7.2,5) {\color{Red} $4$};
\node () at (3,5.2) {\color{Red} $5$};
\node () at (4.25,6) {\color{Red} $6$};
\node () at (5.75,5.3) {\color{Red} $7$};
\node () at (4.7,4.45) {\color{Red} $8$};
\node () at (0.5,9) {\color{Green} $p_1^\mu$};
\node () at (9.5,9) {\color{LimeGreen} $p_2^\mu$};
\node () at (4.4,0.5) {\color{Blue} $q_1^\mu$};
\node () at (5,2.5) {\Large \color{Black} $H$};
\end{tikzpicture}.\nonumber
\end{equation}
We further impose the following constraint on the weights of the soft edges:
\begin{eqnarray}
w(e_3) < w(e_2) = w(e_6) < w(e_1) = w(e_4) = w(e_5) = w(e_7) = w(e_8).
\end{eqnarray}
With these constructions, the weight-ordered subgraphs of $S$ can be automatically defined: $S=\cup_{i=1}^4 S_i$; $S_1 = \varnothing$; $S_2$ consists of $e_3$ only; $S_3$ consists of edges $e_2$, $e_6$ and the vertex they are both incident with; $S_4$ consists of the remaining red edges and vertices. One can then follow the steps 1-3 above and obtain $S_\text{II}^H$ and $S_\text{II}^J$:
\begin{subequations}
\begin{align}
&S_4^H=
\begin{tikzpicture}[baseline=11ex, scale=0.4]
\path (6,6) edge [thick, Red, bend left=20] (7,3.8) {};
\path (6,6) edge [thick, Red, bend right=20] (5,3.8) {};
\node [draw, circle, minimum size=3pt, color=Red, fill=Red, inner sep=0pt, outer sep=0pt] () at (6,6) {};
\node [draw, circle, minimum size=3pt, color=Red, fill=Red, inner sep=0pt, outer sep=0pt] () at (5.3,5) {};
\node () at (7.1,5) {\color{Red} $4$};
\node () at (5.25,5.7) {\color{Red} $7$};
\node () at (4.7,4.45) {\color{Red} $8$};
\end{tikzpicture},
\quad S_4^{J_1}=
\begin{tikzpicture}[baseline=11ex, scale=0.4]
\path (6,6) edge [thick, Red, bend left=20] (7,3.8) {};
\path (6,6) edge [thick, Red, bend right=20] (5,3.8) {};
\node [draw, circle, minimum size=3pt, color=Red, fill=Red, inner sep=0pt, outer sep=0pt] () at (6,6) {};
\node () at (4.8,5) {\color{Red} $1$};
\node () at (7.1,5) {\color{Red} $5$};
\end{tikzpicture},
\quad S_3^{J_1}=
\begin{tikzpicture}[baseline=11ex, scale=0.4]
\path (6,6) edge [thick, Red, bend left=20] (7,3.8) {};
\path (6,6) edge [thick, Red, bend right=20] (5,3.8) {};
\node [draw, circle, minimum size=3pt, color=Red, fill=Red, inner sep=0pt, outer sep=0pt] () at (6,6) {};
\node () at (4.8,5) {\color{Red} $2$};
\node () at (7.1,5) {\color{Red} $6$};
\end{tikzpicture},
\quad S_2^{J_1}=
\begin{tikzpicture}[baseline=9ex, scale=0.4]
\path (5,3.8) edge [thick, Red, bend left=20] (7,3.8) {};
\node () at (6,4.4) {\color{Red} $3$};
\end{tikzpicture}\ ,\\
&S^H=
\begin{tikzpicture}[baseline=11ex, scale=0.4]
\path (6,6) edge [thick, Red, bend left=20] (7,3.8) {};
\path (6,6) edge [thick, Red, bend right=20] (5,3.8) {};
\node [draw, circle, minimum size=3pt, color=Red, fill=Red, inner sep=0pt, outer sep=0pt] () at (6,6) {};
\node [draw, circle, minimum size=3pt, color=Red, fill=Red, inner sep=0pt, outer sep=0pt] () at (5.3,5) {};
\node () at (7.1,5) {\color{Red} $4$};
\node () at (5.25,5.7) {\color{Red} $7$};
\node () at (4.7,4.45) {\color{Red} $8$};
\end{tikzpicture},
\quad S^J = S^{J_1} =
\begin{tikzpicture}[baseline=11ex, scale=0.4]
\path (5.5,7) edge [thick, Red, bend left=10] (6,3.8) {};
\path (5.5,7) edge [thick, Red, bend left=20] (7,3.8) {};
\path (6.3,6) edge [thick, Red, bend left=30] (8,3.8) {};
\path (5.5,7) edge [thick, Red, bend right=10] (4.5,3.8) {};
\node [draw, circle, minimum size=3pt, color=Red, fill=Red, inner sep=0pt, outer sep=0pt] () at (5.5,7) {};
\node [draw, circle, minimum size=3pt, color=Red, fill=Red, inner sep=0pt, outer sep=0pt] () at (6.3,6) {};
\node () at (4.4,5) {\color{Red} $1$};
\node () at (5.7,5) {\color{Red} $5$};
\node () at (6.5,5) {\color{Red} $6$};
\node () at (6.3,6.67) {\color{Red} $2$};
\node () at (8,5) {\color{Red} $3$};
\end{tikzpicture}\ .
\end{align}
\end{subequations}
Below we discuss some general properties of $S_\textup{II}^H$ and $S_\textup{II}^J$.

\begin{corollary}
\label{lemma-onshell_Si_empty_corollary1}
The graphs $S_\textup{II}^H$ and $S_\textup{II}^J$ satisfy the following properties.
\begin{enumerate}
    \item [$\scriptstyle{\textup{\circled{1}}}$] For each $i\in \{1,\dots,M\}$, the graphs $\cup_{i'=i}^M S_{i'}^H \cup H$ and $\cup_{i'=i}^M S_{i'}^J \cup J$ are both connected.
    
    \item [$\scriptstyle{\textup{\circled{2}}}$] Given any minimum spanning tree of $\widetilde{S}$, denoted as $T_S^1$, we have
    \begin{eqnarray}
    \label{eq:lemma10_corollary_Sj_Sh_tree_structures}
    n_{S_\textup{II}^J} = L(\widetilde{S}_\textup{II}^J),\quad n_{S_\textup{II}^H} = L(\widetilde{S}_\textup{II}^H),\quad n_{S_1} = L(\widetilde{S}_1).
    \end{eqnarray}
    
    \item [$\scriptstyle{\textup{\circled{3}}}$] If a soft vertex $v\in S_\textup{II}$ is simultaneously incident with edges from both $S_\textup{II}^H$ and $S_\textup{II}^J$, then $v\in S_\textup{II}^H$. Furthermore, there exists an edge $e_0^H\in S_\textup{II}^H$, such that (1) $v$ is incident with~$e_0^H$; (2) $w(e_0^H)>w(e^J)$ for any $e^J\in S_\textup{II}^J$ that is incident with $v$.
\end{enumerate}
\end{corollary}

\begin{proof}
The first statement is straightforward from the defining properties of $S_\text{II}^H$ and $S_\text{II}^J$: for each $i\in \{1,\dots,M\}$, $S_i^H$ is adjacent to $\cup_{i'>i}^M S_{i'}^H \cup H$ and $S_i^J$ is adjacent to $\cup_{i'>i}^M S_{i'}^J \cup J$. In particular, $S_\text{II}^H\cup H$ and $S_\text{II}^J\cup J$ are connected.

To justify the second statement, first note, based on corollary~\ref{lemma-onshell_sij_configuration_constraints_corollary1}, that for each $i$, the graph $\cup_{i'>i}^M S_{i'}\cup H\cup J$ is connected. This allows us to apply ($\Leftarrow$) of lemma~\ref{lemma-weight_hierarchical_partition_tree_structure} by treating $\Gamma = \widetilde{S}$. By applying this for each $i,j$, we conclude that $\widetilde{s}_{i,j}\cap T_S^1$ is a spanning tree of $\widetilde{s}_{i,j}$.
Then, as we have shown above, for each $i$, both $\cup_{i'=i}^M S_{i'}^H \cup H$ and $\cup_{i'=i}^M S_{i'}^J \cup J$ are connected. We can use this property to apply ($\Rightarrow$) of lemma~\ref{lemma-weight_hierarchical_partition_tree_structure} by treating $\Gamma = \widetilde{S}_\text{II}^H$ and $\widetilde{S}_\text{II}^J$. Finally, this allows us to conclude that $\widetilde{S}_\text{II}^H\cap T(\r)$ and $\widetilde{S}_\text{II}^J\cap T(\r)$ are (minimum) spanning trees of $\widetilde{S}_\text{II}^H$ and $\widetilde{S}_\text{II}^J$, respectively. These observations validate the first two equations in (\ref{eq:lemma10_corollary_Sj_Sh_tree_structures}).

Using the same reasoning above, it is straightforward to derive that $\widetilde{S}_1\cap T_S^1$ is a spanning tree of $\widetilde{S}_1$. Equivalently, $n_{S_1} = L(\widetilde{S}_1)$.

To justify the third statement, we consider the possibility that there exists an edge $e_0^J\in S_\text{II}^J$, such that $e_0^J$ is incident with $v$, and $w(e_H)\leqslant w(e_0^J)$ holds for any other edge $e_H\in S_\text{II}^H$ that is also incident with~$v$. By choosing any one of such $e_H$ and denoting the weight-ordered subgraph of $S$ containing $e_H$ as $s_{i_0,j_0}$, one can show that $v\in \cup_{i'=i_0}^M S_{i'}^J$.\footnote{To see this, suppose $e_*$ is an edge such that (1) $e_*$ is incident with $v$, and (2) $w(e)\leqslant w(e_*)$ for any other edge $e$ that is incident with $v$. Since we have assumed above that $w(e_H)\leqslant w(e_0^J)$ for any $e_H\in S_\text{II}^H$ that is incident with $v$, it follows that $e_*$ must belong to $S_\text{II}^J$. More precisely, $e_*\in \cup_{i'>i_0}^M S_{i'}^J$. The vertex $v$ must be in the same weight-ordered subgraph as $e_*$, so we also have $v\in \cup_{i'>i_0}^M S_{i'}^J$.} Then the subgraph $s_{i_0,j_0}$ is adjacent to $\cup_{i'>i_0}^M S_{i'}^J\cup J$, and by definition, it would have been partitioned as part of $S_{i_0}^J$. This contradicts our assumption that $e_H\in S_\text{II}^H$, thus we have verified the third statement.
\end{proof}

A direct result from corollary~\ref{lemma-onshell_Si_empty_corollary1} is that, eqs.~(\ref{eq:lemma_onshell_leading_U_generic_form}) and (\ref{eq:lemma_onshell_leading_Fp2_generic_form}), which characterize the $\mathcal{U}^{(R)}$ and $\mathcal{F}^{(p^2,R)}$ terms, can be rewritten into the following more precise forms.
\begin{corollary}
\label{lemma-onshell_Si_empty_corollary2}
The $\mathcal{U}^{(R)}$ and $\mathcal{F}^{(p^2,R)}$ terms are characterized by
\begin{subequations}
\allowdisplaybreaks
\begin{align}
    \begin{split}
    \label{eq:lemma12_leading_U_terms_forms}
        \boldsymbol{\mathcal{U}^{(R)}}:&\quad n_H = L(H),\quad n_{J_i}=L(\widetilde{J}_i)\ \ \forall i,\\
        &\quad n_{S^H} = L(\widetilde{S}^H),\quad n_{S^J} = L(\widetilde{S}^J),\quad n_{S_1} = L(\widetilde{S}_1).
    \end{split}\\
    \begin{split}
    \label{eq:lemma12_leading_Fp2_terms_forms}
        \boldsymbol{\mathcal{F}^{(p_i^2,R)}}:&\quad n_H = L(H),\quad n_{J_i}=L(\widetilde{J}_i)+1,\quad n_{J_j}=L(\widetilde{J}_j)\ \ \forall j\neq i,\\
        &\quad n_{S^H} = L(\widetilde{S}^H),\quad n_{S^J} = L(\widetilde{S}^J),\quad n_{S_1} = L(\widetilde{S}_1).
    \end{split}
\end{align}
\end{subequations}
\end{corollary}
\begin{proof}
For each $\mathcal{U}^{(R)}$ or $\mathcal{F}^{(p^2,R)}$ term~$\x^{\r}$, the graph $\widetilde{S}\cap T(\r)$ is a spanning tree of $\widetilde{S}$. Then eq.~(\ref{eq:lemma10_corollary_Sj_Sh_tree_structures}) holds automatically, which is exactly the statement of the corollary.
\end{proof}

We have thus characterized all the $\mathcal{U}^{(R)}$ and $\mathcal{F}^{(p^2,R)}$ terms using the loop numbers of $H$, $\widetilde{J}$, $\widetilde{S}^H$, $\widetilde{S}^J$ and $\widetilde{S}_1$. Our next goal is to characterize the $\mathcal{F}^{(q^2,R)}$ terms using the same parameters. To achieve this, we shall investigate all the possible configuration of their corresponding spanning 2-trees.

\subsubsection{Configurations of the \texorpdfstring{$\boldsymbol{\mathcal{F}^{(q^2,R)}}$}{TEXT} spanning 2-trees}
\label{section-constraints_Fq2R_terms}

So far, we have characterized each given $\mathcal{F}^{(q^2,R)}$ term $\x^{\r}$ using eq.~(\ref{eq:lemma_onshell_leading_Fq2_generic_form}), as part of lemma~\ref{lemma-onshell_subgraphs_tree_structures}. Now, we are particularly interested in the cases where $k\geqslant 1$, namely,
\begin{eqnarray}
\label{eq:onshell_leading_Fq2_generic_form_rewritten}
    n_H +n_J = L(H\cup J) + k+1,\quad n_S=L(\widetilde{S})-k\quad (k=1,2,\dots).
\end{eqnarray}
The equation $n_S=L(\widetilde{S})-k$ implies the existence of $k$ loops in the graph $\widetilde{S}\cap T^2(\r)$; equivalently, there are exactly $k$ paths in $S\cap T^2(\r)$ whose endpoints are in $H\cup J$. Let us refer to these paths as the \emph{principal soft paths}, and use $P_1,\dots,P_k$ to denote them. The general configuration of $T^2(\r)$ can then be described by the principal soft paths, as outlined by the following lemma.
\begin{lemma}
\label{lemma-onshell_Fq2Rterms_kgeq1_tree_structures}
For every $\mathcal{F}^{(q^2,R)}$ term $\x^{\r}$ that is characterized by
\begin{eqnarray}
    n_H +n_J = L(H\cup J) + k+1,\quad n_S=L(\widetilde{S})-k,\quad k\geqslant 1,\nonumber
\end{eqnarray}
one connected component of the spanning 2-tree $T^2(\r)$ must be shown in figure~\ref{figure-onshell_lemma10_generic_configuration}. In other words, it satisfies the following properties.
\begin{enumerate}
    \item It contains all the principal soft paths $P_1,\dots,P_k$. We thus denote it as $t(\r;P)$.
    \item The graph $(H\cup J)\cap t(\r;P)$ has $k+1$ connected components. Two of them are attached by two on-shell external momenta $p_i^\mu$ and $p_j^\mu$ respectively, while the other $k-1$ are not attached by any external momenta.
    \item The $k+1$ components of $(H\cup J)\cap t(\r;P)$ are aligned by the principal soft paths $P_1,\dots,P_k$. In particular, the $k-1$ components not attached by any external momenta are positioned between the two components attached by $p_i^\mu$ and $p_j^\mu$, respectively.
\end{enumerate}
\begin{figure}[t]
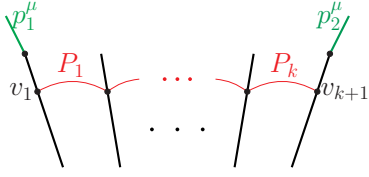

\centering
\include{figs/onshell_lemma10_generic_configuration}
\vspace{-3em}
\caption{The general configuration of $t(\r;P)$ as stated in lemma~\ref{lemma-onshell_Fq2Rterms_kgeq1_tree_structures}. The $k+1$ black line segments represent the components of $(H\cup J)\cap t(\r;P)$, and the $k$ red curves represent the principal soft paths $P_1,\dots,P_k$. Note that we have omitted the branches that can potentially attach to them in the figure.}
\label{figure-onshell_lemma10_generic_configuration}
\end{figure}
\end{lemma}

\begin{proof}
We justify the statements above through a series of observations below.
\begin{itemize}
    \item [(\emph{1})] \emph{if two principal soft paths, $P_a$ and $P_b$, are located in distinct components of $T^2(\r)$, then there is always a spanning 2-tree of $G$ whose weight is smaller than $w(T^2(\r))$.}
    
    For convenience, let us suppose that $P_a$ and $P_b$ are the only principal soft paths (the same analysis applies when more paths are present), which are in distinct components of $T^2(\r)$. We denote the components of $T^2$ containing $P_a$ and $P_b$ as $t(\r;a)$ and $t(\r;b)$, respectively. With this setup, we have $P_a\subset S\cap t(\r;a)$ and $P_b\subset S\cap t(\r;b)$. Both $(H\cup J)\cap t(\r;a)$ and $(H\cup J)\cap t(\r;b)$ have two connected components. We further denote the two components of $(H\cup J)\cap t(\r;a)$ as $\gamma_1^a$ and $\gamma_2^a$, and the components of $(H\cup J)\cap t(\r;b)$ as $\gamma_1^b$ and $\gamma_2^b$.
    
    A key characteristic is that any edge connecting $\gamma_1^a$ and $\gamma_2^a$ must have $w < -1$. To see the reason, consider the alternative: if there were an edge, denoted as $e'$ and connecting $\gamma_1^a$ and $\gamma_2^a$ with $w(e')\geqslant -1$, then adding $e'$ to the graph $T^2(\r)$ would create a loop in $t(\r;a)$. Moreover, this loop would include $P_a$. By removing any edge $e''\in P_a$, the graph $T^2(\r)\cup e' \setminus e''$ would become another spanning 2-tree. This new configuration clearly corresponds to an $\mathcal{F}^{(q^2)}$ term with a weight given by:
    \begin{eqnarray}
    w(T^2(\r)) + w(e'') - w(e') < w(T^2(\r)).
    \end{eqnarray}
    This inequality holds because $w(e'')< -1\leqslant w(e')$, and it contradicts the minimum-weight criterion. Therefore, every edge connecting $\gamma_1^a$ and $\gamma_2^a$ must satisfy $w<-1$. Similarly, the same condition applies to each edge connecting $\gamma_1^b$ and~$\gamma_2^b$ must also satisfy $w<-1$.
    
    For any two vertices $v_1,v_2\in H\cup J$ that are in distinct elements of $\{\gamma_1^a, \gamma_2^a, \gamma_1^b, \gamma_2^b\}$, it follows from the connectedness of $H\cup J$ that there is a path $P_*\subset H\cup J$ joining $v_1$ and $v_2$. All the edges of $P_*$ then satisfy $-1\leqslant w\leqslant 0$, and some of them join distinct components of $(H\cup J)\cap T^2(\r)$. Any one of these edges, according to the analysis above, \emph{cannot} connect $\gamma_1^a$ and $\gamma_2^a$, or connect $\gamma_1^b$ and $\gamma_2^b$. We thus conclude that there are some edges $e^{HJ}_1,\dots,e^{HJ}_n \in H\cup J$, such that
    \begin{itemize}
        \item [$\scriptstyle{\textup{\circled{1}}}$] for each of them, its endpoints are in $\gamma_i^a$ and $\gamma_j^b$ ($i,j\in \{1,2\}$) respectively;
        \item [$\scriptstyle{\textup{\circled{2}}}$] the graph $\gamma_1^a \cup \gamma_2^a \cup \gamma_1^b \cup \gamma_2^b \cup e_1^{HJ} \cup\dots\cup e_n^{HJ}$ is connected.
    \end{itemize}
    The general configuration of $T^2(\r)$ satisfying these conditions is shown in figure~\ref{onshell_Fq2term_multiple_soft_path_possibility}.
    \begin{figure}[t]
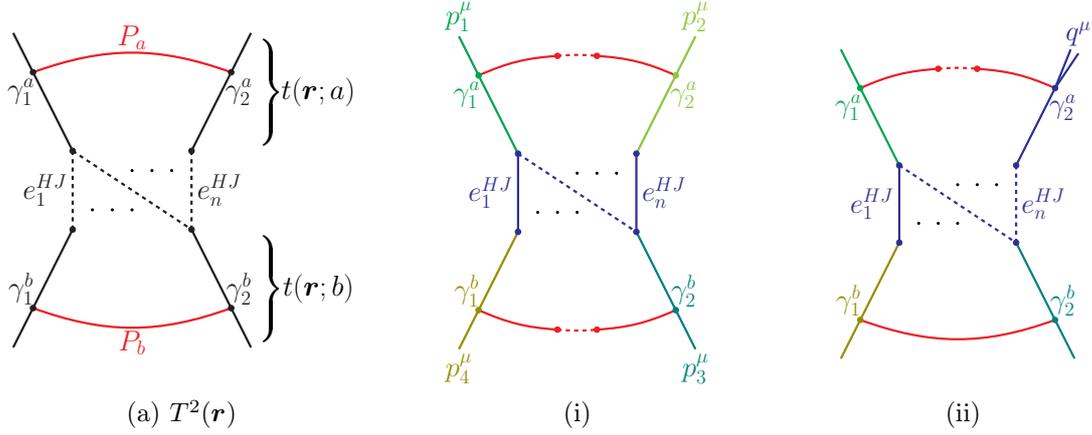

    \centering
    \begin{subfigure}[b]{0.32\textwidth}
    \centering
    \include{figs/onshell_Fq2term_multiple_soft_path_possibility}
    \vspace{-2em}
    \caption{$T^2(\r)$}
    \label{onshell_Fq2term_multiple_soft_path_possibility}
    \end{subfigure}
    \hfill
    \setcounter{subfigure}{0}
    \renewcommand\thesubfigure{\roman{subfigure}}
    \begin{subfigure}[b]{0.32\textwidth}
    \centering
    \include{figs/onshell_Fq2term_multiple_soft_path_modification1}
    \vspace{-3em}
    \caption{}
    \label{onshell_Fq2term_multiple_soft_path_modification1}
    \end{subfigure}
    \hfill
    \begin{subfigure}[b]{0.32\textwidth}
    \centering
    \include{figs/onshell_Fq2term_multiple_soft_path_modification2}
    \vspace{-2em}
    \caption{}
    \label{onshell_Fq2term_multiple_soft_path_modification2}
    \end{subfigure}
    \caption{A graphic illustration of the analysis in the proof of Statement (\emph{1}). (a): the configuration of $T^2(\r)$, one of whose components $t(\r;a)$ contains the soft path $P_a$ and the other $t(\r;b)$ contains the soft path $P_b$, with $P_a$ and $P_b$ both joining some vertices of $H\cup J$. (i): if both $t(\r;a)$ and $t(\r;b)$ are attached by two or more external momenta, one can modify $T^2(\r)$ by adding two edges in $H\cup J$ and deleting two edges in $S$, which are in $P_a$ and $P_b$ respectively, to obtain another $\mathcal{F}^{(q^2)}$ term with a smaller weight. (ii): if either $t(\r;a)$ or $t(\r;b)$ is attached by exactly one external momentum ($q^\mu$), one can add one edge in $H\cup J$ and delete one edge in $P_a$ or $P_b$ to obtain another $\mathcal{F}^{(q^2)}$ term with a smaller weight.}
    \label{figure-onshell_Fq2term_multiple_soft_paths}
    \end{figure}

    However, such an $\x^{\r}$ cannot be a leading term, because we can always find another $\mathcal{F}^{(q^2)}$ term with a smaller weight. If both $t(\r;a)$ and $t(\r;b)$ are attached by two or more external momenta, then one can modify $T^2(\r)$ by deleting two edges in $P_a$ and $P_b$ respectively, and adding two edges in $H\cup J$ (see figure~\ref{onshell_Fq2term_multiple_soft_path_modification1}). If either $t(\r;a)$ or $t(\r;b)$ is attached by exactly one (off-shell) external momentum, we can then modify $T^2(\r)$ by deleting one edge in $S$ and adding one edge in $H\cup J$ (see figure~\ref{onshell_Fq2term_multiple_soft_path_modification2}). In both cases, the weight of the obtained spanning 2-tree is smaller than $w(T^2(\r))$, because $w(e^S)<-1\leqslant w(e^{HJ})$ for any $e^S\in S$ and $e^{HJ}\in H\cup J$.
\end{itemize}
Therefore, we have ruled out the possibility of both components of $T^2(\r)$ containing principal soft paths. In simpler terms, all principal soft paths are within the same component of $T^2(\r)$. This allows us to designate the component of $T^2(\r)$ containing all the principal soft paths as $t(\r;P)$, and the other component, which has no principal soft paths, as $t(\r;\varnothing)$. By definition, $(H\cup J)\cap t(\r;P)$ comprises a total of $k+1$ connected components, which we will denote as $t_1(\r;P),\dots,t_{k+1}(\r;P)$. Namely,
\begin{eqnarray}
\label{eq:onshell_Fq2R_path_component_decomposition1}
    t(r;P)\cup t(\r;\varnothing) = T^2(\r),\qquad \cup_{i=1}^{k+1} t_i(\r;P) = (H\cup J)\cap t(\r;P).
\end{eqnarray}

With these notations, the following observation is immediate: for each $i=1,\dots,k+1$,
\begin{itemize}
    \item [(\emph{2})] \emph{there is an edge $e_i^{HJ}\in H\cup J$, whose endpoints are in $t_i(\r;P)$ and $t(\r;\varnothing)$ respectively.}
    
    This can be seen by considering any path $P_* \subset H\cup J$ whose endpoints are $v_1\in t_i(\r;P)$ and $v_2\in t(\r;\varnothing)$. Recall that in the proof of (\emph{1}) above, we have explained that none of the edges of $P_*$ can connect $t_i(\r;P)$ and $t_j(\r;P)$ (where $j\neq i$), so there must exist an edge $e_i^{HJ}\in P_*\subset H\cup J$, whose endpoints are in $t_i(\r;P)$ and $t(\r;\varnothing)$ respectively.
\end{itemize}

Observations (\emph{1}) and (\emph{2}) further lead to:
\begin{itemize}
    \item [(\emph{3})] \emph{$t(\r;P)$ is attached by on-shell external momenta only.}
    
    If $t(\r;P)$ is attached by an off-shell external momentum $q^\mu$, we can assume, without loss of generality, that $q^\mu$ attaches to $t_1(\r;P)$. We further suppose that the principal soft path $P_1$ connects $t_1(\r;P)$ and another component of $(H\cup J)\cap t(\r;P)$, say, $t_2(\r;P)$ (see figure~\ref{onshell_lemma10_observation3_original}). Then for any $e^S\in P_1$, $t(\r;P)\setminus e^S$ has two connected components: one containing $t_1(\r;P)$ and the other containing $t_2(\r;P)$.
    From (\emph{2}) above, there exists an edge $e_2^{HJ}\in H\cup J$ whose endpoints are in $t_2(\r;P)$ and $t(\r;\varnothing)$ respectively. The spanning 2-tree $T_*^2\equiv (T^2(\r) \cup e_2^{HJ})\setminus e^S$ (see figure~\ref{onshell_lemma10_observation3_comparison}) then corresponds to an $\mathcal{F}^{(q^2)}$ term since the momentum flowing between its components is off shell (more precisely, $q^\mu$). Furthermore,
    \begin{eqnarray}
        w(T_*^2) = w(T^2(\r)) + w(e^S) - w(e_2^{HJ}) < w(T^2(\r))
    \end{eqnarray}
    where we have used $w(e^S) < w(e^{HJ})$. This violates the minimum-weight criterion and rules out the possibility that $q^\mu$ attaches to $t(\r;P)$.
\end{itemize}

\begin{figure}[t]
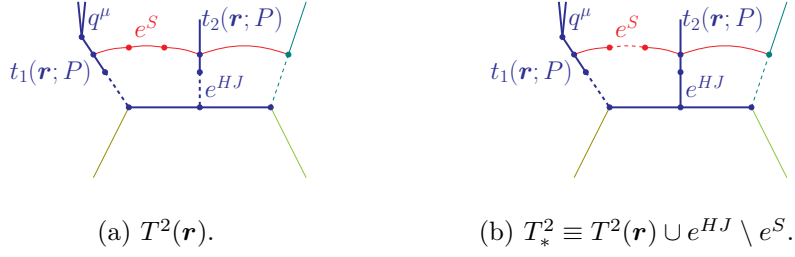

\centering
\begin{subfigure}[b]{0.3\textwidth}
\centering
    \include{figs/onshell_lemma10_observation3_original}
\vspace{-2em}\caption{$T^2(\r)$.}
\label{onshell_lemma10_observation3_original}
\end{subfigure}
\hspace{4em}
\begin{subfigure}[b]{0.3\textwidth}
\centering
    \include{figs/onshell_lemma10_observation3_comparison}
\vspace{-2em}\caption{$T_*^2\equiv T^2(\r)\cup e^{HJ}\setminus e^S$.}
\label{onshell_lemma10_observation3_comparison}
\end{subfigure}
\caption{A graphic illustration of the analysis of (\emph{3}). (a) The spanning 2-tree $T^2(\r)$ where an off-shell external momentum $q^\mu$ attaches to $t(\r;P)$. (b) The spanning 2-tree $T_*^2\equiv T^2(\r)\cup e^{HJ}\setminus e^S$. It can be verified directly that $w(T_*^2)<w(T^2(\r))$.}
\label{figure-onshell_lemma10_observation3}
\end{figure}

Using the same reasoning as mentioned above, we can exclude all configurations where two or more on-shell external momenta are attached to the same $t_i(\r;P)$. Consequently, only on-shell external momenta can attach to $t(\r;P)$, and any two of them must attach to distinct components of $(H\cup J)\cap t(\r;P)$, namely, $t_{i_1}(\r;P)$ and $t_{i_2}(\r;P)$ with $i_1\neq i_2$. Further analysis rules out the possibility that $t(\r;P)$ is attached by three or more on-shell external momenta:
\begin{itemize}
    \item [(\emph{4})] \emph{$t(\r;P)$ is attached by exactly two on-shell external momenta.}
    
    Let us suppose that there are three or more on-shell external momenta $p_1^\mu,p_2^\mu,\dots,p_n^\mu$ ($3\leqslant n\leqslant K$), which respectively attach to $t_1(\r;P),\dots, t_n(\r;P)$. Then there always exists a soft edge $e^S\in t(\r;P)$ and an index $i\in\{ 1,\dots,n \}$, such that $t_i(\r;P)$ is in one component of $t(\r;P)\setminus e^S$, while all the other $t_j(\r;P)$ ($j\neq i$) are in the other. An example with $i=1$ is in figure~\ref{onshell_lemma10_observation4_original}. According to (\emph{2}) above, there exists an edge $e_i^{HJ}\in H\cup J$, whose endpoints are in $t_i(\r;P)$ and $t(\r;\varnothing)$ respectively. The graph $T_*^2\equiv T^2(\r)\cup e_i^{HJ}\setminus e^S$ (see figure~\ref{onshell_lemma10_observation4_comparison}) is then another spanning 2-tree, and the momentum flowing between its components is $\sum_{j\neq i} p_j^\mu$, which must be off shell since there are two or more on-shell external momenta in the sum. $T_*^2$ then corresponds to an $\mathcal{F}^{(q^2)}$ term whose weight is
    \begin{eqnarray}
        w(T_*^2) = w(T^2(\r)) + w(e^S) - w(e_2^{HJ}) < w(T^2(\r)).
    \end{eqnarray}
    This violates the minimum-weight criterion, thus $t(\r;P)$ is attached by exactly two on-shell external momenta.
    \begin{figure}[t]
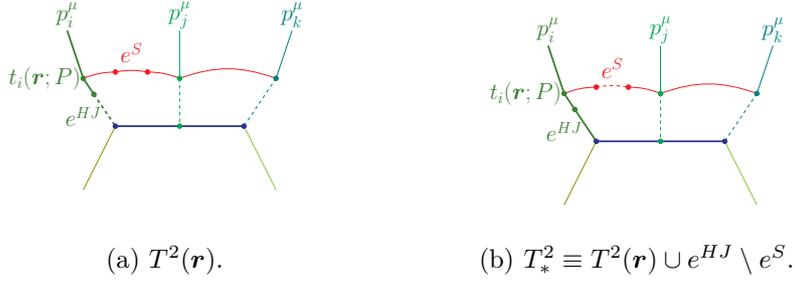

    \centering
    \begin{subfigure}[b]{0.3\textwidth}
    \centering
    \include{figs/onshell_lemma10_observation4_original}
    \vspace{-2em}
    \captionsetup{width=1.05\linewidth}
    \caption{$T^2(\r)$.}
    \label{onshell_lemma10_observation4_original}
    \end{subfigure}
    \hspace{4em}
    \begin{subfigure}[b]{0.3\textwidth}
    \centering
    \include{figs/onshell_lemma10_observation4_comparison}
    \vspace{-2em}
    \caption{$T_*^2\equiv T^2(\r)\cup e^{HJ}\setminus e^S$.}
    \label{onshell_lemma10_observation4_comparison}
    \end{subfigure}
    \caption{A graphic illustration of the analysis of (\emph{4}). (a) The spanning 2-tree $T^2(\r)$ where three on-shell external momenta $p_1^\mu,p_2^\mu,p_3^\mu$ attach to $t(\r;P)$. (b) The spanning 2-tree $T_*^2\equiv T^2(\r)\cup e_1^{HJ}\setminus e^S$. It can be verified directly that $w(T_*^2)<w(T^2(\r))$.}
    \label{figure-onshell_lemma10_observation4}
    \end{figure}
\end{itemize}

In the rest of the proof, we shall denote the two on-shell external momenta attaching to $t(\r;P)$ as $p_1^\mu$ and $p_2^\mu$. Furthermore, we relabel the graphs $t_1(\r;P),\dots, t_{k+1}(\r;P)$ as follows: $t_1(\r;P)$ is attached by $p_1^\mu$, $t_{k+1}(\r;P)$ is attached by $p_2^\mu$, while all the other $k-1$ graphs, $t_i(\r;P)$ ($i=2,\dots,k$), are attached by no external momenta.
Now, consider any edge within the principal soft paths and denote it as $e^S$, i.e. $e^S\in P_1\cup\dots\cup P_k$. After $e^S$ is removed from $t(\r;P)$, the resulting graph has two connected components, each of which contains a subset of $t_1(\r;P),\dots, t_{k+1}(\r;P)$. A key observation is as follows:
\begin{itemize}
    \item [(\emph{5})] \emph{for any edge $e^S$ in the principal soft paths, i.e. $e^S\in P_1\cup\dots\cup P_k$, the external momenta $p_1^\mu$ and $p_2^\mu$ must attach to distinct elements of $t(\r;P)\setminus e^S$.}

    To see this, we consider the contrary, i.e. both $t_1(\r;P)$ and $t_{k+1}(\r;P)$ are simultaneously in one component of $t(\r;P)\setminus e^S$. The other component of $t(\r;P)\setminus e^S$, then, can only contain $t_i(\r;P)$ for some $i\in \{2,\dots,k\}$. According to (\emph{2}) above, there exists some edge $e^{HJ}\in H\cup J$ whose endpoints are respectively in $t_i(\r;P)$ and $t(\r;\varnothing)$. Then the graph
    \begin{eqnarray}
        T_*^2\equiv T^2(\r)\cup e^{HJ} \setminus e^S
    \end{eqnarray}
    is another spanning 2-tree, one of whose components is attached by $p_1^\mu$ and $p_2^\mu$ only. It thus corresponds to an $\mathcal{F}^{(q^2)}$ term whose weight is smaller than $w(T^2(\r))$, violating the minimum-weight criterion. As a result, $t_1(\r;P)$ and $t_{k+1}(\r;P)$ must be contained in distinct components of $t(\r;P)\setminus e^S$ respectively.
\end{itemize}

\begin{figure}[t]
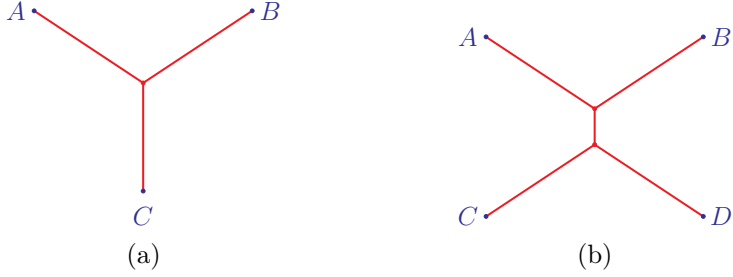

\centering
\begin{subfigure}[b]{0.3\textwidth}
\centering
\include{figs/onshell_lemma10_principal_overlap1}
\vspace{-3em}
\caption{}
\label{onshell_lemma10_principal_overlap1}
\end{subfigure}
\hspace{3em}
\begin{subfigure}[b]{0.3\textwidth}
\centering
\include{figs/onshell_lemma10_principal_overlap2}
\vspace{-3em}
\caption{}
\label{onshell_lemma10_principal_overlap2}
\end{subfigure}
\caption{Possible configurations for $P_a\cap P_b \neq \varnothing$. Note that the vertices $A,B,C$ and $D$ represent different components of $(H\cup J)\cap t(\r;P)$. (a) $P_a$ is the path from $A$ to $C$, and $P_b$ is the path from $B$ to $C$, such that $P_a\cap P_b$ shares an endpoint $C$ with $P_a$ and/or $P_b$; (b)~$P_a$ is the path from $A$ to $C$ and $P_b$ is the path from $B$ to $C$, such that $P_a\cap P_b$ does not share any endpoints with $P_a$ or $P_b$.}
\label{figure-onshell_lemma10_principal_overlap}
\end{figure}
Observation (\emph{5}) implies the following two properties of $t(\r;P)$. First,
\begin{itemize}
    \item [(\emph{5.1})]\emph{$P_a\cap P_b = \varnothing$ for any principal soft paths $P_a$ and $P_b$.}

    In the contrary case where $P_a\cap P_b \neq \varnothing$, whether $P_a\cap P_b$ shares an endpoint with $P_a$ or $P_b$ (as depicted in figure~\ref{onshell_lemma10_principal_overlap1}), or not (see figure~\ref{onshell_lemma10_principal_overlap2}), the following configuration emerges as a subgraph of $t(\r;P)$:
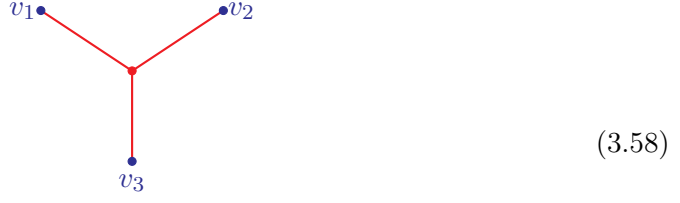
    \begin{equation}
    \label{eq:lemma10_proof_principal_overlap_result}
    \begin{tikzpicture}[baseline=8ex, scale=0.4]
    \draw [thick,color=Red] (2,8) -- (5,6);
    \draw [thick,color=Red] (8,8) -- (5,6);
    \draw [thick,color=Red] (5,3) -- (5,6);
    \node [draw, circle, minimum size=3pt, color=Red, fill=Red, inner sep=0pt, outer sep=0pt] () at (5,6) {};
    \node [draw, circle, minimum size=3pt, color=Blue, fill=Blue, inner sep=0pt, outer sep=0pt] () at (2,8) {};
    \node [draw, circle, minimum size=3pt, color=Blue, fill=Blue, inner sep=0pt, outer sep=0pt] () at (8,8) {};
    \node [draw, circle, minimum size=3pt, color=Blue, fill=Blue, inner sep=0pt, outer sep=0pt] () at (5,3) {};
    \node () at (1.4,8) {\color{Blue} $v_1$};
    \node () at (8.6,8) {\color{Blue} $v_2$};
    \node () at (5,2.3) {\color{Blue} $v_3$};
    \end{tikzpicture}
    \end{equation}
    where the vertices $v_1$, $v_2$ and $v_3$ are in distinct components of $(H\cup J)\cap t(\r;P)$. It then follows that at least one of $v_1,v_2,v_3$ is in $t_i(\r;P)$ for some $i\in \{2,\dots,k\}$, and there exists an edge $e^S\in P_a\cup P_b$ such that one component of $t(\r;P)\setminus e^S$ is attached by no external momenta. Thus $p_1^\mu$ and $p_2^\mu$ attach to the other component, which violates Observation (\emph{5}). Therefore, we must have $P_a\cap P_b = \varnothing$.
\end{itemize}
Another implication is that:
\begin{itemize}
    \item [(\emph{5.2})] \emph{the graphs $t_1(\r;P),\dots, t_{k+1}(\r;P)$ are aligned by the principal soft paths, such that the graphs $t_i(\r;P)$ ($i= 2,\dots,k$) are all between $t_1(\r;P)$ and $t_{k+1}(\r;P)$}.

    To see this, first note that all the components of $(H\cup J)\cap t(\r;P)$ must be aligned, otherwise configuration~(\ref{eq:lemma10_proof_principal_overlap_result}) would appear, which would violate (\emph{5}) above. Then, if some $t_i(\r;P)$ with $i\in \{ 2,\dots,k \}$ are not between $t_1(\r;P)$ and $t_{k+1}(\r;P)$, then there exists some edges $e^S$ such that $p_1^\mu$ and $p_2^\mu$ attach to the same component of $t(\r;P)\setminus e^S$, which would also violate Observation (\emph{5}).
\end{itemize}

To summarize, with Observations (\emph{1})-(\emph{5}) and their implications (\emph{5.1}) and (\emph{5.2}) mentioned above, we have established the validity of all the statements in this lemma. The graph $t(\r;P)$ can be effectively described using figure~\ref{figure-onshell_lemma10_generic_configuration}.
\end{proof}

Upon the insights from lemma~\ref{lemma-onshell_Fq2Rterms_kgeq1_tree_structures}, which comprehensively characterizes the properties of $t(\r;P)$, we next delve into the structure of $t(\r;\varnothing)$. To recap, $t(\r;\varnothing)$ is the other component of $T^2(\r)$, containing no principal soft paths, and attached by all the external momenta of $G$ except $p_i^\mu$ and $p_j^\mu$.
\begin{corollary}
\label{lemma-onshell_Fq2Rterms_kgeq1_tree_structures_corollary1}
    The graph $H\cap t(\r;\varnothing)$ is connected.
\end{corollary}
\begin{proof}
    As $t(\r;\varnothing)$ contains no soft paths connecting vertices of $H\cup J$, it follows that the graph $(H\cup J)\cap t(\r;\varnothing)$ is connected. In order to show the connectedness of $H\cap t(\r;\varnothing)$, we need to demonstrate the absence of the following structures within $t(\r;\varnothing)$:
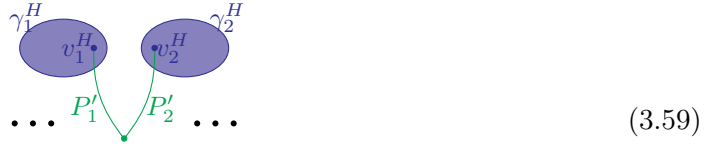
    \begin{equation}
    \label{eq:lemma10_corollary1_multiple_jet_edges_config}
    \begin{tikzpicture}[baseline=8ex, scale=0.4, mydot/.style={circle, fill, inner sep=0.8pt}]
    \node [draw, ellipse, minimum height=2em, minimum width=3em, color=Blue, fill=Blue!50, inner sep=0pt,outer sep=0pt] () at (3,6) {};
    \node [draw, ellipse, minimum height=2em, minimum width=3em, color=Blue, fill=Blue!50, inner sep=0pt,outer sep=0pt] () at (7,6) {};
    
    \path (5,3) edge [color=Green, bend left =20] (4,6) {};
    \path (5,3) edge [color=Green, bend right =20] (6,6) {};
    \node [draw, circle, minimum size=2pt, color=Blue, fill=Blue, inner sep=0pt, outer sep=0pt] () at (4,6) {};
    \node [draw, circle, minimum size=2pt, color=Blue, fill=Blue, inner sep=0pt, outer sep=0pt] () at (6,6) {};
    \node [draw, circle, minimum size=2pt, color=Green, fill=Green, inner sep=0pt, outer sep=0pt] () at (5,3) {};
    
    \path (1,3.5)-- node[mydot, pos=.2] {} node[mydot] {} node[mydot, pos=.8] {}(3,3.5);
    \path (9,3.5)-- node[mydot, pos=.2] {} node[mydot] {} node[mydot, pos=.8] {}(7,3.5);

    \node () at (3.7,4) {\small \color{Green} $P'_1$};
    \node () at (6.2,4) {\small \color{Green} $P'_2$};
    \node () at (3.5,6) {\small \color{Blue} $v_1^H$};
    \node () at (6.6,6) {\small \color{Blue} $v_2^H$};
    \node () at (1.8,7) {\small \color{Blue} $\gamma_1^H$};
    \node () at (8.4,7) {\small \color{Blue} $\gamma_2^H$};
    \end{tikzpicture}
    \end{equation}
    In the forbidden configuration above, two paths $P'_1,P'_2\subset J\cap t(\r;\varnothing)$ connect one jet vertex to two hard vertices, $v_1^H\in \gamma_1^H$ and $v_2^H\in \gamma_2^H$, respectively, where $\gamma_1^H$ and $\gamma_2^H$ represent two components of $H\cap t(\r;\varnothing)$. A key observation is that for any path $P^H\subset H$ connecting $\gamma_1^H$ and $\gamma_2^H$, all its vertices must be in $t(\r;\varnothing)$. To exclude the possibility that some vertices of $P^H$ are in $t(\r;P)$, one can apply the same reasoning of Observation (\emph{1}) in the proof of lemma~\ref{lemma-onshell_Fq2Rterms_kgeq1_tree_structures}, and show that $t(\r;P)\cap P^H\neq \varnothing$ would imply the existence of another $\mathcal{F}^{(q^2)}$ term whose weight is smaller than $w(T^2(\r))$, violating the minimum-weight criterion. As a result, all the vertices of $P^H$ must be in $t(\r;P)$.
    
    Next, from corollary~\ref{lemma-onshell_subgraphs_tree_structures_corollary2}, there exists a path $P^H$ connecting $v_1^H$ and $v_2^H$, whose edges all satisfy $w>-1$. It then follows that there is one particular edge $e_*^H\in P^H$ connecting $\gamma_1^H$ and $\gamma_2^H$, and $w(e_*^H)>-1$ from above. In the graph $t(\r;\varnothing)\cup e_*^H$, there is then a loop containing some edges from $P'_1\cup P'_2$. We take any edge $e^{J}\in P'_1\cup P'_2$, then the graph $T^2(\r)\cup e_*^H\setminus e^{J}$ would have the following structure, in contrast with the one in (\ref{eq:lemma10_corollary1_multiple_jet_edges_config}):
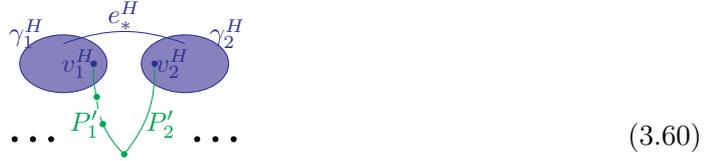
    \begin{equation}
    \label{eq:lemma10_corollary1_multiple_jet_edges_config_comparison}
    \begin{tikzpicture}[baseline=8ex, scale=0.4, mydot/.style={circle, fill, inner sep=0.8pt}]
    \node [draw, ellipse, minimum height=2em, minimum width=3em, color=Blue, fill=Blue!50, inner sep=0pt,outer sep=0pt] () at (3,6) {};
    \node [draw, ellipse, minimum height=2em, minimum width=3em, color=Blue, fill=Blue!50, inner sep=0pt,outer sep=0pt] () at (7,6) {};
    
    \path (3,6.67) edge [color=Blue, bend left =20] (7,6.67) {};
    \path (5,3) edge [color=Green, bend left =10] (4.3,4) {};\path (4.3,4) edge [dashed, color=Green, bend left =10] (4.1,4.9) {};\path (4.1,4.9) edge [color=Green, bend left =10] (4,6) {};
    \path (5,3) edge [color=Green, bend right =20] (6,6) {};
    \node [draw, circle, minimum size=2pt, color=Blue, fill=Blue, inner sep=0pt, outer sep=0pt] () at (4,6) {};
    \node [draw, circle, minimum size=2pt, color=Blue, fill=Blue, inner sep=0pt, outer sep=0pt] () at (6,6) {};
    \node [draw, circle, minimum size=2pt, color=Green, fill=Green, inner sep=0pt, outer sep=0pt] () at (5,3) {};
    \node [draw, circle, minimum size=2pt, color=Green, fill=Green, inner sep=0pt, outer sep=0pt] () at (4.3,4) {};
    \node [draw, circle, minimum size=2pt, color=Green, fill=Green, inner sep=0pt, outer sep=0pt] () at (4.1,4.9) {};
    
    \path (1,3.5)-- node[mydot, pos=.2] {} node[mydot] {} node[mydot, pos=.8] {}(3,3.5);
    \path (9,3.5)-- node[mydot, pos=.2] {} node[mydot] {} node[mydot, pos=.8] {}(7,3.5);

    \node () at (3.7,4) {\small \color{Green} $P'_1$};
    \node () at (6.2,4) {\small \color{Green} $P'_2$};
    \node () at (3.5,6) {\small \color{Blue} $v_1^H$};
    \node () at (6.6,6) {\small \color{Blue} $v_2^H$};
    \node () at (1.8,7) {\small \color{Blue} $\gamma_1^H$};
    \node () at (8.4,7) {\small \color{Blue} $\gamma_2^H$};
    \node () at (5,7.6) {\small \color{Blue} $e_*^H$};
    \end{tikzpicture}
    \end{equation}
Since $w(e^{J})=-1<w(e_*^H)$, the spanning 2-tree $T^2(\r)\cup e_*^H\setminus e^{J}$ corresponds to another $\mathcal{F}^{(q^2)}$ term whose weight is smaller than $w(T^2(\r))$. This contradicts the minimum-weight criterion, forbidding configurations where multiple paths in $J\cap t(\r;\varnothing)$ connect a jet vertex and $H$, as illustrated in (\ref{eq:lemma10_corollary1_multiple_jet_edges_config}). The corollary is thus proved.
\end{proof}

In proving corollary~\ref{lemma-onshell_Fq2Rterms_kgeq1_tree_structures_corollary1}, we have shown that for each given jet vertex $v^J\in t(\r;\varnothing)$, there is a unique path $P\subset J\cap t(\r;\varnothing)$ connecting $v^J$ and $H$. This has the following crucial implication: for each jet $J_l$, the graph $\widetilde{J}_l\cap t(\r;\varnothing)$ is a tree graph. This is because each component of $J_l\cap t(\r;\varnothing)$ is a tree graph, adjacent to a single hard vertex. After all the hard vertices are identified with the auxiliary vertex of $\widetilde{J}_l$, the graph $J_l\cap t(\r;\varnothing)$ becomes a graph containing no loops, which is $\widetilde{J}_l\cap t(\r;\varnothing)$ by definition.

In particular, for $l\neq i,j$, we will later show that $\widetilde{J}_l\cap T^2(\r)$ is a spanning tree of $\widetilde{J}_l$, for any $l\neq i,j$. Now, leveraging corollary~\ref{lemma-onshell_Fq2Rterms_kgeq1_tree_structures_corollary1}, we shall revisit the structure of $t(\r;P)$.
\begin{corollary}
\label{lemma-onshell_Fq2Rterms_kgeq1_tree_structures_corollary2}
    For all the graphs $t_1(\r;P),\dots,t_{k+1}(\r;P)$, suppose $t_1(\r;P)$ is attached by $p_i^\mu$ and adjacent to the principal soft path $P_1$ at vertex $v_1$, while $t_{k+1}(\r;P)$ is attached by $p_j^\mu$ and adjacent to $P_k$ at $v_{k+1}$ (as shown in figure~\ref{figure-onshell_lemma10_generic_configuration}). Then:
    \begin{enumerate}
        \item [$\scriptstyle{\textup{\circled{1}}}$] $t_1(\r;P)\cap J_l =\varnothing = t_{k+1}(\r;P)\cap J_l$ for any $l\neq i,j$;        
        \item [$\scriptstyle{\textup{\circled{2}}}$] $t_a(\r;P)\subset H$ for all the $a=2,\dots,k$;        
        \item [$\scriptstyle{\textup{\circled{3}}}$] $v_1\in J_i\ \Leftrightarrow\ t_1(\r;P)\subset J_i$;\\$v_{k+1}\in J_j\ \Leftrightarrow\ t_{k+1}(\r;P)\subset J_j$.
    \end{enumerate}
\end{corollary}

\begin{proof}
    First, we consider the possibility that $t_1(\r;P)\cap J_l \neq \varnothing$ for a given $l\neq i,j$. From corollary~\ref{lemma-onshell_Fq2Rterms_kgeq1_tree_structures_corollary1}, there is a unique path in $J_l\cap t(\r;\varnothing)$ connecting the external momentum~$p_l^\mu$ and the graph $H\cap t(\r';\varnothing)$. Since $J_l$ is connected, there is an edge $e_1^{J_l}$ whose endpoints are in $t_1(\r;P)$ and $J_l\cap t(\r;\varnothing)$ respectively, and we denote the one in $J_l\cap t(\r;\varnothing)$ as $v^{J_l}$. By construction, there is an edge $e_2^{J_l}\in J_l\cap t(\r;\varnothing)$, such that the graph $t(\r;P)\setminus e_2^{J_l}$ has two connected components, one of which contains $v^{J_l}$ and is attached by $p_l^\mu$. With these notations, the configuration of $T^2(\r)$ is shown in figure~\ref{onshell_lemma10_corollary2_statement1_original}.
    \begin{figure}[t]
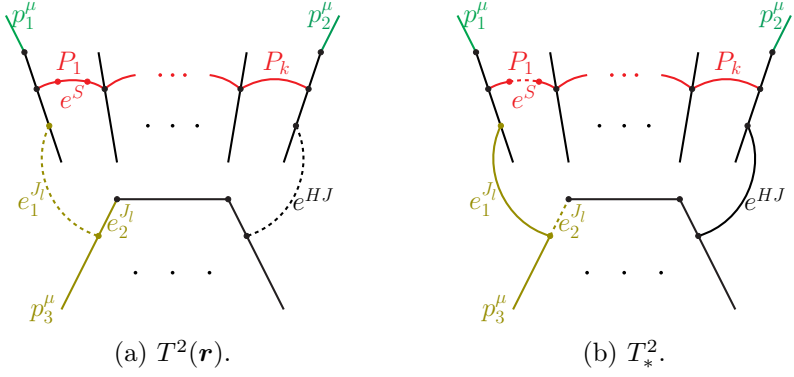

    \centering
    \begin{subfigure}[b]{0.3\textwidth}
    \centering
    \include{figs/onshell_lemma10_corollary2_statement1_original}
    \vspace{-3em}
    \caption{$T^2(\r)$.}
    \label{onshell_lemma10_corollary2_statement1_original}
    \end{subfigure}
    \hspace{3em}
    \begin{subfigure}[b]{0.3\textwidth}
    \centering
    \include{figs/onshell_lemma10_corollary2_statement1_comparison}
    \vspace{-3em}
    \caption{$T_*^2$.}
    \label{onshell_lemma10_corollary2_statement1_comparison}
    \end{subfigure}
    \caption{A graphic illustration of the analysis in proving $\scriptstyle{\textup{\protect\circled{1}}}$ of corollary~\ref{lemma-onshell_Fq2Rterms_kgeq1_tree_structures_corollary2}. Note that both $T^2(\r)$ and $T_*^2$ correspond to $\mathcal{F}^{(q^2)}$ terms, and $T_*^2\equiv T^2(\r)\cup (e_1^{J_l}\cup e^{HJ}) \setminus (e_2^{J_l}\cup e^S)$, where $e_1^{J_l},e_2^{J_l}\in J_l$, $e^S\in P_1\subset S$, and $e^{HJ}\in H\cup J$.}
    \label{figure-onshell_lemma10_corollary2_statement1}
    \end{figure}

    From Observation (\emph{1}) in the proof of lemma~\ref{lemma-onshell_Fq2Rterms_kgeq1_tree_structures}, there is an edge $e^{HJ}$ whose endpoints are in $t_{k+1}(\r;P)$ and $t(\r;\varnothing)$ respectively. By choosing any edge $e^S\in P_1$, we construct the spanning 2-tree
    \begin{eqnarray}
        T_*^2\equiv T^2(\r)\cup (e_1^{J_l}\cup e^{HJ}) \setminus (e_2^{J_l}\cup e^S),
    \end{eqnarray}
    as shown in figure~\ref{onshell_lemma10_corollary2_statement1_comparison}. One component of $T_*^2$ is attached by only $p_i^\mu$ and $p_l^\mu$, thus the total momentum flowing between the components of $T_*^2$ is off shell, corresponding to an $\mathcal{F}^{(q^2)}$ term. The weight of $T^2(\r_*)$ is then
    \begin{eqnarray}
        w(T_*^2) = w(T^2(\r)) + w(e_2^{J_l}) + w(e^S) - w(e_1^{J_l}) - w(e_{}^{HJ}) < w(T^2(\r)),
    \end{eqnarray}
    where we have used $w(e_1^{J_l}) = w(e_2^{J_l})$ and $w(e^S)<-1\leqslant w(e^{HJ})$ by construction. The inequality above violates the minimum-weight criterion, indicating that $t_1(\r;P)\cap J_l = \varnothing$ for any $l\neq i,j$. Similarly, $t_{k+1}(\r;P)\cap J_l = \varnothing$.

    In order to justify statement $\scriptstyle{\textup{\circled{2}}}$, first note that $t_a(\r;P)\cap J_i=\varnothing$, otherwise there would be an edge $e^{J_i}\in J_i$ connecting $t_1(\r;P)$ and $t_a(\r;P)$, leading to a contradiction to the minimum-weight criterion (as explained in deriving Observation (\emph{1}) in the proof of lemma~\ref{lemma-onshell_Fq2Rterms_kgeq1_tree_structures}). Similarly, $t_a(\r;P)\cap J_j=\varnothing$. In order to see that $t_a(\r;P)\cap J_l=\varnothing$ for any $l\neq i,j$, the same analysis in justifying $\scriptstyle{\textup{\circled{1}}}$ applies, which we shall not reiterate here.

    Finally, let us consider statement $\scriptstyle{\textup{\circled{3}}}$. Note that ($\Leftarrow$) is trivial, and by symmetry, we only need to show that $v_1\in J_i\ \Rightarrow\ t_1(\r;P)\subset J_i$. Suppose $v_1\in J_i$ and $t_1(\r;P)\not\subset J_i$ instead, then there must be some hard vertex $v_1^H\in t_1(\r;P)$ (see figure~\ref{onshell_lemma10_corollary2_statement3_original}). Let us denote the connected component of $H\cap T^2(\r)$, which contains $v_1^H$, as $\gamma_1^H$. We then note that the graph $H\cap t(\r;\varnothing)$ is nonempty, because the overall external momentum attaching to $t(\r;P)$ is off shell. It then follows from the connectedness of $H$, that there is always an edge $e_*^H\in H$ such that its endpoints are in $\gamma_1^H$ and $H\cap t(\r;\varnothing)$ respectively. The spanning 2-tree $T^2(\r) \cup e_*^H \setminus e^{J_i}$ (figure~\ref{onshell_lemma10_corollary2_statement3_comparison}), where $e^{J_i}\in J_i\cap t(\r;P)$, then corresponds to another $\mathcal{F}^{(q^2)}$ term, whose weight is smaller than $w(T^2(\r))$. This violation of the minimum-weight criterion implies that $t_1(\r;P)\subset J_i$. Statement $\scriptstyle{\textup{\circled{3}}}$ is then verified.
    \begin{figure}[t]
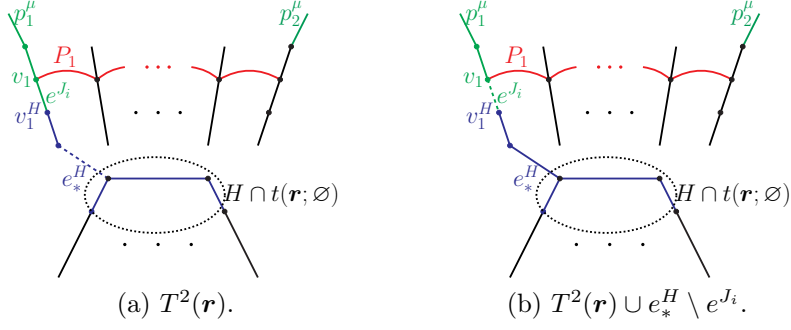

    \centering
    \begin{subfigure}[b]{0.3\textwidth}
    \centering
    \include{figs/onshell_lemma10_corollary2_statement3_original}
    \vspace{-3em}
    \caption{$T^2(\r)$.}
    \label{onshell_lemma10_corollary2_statement3_original}
    \end{subfigure}
    \hspace{3em}
    \begin{subfigure}[b]{0.3\textwidth}
    \centering
    \include{figs/onshell_lemma10_corollary2_statement3_comparison}
    \vspace{-3em}
    \caption{$T^2(\r) \cup e_*^H \setminus e^{J_i}$.}
    \label{onshell_lemma10_corollary2_statement3_comparison}
    \end{subfigure}
    \caption{A graphic illustration of the analysis in proving $\scriptstyle{\textup{\protect\circled{3}}}$ of corollary~\ref{lemma-onshell_Fq2Rterms_kgeq1_tree_structures_corollary2}. Note that both $T^2(\r)$ and $T^2(\r)\cup e_*^H\setminus e^{J_i}$ correspond to $\mathcal{F}^{(q^2)}$ terms, and the weight of $T^2(\r)\cup e_*^H\setminus e^{J_i}$ is smaller than $w(T^2(\r))$.}
    \label{figure-onshell_lemma10_corollary2_statement3}
    \end{figure}
\end{proof}

As we have demonstrated in proving corollary~\ref{lemma-onshell_Fq2Rterms_kgeq1_tree_structures_corollary1}, for each jet vertex $v^J\in t(\r;\varnothing)$ there is a unique path $P\subset J\cap t(\r;\varnothing)$ connecting $v^J$ and $H$. Now that corollary~\ref{lemma-onshell_Fq2Rterms_kgeq1_tree_structures_corollary2} is established, it is evident that for each $l\neq i,j$, the graph $J_l\cap T^2(\r)$ is contained in $t(\r;\varnothing)$, which, after all the hard vertices are identified with an auxiliary vertex, becomes a tree graph. Consequently, $\widetilde{J}_l\cap t(\r;\varnothing)$ is a spanning tree of $\widetilde{J}_l$.

We will now continue our examination of the structure of $T^2(\r)$, by investigating the relationship between its principal soft paths and the graphs $S^H$ and $S^J$. We shall begin with the case where $k\geqslant 2$ in eq.~(\ref{eq:onshell_leading_Fq2_generic_form_rewritten}).

\begin{lemma}
\label{lemma-onshell_Fq2Rterm_principal_path_properties}
For any $\mathcal{F}^{(q^2,R)}$~term $\x^{\r}$ that is characterized by
    \begin{eqnarray}
    n_H +n_J = L(H\cup J) + k+1,\quad n_S=L(\widetilde{S})-k,\quad k\geqslant 2,\nonumber
    \end{eqnarray}
each principal soft path $P_a$ ($a\in \{1,\dots,k\}$) must be contained in $S_\textup{II}$. Moreover,
\begin{enumerate}
    \item [1,] $P_a\subseteq S_\textup{II}^H$ if both endpoints of $P_a$ are in $H$;
    \item [2,] $P_a\cap S_\textup{II}^J$ is nonempty and connected if one endpoint of $P_a$ is in $J$.
\end{enumerate}
\end{lemma}

\begin{proof}
The proof of this lemma is divided into three parts. We begin by demonstrating that $P_1, \dots, P_k$ are contained within $S_\text{II}$. Next, we eliminate all scenarios where $P_a$ connects two hard vertices and $P_a\not \subseteq S_\textup{II}^H$. Finally, we briefly explain that statement \emph{2} can be confirmed using the same technique.

\begin{itemize}
    \item To see that $P_1,P_2,\dots,P_k \subset S_\text{II}$, let us take any $k$ edges $\{ e_1^S,\dots,e_k^S \}$ such that $e_i^S\in P_i$ ($i=1,\dots,k$), and any $k+1$ edges $\{ e_1^{HJ},\dots,e_{k+1}^{HJ} \}$ such that $e_j^{HJ}\in (H\cup J)\setminus T^2(\r)$ ($j=1,\dots,k+1$) whose endpoints are in $t_j(\r;P)$ and $t(\r;\varnothing)$ respectively. The graph
\begin{eqnarray}
T^2(\r)\cup (e_1^{HJ}\cup\dots\cup e_{k+1}^{HJ}) \setminus (e_1^S\cup\dots\cup e_k^S)\nonumber
\end{eqnarray}
is a spanning tree, whose weight must be larger than or equal to $w(T^2(\r))$ since $\x^{\r}$ is leading. We then have
\begin{eqnarray}
w(e_1^S) +\dots+ w(e_k^S) \geqslant w(e_1^{HJ}) +\dots+ w(e_{k+1}^{HJ}).
\label{eq:lemma10_proof_paths_in_Sii_step1}
\end{eqnarray}
If there is some $i_0\in \{1,\dots,k\}$ such that $w(e_{i_0}^S)=-2$, then we can rewrite (\ref{eq:lemma10_proof_paths_in_Sii_step1}) as
\begin{subequations}
    \begin{align}
        \begin{split}
            w(e_1^S) +\dots+ w(e_{i_0-1}^S) &+w(e_{i_0+1}^S) +\dots+ w(e_k^S)\\
            & \geqslant w(e_1^{HJ}) +\dots+ w(e_{k+1}^{HJ}) - w(e_{i_0}^S);
        \end{split}
        \label{eq:lemma10_proof_paths_in_Sii_step2}\\
        \begin{split}
            \Leftrightarrow\quad w(e_1^S) +\dots+ w(e_{i_0-1}^S) &+w(e_{i_0+1}^S) +\dots+ w(e_k^S)\\
            & \geqslant w(e_1^{HJ}) +\dots+ w(e_{k+1}^{HJ}) +2;
        \end{split}
        \label{eq:lemma10_proof_paths_in_Sii_step3}\\
        \begin{split}
            \Rightarrow\quad w(e_1^S) +\dots+ w(e_{i_0-1}^S) &+w(e_{i_0+1}^S) +\dots+ w(e_k^S)\\
            & \geqslant -(k+1) +2 = -(k-1).
        \end{split}
        \label{eq:lemma10_proof_paths_in_Sii_step4}
    \end{align}
\end{subequations}
where (\ref{eq:lemma10_proof_paths_in_Sii_step3}) is from $w(e_{i_0}^S)=-2$, and (\ref{eq:lemma10_proof_paths_in_Sii_step4}) is from $w(e) \geqslant -1$ for any $e\in H\cup J$. We now see a contradiction: every of the $k-1$ terms on the left-hand side of (\ref{eq:lemma10_proof_paths_in_Sii_step4}) is less than $-1$, while the right-hand side is exactly $-(k-1)$. As a result, $w(e_i^S)> -2$ for all $i=1,\dots,k$. Since $e_1^S,\dots,e_k^S$ have been chosen arbitrarily from the principal soft paths, equivalently we have $P_1,\dots,P_k\subset S_\text{II}$.

\item
Let us now consider a principal soft path $P_a$ whose endpoints are both in $H$, and $P_a\cap S_\text{II}^J \neq \varnothing$. We then take any component of $P_a\cap S_\text{II}^J$ and denote it as $P'_a$. By definition, $P'_a\subset P_a$ and it is also a path. From $\scriptstyle{\textup{\circled{3}}}$ of corollary~\ref{lemma-onshell_Si_empty_corollary1}, neither endpoint of~$P'_a$ belongs to~$S_\text{II}^J$. This implies that there are multiple edges in $P'_a$. Otherwise, if $P'_a$ consists of only one edge $e'_a$, it would be adjacent to $H\cup S_\text{II}^H$ rather than $J\cup S_\text{II}^J$, and assigned to $S_\text{II}^H$ from the steps above eq.~(\ref{eq:definition_Si}).

We use $A'$ and $B'$ to denote the endpoints of $P'_a$, and define $\Gamma_a$ as the graph whose vertices and edges can be joined to the internal vertices of $P'_a$ via a path in $t(\r;P)$. In other words, $\Gamma_a$ is the union of~$P'_a$ and the soft branch attaching to $P'_a$ (see figure~\ref{figure-onshell_lemma11_statement1_configuration}).
\begin{figure}[t]
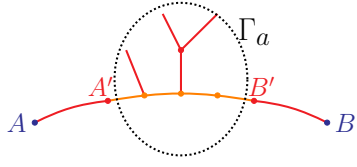

\centering
\include{figs/onshell_lemma11_statement1_configuration}
\vspace{-3em}
\caption{The generic configuration of $P_a$ (joining the vertices $A,B\in H$) and its sub-path $P'_a$ (joining $A',B'\in H\cup S_\text{II}^H$). Note that $P'_a$ consists of multiples edges (marked in orange) that are all in $S_\textup{II}^J$, while $A',B'\notin S_\text{II}^J$. The graph $\Gamma_a$ is enclosed by the dotted curve, which includes $P'_a$ and its soft branch.}
\label{figure-onshell_lemma11_statement1_configuration}
\end{figure}
Furthermore, among all the weights of edges of $\Gamma_a\cap S_\text{II}^J$, we refer to the largest one as $w_*$, and use $\gamma_*$ to denote any connected component of the following subgraph, consisting of:
\begin{enumerate}
    \item edges of $\Gamma_a\cap S_\text{II}^J$ with $w=w_*$;
    \item vertices of $\Gamma_a\cap S_\text{II}^J$, which are incident with some edges with $w=w_*$, but no edges with $w>w_*$.
\end{enumerate}
Without loss of generality, let us assume that $\gamma_*\subset S_\text{II}^{J_l}$ for some $l\in \{1,\dots,K\}$. As shown in $\scriptstyle{\textup{\circled{1}}}$ of corollary~\ref{lemma-onshell_Si_empty_corollary1}, the graph $\cup_{i'=i_0}^M S_{i'}^{J_l}\cup J_l$ is connected, where we have chosen $i_0$ such that $w_{i_0} = w_*$. Consequently, there exists a path, consisting of edges in $\cup_{i'=i_0}^M S_{i'}^{J_l}$, connecting $J$ and $\gamma_*$. It then follows that there is an edge $e_*\in \cup_{i'=i_0}^M S_{i'}^{J_l}$, whose endpoints are in $\gamma_*$ and $T^2(\r)\setminus P'_a$, respectively. Let us denote the endpoint in $T^2(\r)\setminus P'_a$ as $v_*$. If $v_*\in t(\r;P)$, then there is a loop in $t(\r;P)\cup e_*$. If $v_*\in t(\r;\varnothing)$, then $T^2(\r)\cup e_*$ is a spanning tree of $G$.

If $v_*\in t(\r;P)$ (an example is in figure~\ref{onshell_lemma11_kgeq2_P2_intersects_SiiJ_config1}), the loop in $t(\r;P)\cup e_*$ contains some edges in $e_a'\in P'_a$. Then the graph $t(\r;P)\cup e_*\setminus e'_a$ (see figure~\ref{onshell_lemma11_kgeq2_P2_intersects_SiiJ_config1_modified}) is a tree graph attached by $p_i^\mu$ and $p_j^\mu$, which does not conform with figure~\ref{figure-onshell_lemma10_generic_configuration}, because distinct principal soft paths have nonempty intersection here. Such configurations are forbidden from the leading $\mathcal{F}^{(q^2,R)}$ terms, due to Observation (\emph{5}) in the proof of lemma~\ref{lemma-onshell_Fq2Rterms_kgeq1_tree_structures}. Therefore, $t(\r;P)\cup e_*\setminus e_a'$ cannot appear in a leading $\mathcal{F}^{(q^2,R)}$ term, and it follows that,
\begin{eqnarray}
    w(t(\r;P)) + w(e_a') - w(e_*) > w(t(\r;P))\quad\Leftrightarrow\quad w(e_*) < w(e_a') \leqslant w_*.
\end{eqnarray}
This inequality, however, contradicts the condition that $e_*\in \cup_{i'=i_0}^M S_{i'}^{J_l}$.
\begin{figure}[t]
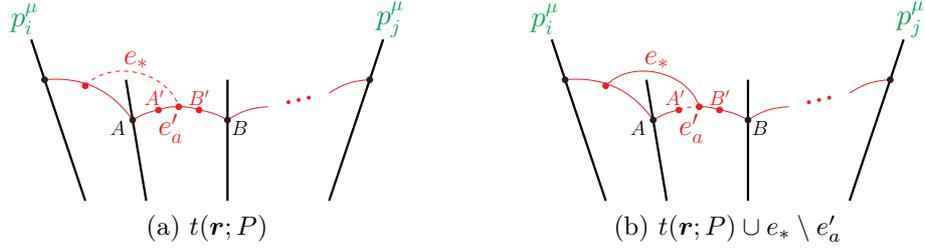

\centering
\begin{subfigure}[b]{0.36\textwidth}
\centering
\include{figs/onshell_lemma11_kgeq2_P2_intersects_SiiJ_config1}
\vspace{-3em}
\caption{$t(\r;P)$}
\label{onshell_lemma11_kgeq2_P2_intersects_SiiJ_config1}
\end{subfigure}
\hspace{3em}
\begin{subfigure}[b]{0.36\textwidth}
\centering
\include{figs/onshell_lemma11_kgeq2_P2_intersects_SiiJ_config1_modified}
\vspace{-3em}
\caption{$t(\r;P)\cup e_*\setminus e'_a$}
\label{onshell_lemma11_kgeq2_P2_intersects_SiiJ_config1_modified}
\end{subfigure}
\caption{The graphs $t(\r;P)$ and $t(\r;P)\cup e_*\setminus e'_a$, which are used in the analysis of the case $l=i$ or $j$ in proving statement \emph{1} of lemma~\ref{lemma-onshell_Fq2Rterm_principal_path_properties}. Here $P_a$ joins two hard vertices $A$ and $B$, and $P'_a\subseteq P_a$, whose endpoints are denoted as $A'$ and $B'$. The edges $e_*\in \cup_{i'=i_0}^M S_{i'}^{J_i}$ and $e'_a\in P'_a$ have been explained in context.}
\label{figure-onshell_lemma11_proof1}
\end{figure}

We then consider the case $v_*\in t(\r;\varnothing)$ (an example is shown in figure~\ref{onshell_lemma11_kgeq2_P2_intersects_SiiJ_config2}). In proving corollary~\ref{lemma-onshell_Fq2Rterms_kgeq1_tree_structures_corollary1}, we have demonstrated that there is a unique path connecting $p_l^\mu$ and $H\cap t(\r;\varnothing)$. So there always exists an edge $e^{J_l}\in J_l$, such that one connected component of $t(\r;\varnothing)\setminus e^{J_l}$, which $e_*$ is adjacent to, is attached only by $p_l^\mu$, while the other component is attached by all the other external momenta of $t(\r;\varnothing)$. In addition, there exists $e'_a\in P'_a$ and $e^{HJ}\in H\cup J$, such that the spanning 2-tree $T^2(\r)\cup (e_*\cup e^{HJ})\setminus (e'_a\cup e^{J_l})$ (see figure~\ref{onshell_lemma11_kgeq2_P2_intersects_SiiJ_config2_modified}) corresponds to an $\mathcal{F}^{(q^2)}$ term $\x^{\r'}$. Note that $T^2(\r)$ does not conform with figure~\ref{figure-onshell_lemma10_generic_configuration} because both its components contain some principal soft paths, so $w(T^2(\r')) > w(T^2(\r))$. Meanwhile, by construction,
\begin{eqnarray}
    w(T^2(\r')) = w(T^2(\r)) + w(e'_a) + w(e^{J_l}) - w(e_*) - w(e^{HJ}) \leqslant w(T^2(\r)),
\end{eqnarray}
where we have used $w(e'_a)\leqslant w(e_*)$ and $w(e^{J_l}) =-1 \leqslant w(e^{HJ})$. This leads to a contradiction.
\begin{figure}[t]
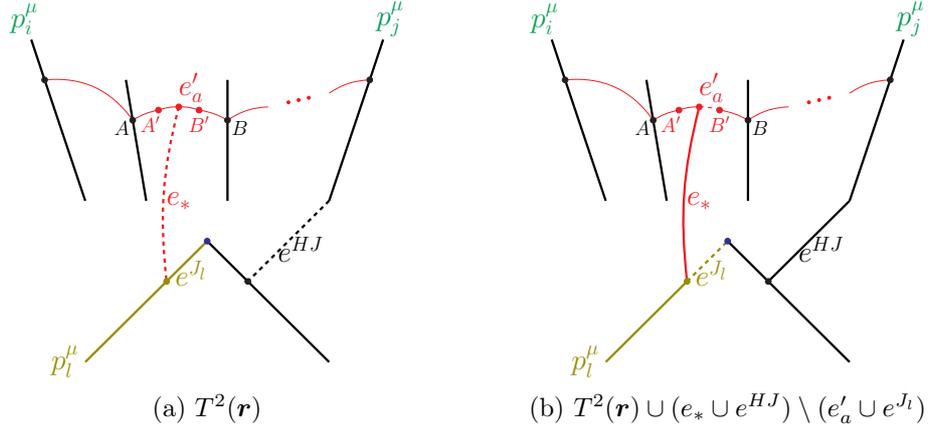

\centering
\begin{subfigure}[b]{0.36\textwidth}
\centering
\include{figs/onshell_lemma11_kgeq2_P2_intersects_SiiJ_config2}
\vspace{-3em}
\caption{$T^2(\r)$}
\label{onshell_lemma11_kgeq2_P2_intersects_SiiJ_config2}
\end{subfigure}
\hspace{3em}
\begin{subfigure}[b]{0.36\textwidth}
\centering
\include{figs/onshell_lemma11_kgeq2_P2_intersects_SiiJ_config2_modified}
\vspace{-3em}
\caption{$T^2(\r)\cup (e_*\cup e^{HJ})\setminus (e'_a\cup e^{J_l})$}
\label{onshell_lemma11_kgeq2_P2_intersects_SiiJ_config2_modified}
\end{subfigure}
\caption{The graphs $T^2(\r)$ and $T^2(\r)\cup (e_*\cup e^{HJ})\setminus (e'_a\cup e^{J_l})$, which are used in the analysis of the case $l\neq i,j$ in proving statement \emph{1} of lemma~\ref{lemma-onshell_Fq2Rterm_principal_path_properties}. Here $P_a$ joins two hard vertices $A$ and $B$, and $P'_a\subseteq P_a$, whose endpoints are denoted as $A'$ and $B'$. The edges $e_*\in \cup_{i'=i_0}^M S_{i'}^{J_l}$, $e'_a\in P'_a$, $e^{J_l}\in J_l$, and $e^{HJ}\in H\cup J$ have been explained in context.}
\label{figure-onshell_lemma11_proof2}
\end{figure}

Therefore, after examining all the possibilities, we have confirmed that the assumption of $P_a\cap S_\text{II}^J \neq \varnothing$ invariably leads to contradictions. Consequently, we have established that $P_a\subseteq S_\text{II}^H$.

\item Finally, we demonstrate that if one endpoint of $P_a$, denoted as $v_1$, is in $J$, then $P_a\cap S_\textup{II}^J$ is a connected path. By definition, the edge of $P_a$ which is incident with $v_1$ belongs to $S_\text{II}^J$, so $P_a\cap S_\textup{II}^J \neq \varnothing$. If $P_a\cap S_\textup{II}^J$ contains another connected component, a path $P'_a\subset P_a$, then from $\scriptstyle{\textup{\circled{3}}}$ of corollary~\ref{lemma-onshell_Si_empty_corollary1}, both endpoints of $P'_a$ are in $H\cup S_\text{II}^H$. Applying the same analysis above, which eliminates the possibility of $P_a\cap S\text{II}^J \neq \varnothing$ when both endpoints of $P_a$ are in $H$, one can justify this statement by contradiction. For the sake of brevity, we shall not reiterate the argument here.
\end{itemize}
We have thus verified all the statements of this lemma.
\end{proof}

We next explore the relationship between the unique principal soft path in $T^2(\r)$ and the graphs $S^H,S^J$, for the case where $k=1$ in eq.~(\ref{eq:onshell_leading_Fq2_generic_form_rewritten}).
\begin{corollary}
\label{lemma-onshell_Fq2Rterm_principal_path_properties_corollary1}
For any $\mathcal{F}^{(q^2,R)}$~term $\x^{\r}$ characterized by
    \begin{eqnarray}
    n_H +n_J = L(H\cup J) + 2,\quad n_S=L(\widetilde{S})-1,\nonumber
    \end{eqnarray}
the (unique) principal soft path $P$ satisfies one of the following:
\begin{enumerate}
    \item [$\scriptstyle{\textup{\circled{1}}}$] if both endpoints of $P$ are in $H$, then $P\subseteq S_\textup{II}^H$;
    \item [$\scriptstyle{\textup{\circled{2}}}$] if one endpoint of $P$ is in $J$ while the other is in $H$, then $P\cap S_\textup{II}^J$ is connected;
    \item [$\scriptstyle{\textup{\circled{3}}}$] if the endpoints of $P$ are in two distinct jets, then $P\cap S_1$ is nonempty and connected. 
\end{enumerate}
\end{corollary}
\begin{proof}
    From lemma~\ref{lemma-onshell_Fq2Rterms_kgeq1_tree_structures}, there are exactly two components of $(H\cup J)\cap t(\r;P)$, each one being attached by one on-shell external momentum. Let us use $t_1(\r;P)$ and $t_2(\r;P)$ to denote the components attached by $p_1^\mu$ and $p_2^\mu$ respectively. All the other external momenta attach to $t(\r;\varnothing)$. Furthermore, there is an edge $e_1^{HJ}\in H\cup J$ connecting $t_1(\r;P)$ and $t(\r;\varnothing)$, and $e_2^{HJ}\in H\cup J$ connecting $t_2(\r;P)$ and $t(\r;\varnothing)$.

    If there is an edge $e^S\in P\cap S_1$, the following two spanning 2-trees
    \begin{eqnarray}
        T_1^2 \equiv T^2(\r)\cup e_2^{HJ}\setminus e^S,\qquad T_2^2 \equiv T^2(\r)\cup e_1^{HJ}\setminus e^S
    \end{eqnarray}
    would corresponds to an $\mathcal{F}^{(p_1^2)}$ term and an $\mathcal{F}^{(p_2^2)}$ term, respectively. Furthermore,
    \begin{align}
    \begin{split}
        &w(T_1^2) = w(T^2(\r))+w(e^S)-w(e_2^{HJ}) + 1 \leqslant w(T^2(\r));\\
        &w(T_2^2) = w(T^2(\r))+w(e^S)-w(e_1^{HJ}) + 1 \leqslant w(T^2(\r)).
    \end{split}
    \end{align}
    where the term $+1$ is from the kinematic contribution of $\mathcal{F}^{(p^2)}$ terms. Note that in deriving the inequalities above, we have used $w(e^S)=-2$ and $w(e_1^{HJ}), w(e_2^{HJ})\geqslant -1$.
    
    Since $\x^{\r}$ is an $\mathcal{F}^{(q^2,R)}$ term, it is clear that $T_1^2$ and $T_2^2$ must correspond to an $\mathcal{F}^{(p_1^2,R)}$ term and an $\mathcal{F}^{(p_2^2,R)}$ term respectively. From \emph{3-2} of corollary~\ref{lemma-onshell_Fp2_vertices_universal_property_corollary2}, the $p_i$ component of any $T^2(\r)$, where $\x^{\r}$ is an $\mathcal{F}^{(p_i^2,R)}$ term, must not contain any hard vertices. It then follows that if one endpoint of $P$ is in $H$, then $P\cap S_1 = \varnothing$, and equivalently, $P\subset S_\text{II}$. The proof of lemma~\ref{lemma-onshell_Fq2Rterm_principal_path_properties} can then be applied to justify $\scriptstyle{\textup{\circled{1}}}$ and $\scriptstyle{\textup{\circled{2}}}$.

    \begin{figure}[t]
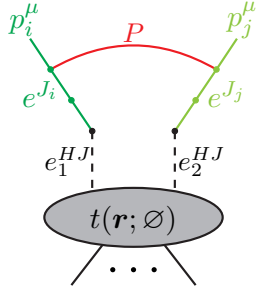

    \centering
    \include{figs/onshell_lemma11_corollary1_statement3_configuration}
    \vspace{-3em}\caption{The general configuration of $T^2(\r)$ with $k=1$ in eq.~(\ref{eq:onshell_leading_Fq2_generic_form_rewritten}), where the unique principal soft path $P$ has endpoints in two distinct jets $J_i$ and $J_j$.}
    \label{figure-onshell_lemma11_corollary1_statement3_configuration}
    \end{figure}
    We then focus on $\scriptstyle{\textup{\circled{3}}}$. For generic $\mathcal{F}^{(q^2,R)}$ term $\x^{\r}$ such that $P$ joins two distinct jets, the general configuration of $T^2(\r)$ is shown in figure~\ref{figure-onshell_lemma11_corollary1_statement3_configuration}. One can then construct another spanning 2-tree
    \begin{eqnarray}
        T^2(\r) \cup (e_1^{HJ}\cup e_2^{HJ}) \setminus (e^{J_i}\cup e^{J_j}),
    \end{eqnarray}
    where $e^{J_i}$ and $e^{J_j}$ are chosen from $J_i$ and $J_j$ respectively, as shown in the figure. The obtained spanning 2-tree, say $T_*^2$, is characterized by
    \begin{eqnarray}
        n_H = L(H),\qquad n_J = L(\widetilde{J})+2,\qquad n_S = L(\widetilde{S})-1.
    \label{eq:onshell_lemma11_corollary1_statement3_step1}
    \end{eqnarray}
    Moreover, the weight of $T_*^2$ is
    \begin{eqnarray}
        w(T_*^2) = w(T^2(\r)) + w(e^{J_i}) + w(e^{J_j}) - w(e_1^{HJ}) - w(e_2^{HJ}) \leqslant w(T^2(\r)),
    \end{eqnarray}
    where we have used $w(e_1^{HJ}), w(e_1^{HJ}) \geqslant -1 = w(e^{J_i}) = w(e^{J_j})$. So $T_*^2$ must also correspond to an $\mathcal{F}^{(q^2,R)}$ term. Let us now compare it with any $\mathcal{U}^{(R)}$ term $\x^{\r'}$, which have already been characterized by eq.~(\ref{eq:lemma_onshell_leading_U_generic_form}). By definition, $w(T_*^2)$ and $w(T^1(\r'))$ are equal, where $w(T_*^2)$ has an extra $-2$ contribution from the weights of two jet edges, while $w(T^1(\r'))$ has an extra contribution, which must also be $-2$, from the weight of one soft edge. It then follows that $P$ contains an edge with $w=-2$, and in other words, $P\cap S_1\neq \varnothing$. 
    
    With this analysis, we can furthermore rewrite eq.~(\ref{eq:onshell_lemma11_corollary1_statement3_step1}) into:
    \begin{eqnarray}
        n_H = L(H),\qquad n_J = L(\widetilde{J})+2,\qquad n_{S_1} = L(\widetilde{S}_1)-1,\qquad n_{S_\text{II}} = L(\widetilde{S}_\text{II}).
    \label{eq:onshell_lemma11_corollary1_statement3_step2}
    \end{eqnarray}
    This implies that $P\cap S_1$ is connected.
\end{proof}

Note that an example illustrating the proof of $\scriptstyle{\textup{\circled{3}}}$ above will be provided later in (\ref{eq:3loop_nonplanar_example_lemma11}).

Finally, we consider $k=0$ in eq.~(\ref{eq:onshell_leading_Fq2_generic_form_rewritten}), namely,
\begin{eqnarray}
    n_H +n_J = L(H\cup J) +1,\quad n_S=L(\widetilde{S}).
\end{eqnarray}
In this case, there are no principal soft paths in $T^2(\r)$. Actually, the equation above can be made more precise as follows.
\begin{corollary}
\label{lemma-onshell_Fq2Rterm_principal_path_properties_corollary2}
    Any leading term with $n_H +n_J = L(H\cup J) + 1$ further satisfies
    \begin{eqnarray}
        n_H = L(H)+1,\qquad n_J=L(\widetilde{J}).\nonumber
    \end{eqnarray}
\end{corollary}
\begin{proof}
    A first observation from eq.~(\ref{eq:removed_edge_number_constraint1}) is that $n_H\geqslant L(H)$, which indicates the following equations:
    \begin{eqnarray}
    \label{eq:lemma11_corollary2_characterization}
        n_H = L(H)+k',\qquad n_J = L(\widetilde{J}) -k'+1 \qquad (k'\geqslant 0).
    \end{eqnarray}

    We next argue that $k'\geqslant 2$ is impossible. To illustrate, let us take $k'=2$. Then $n_H = L(H)+2$ and $n_J = L(\widetilde{J}) -1$ would lead to the following configuration of $T^2(\r)$: $H\cap T^2(\r)$ consists of three connected components $\gamma_1^H$, $\gamma_2^H$, and $\gamma_3^H$, and there are two paths in $J$ connecting a jet vertex $v^J$ and $H$ (see figure~\ref{figure-onshell_lemma11_corollary2}). On one hand, every edge $e^H\in H$ connecting $\gamma_1^H$ and $\gamma_2^H\cup \gamma_3^H$ satisfies $w=-1$, because if $w>-1$, there is always a way to modify the $T^2(\r)$ above into another $\mathcal{F}^{(q^2)}$ term with a smaller weight, thus violating the minimum-weight criterion. On the other hand, according to lemma~\ref{lemma-onshell_subgraphs_tree_structures_corollary2}, there exists a path $P^H\subset H$ connecting $\gamma_1^H$ and $\gamma_2^H$, such that each of its edges satisfies $w>-1$. A contradiction is thus derived, which forbids the possibility of $k'\geqslant 2$ in eq.~(\ref{eq:lemma11_corollary2_characterization}).
    \begin{figure}[t]
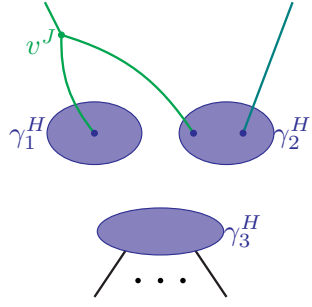

    \centering
    \include{figs/onshell_lemma11_corollary2}
    \vspace{-2em}
    \caption{The generic configuration of $T^2(\r)$, which describes the $\mathcal{F}^{(q^2,R)}$ terms characterized by $n_H = L(H)+2$ and $n_J = L(\widetilde{J}) -1$. The blue blobs labelled by $\gamma_1^H$, $\gamma_2^H$, and $\gamma_3^H$ represent the three connected components of $H\cap T^2(\r)$.}
    \label{figure-onshell_lemma11_corollary2}
    \end{figure}

    Finally, note that $k'=1$ would give rise to $\mathcal{F}^{(p^2,R)}$ terms, which has already been characterized by eq.~(\ref{eq:lemma_onshell_leading_Fp2_generic_form}). So we have $n_H = L(H)+1$ and $n_J=L(\widetilde{J})$ here.
\end{proof}

\subsubsection{Determining the hard and soft weights}
\label{section-determining_hard_soft_weights}

Now we are ready to characterize all the possible $\mathcal{F}^{(q^2,R)}$ terms, as eqs.~(\ref{eq:lemma_onshell_leading_U_generic_form}) and (\ref{eq:lemma_onshell_leading_Fp2_generic_form}). This is the final step before determining the hard and soft weights of a generic region $R$.
\begin{lemma}
\label{lemma-onshell_leading_terms_forms}
The $\mathcal{F}^{(q^2,R)}$ terms are characterized by the following equations.
\begingroup
\allowdisplaybreaks
\begin{align}
    \begin{split}
        & n_H + n_J = L(H\cup J)+k+1,\quad n_S = L(\widetilde{S})-k,\quad k=0,1,\dots.\nonumber
    \end{split}\\
    \begin{split}
    \label{eq:lemma12_Fq2Rterms_forms_kgeq2_possibility1}
        & \underline{k\geqslant 2}:\ \ n_H = L(H)+k-1,\quad n_{J}=L(\widetilde{J})+2,\\
        & \phantom{k\geqslant 2:}\ \ \ n_{S^H} = L(\widetilde{S}^H)-k+2,\quad n_{S^J} = L(\widetilde{S}^J)-2,\quad n_{S_1} = L(\widetilde{S}_1).
    \end{split}\\
    \begin{split}
    \label{eq:lemma12_Fq2Rterms_forms_kgeq2_possibility2}
        & \phantom{k=1}\textup{or}\ \ n_H = L(H)+k,\quad n_{J}=L(\widetilde{J})+1,\\
        & \phantom{k=1:}\ \ \ n_{S^H} = L(\widetilde{S}^H)-k+1,\quad n_{S^J} = L(\widetilde{S}^J)-1,\quad n_{S_1} = L(\widetilde{S}_1).
    \end{split}\\
    \begin{split}
    \label{eq:lemma12_Fq2Rterms_forms_kgeq2_possibility3}
        & \phantom{k=1}\textup{or}\ \ n_H = L(H)+k+1,\quad n_{J}=L(\widetilde{J}),\\
        & \phantom{k=1:}\ \ \ n_{S^H} = L(\widetilde{S}^H)-k,\quad n_{S^J} = L(\widetilde{S}^J),\quad n_{S_1} = L(\widetilde{S}_1).
    \end{split}\\
    \begin{split}
    \label{eq:lemma12_Fq2Rterms_forms_kgeq1_possibility1}
        & \underline{k=1}:\ \ n_H = L(H)+1,\quad n_{J}=L(\widetilde{J})+1,\\
        & \phantom{k=1:}\ \ \ n_{S^H} = L(\widetilde{S}^H),\quad n_{S^J} = L(\widetilde{S}^J)-1,\quad n_{S_1} = L(\widetilde{S}_1).
    \end{split}\\
    \begin{split}
    \label{eq:lemma12_Fq2Rterms_forms_kgeq1_possibility2}
        & \phantom{k=1}\textup{or}\ \ n_H = L(H)+2,\quad n_{J}=L(\widetilde{J}),\\
        & \phantom{k=1:}\ \ \ n_{S^H} = L(\widetilde{S}^H)-1,\quad n_{S^J} = L(\widetilde{S}^J),\quad n_{S_1} = L(\widetilde{S}_1).
    \end{split}\\
    \begin{split}
    \label{eq:lemma12_Fq2Rterms_forms_kgeq1_possibility3}
        & \phantom{k=1}\textup{or}\ \ n_H = L(H),\quad n_{J}=L(\widetilde{J})+2,\\
        & \phantom{k=1:}\ \ \ n_{S^H} = L(\widetilde{S}^H),\quad n_{S^J} = L(\widetilde{S}^J),\quad n_{S_1} = L(\widetilde{S}_1)-1.
    \end{split}\\
    \begin{split}
    \label{eq:lemma12_leading_Fq2_terms_forms_keq0}
        & \underline{k=0}:\ \ n_H = L(H),\quad n_{J}=L(\widetilde{J}),\\
        & \phantom{k=0:}\ \ \ n_{S^H} = L(\widetilde{S}^H),\quad n_{S^J} = L(\widetilde{S}^J),\quad n_{S_1} = L(\widetilde{S}_1).
    \end{split}
\end{align}
\endgroup
\end{lemma}

\begin{proof}
We first consider the cases where $k\geqslant 2$. Recall the general configuration of $t(\r;P)$ as represented by figure~\ref{figure-onshell_lemma10_generic_configuration}. In this configuration, the principal soft paths $P_1,\dots,P_k$ align the $k+1$ components of $(H\cup J)\cap t(\r;P)$, which are labelled $t_1(\r;P),\dots,t_{k+1}(\r;P)$, respectively. Specifically, one endpoint of $P_1$, denoted as $v_1$, is in $t_1(\r;P)$, and one endpoint of $P_k$, denoted as $v_{k+1}$, is in $t_{k+1}(\r;P)$.

According to $\scriptstyle{\textup{\circled{1}}}$ of corollary~\ref{lemma-onshell_Fq2Rterms_kgeq1_tree_structures_corollary2}, we have $t_1(\r;P)\subset H\cup J_i$ and $t_{k+1}(\r;P)\subset H\cup J_j$. Now we discuss the possible positions of $v_1$ and $v_{k+1}$.

\begin{itemize}
    \item If $v_1\in J_i$ and $v_{k+1}\in J_j$, then from $\scriptstyle{\textup{\circled{2}}}$ and $\scriptstyle{\textup{\circled{3}}}$ of corollary~\ref{lemma-onshell_Fq2Rterms_kgeq1_tree_structures_corollary2}, $H\cup t(\r;P)$ has $k-1$ connected components: $t_2(\r;P),\dots,t_k(\r;P)$. Since $H\cap t(\r;\varnothing)$ is connected (corollary~\ref{lemma-onshell_Fq2Rterms_kgeq1_tree_structures_corollary1}), there are $k$ components of $H\cap T^2(\r)$ in total, each of which is a tree graph. As a result,
    \begin{eqnarray}
    \label{eq:proof_onshell_lemma12_kgeq2_step1}
        n_H = L(H) + k - 1.
    \end{eqnarray}
    From $\scriptstyle{\textup{\circled{3}}}$ of corollary~\ref{lemma-onshell_Fq2Rterms_kgeq1_tree_structures_corollary2}, we further know that $t_1(\r;P)\subset J_i$, which is one component of $J_i\cap T^2(\r)$. All the other components of $J_i\cap T^2(\r)$, in contrast, are in $t(\r;\varnothing)$. As we have pointed out below the proof of corollary~\ref{lemma-onshell_Fq2Rterms_kgeq1_tree_structures_corollary1}, $\widetilde{J}_i\cap t(\r;\varnothing)$ is a tree graph. It then follows that the graph $\widetilde{J}_i\cap T^2(\r)$ has two components: one is $t_1(\r;P)$, while the other, which is $\widetilde{J}_i\cap t(\r;\varnothing)$, includes all the edges and vertices of $J_i\cap t(\r;\varnothing)$ as well as the auxiliary vertex. In other words, $\widetilde{J}_i\cap T^2(\r)$ is a spanning 2-tree of $\widetilde{J}_i$. By symmetry, $\widetilde{J}_j\cap T^2(\r)$ is a spanning 2-tree of $\widetilde{J}_j$. We thus have
    \begin{eqnarray}
    \label{eq:proof_onshell_lemma12_kgeq2_step2}
        n_{J_i}=L(\widetilde{J}_i)+1,\quad n_{J_j}=L(\widetilde{J}_j)+1.
    \end{eqnarray}
    For each $l\neq i,j$, as commented below the proof of corollary~\ref{lemma-onshell_Fq2Rterms_kgeq1_tree_structures_corollary2}, the graph $\widetilde{J}_l\cap T^2(\r)$, which is a spanning tree of $\widetilde{J}_l$. We then have
    \begin{eqnarray}
    \label{eq:proof_onshell_lemma12_kgeq2_step3}
        n_{J_l}=L(\widetilde{J}_l)\ \ (l\neq i,j).
    \end{eqnarray}
    Eqs.~(\ref{eq:proof_onshell_lemma12_kgeq2_step2}) and (\ref{eq:proof_onshell_lemma12_kgeq2_step3}) lead to $n_J = L(\widetilde{J})+2$.
    
    According to statement \emph{2} of lemma~\ref{lemma-onshell_Fq2Rterm_principal_path_properties}, the graphs $P_1\cap S_\text{II}^J$ and $P_k\cap S_\text{II}^J$ are both connected and nonempty. In other words, $P_1\cap S_\text{II}^J$ and $P_k\cap S_\text{II}^J$ are sub-paths of $P_1$ and $P_k$, respectively. For the path $P_1\cap S_\text{II}^J$, both its endpoints are in $H\cup S_\text{II}^H$, so it forms a loop in $\widetilde{S}_\text{II}^J\cap T^2(\r)$ after its endpoints are identified with the auxiliary vertex of $\widetilde{S}_\text{II}^J$. The same analysis also applies to the path $P_k\cap S_\text{II}^J$. Therefore, there are two loops in $\widetilde{S}_\text{II}^J\cap T^2(\r)$, implying:
    \begin{eqnarray}
    \label{eq:proof_onshell_lemma12_kgeq2_step4}
        n_{S_\text{II}^J} = L(\widetilde{S}_\text{II}^J)-2.
    \end{eqnarray}
    Meanwhile, all the remaining $k-2$ paths $P_2,\dots, P_{k-1}$ are contained in $S_\text{II}^H$ according to statement~\emph{1} of lemma~\ref{lemma-onshell_Fq2Rterm_principal_path_properties}. Each path has endpoints in $H$ by definition, and becomes a loop in $\widetilde{S}_\text{II}^H\cap T^2(r)$ after its endpoints are identified with the auxiliary vertex of~$\widetilde{S}_\text{II}^H$. We thus have $k-2$ loops in $\widetilde{S}_\text{II}^H\cap T^2(r)$, implying:
    \begin{eqnarray}
    \label{eq:proof_onshell_lemma12_kgeq2_step5}
        n_{S_\text{II}^H} = L(\widetilde{S}_\text{II}^H)-k+2.
    \end{eqnarray}
    Combining eqs.~(\ref{eq:proof_onshell_lemma12_kgeq2_step4}) and (\ref{eq:proof_onshell_lemma12_kgeq2_step5}) with $n_S = L(\widetilde{S}) - k$, we have
    \begin{eqnarray}
    \label{eq:proof_onshell_lemma12_kgeq2_step6}
        n_{S_1} = L(\widetilde{S}) - L(\widetilde{S}_\text{II}^H) - L(\widetilde{S}_\text{II}^J) = L(\widetilde{S}_1),
    \end{eqnarray}
    where we have used the relation $L(\widetilde{S}) = L(\widetilde{S}_\text{II}^H) + L(\widetilde{S}_\text{II}^J) + L(\widetilde{S}_1)$, which can be obtained by using the Euler's formula as in proving eqs.~(\ref{eq:lemma_onshell_subgraphs_loop_relation1}) and (\ref{eq:lemma_onshell_subgraphs_loop_relation2}). So far, all the equations in~(\ref{eq:lemma12_Fq2Rterms_forms_kgeq2_possibility1}) have been shown.

    \item If $v_1\in J_i$ and $v_{k+1}\in H$, then $n_{J_i} = L(\widetilde{J}_i)+1$ still holds as above. In contrast, the graph $J_j\cap t(\r;P)$ is adjacent to $H$ in this case, so after identifying all the hard vertices to the auxiliary vertex of $\widetilde{J}_j$, all the components of $J_j\cap T^2(\r)$ will be in the same component with the auxiliary vertex. The graph $\widetilde{J}_j\cap T^2(\r)$, is then a spanning tree of $\widetilde{J}_j$.
    
    Another difference from above is that $H\cap t_{k+1}(\r;P)\neq \varnothing$, implying that $H\cap T^2(\r)$ has $k+1$ components in total ($k$ in $t(\r;P)$ and one in $t(\r;\varnothing)$), each of which is a tree. So $n_H = L(H)+k$.
    
    Finally, $k-1$ of the principal soft paths $P_2,\dots,P_k$ are contained in $S_\text{II}^H$ from \emph{1} of lemma~\ref{lemma-onshell_Fq2Rterm_principal_path_properties}, each corresponding to a loop in $\widetilde{S}_\text{II}^H\cap T^2(\r)$ as explained above. Meanwhile, from \emph{2} of lemma~\ref{lemma-onshell_Fq2Rterm_principal_path_properties}, $P_1\cap S_\text{II}^J$ is nonempty and connected, which corresponds to a loop in $\widetilde{S}_\text{II}^J\cap T^2(\r)$. So $n_{S^H} = L(\widetilde{S}^H)-k+1$ and $n_{S^J} = L(\widetilde{S}^J)-1$. To summarize,
    \begin{align}
    \begin{split}
        &n_H = L(H)+k,\quad n_{J_i}=L(\widetilde{J}_i)+1,\quad n_{J_l}=L(\widetilde{J}_l)\ \ \forall\ l\neq i,\\
        &n_{S^H} = L(\widetilde{S}^H)-k+1,\quad n_{S^J} = L(\widetilde{S}^J)-1,\quad n_{S_1} = L(\widetilde{S}_1).
    \end{split}
    \end{align}
    which yields eq.~(\ref{eq:lemma12_Fq2Rterms_forms_kgeq2_possibility2}) directly. Note that eq.~(\ref{eq:lemma12_Fq2Rterms_forms_kgeq2_possibility2}) also holds for the case where $v_1\in H$ and $v_{k+1}\in J_j$.

    \item If both $v_1,v_{k+1}\in H$, then for each jet $J_l$ ($l\in \{1,\dots,K\}$), based on the analysis above we have $n_{J_l} = L(\widetilde{J}_l)$. The graph $H\cap T^2(\r)$ now has $k+2$ components in total ($k+1$ in $t(\r;P)$ and one in $t(\r;\varnothing)$). Meanwhile, all the $k$ principal soft paths $P_1,\dots,P_k \subset S_\text{II}^H$, which implies the existence of $k$ loops in $\widetilde{S}_\text{II}^H\cap T^2(\r)$. Equivalently, $n_{S_\text{II}^H} = L(\widetilde{S}_\text{II}^H) -k$. We thus have
    \begin{align}
    \begin{split}
        &n_H = L(H)+k+1,\quad n_{J_l}=L(\widetilde{J}_l)\ \ \forall\ l,\\
        &n_{S^H} = L(\widetilde{S}^H)-k,\quad n_{S^J} = L(\widetilde{S}^J),\quad n_{S_1} = L(\widetilde{S}_1).
    \end{split}
    \end{align}
    This immediately leads to eq.~(\ref{eq:lemma12_Fq2Rterms_forms_kgeq2_possibility3}).
\end{itemize}
So far we have exhausted all the configurations for $k\geqslant 2$.

\bigbreak
For $k=1$, we characterize the $\mathcal{F}^{(q^2,R)}$ terms according to corollary~\ref{lemma-onshell_Fq2Rterm_principal_path_properties_corollary1}. It is worth noticing that if one endpoint of $P$ is in $H$, we can apply $\scriptstyle{\textup{\circled{1}}}$ and $\scriptstyle{\textup{\circled{2}}}$ of corollary~\ref{lemma-onshell_Fq2Rterm_principal_path_properties_corollary1}, which are special cases of \emph{1} and \emph{2} of lemma~\ref{lemma-onshell_Fq2Rterm_principal_path_properties}, respectively. Therefore, by setting $k=1$ in eqs.~(\ref{eq:lemma12_Fq2Rterms_forms_kgeq2_possibility1}) and (\ref{eq:lemma12_Fq2Rterms_forms_kgeq2_possibility2}), we immediately obtain eqs.~(\ref{eq:lemma12_Fq2Rterms_forms_kgeq1_possibility1}) and (\ref{eq:lemma12_Fq2Rterms_forms_kgeq1_possibility2}). If the endpoints of $P$ are in distinct jets, we can apply $\scriptstyle{\textup{\circled{3}}}$ of corollary~\ref{lemma-onshell_Fq2Rterm_principal_path_properties_corollary1}, which indicates that the path $P\cap S_1$ becomes the unique loop in $\widetilde{S}\cap T^2(\r)$ after both its endpoints are identified with the auxiliary vertex. As a result, we have $n_{S_1} = L(\widetilde{S}_1)-1$, $n_{S^H} = L(\widetilde{S}^H)$ and $n_{S^J} = L(\widetilde{S}^J)$.

Furthermore, $t_1(\r;P)\subset J_i$ from $\scriptstyle{\textup{\circled{3}}}$ of corollary~\ref{lemma-onshell_Fq2Rterms_kgeq1_tree_structures_corollary2}, so $\widetilde{J}_i\cap T^2(\r)$ has two connected components, one is $t_1(\r;P)$ and the other, including the auxiliary vertex of $\widetilde{J}_i$, is in $t(\r;\varnothing)$. Consequently, $\widetilde{J}_I\cap T^2(\r)$ is a spanning 2-tree of $\widetilde{J}_i$, which implies that $n_{J_i}=L(\widetilde{J}_i)+1$. Similarly, $n_{J_j}=L(\widetilde{J}_j)+1$. For any $l\neq i,j$, based on the preceding analysis for $k\geqslant 2$, we have $n_{J_l}=L(\widetilde{J}_l)$. These analyses justify eq.~(\ref{eq:lemma12_Fq2Rterms_forms_kgeq1_possibility3}).

\bigbreak
Finally, for $k=0$, the spanning 2-tree $T^2(\r)$ contains no soft paths connecting vertices of $H\cup J$, so $\widetilde{S}\cap T^2(\r)$ is a spanning tree of $\widetilde{S}$. From $\scriptstyle{\textup{\circled{2}}}$ of corollary~\ref{lemma-onshell_Si_empty_corollary1}, the relations $n_{S^H}=L(\widetilde{S}^H)$, $n_{S^J} = L(\widetilde{S}^J)$ and $n_{S_1} = L(\widetilde{S}_1)$ are automatic. Moreover, from corollary~\ref{lemma-onshell_Fq2Rterm_principal_path_properties_corollary2}, the relation $n_H+n_J= L(H) +L(\widetilde{J})+1$ implies $n_H=L(H)+1$ and $n_{J}= L(\widetilde{J})$ directly. The last equation in this lemma, eq.~(\ref{eq:lemma12_leading_Fq2_terms_forms_keq0}), is then derived.
\end{proof}

A straightforward application of lemma~\ref{lemma-onshell_leading_terms_forms} determines the hard and soft weights, as summarized in the following corollary.

\begin{corollary}
\label{lemma-onshell_leading_terms_forms_corollary1}
For each edge $e$, $w(e)=-2$ if $e\in S$; $w(e)=0$ if $e\in H$.
\end{corollary}
\begin{proof}
For any given region vector $\v_R$, by definition, the quantity $\r\cdot \v_R$ is of some fixed value for each $\x^{\r}$ characterized by eqs.~(\ref{eq:lemma12_leading_U_terms_forms}), (\ref{eq:lemma12_leading_Fp2_terms_forms}), and (\ref{eq:lemma12_Fq2Rterms_forms_kgeq2_possibility1})-(\ref{eq:lemma12_leading_Fq2_terms_forms_keq0}). Meanwhile, let us consider another vector $\v'$ defined as
\begin{eqnarray}
\v'\equiv (\underset{N(H)+N(S^H)}{\underbrace{0,\dots,0}},\ \underset{N(J)+N(S^J)}{\underbrace{-1,\dots,-1}}\ ,\ \underset{N(S_1)}{\underbrace{-2,\dots,-2}}\ ;1).
\end{eqnarray}
Namely, each of the edges in $H\cup S^H$ corresponds to entry $0$, each of the edges in $J\cup S^J$ corresponds to entry $-1$, and each of the edges in $S_1$ corresponds to entry $-2$. A key observation is that
\begin{eqnarray}
\r\cdot \v' = -\sum_{i=1}^K n_{J_i} - n_{S^J} - 2n_{S_1} = -L(\widetilde{J}) - L(\widetilde{S}^J) -2L(\widetilde{S}_1).
\end{eqnarray}
The first equality is from the definition of $\v'$. To derive the second equality, we have inserted the values of $n_{J_i}$, $n_{S^J}$ and $n_{S_1}$ in each of the equations (\ref{eq:lemma12_leading_U_terms_forms}), (\ref{eq:lemma12_leading_Fp2_terms_forms}), and (\ref{eq:lemma12_Fq2Rterms_forms_kgeq2_possibility1})-(\ref{eq:lemma12_leading_Fq2_terms_forms_keq0}). It is straightforward to check that the same value, $-L(\widetilde{J}) - L(\widetilde{S}^J) -2L(\widetilde{S}_1)$, is obtained.

This observation $\r\cdot \v' = -L(\widetilde{J}) - L(\widetilde{S}^J) -2L(\widetilde{S}_1)$ implies the violation of the facet criterion unless $\v'= \v_R$, which holds only if the weight of each soft edge is exactly $-2$ (equivalently, $S^H=S^J=\varnothing$), and the weight of each hard edge is exactly $0$. We have thus completed the proof.
\end{proof}

\begin{figure}[t]
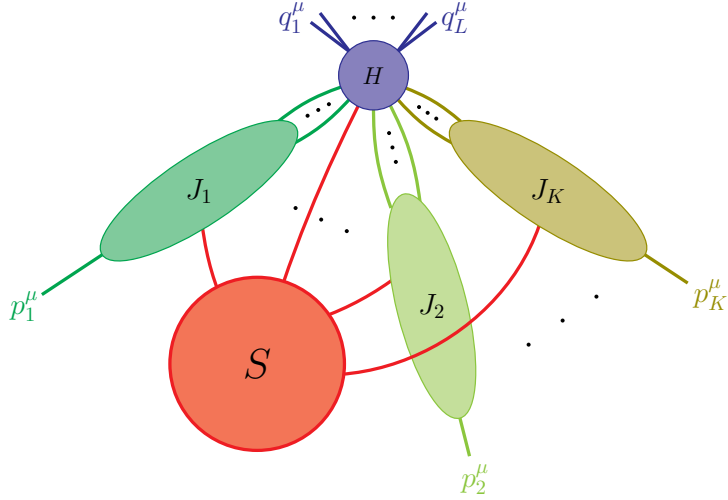

\centering
\include{figs/onshell_region_H_J_S_precise}
\vspace{-3em}
\caption{The picture of a generic region $R$, where $H$, $J$, and $S$ respectively denote the hard, jet, and soft subgraphs of $G$ that are associated with $R$. The dots represent that multiple lines connect the blobs. The hard subgraph $H$ is connected, and attached by all the off-shell external momenta $q_1^\mu,\dots,q_L^\mu$. Each jet subgraph $J_i$ is connected, and attached by the on-shell external momentum $p_i^\mu$. The soft subgraph $S$ can be disconnected, and cannot be attached by any external momenta. Each connected component of $S$ is adjacent to two or more jet subgraphs and possibly the hard subgraph.}
\label{figure-onshell_region_H_J_S_precise}
\end{figure}

\bigbreak
Let us summarize the result obtained from the technical analysis in this subsection. Every region vector $\v_R$ must be of the form
\begin{align}
\label{generic_form_region_vector_onshell}
\begin{split}
    &\v_R= (u_{R,1},u_{R,2},\dots,u_{R,N};1),\qquad u_{R,e}\in \{0,-1,-2\},\\
    &\qquad u_{R,e}=0 \quad\Leftrightarrow\quad e\in H \\
    &\qquad u_{R,e}=-1 \quad\Leftrightarrow\quad e\in J\equiv \cup_{i=1}^K J_i\\
    &\qquad u_{R,e}=-2 \quad\Leftrightarrow\quad e\in S 
\end{split}
\end{align}
where the subgraphs $H,J,S$, defined in section~\ref{section-subgraphs_associated_each_region}, are shown in figure~\ref{figure-onshell_region_H_J_S_precise}. Namely, the hard subgraph $H$ is connected, and attached by all the off-shell external momenta $q_1^\mu,\dots,q_L^\mu$; each jet subgraph $J_i$ is connected, and attached by the on-shell external momentum $p_i^\mu$; the soft subgraph $S$ is attached by no external momenta, and each connected component of $S$ must be adjacent to two or more jet subgraphs.

A key observation is that each region can be interpreted as a specific pinch surface of $G|_{p^2=0}$, where $G|_{p^2=0}$ is the Feynman integral obtained by setting all the $p_i^\mu$ strictly on shell. This correspondence allows us to deduce the set of regions from the set of pinch surfaces, which follow a classical picture due to the Coleman-Norton interpretation.

\subsection{Some examples from a three-loop graph}
\label{section-examples_from_3loop_graph}

\begin{figure}[t]
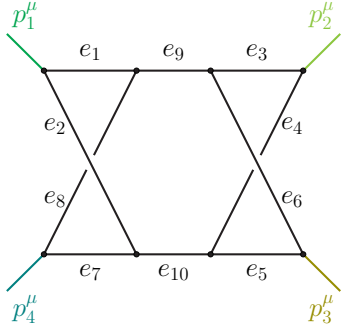

\centering
\include{figs/3loop_nonplanar_graph}
\vspace{-3em}\caption{A three-loop nonplanar graph.}
\label{figure-3loop_nonplanar_graph}
\end{figure}

In this subsection, we provide examples to support the technical proofs presented in section~\ref{section-generic_form_region_rigorous_proof}. We will examine some non-region configurations that contravene lemmas~\ref{lemma-onshell_heavy_in_light_constraint}, \ref{lemma-onshell_hard_subgraph_connected}, \ref{lemma-onshell_sij_configuration_constraints}, \ref{lemma-onshell_Si_empty}, and \ref{lemma-onshell_Fq2Rterm_principal_path_properties}, respectively. Subsequently, we employ the rationale outlined in these respective lemmas to elucidate why these configurations do not qualify as valid regions. All the examples are taken from the same three-loop nonplanar graph, figure~\ref{figure-3loop_nonplanar_graph}. In this graph, the external momenta $p_1^\mu$, $p_2^\mu$, $p_3^\mu$, and $p_4^\mu$ are all approaching the on-shell limit, while the internal edges are labelled by $e_1,\dots,e_{10}$.

First, we consider the following configuration of weight structure, where the weight of each edge is shown.
\begin{equation}
    \label{eq:3loop_nonplanar_example_lemma4and6}
    \begin{tikzpicture}[baseline=10ex, scale=0.3, mydot/.style={circle, fill, inner sep=0.8pt}]
    \draw [thick,color=Green] (1.5,7.5) -- (4,7.5);
    \draw [ultra thick,color=Blue] (4,7.5) -- (6,7.5);
    \draw [thick,color=LimeGreen] (6,7.5) -- (8.5,7.5);
    \draw [thick,color=teal] (1.5,2.5) -- (4,2.5);
    \draw [ultra thick,color=Blue] (4,2.5) -- (6,2.5);
    \draw [thick,color=olive] (6,2.5) -- (8.5,2.5);
    \draw [thick,color=Green] (1.5,7.5) -- (4,2.5);
    \draw [thick,color=olive] (6,7.5) -- (8.5,2.5);
    \draw [thick,color=teal] (4,7.5) -- (2.85,5.2);
    \draw [thick,color=teal] (2.65,4.8) -- (1.5,2.5);
    \draw [thick,color=LimeGreen] (8.5,7.5) -- (7.35,5.2);
    \draw [thick,color=LimeGreen] (7.15,4.8) -- (6,2.5);

    \draw [thick,color=Green] (0.5,8.5) -- (1.5,7.5);
    \draw [thick,color=LimeGreen] (9.5,8.5) -- (8.5,7.5);
    \draw [thick,color=olive] (9.5,1.5) -- (8.5,2.5);
    \draw [thick, color=teal] (0.5,1.5) -- (1.5,2.5);

    \node [draw,circle,minimum size=3pt,color=Green,fill=Green,inner sep=0pt,outer sep=0pt] () at (1.5,7.5) {};
    \node [draw,circle,minimum size=3pt,color=Blue,fill=Blue,inner sep=0pt,outer sep=0pt] () at (4,7.5) {};
    \node [draw,circle,minimum size=3pt,color=Blue,fill=Blue,inner sep=0pt,outer sep=0pt] () at (6,7.5) {};
    \node [draw,circle,minimum size=3pt,color=LimeGreen,fill=LimeGreen,inner sep=0pt,outer sep=0pt] () at (8.5,7.5) {};
    \node [draw,circle,minimum size=3pt,color=teal,fill=teal,inner sep=0pt,outer sep=0pt] () at (1.5,2.5) {};
    \node [draw,circle,minimum size=3pt,color=Blue,fill=Blue,inner sep=0pt,outer sep=0pt] () at (4,2.5) {};
    \node [draw,circle,minimum size=3pt,color=Blue,fill=Blue,inner sep=0pt,outer sep=0pt] () at (6,2.5) {};
    \node [draw,circle,minimum size=3pt,color=olive,fill=olive,inner sep=0pt,outer sep=0pt] () at (8.5,2.5) {};

    \node () at (2.5,8.1) {${\color{Green} -1}$};
    \node () at (7.5,8.1) {${\color{LimeGreen} -1}$};
    \node () at (7.5,1.9) {\large ${\color{olive} -1}$};
    \node () at (2.5,1.9) {\large ${\color{teal} -1}$};
    \node () at (1.3,6) {${\color{Green} -1}$};
    \node () at (1.3,4) {${\color{teal} -1}$};
    \node () at (8.6,6) {${\color{LimeGreen} -1}$};
    \node () at (8.6,4) {${\color{olive} -1}$};
    \node () at (5,8.1) {${\color{Blue} 0}$};
    \node () at (5,1.9) {${\color{Blue} 0}$};
    \end{tikzpicture}
\end{equation}
This configuration does not conform with lemma~\ref{lemma-onshell_heavy_in_light_constraint}, thus must be excluded by its proof. Let us define $\gamma$ as the graph consisting of $e_9$ and its endpoints. From the proof of lemma~\ref{lemma-onshell_heavy_in_light_constraint}, there must be some $\mathcal{F}^{(q^2,R)}$ terms $\x^{\r}$, such that the endpoints of $e_9$ are in distinct components of $T^2(\r)$, respectively. An example of such $\mathcal{F}^{(q^2,R)}$ terms is $(-(p_1+p_4)^2) x_2x_6x_7x_9$ (with weight $-3$). However, there is another $\mathcal{F}^{(q^2)}$ term $(-(p_1+p_4)^2) x_2x_3x_6x_7$ whose weight is smaller ($-4$). Hence, the configuration of (\ref{eq:3loop_nonplanar_example_lemma4and6}) cannot be included in the list of valid regions. It is important to note that the analysis above is a direct application of the proof presented in lemma~\ref{lemma-onshell_heavy_in_light_constraint}.

Alternatively, the configuration in (\ref{eq:3loop_nonplanar_example_lemma4and6}) can be excluded from the proof of lemma~\ref{lemma-onshell_hard_subgraph_connected}. Let us consider the $\mathcal{U}^{(R)}$ term $\x^{\r} = x_2x_6x_7$. In $T^1(\r)$ there is a unique path connecting $e_9$ and $e_{10}$, which consists of $e_3$, $e_4$, and their endpoints. Since $e_4$ is a jet edge, removing it from $T^1(\r)$ results in an $\mathcal{F}^{(p^2,R)}$ term $(-p_3^2)x_2x_4x_6x_7$. However, the corresponding $p_3$ component contains a hard edge $e_{10}$ and hard vertices (endpoints of $e_{10}$), which does not satisfy statement \emph{3-2} of corollary~\ref{lemma-onshell_Fp2_vertices_universal_property_corollary2}.

Next, we consider the following weight structure.
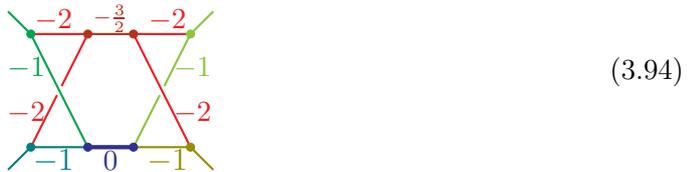
\begin{equation}
    \label{eq:3loop_nonplanar_example_lemma8}
    \begin{tikzpicture}[baseline=10ex, scale=0.3, mydot/.style={circle, fill, inner sep=0.8pt}]
    \draw [thick,color=Red] (1.5,7.5) -- (4,7.5);
    \draw [thick,color=BrickRed] (4,7.5) -- (6,7.5);
    \draw [thick,color=Red] (6,7.5) -- (8.5,7.5);
    \draw [thick,color=teal] (1.5,2.5) -- (4,2.5);
    \draw [ultra thick,color=Blue] (4,2.5) -- (6,2.5);
    \draw [thick,color=olive] (6,2.5) -- (8.5,2.5);
    \draw [thick,color=Green] (1.5,7.5) -- (4,2.5);
    \draw [thick,color=Red] (6,7.5) -- (8.5,2.5);
    \draw [thick,color=Red] (4,7.5) -- (2.85,5.2);
    \draw [thick,color=Red] (2.65,4.8) -- (1.5,2.5);
    \draw [thick,color=LimeGreen] (8.5,7.5) -- (7.35,5.2);
    \draw [thick,color=LimeGreen] (7.15,4.8) -- (6,2.5);

    \draw [thick,color=Green] (0.5,8.5) -- (1.5,7.5);
    \draw [thick,color=LimeGreen] (9.5,8.5) -- (8.5,7.5);
    \draw [thick,color=olive] (9.5,1.5) -- (8.5,2.5);
    \draw [thick, color=teal] (0.5,1.5) -- (1.5,2.5);

    \node [draw,circle,minimum size=3pt,color=Green,fill=Green,inner sep=0pt,outer sep=0pt] () at (1.5,7.5) {};
    \node [draw,circle,minimum size=3pt,color=BrickRed,fill=BrickRed,inner sep=0pt,outer sep=0pt] () at (4,7.5) {};
    \node [draw,circle,minimum size=3pt,color=BrickRed,fill=BrickRed,inner sep=0pt,outer sep=0pt] () at (6,7.5) {};
    \node [draw,circle,minimum size=3pt,color=LimeGreen,fill=LimeGreen,inner sep=0pt,outer sep=0pt] () at (8.5,7.5) {};
    \node [draw,circle,minimum size=3pt,color=teal,fill=teal,inner sep=0pt,outer sep=0pt] () at (1.5,2.5) {};
    \node [draw,circle,minimum size=3pt,color=Blue,fill=Blue,inner sep=0pt,outer sep=0pt] () at (4,2.5) {};
    \node [draw,circle,minimum size=3pt,color=Blue,fill=Blue,inner sep=0pt,outer sep=0pt] () at (6,2.5) {};
    \node [draw,circle,minimum size=3pt,color=olive,fill=olive,inner sep=0pt,outer sep=0pt] () at (8.5,2.5) {};

    \node () at (2.5,8.1) {${\color{Red} -2}$};
    \node () at (7.5,8.1) {${\color{Red} -2}$};
    \node () at (7.5,1.9) {\large ${\color{olive} -1}$};
    \node () at (2.5,1.9) {\large ${\color{teal} -1}$};
    \node () at (1.3,6) {${\color{Green} -1}$};
    \node () at (1.3,4) {${\color{Red} -2}$};
    \node () at (8.6,6) {${\color{LimeGreen} -1}$};
    \node () at (8.6,4) {${\color{Red} -2}$};
    \node () at (5,8.2) {${\color{BrickRed}\scriptstyle -\frac{3}{2}}$};
    \node () at (5,1.9) {${\color{Blue} 0}$};
    \end{tikzpicture}
\end{equation}
This configuration does not conform with the statement of lemma~\ref{lemma-onshell_sij_configuration_constraints}, so we can apply the corresponding proof to exclude it from the list of regions. The point is that, $e_9$ is present in all the leading spanning (2-)trees. If, on the contrary, there is a leading term $\x^{\r}$ such that $e_9\notin T(\r)$, we can apply the analysis of figure~\ref{figure-onshell_lemma8_proof_Fq2} and modify $T(\r)$ into another spanning (2-)tree whose weight is smaller than $w(T(\r))$, which is impossible due to the minimum-weight criterion. The presence of $e_9$, however, violates the facet criterion. So the configuration in (\ref{eq:3loop_nonplanar_example_lemma8}) is not a region.

Then, we consider the following configuration, which does not conform with lemma~\ref{lemma-onshell_Si_empty}.
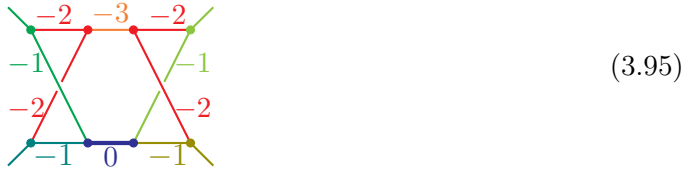
\begin{equation}
    \label{eq:3loop_nonplanar_example_lemma9}
    \begin{tikzpicture}[baseline=10ex, scale=0.3, mydot/.style={circle, fill, inner sep=0.8pt}]
    \draw [thick,color=Red] (1.5,7.5) -- (4,7.5);
    \draw [thick,color=Orange] (4,7.5) -- (6,7.5);
    \draw [thick,color=Red] (6,7.5) -- (8.5,7.5);
    \draw [thick,color=teal] (1.5,2.5) -- (4,2.5);
    \draw [ultra thick,color=Blue] (4,2.5) -- (6,2.5);
    \draw [thick,color=olive] (6,2.5) -- (8.5,2.5);
    \draw [thick,color=Green] (1.5,7.5) -- (4,2.5);
    \draw [thick,color=Red] (6,7.5) -- (8.5,2.5);
    \draw [thick,color=Red] (4,7.5) -- (2.85,5.2);
    \draw [thick,color=Red] (2.65,4.8) -- (1.5,2.5);
    \draw [thick,color=LimeGreen] (8.5,7.5) -- (7.35,5.2);
    \draw [thick,color=LimeGreen] (7.15,4.8) -- (6,2.5);

    \draw [thick,color=Green] (0.5,8.5) -- (1.5,7.5);
    \draw [thick,color=LimeGreen] (9.5,8.5) -- (8.5,7.5);
    \draw [thick,color=olive] (9.5,1.5) -- (8.5,2.5);
    \draw [thick, color=teal] (0.5,1.5) -- (1.5,2.5);

    \node [draw,circle,minimum size=3pt,color=Green,fill=Green,inner sep=0pt,outer sep=0pt] () at (1.5,7.5) {};
    \node [draw,circle,minimum size=3pt,color=Red,fill=Red,inner sep=0pt,outer sep=0pt] () at (4,7.5) {};
    \node [draw,circle,minimum size=3pt,color=Red,fill=Red,inner sep=0pt,outer sep=0pt] () at (6,7.5) {};
    \node [draw,circle,minimum size=3pt,color=LimeGreen,fill=LimeGreen,inner sep=0pt,outer sep=0pt] () at (8.5,7.5) {};
    \node [draw,circle,minimum size=3pt,color=teal,fill=teal,inner sep=0pt,outer sep=0pt] () at (1.5,2.5) {};
    \node [draw,circle,minimum size=3pt,color=Blue,fill=Blue,inner sep=0pt,outer sep=0pt] () at (4,2.5) {};
    \node [draw,circle,minimum size=3pt,color=Blue,fill=Blue,inner sep=0pt,outer sep=0pt] () at (6,2.5) {};
    \node [draw,circle,minimum size=3pt,color=olive,fill=olive,inner sep=0pt,outer sep=0pt] () at (8.5,2.5) {};

    \node () at (2.5,8.1) {${\color{Red} -2}$};
    \node () at (7.5,8.1) {${\color{Red} -2}$};
    \node () at (7.5,1.9) {\large ${\color{olive} -1}$};
    \node () at (2.5,1.9) {\large ${\color{teal} -1}$};
    \node () at (1.3,6) {${\color{Green} -1}$};
    \node () at (1.3,4) {${\color{Red} -2}$};
    \node () at (8.6,6) {${\color{LimeGreen} -1}$};
    \node () at (8.6,4) {${\color{Red} -2}$};
    \node () at (5,8.2) {${\color{Orange} -3}$};
    \node () at (5,1.9) {${\color{Blue} 0}$};
    \end{tikzpicture}
\end{equation}
To see that this configuration does not appear in the list of valid regions, the key point is that $e_9$ is absent in any of the leading spanning (2-)trees. Conversely, if there were a leading term $\x^{\r}$ with $e_9\in T(\r)$, as discussed in the proof of lemma~\ref{lemma-onshell_Si_empty}, it could only be an $\mathcal{F}^{(q^2,R)}$ term. An example of such a term is $(-(p_1+p_2)^2) x_2x_4x_6x_8$. However, the weight $w(T(\r))$ (which is $-6$) is larger than that of the $\mathcal{U}^{(R)}$ term $x_6x_8x_9$ (which is $-7$), hence $e_9\notin T(\r)$ for any leading term $\x^{\r}$, thereby violating the facet criterion. Consequently, the configuration in (\ref{eq:3loop_nonplanar_example_lemma9}) is not a valid region.

Finally, let us consider the following weight structure.
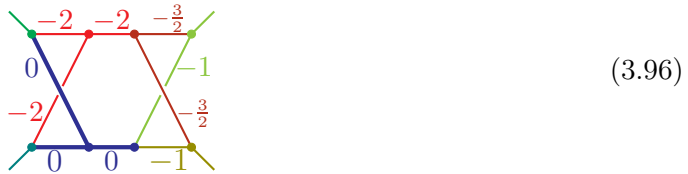
\begin{equation}
    \label{eq:3loop_nonplanar_example_lemma11}
    \begin{tikzpicture}[baseline=10ex, scale=0.3, mydot/.style={circle, fill, inner sep=0.8pt}]
    \draw [thick,color=Red] (1.5,7.5) -- (4,7.5);
    \draw [thick,color=Red] (4,7.5) -- (6,7.5);
    \draw [thick,color=BrickRed] (6,7.5) -- (8.5,7.5);
    \draw [ultra thick,color=Blue] (1.5,2.5) -- (4,2.5);
    \draw [ultra thick,color=Blue] (4,2.5) -- (6,2.5);
    \draw [thick,color=olive] (6,2.5) -- (8.5,2.5);
    \draw [ultra thick,color=Blue] (1.5,7.5) -- (4,2.5);
    \draw [thick,color=BrickRed] (6,7.5) -- (8.5,2.5);
    \draw [thick,color=Red] (4,7.5) -- (2.85,5.2);
    \draw [thick,color=Red] (2.65,4.8) -- (1.5,2.5);
    \draw [thick,color=LimeGreen] (8.5,7.5) -- (7.35,5.2);
    \draw [thick,color=LimeGreen] (7.15,4.8) -- (6,2.5);

    \draw [thick,color=Green] (0.5,8.5) -- (1.5,7.5);
    \draw [thick,color=LimeGreen] (9.5,8.5) -- (8.5,7.5);
    \draw [thick,color=olive] (9.5,1.5) -- (8.5,2.5);
    \draw [thick, color=teal] (0.5,1.5) -- (1.5,2.5);

    \node [draw,circle,minimum size=3pt,color=Green,fill=Green,inner sep=0pt,outer sep=0pt] () at (1.5,7.5) {};
    \node [draw,circle,minimum size=3pt,color=Red,fill=Red,inner sep=0pt,outer sep=0pt] () at (4,7.5) {};
    \node [draw,circle,minimum size=3pt,color=BrickRed,fill=BrickRed,inner sep=0pt,outer sep=0pt] () at (6,7.5) {};
    \node [draw,circle,minimum size=3pt,color=LimeGreen,fill=LimeGreen,inner sep=0pt,outer sep=0pt] () at (8.5,7.5) {};
    \node [draw,circle,minimum size=3pt,color=teal,fill=teal,inner sep=0pt,outer sep=0pt] () at (1.5,2.5) {};
    \node [draw,circle,minimum size=3pt,color=Blue,fill=Blue,inner sep=0pt,outer sep=0pt] () at (4,2.5) {};
    \node [draw,circle,minimum size=3pt,color=Blue,fill=Blue,inner sep=0pt,outer sep=0pt] () at (6,2.5) {};
    \node [draw,circle,minimum size=3pt,color=olive,fill=olive,inner sep=0pt,outer sep=0pt] () at (8.5,2.5) {};

    \node () at (2.5,8.1) {${\color{Red} -2}$};
    \node () at (7.5,8.2) {${\color{BrickRed}\scriptstyle -\frac{3}{2}}$};
    \node () at (7.5,1.9) {\large ${\color{olive} -1}$};
    \node () at (2.5,1.9) {\large ${\color{Blue} 0}$};
    \node () at (1.5,6) {${\color{Blue} 0}$};
    \node () at (1.2,4) {${\color{Red} -2}$};
    \node () at (8.6,6) {${\color{LimeGreen} -1}$};
    \node () at (8.6,4) {${\color{BrickRed}\scriptstyle -\frac{3}{2}}$};
    \node () at (5,8.1) {${\color{Red} -2}$};
    \node () at (5,1.9) {${\color{Blue} 0}$};
    \end{tikzpicture}
\end{equation}
We shall use this configuration to check statement $\scriptstyle{\textup{\circled{3}}}$ of corollary~\ref{lemma-onshell_Fq2Rterm_principal_path_properties_corollary1}. Consider the $\mathcal{F}^{(q^2)}$ term $\x^{\r}\equiv (-(p_1+p_4)^2) x_1x_4x_5x_8$ which is potentially leading. The principal soft path in $T^2(\r)$ consists $e_3$, $e_6$, and their common endpoint. By definition, neither $e_3$ nor $e_6$ belongs to $S_1$, and the weight associated with this spanning 2-tree is $-6$. In contrast, the weight of any $\mathcal{U}^{(R)}$ terms (for example, $x_1x_3x_8$) is $-\frac{11}{2}$. The minimum-weight criterion is thus violated, implying that $(-(p_1+p_4)^2) x_1x_4x_5x_8$ cannot be an $\mathcal{F}^{(q^2,R)}$ term.

These examples presented in this subsection serve as tangible illustrations of the technical and intricate proofs detailed in section~\ref{section-generic_form_region_rigorous_proof}. By delving into these concrete examples, we provide a practical grasp of the underlying complexities of establishing eq.~(\ref{generic_form_region_vector_onshell}).

\subsection{Further constraints on the region vectors}
\label{section-further_restrictions_region_vectors}

Eq.~(\ref{generic_form_region_vector_onshell}), with the subgraphs $H$, $J$, and $S$ depicted in figure~\ref{figure-onshell_region_H_J_S_precise}, provides a necessary condition for the regions appearing in the on-shell expansion. For any region vector satisfying this condition, it has been shown in ref.~\cite{GrdHzgJnsMaSchlk22} that the following two \emph{additional requirements} are needed.\footnote{Aside from these two extra requirements, ref.~\cite{GrdHzgJnsMaSchlk22} also require that every connected component of $S$ must connect at least two different jet subgraphs $J_i$ and $J_j$. But we shall not present this as an ``extra requirement'' here, as we have already derived it in section~\ref{section-generic_form_region_rigorous_proof}. It is worth noticing that in ref.~\cite{GrdHzgJnsMaSchlk22}, the requirements on $H$ and $J$ are stated in terms of \emph{reduced forms} of a graph, and those statements are equivalent to the ones presented here.}
\begin{enumerate}
    \item \emph{The total momentum flowing into each 1VI component of $H$ is off shell.}
    \item \emph{The total momentum flowing into each 1VI component of $\widetilde{J}_i$ is $(p_i+l)^\mu$, where $l^\mu$ is zero or some soft momentum.}
\end{enumerate}
The word ``1VI'' represents \emph{one-vertex irreducible}. We call a graph one-vertex irreducible, if it remains connected after any one of its vertices is removed. In graph theory, 1VI graphs are also referred to as biconnected graphs.

\begin{figure}[t]
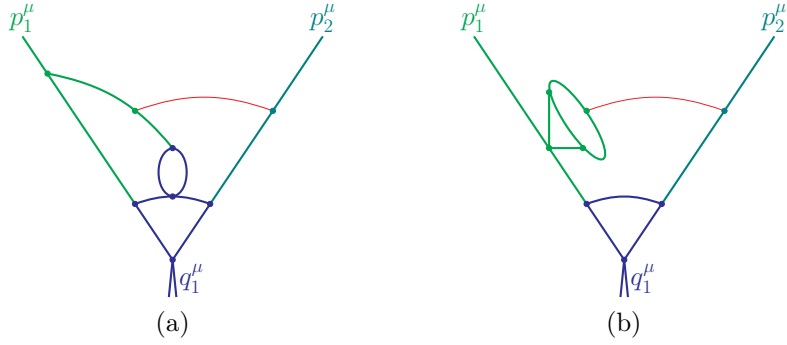

\centering
\begin{subfigure}[b]{0.3\textwidth}
\centering
\include{figs/onshell_configuration_excluded_by_extra1}
\vspace{-3em}
\caption{}
\label{onshell_configuration_excluded_by_extra1}
\end{subfigure}
\hspace{3em}
\begin{subfigure}[b]{0.3\textwidth}
\centering
\include{figs/onshell_configuration_excluded_by_extra2}
\vspace{-3em}
\caption{}
\label{onshell_configuration_excluded_by_extra2}
\end{subfigure}
\caption{Two examples of configurations that satisfy eq.~(\ref{generic_form_region_vector_onshell}) but are excluded from the set of regions, because they do not meet the requirements of $H,J,S$. Figure~(a): the upper hard loop is an individual 1VI component of $H$, and the momentum flowing into it is collinear to $p_1^\mu$. Figure~(b): the union of the two jet loops is an individual 1VI component of $J_1$, and the momentum flowing into it is soft.}
\label{figure-onshell_configuration_excluded_by_extra}
\end{figure}

These extra requirements exclude the configurations in figure~\ref{figure-onshell_configuration_excluded_by_extra} from the list of regions, although they all conform with eq.~(\ref{generic_form_region_vector_onshell}). In figure~\ref{onshell_configuration_excluded_by_extra1}, one of the components of the hard subgraph $H$, which is the upper hard loop, is an individual 1VI component of $H$. The momentum flowing into this component is parallel to $p_1^\mu$, which is on shell. So it does not meet the requirement of $H$. Similarly, in figure~\ref{onshell_configuration_excluded_by_extra2} the union of the two jet loops is an individual 1VI component of $J_1$, and the momentum flowing into it is simply the soft line momentum. So it does not meet the requirement of $J$.

Based on these requirements, a graph-finding algorithm has been implemented in ref.~\cite{GrdHzgJnsMaSchlk22} to generate the list of regions for any wide-angle scattering graph. The algorithm has been specifically implemented in Maple~\cite{git-Maple_file}. Notably, it has been verified that within a scalar theory involving three- and four-point interactions, the program's output precisely aligns with the predictions of the Newton polytope approach, while it often exhibits faster computational performance.

\subsection{Summary}
\label{section-summary_onshell_regions}

In general, the regions appearing in the on-shell expansion feature the hard, collinear and soft modes, whose scaling in momentum space is shown as follows:
\begin{align}
\label{WideAngleMods}
\begin{cases}
\text{hard:} & k_H^\mu\sim Q(1,1,1);\\
\text{collinear to }{p_i}: & k_{C_i}^\mu\sim Q(1,\lambda,\lambda^{1/2});\\
\text{soft:} & k_S^\mu\sim Q(\lambda,\lambda,\lambda).
\end{cases}
\end{align}
Note that in this scaling, we have used the $p_i$ lightcone coordinate, for $i=1,\dots,K$. Namely, $k^\mu = (k\cdot \overline{\beta}_i,\ k\cdot \beta_i,\ k\cdot \beta_{i\perp})$, where $\beta_i^\mu$ is a null vector in the direction of $p_i^\mu$, defined as $\beta_i^\mu=\frac{1}{\sqrt{2}}(1,\boldsymbol{p}_i/|\boldsymbol{p}_i|)$. For each $\beta_i^\mu$, we also define $\overline{\beta}_i^\mu \equiv \frac{1}{\sqrt{2}} (1, -\boldsymbol{p}_i/|\boldsymbol{p}_i|)$, so that $\beta_i \cdot \overline{\beta}_i =1$. Note that the ``soft mode'' in (\ref{WideAngleMods}) is referred to as the ``ultrasoft mode'' in some other literature.

In more detail, each region is identified by a particular partition of the entire graph $G$ into the hard subgraph~$H$, the jet subgraphs~$J_i$ ($i=1,\dots,K$), and the soft subgraph~$S$. The region where all the internal propagators belong to $H$, is referred to as the hard region, while all the other regions, referred to as the infrared regions, all correspond to the pinch surfaces of $G|_{p^2=0}$. Graphically, the infrared regions can be depicted by figure~\ref{figure-onshell_region_H_J_S_precise}: the hard subgraph $H$ is connected, to which all the off-shell external momenta $q_j^\mu$ attach; each nontrivial jet subgraph $J_i$ is connected and adjacent to $H$, and the only external momentum that attaches to $J_i$ is $p_i^\mu$; the soft subgraph $S$ may have several distinct connected components, each of which is adjacent to two or more jets and possibly $H$.

Additionally, we require the following for the hard and jet subgraphs: the total momentum flowing into each 1VI component of $H$ is off shell; meanwhile, the total momentum flowing into each 1VI component of $J_i$ is $(p_i+l)^\mu$, where $l^\mu$ is zero or soft. These extra requirements are from ref.~\cite{GrdHzgJnsMaSchlk22}, aiming to eliminate spurious configurations that conform with figure~\ref{figure-onshell_region_H_J_S_precise} but result in scaleless integrals.

In summary, we have the following “on-shell-expansion region theorem”.
\begin{center}
\framebox{\Longstack[l]{\textbf{On-shell-expansion region theorem}\\\qquad The region vectors appearing in the on-shell expansion of wide-angle scattering are\\
all of the form of $\v_R= (u_{R,1},u_{R,2},\dots,u_{R,N};1)$, such that for each edge $e$,\\
\qquad $\bullet \ \phantom{-}u_{R,e}=0\phantom{1} \quad \Leftrightarrow\quad e\in H$;\\
\qquad $\bullet\ \phantom{0}u_{R,e}=-1 \quad \Leftrightarrow\quad e\in J$;\\
\qquad $\bullet\ \phantom{0}u_{R,e}=-2 \quad \Leftrightarrow\quad e\in S$.\\
The subgraphs $H$, $J$, and $S$ are shown in figure~\ref{figure-onshell_region_H_J_S_precise}, which further satisfy:\\
\qquad (1) the total momentum flowing into each 1VI component of $H$ is off-shell;\\
\qquad (2) the total momentum flowing into each 1VI component of $\widetilde{J}_i$ is $(p_i+l)^\mu$, with $l^\mu$\\
\qquad \phantom{(2) }being either zero or soft.}}
\end{center}

We comment that the on-shell expansion region theorem above provides the complete list of regions to all orders, which is independent of the numerator in momentum representation and the spacetime dimension. For any given theory, the regions contributing to the (sub)$^n$leading power ($n=0,1,\dots$) of the asymptotic expansion can be determined through straightforward power counting. Notably, in four-dimensional gauge theory, regions contributing to the leading power must satisfy the following conditions: (1) no soft parton is attached to $H$; (2) no soft fermion or scalar is attached to $J$; (3) for each jet, the full set of its partons attached to $H$ consists of exactly one parton with physical polarization, and all others being scalar-polarized gauge bosons~\cite{LbyStm78,Stm78I,Stm95book,Stm96lectures,ClsSprStm04}. Violation of any of these conditions would lead to power suppression, implying that the corresponding region only begins to contribute beyond the leading power. It is then straightforward to generalize the Soft-Collinear Effective Theory analyses~\cite{BurStw13lectures, BchBrgFrl15book, Bch18lectures, BurFlmLk00, BurFlmPjlStw01, BurPjlSwt02-1, BurPjlSwt02-2,BnkChpkDhlFdm02,BnkFdm03,RthstStw16EFT} to include power corrections.

\section{Regions in wide-angle scattering: the soft expansion}
\label{section-regions_soft_expansion}

This section centers on another expansion technique for wide-angle scattering, which we refer to as the \emph{soft expansion}. Specifically, we focus on those graphs featuring three types of external momenta: on-shell external momenta $p_i^\mu$ ($i=1,\dots,K$), off-shell external momenta $q_j^\mu$ ($j=1,\dots,L$), and soft external momenta $l_k^\mu$ ($k=1,\dots,M$). Furthermore,
\begin{subequations}
\label{eq:wideangle_soft_kinematics_rewritten}
\begin{align}
    & p_i^2=0\ \ (i=1,\dots,K), \quad q_j^2\sim Q^2\ \ (j=1,\dots,L), \quad l_k^2=0\ \ (k=1,\dots,M),\\
    & p_{i_1}\cdot p_{i_2}\sim Q^2\ \ (i_1\neq i_2), \quad p_i\cdot l_k\sim q_j\cdot l_k\sim \lambda Q^2, \quad l_{k_1}\cdot l_{k_2}\sim \lambda^2 Q^2\ \ (k_1\neq k_2),
\end{align}
\end{subequations}
as the scaling parameter $\lambda\ll 1$. In contrast to the on-shell expansion where $p_i^2\sim \lambda Q^2$, here we require each $p_i^\mu$ to be strictly lightlike, i.e. $p_i^2=0$.

Let us begin by studying the one-loop box graph with external momenta 
$p_1^\mu$, $p_2^\mu$, $q_1^\mu$ and $l_1^\mu$, where $p_1^\mu$ and $p_2^\mu$ are diagonal. From table~\ref{table-oneloop_Sudakov_form_factor_result}, there are four regions in total: the soft region, the collinear-1 region, the collinear-2 region, and the hard region. By coloring the hard, jet, and soft subgraphs in blue, green, and red respectively, we have:
\begin{equation}
\begin{aligned}
\begin{tikzpicture}[baseline=11ex, scale=0.4]
\path (3,7) edge [ultra thick] (7,7) {};
\path (7,7) edge [ultra thick] (7,3) {};
\path (3,3) edge [ultra thick] (7,3) {};
\path (3,3) edge [ultra thick] (3,7) {};
\path (2,8) edge [Green, thick] (3,7) {};
\path (8,8) edge [Red, thick] (7,7) {};
\path (8,2) edge [LimeGreen, thick] (7,3) {};
\path (2,1.8) edge [Blue, ultra thick] (3,3) {};
\path (1.8,2) edge [Blue, ultra thick] (3,3) {};
\node [draw, circle, minimum size=3pt, color=Black, fill=Black, inner sep=0pt, outer sep=0pt] () at (3,3) {};
\node [draw, circle, minimum size=3pt, color=Black, fill=Black, inner sep=0pt, outer sep=0pt] () at (7,3) {};
\node [draw, circle, minimum size=3pt, color=Black, fill=Black, inner sep=0pt, outer sep=0pt] () at (3,7) {};
\node [draw, circle, minimum size=3pt, color=Black, fill=Black, inner sep=0pt, outer sep=0pt] () at (7,7) {};
\node () at (1.5,1.5) {\color{Blue} $q_1^\mu$};
\node () at (8.5,1.5) {\color{LimeGreen} $p_2^\mu$};
\node () at (8.5,8.5) {\color{Red} $l_1^\mu$};
\node () at (1.5,8.5) {\color{Green} $p_1^\mu$};
\end{tikzpicture}
\xrightarrow[]{\text{regions}}
\begin{tikzpicture}[baseline=9ex, scale=0.33]
\path (3,7) edge [Red, thick] (7,7) {};
\path (7,7) edge [Red, thick] (7,3) {};
\path (3,3) edge [LimeGreen, thick] (7,3) {};
\path (3,3) edge [Green, thick] (3,7) {};
\path (2,8) edge [Green, thick] (3,7) {};
\path (8,8) edge [Red, thick] (7,7) {};
\path (8,2) edge [LimeGreen, thick] (7,3) {};
\path (2,1.8) edge [Blue, ultra thick] (3,3) {};
\path (1.8,2) edge [Blue, ultra thick] (3,3) {};
\node [draw, circle, minimum size=3pt, color=Blue, fill=Blue, inner sep=0pt, outer sep=0pt] () at (3,3) {};
\node [draw, circle, minimum size=3pt, color=LimeGreen, fill=LimeGreen, inner sep=0pt, outer sep=0pt] () at (7,3) {};
\node [draw, circle, minimum size=3pt, color=Green, fill=Green, inner sep=0pt, outer sep=0pt] () at (3,7) {};
\node [draw, circle, minimum size=3pt, color=Red, fill=Red, inner sep=0pt, outer sep=0pt] () at (7,7) {};
\end{tikzpicture},\ \ 
\begin{tikzpicture}[baseline=9ex, scale=0.33]
\path (3,7) edge [Green, thick] (7,7) {};
\path (7,7) edge [Green, thick] (7,3) {};
\path (3,3) edge [Blue, ultra thick] (7,3) {};
\path (3,3) edge [Green, thick] (3,7) {};
\path (2,8) edge [Green, thick] (3,7) {};
\path (8,8) edge [Red, thick] (7,7) {};
\path (8,2) edge [LimeGreen, thick] (7,3) {};
\path (2,1.8) edge [Blue, ultra thick] (3,3) {};
\path (1.8,2) edge [Blue, ultra thick] (3,3) {};
\node [draw, circle, minimum size=3pt, color=Blue, fill=Blue, inner sep=0pt, outer sep=0pt] () at (3,3) {};
\node [draw, circle, minimum size=3pt, color=Blue, fill=Blue, inner sep=0pt, outer sep=0pt] () at (7,3) {};
\node [draw, circle, minimum size=3pt, color=Green, fill=Green, inner sep=0pt, outer sep=0pt] () at (3,7) {};
\node [draw, circle, minimum size=3pt, color=Green, fill=Green, inner sep=0pt, outer sep=0pt] () at (7,7) {};
\end{tikzpicture},\ \ 
\begin{tikzpicture}[baseline=9ex, scale=0.33]
\path (3,7) edge [LimeGreen, thick] (7,7) {};
\path (7,7) edge [LimeGreen, thick] (7,3) {};
\path (3,3) edge [LimeGreen, thick] (7,3) {};
\path (3,3) edge [Blue, ultra thick] (3,7) {};
\path (2,8) edge [Green, thick] (3,7) {};
\path (8,8) edge [Red, thick] (7,7) {};
\path (8,2) edge [LimeGreen, thick] (7,3) {};
\path (2,1.8) edge [Blue, ultra thick] (3,3) {};
\path (1.8,2) edge [Blue, ultra thick] (3,3) {};
\node [draw, circle, minimum size=3pt, color=Blue, fill=Blue, inner sep=0pt, outer sep=0pt] () at (3,3) {};
\node [draw, circle, minimum size=3pt, color=LimeGreen, fill=LimeGreen, inner sep=0pt, outer sep=0pt] () at (7,3) {};
\node [draw, circle, minimum size=3pt, color=Blue, fill=Blue, inner sep=0pt, outer sep=0pt] () at (3,7) {};
\node [draw, circle, minimum size=3pt, color=LimeGreen, fill=LimeGreen, inner sep=0pt, outer sep=0pt] () at (7,7) {};
\end{tikzpicture},\ \ 
\begin{tikzpicture}[baseline=9ex, scale=0.33]
\path (3,7) edge [Blue, ultra thick] (7,7) {};
\path (7,7) edge [Blue, ultra thick] (7,3) {};
\path (3,3) edge [Blue, ultra thick] (7,3) {};
\path (3,3) edge [Blue, ultra thick] (3,7) {};
\path (2,8) edge [Green, thick] (3,7) {};
\path (8,8) edge [Red, thick] (7,7) {};
\path (8,2) edge [LimeGreen, thick] (7,3) {};
\path (2,1.8) edge [Blue, ultra thick] (3,3) {};
\path (1.8,2) edge [Blue, ultra thick] (3,3) {};
\node [draw, circle, minimum size=3pt, color=Blue, fill=Blue, inner sep=0pt, outer sep=0pt] () at (3,3) {};
\node [draw, circle, minimum size=3pt, color=Blue, fill=Blue, inner sep=0pt, outer sep=0pt] () at (7,3) {};
\node [draw, circle, minimum size=3pt, color=Blue, fill=Blue, inner sep=0pt, outer sep=0pt] () at (3,7) {};
\node [draw, circle, minimum size=3pt, color=Blue, fill=Blue, inner sep=0pt, outer sep=0pt] () at (7,7) {};
\end{tikzpicture}.
\end{aligned}
\label{eq:threshold_motivation_box_4regions}
\end{equation}
Like the on-shell expansion, each region above can be linked to a pinch surface of $G|_{l^\mu=0}$. We next adjust the positions of the external momenta such that $p_1^\mu$ and $l_1^\mu$ are diagonal. In this case, only the collinear-2 region and the hard region are relevant:
\begin{equation}
\begin{aligned}
\begin{tikzpicture}[baseline=11ex, scale=0.4]
\path (3,7) edge [ultra thick] (7,7) {};
\path (7,7) edge [ultra thick] (7,3) {};
\path (3,3) edge [ultra thick] (7,3) {};
\path (3,3) edge [ultra thick] (3,7) {};
\path (2,8) edge [Green, thick] (3,7) {};
\path (8,8) edge [LimeGreen, thick] (7,7) {};
\path (8,2) edge [Red, thick] (7,3) {};
\path (2,1.8) edge [Blue, ultra thick] (3,3) {};
\path (1.8,2) edge [Blue, ultra thick] (3,3) {};
\node [draw, circle, minimum size=3pt, color=Black, fill=Black, inner sep=0pt, outer sep=0pt] () at (3,3) {};
\node [draw, circle, minimum size=3pt, color=Black, fill=Black, inner sep=0pt, outer sep=0pt] () at (7,3) {};
\node [draw, circle, minimum size=3pt, color=Black, fill=Black, inner sep=0pt, outer sep=0pt] () at (3,7) {};
\node [draw, circle, minimum size=3pt, color=Black, fill=Black, inner sep=0pt, outer sep=0pt] () at (7,7) {};
\node () at (1.5,1.5) {\color{Blue} $q_1^\mu$};
\node () at (8.5,1.5) {\color{Red} $l_1^\mu$};
\node () at (8.5,8.5) {\color{LimeGreen} $p_2^\mu$};
\node () at (1.5,8.5) {\color{Green} $p_1^\mu$};
\end{tikzpicture}
\xrightarrow[]{\text{regions}}
\begin{tikzpicture}[baseline=9ex, scale=0.33]
\path (3,7) edge [LimeGreen, thick] (7,7) {};
\path (7,7) edge [LimeGreen, thick] (7,3) {};
\path (3,3) edge [LimeGreen, thick] (7,3) {};
\path (3,3) edge [Blue, ultra thick] (3,7) {};
\path (2,8) edge [Green, thick] (3,7) {};
\path (8,8) edge [LimeGreen, thick] (7,7) {};
\path (8,2) edge [Red, thick] (7,3) {};
\path (2,1.8) edge [Blue, ultra thick] (3,3) {};
\path (1.8,2) edge [Blue, ultra thick] (3,3) {};
\node [draw, circle, minimum size=3pt, color=Blue, fill=Blue, inner sep=0pt, outer sep=0pt] () at (3,3) {};
\node [draw, circle, minimum size=3pt, color=LimeGreen, fill=LimeGreen, inner sep=0pt, outer sep=0pt] () at (7,3) {};
\node [draw, circle, minimum size=3pt, color=Blue, fill=Blue, inner sep=0pt, outer sep=0pt] () at (3,7) {};
\node [draw, circle, minimum size=3pt, color=LimeGreen, fill=LimeGreen, inner sep=0pt, outer sep=0pt] () at (7,7) {};
\end{tikzpicture},\ \ 
\begin{tikzpicture}[baseline=9ex, scale=0.33]
\path (3,7) edge [Blue, ultra thick] (7,7) {};
\path (7,7) edge [Blue, ultra thick] (7,3) {};
\path (3,3) edge [Blue, ultra thick] (7,3) {};
\path (3,3) edge [Blue, ultra thick] (3,7) {};
\path (2,8) edge [Green, thick] (3,7) {};
\path (8,8) edge [LimeGreen, thick] (7,7) {};
\path (8,2) edge [Red, thick] (7,3) {};
\path (2,1.8) edge [Blue, ultra thick] (3,3) {};
\path (1.8,2) edge [Blue, ultra thick] (3,3) {};
\node [draw, circle, minimum size=3pt, color=Blue, fill=Blue, inner sep=0pt, outer sep=0pt] () at (3,3) {};
\node [draw, circle, minimum size=3pt, color=Blue, fill=Blue, inner sep=0pt, outer sep=0pt] () at (7,3) {};
\node [draw, circle, minimum size=3pt, color=Blue, fill=Blue, inner sep=0pt, outer sep=0pt] () at (3,7) {};
\node [draw, circle, minimum size=3pt, color=Blue, fill=Blue, inner sep=0pt, outer sep=0pt] () at (7,7) {};
\end{tikzpicture}.
\end{aligned}
\label{eq:threshold_motivation_box_2regions}
\end{equation}
As validated in ref.~\cite{AnstsDhrDltHzgmMstbg13}, the summation of contributions from the regions in (\ref{eq:threshold_motivation_box_4regions}) and (\ref{eq:threshold_motivation_box_2regions}) accurately reproduces the corresponding original Feynman integrals. From these observations, it is natural to propose the following. First, similar to the regions in the on-shell expansion, each region in the soft expansion involves only the hard, collinear, and soft modes, which are in correspondence with the pinch surfaces of $G|_{l^\mu=0}$. Second, it is anticipated that additional requirements govern the hard, jet, and soft subgraphs, and configurations failing to meet these requirements would correspond to scaleless integrals. We will study these aspects in detail soon, with a specific focus on uncovering the requirements placed on the subgraphs.

We begin in section~\ref{section-motivation_from_examples} with some examples of both regions and non-region configurations, which serve as the basis for formulating a comprehensive prescription governing the regions. Moving forward to section~\ref{section-general_prescription_threshold}, we consolidate insights from the examples and propose a general prescription asserting that each region involves solely the hard, collinear, and soft modes, aligning with the principles outlined in the on-shell-expansion region theorem. Section~\ref{section-leading_terms_infrared_regions} delves into the characterization of the leading terms of the Lee-Pomeransky polynomial for a generic region. Then in section~\ref{section-further_restrictions_region_vectors_threshold}, we establish a set of requirements for the subgraphs $H$, $J$, and $S$, demonstrating their necessity and sufficiency for the regions of the soft expansion. The results of these analyses are summarized in section~\ref{section-summary_threshold} and referred to as the soft-expansion region proposition. Finally, in section~\ref{section-extending_results_additional_expansions}, we discuss how these findings can be seen as a generalization of the results from the on-shell expansion discussed in section~\ref{section-regions_onshell}. We also explore the further extension of these results when incorporating additional expansions.

\subsection{Motivations from examples}
\label{section-motivation_from_examples}

Let us first consider the Sudakov form factor graphs, which are relevant in some specific decay processes: an off-shell particle decays into two lightlike particles travelling in two distinct directions, accompanied by some soft radiation carrying momenta $l_1^\mu$, $l_2^\mu$, etc. According to eq.~(\ref{eq:wideangle_soft_kinematics_rewritten}), the virtuality of each lightlike or soft external momentum is precisely zero. Four examples of regions without (internal) soft propagators are shown below. Next to each figure, we have included a ``\greencheckmark[ForestGreen]'' to emphasize that it is a valid region.
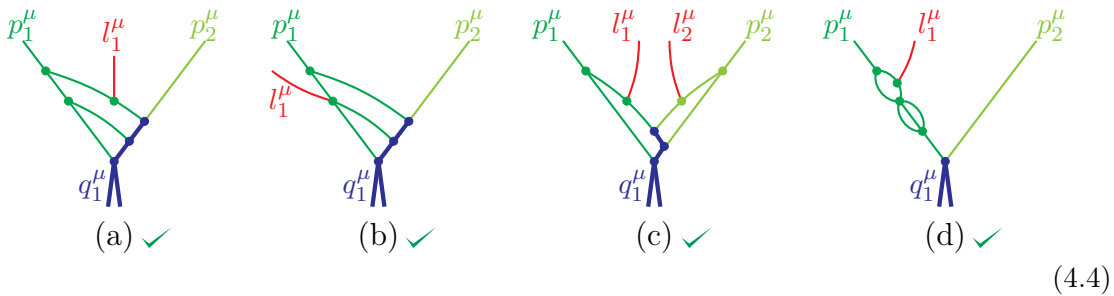
\begin{equation}
\begin{aligned}
\begin{tikzpicture}[baseline=11ex, scale=0.4]
\path (2,7) edge [Green, thick] (5,3) {};
\path (5,6.5) edge [Red, thick] (5,5) {};
\path (2.75,6) edge [Green, thick, bend left = 10] (6,4.33) {};
\path (3.5,5) edge [Green, thick, bend left = 10] (5.5,3.67) {};
\path (8,7) edge [LimeGreen, thick] (6,4.33) {};
\path (6,4.33) edge [Blue, ultra thick] (5,3) {};
\path (4.8,1.5) edge [Blue, ultra thick] (5,3) {};
\path (5.2,1.5) edge [Blue, ultra thick] (5,3) {};
\node [draw, circle, minimum size=3pt, color=Green, fill=Green, inner sep=0pt, outer sep=0pt] () at (2.75,6) {};
\node [draw, circle, minimum size=3pt, color=Blue, fill=Blue, inner sep=0pt, outer sep=0pt] () at (5.5,3.67) {};
\node [draw, circle, minimum size=3pt, color=Green, fill=Green, inner sep=0pt, outer sep=0pt] () at (3.5,5) {};
\node [draw, circle, minimum size=3pt, color=Blue, fill=Blue, inner sep=0pt, outer sep=0pt] () at (6,4.33) {};
\node [draw, circle, minimum size=3pt, color=Green, fill=Green, inner sep=0pt, outer sep=0pt] () at (5,5) {};
\node [draw, circle, minimum size=3pt, color=Blue, fill=Blue, inner sep=0pt, outer sep=0pt] () at (5,3) {};
\node () at (4.3,2.2) {\color{Blue} $q_1^\mu$};
\node () at (5,7.2) {\color{Red} $l_1^\mu$};
\node () at (8,7.5) {\color{LimeGreen} $p_2^\mu$};
\node () at (2,7.5) {\color{Green} $p_1^\mu$};
\node () at (5,0.5) {\large (a)};
\node () at (6.4,2.4) {$\greencheckmark[ForestGreen]$};
\end{tikzpicture}\quad
\begin{tikzpicture}[baseline=11ex, scale=0.4]
\path (2,7) edge [Green, thick] (5,3) {};
\path (1.5,6) edge [Red, thick, bend right = 10] (3.5,5) {};
\path (2.75,6) edge [Green, thick, bend left = 10] (6,4.33) {};
\path (3.5,5) edge [Green, thick, bend left = 10] (5.5,3.67) {};
\path (8,7) edge [LimeGreen, thick] (6,4.33) {};
\path (6,4.33) edge [Blue, ultra thick] (5,3) {};
\path (4.8,1.5) edge [Blue, ultra thick] (5,3) {};
\path (5.2,1.5) edge [Blue, ultra thick] (5,3) {};
\node [draw, circle, minimum size=3pt, color=Green, fill=Green, inner sep=0pt, outer sep=0pt] () at (2.75,6) {};
\node [draw, circle, minimum size=3pt, color=Blue, fill=Blue, inner sep=0pt, outer sep=0pt] () at (5.5,3.67) {};
\node [draw, circle, minimum size=3pt, color=Green, fill=Green, inner sep=0pt, outer sep=0pt] () at (3.5,5) {};
\node [draw, circle, minimum size=3pt, color=Blue, fill=Blue, inner sep=0pt, outer sep=0pt] () at (6,4.33) {};
\node [draw, circle, minimum size=3pt, color=Blue, fill=Blue, inner sep=0pt, outer sep=0pt] () at (5,3) {};
\node () at (4.3,2.2) {\color{Blue} $q_1^\mu$};
\node () at (1.8,5) {\color{Red} $l_1^\mu$};
\node () at (8,7.5) {\color{LimeGreen} $p_2^\mu$};
\node () at (2,7.5) {\color{Green} $p_1^\mu$};
\node () at (5,0.5) {\large (b)};
\node () at (6.4,2.4) {$\greencheckmark[ForestGreen]$};
\end{tikzpicture}\quad
\begin{tikzpicture}[baseline=11ex, scale=0.4]
\path (2,7) edge [Green, thick] (5,3) {};
\path (2.75,6) edge [Green, thick, bend left = 10] (5,4) {};
\path (4.5,7) edge [Red, thick, bend left = 10] (4.1,5) {};
\path (5.5,7) edge [Red, thick, bend right = 10] (5.9,5) {};
\path (5,4) edge [LimeGreen, thick, bend left = 10] (7.25,6) {};
\path (8,7) edge [LimeGreen, thick] (5,3) {};
\path (5.33,3.5) edge [Blue, ultra thick] (5,4) {};
\path (5.33,3.5) edge [Blue, ultra thick] (5,3) {};
\path (4.8,1.5) edge [Blue, ultra thick] (5,3) {};
\path (5.2,1.5) edge [Blue, ultra thick] (5,3) {};
\node [draw, circle, minimum size=3pt, color=Green, fill=Green, inner sep=0pt, outer sep=0pt] () at (2.75,6) {};
\node [draw, circle, minimum size=3pt, color=LimeGreen, fill=LimeGreen, inner sep=0pt, outer sep=0pt] () at (7.25,6) {};
\node [draw, circle, minimum size=3pt, color=Green, fill=Green, inner sep=0pt, outer sep=0pt] () at (4.1,5) {};
\node [draw, circle, minimum size=3pt, color=LimeGreen, fill=LimeGreen, inner sep=0pt, outer sep=0pt] () at (5.9,5) {};
\node [draw, circle, minimum size=3pt, color=Blue, fill=Blue, inner sep=0pt, outer sep=0pt] () at (5,3) {};
\node [draw, circle, minimum size=3pt, color=Blue, fill=Blue, inner sep=0pt, outer sep=0pt] () at (5,4) {};
\node [draw, circle, minimum size=3pt, color=Blue, fill=Blue, inner sep=0pt, outer sep=0pt] () at (5.33,3.5) {};
\node () at (4.3,2.2) {\color{Blue} $q_1^\mu$};
\node () at (4,7.5) {\color{Red} $l_1^\mu$};
\node () at (6,7.5) {\color{Red} $l_2^\mu$};
\node () at (8.5,7.5) {\color{LimeGreen} $p_2^\mu$};
\node () at (1.5,7.5) {\color{Green} $p_1^\mu$};
\node () at (5,0.5) {\large (c)};
\node () at (6.4,2.4) {$\greencheckmark[ForestGreen]$};
\end{tikzpicture}\quad
\begin{tikzpicture}[baseline=11ex, scale=0.4]
\path (2,7) edge [Green, thick] (2.75,6) {};
\path (3.5,5) edge [Green, thick] (5,3) {};
\path (3.42,5.6) edge [Red, thick, bend right =10] (4,7) {};
\path (2.75,6) edge [Green, thick, bend right = 60] (3.5,5) {};
\path (2.75,6) edge [Green, thick, bend left = 60] (3.5,5) {};
\path ((4.25,4) edge [Green, thick, bend right = 60] (3.5,5) {};
\path (4.25,4) edge [Green, thick, bend left = 60] (3.5,5) {};
\path (8,7) edge [LimeGreen, thick] (5,3) {};
\path (4.8,1.5) edge [Blue, ultra thick] (5,3) {};
\path (5.2,1.5) edge [Blue, ultra thick] (5,3) {};
\node [draw, circle, minimum size=3pt, color=Green, fill=Green, inner sep=0pt, outer sep=0pt] () at (2.75,6) {};
\node [draw, circle, minimum size=3pt, color=Green, fill=Green, inner sep=0pt, outer sep=0pt] () at (3.42,5.6) {};
\node [draw, circle, minimum size=3pt, color=Green, fill=Green, inner sep=0pt, outer sep=0pt] () at (3.5,5) {};
\node [draw, circle, minimum size=3pt, color=Green, fill=Green, inner sep=0pt, outer sep=0pt] () at (4.25,4) {};
\node [draw, circle, minimum size=3pt, color=Blue, fill=Blue, inner sep=0pt, outer sep=0pt] () at (5,3) {};
\node () at (4.3,2.2) {\color{Blue} $q_1^\mu$};
\node () at (4.5,7.5) {\color{Red} $l_1^\mu$};
\node () at (8.5,7.5) {\color{LimeGreen} $p_2^\mu$};
\node () at (1.5,7.5) {\color{Green} $p_1^\mu$};
\node () at (5,0.5) {\large (d)};
\node () at (6.4,2.4) {$\greencheckmark[ForestGreen]$};
\end{tikzpicture}
\end{aligned}
\label{eq:threshold_motivation_Sudakov_collinear_regions}
\end{equation}
In (a) and (b) above, the jet $J_1$ is nontrivial and there are two loops in the contracted jet subgraph $\widetilde{J}_1$, while the jet $J_2$ is trivial (consisting of the external momentum $p_2^\mu$ only). The soft momentum $l_1^\mu$ is attached to $J_1$ directly. In (c) there are two nontrivial jets $J_1$ and $J_2$, which are attached by $l_1^\mu$ and $l_2^\mu$ respectively. In (d), $J_1$ is nontrivial, and $\widetilde{J}_1$ consists of multiple 1VI components. The external momentum $l_1^\mu$ is attached to the ``outermost'' 1VI component, i.e. the one attached by $p_1^\mu$.

In contrast, the examples below are non-region configurations, each of which can be obtained by slightly modifying the regions above. Next to each figure, we have included a ``\crossmark[Red]'' to emphasize that it is \emph{not} a valid region.
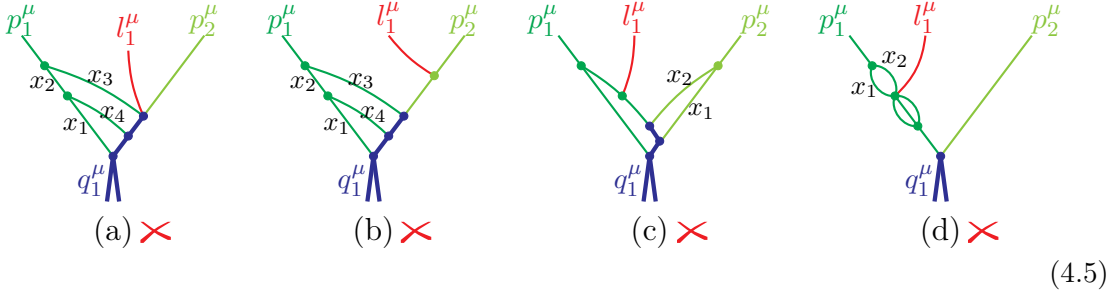
\begin{equation}
\begin{aligned}
\begin{tikzpicture}[baseline=11ex, scale=0.4]
\path (2,7) edge [Green, thick] (5,3) {};
\path (5.5,6.5) edge [Red, thick, bend right =10] (6,4.33) {};
\path (2.75,6) edge [Green, thick, bend left = 10] (6,4.33) {};
\path (3.5,5) edge [Green, thick, bend left = 10] (5.5,3.67) {};
\path (8,7) edge [LimeGreen, thick] (6,4.33) {};
\path (6,4.33) edge [Blue, ultra thick] (5,3) {};
\path (4.8,1.5) edge [Blue, ultra thick] (5,3) {};
\path (5.2,1.5) edge [Blue, ultra thick] (5,3) {};
\node [draw, circle, minimum size=3pt, color=Green, fill=Green, inner sep=0pt, outer sep=0pt] () at (2.75,6) {};
\node [draw, circle, minimum size=3pt, color=Blue, fill=Blue, inner sep=0pt, outer sep=0pt] () at (5.5,3.67) {};
\node [draw, circle, minimum size=3pt, color=Green, fill=Green, inner sep=0pt, outer sep=0pt] () at (3.5,5) {};
\node [draw, circle, minimum size=3pt, color=Blue, fill=Blue, inner sep=0pt, outer sep=0pt] () at (6,4.33) {};
\node [draw, circle, minimum size=3pt, color=Blue, fill=Blue, inner sep=0pt, outer sep=0pt] () at (5,3) {};
\node () at (4.3,2.2) {\color{Blue} $q_1^\mu$};
\node () at (5.5,7.2) {\color{Red} $l_1^\mu$};
\node () at (8,7.5) {\color{LimeGreen} $p_2^\mu$};
\node () at (2,7.5) {\color{Green} $p_1^\mu$};
\node () at (3.8,4) {\small $x_1$};
\node () at (2.7,5.4) {\small $x_2$};
\node () at (4.6,5.6) {\small $x_3$};
\node () at (5,4.4) {\small $x_4$};
\node () at (5,0.5) {\large (a)};
\node () at (6.4,2.4) {$\crossmark[Red]$};
\end{tikzpicture}\quad
\begin{tikzpicture}[baseline=11ex, scale=0.4]
\path (2,7) edge [Green, thick] (5,3) {};
\path (5.5,7) edge [Red, thick, bend right = 10] (7,5.67) {};
\path (2.75,6) edge [Green, thick, bend left = 10] (6,4.33) {};
\path (3.5,5) edge [Green, thick, bend left = 10] (5.5,3.67) {};
\path (8,7) edge [LimeGreen, thick] (6,4.33) {};
\path (6,4.33) edge [Blue, ultra thick] (5,3) {};
\path (4.8,1.5) edge [Blue, ultra thick] (5,3) {};
\path (5.2,1.5) edge [Blue, ultra thick] (5,3) {};
\node [draw, circle, minimum size=3pt, color=Green, fill=Green, inner sep=0pt, outer sep=0pt] () at (2.75,6) {};
\node [draw, circle, minimum size=3pt, color=Blue, fill=Blue, inner sep=0pt, outer sep=0pt] () at (5.5,3.67) {};
\node [draw, circle, minimum size=3pt, color=Green, fill=Green, inner sep=0pt, outer sep=0pt] () at (3.5,5) {};
\node [draw, circle, minimum size=3pt, color=Blue, fill=Blue, inner sep=0pt, outer sep=0pt] () at (6,4.33) {};
\node [draw, circle, minimum size=3pt, color=Blue, fill=Blue, inner sep=0pt, outer sep=0pt] () at (5,3) {};
\node [draw, circle, minimum size=3pt, color=LimeGreen, fill=LimeGreen, inner sep=0pt, outer sep=0pt] () at (7,5.67) {};
\node () at (4.3,2.2) {\color{Blue} $q_1^\mu$};
\node () at (5.5,7.6) {\color{Red} $l_1^\mu$};
\node () at (8,7.5) {\color{LimeGreen} $p_2^\mu$};
\node () at (2,7.5) {\color{Green} $p_1^\mu$};
\node () at (3.8,4) {\small $x_1$};
\node () at (2.7,5.4) {\small $x_2$};
\node () at (4.6,5.5) {\small $x_3$};
\node () at (5,4.4) {\small $x_4$};
\node () at (5,0.5) {\large (b)};
\node () at (6.4,2.4) {$\crossmark[Red]$};
\end{tikzpicture}\quad
\begin{tikzpicture}[baseline=11ex, scale=0.4]
\path (2,7) edge [Green, thick] (5,3) {};
\path (2.75,6) edge [Green, thick, bend left = 10] (5,4) {};
\path (4.5,7) edge [Red, thick, bend left = 10] (4.1,5) {};
\path (5,4) edge [LimeGreen, thick, bend left = 10] (7.25,6) {};
\path (8,7) edge [LimeGreen, thick] (5,3) {};
\path (5.33,3.5) edge [Blue, ultra thick] (5,4) {};
\path (5.33,3.5) edge [Blue, ultra thick] (5,3) {};
\path (4.8,1.5) edge [Blue, ultra thick] (5,3) {};
\path (5.2,1.5) edge [Blue, ultra thick] (5,3) {};
\node [draw, circle, minimum size=3pt, color=Green, fill=Green, inner sep=0pt, outer sep=0pt] () at (2.75,6) {};
\node [draw, circle, minimum size=3pt, color=LimeGreen, fill=LimeGreen, inner sep=0pt, outer sep=0pt] () at (7.25,6) {};
\node [draw, circle, minimum size=3pt, color=Green, fill=Green, inner sep=0pt, outer sep=0pt] () at (4.1,5) {};
\node [draw, circle, minimum size=3pt, color=Blue, fill=Blue, inner sep=0pt, outer sep=0pt] () at (5,3) {};
\node [draw, circle, minimum size=3pt, color=Blue, fill=Blue, inner sep=0pt, outer sep=0pt] () at (5,4) {};
\node [draw, circle, minimum size=3pt, color=Blue, fill=Blue, inner sep=0pt, outer sep=0pt] () at (5.33,3.5) {};
\node () at (4.3,2.2) {\color{Blue} $q_1^\mu$};
\node () at (4.5,7.6) {\color{Red} $l_1^\mu$};
\node () at (8.5,7.5) {\color{LimeGreen} $p_2^\mu$};
\node () at (1.5,7.5) {\color{Green} $p_1^\mu$};
\node () at (6.7,4.5) {\small $x_1$};
\node () at (6,5.6) {\small $x_2$};
\node () at (5,0.5) {\large (c)};
\node () at (6.4,2.4) {$\crossmark[Red]$};
\end{tikzpicture}\quad
\begin{tikzpicture}[baseline=11ex, scale=0.4]
\path (2,7) edge [Green, thick] (2.75,6) {};
\path (3.5,5) edge [Green, thick] (5,3) {};
\path (3.5,5) edge [Red, thick, bend right = 20] (4.5,7) {};
\path (2.75,6) edge [Green, thick, bend right = 60] (3.5,5) {};
\path (2.75,6) edge [Green, thick, bend left = 60] (3.5,5) {};
\path ((4.25,4) edge [Green, thick, bend right = 60] (3.5,5) {};
\path (4.25,4) edge [Green, thick, bend left = 60] (3.5,5) {};
\path (8,7) edge [LimeGreen, thick] (5,3) {};
\path (4.8,1.5) edge [Blue, ultra thick] (5,3) {};
\path (5.2,1.5) edge [Blue, ultra thick] (5,3) {};
\node [draw, circle, minimum size=3pt, color=Green, fill=Green, inner sep=0pt, outer sep=0pt] () at (2.75,6) {};
\node [draw, circle, minimum size=3pt, color=Green, fill=Green, inner sep=0pt, outer sep=0pt] () at (3.5,5) {};
\node [draw, circle, minimum size=3pt, color=Green, fill=Green, inner sep=0pt, outer sep=0pt] () at (4.25,4) {};
\node [draw, circle, minimum size=3pt, color=Blue, fill=Blue, inner sep=0pt, outer sep=0pt] () at (5,3) {};
\node () at (4.3,2.2) {\color{Blue} $q_1^\mu$};
\node () at (4.5,7.6) {\color{Red} $l_1^\mu$};
\node () at (8.5,7.5) {\color{LimeGreen} $p_2^\mu$};
\node () at (1.5,7.5) {\color{Green} $p_1^\mu$};
\node () at (2.5,5.2) {\small $x_1$};
\node () at (3.5,6.2) {\small $x_2$};
\node () at (5,0.5) {\large (d)};
\node () at (6.4,2.4) {$\crossmark[Red]$};
\end{tikzpicture}
\end{aligned}
\label{eq:threshold_motivation_Sudakov_collinear_nonregions}
\end{equation}
In configurations (a)-(c) above, there is a nontrivial jet that is not attached by any soft external momenta, namely, $J_1$ in (a) and (b), and $J_2$ in (c). In configuration (d), the only nontrivial jet is $J_1$, which is attached by the soft momentum $l_1^\mu$. Note that $l_1^\mu$ does not attach to any internal vertex of the ``outermost'' 1VI component of $\widetilde{J}_1$ (the one is attached by $p_1^\mu$), differing from the case of (d) in (\ref{eq:threshold_motivation_Sudakov_collinear_regions}).

To understand why these configurations lead to scaleless integrals, one approach is to explore the homogeneity properties of their associated leading polynomials $\mathcal{P}^{(R)}(\x;\s)$. Using the parameterization in (\ref{eq:threshold_motivation_Sudakov_collinear_nonregions}), one can check that for both (a) and (b), $\mathcal{P}^{(R)}(\x;\s)$ is quadratic in $x_1,x_2,x_3,x_4$; for both (c) and (d), $\mathcal{P}^{(R)}(\x;\s)$ is linear in $x_1,x_2$. Each homogeneity property restricts the overall dimension of the lower face $f_R$, making \hbox{$\text{dim}f_R<N$} and the corresponding Feynman integrals scaleless, as explained in the facet criterion.

In fact, prior to any expansion, the original Feynman integral of (d) is already scaleless, because the total momentum flowing into one 1VI component of $G$ is exactly $p_i^\mu$, and $p_i^2=0$. So we shall not study such cases. In other words, we will assume that the total momentum flowing into each given 1VI component of $G$ has a nonzero virtuality.

Let us now propose the first rule for the regions from these examples: \emph{if a region does not feature any internal soft propagators, then every nontrivial jet must be attached by some soft external momenta.}

We next examine some examples of regions with soft propagators.
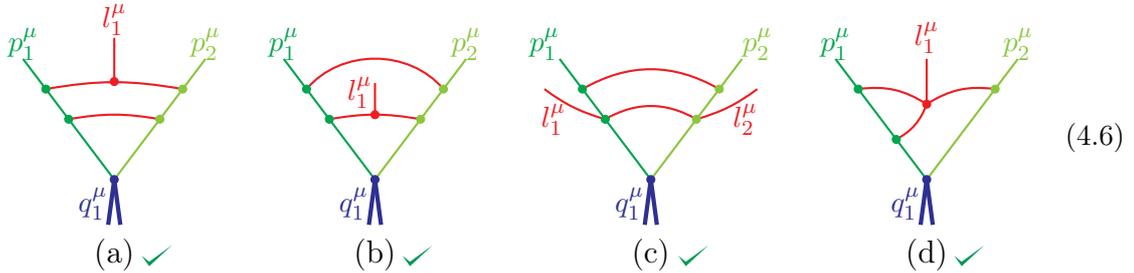
\begin{equation}
\begin{aligned}
\begin{tikzpicture}[baseline=11ex, scale=0.4]
\path (2,7) edge [Green, thick] (5,3) {};
\path (5,7.7) edge [Red, thick] (5,6.25) {};
\path (2.75,6) edge [Red, thick, bend left = 10] (7.25,6) {};
\path (3.5,5) edge [Red, thick, bend left = 10] (6.5,5) {};
\path (8,7) edge [LimeGreen, thick] (5,3) {};
\path (4.8,1.5) edge [Blue, ultra thick] (5,3) {};
\path (5.2,1.5) edge [Blue, ultra thick] (5,3) {};
\node [draw, circle, minimum size=3pt, color=Green, fill=Green, inner sep=0pt, outer sep=0pt] () at (2.75,6) {};
\node [draw, circle, minimum size=3pt, color=LimeGreen, fill=LimeGreen, inner sep=0pt, outer sep=0pt] () at (7.25,6) {};
\node [draw, circle, minimum size=3pt, color=Green, fill=Green, inner sep=0pt, outer sep=0pt] () at (3.5,5) {};
\node [draw, circle, minimum size=3pt, color=LimeGreen, fill=LimeGreen, inner sep=0pt, outer sep=0pt] () at (6.5,5) {};
\node [draw, circle, minimum size=3pt, color=Red, fill=Red, inner sep=0pt, outer sep=0pt] () at (5,6.25) {};
\node [draw, circle, minimum size=3pt, color=Blue, fill=Blue, inner sep=0pt, outer sep=0pt] () at (5,3) {};
\node () at (4.3,2.2) {\color{Blue} $q_1^\mu$};
\node () at (5,8.3) {\color{Red} $l_1^\mu$};
\node () at (8,7.5) {\color{LimeGreen} $p_2^\mu$};
\node () at (2,7.5) {\color{Green} $p_1^\mu$};
\node () at (5,0.5) {\large (a)};
\node () at (6.4,2.4) {$\greencheckmark[ForestGreen]$};
\end{tikzpicture}\quad
\begin{tikzpicture}[baseline=11ex, scale=0.4]
\path (2,7) edge [Green, thick] (5,3) {};
\path (5,6.2) edge [Red, thick] (5,5.15) {};
\path (2.75,6) edge [Red, thick, bend left = 50] (7.25,6) {};
\path (3.5,5) edge [Red, thick, bend left = 10] (6.5,5) {};
\path (8,7) edge [LimeGreen, thick] (5,3) {};
\path (4.8,1.5) edge [Blue, ultra thick] (5,3) {};
\path (5.2,1.5) edge [Blue, ultra thick] (5,3) {};
\node [draw, circle, minimum size=3pt, color=Green, fill=Green, inner sep=0pt, outer sep=0pt] () at (2.75,6) {};
\node [draw, circle, minimum size=3pt, color=LimeGreen, fill=LimeGreen, inner sep=0pt, outer sep=0pt] () at (7.25,6) {};
\node [draw, circle, minimum size=3pt, color=Green, fill=Green, inner sep=0pt, outer sep=0pt] () at (3.5,5) {};
\node [draw, circle, minimum size=3pt, color=LimeGreen, fill=LimeGreen, inner sep=0pt, outer sep=0pt] () at (6.5,5) {};
\node [draw, circle, minimum size=3pt, color=Red, fill=Red, inner sep=0pt, outer sep=0pt] () at (5,5.15) {};
\node [draw, circle, minimum size=3pt, color=Blue, fill=Blue, inner sep=0pt, outer sep=0pt] () at (5,3) {};
\node () at (4.3,2.2) {\color{Blue} $q_1^\mu$};
\node () at (4.5,5.9) {\color{Red} $l_1^\mu$};
\node () at (8,7.5) {\color{LimeGreen} $p_2^\mu$};
\node () at (2,7.5) {\color{Green} $p_1^\mu$};
\node () at (5,0.5) {\large (b)};
\node () at (6.4,2.4) {$\greencheckmark[ForestGreen]$};
\end{tikzpicture}\quad
\begin{tikzpicture}[baseline=11ex, scale=0.4]
\path (2,7) edge [Green, thick] (5,3) {};
\path (1.5,6) edge [Red, thick, bend right = 10] (3.5,5) {};
\path (8.5,6) edge [Red, thick, bend left = 10] (6.5,5) {};
\path (2.75,6) edge [Red, thick, bend left = 30] (7.25,6) {};
\path (3.5,5) edge [Red, thick, bend left = 30] (6.5,5) {};
\path (8,7) edge [LimeGreen, thick] (5,3) {};
\path (4.8,1.5) edge [Blue, ultra thick] (5,3) {};
\path (5.2,1.5) edge [Blue, ultra thick] (5,3) {};
\node [draw, circle, minimum size=3pt, color=Green, fill=Green, inner sep=0pt, outer sep=0pt] () at (2.75,6) {};
\node [draw, circle, minimum size=3pt, color=LimeGreen, fill=LimeGreen, inner sep=0pt, outer sep=0pt] () at (7.25,6) {};
\node [draw, circle, minimum size=3pt, color=Green, fill=Green, inner sep=0pt, outer sep=0pt] () at (3.5,5) {};
\node [draw, circle, minimum size=3pt, color=LimeGreen, fill=LimeGreen, inner sep=0pt, outer sep=0pt] () at (6.5,5) {};
\node [draw, circle, minimum size=3pt, color=Blue, fill=Blue, inner sep=0pt, outer sep=0pt] () at (5,3) {};
\node () at (4.3,2.2) {\color{Blue} $q_1^\mu$};
\node () at (1.8,5) {\color{Red} $l_1^\mu$};
\node () at (8.1,5) {\color{Red} $l_2^\mu$};
\node () at (8.5,7.5) {\color{LimeGreen} $p_2^\mu$};
\node () at (1.5,7.5) {\color{Green} $p_1^\mu$};
\node () at (5,0.5) {\large (c)};
\node () at (6.4,2.4) {$\greencheckmark[ForestGreen]$};
\end{tikzpicture}\quad
\begin{tikzpicture}[baseline=11ex, scale=0.4]
\path (2,7) edge [Green, thick] (5,3) {};
\path (5,7) edge [Red, thick] (5,5.5) {};
\path (2.75,6) edge [Red, thick, bend left = 20] (5,5.5) {};
\path (4,4.33) edge [Red, thick, bend right = 20] (5,5.5) {};
\path (7.25,6) edge [Red, thick, bend right = 20] (5,5.5) {};
\path (8,7) edge [LimeGreen, thick] (5,3) {};
\path (4.8,1.5) edge [Blue, ultra thick] (5,3) {};
\path (5.2,1.5) edge [Blue, ultra thick] (5,3) {};
\node [draw, circle, minimum size=3pt, color=Green, fill=Green, inner sep=0pt, outer sep=0pt] () at (2.75,6) {};
\node [draw, circle, minimum size=3pt, color=LimeGreen, fill=LimeGreen, inner sep=0pt, outer sep=0pt] () at (7.25,6) {};
\node [draw, circle, minimum size=3pt, color=Green, fill=Green, inner sep=0pt, outer sep=0pt] () at (4,4.33) {};
\node [draw, circle, minimum size=3pt, color=Red, fill=Red, inner sep=0pt, outer sep=0pt] () at (5,5.5) {};
\node [draw, circle, minimum size=3pt, color=Blue, fill=Blue, inner sep=0pt, outer sep=0pt] () at (5,3) {};
\node () at (4.3,2.2) {\color{Blue} $q_1^\mu$};
\node () at (5,7.7) {\color{Red} $l_1^\mu$};
\node () at (8,7.5) {\color{LimeGreen} $p_2^\mu$};
\node () at (2,7.5) {\color{Green} $p_1^\mu$};
\node () at (5,0.5) {\large (d)};
\node () at (6.4,2.4) {$\greencheckmark[ForestGreen]$};
\end{tikzpicture}
\end{aligned}
\label{eq:threshold_motivation_Sudakov_soft_regions}
\end{equation}
In (a) and (b), there are two soft components, and the only soft external momentum $l_1^\mu$ directly attaches to either one of them. In (c), neither components of $S$ is attached by any soft external momenta, while $J_1$ and $J_2$ are attached by $l_1^\mu$ and $l_2^\mu$ respectively. In (d), there is a unique connected component of $S$, which the soft momentum $l^\mu$ directly attaches to.

In contrast, below are some non-region configurations, which can be obtained by slightly modifying the regions above.
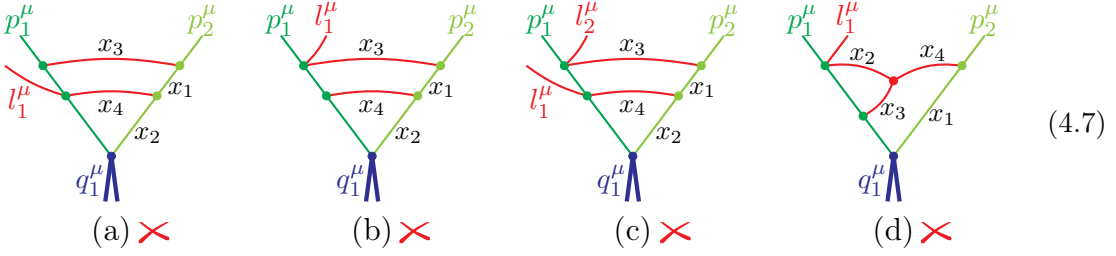
\begin{equation}
\begin{aligned}
\begin{tikzpicture}[baseline=11ex, scale=0.4]
\path (2,7) edge [Green, thick] (5,3) {};
\path (1.5,6) edge [Red, thick, bend right = 10] (3.5,5) {};
\path (2.75,6) edge [Red, thick, bend left = 10] (7.25,6) {};
\path (3.5,5) edge [Red, thick, bend left = 10] (6.5,5) {};
\path (8,7) edge [LimeGreen, thick] (5,3) {};
\path (4.8,1.5) edge [Blue, ultra thick] (5,3) {};
\path (5.2,1.5) edge [Blue, ultra thick] (5,3) {};
\node [draw, circle, minimum size=3pt, color=Green, fill=Green, inner sep=0pt, outer sep=0pt] () at (2.75,6) {};
\node [draw, circle, minimum size=3pt, color=LimeGreen, fill=LimeGreen, inner sep=0pt, outer sep=0pt] () at (7.25,6) {};
\node [draw, circle, minimum size=3pt, color=Green, fill=Green, inner sep=0pt, outer sep=0pt] () at (3.5,5) {};
\node [draw, circle, minimum size=3pt, color=LimeGreen, fill=LimeGreen, inner sep=0pt, outer sep=0pt] () at (6.5,5) {};
\node [draw, circle, minimum size=3pt, color=Blue, fill=Blue, inner sep=0pt, outer sep=0pt] () at (5,3) {};
\node () at (4.3,2.2) {\color{Blue} $q_1^\mu$};
\node () at (2,4.8) {\color{Red} $l_1^\mu$};
\node () at (8,7.5) {\color{LimeGreen} $p_2^\mu$};
\node () at (2,7.5) {\color{Green} $p_1^\mu$};
\node () at (7.3,5.2) {\small $x_1$};
\node () at (6.2,3.7) {\small $x_2$};
\node () at (5,6.6) {\small $x_3$};
\node () at (5,4.6) {\small $x_4$};
\node () at (5,0.5) {\large (a)};
\node () at (6.4,2.4) {$\crossmark[Red]$};
\end{tikzpicture}\quad
\begin{tikzpicture}[baseline=11ex, scale=0.4]
\path (2,7) edge [Green, thick] (5,3) {};
\path (2.75,6) edge [Red, thick, bend right = 10] (3.5,7) {};
\path (2.75,6) edge [Red, thick, bend left = 10] (7.25,6) {};
\path (3.5,5) edge [Red, thick, bend left = 10] (6.5,5) {};
\path (8,7) edge [LimeGreen, thick] (5,3) {};
\path (4.8,1.5) edge [Blue, ultra thick] (5,3) {};
\path (5.2,1.5) edge [Blue, ultra thick] (5,3) {};
\node [draw, circle, minimum size=3pt, color=Green, fill=Green, inner sep=0pt, outer sep=0pt] () at (2.75,6) {};
\node [draw, circle, minimum size=3pt, color=LimeGreen, fill=LimeGreen, inner sep=0pt, outer sep=0pt] () at (7.25,6) {};
\node [draw, circle, minimum size=3pt, color=Green, fill=Green, inner sep=0pt, outer sep=0pt] () at (3.5,5) {};
\node [draw, circle, minimum size=3pt, color=LimeGreen, fill=LimeGreen, inner sep=0pt, outer sep=0pt] () at (6.5,5) {};
\node [draw, circle, minimum size=3pt, color=Blue, fill=Blue, inner sep=0pt, outer sep=0pt] () at (5,3) {};
\node () at (4.3,2.2) {\color{Blue} $q_1^\mu$};
\node () at (3.5,7.6) {\color{Red} $l_1^\mu$};
\node () at (8,7.5) {\color{LimeGreen} $p_2^\mu$};
\node () at (2,7.5) {\color{Green} $p_1^\mu$};
\node () at (7.3,5.2) {\small $x_1$};
\node () at (6.2,3.7) {\small $x_2$};
\node () at (5,6.6) {\small $x_3$};
\node () at (5,4.6) {\small $x_4$};
\node () at (5,0.5) {\large (b)};
\node () at (6.4,2.4) {$\crossmark[Red]$};
\end{tikzpicture}\quad
\begin{tikzpicture}[baseline=11ex, scale=0.4]
\path (2,7) edge [Green, thick] (5,3) {};
\path (1.5,6) edge [Red, thick, bend right = 10] (3.5,5) {};
\path (2.75,6) edge [Red, thick, bend right = 10] (3.5,7) {};
\path (2.75,6) edge [Red, thick, bend left = 10] (7.25,6) {};
\path (3.5,5) edge [Red, thick, bend left = 10] (6.5,5) {};
\path (8,7) edge [LimeGreen, thick] (5,3) {};
\path (4.8,1.5) edge [Blue, ultra thick] (5,3) {};
\path (5.2,1.5) edge [Blue, ultra thick] (5,3) {};
\node [draw, circle, minimum size=3pt, color=Green, fill=Green, inner sep=0pt, outer sep=0pt] () at (2.75,6) {};
\node [draw, circle, minimum size=3pt, color=LimeGreen, fill=LimeGreen, inner sep=0pt, outer sep=0pt] () at (7.25,6) {};
\node [draw, circle, minimum size=3pt, color=Green, fill=Green, inner sep=0pt, outer sep=0pt] () at (3.5,5) {};
\node [draw, circle, minimum size=3pt, color=LimeGreen, fill=LimeGreen, inner sep=0pt, outer sep=0pt] () at (6.5,5) {};
\node [draw, circle, minimum size=3pt, color=Blue, fill=Blue, inner sep=0pt, outer sep=0pt] () at (5,3) {};
\node () at (4.3,2.2) {\color{Blue} $q_1^\mu$};
\node () at (2,4.8) {\color{Red} $l_1^\mu$};
\node () at (3.5,7.6) {\color{Red} $l_2^\mu$};
\node () at (8,7.5) {\color{LimeGreen} $p_2^\mu$};
\node () at (2,7.5) {\color{Green} $p_1^\mu$};
\node () at (7.3,5.2) {\small $x_1$};
\node () at (6.2,3.7) {\small $x_2$};
\node () at (5,6.6) {\small $x_3$};
\node () at (5,4.6) {\small $x_4$};
\node () at (5,0.5) {\large (c)};
\node () at (6.4,2.4) {$\crossmark[Red]$};
\end{tikzpicture}\quad
\begin{tikzpicture}[baseline=11ex, scale=0.4]
\path (2,7) edge [Green, thick] (5,3) {};
\path (2.75,6) edge [Red, thick] (3.5,7) {};
\path (2.75,6) edge [Red, thick, bend left = 20] (5,5.5) {};
\path (4,4.33) edge [Red, thick, bend right = 20] (5,5.5) {};
\path (7.25,6) edge [Red, thick, bend right = 20] (5,5.5) {};
\path (8,7) edge [LimeGreen, thick] (5,3) {};
\path (4.8,1.5) edge [Blue, ultra thick] (5,3) {};
\path (5.2,1.5) edge [Blue, ultra thick] (5,3) {};
\node [draw, circle, minimum size=3pt, color=Green, fill=Green, inner sep=0pt, outer sep=0pt] () at (2.75,6) {};
\node [draw, circle, minimum size=3pt, color=LimeGreen, fill=LimeGreen, inner sep=0pt, outer sep=0pt] () at (7.25,6) {};
\node [draw, circle, minimum size=3pt, color=Green, fill=Green, inner sep=0pt, outer sep=0pt] () at (4,4.33) {};
\node [draw, circle, minimum size=3pt, color=Red, fill=Red, inner sep=0pt, outer sep=0pt] () at (5,5.5) {};
\node [draw, circle, minimum size=3pt, color=Blue, fill=Blue, inner sep=0pt, outer sep=0pt] () at (5,3) {};
\node () at (4.3,2.2) {\color{Blue} $q_1^\mu$};
\node () at (3.5,7.6) {\color{Red} $l_1^\mu$};
\node () at (8,7.5) {\color{LimeGreen} $p_2^\mu$};
\node () at (2,7.5) {\color{Green} $p_1^\mu$};
\node () at (6.6,4.3) {\small $x_1$};
\node () at (4,6.3) {\small $x_2$};
\node () at (5,4.5) {\small $x_3$};
\node () at (6.3,6.4) {\small $x_4$};
\node () at (5,0.5) {\large (d)};
\node () at (6.4,2.4) {$\crossmark[Red]$};
\end{tikzpicture}
\end{aligned}
\label{eq:threshold_motivation_Sudakov_soft_nonregions}
\end{equation}
In each configuration above, all the soft external momenta attach to exactly one of the two jets. Similar to the analysis below (\ref{eq:threshold_motivation_Sudakov_collinear_nonregions}), it is straightforward to understand why the associated Feynman integrals do not contribute from the homogeneity properties of the leading polynomial $\mathcal{P}^{(R)}(\x;\s)$. One can actually check that all the leading terms for (a)-(d) are quadratic in the parameters $x_1,x_2,x_3,x_4$. It then follows that $\text{dim}f_R< N$, and the corresponding integrals are scaleless.

Let us now propose the second rule from these examples: \emph{if there are exactly two jets and the soft subgraph is not empty, then either there is one soft component attached by some soft external momenta, or both jets are attached by some soft external momenta.}

We then focus on more complicated configurations of wide-angle scattering that feature more than two jets. Below are some examples of valid regions in the $2\to 2$ wide-angle scattering.
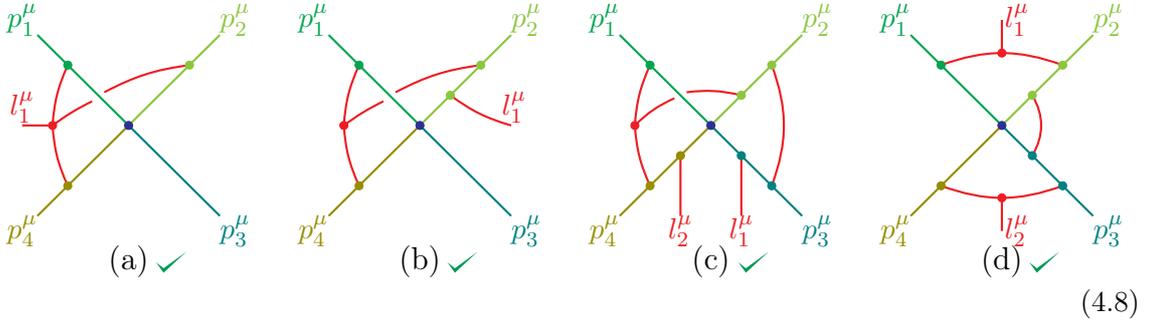
\begin{equation}
\begin{aligned}
\begin{tikzpicture}[baseline=11ex, scale=0.4]
\path (2,8) edge [Green, thick] (5,5) {};
\path (8,8) edge [LimeGreen, thick] (5,5) {};
\path (8,2) edge [teal, thick] (5,5) {};
\path (2,2) edge [olive, thick] (5,5) {};
\path (2.5,5) edge [Red, thick] (1.5,5) {};
\path (2.5,5) edge [Red, thick, bend left = 10] (3,7) {};
\path (2.5,5) edge [Red, thick, bend left = 5] (3.8,5.8) {};
\path (4.2,6.1) edge [Red, thick, bend left = 10] (7,7) {};
\path (2.5,5) edge [Red, thick, bend right = 10] (3,3) {};
\node [draw, circle, minimum size=3pt, color=Green, fill=Green, inner sep=0pt, outer sep=0pt] () at (3,7) {};
\node [draw, circle, minimum size=3pt, color=LimeGreen, fill=LimeGreen, inner sep=0pt, outer sep=0pt] () at (7,7) {};
\node [draw, circle, minimum size=3pt, color=olive, fill=olive, inner sep=0pt, outer sep=0pt] () at (3,3) {};
\node [draw, circle, minimum size=3pt, color=Red, fill=Red, inner sep=0pt, outer sep=0pt] () at (2.5,5) {};
\node [draw, circle, minimum size=3pt, color=Blue, fill=Blue, inner sep=0pt, outer sep=0pt] () at (5,5) {};
\node () at (1.5,5.6) {\color{Red} $l_1^\mu$};
\node () at (1.5,8.5) {\color{Green} $p_1^\mu$};
\node () at (8.5,8.5) {\color{LimeGreen} $p_2^\mu$};
\node () at (8.5,1.5) {\color{teal} $p_3^\mu$};
\node () at (1.5,1.5) {\color{olive} $p_4^\mu$};
\node () at (5,0.5) {\large (a)};
\node () at (6.4,2.4) {$\greencheckmark[ForestGreen]$};
\end{tikzpicture}\quad
\begin{tikzpicture}[baseline=11ex, scale=0.4]
\path (2,8) edge [Green, thick] (5,5) {};
\path (8,8) edge [LimeGreen, thick] (5,5) {};
\path (8,2) edge [teal, thick] (5,5) {};
\path (2,2) edge [olive, thick] (5,5) {};
\path (6,6) edge [Red, thick, bend right = 10] (8,5) {};
\path (2.5,5) edge [Red, thick, bend left = 10] (3,7) {};
\path (2.5,5) edge [Red, thick, bend left = 5] (3.8,5.8) {};
\path (4.2,6.1) edge [Red, thick, bend left = 10] (7,7) {};
\path (2.5,5) edge [Red, thick, bend right = 10] (3,3) {};
\node [draw, circle, minimum size=3pt, color=Green, fill=Green, inner sep=0pt, outer sep=0pt] () at (3,7) {};
\node [draw, circle, minimum size=3pt, color=LimeGreen, fill=LimeGreen, inner sep=0pt, outer sep=0pt] () at (7,7) {};
\node [draw, circle, minimum size=3pt, color=LimeGreen, fill=LimeGreen, inner sep=0pt, outer sep=0pt] () at (6,6) {};
\node [draw, circle, minimum size=3pt, color=olive, fill=olive, inner sep=0pt, outer sep=0pt] () at (3,3) {};
\node [draw, circle, minimum size=3pt, color=Red, fill=Red, inner sep=0pt, outer sep=0pt] () at (2.5,5) {};
\node [draw, circle, minimum size=3pt, color=Blue, fill=Blue, inner sep=0pt, outer sep=0pt] () at (5,5) {};
\node () at (8.1,5.6) {\color{Red} $l_1^\mu$};
\node () at (1.5,8.5) {\color{Green} $p_1^\mu$};
\node () at (8.5,8.5) {\color{LimeGreen} $p_2^\mu$};
\node () at (8.5,1.5) {\color{teal} $p_3^\mu$};
\node () at (1.5,1.5) {\color{olive} $p_4^\mu$};
\node () at (5,0.5) {\large (b)};
\node () at (6.4,2.4) {$\greencheckmark[ForestGreen]$};
\end{tikzpicture}\quad
\begin{tikzpicture}[baseline=11ex, scale=0.4]
\path (2,8) edge [Green, thick] (5,5) {};
\path (8,8) edge [LimeGreen, thick] (5,5) {};
\path (8,2) edge [teal, thick] (5,5) {};
\path (2,2) edge [olive, thick] (5,5) {};
\path (6,4) edge [Red, thick] (6,2) {};
\path (4,4) edge [Red, thick] (4,2) {};
\path (7,7) edge [Red, thick, bend left = 20] (7,3) {};
\path (2.5,5) edge [Red, thick, bend left = 10] (3,7) {};
\path (2.5,5) edge [Red, thick, bend left = 10] (3.8,5.9) {};
\path (4.2,6.1) edge [Red, thick, bend left = 10] (6,6) {};
\path (2.5,5) edge [Red, thick, bend right = 10] (3,3) {};
\node [draw, circle, minimum size=3pt, color=Green, fill=Green, inner sep=0pt, outer sep=0pt] () at (3,7) {};
\node [draw, circle, minimum size=3pt, color=LimeGreen, fill=LimeGreen, inner sep=0pt, outer sep=0pt] () at (7,7) {};
\node [draw, circle, minimum size=3pt, color=LimeGreen, fill=LimeGreen, inner sep=0pt, outer sep=0pt] () at (6,6) {};
\node [draw, circle, minimum size=3pt, color=teal, fill=teal, inner sep=0pt, outer sep=0pt] () at (7,3) {};
\node [draw, circle, minimum size=3pt, color=teal, fill=teal, inner sep=0pt, outer sep=0pt] () at (6,4) {};
\node [draw, circle, minimum size=3pt, color=olive, fill=olive, inner sep=0pt, outer sep=0pt] () at (3,3) {};
\node [draw, circle, minimum size=3pt, color=olive, fill=olive, inner sep=0pt, outer sep=0pt] () at (4,4) {};
\node [draw, circle, minimum size=3pt, color=Red, fill=Red, inner sep=0pt, outer sep=0pt] () at (2.5,5) {};
\node [draw, circle, minimum size=3pt, color=Blue, fill=Blue, inner sep=0pt, outer sep=0pt] () at (5,5) {};
\node () at (4,1.5) {\color{Red} $l_2^\mu$};
\node () at (6,1.5) {\color{Red} $l_1^\mu$};
\node () at (1.5,8.5) {\color{Green} $p_1^\mu$};
\node () at (8.5,8.5) {\color{LimeGreen} $p_2^\mu$};
\node () at (8.5,1.5) {\color{teal} $p_3^\mu$};
\node () at (1.5,1.5) {\color{olive} $p_4^\mu$};
\node () at (5,0.5) {\large (c)};
\node () at (6.4,2.4) {$\greencheckmark[ForestGreen]$};
\end{tikzpicture}\quad
\begin{tikzpicture}[baseline=11ex, scale=0.4]
\path (2,8) edge [Green, thick] (5,5) {};
\path (8,8) edge [LimeGreen, thick] (5,5) {};
\path (8,2) edge [teal, thick] (5,5) {};
\path (2,2) edge [olive, thick] (5,5) {};
\path (5,7.4) edge [Red, thick] (5,8.5) {};
\path (5,2.6) edge [Red, thick] (5,1.5) {};
\path (3,7) edge [Red, thick, bend left = 20] (7,7) {};
\path (6,6) edge [Red, thick, bend left = 30] (6,4) {};
\path (3,3) edge [Red, thick, bend right = 20] (7,3) {};
\node [draw, circle, minimum size=3pt, color=Green, fill=Green, inner sep=0pt, outer sep=0pt] () at (3,7) {};
\node [draw, circle, minimum size=3pt, color=LimeGreen, fill=LimeGreen, inner sep=0pt, outer sep=0pt] () at (7,7) {};
\node [draw, circle, minimum size=3pt, color=LimeGreen, fill=LimeGreen, inner sep=0pt, outer sep=0pt] () at (6,6) {};
\node [draw, circle, minimum size=3pt, color=teal, fill=teal, inner sep=0pt, outer sep=0pt] () at (7,3) {};
\node [draw, circle, minimum size=3pt, color=teal, fill=teal, inner sep=0pt, outer sep=0pt] () at (6,4) {};
\node [draw, circle, minimum size=3pt, color=olive, fill=olive, inner sep=0pt, outer sep=0pt] () at (3,3) {};
\node [draw, circle, minimum size=3pt, color=Blue, fill=Blue, inner sep=0pt, outer sep=0pt] () at (5,5) {};
\node [draw, circle, minimum size=3pt, color=Red, fill=Red, inner sep=0pt, outer sep=0pt] () at (5,7.4) {};
\node [draw, circle, minimum size=3pt, color=Red, fill=Red, inner sep=0pt, outer sep=0pt] () at (5,2.6) {};
\node () at (5.5,8.5) {\color{Red} $l_1^\mu$};
\node () at (5.5,1.5) {\color{Red} $l_2^\mu$};
\node () at (1.5,8.5) {\color{Green} $p_1^\mu$};
\node () at (8.5,8.5) {\color{LimeGreen} $p_2^\mu$};
\node () at (8.5,1.5) {\color{teal} $p_3^\mu$};
\node () at (1.5,1.5) {\color{olive} $p_4^\mu$};
\node () at (5,0.5) {\large (d)};
\node () at (6.4,2.4) {$\greencheckmark[ForestGreen]$};
\end{tikzpicture}
\end{aligned}
\label{eq:threshold_motivation_scattering_soft_regions}
\end{equation}
In (a) and (b) above, the only connected component of $S$, which contains all the internal soft propagators, is adjacent to $J_1$, $J_2$, and $J_3$. The soft external momentum $l_1^\mu$ attaches to $S$ in (a) and to $J_2$ in (b). In (c), $S$ consists of two connected components, one adjacent to $J_1,J_2,J_4$ and the other adjacent to $J_2,J_4$, meanwhile the two soft external momenta $l_1^\mu$ and $l_2^\mu$ attach to $J_3$ and $J_4$ respectively. In (d), $S$ consists of three connected components, two of which are attached by $l_1^\mu$ and $l_2^\mu$, respectively.

Note that in a valid region where a given soft component is adjacent to three or more jets, it is possible that the soft external momenta attach to only one of these jets in a valid region (see (b) above). This is a key difference from the regions in Sudakov form factors (in contrast, refer to the non-region examples in (\ref{eq:threshold_motivation_Sudakov_soft_nonregions})). Therefore, those soft components that are adjacent to three or more jets should be treated differently from those soft components adjacent to exactly two jets.

In contrast, the following configurations are not valid regions:
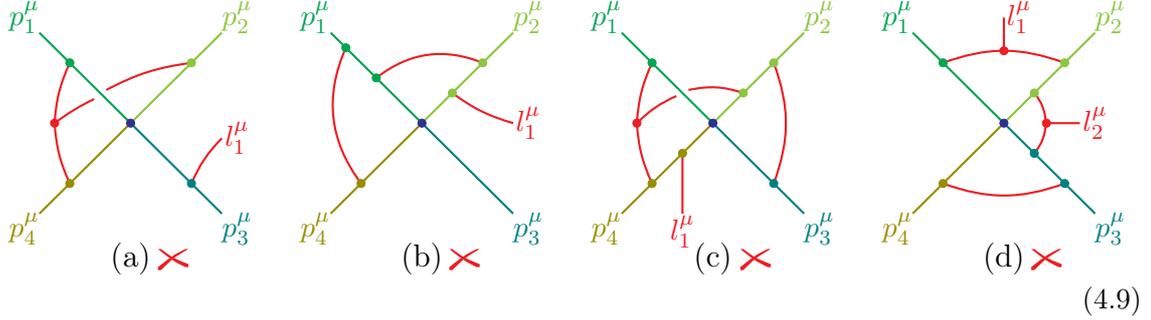
\begin{equation}
\begin{aligned}
\begin{tikzpicture}[baseline=11ex, scale=0.4]
\path (2,8) edge [Green, thick] (5,5) {};
\path (8,8) edge [LimeGreen, thick] (5,5) {};
\path (8,2) edge [teal, thick] (5,5) {};
\path (2,2) edge [olive, thick] (5,5) {};
\path (7,3) edge [Red, thick, bend left =10] (8,4.5) {};
\path (2.5,5) edge [Red, thick, bend left = 10] (3,7) {};
\path (2.5,5) edge [Red, thick, bend left = 5] (3.8,5.8) {};
\path (4.2,6.1) edge [Red, thick, bend left = 10] (7,7) {};
\path (2.5,5) edge [Red, thick, bend right = 10] (3,3) {};
\node [draw, circle, minimum size=3pt, color=Green, fill=Green, inner sep=0pt, outer sep=0pt] () at (3,7) {};
\node [draw, circle, minimum size=3pt, color=LimeGreen, fill=LimeGreen, inner sep=0pt, outer sep=0pt] () at (7,7) {};
\node [draw, circle, minimum size=3pt, color=olive, fill=olive, inner sep=0pt, outer sep=0pt] () at (3,3) {};
\node [draw, circle, minimum size=3pt, color=teal, fill=teal, inner sep=0pt, outer sep=0pt] () at (7,3) {};
\node [draw, circle, minimum size=3pt, color=Red, fill=Red, inner sep=0pt, outer sep=0pt] () at (2.5,5) {};
\node [draw, circle, minimum size=3pt, color=Blue, fill=Blue, inner sep=0pt, outer sep=0pt] () at (5,5) {};
\node () at (8.5,4.5) {\color{Red} $l_1^\mu$};
\node () at (1.5,8.5) {\color{Green} $p_1^\mu$};
\node () at (8.5,8.5) {\color{LimeGreen} $p_2^\mu$};
\node () at (8.5,1.5) {\color{teal} $p_3^\mu$};
\node () at (1.5,1.5) {\color{olive} $p_4^\mu$};
\node () at (5,0.5) {\large (a)};
\node () at (6.4,2.4) {$\crossmark[Red]$};
\end{tikzpicture}\quad
\begin{tikzpicture}[baseline=11ex, scale=0.4]
\path (2,8) edge [Green, thick] (5,5) {};
\path (8,8) edge [LimeGreen, thick] (5,5) {};
\path (8,2) edge [teal, thick] (5,5) {};
\path (2,2) edge [olive, thick] (5,5) {};
\path (6,6) edge [Red, thick, bend right = 10] (8,5) {};
\path (3,3) edge [Red, thick, bend left = 30] (2.5,7.5) {};
\path (3.5,6.5) edge [Red, thick, bend left = 30] (7,7) {};
\node [draw, circle, minimum size=3pt, color=Green, fill=Green, inner sep=0pt, outer sep=0pt] () at (2.5,7.5) {};
\node [draw, circle, minimum size=3pt, color=Green, fill=Green, inner sep=0pt, outer sep=0pt] () at (3.5,6.5) {};
\node [draw, circle, minimum size=3pt, color=LimeGreen, fill=LimeGreen, inner sep=0pt, outer sep=0pt] () at (7,7) {};
\node [draw, circle, minimum size=3pt, color=LimeGreen, fill=LimeGreen, inner sep=0pt, outer sep=0pt] () at (6,6) {};
\node [draw, circle, minimum size=3pt, color=olive, fill=olive, inner sep=0pt, outer sep=0pt] () at (3,3) {};
\node [draw, circle, minimum size=3pt, color=Blue, fill=Blue, inner sep=0pt, outer sep=0pt] () at (5,5) {};
\node () at (8.5,5) {\color{Red} $l_1^\mu$};
\node () at (1.5,8.5) {\color{Green} $p_1^\mu$};
\node () at (8.5,8.5) {\color{LimeGreen} $p_2^\mu$};
\node () at (8.5,1.5) {\color{teal} $p_3^\mu$};
\node () at (1.5,1.5) {\color{olive} $p_4^\mu$};
\node () at (5,0.5) {\large (b)};
\node () at (6.4,2.4) {$\crossmark[Red]$};
\end{tikzpicture}\quad
\begin{tikzpicture}[baseline=11ex, scale=0.4]
\path (2,8) edge [Green, thick] (5,5) {};
\path (8,8) edge [LimeGreen, thick] (5,5) {};
\path (8,2) edge [teal, thick] (5,5) {};
\path (2,2) edge [olive, thick] (5,5) {};
\path (4,4) edge [Red, thick] (4,2) {};
\path (7,7) edge [Red, thick, bend left = 20] (7,3) {};
\path (2.5,5) edge [Red, thick, bend left = 10] (3,7) {};
\path (2.5,5) edge [Red, thick, bend left = 10] (3.8,5.9) {};
\path (4.2,6.1) edge [Red, thick, bend left = 15] (6,6) {};
\path (2.5,5) edge [Red, thick, bend right = 10] (3,3) {};
\node [draw, circle, minimum size=3pt, color=Green, fill=Green, inner sep=0pt, outer sep=0pt] () at (3,7) {};
\node [draw, circle, minimum size=3pt, color=LimeGreen, fill=LimeGreen, inner sep=0pt, outer sep=0pt] () at (7,7) {};
\node [draw, circle, minimum size=3pt, color=LimeGreen, fill=LimeGreen, inner sep=0pt, outer sep=0pt] () at (6,6) {};
\node [draw, circle, minimum size=3pt, color=teal, fill=teal, inner sep=0pt, outer sep=0pt] () at (7,3) {};
\node [draw, circle, minimum size=3pt, color=olive, fill=olive, inner sep=0pt, outer sep=0pt] () at (3,3) {};
\node [draw, circle, minimum size=3pt, color=olive, fill=olive, inner sep=0pt, outer sep=0pt] () at (4,4) {};
\node [draw, circle, minimum size=3pt, color=Red, fill=Red, inner sep=0pt, outer sep=0pt] () at (2.5,5) {};
\node [draw, circle, minimum size=3pt, color=Blue, fill=Blue, inner sep=0pt, outer sep=0pt] () at (5,5) {};
\node () at (4,1.4) {\color{Red} $l_1^\mu$};
\node () at (1.5,8.5) {\color{Green} $p_1^\mu$};
\node () at (8.5,8.5) {\color{LimeGreen} $p_2^\mu$};
\node () at (8.5,1.5) {\color{teal} $p_3^\mu$};
\node () at (1.5,1.5) {\color{olive} $p_4^\mu$};
\node () at (5,0.5) {\large (c)};
\node () at (6.4,2.4) {$\crossmark[Red]$};
\end{tikzpicture}\quad
\begin{tikzpicture}[baseline=11ex, scale=0.4]
\path (2,8) edge [Green, thick] (5,5) {};
\path (8,8) edge [LimeGreen, thick] (5,5) {};
\path (8,2) edge [teal, thick] (5,5) {};
\path (2,2) edge [olive, thick] (5,5) {};
\path (5,7.4) edge [Red, thick] (5,8.5) {};
\path (6.4,5) edge [Red, thick] (7.5,5) {};
\path (3,7) edge [Red, thick, bend left = 20] (7,7) {};
\path (6,6) edge [Red, thick, bend left = 40] (6,4) {};
\path (3,3) edge [Red, thick, bend right = 20] (7,3) {};
\node [draw, circle, minimum size=3pt, color=Green, fill=Green, inner sep=0pt, outer sep=0pt] () at (3,7) {};
\node [draw, circle, minimum size=3pt, color=LimeGreen, fill=LimeGreen, inner sep=0pt, outer sep=0pt] () at (7,7) {};
\node [draw, circle, minimum size=3pt, color=LimeGreen, fill=LimeGreen, inner sep=0pt, outer sep=0pt] () at (6,6) {};
\node [draw, circle, minimum size=3pt, color=teal, fill=teal, inner sep=0pt, outer sep=0pt] () at (7,3) {};
\node [draw, circle, minimum size=3pt, color=teal, fill=teal, inner sep=0pt, outer sep=0pt] () at (6,4) {};
\node [draw, circle, minimum size=3pt, color=olive, fill=olive, inner sep=0pt, outer sep=0pt] () at (3,3) {};
\node [draw, circle, minimum size=3pt, color=Blue, fill=Blue, inner sep=0pt, outer sep=0pt] () at (5,5) {};
\node [draw, circle, minimum size=3pt, color=Red, fill=Red, inner sep=0pt, outer sep=0pt] () at (5,7.4) {};
\node [draw, circle, minimum size=3pt, color=Red, fill=Red, inner sep=0pt, outer sep=0pt] () at (6.4,5) {};
\node () at (5.5,8.5) {\color{Red} $l_1^\mu$};
\node () at (8,5) {\color{Red} $l_2^\mu$};
\node () at (1.5,8.5) {\color{Green} $p_1^\mu$};
\node () at (8.5,8.5) {\color{LimeGreen} $p_2^\mu$};
\node () at (8.5,1.5) {\color{teal} $p_3^\mu$};
\node () at (1.5,1.5) {\color{olive} $p_4^\mu$};
\node () at (5,0.5) {\large (d)};
\node () at (6.4,2.4) {$\crossmark[Red]$};
\end{tikzpicture}
\end{aligned}
\label{eq:threshold_motivation_scattering_soft_nonregions}
\end{equation}
In configuration (a) above, the unique soft component with internal edges is adjacent to $J_1$, $J_2$ and $J_4$, while the only external momentum $l_1^\mu$ attaches to $J_3$. In configuration (b), there are two soft components, each adjacent to exactly two jets, while $l_1^\mu$ attaches to $J_2$. In configuration (c), there are two soft components, and the only soft external momentum attaches to $J_4$. Finally, in configuration (d) there are three soft components, one being adjacent to $J_1,J_2$ and attached by $l_1^\mu$, while another being adjacent to $J_2,J_3$ and attached by $l_2^\mu$.

By comparing the examples in (\ref{eq:threshold_motivation_scattering_soft_regions}) and (\ref{eq:threshold_motivation_scattering_soft_nonregions}), we propose the third rule: \emph{for any soft component $S_{ij}$ that is adjacent to two jets $J_i$ and $J_j$, either $S_{ij}$ is attached by some soft external momenta, or both $J_i$ and $J_j$ are attached by some soft momenta not from $S_{ij}$; for any soft component $S_{ijk}$ that is adjacent to three} (\emph{or more}) \emph{jets $J_i$, $J_j$, and $J_k$, either $S_{ijk}$ is attached by some soft external momenta, or some of $J_i$, $J_j$, and $J_k$ are attached by soft momenta not from $S_{ijk}$.}

Later in this section, all the three rules above will be verified to all orders.

\subsection{The generic form of a region vector}
\label{section-general_prescription_threshold}

\begin{figure}[t]
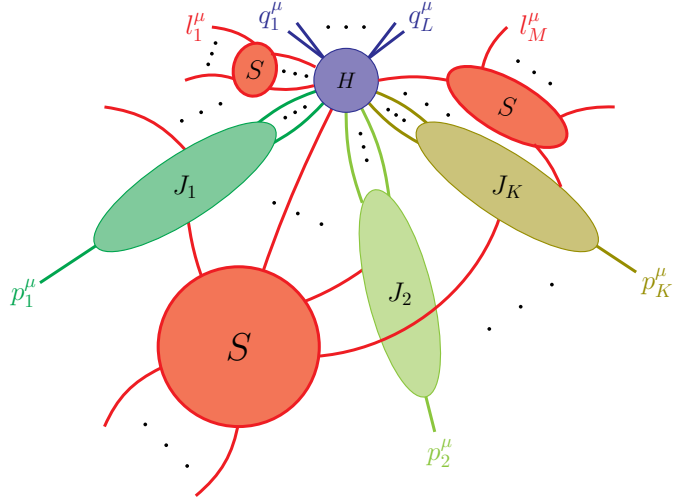

\centering
\include{figs/threshold_region_H_J_S_precise}
\vspace{-3em}
\caption{The configuration of a generic region $R$ in the soft expansion. The entire graph $G$ is divided into the hard subgraph $H$, the jet subgraphs $J_1,\dots,J_K$, and the soft subgraph $S$. Note that $H, J_1,\dots, J_K$ are all connected, and the soft external momenta $l_1^\mu,\dots,l_M^\mu$ can attach to any vertex of~$G$. In contrast to the regions in the on-shell expansion (figure~\ref{figure-onshell_region_H_J_S_precise}), some soft components may be adjacent to only one, or even zero jet subgraph.}
\label{figure-threshold_region_H_J_S_precise}
\end{figure}

As a key observation from the examples above, the list of regions for each given graph consists of one hard region and several infrared regions. Furthermore, all the infrared regions have a correspondence with the pinch surfaces of $G|_{l^\mu=0}$, where we have set all the on-shell external momenta $p_i^\mu$ to be strictly lightlike, and all the soft external momenta to be zero. Namely, the region vectors are of the following form
\begin{eqnarray}
    \v_R= (u_{R,1},u_{R,2},\dots,u_{R,N};1),
\end{eqnarray}
where $u_{R,e}=0$ if $e\in H$, $u_{R,e}=-1$ if $e\in J$, and $u_{R,e}=-2$ if $e\in S$, with the hard, jet, and soft subgraphs shown in figure~\ref{figure-threshold_region_H_J_S_precise}.

A similar property holds for the regions in the on-shell expansion, as we have proved in section~\ref{section-generic_form_region_rigorous_proof} as part of the on-shell-expansion region theorem. We expect that another proof can be established for the soft expansion. However, rather than dedicating an extensive subsection to a rigorous proof, we will only present this observation as the following proposition.
\begin{proposition}
\label{proposition-threshold_expansion_vector_generic_form}
Any region vector $\v_R$ in the soft expansion can be expressed as:
\begin{eqnarray}
    && \v_R= (u_{R,1},u_{R,2},\dots,u_{R,N};1),\text{ with} \nonumber\\
    && u_{R,e}=0\ \Leftrightarrow\ e\in H,\qquad u_{R,e}=-1\ \Leftrightarrow\ e\in J,\qquad u_{R,e}=-2\ \Leftrightarrow\ e\in S, \nonumber
\end{eqnarray}
where the subgraphs $H,J,S$ are shown in figure~\ref{figure-threshold_region_H_J_S_precise}, satisfying the following conditions.
\begin{itemize}
    \item The hard subgraph $H$ is connected and attached by all the off-shell external momenta $q_1^\mu,\dots,q_L^\mu$. It is possible that some soft external momenta attach to $H$.
    \item Each jet subgraph $J_i$ ($i\in \{1,\dots,K\}$) is connected and attached by the on-shell external momentum $p_i^\mu$. It is possible that some soft external momenta attach to $J_i$.
\end{itemize}
\end{proposition}

In contrast to the general configuration of regions in the on-shell expansion, here we refrain from imposing the constraint on $S$ that each its component must be adjacent to two or more jets. This difference arises from the recognition that the configurations depicted below are valid regions in the soft expansion:
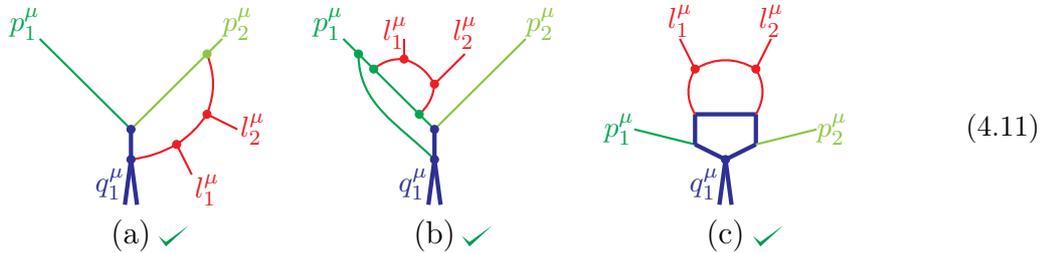
\begin{equation}
\begin{aligned}
\begin{tikzpicture}[baseline=11ex, scale=0.4]
\path (2,7) edge [Green, thick] (5,4) {};
\path (8,7) edge [LimeGreen, thick] (5,4) {};
\path (5,3) edge [Red, thick, bend right = 10] (6.5,3.5) {};
\path (6.5,3.5) edge [Red, thick, bend right = 10] (7.5,4.5) {};
\path (7.5,4.5) edge [Red, thick, bend right = 20] (7.5,6.5) {};
\path (7.5,4.5) edge [Red, thick] (8.5,4) {};
\path (6.5,3.5) edge [Red, thick] (7,2.5) {};
\path (5,3) edge [Blue, ultra thick] (5,4) {};
\path (4.8,1.5) edge [Blue, ultra thick] (5,3) {};
\path (5.2,1.5) edge [Blue, ultra thick] (5,3) {};
\node [draw, circle, minimum size=3pt, color=LimeGreen, fill=LimeGreen, inner sep=0pt, outer sep=0pt] () at (7.5,6.5) {};
\node [draw, circle, minimum size=3pt, color=Red, fill=Red, inner sep=0pt, outer sep=0pt] () at (6.5,3.5) {};
\node [draw, circle, minimum size=3pt, color=Red, fill=Red, inner sep=0pt, outer sep=0pt] () at (7.5,4.5) {};
\node [draw, circle, minimum size=3pt, color=Blue, fill=Blue, inner sep=0pt, outer sep=0pt] () at (5,3) {};
\node [draw, circle, minimum size=3pt, color=Blue, fill=Blue, inner sep=0pt, outer sep=0pt] () at (5,4) {};
\node () at (4.3,2.2) {\color{Blue} $q_1^\mu$};
\node () at (7.5,2) {\color{Red} $l_1^\mu$};
\node () at (9,4) {\color{Red} $l_2^\mu$};
\node () at (8.5,7.5) {\color{LimeGreen} $p_2^\mu$};
\node () at (1.5,7.5) {\color{Green} $p_1^\mu$};
\node () at (5,0.5) {\large (a)};
\node () at (6.4,2.4) {$\greencheckmark[ForestGreen]$};
\end{tikzpicture}\quad
\begin{tikzpicture}[baseline=11ex, scale=0.4]
\path (2,7) edge [Green, thick] (5,4) {};
\path (8,7) edge [LimeGreen, thick] (5,4) {};
\path (2.5,6.5) edge [Green, thick, out=-90, in=135] (5,3) {};
\path (3,6) edge [Red, thick, bend left = 20] (4,6.33) {};
\path (4,6.33) edge [Red, thick, bend left = 20] (5,5.5) {};
\path (5,5.5) edge [Red, thick, bend left = 20] (4.5,4.5) {};
\path (4,7) edge [Red, thick] (4,6.33) {};
\path (6,6.5) edge [Red, thick] (5,5.5) {};
\path (5,3) edge [Blue, ultra thick] (5,4) {};
\path (4.8,1.5) edge [Blue, ultra thick] (5,3) {};
\path (5.2,1.5) edge [Blue, ultra thick] (5,3) {};
\node [draw, circle, minimum size=3pt, color=Green, fill=Green, inner sep=0pt, outer sep=0pt] () at (2.5,6.5) {};
\node [draw, circle, minimum size=3pt, color=Green, fill=Green, inner sep=0pt, outer sep=0pt] () at (3,6) {};
\node [draw, circle, minimum size=3pt, color=Green, fill=Green, inner sep=0pt, outer sep=0pt] () at (4.5,4.5) {};
\node [draw, circle, minimum size=3pt, color=Red, fill=Red, inner sep=0pt, outer sep=0pt] () at (4,6.33) {};
\node [draw, circle, minimum size=3pt, color=Red, fill=Red, inner sep=0pt, outer sep=0pt] () at (5,5.5) {};
\node [draw, circle, minimum size=3pt, color=Blue, fill=Blue, inner sep=0pt, outer sep=0pt] () at (5,3) {};
\node [draw, circle, minimum size=3pt, color=Blue, fill=Blue, inner sep=0pt, outer sep=0pt] () at (5,4) {};
\node () at (4.3,2.2) {\color{Blue} $q_1^\mu$};
\node () at (3.7,7.2) {\color{Red} $l_1^\mu$};
\node () at (6,7.2) {\color{Red} $l_2^\mu$};
\node () at (8.5,7.5) {\color{LimeGreen} $p_2^\mu$};
\node () at (1.5,7.5) {\color{Green} $p_1^\mu$};
\node () at (5,0.5) {\large (b)};
\node () at (6.4,2.4) {$\greencheckmark[ForestGreen]$};
\end{tikzpicture}\quad
\begin{tikzpicture}[baseline=11ex, scale=0.4]
\path (2,4) edge [Green, thick] (4,3.5) {};
\path (4,3.5) edge [Blue, ultra thick] (5,3) {};
\path (6,3.5) edge [Blue, ultra thick] (5,3) {};
\path (4,3.5) edge [Blue, ultra thick] (4,4.5) {};
\path (4,4.5) edge [Blue, ultra thick] (6,4.5) {};
\path (6,4.5) edge [Blue, ultra thick] (6,3.5) {};
\path (3.5,7) edge [Red, thick] (4,6) {};
\path (6.5,7) edge [Red, thick] (6,6) {};
\path (4,4.5) edge [Red, thick, bend left =30] (4,6) {};
\path (4,6) edge [Red, thick, bend left =30] (6,6) {};
\path (6,6) edge [Red, thick, bend left =30] (6,4.5) {};
\path (8,4) edge [LimeGreen, thick] (6,3.5) {};
\path (4.8,1.5) edge [Blue, ultra thick] (5,3) {};
\path (5.2,1.5) edge [Blue, ultra thick] (5,3) {};
\node [draw, circle, minimum size=3pt, color=Red, fill=Red, inner sep=0pt, outer sep=0pt] () at (4,6) {};
\node [draw, circle, minimum size=3pt, color=Red, fill=Red, inner sep=0pt, outer sep=0pt] () at (6,6) {};
\node [draw, circle, minimum size=3pt, color=Blue, fill=Blue, inner sep=0pt, outer sep=0pt] () at (5,3) {};
\node () at (4.3,2.2) {\color{Blue} $q_1^\mu$};
\node () at (3.5,7.6) {\color{Red} $l_1^\mu$};
\node () at (6.5,7.6) {\color{Red} $l_2^\mu$};
\node () at (8.5,4) {\color{LimeGreen} $p_2^\mu$};
\node () at (1.5,4) {\color{Green} $p_1^\mu$};
\node () at (5,0.5) {\large (c)};
\node () at (6.4,2.4) {$\greencheckmark[ForestGreen]$};
\end{tikzpicture}
\end{aligned}
\label{eq:threshold_motivation_soft_interacting_less_than_two_jets}
\end{equation}
In each of these regions, there exists a soft component that is adjacent to only one, or even zero, jet subgraph. As another feature, each of these soft components is attached by two soft external momenta. This observation serves as motivation for us to derive further constraints on the subgraphs, as we will delve into in section~\ref{section-further_restrictions_region_vectors_threshold}.

Although a rigorous proof of proposition~\ref{proposition-threshold_expansion_vector_generic_form} is currently unavailable, it is worth noting that the approach taken in proving lemmas~\ref{lemma-onshell_basic_weight_structure_Uterm}-\ref{lemma-onshell_leading_terms_forms} can be instrumental. Extending these lemmas from the on-shell expansion to the soft expansion may pose challenges, but it will be sufficient to extend only some of them. For example, the proof of lemma~\ref{lemma-onshell_basic_weight_structure_Uterm} involves relating a given $\mathcal{U}^{(R)}$ term $\x^{\r_1}$ to an $\mathcal{F}^{(p_i^2)}$ term $\x^{\r_2}$ via $T^2(\r_2) = T^1(\r_1)\setminus e$, where $e$ is \emph{any} edge in the $p_i$ trunk of $T^1(\r_1)$. However, in the soft expansion, such a $T^2(\r_2)$ may not exist, as $T^1(\r_1)\setminus e$ does not necessarily correspond to an $\mathcal{F}^{(p_i^2)}$ term in the soft expansion. This discrepancy arises because, if one component of $T^2$ is connected only by $p_i^\mu$, then the corresponding term carries an overall factor of $p_i^2=0$. Nevertheless, once the extension of lemma~\ref{lemma-onshell_basic_weight_structure_Uterm} is accomplished, extending lemmas~\ref{lemma-onshell_Fp2_external_less_equal_minusone}-\ref{lemma-onshell_heavy_in_light_constraint} becomes trivial. The proofs of these lemmas rely solely on the use of lemma~\ref{lemma-onshell_basic_weight_structure_Uterm} and can be automatically extended. We will leave this as a task for future researchers interested in pursuing this line of study.

With proposition~\ref{proposition-threshold_expansion_vector_generic_form} established, we will next explore how the soft components connect $H\cup J$. Understanding this is crucial for excluding those configurations that conform with proposition~\ref{proposition-threshold_expansion_vector_generic_form} while yielding scaleless integrals.

\subsection{Leading terms in the infrared regions}
\label{section-leading_terms_infrared_regions}

In this subsection, we investigate the general form of the leading terms associated with a given region. We start by defining the contracted soft and jet graphs in a certain region. The contracted soft subgraph $\widetilde{S}$ is obtained from $S$ by contracting $H\cup J$ into a vertex, and identify it with a new soft vertex which we refer to as the \emph{auxiliary }(\emph{soft})\emph{ vertex}. Similarly, $\widetilde{J}_i$ is obtained from $J_i$ by contracting $H$ to a vertex, and identify it with a new vertex of $J_i$, the \emph{auxiliary }(\emph{$J_i$})\emph{ vertex}.

\begin{figure}[t]
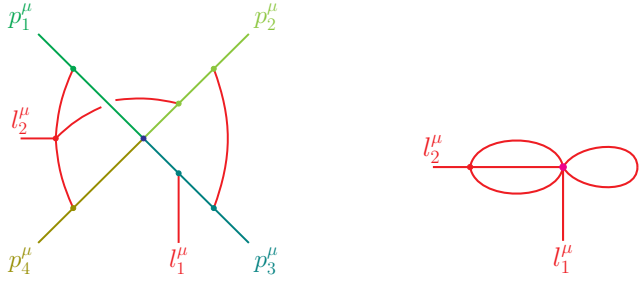

\centering
\begin{subfigure}[b]{0.25\textwidth}
\centering
\include{figs/soft_expansion_contracted_subgraphs_Feynman}
\label{soft_expansion_contracted_subgraphs_Feynman}
\end{subfigure}
\hspace{2em}\qquad
\begin{subfigure}[b]{0.2\textwidth}
\centering
\include{figs/soft_expansion_contracted_subgraphs_soft}
\label{soft_expansion_contracted_subgraphs_soft}
\end{subfigure}
\vspace{-2em}\caption{An example of wide-angle scattering graph with soft external momenta, and its corresponding contracted soft subgraph $\widetilde{S}$.}
\label{figure-soft_expansion_contracted_subgraphs_example}
\end{figure}

In contrast to the contracted soft subgraphs defined for the on-shell expansion, here $\widetilde{S}$ is attached by all the external soft momenta. An example is shown in figure~\ref{figure-soft_expansion_contracted_subgraphs_example}, where $l_1^\mu$ attaches to $J_3$ and $l_2^\mu$ attaches to $S$, and they both belong to $\widetilde{S}$.

We next rewrite the Symanzik polynomial $\mathcal{F}(\x;\s)$ as
\begin{eqnarray}
    \mathcal{F}(\x;\s) \equiv \sum_{i=1}^K \mathcal{F}^{(p_i^2)}(\x;\s) + \mathcal{F}^{(q^2)}(\x;\s) + \mathcal{F}^{(l^2)}(\x;\s),
\end{eqnarray}
where the kinematic factor of each $\mathcal{F}^{(p_i^2)}$ term is of the form $(p_i+l)^2$, with $l^\mu$ being a nonzero soft momentum. Similarly, for each $\mathcal{F}^{(q^2)}$ term the kinematic factor is $q^2\sim Q^2$, with $Q$ being a large scale, and for each $\mathcal{F}^{(l^2)}$ term the kinematic factor is $l^2$, where $l^\mu$ is soft and $l^2\neq 0$. For each given region $R$, the corresponding leading polynomials are denoted as $\mathcal{F}^{(p_i^2,R)}(\x;\s)$, $\mathcal{F}^{(q^2,R)}(\x;\s)$, and $\mathcal{F}^{(l^2,R)}(\x;\s)$, respectively.

Recall that for any subgraph $\gamma\subset G$ and any spanning tree $T^1$ or spanning 2-tree $T^2$, we use $n_{\gamma}^{}$ to denote the number of edges in $\gamma \setminus T^1$ or $\gamma\setminus T^2$. A first observation is that for any $\mathcal{U}$, $\mathcal{F}^{(p^2)}$, and $\mathcal{F}^{(q^2)}$ terms,
\begin{eqnarray}
n_H^{} \geqslant L(H),\qquad\quad n_S^{} \leqslant L(\widetilde{S}).
\label{eq:threshold_leading_terms_fundamental_constraints}
\end{eqnarray}
The first inequality holds because we need to remove at least $L(H)$ edges from $G$ to make $H\cap T$ free from loops, which can be seen as a special case of the general property (\ref{eq:removed_edge_number_constraint1}). The second inequality can be proved by contradiction. If $n_S>L(\widetilde{S})$, the graph $\widetilde{S}\cap T$ is disconnected, and the auxiliary vertex belongs to one connected component. The other component of $\widetilde{S}\cap T$ consists solely of soft vertices and edges, so it cannot be attached by any of $p_1^\mu,\dots,p_K^\mu,q_1^\mu,\dots,q_L^\mu$. Such a configuration does not contribute to the polynomial $\mathcal{U}+\mathcal{F}^{(p^2)}+\mathcal{F}^{(q^2)}$.

In ref.~\cite{GrdHzgJnsMaSchlk22}, the $\mathcal{U}^{(R)}$, $\mathcal{F}^{(p^2,R)}$, and $\mathcal{F}^{(q^2,R)}$ terms in the on-shell expansion are characterized by specific values of $n_H$, $n_{J_i}$, and $n_S$. This characterization relies on $n_H^{} \geqslant L(H)$ and $n_S^{} \leqslant L(\widetilde{S})$, which are identical to (\ref{eq:threshold_leading_terms_fundamental_constraints}) above. Therefore, for the soft expansion here, we can apply the same analysis to characterize the $\mathcal{U}^{(R)}$, $\mathcal{F}^{(p^2,R)}$, and $\mathcal{F}^{(q^2,R)}$ polynomials in terms of $n_H$, $n_{J_i}$, and $n_S$. 
Instead of repeating the entire analysis, for simplicity, we will only list the corresponding results in this subsection. First, for each $\mathcal{U}^{(R)}$ term $\x^{\r}$, the graphs $H\cap T^1(\r)$, $\widetilde{J}_i\cap T^1(\r)$ ($j=1,\dots,K$) and $\widetilde{S}_k\cap T^1(\r)$ ($k=1,\dots,\mathfrak{N}$, where $\mathfrak{N}$ is the number of connected components of $S$, and each $S_k$ is one connected component) are spanning trees of $H$, $\widetilde{J}_i$, and $\widetilde{S}_k$ respectively. Namely,
\begin{eqnarray}
    \mathcal{U}^{(R)}(\x):&&\ n_H^{} = L(H),\nonumber \\
    &&\ n_{J_i}^{} = L(\widetilde{J}_i) \quad (i=1,\dots,K), \nonumber\\
    &&\ n_{S_k}^{}= L(\widetilde{S}_k) \quad (k=1,\dots,\mathfrak{N}).
\label{eq:threshold_minimal_u_term_condition_precise}
\end{eqnarray}
Reversely, by combining any chosen spanning trees of $H$, $\widetilde{J}_i$, and $\widetilde{S}_k$, the result is a (minimum) spanning tree of $G$, which corresponds to a $\mathcal{U}^{(R)}$ term. Equivalently, let us use $\x^{[H]}$, $\x^{[J_i]}$, and $\x^{[S_k]}$ to denote the Lee-Pomeransky parameters associated with the edges of $H$, $J_i$, and $S_k$ respectively, then the polynomial $\mathcal{U}^{(R)}(\x)$ can be factorized as:
\begin{eqnarray}
    \mathcal{U}^{(R)}(\x) = \mathcal{U}_H(\x^{[H]}) \cdot \Big( \prod_{i=1}^K \mathcal{U}_{J_i}(\x^{[J_i]}) \Big) \cdot \Big( \prod_{k=1}^n \mathcal{U}_S(\x^{[S_k]}) \Big),
\label{eq:threshold_leading_Uterms_factorize}
\end{eqnarray}
where $\mathcal{U}_H(\x^{[H]})$, $\mathcal{U}_{J_i}(\x^{[J_i]})$ and $\mathcal{U}_{S_k}(\x^{[S_k]})$ are the polynomials containing only $\x^{[H]}$, $\x^{[J_i]}$, and $\x^{[S_k]}$, and corresponding to the spanning trees of $H$, $\widetilde{J}_i$, and $\widetilde{S}_k$ respectively.

Next, for each $\mathcal{F}^{(p_i^2,R)}$ term $\x^{\r}$, one component of $T^2(\r)$ is attached by $p^\mu$ and some soft external momenta. The graphs $H\cap T^2(\r)$, $\widetilde{J}_j\cap T^2(\r)$ ($j\neq i$), and $\widetilde{S}_k \cap T^2(\r)$ ($k=1,\dots,\mathfrak{N}$) are spanning trees of $H$, $\widetilde{J}_j$, and $\widetilde{S}_k$ respectively, while $\widetilde{J}_i\cap T^1(\r)$ is a spanning 2-tree of $\widetilde{J}_i$. Namely,
\begin{eqnarray}
    \mathcal{F}^{(p_i^2,R)}(\x;\s):
    &&\ n_H^{} = L(H),\nonumber \\
    &&\ n_{J_i}^{} = L(\widetilde{J}_i)+1,\quad n_{J_j}^{} = L(\widetilde{J}_j)\quad (\forall j\neq i), \nonumber\\
    &&\ n_{S_k}^{}= L(\widetilde{S}_k)\quad (k=1,\dots,\mathfrak{N}).
\label{eq:threshold_minimal_fpi2_term_condition_precise}
\end{eqnarray}

We further rewrite $\mathcal{F}^{(q^2,R)} = \mathcal{F}_\text{I}^{(q^2,R)} + \mathcal{F}_\text{II}^{(q^2,R)}$. For each $\mathcal{F}_\text{I}^{(q^2,R)}$ term $\x^{\r}$, the graphs $\widetilde{J}_i\cap T^2(\r)$ ($j=1,\dots,K$) and $\widetilde{S}_k\cap T^2(\r)$ ($k=1,\dots,\mathfrak{N}$) are spanning trees of $\widetilde{J}_i$ and $\widetilde{S}_k$ respectively, and $H\cap T^2(\r)$ is a spanning 2-tree of $H$. Namely,
\begin{eqnarray}
    \mathcal{F}_\text{I}^{(q^2,R)}(\x;\s):
    &&\ n_H^{} = L(H)+1,\nonumber \\
    &&\ n_{J_i}^{} = L(\widetilde{J}_i)\quad (i=1,\dots,K), \nonumber\\
    &&\ n_{S_k}^{}= L(\widetilde{S}_k)\quad (k=1,\dots,\mathfrak{N}).
\label{eq:threshold_minimal_fq2i_term_condition_precise}
\end{eqnarray}

For each $\mathcal{F}_\text{II}^{(q^2,R)}$ term $\x^{\r}$, one component of $T^2(\r)$ is attached by two on-shell external momenta $p_i^\mu$ and $p_j^\mu$, and possibly some other soft external momenta. The graphs $H\cap T^2(\r)$ and $\widetilde{J}_l\cap T^2(\r)$ ($l\neq i,j$) are spanning trees of $H$ and $\widetilde{J}_l$ respectively, while $\widetilde{J}_i\cap T^2(\r)$ and $\widetilde{J}_j\cap T^2(\r)$ are spanning 2-trees of $\widetilde{J}_i$ and $\widetilde{J}_j$ respectively. Furthermore, there is a soft component $S_{k[i,j]}$, which is adjacent to both $J_i$ and $J_j$, and $S_{k[i,j]} \cap T^2(\r)$ contains a path joining $J_i$ and $J_j$. Equivalently, $\widetilde{S}_{k[i,j]} \cap T^2(\r)$ contains a loop. Meanwhile, all the other $\widetilde{S}_k\cap T^2(\r)$ ($k\neq k[i,j]$) are spanning trees of $\widetilde{S}_k$. We then have
\begin{eqnarray}
    \mathcal{F}_\text{II}^{(q^2,R)}(\x;\s):
    &&\ n_H^{} = L(H),\nonumber \\
    &&\ n_{J_{i}}^{} = L(\widetilde{J}_i)+1,\quad n_{J_j}^{} = L(\widetilde{J}_j)+1,\quad n_{J_l}=L(\widetilde{J}_l)\ \ (\forall l\neq i,j),\nonumber\\
    &&\ n_{S_{k[i,j]}}^{}= L(\widetilde{S}_{k[i,j]})-1,\quad n_{S_k}^{}= L(\widetilde{S}_k)\ \ (\forall k\neq k[i,j]).
\label{eq:threshold_minimal_fq2ii_term_condition_precise}
\end{eqnarray}

It is worth noticing that the $\mathcal{U}^{(R)}$, $\mathcal{F}^{(p_i^2,R)}$ and $\mathcal{F}^{(q^2)}$ terms are all of the same weight
\begin{eqnarray}
\label{eq:soft_expansion_leading_weight}
    -L(\widetilde{J})-2L(\widetilde{S}),
\end{eqnarray}
which can be directly verified from eqs.~(\ref{eq:threshold_minimal_u_term_condition_precise}) and (\ref{eq:threshold_minimal_fpi2_term_condition_precise})-(\ref{eq:threshold_minimal_fq2ii_term_condition_precise}). We stress once more, that the results above can be obtained by directly applying the on-shell expansion analysis given in ref.~\cite{GrdHzgJnsMaSchlk22}.

We still need to characterize the $\mathcal{F}^{(l^2,R)}$ terms, which do not satisfy (\ref{eq:threshold_leading_terms_fundamental_constraints}), and confirm that they also correspond to the weight in (\ref{eq:soft_expansion_leading_weight}). For each given $\mathcal{F}^{(l^2)}$ term $\x^{\r}$, the kinematic contribution is $+2$. Also, the two components of $T^2(\r)$ can be seen to be separated by a cut going through a set of edges, where two or more soft external momenta are on one side of the cut, and all the other external momenta are on the other. Three examples of such cuts are shown in figures~\ref{figure-threshold_Fl2_term_cuts}, where all the cut edges belong to $S$ in figure~\ref{threshold_Fl2_term_cut1}, belong to $S\cup J_1$ in figure~\ref{threshold_Fl2_term_cut2}, and belong to $S\cup J_1\cup J_2$ in figure~\ref{threshold_Fl2_term_cut3}. In terms of the numbers $n_H$, $n_{J_i}$ and $n_S$, we have:
\begin{figure}[t]
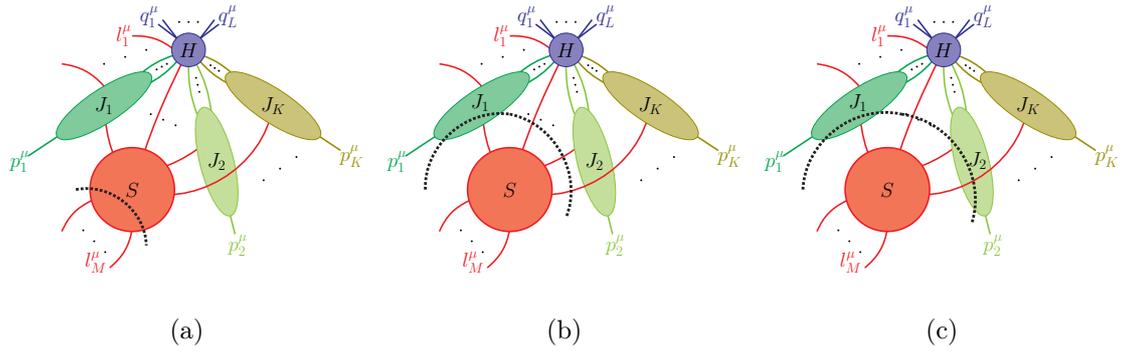

\centering
\begin{subfigure}[b]{0.32\textwidth}
\centering
\include{figs/threshold_Fl2_term_cut1}
\vspace{-2em}
\caption{}
\label{threshold_Fl2_term_cut1}
\end{subfigure}
\begin{subfigure}[b]{0.32\textwidth}
\centering
\include{figs/threshold_Fl2_term_cut2}
\vspace{-2em}
\caption{}
\label{threshold_Fl2_term_cut2}
\end{subfigure}
\begin{subfigure}[b]{0.32\textwidth}
\centering
\include{figs/threshold_Fl2_term_cut3}
\vspace{-2em}
\caption{}
\label{threshold_Fl2_term_cut3}
\end{subfigure}
\caption{Examples of the cuts that separate the two components of $T^2(\r)$. In each figure the cut is represented by the dotted curve, and one side of the cut is attached by soft external momenta only. (a) All the cut edges belong to $S$. (b) The cut edges belong to $S\cup J_1$. (c) The cut edges belong to $S\cup J_1\cup J_2$.}
\label{figure-threshold_Fl2_term_cuts}
\end{figure}
\begin{subequations}
    \begin{align}
    \text{figure }\ref{threshold_Fl2_term_cut1}\text{:}&\qquad n_H^{} = L(H),\qquad n_{J} = L(\widetilde{J}),\phantom{+1}\qquad n_S= L(\widetilde{S})+1.
    \label{eq:threshold_fl2_term_condition1}\\
    \text{figure }\ref{threshold_Fl2_term_cut2}\text{:}&\qquad n_H^{} = L(H),\qquad n_{J} = L(\widetilde{J})+1,\qquad n_S= L(\widetilde{S}).
    \label{eq:threshold_fl2_term_condition2}\\
    \text{figure }\ref{threshold_Fl2_term_cut3}\text{:}&\qquad n_H^{} = L(H),\qquad n_{J} = L(\widetilde{J})+2,\qquad n_S= L(\widetilde{S})-1.
    \label{eq:threshold_fl2_term_condition3}
    \end{align}
\end{subequations}

Note that there are other options of cuts to obtain an $\mathcal{F}^{(l^2,R)}$ term, but it suffices to observe from the examples above, that the minimum weight of $\mathcal{F}^{(l^2)}$ is achieved when all the cut edges belong to one connected component of $S$. The existence of any cut edges in $H\cup J$ would increase the overall weight. As a result, figure~\ref{threshold_Fl2_term_cut1} describes the $\mathcal{F}^{(l^2,R)}$ terms, and we have
\begin{eqnarray}
    \mathcal{F}^{(l^2,R)}(\x;\s):
    &&\ n_H^{} = L(H),\nonumber \\
    &&\ n_{J_i}^{} = L(\widetilde{J}_i)\quad (i=1,\dots,K), \nonumber\\
    &&\ n_{S_j}^{}= L(\widetilde{S}_j)+1,\quad n_{S_k}^{}= L(\widetilde{S}_k)\quad (k\neq j).
\label{eq:threshold_minimal_fl2_term_condition_precise}
\end{eqnarray}
It is also straightforward to check that these $\mathcal{F}^{(l^2,R)}$ terms also correspond to the weight $-L(\widetilde{J})-2L(\widetilde{S})$ in (\ref{eq:soft_expansion_leading_weight}). Thus they are indeed leading terms.

To summarize, eqs.~(\ref{eq:threshold_minimal_u_term_condition_precise}), (\ref{eq:threshold_minimal_fpi2_term_condition_precise})-(\ref{eq:threshold_minimal_fq2ii_term_condition_precise}), and (\ref{eq:threshold_minimal_fl2_term_condition_precise}) characterize the complete set of leading terms in the soft expansion. Based on these equations, we shall investigate additional constraints on the hard, jet, and soft subgraphs.

\subsection{Further constraints on the region vectors}
\label{section-further_restrictions_region_vectors_threshold}

In section~\ref{section-general_prescription_threshold}, we have proposed that all the regions in the soft expansion can be depicted by figure~\ref{figure-threshold_region_H_J_S_precise}. To comprehensively define the framework of this representation, it is imperative to eliminate spurious configurations that conform with figure~\ref{figure-threshold_region_H_J_S_precise} but result in scaleless integrals. In this subsection, we shall first present the requirements of the hard, jet, and soft subgraphs, and then demonstrate that these requirements are both necessary and sufficient for a configuration of figure~\ref{figure-threshold_region_H_J_S_precise} to qualify as a region in the soft expansion.

\subsubsection{Statement of the requirements}
\label{section-statement_requirements}
Let us start with some basic concepts. Given any nontrivial jet $J_i$, we term it \emph{regular}, if each 1VI component of $\widetilde{J}_i$ depends on both $p_i^\mu$ and some \emph{nonzero} soft momentum $l^\mu$.
\begin{figure}[t]
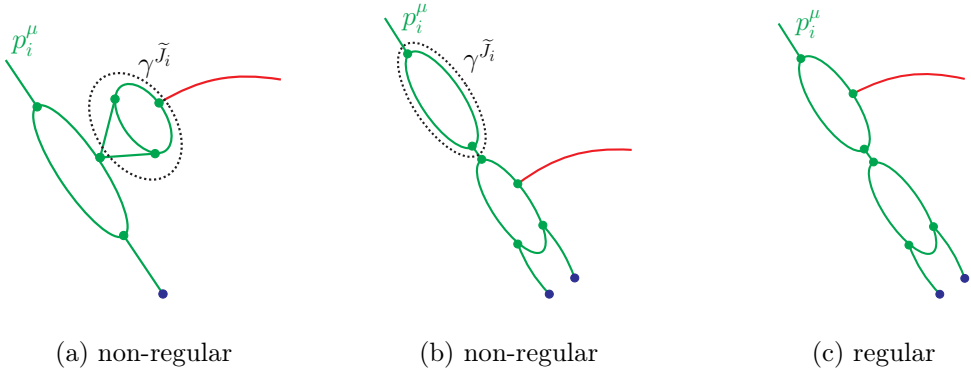

\centering
\begin{subfigure}[b]{0.25\textwidth}
\centering
\include{figs/regular_nonregular_jet_example1}
\vspace{-2em}
\caption{non-regular}
\label{regular_nonregular_jet_example1}
\end{subfigure}
\hspace{2em}
\begin{subfigure}[b]{0.25\textwidth}
\centering
\include{figs/regular_nonregular_jet_example2}
\vspace{-2em}
\caption{non-regular}
\label{regular_nonregular_jet_example2}
\end{subfigure}
\hspace{2em}
\begin{subfigure}[b]{0.25\textwidth}
\centering
\include{figs/regular_nonregular_jet_example3}
\vspace{-2em}
\caption{regular}
\label{regular_nonregular_jet_example3}
\end{subfigure}
\caption{Two examples of non-regular and one example of regular jets. (a) $J_i$ is not regular, because the 1VI components which is encircled by the dotted curve does not depend on $p_i^\mu$. (b) $J_i$ is non-regular because the 1VI components which is encircled by the dotted curve does not depend on any soft momentum $l^\mu$. (c) $J_i$ is regular.}
\label{figure-regular_nonregular_jet_examples}
\end{figure}

Figure~\ref{figure-regular_nonregular_jet_examples} provides some examples of regular and non-regular $J_i$. In particular, the $\gamma^{\widetilde{J}_i}$ specified in (a) does not depend on $p_i^\mu$, and the $\gamma^{\widetilde{J}_i}$ specified in (b) does not depend on any soft momentum $l^\mu$. Note that the topology in figure~\ref{figure-regular_nonregular_jet_examples}(b) leads to a scaleless Feynman integral before any expansion is performed, so we will not consider it in the upcoming analyses. In contrast, the jet $J_i$ in (c) is regular, because each 1VI component of $\widetilde{J}_i$ depends on both $p_i^\mu$ and some nonzero soft momentum $l^\mu$.

\begin{figure}[t]
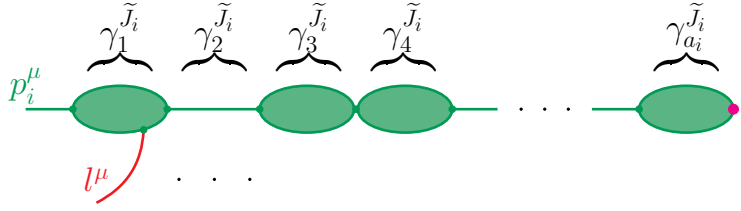

\centering
\include{figs/threshold_jet_1VI_labelling}
\vspace{-3em}
\caption{The configuration of $\widetilde{J}_i$ where $J_i$ is regular. The pink vertex is the auxiliary $J_i$ vertex. The 1VI components of $\widetilde{J}_i$ are labelled as $\gamma_1^{\widetilde{J}_i},\dots,\gamma_{a_i}^{\widetilde{J}_i}$, among which the green blobs represent nontrivial 1VI components (for example, $\gamma_1^{\widetilde{J}_i}$), and the internal edges represent trivial 1VI components (for example, $\gamma_2^{\widetilde{J}_i}$). Note that $\gamma_1^{\widetilde{J}_i}$ must be attached by some soft momentum, which we denote as $l^\mu$.}
\label{figure-threshold_jet_1VI_labelling}
\end{figure}

The general configuration of $\widetilde{J}_i$, with $J_i$ being regular, is depicted by figure~\ref{figure-threshold_jet_1VI_labelling}, where all the 1VI components align with the external momentum $p_i^\mu$. This configuration further allows us to label the 1VI components of $\widetilde{J}_i$ based on their distances from the external momentum $p_i^\mu$: we denote the ``outermost'' 1VI component of $\widetilde{J}_i$, which is attached by $p_i^\mu$, as $\gamma_{1}^{\widetilde{J}_i}$; we denote the one adjacent to $\gamma_{1}^{\widetilde{J}_i}$ as $\gamma_{2}^{J_i}$, and so on. In this way, the 1VI components of $\widetilde{J}_i$ are denoted by $\gamma_1^{\widetilde{J}_i}, \dots, \gamma_{a_i}^{\widetilde{J}_i}$, with $a_i$ the number of 1VI components of $\widetilde{J}_i$. In particular, $\gamma_1^{\widetilde{J}_i}$ is either attached by some soft external momenta, or adjacent to some internal soft propagators, whose total momentum is $l^\mu$ as shown in figure~\ref{figure-threshold_jet_1VI_labelling}.

\begin{figure}[t]
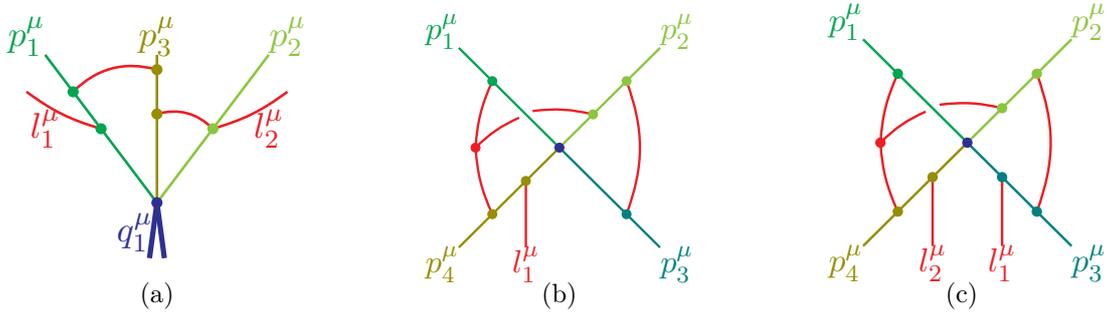

\centering
\begin{subfigure}[b]{0.3\textwidth}
\centering
\include{figs/IRcompatible_nonIRcompatible_jet_example1}
\vspace{-1em}
\caption{}
\label{IRcompatible_nonIRcompatible_jet_example1}
\end{subfigure}
\hfill
\begin{subfigure}[b]{0.3\textwidth}
\centering
\include{figs/IRcompatible_nonIRcompatible_jet_example2}
\vspace{-2em}
\caption{}
\label{IRcompatible_nonIRcompatible_jet_example2}
\end{subfigure}
\hfill
\begin{subfigure}[b]{0.3\textwidth}
\centering
\include{figs/IRcompatible_nonIRcompatible_jet_example3}
\vspace{-2em}
\caption{}
\label{IRcompatible_nonIRcompatible_jet_example3}
\end{subfigure}
\caption{Examples of non-region configurations where some jets are not soft compatible, and regions where all the jets are soft compatible. (a): $J_1$ and $J_2$ are soft-compatible jets while $J_3$ is not, because $J_3$ is neither joined to a soft external momentum via a soft path, nor joined to $J_1$ or $J_2$ via a soft component that is adjacent to three or more jets. (b): $J_1,J_2,J_4$ are soft-compatible jets while $J_3$ is not, because of the same reason as (a). In (c) all the jets $J_1$-$J_4$ are soft compatible.}
\label{figure-IRcompatible_nonIRcompatible_jet_examples}
\end{figure}

The notion of (non-)regular jets describes the intrinsic structure of an individual jet. To elucidate how jets interact with each other through the soft subgraph, we next introduce the notion of soft-compatible jets. We say that $J_i$ is \emph{soft compatible} if it satisfies \emph{either} of the following two conditions:
\begin{enumerate}
    \item [(1)] there is a soft path connecting a soft external momentum and $J_i$;
    \item [(2)] there exists a connected component of $S$ adjacent to three or more jets, including $J_i$ and another jet that has been confirmed soft compatible.
\end{enumerate}
Let us examine the concrete examples in figure~\ref{figure-IRcompatible_nonIRcompatible_jet_examples} to better understand this concept. In figure~\ref{IRcompatible_nonIRcompatible_jet_example1}, both $J_1$ and $J_2$ are considered soft compatible because they are both (directly) attached by some soft external momenta. In contrast, $J_3$ is not soft compatible because it satisfies neither (1) nor (2) above: no soft path connects $l_1^\mu$ or $l_2^\mu$ to $J_3$, and no soft component is adjacent to three or more jets. Similarly, in figure~\ref{IRcompatible_nonIRcompatible_jet_example2} the jets $J_1$, $J_2$, and $J_4$ are soft compatible. First of all, $J_4$ is soft compatible because it is attached by a soft external momentum $l_1^\mu$. Then we observe that there is a soft component simultaneously adjacent to $J_1$, $J_2$, and $J_4$, so $J_1$ and $J_2$ are also soft compatible. The jets $J_3$, in contrast, satisfies neither (1) nor (2) above, thus is not soft compatible. However, if we allow another soft external momentum to attach to $J_3$, as shown in figure~\ref{IRcompatible_nonIRcompatible_jet_example3}, $J_3$ also turns into a soft-compatible jet.

Indeed, one can verify that for each valid region presented in section~\ref{section-motivation_from_examples}, all the jets are both regular and soft compatible; for each non-region configuration, there exist some jets which are either non regular or non soft compatible.

Based on these concepts, we propose the following extra requirements of $H$, $J$, and $S$ in the soft expansion.
\begin{enumerate}
    \item \emph{The total momentum flowing into each 1VI component of $H$ is off shell.}
    \item \emph{Each jet is regular and soft compatible.}
    \item \emph{If a soft component is attached by zero or one soft external momentum, then it must be adjacent to two or more jets.}
\end{enumerate}
Note that the requirement of $H$ is identical to that for the on-shell expansion, as stated in section~\ref{section-further_restrictions_region_vectors}. The requirement of $S$ is based on the examples in (\ref{eq:threshold_motivation_soft_interacting_less_than_two_jets}), indicating that it is possible for a soft component to be adjacent to zero or one jet, provided it is attached by two or more soft external momenta.

In the remaining of this subsection, we prove that these requirements are necessary and sufficient for any configuration in figure~\ref{figure-threshold_region_H_J_S_precise} to be a valid region in the soft expansion.

\subsubsection{Proof of the necessity}
\label{section-proof_necessity}

The requirement of $H$ above is precisely the same for the on-shell expansion. As mentioned in section~\ref{section-further_restrictions_region_vectors}, its necessity has been demonstrated in ref.~\cite{GrdHzgJnsMaSchlk22}. The same analysis can be applied to the soft expansion, which we do not repeat here.

If the requirement of $S$ is not met, i.e., one component of $S$, say $S_0$, is attached by zero or one external momentum meanwhile adjacent to zero or one jet, then one can check from eqs.~(\ref{eq:threshold_minimal_u_term_condition_precise}), (\ref{eq:threshold_minimal_fpi2_term_condition_precise})-(\ref{eq:threshold_minimal_fq2ii_term_condition_precise}), and (\ref{eq:threshold_minimal_fl2_term_condition_precise}) that $n_{S_0}^{} = L(\widetilde{S}_0)$ holds for all the $\mathcal{P}^{(R)}$ terms. This contradicts the facet criterion, thus the requirement of $S$ is also necessary.

If a given jet $J_i$ is not regular, then by definition, one 1VI component of $\widetilde{J}_i$ is either independent of $p_i^\mu$ or soft momenta. Let us denote this component as $\gamma^{\widetilde{J}_i}$. If it does not depend on $p_i^\mu$ (as in figure~\ref{regular_nonregular_jet_example1}), then one can deduce $\text{dim}f_R<N$, which further implies that $f_R$ is not a facet of the Newton polytope~\cite{GrdHzgJnsMaSchlk22}. If it does not depend on any soft momenta (as in figure~\ref{regular_nonregular_jet_example2}), then the configuration of $\gamma^{\widetilde{J}_i}$ would lead to a scaleless integral in the original graph, which is a trivial case we shall not consider. Therefore, we have proved by contradiction that all the jets must be regular.

It remains to show that each jet must be soft compatible as well. To do this, we take any non-soft-compatible jet $J_i$, and focus on those jets $J_{i'}$ satisfying the following property: there exist some soft components $S_{i_1},\dots,S_{i_n}$ and jets $J_{i_1},\dots,J_{i_{n-1}}$ for a positive integer~$n$, such that
\begin{enumerate}
    \item [(1)] each $S_{i_j}$ ($j\in \{1,\dots,n\}$) is adjacent to three or more jets;
    \item [(2)] the jets adjacent to $S_{i_1}$ include $J_i$ and $J_{i_1}$, the jets adjacent to $S_{i_k}$ include $J_{i_{k-1}}$ and $J_{i_{k}}$ ($k=2,\dots,n-1$), and the jets adjacent to $S_{i_n}$ include $J_{i_{n-1}}$ and $J_{i'}$.
\end{enumerate}
For example, for the non-soft-compatible jet $J_1$ in (a) of (\ref{eq:threshold_motivation_scattering_soft_nonregions}), both $J_2$ and $J_4$ are in the associated set $\{J_{i'}\}$ defined above, because there is a soft component $S_1$ that is adjacent to three jets, including both $J_2$ ($J_4$) and $J_1$. (In this case, the value of $n$ above is equal to~$1$.) For the non-soft-compatible jet $J_1$ in (b) of (\ref{eq:threshold_motivation_scattering_soft_nonregions}), in contrast, the associated set $\{J_{i'}\}$ is empty, because there are no soft components adjacent to three or more jets in the figure.

\begin{figure}[t]
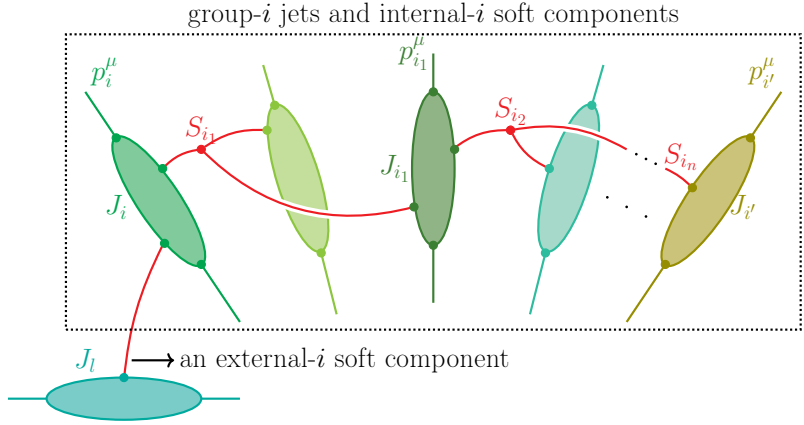

\centering
\include{figs/group_i_jets}
\vspace{-3em}
\caption{A graphic illustration of the group-$i$ jets and the internal-$i$ soft components, which are all enclosed by the dotted rectangle. For any $J_{i'}$ in this rectangle, there is a set of soft components $S_{i_1},\dots,S_{i_n}$ and a set of jets $J_{i_1},\dots,J_{i_{n-1}}$ connecting $J_i$ and $J_{i'}$. The soft components $S_{i_1},\dots,S_{i_n}$, each adjacent to three or more jets, are the internal-$i$ soft components, while all those soft components connecting a group-$i$ jet and a non-group-$i$ jet are the external-$i$ soft components.}
\label{figure-group_i_jets}
\end{figure}

For the most general case of a non-soft-compatible jet $J_i$, all its associated jets $J_{i'}$, and (one of) the remaining jets $J_l$ are depicted in figure~\ref{figure-group_i_jets}. For convenience, let us call the jets in $J_i\cup \{J_{i'}\}$ the \emph{group-$i$ jets}, call the soft components that are solely adjacent to the group-$i$ jets the \emph{internal-$i$ soft components}, and call the soft components, each of which is adjacent to one group-$i$ jet and a non-group-$i$ jet, the \emph{external-$i$ soft components}. One can verify that the configuration of figure~\ref{figure-group_i_jets} features the following properties. First, none of the group-$i$ jets is soft compatible, otherwise $J_i$ would also be soft compatible by definition. Second, each soft external momenta must not attach to any group-$i$ jet, any external-$i$ soft component, or any internal-$i$ soft component, otherwise the group-$i$ jets would become soft compatible, contradicting the property above. Third, any external-$i$ soft component is adjacent to exactly two jets, otherwise, if it is adjacent to three or more jets, all the jets it is adjacent to would be included in the set of group-$i$ jets by definition.

We now claim that in addition to the region vector $\v_R$, the following vector $\v'_R$ is also normal to the lower face $f_R$:
\begin{eqnarray}
\v'_R = &&\Big (\underset{N(H)}{\underbrace{0,\dots,0}}\ , \underset{\sum_{\text{non-group-}i}N(J)}{\underbrace{0,\dots,0}}, \underset{\sum_{\text{group-}i}N(J)}{\underbrace{1,\dots,1}},\nonumber\\
&&\hspace{3.5cm}\underset{\sum_{\text{external-}i}N(S)}{\underbrace{1,\dots,1}}, \underset{\sum_{\text{internal-}i}N(S)}{\underbrace{2,\dots,2}}, \underset{\sum_{\text{other}}N(S)}{\underbrace{0,\dots,0}}\ ;0 \Big ).
\label{eq:soft_expansion_non_soft_compatible_extra_vector}
\end{eqnarray}
Note that the sum $\sum_{\text{non-group-}i}$ is only for the non-group-$i$ jets, so $\sum_{\text{non-group-}i}N(J)$ denotes precisely the number of edges of the non-group-$i$ jets. The other notations, $\sum_{\text{group-}i}$, $\sum_{\text{external-}i}$, $\sum_{\text{internal-}i}$, and $\sum_{\text{other}}$, are defined similarly. Below we shall verify
\begin{eqnarray}
\label{eq:non_soft_compatible_extra_vector_inner_product}
    \v'_R\cdot \r = \sum_{\text{group-}i} L(\widetilde{J}) + \sum_{\text{external-}i} L(\widetilde{S}) + 2\sum_{\text{internal-}i} L(\widetilde{S})
\end{eqnarray}
for each leading term $\x^{\r}$, which suffices to show that $\v'_R$ is normal to $f_R$.

First, eq.~(\ref{eq:non_soft_compatible_extra_vector_inner_product}) holds when $\x^{\r}$ is a $\mathcal{U}^{(R)}$ term as characterized by eq.~(\ref{eq:threshold_minimal_u_term_condition_precise}). This is because among all the $N+1$ entries of the vector $\r$, only those associated to the group-$i$ jets, the external-$i$ soft components, and the internal-$i$ soft components can contribute to $\v'_R\cdot \r$, according to (\ref{eq:soft_expansion_non_soft_compatible_extra_vector}). Meanwhile, for each group-$i$ jet $J_a$, there are exactly $L(\widetilde{J}_a)$ nonzero entries of $\r$, each of which is $1$. So we have $\sum_{\text{group-}i} L(\widetilde{J})$ as the contribution to $\v'_R\cdot \r$ from the group-$i$ jets. Similarly, the terms $\sum_{\text{external-}i} L(\widetilde{S})$ and $2\sum_{\text{internal-}i} L(\widetilde{S})$ are also in $\v'_R\cdot \r$. Thus, we have verified eq.~(\ref{eq:non_soft_compatible_extra_vector_inner_product}) for $\x^{\r}$ being a $\mathcal{U}^{(R)}$ term.

Next, for $\x^{\r}$ being an $\mathcal{F}^{(p^2,R)}$ term, from eq.~(\ref{eq:threshold_minimal_fpi2_term_condition_precise}) there exists one jet $J_j$ satisfying $n_{J_j} = L(\widetilde{J}_j)+1$. From the defining property of $\mathcal{F}^{(p^2,R)}$ terms, one component of the spanning 2-tree $T^2(\r)$ is attached by the momentum $p_j^\mu$ and some soft external momenta, which implies the existence of some soft paths connecting these soft external momenta and $J_i$. Then $J_i$ is soft compatible by definition, thus cannot be a group-$i$ jet. In other words, for each group-$i$ jet $J_a$, we still have $n_{J_a} = L(\widetilde{J}_a)$ as in the previous paragraph. So the reasoning in the paragraph above applies identically, and eq.~(\ref{eq:non_soft_compatible_extra_vector_inner_product}) holds for $\x^{\r}$ being an $\mathcal{F}^{(p^2,R)}$ term.

Then we consider the $\mathcal{F}^{(q^2,R)}$ terms. For $\x^{\r}$ being an $\mathcal{F}_\text{I}^{(p^2,R)}$ term as characterized by eq.~(\ref{eq:threshold_minimal_fq2i_term_condition_precise}), the only difference from the $\mathcal{U}^{(R)}$ terms lies in the number $n_H$. However, it is clear from (\ref{eq:soft_expansion_non_soft_compatible_extra_vector}) that those entries of $\r$ associated with $H$ do not contribute to $\v'_R\cdot \r$. It then follows that $\v'_R\cdot \r$ should be of the same value in (\ref{eq:non_soft_compatible_extra_vector_inner_product}), regardless of whether $\x^{\r}$ is a $\mathcal{U}^{(R)}$ term or an $\mathcal{F}_\text{I}^{(q^2,R)}$ term.

For $\x^{\r}$ being an $\mathcal{F}_\text{II}^{(p^2,R)}$ term, from eq.~(\ref{eq:threshold_minimal_fq2ii_term_condition_precise}) there is one soft component $S_{k[j,l]}$ satisfying $n_{S_{k[j,l]}}^{}= L(\widetilde{S}_{k[j,l]})-1$, and two associated jets $J_j$ and $J_l$ satisfying $n_{J_j}=L(\widetilde{J}_j)+1$ and $n_{J_l}=L(\widetilde{J}_l)+1$, respectively. If $S_{k[j,l]}$ is an internal-$i$ soft component, both $J_j$ and $J_l$ must be group-$i$ jets. The contribution to $\v'_R\cdot \r$ from the internal-$i$ soft components is then $2(\sum_{\text{internal-}i} L(\widetilde{S})-1)$, while the contribution from the group-$i$ jets is $\sum_{\text{group-}i} L(\widetilde{J})+2$. Meanwhile, the contribution from the external-$i$ soft components is $\sum_{\text{external-}i} L(\widetilde{S})$. The sum over these three contributions reproduces exactly the right-hand side of eq.~(\ref{eq:non_soft_compatible_extra_vector_inner_product}). If $S_{k[j,l]}$ is an external-$i$ soft component, then exactly one of $J_j$ and $J_l$ is a group-$i$ jet. If $S_{k[j,l]}$ is neither an internal- or an external-$i$ soft component, then $J_j$ and $J_l$ are both non-group-$i$ jets. Using similar reasoning as above, one can verify that eq.~(\ref{eq:non_soft_compatible_extra_vector_inner_product}) holds for these two cases as well.

Finally, we consider the possibility that $\x^{\r}$ is an $\mathcal{F}^{(l^2,R)}$ term. By comparing eqs.~(\ref{eq:threshold_minimal_u_term_condition_precise}) and (\ref{eq:threshold_minimal_fl2_term_condition_precise}), the only difference between the $\mathcal{U}^{(R)}$ and $\mathcal{F}^{(l^2,R)}$ terms lies in one soft component $S_j$: $n_{S_j} = L(\widetilde{S}_j)$ for $\mathcal{U}^{(R)}$ terms while $n_{S_j} = L(\widetilde{S}_j)+1$ for $\mathcal{F}^{(l^2,R)}$ terms. Furthermore, $S_j$ must be attached by at least two soft external momenta, so $S_j$ can neither be an internal- or external-$i$ soft component. Then $\v'_R\cdot \r$ is of the same value, regardless of whether $\x^{\r}$ is a $\mathcal{U}^{(R)}$ term or an $\mathcal{F}^{(l^2,R)}$ term, and eq.~(\ref{eq:non_soft_compatible_extra_vector_inner_product}) must also hold.

To conclude, we have shown that all the jets must be soft compatible. Those extra requirements of $H$ and $J$, as stated at the end of section~\ref{section-statement_requirements}, are necessary for the regions in the soft expansion.

\subsubsection{Proof of the sufficiency}
\label{section-proof_sufficiency}

Proving the sufficiency of the requirements is equivalent to showing $\text{dim}f_R= N$ when both the requirements of $H$ and $J$ are satisfied. To achieve this goal, we aim to identify the following $N(H)+N(J)+N(S) = N$ vectors in $f_R$:
\begin{subequations}
\label{basis_vectors_infrared_facet}
    \begin{align}
        &\Delta \boldsymbol{r}_{[H]}^{} \equiv (\underset{N(H)}{\underbrace{0,\dots,0,1,0,\dots, 0}}\ ,\underset{N(J)+N(S)}{\underbrace{0, \dots, 0}};0),
        \label{basis_vectors_infrared_facet_hard}\\
        &\Delta \boldsymbol{r}_{[J]}^{} \equiv (\underset{N(H)}{\underbrace{0, \dots, 0}}\ ,\ \underset{N(J)}{\underbrace{0\dots,0,1,0,\dots 0}}\ ,\underset{N(S)}{\underbrace{0, \dots, 0}};1),
        \label{basis_vectors_infrared_facet_jet}\\
        &\Delta \boldsymbol{r}_{[S]}^{} \equiv (\underset{N(H)+N(J)}{\underbrace{0, \dots, 0}},\ \underset{N(S)}{\underbrace{0,\dots,0,1,0,\dots, 0}};2),
        \label{basis_vectors_infrared_facet_soft}
    \end{align}
\end{subequations}
where we have used a specific parameterization of the edges such that the first $N(H)$ parameters correspond to the hard edges, the next $N(J)$ correspond to the jet edges, and the final $N(S)$ correspond to the soft edges. One can check that the vectors in eq.~(\ref{basis_vectors_infrared_facet}) are linearly independent of each other and all perpendicular to $\boldsymbol{v}_R$. Therefore, it suffices to claim that $f_R$ is a facet once all the vectors above are identified in $f_R$. The complete analysis will be provided through lemmas~\ref{lemma-threshold_Uterms_space_dimensionality}-\ref{lemma-threshold_soft_vector_existence} below.

To begin with, we study the basis vectors of the parameter space, which is generated by the spanning trees of a given nontrivial 1VI graph $\gamma$. We call a 1VI graph $\gamma$ \emph{nontrivial} if it contains one or more loops.

\begin{lemma}
\label{lemma-threshold_Uterms_space_dimensionality}
For any nontrivial 1VI graph $\gamma$, the parameter subspace $W_\gamma$, which contains all the points corresponding to the spanning trees of $\gamma$, satisfies
\begin{eqnarray}
\textup{dim}(W_\gamma)= N(\gamma)-1,
\label{eq:threshold_Uterms_space_dimensionality}
\end{eqnarray}
where $N(\gamma)$ is the number of propagators of $\gamma$. Furthermore, the $(N(\gamma)-1)$ independent basis vectors of $W_\gamma$ can be expressed as:
\begin{eqnarray}
\Big ( \underset{N(\gamma)}{\underbrace{1,-1,0,\dots,0}};0 \Big ),\ \Big ( \underset{N(\gamma)}{\underbrace{1,0,-1,0,\dots,0}};0 \Big ),\ \dots,\ \Big ( \underset{N(\gamma)}{\underbrace{1,0,\dots,0,-1}};0 \Big ).
\label{Nminus1_basis_vector_U}
\end{eqnarray}
\end{lemma}
This lemma is also stated and proved (as theorem 1) in ref.~\cite{GrdHzgJnsMaSchlk22}, and for simplicity, we shall not repeat its proof here. As an essential application of this lemma, we show that all the $N(H)$ vectors $\Delta\r_{[H]}$ can be found in $f_R$ if the requirement of $H$ is satisfied.

\begin{lemma}
If the momentum flowing into each 1VI component of $H$ is off shell, then all the vectors $\Delta \boldsymbol{r}_{[H]}^{}$ are in $f_R$.
\end{lemma}

\begin{proof}
Let us denote the 1VI components of $H$ as $\gamma_1^H,\dots,\gamma_{a(H)}^H$, where $a(H)$ is the total number of 1VI components of $H$.

From the factorization property of $\mathcal{U}^{(R)}(\x)$, eq.~(\ref{eq:threshold_leading_Uterms_factorize}), a $\mathcal{U}^{(R)}$ term can be obtained by combining any chosen spanning trees of $H$, $\widetilde{J}_1,\dots,\widetilde{J}_K$, and $\widetilde{S}$. Let us consider a set of $\mathcal{U}^{(R)}$ terms, which share the same spanning trees of $\widetilde{J}_1,\dots,\widetilde{J}_K$, and $\widetilde{S}$, and each of them has a distinct spanning tree of $H$. According to lemma~\ref{lemma-threshold_Uterms_space_dimensionality}, the following vectors are in $f_R$:
\begin{eqnarray}
\Big (\underset{N(\gamma_1^H)+\dots+N(\gamma_{j-1}^H)}{\underbrace{0,\dots,0}}, \underset{N(\gamma_j^H)}{\underbrace{1,0,\dots,0,-1,0,\dots,0}},\underset{N(\gamma_{j+1}^H)+\dots+N(\gamma_{n}^H)}{\underbrace{0,\dots,0}},\underset{N(J)+N(S)}{\underbrace{0,\dots,0}};0 \Big ).\qquad
\label{eq:threshold_hardgraph_basis_vectors_U_polynomial}
\end{eqnarray}
where $j\in \{1,\dots,a(H)\}$ such that $\gamma_j^H$ is nontrivial. Please note that we have opted for a specific parameterization for the initial $N(H)$ entries: the initial $N(\gamma_1^H)$ among them correspond to the edges of $\gamma_1^H$, the subsequent $N(\gamma_2^H)$ correspond to the edges of $\gamma_2^H$, and so on. Each vector of (\ref{eq:threshold_hardgraph_basis_vectors_U_polynomial}) contains two nonzero entries: one is always~$1$, linked to a particular edge of $\gamma_j^H$, while the other is always~$-1$, linked to any of the remaining edges of $\gamma_j^H$.

It then follows that the total number of vectors in (\ref{eq:threshold_hardgraph_basis_vectors_U_polynomial}) is
\begin{eqnarray}
{\sum}'_j \Big( N(\gamma_j^H)-1 \Big) = \sum_{j=1}^{a(H)} \Big( N(\gamma_j^H)-1 \Big)  = N(H) - a(H).
\label{eq:threshold_Uterm_difference_Hvector_number}
\end{eqnarray}
The symbol ${\sum}'$ sums over those $j$ such that $\gamma_j^H$ is nontrivial. To derive the first equality, note that the difference between the left-hand side and the right-hand side arises from the terms $N(\gamma_j^H)-1$ where $\gamma_j^H$ is any trivial 1VI component of $H$; however, such terms do not contribute because they each satisfy $N(\gamma_j^H)-1=0$. To derive the second equality, we have used $\sum_{j=1}^{a(H)} N(\gamma_j^H) = N(H)$.

We have thus found $N(H)-a(H)$ linearly independent vectors in $f_R$. Now, we need to find another $a(H)$ vectors, which are linear combinations of $\Delta \boldsymbol{r}_{[H]}$ but are independent of those in (\ref{eq:threshold_hardgraph_basis_vectors_U_polynomial}). To do this, for each $j\in \{1,\dots,a(H)\}$, we consider a $\mathcal{U}^{(R)}$ term $\x^{\r_1}$ and an $\mathcal{F}_\text{I}^{(q^2,R)}$ term $\x^{\r_2}$, such that there is an edge $e\in \gamma_j^H \cap T^1(\r_1)$ with $T^1(\r_1)\setminus e = T^2(\r_2)$. (Note that since the total momentum flowing into $\gamma_j^H$ is off-shell, such an edge $e$ must exist.) The vector $\r_2 -\r_1$, which is in $f_R$ by definition, takes the form:
\begin{equation}
\label{eq:threshold_hardgraph_basis_vectors_UandFq2}
    \Big (\underset{N(\gamma_1^H)+\dots+N(\gamma_{j-1}^H)}{\underbrace{0,\dots,0}}, \underset{N(\gamma_j^H)}{\underbrace{0,\dots,0,1,0,\dots,0}},\underset{N(\gamma_{j+1}^H)+\dots+N(\gamma_n^H)}{\underbrace{0,\dots,0}},\underset{N(J)+N(S)}{\underbrace{0,\dots,0}};0 \Big ),
\end{equation}
where the only nonzero entry $1$ is associated with the edge $e$ above. Since $j\in \{1,\dots,a(H)\}$, we obtain another $a(H)$ vectors, each following the form of (\ref{eq:threshold_hardgraph_basis_vectors_UandFq2}). These vectors are bound to be independent of each other, and also independent of those in (\ref{eq:threshold_hardgraph_basis_vectors_U_polynomial}). In total, we have found $N(H)$ independent vectors in $f_R$, which are equivalent to $\Delta \boldsymbol{r}_{[H]}^{}$ in (\ref{basis_vectors_infrared_facet_hard}).
\end{proof}

We next consider the $N(J)$ vectors $\Delta\r_{[J]}$. For each $i\in \{1,\dots,K\}$, we use $\Delta\r_{[J_i]}$ to denote the $N(J_i)$ vectors of the following form:
\begin{eqnarray}
\label{basis_vectors_infrared_facet_jet_i}
    \Delta \boldsymbol{r}_{[J_i]}^{} \equiv (\underset{N(H)}{\underbrace{0, \dots, 0}}\ ,\underset{N(J_1)+\dots+N(J_{i-1})}{\underbrace{0, \dots, 0}},\underset{N(J_i)}{\underbrace{0\dots,0,1,0,\dots 0}}\ ,\underset{N(J_{i+1})+\dots+N(J_K)}{\underbrace{0, \dots, 0}},\underset{N(S)}{\underbrace{0, \dots, 0}}\ ;-1),\nonumber\\
\end{eqnarray}
where the only nonzero entry $1$ can be associated with any edge of $J_i$. A first observation, as summarized below, is that if $J_i$ is a regular jet, then the existence of one $\Delta \boldsymbol{r}_{[J_i]}^{}$ in $f_R$ implies the existence of the remaining $N(J_i)-1$.
\begin{lemma}
\label{lemma-threshold_regular_jet_vectors_from_one}
For any regular jet $J_i$, if one of the $N(J_i)$ vectors $\Delta\r_{[J_i]}$ is in $f_R$, then the remaining $N(J_i)-1$ vectors are also in $f_R$.
\end{lemma}

\begin{proof}
We first consider a set of $\mathcal{U}^{(R)}$ terms that share the same spanning trees of $H$, $\widetilde{J}_1, \dots, \widetilde{J}_{i-1}, \widetilde{J}_{i+1}, \widetilde{J}_K$ and $\widetilde{S}$, but have distinct spanning trees of $\widetilde{J}_i$. Lemma~\ref{lemma-threshold_Uterms_space_dimensionality} then indicates that the following vectors are in $f_R$: 
\begin{eqnarray}
&&\Big (\underset{N(H)}{\underbrace{0,\dots,0}}\ , \underset{N(J_1)+\dots+N(J_{i-1})}{\underbrace{0,\dots,0}}, \underset{N(\gamma_i^{\widetilde{J}_i})+\dots+N(\gamma_{j-1}^{\widetilde{J}_i})}{\underbrace{0,\dots,0}},\  \underset{N(\gamma_j^{\widetilde{J}_i})}{\underbrace{1,0,\dots,0,-1,0,\dots,0}}\ ,\nonumber\\
&&\hspace{4.5cm}\underset{N(\gamma_{j+1}^{\widetilde{J}_i})+\dots+N(\gamma_{a_i}^{\widetilde{J}_i})}{\underbrace{0,\dots,0}}, \underset{N(J_{i+1})+\dots+N(J_K)}{\underbrace{0,\dots,0}},\ \underset{N(S)}{\underbrace{0,\dots,0}}\ ;0 \Big ),
\label{eq:threshold_jetgraph_basis_vectors_U_polynomial}
\end{eqnarray}
where $a_i$ is the number of 1VI components of $\widetilde{J}_i$, and $j\in \{1,\dots,a_i\}$ such that $\gamma_j^{\widetilde{J}_i}$ is a nontrivial 1VI component of $\widetilde{J}_i$. Similar to the vectors in (\ref{eq:threshold_hardgraph_basis_vectors_U_polynomial}), each vector above has two nonzero entries: one is always~$1$, linked to a particular edge of $\gamma_j^{\widetilde{J}_i}$, while the other is always~$-1$, linked to any of the remaining edges of $\gamma_j^{\widetilde{J}_i}$. The total number of these linearly independent vectors is then
\begin{eqnarray}
{\sum}'_j \Big( N(\gamma_j^{\widetilde{J}_i})-1 \Big)= \sum_{j=1}^{a_i} \Big( N(\gamma_j^{\widetilde{J}_i})-1 \Big)= N(J_i) -a_i,
\label{eq:threshold_Uterm_difference_Jivector_number}
\end{eqnarray}
where ${\sum}'_j$ sums over those $j$ such that $\gamma_j^{\widetilde{J}_i}$ is nontrivial, and we have used the identity $N(J_i)= \sum_{j=1}^{a_i} N(\gamma_j^{\widetilde{J}_i})$. The derivation of the right-hand side of eq.~(\ref{eq:threshold_Uterm_difference_Jivector_number}) is similar to that of eq.~(\ref{eq:threshold_Uterm_difference_Hvector_number}).

As we have assumed, one of the $N(J_i)$ vectors $\Delta\r_{[J_i]}$ already exists in $f_R$. In the special case of $a_i=1$, this vector and those in (\ref{eq:threshold_Uterm_difference_Jivector_number}) form $N(J_i)$ independent vectors that are equivalent to $\Delta\r_{[J_i]}$. For more general case of $a_i>1$, we still need to show that $f_R$ contains another $a_i$ vectors, which are independent from those in (\ref{eq:threshold_jetgraph_basis_vectors_U_polynomial}).

Since $J_i$ is regular, the outermost 1VI component of $\widetilde{J}_i$ (which we have denoted as $\gamma_1^{\widetilde{J}_i}$) is either attached by a soft external momentum or an internal soft edge, as explained in section~\ref{section-statement_requirements} (also see figure~\ref{figure-threshold_jet_1VI_labelling}).
\begin{itemize}
    \item If $\gamma_1^{\widetilde{J}_i}$ is attached by a soft external momentum $l^\mu$, we consider any $\mathcal{U}^{(R)}$ term $\x^{\r_1}$ such that $l^\mu$ attaches to $\gamma_1^{\widetilde{J}_i}\cap T^1(\r)$. Then for each $j\in \{1,\dots,a_i\}$, there is an $\mathcal{F}^{(p_i^2,R)}$ term $\x^{\r_2}$ such that $T^2(\r_2) = T^1(\r_1) \setminus e$, where $e$ is an edge in $\gamma_j^{\widetilde{J}_i}\cap T^1(\r_1)$. The vector $\r_2-\r_1$, which is in $f_R$, is then
    \begin{eqnarray}
    \label{eq:lemma3_threshold_vectors_to_find_remaining}
    \r_2-\r_1 = \big( 0,\dots,0,\underset{e}{\underset{\downarrow}{1}},0,\dots,0; 1 \big),
    \end{eqnarray}
    where the first nonzero entry $1$ is associated with the edge $e$. Since $j$ can be chosen arbitrarily from $\{1,\dots,a_i\}$, there are $a_i$ such vectors. All these vectors are in (\ref{basis_vectors_infrared_facet_jet_i}), meanwhile linearly independent of those in (\ref{eq:threshold_jetgraph_basis_vectors_U_polynomial}).
\end{itemize}

In the cases where $\gamma_1^{\widetilde{J}_i}$ is attached by only internal soft edges, we denote the connected component of $S$ containing this particular soft edge as $S_1$, and denote another jet $S_1$ is adjacent to by $J_k$. (Recall that from proposition~\ref{proposition-threshold_expansion_vector_generic_form}, each soft component is adjacent to two or more jets.)

\begin{itemize}
    \item If $S_1$ itself is attached by a soft external momentum $l^\mu$, then there is a soft path $P^S$ joining $l^\mu$ and $\gamma_1^{\widetilde{J}_i}$. We can then consider a $\mathcal{U}^{(R)}$ term $\x^{\r_1}$ such that $P^S\subset T^1(\r)$. Then we can apply the same analysis as the case above, and obtain another $a_i$ linearly independent vectors, which are in the same form as (\ref{eq:lemma3_threshold_vectors_to_find_remaining}).
    
    \item If no soft external momenta attach to $S_1$, then let us consider a set of $\mathcal{F}_\text{II}^{(q^2,R)}$ terms $\x^{\r_1}, \x^{\r_2}, \dots, \x^{\r_{a_i}}$ such that each of them is characterized by:
    \begin{align}
    \label{eq:threshold_proof_sufficiency_jet_vectors_step1}
    \begin{split}
    &n_H = L(H), \\
    &n_{J_i}=L(\widetilde{J}_i)+1,\quad n_{J_k}=L(\widetilde{J}_k)+1,\quad n_{J_l}=L(\widetilde{J}_l)\ \ (\forall l\neq i,k),\\
    &n_{S_1} = L(\widetilde{S}_1)-1,\quad n_{S_m} = L(\widetilde{S}_m)\ \ (\forall m\neq 1).
    \end{split}
    \end{align}
    Moreover, we require that there is a path in $S_1$ connecting $\gamma_1^{\widetilde{J}_i}$ and $J_k$, and for any $b_1,b_2\in \{1,\dots,a_i\}$, there exists $e_{b_1}\in \gamma_{b_1}^{\widetilde{J}_i}$ and $e_{b_2}\in \gamma_{b_2}^{\widetilde{J}_i}$, such that
    \begin{eqnarray}
    \label{eq:threshold_proof_sufficiency_jet_vectors_step2}
        T^2(\r_{b_1})\cup e_{b_1} = T^2(\r_{b_2})\cup e_{b_2}.
    \end{eqnarray}
    In other words, we have
    \begin{subequations}
        \begin{align}
            &H\cap T^2(\r_{b_1}) = H\cap T^2(\r_{b_2}),\\
            &\widetilde{J}_l\cap T^2(\r_{b_1}) = \widetilde{J}_l\cap T^2(\r_{b_2})\ \ (l\neq i),\\
            &\widetilde{S}\cap T^2(\r_{b_1}) = \widetilde{S}\cap T^2(\r_{b_2}),
        \end{align}
    \end{subequations}
    meanwhile $(\widetilde{J}_i\cap T^2(\r_{b_1}))\cup e_{b_1} = (\widetilde{J}_i\cap T^2(\r_{b_2}))\cup e_{b_2}$. Note that such a set of $\mathcal{F}_\text{II}^{(q^2,R)}$ terms $\{ \x^{\r_1}, \x^{\r_2}, \dots, \x^{\r_{a_i}} \}$ always exist. An example with $i=1$ and $a_i=3$ is shown in figure~\ref{figure-threshold_lemma3_proof_examples}.
\begin{figure}[t]
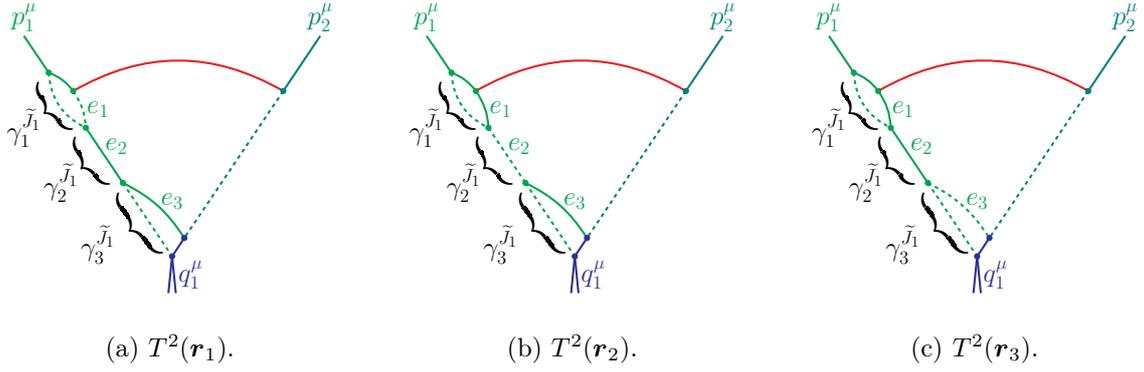

\centering
\begin{subfigure}[b]{0.3\textwidth}
\centering
\include{figs/threshold_lemma3_proof_example1}
\vspace{-2em}
\caption{$T^2(\r_1)$.}
\label{threshold_lemma3_proof_example1}
\end{subfigure}
\hfill
\begin{subfigure}[b]{0.3\textwidth}
\centering
\include{figs/threshold_lemma3_proof_example2}
\vspace{-2em}
\caption{$T^2(\r_2)$.}
\label{threshold_lemma3_proof_example2}
\end{subfigure}
\hfill
\begin{subfigure}[b]{0.3\textwidth}
\centering
\include{figs/threshold_lemma3_proof_example3}
\vspace{-2em}
\caption{$T^2(\r_3)$.}
\label{threshold_lemma3_proof_example3}
\end{subfigure}
\caption{A graphic illustration of the $\mathcal{F}^{(q^2,R)}$ terms whose corresponding spanning 2-trees are $T^2(\r_1), \dots, T^2(\r_{a_i})$. Here we have taken $i=1$ and $a_i=3$. These examples can be thought of as embedded in a larger graph with more internal propagators and external momenta. Eq.~(\ref{eq:threshold_proof_sufficiency_jet_vectors_step2}) clearly follows from these examples.}
\label{figure-threshold_lemma3_proof_examples}
\end{figure}
    
    With all these constructions above, the vector $\r_{b_1}-\r_{b_2}$ is then
    \begin{eqnarray}
    \label{eq:threshold_sufficiency_proof_jet_vectors_rb1rb2}
    \r_{b_1}-\r_{b_2} = \big( 0,\dots,0, \underset{e_{b_1}}{\underset{\downarrow}{1}}, 0,\dots,0,\underset{e_{b_2}}{\underset{\downarrow}{-1}}, 0,\dots,0;0 \big).
    \end{eqnarray}
    As indicated above, the entries $1$ and $-1$ respectively correspond to the edges $e_{b_1}$ and $e_{b_2}$, which are in distinct 1VI components of $\widetilde{J}_i$, so each of these vectors is linearly independent of those in (\ref{eq:threshold_jetgraph_basis_vectors_U_polynomial}). Since $b_1$ and $b_2$ can be chosen freely in $\{ 1,\dots,a_i \}$, the total number of independent vectors in (\ref{eq:threshold_sufficiency_proof_jet_vectors_rb1rb2}) is $a_i-1$. As we have also assumed, $f_R$ already contains one of the vectors $\Delta\r_{[J_i]}$. Together with these $a_i-1$ vectors, we obtain all the $a_i$ vectors that are independent of those in (\ref{eq:threshold_jetgraph_basis_vectors_U_polynomial}).
\end{itemize}

To summarize, we have discussed all the possible configurations of $\gamma_1^{\widetilde{J}_i}$ and demonstrated that another $a_i$ vectors, which are independent from each other and also independent from those in (\ref{eq:threshold_jetgraph_basis_vectors_U_polynomial}), exist in $f_R$. The lemma is thus proved.
\end{proof}

From this lemma, one can directly show that all the $N(J_i)$ vectors $\Delta\r_{[J_i]}$ are in $f_R$, if a soft path can ``conduct'' a soft external momentum to $J_i$.

\begin{corollary}
\label{lemma-threshold_regular_jet_vectors_from_one_corollary1}
For a given regular jet $J_i$, all the $N(J_i)$ vectors $\Delta \r_{[J_i]}$ are in $f_R$ if there is a path $P\subset S$ connecting $J_i$ and a soft external momentum $l^\mu$.
\end{corollary}
\begin{proof}
    We consider any $\mathcal{U}^{(R)}$ term $\x^{\r_1}$ such that $P\subset T^1(\r)$. Then there exists an edge $e\in J_i\cap T^1(\r)$ such that the external momenta attaching to one component of the spanning 2-tree $T^1(\r)\setminus e$ is $(p_i+l)^\mu$. This spanning 2-tree then corresponds to an $\mathcal{F}^{(p_i^2,R)}$ term $\x^{\r_2}$, and the vector $\r_1-\r_2$ is
    \begin{eqnarray}
    \r_1-\r_2 = \big( 0,\dots,0,1,0,\dots,0; 1 \big),
    \end{eqnarray}
    where the nonzero entry $1$ is associated with the edge $e$. Thus one of the $N(J_i)$ vectors $\Delta\r_{[J_i]}$, which is above, is in $f_R$. Since $J_i$ is regular, on applying lemma~\ref{lemma-threshold_regular_jet_vectors_from_one} we know that all the $N(J_i)$ vectors $\Delta \r_{[J_i]}$ are in $f_R$.
\end{proof}

Let us note that corollary~\ref{lemma-threshold_regular_jet_vectors_from_one_corollary1} is a sufficient condition for the existence of all the $N(J_i)$ vectors $\Delta\r_{[J_i]}$. It is not necessary. For example, the jets $J_1,J_4$ of example (b), and the jets $J_1,J_2$ of example (c) in (\ref{eq:threshold_motivation_scattering_soft_regions}) do not satisfy the conditions that there is a path $P\subset S$ joining $J_i$ and an soft external momentum, but for each of them, all the $N(J_i)$ vectors $\Delta\r_{[J_i]}$ are in the $f_R$ as $R$ is a valid region. Nevertheless, these jets satisfy another property: they are all adjacent to a soft component $S'$ that is adjacent to three or more jets in total. The following lemma is motivated from this observation.

\begin{lemma}
\label{lemma-threshold_IRcompatible_jet_vector_conduction}
Suppose all the $N(J_a)$ vectors $\Delta\r_{[J_a]}$ are in $f_R$ for a given jet $J_a$, then all the $N(J_b)$ vectors $\Delta\r_{[J_b]}$ are also in $f_R$, provided that both the following conditions are satisfied:
\begin{itemize}
    \item [(1)] $J_b$ is a regular jet;
    \item [(2)] one component of the soft subgraph is adjacent to three or more jets, including $J_a$ and~$J_b$.
\end{itemize}
\end{lemma}

\begin{proof}
Let us suppose that $S'$ is adjacent to exactly three jets: $J_a$, $J_b$, and another jet $J_c$. (The same analysis below applies when $S'$ is adjacent to four or more jets.) We then consider a special spanning 2-tree of $G$, say $T^2$: one component of $T^2$ is attached by the external momenta $p_a^\mu, p_b^\mu, p_c^\mu$, and the graphs $\widetilde{J}_a\cap T^2$, $\widetilde{J}_b\cap T^2$, and $\widetilde{J}_c\cap T^2$ are spanning 2-trees of $\widetilde{J}_a$, $\widetilde{J}_b$, and $\widetilde{J}_c$ respectively; meanwhile, for any two jets of $J_a,J_b,J_c$, there is a soft path in $S'\cap T^2$ that joins them, which means that the graph $\widetilde{S}'\cap T^2$ contains two loops (see figure~\ref{figure-threshold_lemma4_Sprime_configuration}). $T^2$ is then characterized by
\begin{eqnarray}
    &&\ n_H^{} = L(H),\nonumber \\
    &&\ n_{J_{a}}^{} = L(\widetilde{J}_a)+1,\quad n_{J_b}^{} = L(\widetilde{J}_b)+1,\quad n_{J_c}^{} = L(\widetilde{J}_c)+1,\quad n_{J_d}=L(\widetilde{J}_d)\ \ (\forall d\neq a,b,c),\nonumber\\
    &&\ n_{S'}^{}= L(\widetilde{S}')-2,\quad n_{S_k}^{}= L(\widetilde{S}_k)\ \ (S_k\text{ is any other component of }S).
\end{eqnarray}
Although $T^2$ does \emph{not} correspond to any leading terms, we will soon construct certain $\mathcal{F}^{(q^2,R)}$ and $\mathcal{U}^{(R)}$ terms by modifying~$T^2$.
\begin{figure}[t]
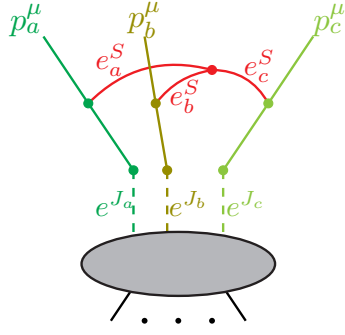

\centering
\include{figs/threshold_lemma4_Sprime_configuration}
\caption{An example of the spanning tree $T^2$, one of whose components is attached by the on-shell external momenta $p_a^\mu, p_b^\mu, p_c^\mu$, and there are some soft paths in $S'$ joining $J_a,J_b,J_c$. The lower grey blob represents the other component of $T^2$.}
\label{figure-threshold_lemma4_Sprime_configuration}
\end{figure}

Since $\widetilde{J}_a\cap T^2$, $\widetilde{J}_b\cap T^2$ and $\widetilde{J}_c\cap T^2$ are spanning 2-trees of $\widetilde{J}_a$, $\widetilde{J}_b$ and $\widetilde{J}_c$ respectively, there exists three jet edges $e^{J_a}\in J_a\setminus T^2$, $e^{J_b}\in J_b\setminus T^2$ and $e^{J_c}\in J_c\setminus T^2$, such that they each connect the two components of $T^2$. We further identify three soft edges $e_a^S, e_b^S, e_c^S \in S'\cap T^2$, such that $(S'\cap T^2)\setminus e_a^S$ connects $J_b$ and $J_c$ but not $J_a$, $(S'\cap T^2)\setminus e_b^S$ connects $J_a$ and $J_c$ but not $J_b$, and $(S'\cap T^2)\setminus e_c^S$ connects $J_a$ and $J_b$ but not $J_c$. These special edges have been marked in figure~\ref{figure-threshold_lemma4_Sprime_configuration}.

We can then construct the following three $\mathcal{F}^{(q^2,R)}$ terms $\x^{\r_{bc}}$, $\x^{\r_{ac}}$, and $\x^{\r_{ab}}$:
\begin{itemize}
    \item $\x^{\r_{bc}}$, with $T^2(\r_{bc}) = T^2\cup e^{J_a}\setminus e_a^S$ (see figure~\ref{threshold_lemma4_Fq2term_a});
    \item $\x^{\r_{ac}}$, with $T^2(\r_{ac}) = T^2\cup e^{J_b}\setminus e_b^S$ (see figure~\ref{threshold_lemma4_Fq2term_b});
    \item $\x^{\r_{ab}}$, with $T^2(\r_{ab}) = T^2\cup e^{J_c}\setminus e_c^S$ (see figure~\ref{threshold_lemma4_Fq2term_c}).
\end{itemize}
These terms are all in $\mathcal{F}^{(q^2,R)}$ because they conform with eq.~(\ref{eq:threshold_minimal_fq2ii_term_condition_precise}). Namely, for $\r = \r_{bc}, \r_{ac}, \r_{ab}$, one component of $T^2(\r)$ is attached by two on-shell external momenta, and contains a soft path connecting the corresponding jets.
\begin{figure}[t]
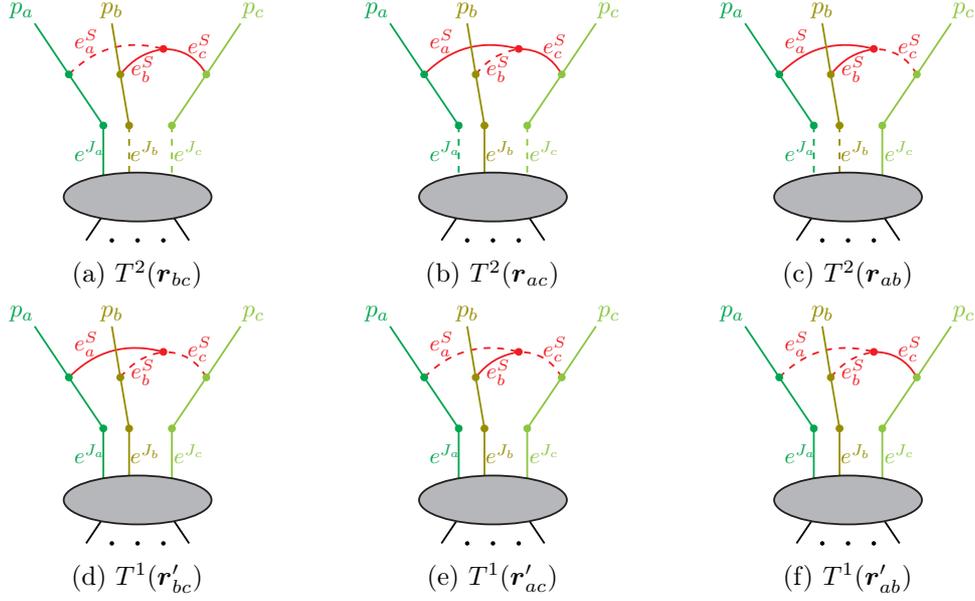

\centering
\begin{subfigure}[b]{0.25\textwidth}
\centering
\include{figs/threshold_lemma4_Fq2term_a}
\vspace{-1em}\caption{$T^2(\r_{bc})$}
\label{threshold_lemma4_Fq2term_a}
\end{subfigure}
\qquad
\begin{subfigure}[b]{0.25\textwidth}
\centering
\include{figs/threshold_lemma4_Fq2term_b}
\vspace{-1em}\caption{$T^2(\r_{ac})$}
\label{threshold_lemma4_Fq2term_b}
\end{subfigure}
\qquad
\begin{subfigure}[b]{0.25\textwidth}
\centering
\include{figs/threshold_lemma4_Fq2term_c}
\vspace{-1em}\caption{$T^2(\r_{ab})$}
\label{threshold_lemma4_Fq2term_c}
\end{subfigure}
\\
\begin{subfigure}[b]{0.25\textwidth}
\centering
\include{figs/threshold_lemma4_Uterm_d}
\vspace{-1em}\caption{$T^1(\r'_{bc})$}
\label{threshold_lemma4_Uterm_d}
\end{subfigure}
\qquad
\begin{subfigure}[b]{0.25\textwidth}
\centering
\include{figs/threshold_lemma4_Uterm_e}
\vspace{-1em}\caption{$T^1(\r'_{ac})$}
\label{threshold_lemma4_Uterm_e}
\end{subfigure}
\qquad
\begin{subfigure}[b]{0.25\textwidth}
\centering
\include{figs/threshold_lemma4_Uterm_f}
\vspace{-1em}\caption{$T^1(\r'_{ab})$}
\label{threshold_lemma4_Uterm_f}
\end{subfigure}
\captionsetup{margin = 2cm}
\caption{The spanning (2-)trees corresponding to the terms $\x^{\r_{bc}}$, $\x^{\r_{ac}}$, $\x^{\r_{ab}}$, $\x^{\r'_{bc}}$, $\x^{\r'_{ac}}$, and $\x^{\r'_{ab}}$, respectively.}
\label{figure-threshold_lemma4_UandFq2_terms}
\end{figure}
In addition, we construct the following three $\mathcal{U}^{(R)}$ terms $\x^{\r'_{bc}}$, $\x^{\r'_{ac}}$, and $\x^{\r'_{ab}}$:
\begin{itemize}
    \item $\x^{\r'_{bc}}$, with $T^1(\r') = T^2\cup (e^{J_a}\cup e^{J_b}\cup e^{J_c}) \setminus (e_b^S\cup e_c^S)$ (see figure~\ref{threshold_lemma4_Uterm_d});
    \item $\x^{\r'_{ac}}$, with $T^1(\r'_{ac}) = T^2\cup (e^{J_a}\cup e^{J_b}\cup e^{J_c}) \setminus (e_a^S\cup e_c^S)$ (see figure~\ref{threshold_lemma4_Uterm_e});
    \item $\x^{\r'_{ab}}$, with $T^1(\r'_{ab}) = T^2\cup (e^{J_a}\cup e^{J_b}\cup e^{J_c}) \setminus (e_a^S\cup e_b^S)$ (see figure~\ref{threshold_lemma4_Uterm_f}).
\end{itemize}
It is direct to check that they are all in $\mathcal{U}^{(R)}$ because they conform with eq.~(\ref{eq:threshold_minimal_u_term_condition_precise}). We now observe the following relations:
\begin{subequations}
\label{eq:threshold_lemma4_points_differences}
    \begin{align}
        \r_{ac} -\r'_{bc} &= \big( 0,\dots,0, \underset{e^{J_a}}{\underset{\downarrow}{1}}, 0,\dots,0, \underset{e^{J_c}}{\underset{\downarrow}{1}}, 0,\dots,0, \underset{e_c^S}{\underset{\downarrow}{-1}}, 0,\dots,0;0 \big);
        \label{eq:threshold_lemma4_points_difference_1}
        \\
        \r_{bc} -\r'_{ac} &= \big( 0,\dots,0, \underset{e^{J_b}}{\underset{\downarrow}{1}}, 0,\dots,0, \underset{e^{J_c}}{\underset{\downarrow}{1}}, 0,\dots,0, \underset{e_c^S}{\underset{\downarrow}{-1}}, 0,\dots,0;0 \big).
        \label{eq:threshold_lemma4_points_difference_2}
    \end{align}
\end{subequations}
On the right-hand side above, we have indicated that each nonzero entry is linked to a certain edge of $G$. For example, in the expression of $\r_{ac} -\r'_{bc}$, the first $1$ is linked to $e^{J_a}$, the second $1$ is linked to $e^{J_c}$, and $-1$ is linked to $e_c^S$.

Since $\r_{ac}$, $\r'_{bc}$, $\r_{bc}$ and $\r'_{ac}$ are all leading terms, both $\r_{ac} - \r'_{bc}$ and $\r_{bc} - \r'_{ac}$ are in $f_R$. In addition, the vector
\begin{eqnarray}
\delta \r_a\equiv \big( 0,\dots,0, \underset{e^{J_a}}{\underset{\downarrow}{1}}, 0,\dots,0;1 \big)
\end{eqnarray}
is also in $f_R$ since it is precisely one of the $N(J_a)$ vectors $\Delta\r_{[J_a]}$. So the following vector
\begin{eqnarray}
-\left(\r_{ac} -\r'_{bc}\right) + \left(\r_{bc} -\r'_{ac}\right) +\delta \r_a  = \big( 0,\dots,0, \underset{e^{J_b}}{\underset{\downarrow}{1}}, 0,\dots,0;1 \big)
\end{eqnarray}
is also in $f_R$. Since $J_b$ is a regular jet, a direct application of lemma~\ref{lemma-threshold_regular_jet_vectors_from_one} indicates that all the $N(J_b)$ vectors $\Delta\r_{J_b}$ are in $f_R$. The lemma is thus proved.
\end{proof}

By combining lemma~\ref{lemma-threshold_IRcompatible_jet_vector_conduction} and the concept of soft-compatible jets, one can show the existence of all the $N(J)$ vectors $\Delta\r_{[J]}$ in (\ref{basis_vectors_infrared_facet_jet}).
\begin{corollary}
\label{lemma-threshold_IRcompatible_jet_vector_conduction_corollary1}
    All the $N(J)$ vectors $\Delta\r_{[J]}$ are in $f_R$ if every jet is both regular and soft compatible.
\end{corollary}
\begin{proof}
    If all the jets are soft compatible, by definition some of them must be joined to some soft external momenta via paths in $S$. Let us denote those jets as $J_1,\dots,J_{n_1}$. Since all the jets are regular, from corollary~\ref{lemma-threshold_regular_jet_vectors_from_one_corollary1} the vectors $\Delta\r_{[J_1]},\dots, \Delta\r_{[J_{n_1}]}$ are all in $f_R$. Let us then consider another set of jets $J_{n_1+1},\dots,J_{n_2}$ (with $n_2>n_1$): for each of them, say $J_i$, there exists another jet $J_j\in \{J_1,\dots,J_{n_1}\}$ and a soft component $S'$, such that $S'$ is adjacent to three or more jets including $J_i$ and $J_j$. From lemma~\ref{lemma-threshold_IRcompatible_jet_vector_conduction}, all the vectors $\Delta\r_{[J_{n_1+1}]},\dots,\Delta\r_{[J_{n_2}]}$ are in $f_R$ as well. Since all the jets are soft compatible, we can repeat this procedure until all the vectors $\Delta\r_{[J]}$ are identified in $f_R$.
\end{proof}

Finally, we aim to show that all the $N(S)$ vectors $\Delta\r_{[S]}$ exist in $f_R$. To this end, we first classify the components of $S$ into the following two types: (a) those adjacent to two or more jets, and (b) those adjacent to zero or one jet while attached by two or more soft external momenta. We denote the union of the type-(a) soft components by $S^{(a)}$ and the type-(b) ones by $S^{(b)}$. It is straightforward to see that each of those vectors $\Delta\r_{[S]}$, where the entry $1$ corresponds to a soft edge in $\gamma_S^{(b)}$, is present in $f_R$. In fact, these vectors can be obtained directly from $\r_2-\r_1$, where $\r_2$ corresponds to an $\mathcal{F}^{(l^2,R)}$ term and $\r_1$ corresponds to a $\mathcal{U}^{(R)}$ term, such that there exists an edge $e\in \gamma_S^{(b)}$, with $T^1(\r_2) = T^1(\r_1)\setminus e$.

To see that those vectors $\Delta\r_{[S]}$ with the entry $1$ corresponding to an edge in $S^{(a)}$ are present in $f_R$, we employ the following lemma.
\begin{lemma}
If every jet is both regular and soft compatible, then the vectors $\Delta \boldsymbol{r}_{[S]}^{}$ are all in $f_R$.
\label{lemma-threshold_soft_vector_existence}
\end{lemma}
\begin{proof}
Let us take any soft component of $S^{(a)}$ and denote it as $S_1^{(a)}$. By definition, $\widetilde{S}_1^{(a)}$ is a 1VI graph. We then consider a set of $\mathcal{U}^{(R)}$ terms that share the same spanning trees of $H$, $\widetilde{J}_1,\dots,\widetilde{J}_K, \widetilde{S}_i$ (with $S_i\neq S_1^{(a)}$), while have distinct spanning tree of $\widetilde{S}_1^{(a)}$. Again, from lemma~\ref{lemma-threshold_Uterms_space_dimensionality}, the following $N(S_1^{(a)})-1$ vectors are in $f_R$:
\begin{eqnarray}
\Big (\underset{N(H)+N(J)}{\underbrace{0,\dots,0}}, \underset{N(S_1^{(a)})}{\underbrace{1,0,\dots,0,-1,0,\dots,0}},\underset{N(S)-N(S_1^{(a)})}{\underbrace{0,\dots,0}};0 \Big ),
\label{pinch_softgraph_basis_vectors_U_polynomial}
\end{eqnarray}
where we have parameterized the soft edges in a way that the first $N(S_1^{(a)})$ of them are all in $S_1^{(a)}$. Note that in the expression above, the entry $1$ is fixed as the first one of all the $N(S_1)$ entries linked to $S_1$, while $-1$ can be anywhere else. Thus the total number of vectors in (\ref{pinch_softgraph_basis_vectors_U_polynomial}) is $N(S_1^{(a)})-1$.

For any given $\mathcal{U}^{(R)}$ term $\x^{\r_1}$, we denote the corresponding spanning tree by $T^1$. As $S_1^{(a)}$ is a type-(a) soft component, we assume that it is adjacent to the jets $J_i$, $J_j$, and possibly some other jets. Then there is an edge $e_0\in S_1^{(a)}\setminus T^1$, such that there exist two edges $e_i\in J_i$ and $e_j\in J_j$ in the unique loop of $T^1\cup e_0$. The graph $T^2\equiv T^1\cup e_0 \setminus (e_i\cup e_j)$ is a spanning 2-tree, which, in line with eq.~(\ref{eq:threshold_minimal_fq2ii_term_condition_precise}), corresponds to an $\mathcal{F}_\text{II}^{(q^2,R)}$ term $\x^{\r_2}$. The vector \hbox{$\delta\boldsymbol{r}'\equiv \boldsymbol{r}_1-\boldsymbol{r}_2$} then reads
\begin{eqnarray}
\delta \boldsymbol{r}' = \big( 0,\dots,0, \underset{e_i}{\underset{\downarrow}{-1}}, 0,\dots,0, \underset{e_j}{\underset{\downarrow}{-1}}, 0,\dots,0, \underset{e_0}{\underset{\downarrow}{1}}, 0,\dots,0;0 \big),
\end{eqnarray}
Since every jet is both regular and soft compatible, from corollary~\ref{lemma-threshold_IRcompatible_jet_vector_conduction_corollary1}, all the $N(J_i)$ vectors $\delta \boldsymbol{r}_{[J]}$ are in $f_R$, which certainly include the following two vectors:
\begin{eqnarray}
\delta \boldsymbol{r}_i = \big( 0,\dots,0, \underset{e_i}{\underset{\downarrow}{1}}, 0,\dots,0;1 \big),\qquad \delta \boldsymbol{r}_j = \big( 0,\dots,0, \underset{e_j}{\underset{\downarrow}{1}}, 0,\dots,0;1 \big).
\end{eqnarray}
The vector $\delta \boldsymbol{r}\equiv \delta \boldsymbol{r}'+ \delta \boldsymbol{r}_{i}^{} + \delta \boldsymbol{r}_{j}^{}$ is then
\begin{eqnarray}
\delta \boldsymbol{r} = \big( 0,\dots,0, \underset{e_0}{\underset{\downarrow}{1}}, 0,\dots,0;2 \big).
\label{pinch_softgraph_basis_vectors_UminusF_polynomial}
\end{eqnarray}
This vector $\delta\r$, together with those vectors in eq.~(\ref{pinch_softgraph_basis_vectors_U_polynomial}), form a set of $N(S_1^{(a)})$ vectors that are independent from each other. Since $S_1^{(a)}$ is chosen as any type-(a) soft component, combining with the argument above this lemma, we deduce that all the $N(S)$ vectors $\Delta\r_{[S]}$ exist in $f_R$.
\end{proof}

\subsection{Summary}
\label{section-summary_threshold}

The regions appearing in the soft expansion are similar to those of the on-shell expansion. Specifically, each region is identified by a particular partition of the entire graph $G$ into the hard subgraph~$H$, the jet subgraphs~$J_i$ ($i=1,\dots,K$), and the soft subgraph~$S$ (the momentum scaling of the modes are shown in~(\ref{WideAngleMods})). The special region where all the propagators belong to $H$ is called the hard region, and each of the other regions, called the infrared regions, are in one-to-one correspondence with the pinch surfaces of $G|_{l^\mu=0}$. This indicates that the general configuration of infrared regions can be depicted by figure~\ref{figure-threshold_region_H_J_S_precise}: the hard subgraph $H$ is connected, to which all the off-shell external momenta $q_j^\mu$ and possibly some soft external momenta attach; each nontrivial jet subgraph $J_i$ is connected and adjacent to $H$, attached by $p_i^\mu$ and possibly some soft external momenta; the soft subgraph $S$ may have several distinct connected components, each of which must be adjacent to multiple jets and possibly $H$.

Similar to the on-shell expansion regions, additional requirements for the subgraph $H,J,S$ are needed for the soft expansion regions. These requirements, stated in section~\ref{section-statement_requirements}, aim to exclude those configurations that conform with figure~\ref{figure-threshold_region_H_J_S_precise} but yield scaleless integrals. In the soft expansion, the requirement of $H$ is the same as that in the on-shell expansion, while the requirements of $J$ and $S$ are different. We require each jet to be both regular and soft compatible, and allow each component of $S$ to be adjacent to zero or one jet, provided it is attached by two or more soft external momenta.

In summary, we have the following ``soft-expansion region proposition''.
\begin{center}
\framebox{\Longstack[l]{\textbf{Soft-expansion region proposition}\\\qquad The region vectors appearing in the soft expansion of wide-angle scattering are\\
all of the form of $\v_R= (u_{R,1},u_{R,2},\dots,u_{R,N};1)$, such that for each edge $e$,\\
\qquad $\bullet \ \phantom{-}u_{R,e}=0\phantom{1} \quad \Leftrightarrow\quad e\in H$;\\
\qquad $\bullet\ \phantom{0}u_{R,e}=-1 \quad \Leftrightarrow\quad e\in J$;\\
\qquad $\bullet\ \phantom{0}u_{R,e}=-2 \quad \Leftrightarrow\quad e\in S$.\\
The subgraphs $H,J,S$ are shown in figure~\ref{figure-threshold_region_H_J_S_precise}, which further satisfy the following.\\
\qquad (1) Requirement of $H$: the momentum flowing into each 1VI component of $H$ is\\ \qquad \phantom{(1) Requirement of $H$:} off shell.\\
\qquad (2) Requirement of $J$: each jet $J_i$ is both regular and soft compatible.\\
\qquad (3) Requirement of $S$: if a soft component is attached by zero or one soft exter-\\ \qquad \phantom{(1) Requirement of $H$:} nal momentum, then it is adjacent to two or more jets.}}
\end{center}
We emphasize that the only conjecture above is proposition~\ref{proposition-threshold_expansion_vector_generic_form}, that figure~\ref{figure-threshold_region_H_J_S_precise} characterizes all the regions in the soft expansion. Once this statement is proved, the soft-expansion region proposition can be rephrased as the ``soft-expansion region theorem''. We will leave such a proof to future work.

Similar to the on-shell-expansion region theorem studied in section~\ref{section-regions_onshell}, the proposition above provides the entire list of regions to all orders, independent of numerators in momentum representation and spacetime dimensions. In any theory, the order of a regions can be obtained by straightforward power counting. In four-dimensional gauge theory, in particular, those regions contributing to the leading power are the \emph{leading pinch surfaces} in the Collins-Soper-Sterman formalism~\cite{ClsSprStm04}, with the following properties: no soft partons are attached to $H$; no soft fermions or scalars are attached to $J$; for each jet, the set of its partons attached to $H$ consists of exactly one physical parton, with all others being scalar-polarized gauge bosons. Regions without these features are subleading.

\begin{figure}[t]
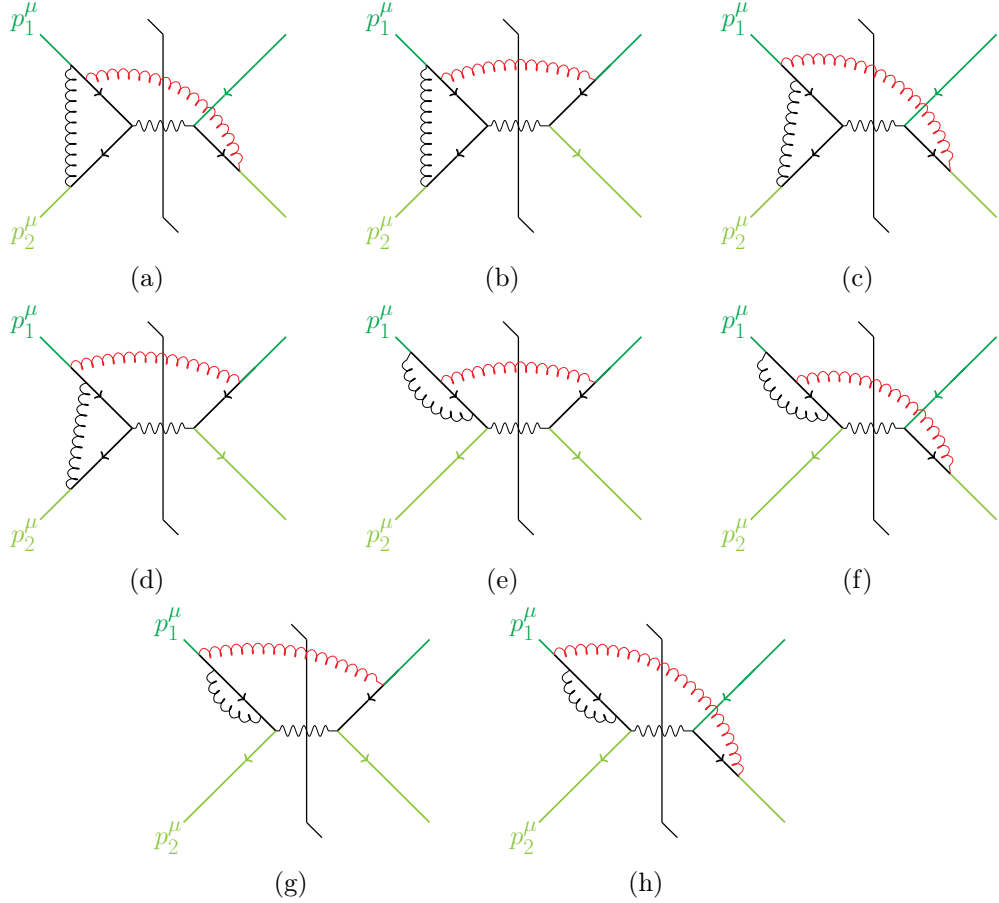

\centering
\begin{subfigure}[b]{0.25\textwidth}
\centering
\include{figs/soft_expansion_theorem_application_DY_1}
\vspace{-3em}\caption{}
\label{soft_expansion_theorem_application_DY_1}
\end{subfigure}
\qquad
\begin{subfigure}[b]{0.25\textwidth}
\centering
\include{figs/soft_expansion_theorem_application_DY_2}
\vspace{-3em}\caption{}
\label{soft_expansion_theorem_application_DY_2}
\end{subfigure}
\qquad
\begin{subfigure}[b]{0.25\textwidth}
\centering
\include{figs/soft_expansion_theorem_application_DY_3}
\vspace{-3em}\caption{}
\label{soft_expansion_theorem_application_DY_3}
\end{subfigure}
\\
\begin{subfigure}[b]{0.25\textwidth}
\centering
\include{figs/soft_expansion_theorem_application_DY_4}
\vspace{-3em}\caption{}
\label{soft_expansion_theorem_application_DY_4}
\end{subfigure}
\qquad
\begin{subfigure}[b]{0.25\textwidth}
\centering
\include{figs/soft_expansion_theorem_application_DY_5}
\vspace{-3em}\caption{}
\label{soft_expansion_theorem_application_DY_5}
\end{subfigure}
\qquad
\begin{subfigure}[b]{0.25\textwidth}
\centering
\include{figs/soft_expansion_theorem_application_DY_6}
\vspace{-3em}\caption{}
\label{soft_expansion_theorem_application_DY_6}
\end{subfigure}
\\
\begin{subfigure}[b]{0.25\textwidth}
\centering
\include{figs/soft_expansion_theorem_application_DY_7}
\vspace{-3em}\caption{}
\label{soft_expansion_theorem_application_DY_7}
\end{subfigure}
\qquad
\begin{subfigure}[b]{0.25\textwidth}
\centering
\include{figs/soft_expansion_theorem_application_DY_8}
\vspace{-3em}\caption{}
\label{soft_expansion_theorem_application_DY_8}
\end{subfigure}
\qquad
\phantom{\begin{subfigure}[b]{0.25\textwidth}
\centering
\include{figs/soft_expansion_theorem_application_DY_9}
\label{soft_expansion_theorem_application_DY_9}
\end{subfigure}}
\caption{The abelian-like cut graphs contributing to the NNLO Drell–Yan cross section in ref.~\cite{BncLnMgnVnzWht14}, where $p_1^\mu$ and $p_2^\mu$ are two in two distinct lightlike directions, and the real gluon (marked in red) is soft. Note that those graphs obtained by interchanging $p_1$ and $p_2$ and/or complex conjugation are not shown here.}
\label{figure-soft_expansion_theorem_application_DY}
\end{figure}

The soft-expansion region proposition finds application in the computation of QCD soft currents and splitting functions. For instance, it was recently employed in the study of single-soft emission, as demonstrated in ref.~\cite{HzgMaMstlbgSrsh23}. Furthermore, its utility extends to the realm of cross sections. In a study like ref.~\cite{BncLnMgnVnzWht14}, investigating the single soft gluon emission\footnote{later extended to double soft emission in ref.~\cite{BjtAbsSngVnzWht18}} in the NNLO Drell–Yan cross section, the abelian-like cut graphs depicted in figure~\ref{figure-soft_expansion_theorem_application_DY} are involved in the calculation. By applying the MoR, the leading-power and next-to-leading-power contributions for each figure can be computed, where the relevant regions can be directly identified using the soft-expansion region proposition. For example, figure~\ref{soft_expansion_theorem_application_DY_1} exhibits contributions from the hard and collinear-1 regions, while omitting contributions from the collinear-2 or soft regions.

The complete set of regions for the graphs in figure~\ref{figure-soft_expansion_theorem_application_DY} is shown in table~\ref{table-soft_expansion_theorem_application_DY}. It is important to note that the collinear-2 regions, pertinent to graphs involving $p_1 \leftrightarrow p_2$, are not included in this table. This result aligns with the findings in ref.~\cite{BncLnMgnVnzWht14}.
\begin{table}[t]
\begin{center}
\begin{tabular}{ |c||c|c|c||c|c|c|c| } 
\hline
    & $H$ & $C_1$ & $S$ & & $H$ & $C_1$ & $S$ \\ \hline
    (a) & $\greencheckmark[Green]$ & $\greencheckmark[Green]$ & & (e) &  & $\greencheckmark[Green]$ & \\ \hline
    (b) & $\greencheckmark[Green]$ & $\greencheckmark[Green]$ & & (f) &  & $\greencheckmark[Green]$ & \\ \hline
    (c) & $\greencheckmark[Green]$ & $\greencheckmark[Green]$ & & (g) &  & $\greencheckmark[Green]$ & \\ \hline
    (d) & $\greencheckmark[Green]$ & $\greencheckmark[Green]$ & & (h) &  & $\greencheckmark[Green]$ & \\ \hline
\end{tabular}
\captionsetup{justification=centering}
\caption{The regions that are relevant in the cut graphs in figure~\ref{figure-soft_expansion_theorem_application_DY}.}
\label{table-soft_expansion_theorem_application_DY}
\end{center}
\end{table}

\subsection{Extending the results to incorporate additional expansions}
\label{section-extending_results_additional_expansions}

\begin{figure}[t]
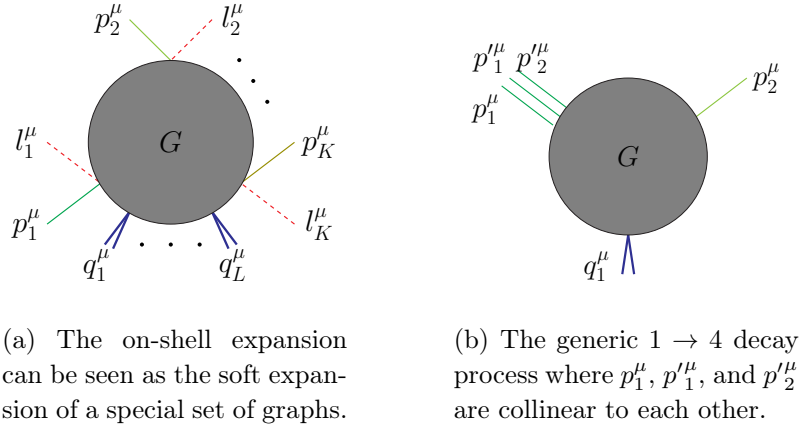

\centering
\begin{subfigure}[b]{0.3\textwidth}
\centering
\include{figs/special_soft_expansion}
\vspace{-2em}
\caption{The on-shell expansion can be seen as the soft expansion of a special set of graphs.}
\label{special_soft_expansion}
\end{subfigure}
\hspace{3em}
\begin{subfigure}[b]{0.3\textwidth}
\centering
\include{figs/generic_collinear_expansion}
\vspace{-2em}
\caption{The generic $1\to 4$ decay process where $p_1^\mu$, ${p'}_1^\mu$, and ${p'}_2^\mu$ are collinear to each other.}
\label{generic_collinear_expansion}
\end{subfigure}
\caption{Graphic illustrations of some applications of the soft expansion analysis. Figure~(a): the soft expansion of a special set of graphs, where the lightlike external momenta $p_1^\mu,\dots,p_K^\mu$ and the soft external momenta $l_1^\mu,\dots,l_K^\mu$ form pairs, each entering the same vertex of $G$. Figure~(b): an example of the decay process where some outgoing momenta are collinear to each other. Here, we have a $1\to 4$ decay process with $p_1^\mu \parallel {p'}_1^\mu \parallel {p'}_2^\mu$.}
\end{figure}

It is worth noticing that the on-shell expansion, which we have studied in detail in section~\ref{section-regions_onshell}, is intimately related to the soft expansion as we study in this section. The relations are from the external kinematics in their definitions, (\ref{eq:wideangle_onshell_kinematics}) and (\ref{eq:wideangle_soft_kinematics}). In the soft expansion, each $p_i$ is defined to be \emph{exactly on shell}, i.e., $(p_i^\text{soft})^2=0$; furthermore, we have $(p_i^\text{soft}+l_k)^2 = 2p_i^\text{soft}\cdot l_k \sim \lambda Q^2$ for any soft momentum $l_k$. In contrast, in the on-shell expansion, each $p_i$ is \emph{close to a lightcone} instead, i.e., $(p_i^\text{onshell})^2\sim \lambda Q^2\neq 0$, thus such $p_i^\text{onshell}$ can be seen as $p_i^\text{soft}+l_k$. As a result, the kinematics of the on-shell expansion is equivalent to the kinematics of the soft expansion regarding a special set of graphs. These graphs are drawn in figure~\ref{special_soft_expansion}, where the on-shell external momenta $\{p_i^\mu\}$ and soft external momenta $\{l_i^\mu\}$ form pairs, each entering the same vertex of $G$. Then the terms in the $\mathcal{F}$ polynomial can be classified into the following two types:
\begin{itemize}
    \item the $\mathcal{F}^{(p_i^2)}$ terms with $i\in \{1,\dots,K\}$, whose kinematic factor is $-(p_i+l_i)^2=-2p_i\cdot l_i$, scaling as $\lambda Q^2$;
    \item the $\mathcal{F}^{(q^2)}$ terms, whose kinematic factors contain terms of the form $q_j^2$, $p_i\cdot p_k$, or $p_i\cdot q_j$, scaling as $Q^2$.
\end{itemize}
Note that no $\mathcal{F}$ terms can have kinematic factors in the form of $l_i\cdot l_j$, because by construction, each $l_i^\mu$ is always accompanied by a $p_i^\mu$. As these $\mathcal{F}$ terms above one-to-one corresponds to those in the on-shell expansion, it then follows that the soft expansion can be seen as a generalization of the on-shell expansion.

From this observation, one can also verify that for the special case shown in figure~\ref{special_soft_expansion}, the statement of the soft-expansion region proposition is identical to that of the on-shell-expansion region theorem.

Building on the same idea, the regions in any expansion that combines the on-shell and soft expansions for wide-angle scattering can be obtained in a similar manner.\footnote{Note that such generalizations considers massless Feynman graphs only. With the inclusion of internal masses, the mode structure could become much more complicated. We will study one particular expansion with massive propagators (the heavy-to-light decay) in section~\ref{section-regions_mass_expansion} and investigate the relevant modes.} Moreover, the concept of identifying regions in the soft expansion can be applied to certain processes where some lightlike external momenta are collinear to each other. As an example, consider the one-to-four decay process where three outgoing momenta are in the same direction, as illustrated in figure~\ref{generic_collinear_expansion}. For more general cases, we may have off-shell external momenta $q_1^\mu,\dots,q_L^\mu$, on-shell external momenta $p_1^\mu,\dots,p_K^\mu$ pointing in different directions, and additional on-shell external momenta $\{ {p'}_k^\mu \}$, each of which collinear to one of $\{p_1^\mu,\dots,p_K^\mu\}$. The kinematics are then:
\begin{subequations}
\label{eq:wideangle_collinear_kinematics}
\begin{align}
    & p_i^2=0={p'}_k^2\ \ (\forall i,k), \quad q_j^2\sim Q^2\ \ (\forall j),\quad p_{i_1}\cdot p_{i_2}\sim Q^2\ \ (\forall i_1\neq i_2),\\
    & p_i\cdot {p'}_k\sim \lambda Q^2\ \textup{ for }{p'}_k \parallel p_i^\mu,\qquad p_i\cdot {p'}_k\sim Q^2\ \textup{ for }{p'}_k\nparallel p_i^\mu.
\end{align}
\end{subequations}
Expanding in the small parameter $\lambda$ defines this ``collinear expansion''~\cite{EbtMstbgVita20collinear,EbtMstbgVita20transverse,EbtMstbgVita20Njettiness,EbtMstbgVita21TMD}. In this context, the general configuration of the regions is similar to figure~\ref{figure-threshold_region_H_J_S_precise}, where no soft external momenta exist, and each jet is attached by one or more lightlike momenta that are collinear to each other. In line with the soft-expansion region proposition, the extra requirement of the jets needs to be adjusted as follows.
\begin{enumerate}
    \item Each jet is regular: for any $i\in \{1,\dots,K\}$, the momentum flowing into each 1VI component of $\widetilde{J}_i$ scales as $Q(1,\lambda,\lambda^{1/2})$ and has a nonzero virtuality.
    \item Each jet is ``collinear compatible'': for any $i\in \{1,\dots,K\}$, either $J_i$ is attached by two or more lightlike external momenta (all collinear to $p_i^\mu$), or there exists a connected component of $S$ adjacent to three or more jets, including $J_i$ and another jet that has been confirmed collinear compatible.
\end{enumerate}

Note that in eq.~(\ref{eq:wideangle_collinear_kinematics}), we have implicitly considered the ``timelike collinear limit'' only. That is, those external momenta collinear to each other are all incoming, or all outgoing. For the ``spacelike collinear limit'', where some incoming momenta are collinear to some outgoing momenta, there are possibly regions ``hidden'' inside the Newton polytope, which feature momenta in the Glauber mode. The identification of such regions requires a proper change of parameters and dissection of the original Newton polytope into distinct sectors~\cite{GrdHzgJnsMa24}.

In conclusion, the techniques used in this section can be applied to a wide range of expansions. Detailed discussions regarding any of these expansions will be deferred to future works.

\section{Regions in heavy-to-light decays: the mass expansion}
\label{section-regions_mass_expansion}

In this section, we study the regions in the mass expansion of the heavy-to-light decay processes, figure~\ref{figure-problem_to_study_mass}. Here, the external momenta are $Q^\mu$, $P^\mu$, and $p^\mu$, which satisfy the conditions (the parameter $\lambda \ll 1$):
\begin{eqnarray}
\label{eq:decay_mass_kinematics_rewritten}
    P^2=M^2\sim Q^2,\quad p^2=m^2\sim \lambda Q^2,\quad P\cdot p\sim Q^2.
\end{eqnarray}
Throughout this section, we will use a lightcone coordinate in which any four-vector $v^\mu$ can be decomposed as $v^\mu = (v\cdot \overline{\beta})\beta^\mu+(v\cdot\beta)\overline{\beta}^\mu+(v\cdot \beta_\perp)\beta_\perp^\mu$, with $\beta^\mu$ a lightlike unit vector in the $p$-direction, i.e. $\beta^2=0$, $p\cdot \overline{\beta}\sim Q$, $p\cdot \beta\sim \lambda Q$, and $p\cdot \beta_\perp\sim \lambda^{1/2}Q$.

In figure~\ref{figure-problem_to_study_mass}, all the edges in the (upper) double line are of the mass $M$, all the edges in the (lower) single line are of the mass $m$, while all the remaining dashed edges are massless. For convenience, let us refer to the union of the mass-$M$ edges as the \emph{$M$ boundary} of the entire Feynman graph $G$ and the union of the mass-$m$ edges as the \emph{$m$ boundary} of $G$.

In a given region, each line momentum has a specific scaling, and we call its corresponding edge in a certain \emph{mode}. For example, any edge carrying the momentum $l_H^\mu\sim Q(1,1,1)$ is in the hard mode, any edge carrying $l_C^\mu\sim Q(1,\lambda,\lambda^{1/2})$ is in the collinear mode, and any edge carrying $l_S^\mu\sim Q(\lambda,\lambda,\lambda)$ is in the soft mode. These three modes are named in compatible with sections~\ref{section-regions_onshell} and \ref{section-regions_soft_expansion}, where we stated that all the regions for the on-shell expansion and the soft expansion involve exclusively the hard, collinear, and soft modes.

For the mass expansion, the mode structure can be much richer. As we will see in the upcoming subsections, all of the following modes can be relevant for the regions of a given graph.
\begin{itemize}
    \item The hard mode: $l_H^\mu\sim Q(1,1,1)$.
    \item The collinear mode: $l_C^\mu \sim Q (1, \lambda, \lambda^{1/2})$.
    \item The soft mode: $l_S^\mu\sim Q (\lambda, \lambda, \lambda)$.
    \item The collinear$\cdot$soft mode: $l_{C\cdot S}^\mu\sim Q (\lambda, \lambda^{2}, \lambda^{3/2})$.
    \item The semihard mode: $l_{sH}^\mu\sim Q (\lambda^{1/2}, \lambda^{1/2}, \lambda^{1/2})$.
    \item The semihard$\cdot$collinear mode: $l_{sH\cdot C}^\mu\sim Q (\lambda^{1/2}, \lambda^{3/2}, \lambda)$.
    \item The semihard$\cdot$soft mode: $l_{sH\cdot S}^\mu\sim Q (\lambda^{3/2}, \lambda^{3/2}, \lambda^{3/2})$.
    \item The semihard$\cdot$collinear$\cdot$soft mode: $l_{sH\cdot C\cdot S}^\mu\sim Q (\lambda^{3/2}, \lambda^{5/2}, \lambda^2)$.
    \item The semicollinear mode: $l_{sC}^\mu\sim Q (1, \lambda^{1/2}, \lambda^{1/4})$.
    \item The semihard$\cdot$semicollinear mode: $l_{sH\cdot sC}^\mu\sim Q (\lambda^{1/2}, \lambda, \lambda^{3/4})$.
    \item[]\dots
\end{itemize}
It is worth noticing that some of these modes have been observed, and their contributions to the MoR have been evaluated, for example, in ref.~\cite{Smn99}. They encompass the hard, semihard, soft, collinear, soft$\cdot$collinear, soft${}^2\cdot$collinear, and so on, as proposed in ref.~\cite{Smn02book}. As we will demonstrate here, at the level of three loops or more, additional modes such as semicollinear, semihard$\cdot$semicollinear, etc., can also be relevant.

In addition to identifying the modes of a generic region $R$, we will propose a set of constraints describing how the subgraphs of $G$ connect to each other in $R$. This not only allows for the immediate exclusion of configurations leading to scaleless integrals from the list of regions, but also has the potential to aid in the construction of a graph-finding algorithm for obtaining the regions without constructing the Newton polytopes.

In section~\ref{section-modes_in_mass_expansions}, we will begin with some examples featuring representatives of the modes mentioned above. Inspired by these examples, we classify the regions into certain types and propose the general prescription for each type in section~\ref{section-characterizing_regions}. Particularly for planar graphs, the prescription is equivalent to a ``terrace formalism'', revealing the regions through the construction of valid ``terraces'', as demonstrated in section~\ref{section-formalism_describing_regions_planar_graphs}.

\subsection{Modes in the mass expansions}
\label{section-modes_in_mass_expansions}

We shall soon delve into specific examples to deduce the relevant modes in each given region. A primary clue for this lies in the relation between the scaling of the momentum off-shellness and the scaling of the Lee-Pomeransky parameter, as expressed in eq.~(\ref{eq:LP_offshellness_relation}). To elaborate, an entry of $-a$ in the region vector implies that the corresponding line momentum satisfies $l_e^2 - m_e^2 \sim \lambda^a Q$. In the context of the on-shell expansion and the soft expansion, which we have previously examined for massless graphs, the mode of each edge is determined once the off-shellness (equal to the virtuality) of its line momentum is known. This is because the hard, collinear, and soft modes, corresponding to the entries with \emph{distinct} values $0$, $-1$, and $-2$ in a given region vector, constitute a complete set of modes relevant to the expansion.

However, in the case of the mass expansion we are currently exploring, where the mode structure can be more intricate, relying solely on Eq.~(\ref{eq:LP_offshellness_relation}) may not suffice. For instance, besides the collinear mode, an entry of $-1$ in the region vector may also correspond to the semihard mode. In such cases, we must employ additional tools like the momentum-conservation law, the Landau equations, and so forth, to determine each mode, as we will elucidate below.

\begin{figure}[t]
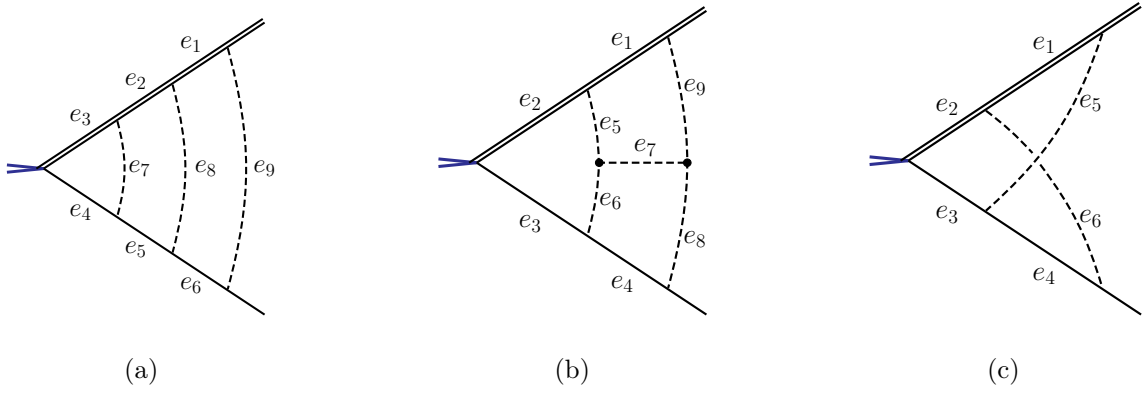

\centering
\begin{subfigure}[b]{0.25\textwidth}
\centering
\include{figs/mass_expansion_example_graph1}
\vspace{-2em}
\caption{}
\label{mass_expansion_example_graph1}
\end{subfigure}
\hfill
\begin{subfigure}[b]{0.25\textwidth}
\centering
\include{figs/mass_expansion_example_graph2}
\vspace{-2em}
\caption{}
\label{mass_expansion_example_graph2}
\end{subfigure}
\hfill
\begin{subfigure}[b]{0.25\textwidth}
\centering
\include{figs/mass_expansion_example_graph3}
\vspace{-2em}
\caption{}
\label{mass_expansion_example_graph3}
\end{subfigure}
\caption{Three examples of the mass expansion in the Sudakov limit, where (a) and (b) are three-loop planar graphs, and (c) is a two-loop nonplanar graph.}
\label{figure-mass_expansion_example_graphs}
\end{figure}

To begin with, let us consider the region vectors corresponding to figures~\ref{mass_expansion_example_graph1}-\ref{mass_expansion_example_graph3}, as listed in table~\ref{table-mass_expansion_example_graphs_regions}. As is manifested, we have classified these region vectors into four sets $S_1$, $S_2$, $S_3$, and $S_4$. The reason for this classification will soon be seen from the interpretation of these regions in momentum space.
\begin{table}[t]
\begin{center}
\begin{tabular}{ |c||c|c|c| } 
\hline
    & figure~\ref{mass_expansion_example_graph1} & figure~\ref{mass_expansion_example_graph2} & figure~\ref{mass_expansion_example_graph3} \\ \hline
    $S_1$ & \textsf{(0,0,0,0,0,0,0,0,0;1)} & \textsf{(0,0,0,0,0,0,0,0,0;1)} & \textsf{(0,0,0,0,0,0;1)} \\ \hline
    \multirow{9}{1em}{$S_2$} & \textsf{(0,0,0,0,0,-1,0,0,-1;1)} & \textsf{(0,0,0,-1,0,0,0,-1,0;1)} & \textsf{(0,0,0,-1,-1,0;1)} \\ 
     & \textsf{(0,0,0,0,-1,-1,0,-1,-1;1)} & \textsf{(0,0,0,-1,0,0,-1,-1,-1;1)} & \textsf{(0,0,-1,-1,-1,-1;1)} \\ 
     & \textsf{(0,0,0,-1,-1,-1,-1,-1,-1;1)} & \textsf{(0,0,-1,-1,-1,0,0,-1,0;1)} & \textsf{(-1,0,-1,-1,-1,-2;1)} \\ 
     & \textsf{(-1,0,0,0,-1,-1,0,-1,-2;1)} & \textsf{(0,0,-1,-1,-1,-1,-1,-1,-1;1)} & \\ 
     & \textsf{(-1,0,0,-1,-1,-1,-1,-1,-2;1)} & \textsf{(-1,0,0,-1,0,0,-1,-1,-2;1)} & \\
     & \textsf{(-1,-1,0,-1,-1,-1,-1,-2,-2;1)} & \textsf{(-1,0,-1,-1,-1,-1,-1,-1,-2;1)} & \\ 
     & \textsf{(-1,-1,0,-1,-1,-2,-1,-2,-3;1)} & \textsf{(-1,0,-1,-1,-1,-1,-2,-2,-2;1)} & \\
     &  & \textsf{(-1,-1,-1,-1,-1,-2,-1,-1,-2;1)} & \\
     &  & \textsf{(-1,0,-1,-2,-1,-1,-2,-3,-2;1)} & \\ \hline
    \multirow{4}{1em}{$S_3$} & \textsf{(0,0,0,0,-1,0,0,0,0;1)} & \textsf{(0,0,-1,0,0,0,0,0,0;1)} & \textsf{(0,0,-1,0,0,0;1)} \\
     & \textsf{(0,0,0,-1,0,0,0,0,0;1)} & & \\
     & \textsf{(0,0,0,-1,0,-1,0,0,-1;1)} & & \\
     & \textsf{(0,0,0,-1,-1,0,-1,0,0;1)} & & \\ \hline 
    \text{$S_4$} & & & \textsf{(-}$\frac{\textsf{1}}{\textsf{2}}$\textsf{,0,-1,-1,-1,-}$\frac{\textsf{1}}{\textsf{2}}$\textsf{;1)}\\
    \hline
\end{tabular}
\captionsetup{justification=centering}
\caption{The region vectors corresponding to figures~\ref{mass_expansion_example_graph1}, \ref{mass_expansion_example_graph2}, and~\ref{mass_expansion_example_graph3} respectively, which are classified into the sets $S_1$, $S_2$, $S_3$, and $S_4$ based on their associated modes.}
\label{table-mass_expansion_example_graphs_regions}
\end{center}
\end{table}

First, for each vector in $S_1$, the first $N$ entries (recall that $N$ is the number of edges) are exactly $0$, which implies that the off-shellness of each propagator is $\mathcal{O}(Q^2\lambda^0) = \mathcal{O}(Q^2)$. In other words, each edges is in the \emph{hard mode} ($H$), with the scaling of its line momentum
\begin{eqnarray}
\label{eq:mass_expansion_hard_mode}
l_H^\mu \sim Q(1,1,1).
\end{eqnarray}
The corresponding region is called the hard region. We emphasize that the hard regions here are identical to those in the on-shell expansion and the soft expansion.

\begin{figure}[t]
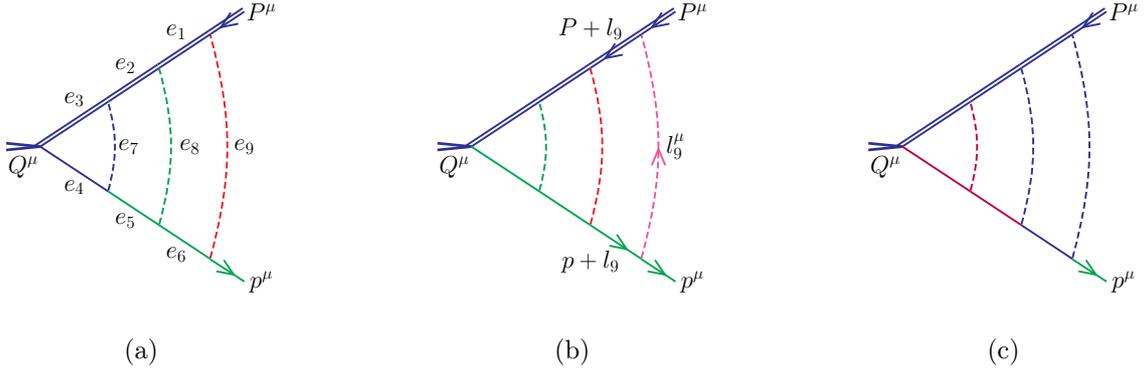

\centering
\begin{subfigure}[b]{0.25\textwidth}
\centering
\include{figs/mass_expansion_example_region1_momentum}
\vspace{-2em}
\caption{}
\label{mass_expansion_example_region1_momentum}
\end{subfigure}
\hfill
\begin{subfigure}[b]{0.25\textwidth}
\centering
\include{figs/mass_expansion_example_region2_momentum}
\vspace{-2em}
\caption{}
\label{mass_expansion_example_region2_momentum}
\end{subfigure}
\hfill
\begin{subfigure}[b]{0.25\textwidth}
\centering
\include{figs/mass_expansion_example_region3_momentum}
\vspace{-2em}
\caption{}
\label{mass_expansion_example_region3_momentum}
\end{subfigure}
\caption{The regions corresponding to some vectors in table~\ref{table-mass_expansion_example_graphs_regions}, where we color every hard-mode edge in \textbf{\color{Blue}blue}, every collinear-mode edge in \textbf{\color{Green}green}, every soft-mode edge in \textbf{\color{Red}red}, every soft$\cdot$collinear edge in \textbf{\color{Rhodamine}rhodamine}, and every semihard edge in \textbf{\color{Purple}purple}. (a): the region associated with $\textsf{(-1,0,0,0,-1,-1,0,-1,-2;1)}$. (b): the region associated with $\textsf{(-1,-1,0,-1,-1,-2,-1,-2,-3;1)}$. (c): the region associated with $\textsf{(0,0,0,-1,-1,0,-1,0,0;1)}$.}
\label{figure-mass_expansion_example_regions_momentum}
\end{figure}

Next, we examine the vectors in $S_2$. To start with, we specifically focus on one subset of $S_2$ whose vectors have entries $0$, $-1$, and $-2$ only. We assert that the edges associated with $0$ are in the hard mode, as defined above; the edges associated with $-1$ and $-2$ are in the \emph{collinear mode} ($C$) and the \emph{soft mode} ($S$) respectively, whose associated momenta scale as:
\begin{eqnarray}
\label{eq:mass_expansion_collinear_and_soft_mode}
    l_C^\mu\sim Q(1,\lambda,\lambda^{1/2}),\qquad l_S^\mu\sim Q(\lambda,\lambda,\lambda).
\end{eqnarray}
One example is the vector $\textsf{(-1,0,0,0,-1,-1,0,-1,-2;1)}$ in the first column of table~\ref{table-mass_expansion_example_graphs_regions}, whose corresponding region is depicted in figure~\ref{mass_expansion_example_region1_momentum} with the hard, collinear, and soft modes colored in blue, green, and red, respectively. It is straightforward to check that this assignment of the modes is consistent with the scalings of the Lee-Pomeransky parameters and momentum conservation. Furthermore, each region follows the same picture as a pinch surface of $G|_{p^2=m^2=0}$.

For those vectors in $S_2$ including the entry $-3$, one example is $\textsf{(-1,-1,0,-1,-1,-2,-1,-2,-3;1)}$ which is in the first column of table~\ref{table-mass_expansion_example_graphs_regions}. The entry $-3$ corresponds to a massless edge $e_9$. In momentum space, it implies that the line momentum $l_9^\mu$ is in the \emph{collinear$\cdot$soft mode} ($CS$), with the following scaling:
\begin{eqnarray}
\label{eq:mass_expansion_collinearsoft_mode}
    l_{CS}^\mu \sim Q (\lambda, \lambda^{2}, \lambda^{3/2}).
\end{eqnarray}
It is straightforward to check that with this interpretation, the off-shellnesses of all the other line momenta are consistent with the vector entries. For example, part of the momentum flow corresponding to the region vector $\textsf{(-1,-1,0,-1,-1,-2,-1,-2,-3;1)}$ in the first column of table~\ref{table-mass_expansion_example_graphs_regions} is shown in figure~\ref{mass_expansion_example_region2_momentum}. The off-shellnesses of $l_1^\mu$ and $l_6^\mu$ are, respectively,
\begin{subequations}
\label{eq:mass_expansion_examples123_collinearsoft_offshellness}
    \begin{align}
        &l_1^2-M^2 = (P+l_9)^2-M^2 = 2P\cdot l_9 +l_9^2 \sim \lambda Q^2;
        \label{eq:mass_expansion_examples123_collinearsoft_offshellness_e1}\\
        &l_6^2-m^2 = (p+l_9)^2-m^2 = 2p\cdot l_9 +l_9^2 \sim \lambda^2 Q^2.
        \label{eq:mass_expansion_examples123_collinearsoft_offshellness_e6}
    \end{align}
\end{subequations}
They align with the entries $-1$ and $-2$ of the region vector. Thus, we have completed the interpretation of all vectors in $S_2$ in momentum space as described above.

Moving on, we examine the vectors in $S_3$. Let us take $\textsf{(0,0,0,0,-1,0,0,0,0;1)}$ from the first column of table~\ref{table-mass_expansion_example_graphs_regions} as one example, where the momentum off-shellness associated with $e_5$ is $\mathcal{O}(\lambda Q^2)$, while those associated with the other edges are $\mathcal{O}(Q^2)$. In momentum space, this implies that $e_5$ is in the \emph{semihard mode} ($sH$):
\begin{eqnarray}
\label{eq:mass_expansion_semihard_mode}
    l_{sH}^\mu\sim Q (\lambda^{1/2}, \lambda^{1/2}, \lambda^{1/2}).
\end{eqnarray}
Meanwhile, all the other edges are in the hard mode. In this configuration, the semihard edge is \emph{exclusively} adjacent to the hard subgraph. Note that the scaling of a semihard momentum is exactly the geometric mean of the scalings of a hard momentum and a soft momentum.

In fact, at higher-loop orders, the semihard, hard, collinear, and soft modes can be simultaneously present in a region, as one can see from some examples in appendix~\ref{appendix-mass_expansion_examples}. It is worth noting that in Soft-Collinear Effective Theory (SCET)~\cite{BurStw13lectures, BchBrgFrl15book, Bch18lectures, BurFlmLk00, BurFlmPjlStw01, BurPjlSwt02-1, BurPjlSwt02-2,BnkChpkDhlFdm02,BnkFdm03,RthstStw16EFT}, the soft mode is a feature of the SCET$_\text{I}$ type theories while the semihard mode is a feature of the SCET$_\text{II}$ type theories. From the observation above, both SCET$_\text{I}$ and SCET$_\text{II}$ can be relevant in the mass expansion of the heavy-to-light decays. This is also consistent with the region analysis in specific calculations, as demonstrated in, for example, ref.~\cite{EglGndgSgnUrch19}, where the two-loop contribution to the heavy-to-light decays has been evaluated using the MoR.

One may wonder whether $e_5$ can be interpreted as a collinear-mode edge instead, since such an assignment of the modes is still consistent with both eq.~(\ref{eq:LP_offshellness_relation}) and momentum conservation. However, the resulting configuration does not conform with any pinch surfaces of $G|_{p^2=m^2=0}$. In other words, as $m^2/M^2\to 0$, no pinch surfaces can have a collinear subgraph adjacent solely to the hard subgraph and attached by no external momenta.\footnote{One can see this from the Coleman-Norton interpretation, which leads to $\alpha_5 l_5^\mu +\sum_{i=2,7,8}\alpha_i l_i^\mu = 0$ for this example, where $\alpha_i$ is the Feynman parameter associated with $e_i$. If $l_5^\mu$ is lightlike in one direction, then in the limit $m^2/M^2\to 0$, the second term vanishes for all $\mu$, while the first term is $\mathcal{O}(Q)$ when $\mu$ corresponds to the large component of $l_5$. Thus, the configuration with $l_5^\mu$ in the collinear mode and all the other $l_i^\mu$ in the hard mode does not conform with any solution of the Landau equations.} As a consequence, for the vectors in $S_3$, it is the semihard mode rather than the collinear mode that is exclusively adjacent to the hard subgraph.

As another example, the region vector $\textsf{(0,0,0,-1,-1,0,-1,0,0;1)}$ for figure~\ref{mass_expansion_example_graph1} corresponds to the region where the momenta $e_4$, $e_5$, and $e_7$ are simultaneously in the semihard mode while all the other line momenta are in the hard mode, as depicted in figure~\ref{mass_expansion_example_region3_momentum}.

Finally, we consider the set $S_4$, which is empty for figures~\ref{mass_expansion_example_graph1} and \ref{mass_expansion_example_graph2}, and consists of exactly one vector \textsf{(-}$\frac{\textsf{1}}{\textsf{2}}$\textsf{,0,-1,-1,-1,-}$\frac{\textsf{1}}{\textsf{2}}$\textsf{,1)} for figure~\ref{mass_expansion_example_graph3}. Similar to our analysis for the $S_3$ vectors, here the edges $e_3$, $e_4$, and $e_5$ are all in the semihard mode, which induces the following scaling and off-shellnesses of the other three line momenta:
\begin{subequations}
    \begin{align}
        &l_1^\mu = (l_5+P)^\mu \sim Q(1,1,1);\hspace{0.9cm}\quad l_1^2 - M^2 = 2P\cdot l_5 + l_5^2 \sim \lambda^{1/2} Q^2;
        \label{eq:mass_expansion_examples3_semicollinear_edge1_offshellness}\\
        &l_6^\mu = (l_4+p)^\mu \sim Q(1,\lambda^{1/2},\lambda^{1/2});\quad l_6^2 \sim \lambda^{1/2} Q^2;
        \label{eq:mass_expansion_examples3_semicollinear_edge6_offshellness}\\
        &l_2^\mu = (l_1+l_6)^\mu \sim Q(1,1,1);\hspace{0.9cm}\quad l_2^2 - M^2 = 2P\cdot (l_5+l_6) + (l_5+l_6)^2 \sim \lambda^0 Q^2.
        \label{eq:mass_expansion_examples3_semicollinear_edge2_offshellness}
    \end{align}
\end{subequations}
In the calculation in eq.~(\ref{eq:mass_expansion_examples3_semicollinear_edge2_offshellness}) above, we have used the relation $l_2^\mu = (l_1+l_6)^\mu = (P+l_5+l_6)^\mu$, so $l_2^2-M^2 = 2P\cdot (l_5+l_6) + (l_5+l_6)^2$. From above, the momentum off-shellnesses agree with the scaling of the Lee-Pomeransky parameters. The edges $e_1$ and $e_2$ are both in the hard mode, while $e_6$ is in the \emph{semicollinear mode} ($sC$), whose line momentum is defined as
\begin{eqnarray}
\label{eq:mass_expansion_semicollinear_mode}
    l_{sC}^\mu\sim Q(1,\lambda^{1/2},\lambda^{1/4}).
\end{eqnarray}
It then follows that, the scaling of a softcollinear momentum is the geometric mean of the scalings of a collinear momentum and a hard momentum. Note that this definition slightly differs from eq.~(\ref{eq:mass_expansion_examples3_semicollinear_edge6_offshellness}), as the magnitude of $l_{6}^\perp$ is $\mathcal{O}(Q\lambda^{1/2})$ instead of $\mathcal{O}(Q\lambda^{1/4})$. This smallness of $l_{6}^\perp$ is accidental: momentum conservation fixes $l_{6}^\perp= l_4^\perp + p^\perp$, thus $l_4^\perp\sim p^\perp\sim Q\lambda^{1/2}$ here. If there is an independent semicollinear loop instead, its scaling would be identical to eq.~(\ref{eq:mass_expansion_semicollinear_mode}). We will see such an example below.

Let us examine a few more examples with semicollinear-mode edges at higher-loop orders. First, we consider the region vector
\begin{eqnarray}
\label{eq:mass_expansion_example4_region_vector}
    (-\frac{1}{2}, -\frac{1}{2}, 0, 0, -1, -1, -\frac{1}{2}, -\frac{3}{2}, -1, -\frac{1}{2}, -\frac{1}{2}, -2; 1),
\end{eqnarray}
which is from the mass expansion of a four-loop graph (figure~\ref{mass_expansion_examples_nonplanar_1}) in appendix~\ref{appendix-mass_expansion_examples}. The edges $e_{1,2,3,4}$ are in the hard mode, $e_{5,6,9}$ are in the semihard mode, $e_{7,10,11}$ are in the semicollinear mode, $e_8$ is in the collinear mode, while $e_{12}$ is in the \emph{semihard$\cdot$collinear mode} ($sH C$), which in general is defined as
\begin{eqnarray}
\label{eq:mass_expansion_semihard_times_collinear_mode}
    l_{sH C}^\mu \sim Q(\lambda^{1/2},\lambda^{3/2},\lambda^1).
\end{eqnarray}
It follows that $l_{sH C}^2\sim \lambda^2 Q^2$, $(P+l_{sH C})^2-M^2 \sim \lambda^{1/2} Q^2$, and $(p+l_{sH C})^2-m^2 \sim \lambda^{3/2} Q^2$. The scaling successfully explains the $1^{\text{st}}$, $8^{\text{th}}$, and $12^{\text{th}}$ entries of the region vector above, which are $-\frac{1}{2}$, $-\frac{3}{2}$, and $-2$, respectively.

Let us leave two short comments regarding the modes in the example above. First, the assignment of modes above is the only way of simultaneously being consistent with the parameter scaling, eq.~(\ref{eq:LP_offshellness_relation}), and the conservation of momenta. Second, as there exists a semicollinear loop, the scaling of this loop momentum follows precisely eq.~(\ref{eq:mass_expansion_semicollinear_mode}), in contrast to the case of a single semicollinear edge in eq.~(\ref{eq:mass_expansion_examples3_semicollinear_edge6_offshellness}).

More modes can emerge as the loop number increases. Let us consider a five-loop graph, figure~\ref{mass_expansion_examples_nonplanar_5} in appendix~\ref{appendix-mass_expansion_examples}. As has been manifested, there are 104 regions in total for this graph, one of which is 
\begin{eqnarray}
\label{eq:mass_expansion_example5_region_vector}
    (-\frac{3}{2}, -\frac{1}{2}, -\frac{1}{2}, 0, 0, -1, -1, -1, -\frac{3}{2}, -\frac{3}{2}, -1, -\frac{1}{2}, -1, -2, -3 ; 1).
\end{eqnarray}
In the momentum configuration of this region, the edges $e_{1,2,3,4,5}$ are in the hard mode, $e_{6,7,11}$ are in the semihard mode, $e_{8,9,10,13}$ are in the collinear mode, $e_{11}$ is in the semicollinear mode, $e_{14}$ is in the semihard$\cdot$collinear mode, while $e_{15}$ is in the \emph{semihard$\cdot$soft mode} ($sHS$), the scaling of whose momentum is defined as:
\begin{eqnarray}
\label{eq:mass_expansion_semihard_times_soft_mode}
    l_{sH S}^\mu \sim Q(\lambda^{3/2},\lambda^{3/2},\lambda^{3/2}).
\end{eqnarray}
We then have $l_{sH S}^2\sim \lambda^3 Q^2$, $(P+l_{sH S})^2 -M^2 \sim \lambda^{3/2} Q^2$, and $(p+l_{sH S})^2 -m^2 \sim \lambda^{3/2} Q^2$. The scaling explains that the $1^{\text{st}}$, $10^{\text{th}}$, and $15^{\text{th}}$ entries of the region vector are $-\frac{3}{2}$, $-\frac{3}{2}$, and $-3$, respectively.

Another region vector of the same graph (figure~\ref{mass_expansion_examples_nonplanar_5}) is
\begin{eqnarray}
\label{eq:mass_expansion_example6_region_vector}
    (-\frac{1}{2}, -\frac{1}{2}, -\frac{1}{2}, 0, 0, -1, -1, -1, -1, -\frac{3}{2}, -1, -\frac{1}{2}, -1, -\frac{3}{2}, -2 ; 1).
\end{eqnarray}
The edges $e_{1,2,3,4,5}$ are in the hard mode, $e_{6,7,11}$ are in the semihard mode, $e_{8,9,10,13}$ are in the collinear mode, $e_{12}$ is in the semicollinear mode, $e_{15}$ is in the semihard$\cdot$collinear mode, while $e_{14}$ is in the \emph{semihard$\cdot$semicollinear mode} ($sHsC$), which in general is defined as
\begin{eqnarray}
\label{eq:mass_expansion_semihard_times_semicollinear_mode}
    l_{sH sC}^\mu \sim Q(\lambda^{1/2},\lambda^1,\lambda^{3/4}).
\end{eqnarray}
We then have $l_{sH sC}^2\sim \lambda^{3/2} Q^2$ and $(p+l_{sH sC})^2 -m^2 \sim \lambda Q^2$. The scaling explains that the $9^{\text{th}}$ and $14^{\text{th}}$ entries of the region vector are $-1$ and $-\frac{3}{2}$, respectively.

So far, we have delved into a range of examples that have unveiled a remarkably diverse mode structure within a given region. The richness of the modes underscores the complexity of mass expansions. In order to fully characterize the regions, it is worth highlighting two pressing questions that have emerged from our exploration above. First, \emph{how can we establish a general prescription for identifying the modes present in any given graph}? Second, \emph{the subgraphs where specific modes are involved appear to interconnect in certain ways; what constraints govern their connections}? These intriguing inquiries form the focal point of the next subsection, where we will endeavor to provide answers and shed further light on the intricacies of mass expansions.

\subsection{Characterizing the regions}
\label{section-characterizing_regions}

Before characterizing the regions in the mass expansion of a heavy-to-light decay process, we shall classify the regions into certain types. To do this, a few concepts are needed. Recall that the external momenta, $P^\mu$, $Q^\mu$, and $p^\mu$, all have some large components at $\mathcal{O}(Q)$. For any given region that is consistent with momentum conservation, some internal edges of $G$ must also carry momenta with $\mathcal{O}(Q)$ components, and the union of these edges and their endpoints is a subgraph of $G$, connecting all the three external momenta. We denote this subgraph as $\Gamma_\text{LMF}^{}$, where ``LMF'' refers to the \emph{large momentum flow} associated with the region.

\begin{figure}[t]
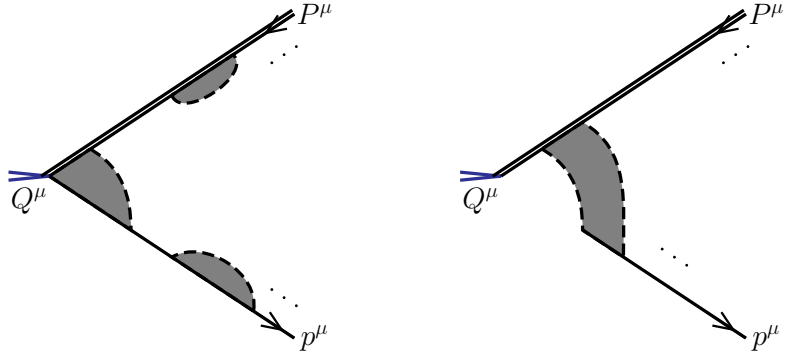

\centering
\begin{subfigure}[b]{0.3\textwidth}
\centering
\include{figs/mass_large_momentum_flow_typeI}
\vspace{-2em}
\caption{The LMF including all the massive edges.}
\label{mass_large_momentum_flow_typeI}
\end{subfigure}
\hspace{3em}
\begin{subfigure}[b]{0.3\textwidth}
\centering
\include{figs/mass_large_momentum_flow_typeII}
\vspace{-2em}
\caption{The LMF excluding some massive edges.}
\label{mass_large_momentum_flow_typeII}
\end{subfigure}
\caption{The generic configuration of the LMF in the regions of the mass expansion. Those edges with momenta not having $\mathcal{O}(Q)$ components are not shown in the figure.}
\label{figure-mass_large_momentum_flow_types}
\end{figure}

As we summarize from the examples in section~\ref{section-modes_in_mass_expansions} and appendix~\ref{appendix-mass_expansion_examples}, as well as those not presented in this manuscript, the generic $\Gamma_\text{LMF}^{}$ can be depicted by either configuration in figure~\ref{figure-mass_large_momentum_flow_types}. In figures~\ref{mass_large_momentum_flow_typeI}, \emph{all} the massive edges, and possibly some massless edges, are included in $\Gamma_\text{LMF}^{}$. In contrast, in figures~\ref{mass_large_momentum_flow_typeII}, some massive edges are excluded from $\Gamma_\text{LMF}^{}$. Based on this difference, we classify the regions in mass expansion into two types: a region is called a \emph{type-I region} if its associated $\Gamma_\text{LMF}^{}$ contains all the massive edges (figures~\ref{mass_large_momentum_flow_typeI}); it is called a \emph{type-II region} if some massive edges are not in its associated $\Gamma_\text{LMF}^{}$ (figures~\ref{mass_large_momentum_flow_typeII}).

Each large momentum flow region can be envisioned with the following phenomenological picture. The heavy particle carries the momentum $P^\mu$ and, at some point(s), emits an off-shell momentum $Q^\mu$. The remaining momentum $p^\mu$ enters the light particle, possibly transferred through some massless gauge bosons. From this picture, it is natural to define the \emph{$p$ branch} of $\Gamma_\text{LMF}^{}$, which includes:
\begin{itemize}
    \item the external momentum $p^\mu$;
    \item all the mass-$m$ and massless edges in $\Gamma_\text{LMF}^{}$, and all the mass-$M$ edges in $\Gamma_\text{LMF}^{}$ that are adjacent to two massless edges in $\Gamma_\text{LMF}^{}$;
    \item all the endpoints of the edges above.
\end{itemize} 
The graph obtained by removing the $p$ branch from $\Gamma_\text{LMF}^{}$ has two disconnected components, which are attached by $P^\mu$ and $Q^\mu$ respectively. Accordingly, we refer to them as the \emph{$P$ branch} and \emph{$Q$ branch} of $\Gamma_\text{LMF}^{}$.

A key feature is that type-II regions contain the semihard-mode edges, while type-I regions do not. Furthermore, within a type-II region, all mass-$m$ edges not belonging to $\Gamma_\text{LMF}^{}$ are in the semihard mode. For each type-II region, we define its semihard subgraph $\Gamma_{sH}$ to encompass all semihard-mode edges and the vertices incident only with semihard-mode edges. Further, we introduce the subgraph $\widehat{\Gamma}_{sH}$ as the union of those components of $\Gamma_{sH}$, each of which contains some mass-$m$ edges. In general, $\widehat{\Gamma}_{sH}$ may consist of multiple connected components. One example is the region vector \textsf{(0,0,0,0,0,0,-1,0,-1,0,0,0,0,0,0;1)} corresponding to figure~\ref{mass_expansion_examples_planar_1}, as presented in appendix~\ref{appendix-mass_expansion_examples}.

With these concepts, type-II regions can be further classified based on the connection between $\widehat{\Gamma}_{sH}$ and $\Gamma_\text{LMF}^{}$. For each type-II region, if $\widehat{\Gamma}_{sH}$ is not adjacent to the $P$ branch of $\Gamma_\text{LMF}^{}$, we refer to it as a \emph{type-IIA region}; otherwise ($\widehat{\Gamma}_{sH}$ is adjacent to the $P$ branch), we refer to it as a \emph{type-IIB region}. From this definition, type-IIB regions occur only for nonplanar graphs. We will demonstrate this through some examples later.

Revisiting table~\ref{table-mass_expansion_example_graphs_regions}, all the regions in the sets $S_1$ and $S_2$ are type-I regions, all the regions in $S_3$ are type-IIA regions, while the single region in $S_4$ is a type-IIB region. Three additional examples of type-IIB regions are presented in (\ref{eq:mass_expansion_example4_region_vector}), (\ref{eq:mass_expansion_example5_region_vector}), and (\ref{eq:mass_expansion_example6_region_vector}), which have been analyzed above. To summarize from these examples, the absence of the semihard mode is a feature of type-I regions, the presence of the semihard mode together with the absence of the semicollinear mode is a feature of type-IIA regions, and the presence of both the semihard and semicollinear modes, and possibly the semihard$\cdot$collinear, semihard$\cdot$soft modes, etc., is a feature of type-IIB regions.

\subsubsection{Some key concepts}
\label{section-some_key_concepts}

Before investigating more structures about the type-I, type-IIA, and type-IIB regions, let us introduce some key concepts regarding the subgraphs of $G$ that are associated with certain modes. These concepts will be repeatedly used when we characterize a generic region with its relevant modes.

First, we define the \emph{on-shellness degree} (OSD) of an $e$ as the number $b$, where $\lambda^b Q^2$ is the scaling of its momentum off-shellness. In other words, the closer $l_e^\mu$ is to its mass shell, the larger $b$ it has. It then follows that for $e$ being massless, its OSD is $0$ if it is hard, its OSD is $1$ if it is collinear, and so on. For $e$ being massive, its OSD is then linked to the scaling of $(P+l_e)^2-M^2 = 2P\cdot l_e$ or $(p+l_e)^2-m^2 = 2p\cdot l_e$. For example, the OSDs of $e_1$ and $e_6$ in figure~\ref{mass_expansion_example_region2_momentum} are $1$ and $2$ respectively, as a result of eq.~(\ref{eq:mass_expansion_examples123_collinearsoft_offshellness}).

We next develop the concept of mode sensitive subgraphs of $G$. To begin with, we define the \emph{hard-sensitive subgraph} $\gamma_H^{}$ to consist of those edges whose off-shellnesses are $\mathcal{O}(Q^2)$ and all their endpoints. In particular, the vertex $O$, through which the momentum $Q^\mu$ enters, is always considered to belong to $\gamma_H^{}$.

For each other mode $X\neq H$, we define the associated \emph{$X$-sensitive subgraph}, $\gamma_{X}^{}$, to include the following elements:
\begin{itemize}
    \item the massless edges whose momenta are in the $X$ mode,
    \item the massive edges of $G\setminus \gamma_H^{}$, whose momenta are in the form of either $(P+l_{X})^\mu$ or $(p+l_{X})^\mu$, with $l_{X}^\mu$ representing a momentum in the $X$ mode,
    \item the vertices $v$ of $G\setminus \gamma_H^{}$ such that, for all the edges incident with $v$, those with the smallest OSD all belong to $\gamma_{X}^{}$.
\end{itemize}
This concept allows us to partition the entire graph $G$ into distinct mode sensitive subgraphs according to the mode structure in any given region. As one example, for the (type-I) region in figure~\ref{mass_expansion_example_region1_momentum}, the hard sensitive subgraph $\gamma_H^{}$ includes $e_2$, $e_3$, $e_4$, $e_7$, and all their endpoints; the collinear sensitive subgraph $\gamma_C^{}$ includes $e_5$, $e_8$, and the vertex incident with both $e_5$ and $e_8$; the soft sensitive subgraph $\gamma_S^{}$ includes all the remaining vertices and edges. As another example, the region \textsf{(-1,0,-1,-2,-1,-1,-2,-3,-2;1)} of figure~\ref{mass_expansion_example_graph2} is:
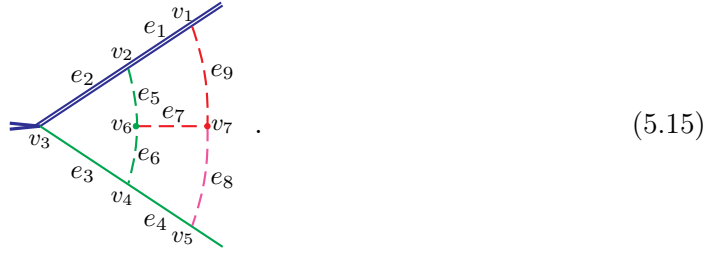
\begin{equation}
\begin{aligned}
\begin{tikzpicture}[baseline=11ex, scale =0.4]
\draw [very thick, Blue] (1,5.1) edge (2,5) node [] {};
\draw [very thick, Blue] (1,4.9) edge (2,5) node [] {};
\draw [thick, Blue] (1.8,5) edge (7.95,9.1) node [] {};
\draw [thick, Blue] (2,5) edge (8,9) node [] {};
\draw [thick, Green] (2,5) edge (8,1) node [] {};
\draw (4.9,6.9) edge [thick, Green, dash pattern=on 2mm off 1mm, bend left = 15] (4.9,3.1) {};
\draw (7,8.3) edge [thick, Red, dash pattern=on 2mm off 1mm, bend left = 10] (7.5,5) {};
\draw (7.5,5) edge [thick, Rhodamine, dash pattern=on 2mm off 1mm, bend left = 10] (7,1.7) {};
\draw (5.15,5) edge [thick, Red, dash pattern=on 2mm off 1mm] (7.5,5) {};
\node [draw, Green, circle, minimum size=2pt, fill, inner sep=0pt, outer sep=0pt] () at (5.15,5) {};
\node [draw, Red, circle, minimum size=2pt, fill, inner sep=0pt, outer sep=0pt] () at (7.5,5) {};
\node () at (5.8,8.2) {\small $e_1$};
\node () at (3.4,6.6) {\small $e_2$};
\node () at (5.8,1.8) {\small $e_4$};
\node () at (3.4,3.4) {\small $e_3$};
\node () at (5.6,6) {\small $e_5$};
\node () at (5.6,4) {\small $e_6$};
\node () at (6.4,5.4) {\small $e_7$};
\node () at (8,3.2) {\small $e_8$};
\node () at (8,6.8) {\small $e_9$};
\node () at (6.7,8.7) {\footnotesize $v_1$};
\node () at (4.7,7.4) {\footnotesize $v_2$};
\node () at (2,4.5) {\footnotesize $v_3$};
\node () at (4.7,2.6) {\footnotesize $v_4$};
\node () at (6.7,1.3) {\footnotesize $v_5$};
\node () at (4.7,5) {\footnotesize $v_6$};
\node () at (8,5) {\footnotesize $v_7$};
\end{tikzpicture}
\end{aligned}\ .
\end{equation}
In particular, note that the edges $e_7$ and $e_9$ are in the soft mode, while $e_8$ is in the soft$\cdot$collinear mode. The mode sensitive graphs are then:
\begin{align}
    \begin{split}
        &\gamma_H^{} = (\{v_2,v_3\},\{e_2\});\\
        &\gamma_C^{} = (\{v_4,v_6\},\{e_3,e_5,e_6\});\\
        &\gamma_S^{} = (\{v_1,v_7\},\{e_1,e_7,e_9\});\\
        &\gamma_{CS}^{} = (\{v_5\},\{e_4,e_8\}).
    \end{split}
\end{align}

We note that for each mode $X$, the OSDs of the mass-$M$, mass-$m$, and massless edges are fixed for $\gamma_X^{}$, and $G = \sqcup_{X} \gamma_X^{}$. For any subgraph of $G$, we will use $w^{(M)}$, $w^{(m)}$, and $w^{(0)}$ to denote the largest OSDs of all its mass-$M$, mass-$m$, and massless edges respectively, as we will see later in the text.

With these concepts, we are ready to propose the general prescriptions for the type-I, type-IIA, and type-IIB regions.

\subsubsection{Type-I regions}
\label{section-typeI_regions}

In any type-I region, all the relevant modes are of the form ``collinear${}^{n_1}\cdot$soft${}^{n_2}$'' (denoted as $C^{n_1} S^{n_2}$), with $n_1\in\{0,1\}$ and $n_2\in \mathbb{N} = \{0,1,2,\dots\}$. The scaling of a line momentum in the $C^{n_1} S^{n_2}$ mode is:
\begin{eqnarray}
\label{eq:mass_expansion_collinear_power_n1_soft_power_n2_mode}
    l_{C^{n_1} S^{n_2}}^\mu\sim Q(\lambda^{n_2}, \lambda^{n_1+n_2}, \lambda^{n_1/2+n_2}).
\end{eqnarray}
In particular, $(n_1,n_2)=(0,0)$ corresponds to the hard mode, $(n_1,n_2)=(1,0)$ corresponds to the collinear mode, and $(n_1,n_2)=(0,1)$ corresponds to the soft mode. From above, $l_{C^{n_1} S^{n_2}}^2\sim \lambda^{n_1+2n_2} Q^2$, and we can rename each mode $C^{n_1} S^{n_2}$ as $X_i$ with $i=n_1+2n_2$. In this way, $X_0$ is the hard mode, $X_1$ is the collinear mode, $X_2$ is the soft mode, etc.

\begin{figure}[t]
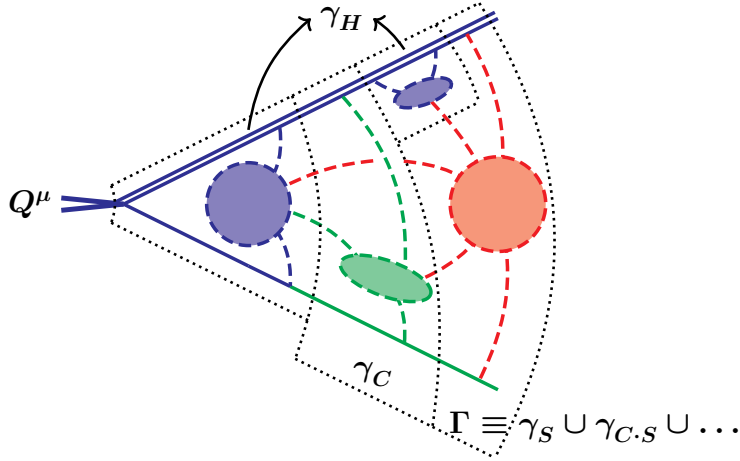

\centering
\include{figs/mass_typeI_general_configuration}
\vspace{-3em}\caption{The general configuration of type-I regions, where certain mode sensitive subgraphs have been encircled by dotted curves. Each dashed line represents an arbitrary number of connections in the general case.}
\label{figure-mass_typeI_general_configuration}
\end{figure}

The general configuration of type-I regions can be depicted by figure~\ref{figure-mass_typeI_general_configuration}, where we have used dotted curves to encircle the subgraphs $\gamma_H^{}$, $\gamma_C^{}$, and $\gamma_{S}^{}\cup \gamma_{C\cdot S}^{}\cup \dots$, respectively. The hard sensitive subgraph $\gamma_H^{}$ always includes the vertex $O$ (which $Q^\mu$ enters), and each component of $\gamma_H^{}$ must contain some mass-$M$ edges. Similarly, each component of the collinear sensitive subgraph $\gamma_{C}^{}$ contains some mass-$m$ edges.

The remaining part of $G$, which is $\Gamma\equiv \gamma_{S}^{}\cup \gamma_{CS}^{}\cup \dots$, is adjacent to $\gamma_C^{}$, containing edges in the soft mode, the collinear$\cdot$soft mode, etc. The configuration of the modes within $\Gamma$ is constrained by three additional requirements, which remove those configurations conforming with figure~\ref{figure-mass_typeI_general_configuration} but leading to scaleless integrals. To explore these requirements, we shall use the following notations in the upcoming text:
\begin{eqnarray}
    G_i \equiv \bigcup_{j=0}^i \gamma_{X_j}^{}.
\end{eqnarray}
Equivalently, $G_{n_1+2n_2} = \gamma_H^{}\cup \gamma_C^{}\cup \dots \cup \gamma_{C^{n_1} S^{n_2}}^{}$ for each $n_1=0,1$ and $n_2\in \mathbb{N}$. In particular, $G_0 = \gamma_H^{}$, $G_1 = \gamma_H^{}\cup \gamma_C^{}$, $G_2 = \gamma_H^{}\cup \gamma_C^{}\cup \gamma_S^{}$, and so on.

\bigbreak
\emph{Requirement 1: for each $i\geqslant 2$, the graph $G_i$ satisfies}
\begin{subequations}
\label{eq:balance_conditions}
    \begin{align}
        &w_i^{(M)} + w_i^{(m)} = w_i^{(0)},\label{eq:balance_condition1}\\
        &w_{i-1}^{(M)} = w_{i}^{(M)}\ \&\ w_{i-1}^{(m)} < w_{i}^{(m)},\quad \text{ or }\quad w_{i-1}^{(M)} < w_{i}^{(M)}\ \&\ w_{i-1}^{(m)} = w_{i}^{(m)},\label{eq:balance_condition2}
    \end{align}
\end{subequations}
\emph{where $w_i^{(M)}$, $w_i^{(m)}$, and~$w_i^{(0)}$ are equal to the largest OSDs of the mass-$M$, mass-$m$, and massless edges in the graph $G_i$, respectively. In addition, $w_1^{(M)}=0$ and $w_1^{(m)}=1$.}
\bigbreak
One implication of this requirement is that the ordered presence of modes in $\Gamma$. For the general case that $\Gamma\neq \varnothing$, let us suppose that the massless edges with the smallest OSD are in the mode $X_i = C^{n_1} S^{n_2}$ (for some $n_1\in\{0,1\}$ and $n_2\in \mathbb{N}$), whose momentum scaling is described by (\ref{eq:mass_expansion_collinear_power_n1_soft_power_n2_mode}). Then, we have
\begin{subequations}
    \begin{align}
        & l_{X_i}^2 \sim \lambda^{n_1+2n_2} Q^2,
        \label{eq:CmSn_offshellness_massless}\\
        & (P+l_{X_i})^2-M^2 = 2P\cdot l_{X_i} +l_{X_i}^2 \sim \lambda^{n_2} Q^2,
        \label{eq:CmSn_offshellness_massive_M}\\
        & (p+l_{X_i})^2-m^2 = 2p\cdot l_{X_i} +l_{X_i}^2 \sim \lambda^{n_1+n_2} Q^2.
        \label{eq:CmSn_offshellness_massive_m}
    \end{align}
\end{subequations}
The numbers $i$, $w_i^{(M)}$, $w_i^{(m)}$, and $w_i^{(0)}$ can be deduced from these relations. By definition, $w_i^{(0)} = n_1+2n_2$. If there are mass-$M$ edges in $\gamma_{X_i}^{}$, then $w_i^{(M)} = n_2$, otherwise, all the mass-$M$ edges of $G_i$ are in $\gamma_H^{}\cup \gamma_C^{}$, which implies $w_i^{(M)} = 0$. Similarly, if there are mass-$m$ edges in $\gamma_{X_i}^{}$, then $w_i^{(m)} = n_1+n_2$, otherwise $w_i^{(m)} = 1$. It is straightforward to check that the only possibility satisfying both eqs.~(\ref{eq:balance_condition1}) and (\ref{eq:balance_condition2}) is $i=2$, or equivalently, $(n_1,n_2)=(0,1)$. In other words, if $\Gamma\neq \varnothing$, then $\gamma_S^{}\neq \varnothing$. Furthermore, eq.~(\ref{eq:balance_condition1}) implies that $\gamma_S^{}$ must contain mass-$M$ edges.

If $\Gamma\setminus \gamma_S^{} \neq \varnothing$, one can continue the analysis above and deduce that $\gamma_{C S}^{}\neq \varnothing$, which must contain mass-$m$ edges. If $\Gamma\setminus \gamma_S^{} \setminus \gamma_{CS}^{} \neq \varnothing$, then $\gamma_{S^2}^{}\neq \varnothing$, which must contain mass-$M$ edges, and so on. In this way, it is clear that \emph{the modes of $\Gamma$ in figure~\ref{figure-mass_typeI_general_configuration} appear in the following order}:
\begin{eqnarray}
\label{eq:mass_typeI_requirement1_implication_ordering}
    S,\ C S,\ S^2,\ C S^2,\ S^3,\ \dots.
\end{eqnarray}
\emph{Moreover, $\gamma_{S^n}^{}$ must contain mass-$M$ edges, and $\gamma_{C S^n}^{}$ must contain mass-$m$ edges, for each $n\in\{1,2,\dots\}$. As a result, the values of $(w_i^{(0)}, w_i^{(M)}, w_i^{(m)})$ are fixed for each $i$:}
\begin{eqnarray}
    (w_i^{(0)}, w_i^{(M)}, w_i^{(m)})&=& (2,1,1)\quad \text{for }i=2;\nonumber\\
    &\ & (3,1,2)\quad \text{for }i=3;\nonumber\\
    &\ & (4,2,2)\quad \text{for }i=4;\nonumber\\
    &\ & (5,2,3)\quad \text{for }i=5;\nonumber\\
    &\ & (6,3,3)\quad \text{for }i=6;\nonumber\\
    &\ & \qquad\dots.\nonumber
\end{eqnarray}

Below we show some configurations, each of which either contradicts figure~\ref{figure-mass_typeI_general_configuration} or violates Requirement 1 above. Here we color the massless edges such that the {\color{Blue}blue} ones are in the hard mode, the {\color{Green}green} ones are in the collinear mode, the {\color{Red}red} ones are in the soft mode, the {\color{Rhodamine}rhodamine} ones are in the collinear$\cdot$soft mode, and the {\color{Orange}orange} ones are in the soft${}^2$ mode. (We have also specified each massless $CS$ edge and $S^2$ edge for clarity.) This is sufficient to uniquely determine the region, so the massive edges are not colored here.
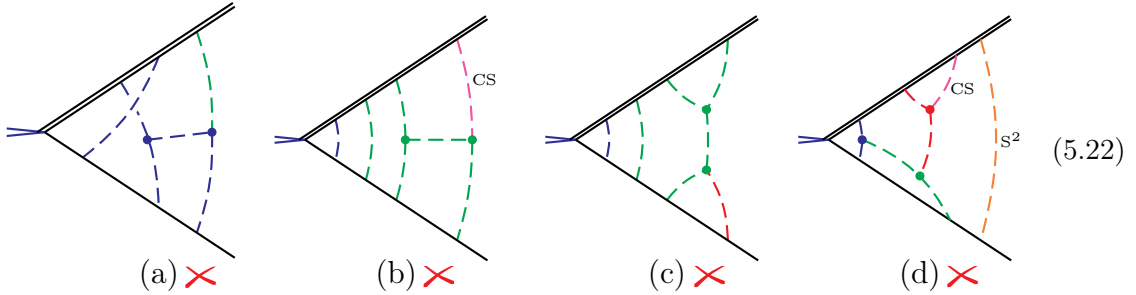
\begin{equation}
\begin{aligned}
\begin{tikzpicture}[baseline=11ex, scale =0.5]

\draw (4,6.33-0.8) edge [thick, Blue, dash pattern=on 2mm off 1mm, bend left = 15] (5,3-0.8) {};
\draw (6,7.67-0.8) edge [thick, Green, dash pattern=on 2mm off 1mm, bend left = 10] (6.4,5-0.8) {};
\draw (6.4,5-0.8) edge [thick, Blue, dash pattern=on 2mm off 1mm, bend left = 10] (6,2.33-0.8) {};
\draw (4.7,4.8-0.8) edge [thick, Blue, dash pattern=on 2mm off 1mm] (6.4,5-0.8) {};
\draw (5,7-0.8) edge [thick, Blue, draw=white, double=white, double distance=3pt, bend left = 15] (3,4.33-0.8) {}; \draw (5,7-0.8) edge [thick, Blue, dash pattern=on 2mm off 1mm, bend left = 15] (3,4.33-0.8) {};
\draw [thick, Blue] (1,5.1-0.8) edge (2,5-0.8) node [] {};
\draw [thick, Blue] (1,4.9-0.8) edge (2,5-0.8) node [] {};
\draw [thick] (1.8,5-0.8) edge (6.95,8.43-0.8) node [] {};
\draw [thick] (2,5-0.8) edge (7,8.33-0.8) node [] {};
\draw [thick] (2,5-0.8) edge (7,1.67-0.8) node [] {};
\node [draw, Blue, circle, minimum size=3pt, fill, inner sep=0pt, outer sep=0pt] () at (4.7,4.8-0.8) {};
\node [draw, Blue, circle, minimum size=3pt, fill, inner sep=0pt, outer sep=0pt] () at (6.4,5-0.8) {};
\node () at (5,1-0.6) {\large (a)};
\node () at (6.1,2.5-0.6) {$\crossmark[Red]$};
\end{tikzpicture}\quad
\begin{tikzpicture}[baseline=11ex, scale =0.4]

\draw (3,5.67) edge [Blue, thick, dash pattern=on 2mm off 1mm, bend left = 15] (3,4.33) {};
\draw (4,6.33) edge [Green, thick, dash pattern=on 2mm off 1mm, bend left = 15] (4,3.67) {};
\draw (5,7) edge [Green, thick, dash pattern=on 2mm off 1mm, bend left = 8] (5.3,5) {};
\draw (5.3,5) edge [Green, thick, dash pattern=on 2mm off 1mm, bend left = 8] (5,3) {};
\draw (5.3,5) edge [Green, thick, dash pattern=on 2mm off 1mm] (7.5,5) {};
\draw (7,8.33) edge [Rhodamine, thick, dash pattern=on 2mm off 1mm, bend left = 8] (7.5,5) {};
\draw (7.5,5) edge [Green, thick, dash pattern=on 2mm off 1mm, bend left = 8] (7,1.67) {};

\draw [thick, Blue] (1,5.1) edge (2,5) node [] {};
\draw [thick, Blue] (1,4.9) edge (2,5) node [] {};
\draw [thick] (1.8,5) edge (7.95,9.1) node [] {};
\draw [thick] (2,5) edge (8,9) node [] {};
\draw [thick] (2,5) edge (8,1) node [] {};

\node [draw, circle, minimum size=3pt, color=Green, fill=Green, inner sep=0pt, outer sep=0pt] () at (5.3,5) {};
\node [draw, circle, minimum size=3pt, color=Green, fill=Green, inner sep=0pt, outer sep=0pt] () at (7.5,5) {};

\node () at (7.9,7) {\tiny CS};
\node () at (5,0.5) {\large (b)};
\node () at (6.4,2.4) {$\crossmark[Red]$};
\end{tikzpicture}\quad
\begin{tikzpicture}[baseline=11ex, scale =0.4]

\draw (3,5.67) edge [Blue, thick, dash pattern=on 2mm off 1mm, bend left = 15] (3,4.33) {};
\draw (4,6.33) edge [Green, thick, dash pattern=on 2mm off 1mm, bend left = 15] (4,3.67) {};
\draw (5,7) edge [Green, thick, dash pattern=on 2mm off 1mm, bend right = 15] (6.3,6) {};
\draw (5,3) edge [Green, thick, dash pattern=on 2mm off 1mm, bend left = 15] (6.3,4) {};
\draw (6.3,6) edge [Green, thick, dash pattern=on 2mm off 1mm, bend left = 5] (6.3,4) {};
\draw (7,8.33) edge [Green, thick, dash pattern=on 2mm off 1mm, bend left = 15] (6.3,6) {};
\draw (7,1.67) edge [Red, thick, dash pattern=on 2mm off 1mm, bend right = 15] (6.3,4) {};

\draw [thick, Blue] (1,5.1) edge (2,5) node [] {};
\draw [thick, Blue] (1,4.9) edge (2,5) node [] {};
\draw [thick] (1.8,5) edge (7.95,9.1) node [] {};
\draw [thick] (2,5) edge (8,9) node [] {};
\draw [thick] (2,5) edge (8,1) node [] {};

\node [draw, circle, minimum size=3pt, color=Green, fill=Green, inner sep=0pt, outer sep=0pt] () at (6.3,6) {};
\node [draw, circle, minimum size=3pt, color=Green, fill=Green, inner sep=0pt, outer sep=0pt] () at (6.3,4) {};

\node () at (5,0.5) {\large (c)};
\node () at (6.4,2.4) {$\crossmark[Red]$};
\end{tikzpicture}\quad
\begin{tikzpicture}[baseline=11ex, scale =0.4]
\draw [thick, Blue] (1,5.1) edge (2,5) node [] {};
\draw [thick, Blue] (1,4.9) edge (2,5) node [] {};
\draw [thick] (1.8,5) edge (7.95,9.1) node [] {};
\draw [thick] (2,5) edge (8,9) node [] {};
\draw [thick] (2,5) edge (8,1) node [] {};

\draw (7,8.33) edge [Orange, thick, dash pattern=on 2mm off 1mm, bend left = 15] (7,1.67) {};
\draw (5.33,6) edge [Red, thick, dash pattern=on 2mm off 1mm, bend left = 15] (5,3.8) {};
\draw (3.1,5) edge [Green, thick, dash pattern=on 2mm off 1mm, bend left = 20] (6,2.33) {};
\draw (3,5.67) edge [Blue, thick, dash pattern=on 2mm off 1mm, bend left = 15] (3,4.33) {};
\draw (4.5,6.67) edge [Red, thick, dash pattern=on 2mm off 1mm, bend right = 15] (5.33,6) {};
\draw (6.2,7.8) edge [Rhodamine, thick, dash pattern=on 2mm off 1mm, bend left = 15] (5.33,6) {};

\node [draw, circle, minimum size=3pt, color=Red, fill=Red, inner sep=0pt, outer sep=0pt] () at (5.33,6) {};
\node [draw, circle, minimum size=3pt, color=Blue, fill=Blue, inner sep=0pt, outer sep=0pt] () at (3.1,5) {};
\node [draw, circle, minimum size=3pt, color=Green, fill=Green, inner sep=0pt, outer sep=0pt] () at (5,3.8) {};
\node () at (6.4,6.7) {\tiny CS};
\node () at (8,5) {\tiny S${}^2$};
\node () at (5,0.5) {\large (d)};
\node () at (6.4,2.4) {$\crossmark[Red]$};
\end{tikzpicture}
\end{aligned}
\label{eq:mass_typeI_requirement1_nonregion_examples}
\end{equation}
The first configuration (a) involves only the hard and the collinear modes, which, however, is not consistent with the configuration of figure~\ref{figure-mass_typeI_general_configuration} because the collinear sensitive subgraph $\gamma_C^{}$ contains no mass-$m$ edges. In contrast, configuration (b) conforms with figure~\ref{figure-mass_typeI_general_configuration} but violates Requirement 1: as explained above, $\gamma_S^{}\neq \varnothing$ if $\Gamma\neq \varnothing$, but here one $CS$-mode edge is present while the soft-mode edges are absent.

Configuration (c) also violates Requirement 1, because one implication of Requirement 1 is that $\gamma_S^{}$ must contain some mass-$M$ edges, which is not satisfied here. One can also see this from eq.~(\ref{eq:balance_condition1}). In the graph $G_2 = \gamma_H^{}\cup \gamma_J^{}\cup \gamma_S^{} =G$, the momentum of each mass-$M$ edge is either $(P+l_C^{})^\mu$ or $(P+l_H^{})^\mu$, with $l_H^\mu$ and $l_C^\mu$ in the hard mode and collinear mode respectively. Thus, their off-shellness all scale as $\lambda^0 Q^2$, and $w_2^{(M)}=0$. Following the same analysis, we have $w_2^{(m)}=1$ and $w_2^{(0)}=2$, and as a result, eq.~(\ref{eq:balance_condition1}) is not satisfied.

Similarly, Requirement 1 is violated in the configuration (d) because $\gamma_{CS}^{}$ does not contain mass-$m$ edges. Alternatively, consider the graph $G_3 = \gamma_H^{}\cup \gamma_J^{}\cup \gamma_S^{}\cup \gamma_{C S}^{}$:
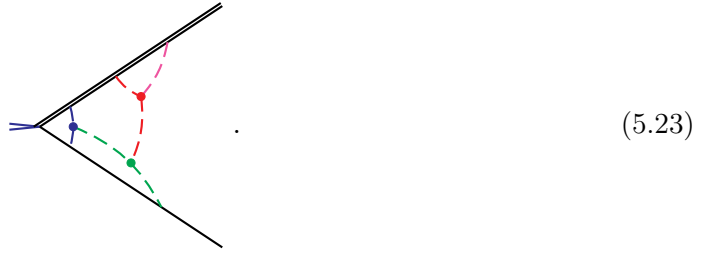
\begin{equation}
\begin{aligned}
\begin{tikzpicture}[baseline=11ex, scale =0.4]
\draw [thick, Blue] (1,5.1) edge (2,5) node [] {};
\draw [thick, Blue] (1,4.9) edge (2,5) node [] {};
\draw [thick] (1.8,5) edge (7.95,9.1) node [] {};
\draw [thick] (2,5) edge (8,9) node [] {};
\draw [thick] (2,5) edge (8,1) node [] {};

\draw (5.33,6) edge [Red, thick, dash pattern=on 2mm off 1mm, bend left = 15] (5,3.8) {};
\draw (3.1,5) edge [Green, thick, dash pattern=on 2mm off 1mm, bend left = 20] (6,2.33) {};
\draw (3,5.67) edge [Blue, thick, dash pattern=on 2mm off 1mm, bend left = 15] (3,4.33) {};
\draw (4.5,6.67) edge [Red, thick, dash pattern=on 2mm off 1mm, bend right = 15] (5.33,6) {};
\draw (6.2,7.8) edge [Rhodamine, thick, dash pattern=on 2mm off 1mm, bend left = 15] (5.33,6) {};

\node [draw, circle, minimum size=3pt, color=Red, fill=Red, inner sep=0pt, outer sep=0pt] () at (5.33,6) {};
\node [draw, circle, minimum size=3pt, color=Blue, fill=Blue, inner sep=0pt, outer sep=0pt] () at (3.1,5) {};
\node [draw, circle, minimum size=3pt, color=Green, fill=Green, inner sep=0pt, outer sep=0pt] () at (5,3.8) {};
.\end{tikzpicture}
\end{aligned}\ .
\end{equation}
Note that the line momenta of the mass-$M$ edges are of the forms $(P+l_{C S}^{})^\mu$, $(P+l_{S}^{})^\mu$, or $(P+l_{H}^{})^\mu$, whose off-shellnesses are either $\lambda Q^2$ or $Q^2$ from eq.~(\ref{eq:CmSn_offshellness_massive_M}). So we have $w_3^{(M)}=1$. Similarly, $w_3^{(m)}=1$ and $w_3^{(0)}=3$, and eq.~(\ref{eq:balance_condition1}) is not satisfied.

Through proper adjustments of each configuration in (\ref{eq:mass_typeI_requirement1_nonregion_examples}), the following valid regions in the mass expansion can be obtained.
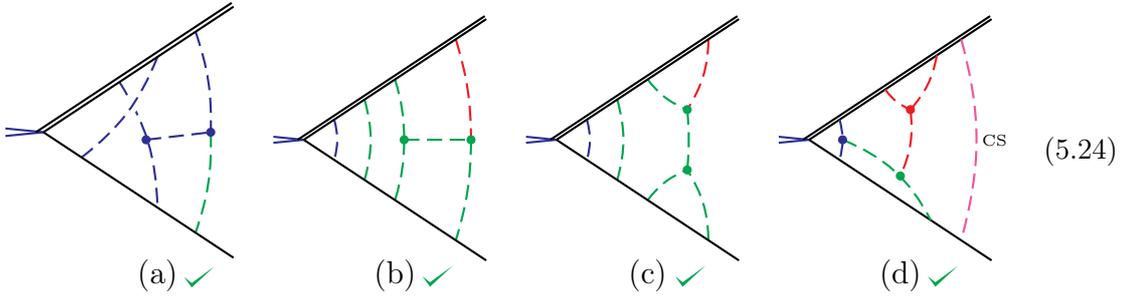
\begin{equation}
\begin{aligned}
\begin{tikzpicture}[baseline=11ex, scale =0.5]

\draw (4,6.33-0.8) edge [thick, Blue, dash pattern=on 2mm off 1mm, bend left = 15] (5,3-0.8) {};
\draw (6,7.67-0.8) edge [thick, Blue, dash pattern=on 2mm off 1mm, bend left = 10] (6.4,5-0.8) {};
\draw (6.4,5-0.8) edge [thick, Green, dash pattern=on 2mm off 1mm, bend left = 10] (6,2.33-0.8) {};
\draw (4.7,4.8-0.8) edge [thick, Blue, dash pattern=on 2mm off 1mm] (6.4,5-0.8) {};
\draw (5,7-0.8) edge [thick, Blue, draw=white, double=white, double distance=3pt, bend left = 15] (3,4.33-0.8) {}; \draw (5,7-0.8) edge [thick, Blue, dash pattern=on 2mm off 1mm, bend left = 15] (3,4.33-0.8) {};
\draw [thick, Blue] (1,5.1-0.8) edge (2,5-0.8) node [] {};
\draw [thick, Blue] (1,4.9-0.8) edge (2,5-0.8) node [] {};
\draw [thick] (1.8,5-0.8) edge (6.95,8.43-0.8) node [] {};
\draw [thick] (2,5-0.8) edge (7,8.33-0.8) node [] {};
\draw [thick] (2,5-0.8) edge (7,1.67-0.8) node [] {};
\node [draw, Blue, circle, minimum size=3pt, fill, inner sep=0pt, outer sep=0pt] () at (4.7,4.8-0.8) {};
\node [draw, Blue, circle, minimum size=3pt, fill, inner sep=0pt, outer sep=0pt] () at (6.4,5-0.8) {};
\node () at (5,1-0.6) {\large (a)};
\node () at (6.1,2.5-0.6) {$\greencheckmark[Green]$};
\end{tikzpicture}\quad
\begin{tikzpicture}[baseline=11ex, scale =0.4]

\draw (3,5.67) edge [Blue, thick, dash pattern=on 2mm off 1mm, bend left = 15] (3,4.33) {};
\draw (4,6.33) edge [Green, thick, dash pattern=on 2mm off 1mm, bend left = 15] (4,3.67) {};
\draw (5,7) edge [Green, thick, dash pattern=on 2mm off 1mm, bend left = 8] (5.3,5) {};
\draw (5.3,5) edge [Green, thick, dash pattern=on 2mm off 1mm, bend left = 8] (5,3) {};
\draw (5.3,5) edge [Green, thick, dash pattern=on 2mm off 1mm] (7.5,5) {};
\draw (7,8.33) edge [Red, thick, dash pattern=on 2mm off 1mm, bend left = 8] (7.5,5) {};
\draw (7.5,5) edge [Green, thick, dash pattern=on 2mm off 1mm, bend left = 8] (7,1.67) {};

\draw [thick, Blue] (1,5.1) edge (2,5) node [] {};
\draw [thick, Blue] (1,4.9) edge (2,5) node [] {};
\draw [thick] (1.8,5) edge (7.95,9.1) node [] {};
\draw [thick] (2,5) edge (8,9) node [] {};
\draw [thick] (2,5) edge (8,1) node [] {};

\node [draw, circle, minimum size=3pt, color=Green, fill=Green, inner sep=0pt, outer sep=0pt] () at (5.3,5) {};
\node [draw, circle, minimum size=3pt, color=Green, fill=Green, inner sep=0pt, outer sep=0pt] () at (7.5,5) {};

\node () at (5,0.5) {\large (b)};
\node () at (6.4,2.4) {$\greencheckmark[Green]$};
\end{tikzpicture}\quad
\begin{tikzpicture}[baseline=11ex, scale =0.4]

\draw (3,5.67) edge [Blue, thick, dash pattern=on 2mm off 1mm, bend left = 15] (3,4.33) {};
\draw (4,6.33) edge [Green, thick, dash pattern=on 2mm off 1mm, bend left = 15] (4,3.67) {};
\draw (5,7) edge [Green, thick, dash pattern=on 2mm off 1mm, bend right = 15] (6.3,6) {};
\draw (5,3) edge [Green, thick, dash pattern=on 2mm off 1mm, bend left = 15] (6.3,4) {};
\draw (6.3,6) edge [Green, thick, dash pattern=on 2mm off 1mm, bend left = 5] (6.3,4) {};
\draw (7,8.33) edge [Red, thick, dash pattern=on 2mm off 1mm, bend left = 15] (6.3,6) {};
\draw (7,1.67) edge [Green, thick, dash pattern=on 2mm off 1mm, bend right = 15] (6.3,4) {};

\draw [thick, Blue] (1,5.1) edge (2,5) node [] {};
\draw [thick, Blue] (1,4.9) edge (2,5) node [] {};
\draw [thick] (1.8,5) edge (7.95,9.1) node [] {};
\draw [thick] (2,5) edge (8,9) node [] {};
\draw [thick] (2,5) edge (8,1) node [] {};

\node [draw, circle, minimum size=3pt, color=Green, fill=Green, inner sep=0pt, outer sep=0pt] () at (6.3,6) {};
\node [draw, circle, minimum size=3pt, color=Green, fill=Green, inner sep=0pt, outer sep=0pt] () at (6.3,4) {};

\node () at (5,0.5) {\large (c)};
\node () at (6.4,2.4) {$\greencheckmark[Green]$};
\end{tikzpicture}\quad
\begin{tikzpicture}[baseline=11ex, scale =0.4]
\draw [thick, Blue] (1,5.1) edge (2,5) node [] {};
\draw [thick, Blue] (1,4.9) edge (2,5) node [] {};
\draw [thick] (1.8,5) edge (7.95,9.1) node [] {};
\draw [thick] (2,5) edge (8,9) node [] {};
\draw [thick] (2,5) edge (8,1) node [] {};

\draw (7,8.33) edge [Rhodamine, thick, dash pattern=on 2mm off 1mm, bend left = 15] (7,1.67) {};
\draw (5.33,6) edge [Red, thick, dash pattern=on 2mm off 1mm, bend left = 15] (5,3.8) {};
\draw (3.1,5) edge [Green, thick, dash pattern=on 2mm off 1mm, bend left = 20] (6,2.33) {};
\draw (3,5.67) edge [Blue, thick, dash pattern=on 2mm off 1mm, bend left = 15] (3,4.33) {};
\draw (4.5,6.67) edge [Red, thick, dash pattern=on 2mm off 1mm, bend right = 15] (5.33,6) {};
\draw (6.2,7.8) edge [Red, thick, dash pattern=on 2mm off 1mm, bend left = 15] (5.33,6) {};

\node [draw, circle, minimum size=3pt, color=Red, fill=Red, inner sep=0pt, outer sep=0pt] () at (5.33,6) {};
\node [draw, circle, minimum size=3pt, color=Blue, fill=Blue, inner sep=0pt, outer sep=0pt] () at (3.1,5) {};
\node [draw, circle, minimum size=3pt, color=Green, fill=Green, inner sep=0pt, outer sep=0pt] () at (5,3.8) {};
\node () at (8.1,5) {\tiny CS};
\node () at (5,0.5) {\large (d)};
\node () at (6.4,2.4) {$\greencheckmark[Green]$};
\end{tikzpicture}
\end{aligned}
\label{eq:mass_typeI_requirement1_region_examples}
\end{equation}
One can verify that each configuration above admits both figure~\ref{figure-mass_typeI_general_configuration} and Requirement~1. In fact, using the parameterization of edges in appendix~\ref{appendix-mass_expansion_examples}, the configurations (a)-(d) correspond to the region vectors \textsf{(0,0,0,0,0,-1,0,0,0,0,0,-1;1)}, \textsf{(-1,0,0,0,0,-1,-1,-1,0,-1,-1,-1,-1,-2,-1;1)}, \textsf{(-1,0,0,0,0,-1,-1,-1,0,-1,-1,-1,-1,-2,-1;1)}, and \textsf{(-1,-1,-1,0,0,-1,-2,0,0,-1,-1,-2,-2,-2,-3;1)}, respectively.
\bigbreak
\emph{Requirement 2: for each $i\in\mathbb{N}$, the graph $G_{i+2}/G_i$ is one-vertex irreducible.}
\bigbreak
Note that $G_j$ in general can be disconnected, and to obtain $G_{i+2}/G_i$ from $G_{i+2}$, one contracts \emph{each component of $G_i$} to a vertex. This requirement aims to further rule out those configurations in which ``the $X_i$ and $X_{i+1}$ modes do not interact''. For example,
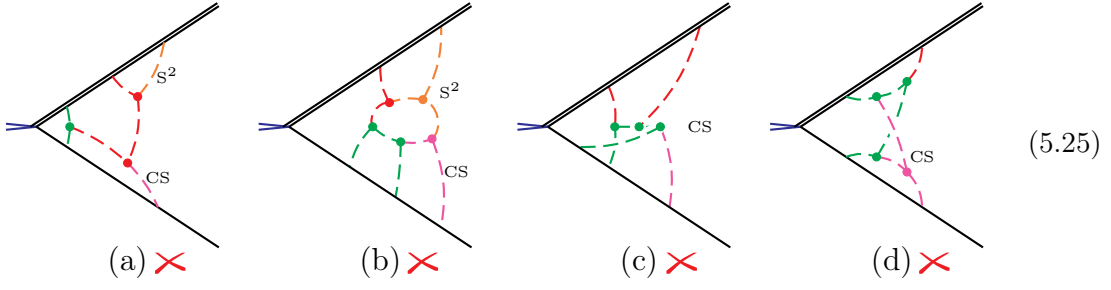
\begin{equation}
\begin{aligned}
\begin{tikzpicture}[baseline=11ex, scale =0.4]
\draw [thick, Blue] (1,5.1) edge (2,5) node [] {};
\draw [thick, Blue] (1,4.9) edge (2,5) node [] {};
\draw [thick] (1.8,5) edge (7.95,9.1) node [] {};
\draw [thick] (2,5) edge (8,9) node [] {};
\draw [thick] (2,5) edge (8,1) node [] {};
\draw (5.33,6) edge [thick, Red, dash pattern=on 2mm off 1mm, bend left = 15] (5,3.8) {};
\draw (3.1,5) edge [thick, Red, dash pattern=on 2mm off 1mm, bend left = 10] (5,3.8) {};
\draw (5,3.8) edge [thick, Rhodamine, dash pattern=on 2mm off 1mm, bend left = 10] (6,2.33) {};
\draw (3,5.67) edge [thick, Green, dash pattern=on 2mm off 1mm, bend left = 15] (3,4.33) {};
\draw (4.5,6.67) edge [thick, Red, dash pattern=on 2mm off 1mm, bend right = 15] (5.33,6) {};
\draw (6.2,7.8) edge [thick, Orange, dash pattern=on 2mm off 1mm, bend left = 15] (5.33,6) {};
\node [draw, circle, Red, minimum size=3pt, fill, inner sep=0pt, outer sep=0pt] () at (5.33,6) {};
\node [draw, Green, circle, minimum size=3pt, fill, inner sep=0pt, outer sep=0pt] () at (3.1,5) {};
\node [draw, circle, Red, minimum size=3pt, fill, inner sep=0pt, outer sep=0pt] () at (5,3.8) {};
\node () at (6,3.3) {\tiny CS};
\node () at (6.3,6.7) {\tiny S$^2$};
\node () at (5,0.5) {\large (a)};
\node () at (6.4,2.4) {$\crossmark[Red]$};
\end{tikzpicture}\quad
\begin{tikzpicture}[baseline=11ex, scale =0.4]
\draw (6.4,5.9) edge [thick, Orange, dash pattern=on 2mm off 1mm, bend right = 15] (7,8.33) {};
\draw (5.3,5.8) edge [thick, Red, dash pattern=on 2mm off 1mm, bend left = 15] (5,7) {};
\draw (4.75,5) edge [thick, Green, dash pattern=on 2mm off 1mm, bend right = 15] (4,3.67) {};
\draw (5.67,4.5) edge [thick, Green, dash pattern=on 2mm off 1mm, bend left = 5] (5.5,2.67) {};
\draw (6.7,4.6) edge [thick, Rhodamine, dash pattern=on 2mm off 1mm, bend left = 15] (7,1.67) {};
\draw [thick, Blue] (1,5.1) edge (2,5) node [] {};
\draw [thick, Blue] (1,4.9) edge (2,5) node [] {};
\draw [thick] (1.8,5) edge (7.95,9.1) node [] {};
\draw [thick] (2,5) edge (8,9) node [] {};
\draw [thick] (2,5) edge (8,1) node [] {};
\draw (5.3,5.8) edge [thick, Red, dash pattern=on 2mm off 1mm, bend right = 60] (4.75,5) {};
\draw (4.75,5) edge [thick, Green, dash pattern=on 2mm off 1mm, bend right = 15] (5.67,4.5) {};
\draw (5.67,4.5) edge [thick, Rhodamine, dash pattern=on 2mm off 1mm, bend right = 15] (6.7,4.6) {};
\draw (6.7,4.6) edge [thick, Orange, dash pattern=on 2mm off 1mm, bend right = 60] (6.4,5.9) {};
\draw (6.4,5.9) edge [thick, Orange, dash pattern=on 2mm off 1mm, bend right = 15] (5.3,5.8) {};
\node [draw, Orange, circle, minimum size=3pt, fill, inner sep=0pt, outer sep=0pt] () at (6.4,5.9) {};
\node [draw, Red, circle, minimum size=3pt, fill, inner sep=0pt, outer sep=0pt] () at (5.3,5.8) {};
\node [draw, Green, circle, minimum size=3pt, fill, inner sep=0pt, outer sep=0pt] () at (4.75,5) {};
\node [draw, Green, circle, minimum size=3pt, fill, inner sep=0pt, outer sep=0pt] () at (5.67,4.5) {};
\node [draw, Rhodamine, circle, minimum size=3pt, fill, inner sep=0pt, outer sep=0pt] () at (6.7,4.6) {};
\node () at (7.5,3.5) {\tiny CS};
\node () at (7.3,6.2) {\tiny S$^2$};
\node () at (5,0.5) {\large (b)};
\node () at (6.4,2.4) {$\crossmark[Red]$};
\end{tikzpicture}\quad
\begin{tikzpicture}[baseline=11ex, scale =0.4]
\draw (4,6.33) edge [thick, Red, dash pattern=on 2mm off 1mm, bend left = 15] (4.2,5) {};
\draw (4.2,5) edge [thick, Green, dash pattern=on 2mm off 1mm, bend left = 15] (4,3.67) {};
\draw (5.7,5) edge [thick, Rhodamine, dash pattern=on 2mm off 1mm, bend left = 15] (6,2.33) {};
\draw (4.2,5) edge [thick, Green, dash pattern=on 2mm off 1mm] (5.7,5) {};
\draw (5.7,5) edge [thick, draw=white, double=white, double distance=3pt, bend left = 15] (3,4.33) {}; \draw (5.7,5) edge [thick, Green, dash pattern=on 2mm off 1mm, bend left = 15] (3,4.33) {};
\draw (7,8.33) edge [thick, draw=white, double=white, double distance=3pt, bend right = 5] (5,5) {}; \draw (7,8.33) edge [thick, Red, dash pattern=on 2mm off 1mm, bend left = 15] (5,5) {};
\node [draw, Green, circle, minimum size=3pt, fill, inner sep=0pt, outer sep=0pt] () at (4.2,5) {};
\node [draw, Green, circle, minimum size=3pt, fill, inner sep=0pt, outer sep=0pt] () at (5,5) {};
\node [draw, Green, circle, minimum size=3pt, fill, inner sep=0pt, outer sep=0pt] () at (5.7,5) {};
\draw [thick, Blue] (1,5.1) edge (2,5) node [] {};
\draw [thick, Blue] (1,4.9) edge (2,5) node [] {};
\draw [thick] (1.8,5) edge (7.95,9.1) node [] {};
\draw [thick] (2,5) edge (8,9) node [] {};
\draw [thick] (2,5) edge (8,1) node [] {};
\node () at (7,5) {\tiny CS};
\node () at (5,0.5) {\large (c)};
\node () at (6.4,2.4) {$\crossmark[Red]$};
\end{tikzpicture}\quad
\begin{tikzpicture}[baseline=11ex, scale =0.4]
\draw [thick, Blue] (1,5.1) edge (2,5) node [] {};
\draw [thick, Blue] (1,4.9) edge (2,5) node [] {};
\draw [thick] (1.8,5) edge (7.95,9.1) node [] {};
\draw [thick] (2,5) edge (8,9) node [] {};
\draw [thick] (2,5) edge (8,1) node [] {};
\draw (4.5,4) edge [thick, Green, dash pattern=on 2mm off 1mm, bend right = 10] (5.5,6.5) {};
\draw (4.5,6) edge [thick, draw=white, double=white, double distance=3pt, bend left = 10] (5.5,3.5) {}; \draw (4.5,6) edge [thick, Rhodamine, dash pattern=on 2mm off 1mm, bend left = 10] (5.5,3.5) {};
\draw (3.5,6) edge [thick, Green, dash pattern=on 2mm off 1mm, bend right = 20] (4.5,6) {};
\draw (4.5,6) edge [thick, Green, dash pattern=on 2mm off 1mm, bend right = 20] (5.5,6.5) {};
\draw (5.5,6.5) edge [thick, Red, dash pattern=on 2mm off 1mm, bend right = 20] (6,7.67) {};
\draw (3.5,4) edge [thick, Green, dash pattern=on 2mm off 1mm, bend left = 20] (4.5,4) {};
\draw (4.5,4) edge [thick, Rhodamine, dash pattern=on 2mm off 1mm, bend left = 20] (5.5,3.5) {};
\draw (5.5,3.5) edge [thick, Rhodamine, dash pattern=on 2mm off 1mm, bend left = 20] (6,2.33) {};
\node [draw, Green, circle, minimum size=3pt, fill, inner sep=0pt, outer sep=0pt] () at (4.5,6) {};
\node [draw, Green, circle, minimum size=3pt, fill, inner sep=0pt, outer sep=0pt] () at (4.5,4) {};
\node [draw, Green, circle, minimum size=3pt, fill, inner sep=0pt, outer sep=0pt] () at (5.5,6.5) {};
\node [draw, Rhodamine, circle, minimum size=3pt, fill, inner sep=0pt, outer sep=0pt] () at (5.5,3.5) {};
\node () at (6,4) {\tiny CS};
\node () at (5,0.5) {\large (d)};
\node () at (6.4,2.4) {$\crossmark[Red]$};
\end{tikzpicture}
\end{aligned}
\label{eq:mass_typeI_requirement2_nonregion_examples}
\end{equation}
Note that all these configurations meet Requirement 1; the reason why they are not regions is that $G_4/G_2$ is not 1VI in (a), while $G_3/G_1$ is not 1VI in (b), (c), and (d), as can be checked directly. However, one can modify them into the following valid regions:
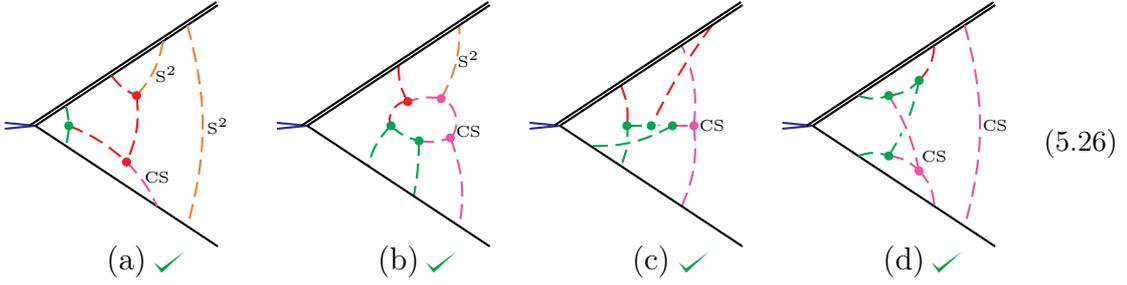
\begin{equation}
\begin{aligned}
\begin{tikzpicture}[baseline=11ex, scale =0.4]
\draw [thick, Blue] (1,5.1) edge (2,5) node [] {};
\draw [thick, Blue] (1,4.9) edge (2,5) node [] {};
\draw [thick] (1.8,5) edge (7.95,9.1) node [] {};
\draw [thick] (2,5) edge (8,9) node [] {};
\draw [thick] (2,5) edge (8,1) node [] {};
\draw (7,8.33) edge [thick, Orange, dash pattern=on 2mm off 1mm, bend left = 15] (7,1.67) {};
\draw (5.33,6) edge [thick, Red, dash pattern=on 2mm off 1mm, bend left = 15] (5,3.8) {};
\draw (3.1,5) edge [thick, Red, dash pattern=on 2mm off 1mm, bend left = 10] (5,3.8) {};
\draw (5,3.8) edge [thick, Rhodamine, dash pattern=on 2mm off 1mm, bend left = 10] (6,2.33) {};
\draw (3,5.67) edge [thick, Green, dash pattern=on 2mm off 1mm, bend left = 15] (3,4.33) {};
\draw (4.5,6.67) edge [thick, Red, dash pattern=on 2mm off 1mm, bend right = 15] (5.33,6) {};
\draw (6.2,7.8) edge [thick, Orange, dash pattern=on 2mm off 1mm, bend left = 15] (5.33,6) {};
\node [draw, circle, Red, minimum size=3pt, fill, inner sep=0pt, outer sep=0pt] () at (5.33,6) {};
\node [draw, Green, circle, minimum size=3pt, fill, inner sep=0pt, outer sep=0pt] () at (3.1,5) {};
\node [draw, circle, Red, minimum size=3pt, fill, inner sep=0pt, outer sep=0pt] () at (5,3.8) {};
\node () at (6,3.3) {\tiny CS};
\node () at (6.3,6.7) {\tiny S$^2$};
\node () at (8,5) {\tiny S$^2$};
\node () at (5,0.5) {\large (a)};
\node () at (6.4,2.4) {$\greencheckmark[Green]$};
\end{tikzpicture}\quad
\begin{tikzpicture}[baseline=11ex, scale =0.4]
\draw (6.4,5.9) edge [thick, Orange, dash pattern=on 2mm off 1mm, bend right = 15] (7,8.33) {};
\draw (5.3,5.8) edge [thick, Red, dash pattern=on 2mm off 1mm, bend left = 15] (5,7) {};
\draw (4.75,5) edge [thick, Green, dash pattern=on 2mm off 1mm, bend right = 15] (4,3.67) {};
\draw (5.67,4.5) edge [thick, Green, dash pattern=on 2mm off 1mm, bend left = 5] (5.5,2.67) {};
\draw (6.7,4.6) edge [thick, Rhodamine, dash pattern=on 2mm off 1mm, bend left = 15] (7,1.67) {};
\draw [thick, Blue] (1,5.1) edge (2,5) node [] {};
\draw [thick, Blue] (1,4.9) edge (2,5) node [] {};
\draw [thick] (1.8,5) edge (7.95,9.1) node [] {};
\draw [thick] (2,5) edge (8,9) node [] {};
\draw [thick] (2,5) edge (8,1) node [] {};
\draw (5.3,5.8) edge [thick, Red, dash pattern=on 2mm off 1mm, bend right = 60] (4.75,5) {};
\draw (4.75,5) edge [thick, Green, dash pattern=on 2mm off 1mm, bend right = 15] (5.67,4.5) {};
\draw (5.67,4.5) edge [thick, Rhodamine, dash pattern=on 2mm off 1mm, bend right = 15] (6.7,4.6) {};
\draw (6.7,4.6) edge [thick, Rhodamine, dash pattern=on 2mm off 1mm, bend right = 60] (6.4,5.9) {};
\draw (6.4,5.9) edge [thick, Rhodamine, dash pattern=on 2mm off 1mm, bend right = 15] (5.3,5.8) {};
\node [draw, Rhodamine, circle, minimum size=3pt, fill, inner sep=0pt, outer sep=0pt] () at (6.4,5.9) {};
\node [draw, Red, circle, minimum size=3pt, fill, inner sep=0pt, outer sep=0pt] () at (5.3,5.8) {};
\node [draw, Green, circle, minimum size=3pt, fill, inner sep=0pt, outer sep=0pt] () at (4.75,5) {};
\node [draw, Green, circle, minimum size=3pt, fill, inner sep=0pt, outer sep=0pt] () at (5.67,4.5) {};
\node [draw, Rhodamine, circle, minimum size=3pt, fill, inner sep=0pt, outer sep=0pt] () at (6.7,4.6) {};
\node () at (7.3,4.8) {\tiny CS};
\node () at (7.3,7.2) {\tiny S$^2$};
\node () at (5,0.5) {\large (b)};
\node () at (6.4,2.4) {$\greencheckmark[Green]$};
\end{tikzpicture}\quad
\begin{tikzpicture}[baseline=11ex, scale =0.4]
\draw (4,6.33) edge [thick, Red, dash pattern=on 2mm off 1mm, bend left = 15] (4.2,5) {};
\draw (4.2,5) edge [thick, Green, dash pattern=on 2mm off 1mm, bend left = 15] (4,3.67) {};
\draw (6,7.67) edge [thick, Rhodamine, dash pattern=on 2mm off 1mm, bend left = 15] (6.4,5) {};
\draw (6.4,5) edge [thick, Rhodamine, dash pattern=on 2mm off 1mm, bend left = 15] (6,2.33) {};
\draw (4.2,5) edge [thick, Green, dash pattern=on 2mm off 1mm] (5.7,5) {};
\draw (5.7,5) edge [thick, Rhodamine, dash pattern=on 2mm off 1mm] (6.4,5) {};
\draw (5.7,5) edge [thick, draw=white, double=white, double distance=3pt, bend left = 15] (3,4.33) {}; \draw (5.7,5) edge [thick, Green, dash pattern=on 2mm off 1mm, bend left = 15] (3,4.33) {};
\draw (7,8.33) edge [thick, draw=white, double=white, double distance=3pt, bend right = 5] (5,5) {}; \draw (7,8.33) edge [thick, Red, dash pattern=on 2mm off 1mm, bend right = 5] (5,5) {};
\node [draw, Green, circle, minimum size=3pt, fill, inner sep=0pt, outer sep=0pt] () at (4.2,5) {};
\node [draw, Green, circle, minimum size=3pt, fill, inner sep=0pt, outer sep=0pt] () at (5,5) {};
\node [draw, Green, circle, minimum size=3pt, fill, inner sep=0pt, outer sep=0pt] () at (5.7,5) {};
\node [draw, Rhodamine, circle, minimum size=3pt, fill, inner sep=0pt, outer sep=0pt] () at (6.4,5) {};
\draw [thick, Blue] (1,5.1) edge (2,5) node [] {};
\draw [thick, Blue] (1,4.9) edge (2,5) node [] {};
\draw [thick] (1.8,5) edge (7.95,9.1) node [] {};
\draw [thick] (2,5) edge (8,9) node [] {};
\draw [thick] (2,5) edge (8,1) node [] {};
\node () at (7,5) {\tiny CS};
\node () at (5,0.5) {\large (c)};
\node () at (6.4,2.4) {$\greencheckmark[Green]$};
\end{tikzpicture}\quad
\begin{tikzpicture}[baseline=11ex, scale =0.4]
\draw [thick, Blue] (1,5.1) edge (2,5) node [] {};
\draw [thick, Blue] (1,4.9) edge (2,5) node [] {};
\draw [thick] (1.8,5) edge (7.95,9.1) node [] {};
\draw [thick] (2,5) edge (8,9) node [] {};
\draw [thick] (2,5) edge (8,1) node [] {};
\draw (4.5,4) edge [thick, Green, dash pattern=on 2mm off 1mm, bend right = 10] (5.5,6.5) {};
\draw (4.5,6) edge [thick, draw=white, double=white, double distance=3pt, bend left = 10] (5.5,3.5) {}; \draw (4.5,6) edge [thick, Rhodamine, dash pattern=on 2mm off 1mm, bend left = 10] (5.5,3.5) {};
\draw (7,8.33) edge [thick, Rhodamine, dash pattern=on 2mm off 1mm, bend left = 15] (7,1.67) {};
\draw (3.5,6) edge [thick, Green, dash pattern=on 2mm off 1mm, bend right = 20] (4.5,6) {};
\draw (4.5,6) edge [thick, Green, dash pattern=on 2mm off 1mm, bend right = 20] (5.5,6.5) {};
\draw (5.5,6.5) edge [thick, Red, dash pattern=on 2mm off 1mm, bend right = 20] (6,7.67) {};
\draw (3.5,4) edge [thick, Green, dash pattern=on 2mm off 1mm, bend left = 20] (4.5,4) {};
\draw (4.5,4) edge [thick, Rhodamine, dash pattern=on 2mm off 1mm, bend left = 20] (5.5,3.5) {};
\draw (5.5,3.5) edge [thick, Rhodamine, dash pattern=on 2mm off 1mm, bend left = 20] (6,2.33) {};
\node [draw, Green, circle, minimum size=3pt, fill, inner sep=0pt, outer sep=0pt] () at (4.5,6) {};
\node [draw, Green, circle, minimum size=3pt, fill, inner sep=0pt, outer sep=0pt] () at (4.5,4) {};
\node [draw, Green, circle, minimum size=3pt, fill, inner sep=0pt, outer sep=0pt] () at (5.5,6.5) {};
\node [draw, Rhodamine, circle, minimum size=3pt, fill, inner sep=0pt, outer sep=0pt] () at (5.5,3.5) {};
\node () at (6,4) {\tiny CS};
\node () at (8,5) {\tiny CS};
\node () at (5,0.5) {\large (d)};
\node () at (6.4,2.4) {$\greencheckmark[Green]$};
\end{tikzpicture}
\end{aligned}
\label{eq:mass_typeI_requirement2_region_examples}
\end{equation}
It is straightforward to check that $G_{i+2}/G_i$ is 1VI for each $i\in \mathbb{N}$. In fact, the four regions above correspond to the vectors \textsf{(-2,-2,-1,0,-1,-2,-2,-1,-1,-2,-2,-4,-2,-3,-4;1)}, \textsf{(-2,-1,-1,-1,-2,-2,-2,-1,-4,-3,-1,-1,-3,-3,-3;1)}, \textsf{(-1,-1,-1,-1,-1,-2,-2,-1,-1,-1,-1,-2,-3,-3,-3;1)}, and \textsf{(-1,-1,0,-1,-2,-2,-2,-1,-1,-1,-3,-3,-1,-3,-3;1)} respectively, as presented in appendix~\ref{appendix-mass_expansion_examples}.

To introduce the third requirement, we rewrite each mode sensitive graph as the union of three subgraphs:
\begin{eqnarray}
    \gamma_X^{} = \gamma_X^{(0)} \cup \gamma_X^{(M)} \cup \gamma_X^{(m)},
\end{eqnarray}
with $\gamma_X^{(M)}$ consisting of all the mass-$M$ edges and their endpoints, $\gamma_X^{(m)}$ consisting of all the mass-$M$ edges and their endpoints, and $\gamma_X^{(0)}$ consisting of all the massless edges and their endpoints. Then, in addition to Requirements 1 and 2 above, we further require the following.

\bigbreak
\emph{Requirement 3: for $n\in \mathbb{N}_+$, each component of $\gamma_{S^n}^{(0)}$ is adjacent to $\gamma_{S^n}^{(M)}$, and each component of  $\gamma_{CS^n}^{(0)}$ is adjacent to $\gamma_{CS^n}^{(m)}$.}
\bigbreak
Let us illustrate this from the following examples:
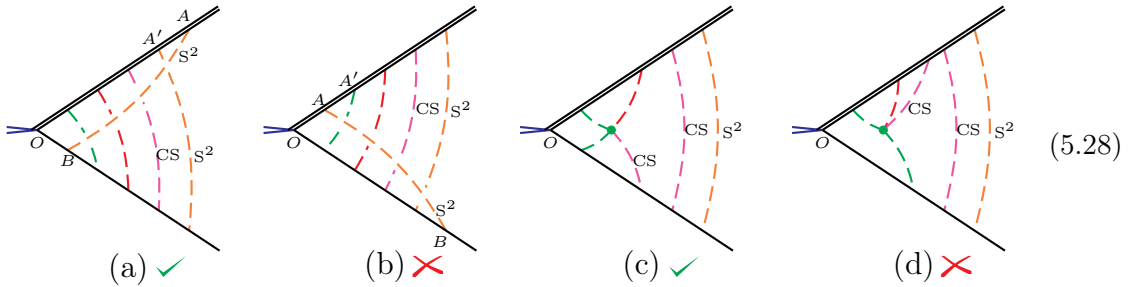
\begin{equation}
\begin{aligned}
\begin{tikzpicture}[baseline=11ex, scale =0.4]
\draw (3,5.67) edge [thick, Green, dash pattern=on 2mm off 1mm, bend left = 15] (4,3.67) {};
\draw (4,6.33) edge [thick, Red, dash pattern=on 2mm off 1mm, bend left = 15] (5,3) {};
\draw (5,7) edge [thick, Rhodamine, dash pattern=on 2mm off 1mm, bend left = 15] (6,2.33) {};
\draw (6,7.67) edge [thick, Orange, dash pattern=on 2mm off 1mm, bend left = 15] (7,1.67) {};
\draw (7,8.33) edge [thick, draw=white, double=white, double distance=3pt, bend left = 15] (3,4.33) {}; \draw (7,8.33) edge [thick, Orange, dash pattern=on 2mm off 1mm, bend left = 15] (3,4.33) {};
\draw [thick, Blue] (1,5.1) edge (2,5) node [] {};
\draw [thick, Blue] (1,4.9) edge (2,5) node [] {};
\draw [thick] (1.8,5) edge (7.95,9.1) node [] {};
\draw [thick] (2,5) edge (8,9) node [] {};
\draw [thick] (2,5) edge (8,1) node [] {};
\node () at (2,4.6) {\tiny $O$};
\node () at (6.8,8.8) {\tiny $A$};
\node () at (5.8,8.1) {\tiny $A'$};
\node () at (3,4) {\tiny $B$};
\node () at (7,7.5) {\tiny S$^2$};
\node () at (6.3,4.2) {\tiny CS};
\node () at (7.5,4.2) {\tiny S$^2$};
\node () at (5,1-0.6) {\large (a)};
\node () at (6.4,2.4) {$\greencheckmark[Green]$};
\end{tikzpicture}\quad
\begin{tikzpicture}[baseline=11ex, scale =0.4]
\draw (4,6.33) edge [thick, Green, dash pattern=on 2mm off 1mm, bend left = 15] (3,4.33) {};
\draw (5,7) edge [thick, Red, dash pattern=on 2mm off 1mm, bend left = 15] (4,3.67) {};
\draw (6,7.67) edge [thick, Rhodamine, dash pattern=on 2mm off 1mm, bend left = 15] (5,3) {};
\draw (7,8.33) edge [thick, Orange, dash pattern=on 2mm off 1mm, bend left = 15] (6,2.33) {};
\draw (3,5.67) edge [thick, draw=white, double=white, double distance=3pt, bend left = 15] (7,1.67) {}; \draw (3,5.67) edge [thick, Orange, dash pattern=on 2mm off 1mm, bend left = 15] (7,1.67) {};
\draw [thick, Blue] (1,5.1) edge (2,5) node [] {};
\draw [thick, Blue] (1,4.9) edge (2,5) node [] {};
\draw [thick] (1.8,5) edge (7.95,9.1) node [] {};
\draw [thick] (2,5) edge (8,9) node [] {};
\draw [thick] (2,5) edge (8,1) node [] {};
\node () at (2,4.6) {\tiny $O$};
\node () at (2.8,6) {\tiny $A$};
\node () at (3.8,6.7) {\tiny $A'$};
\node () at (6.8,1.3) {\tiny $B$};
\node () at (6.4,5.7) {\tiny CS};
\node () at (7.5,5.7) {\tiny S$^2$};
\node () at (7,2.4) {\tiny S$^2$};
\node () at (5,0.5) {\large (b)};
\node () at (6.4,2.4) {$\crossmark[Red]$};
\end{tikzpicture}\quad
\begin{tikzpicture}[baseline=11ex, scale =0.4]
\draw (3,5.67) edge [thick, Green, dash pattern=on 2mm off 1mm, bend right = 15] (4,5) {};
\draw (4,5) edge [thick, Green, dash pattern=on 2mm off 1mm, bend left = 15] (3,4.33) {};
\draw (5,7) edge [thick, Red, dash pattern=on 2mm off 1mm, bend left = 15] (4,5) {};
\draw (4,5) edge [thick, Rhodamine, dash pattern=on 2mm off 1mm, bend left = 15] (5,3) {};
\draw (6,7.67) edge [thick, Rhodamine, dash pattern=on 2mm off 1mm, bend left = 15] (6,2.33) {};
\draw (7,8.33) edge [thick, Orange, dash pattern=on 2mm off 1mm, bend left = 15] (7,1.67) {};
\draw [thick, Blue] (1,5.1) edge (2,5) node [] {};
\draw [thick, Blue] (1,4.9) edge (2,5) node [] {};
\draw [thick] (1.8,5) edge (7.95,9.1) node [] {};
\draw [thick] (2,5) edge (8,9) node [] {};
\draw [thick] (2,5) edge (8,1) node [] {};
\node [draw, Green, circle, minimum size=3pt, fill, inner sep=0pt, outer sep=0pt] () at (4,5) {};
\node () at (2,4.6) {\tiny $O$};
\node () at (5.1,4) {\tiny CS};
\node () at (6.8,5) {\tiny CS};
\node () at (8,5) {\tiny S$^2$};
\node () at (5,0.5) {\large (c)};
\node () at (6.4,2.4) {$\greencheckmark[Green]$};
\end{tikzpicture}\quad
\begin{tikzpicture}[baseline=11ex, scale =0.4]
\draw (3,5.67) edge [thick, Green, dash pattern=on 2mm off 1mm, bend right = 15] (4,5) {};
\draw (4,5) edge [thick, Green, dash pattern=on 2mm off 1mm, bend left = 15] (5,3) {};
\draw (4.5,6.67) edge [thick, Red, dash pattern=on 2mm off 1mm, bend left = 15] (4,5) {};
\draw (5.5,7.33) edge [thick, Rhodamine, dash pattern=on 2mm off 1mm, bend left = 15] (4,5) {};
\draw (6,7.67) edge [thick, Rhodamine, dash pattern=on 2mm off 1mm, bend left = 15] (6,2.33) {};
\draw (7,8.33) edge [thick, Orange, dash pattern=on 2mm off 1mm, bend left = 15] (7,1.67) {};
\draw [thick, Blue] (1,5.1) edge (2,5) node [] {};
\draw [thick, Blue] (1,4.9) edge (2,5) node [] {};
\draw [thick] (1.8,5) edge (7.95,9.1) node [] {};
\draw [thick] (2,5) edge (8,9) node [] {};
\draw [thick] (2,5) edge (8,1) node [] {};
\node [draw, Green, circle, minimum size=3pt, fill, inner sep=0pt, outer sep=0pt] () at (4,5) {};
\node () at (2,4.6) {\tiny $O$};
\node () at (5.2,5.7) {\tiny CS};
\node () at (6.8,5) {\tiny CS};
\node () at (8,5) {\tiny S$^2$};
\node () at (5,0.5) {\large (d)};
\node () at (6.4,2.4) {$\crossmark[Red]$};
\end{tikzpicture}
\end{aligned}
\label{eq:mass_typeI_requirement3_examples}
\end{equation}
It is straightforward to verify that Requirement 3 is satisfied for both (a) and (c) but neither (b) nor (d) above. For example, in region (a) one component of $\gamma_{S^2}^{(0)}$ is precisely the edge $e_{AB}^{}$, which is adjacent to $e_{AA'}^{}\in \gamma_{S^2}^{(M)}$. In contrast, in configuration (b), the edge $e_{A'B'}\in \gamma_{S^2}^{(0)}$ is not adjacent to any edge of $\gamma_{S^2}^{(M)}$. (Notice that $e_{OA}^{},e_{AA'}^{}\in \gamma_{H}$ instead.) Similarly, one can verify that region (c) meets Requirement 3 while configuration (d) does not. In fact, the region vectors corresponding to (a) and (c), which are \textsf{(-2,-2,-1,-1,0,-1,-1,-1,-2,-2,-4,-1,-2,-3,-4;1)} and \textsf{(-2,-1,-1,0,-1,-2,-2,-2,-1,-1,-2,-3,-3,-4;1)} respectively, are both presented in the lists of region vectors in appendix~\ref{appendix-mass_expansion_examples}.

We propose that Requirements 1-3 are necessary and sufficient for any configuration of figure~\ref{figure-mass_typeI_general_configuration} to be a valid type-I region in the mass expansion.
\begin{proposition}
\label{proposition-mass_expansion_typeI}
    Every type-I region can be depicted by figure~\ref{figure-mass_typeI_general_configuration}, and Requirements 1-3 must be satisfied. Conversely, every configuration of figure~\ref{figure-mass_typeI_general_configuration} with Requirements 1-3 satisfied is a type-I region.
\end{proposition}
Moving on, we will now study the type-IIA regions.

\subsubsection{Type-IIA regions}
\label{section-typeIIA_regions}

\begin{figure}[t]
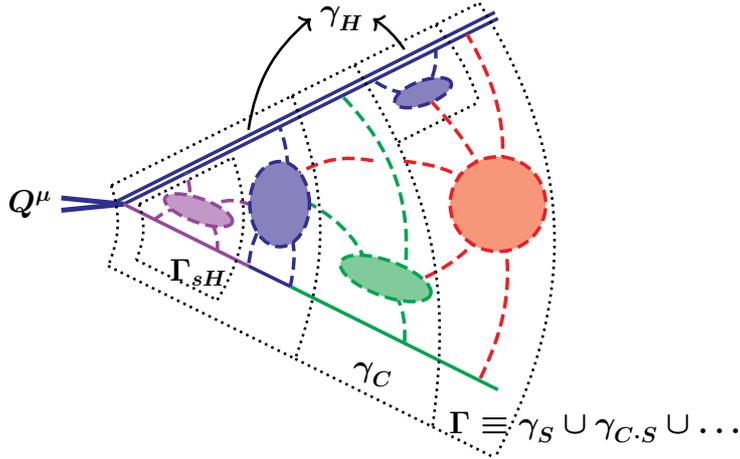

\centering
\include{figs/mass_typeIIA_general_configuration}
\vspace{-3em}\caption{The general configuration of type-IIA regions, where certain mode sensitive subgraphs have been encircled by dotted curves. In comparison with the type-I regions shown in figure~\ref{figure-mass_typeI_general_configuration}, here is a nonempty semihard subgraph $\gamma_{sH}^{}$, which is adjacent to $\gamma_{H}^{}$ only and colored in \textbf{\color{Purple}purple}.}
\label{figure-mass_typeIIA_general_configuration}
\end{figure}

In comparison with type-I regions, the only additional mode in type-IIA regions is the semihard mode. The configuration of a generic type-IIA region, in terms of the mode sensitive subgraphs, is shown in figure~\ref{figure-mass_typeIIA_general_configuration}. In particular, the semihard sensitive subgraph $\gamma_{sH}^{}$ is adjacent to $\gamma_H^{}$ but not to any other subgraphs, and each connected component of $\gamma_{sH}^{}$ contains some mass-$m$ edges and possibly massless edges, but not any mass-$M$ edges.

Constraints on the subgraphs of figure~\ref{figure-mass_typeIIA_general_configuration} are needed for a valid type-IIA region. First, Requirements 1-3, which apply to the type-I regions, are still necessary. Furthermore, there is an additional constraint on the hard sensitive subgraph $\gamma_{H}^{}$. To describe it, let us define $\overline{\gamma}_H^{}$ as the union of $\gamma_{H}^{}$ and all the massive edges whose endpoints are not in $\gamma_{sH}^{}$. Then the extra requirement states as follows.

\emph{Requirement 4: for any edge $e\in \gamma_H^{}$ and any vertex $v\in \gamma_H^{}$, each of the graphs $\overline{\gamma}_H^{} \setminus e$ and $\overline{\gamma}_H^{} \setminus v$ should either be connected or have two connected components, one of which is solely attached by $P^\mu$ or $Q^\mu$.}

One can check that all the type-IIA regions listed in appendix~\ref{appendix-mass_expansion_examples} satisfy this requirement. In contrast, the following configurations do not admit this requirement, hence should not be valid type-IIA regions:
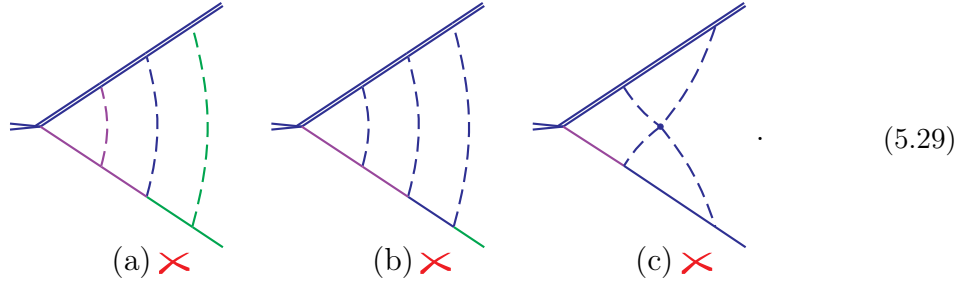
\begin{equation}
\begin{aligned}
\begin{tikzpicture}[baseline=11ex, scale =0.4]

\draw [thick, Blue] (1,5.1) edge (2,5) node [] {};
\draw [thick, Blue] (1,4.9) edge (2,5) node [] {};

\draw (4,3.7) edge [thick, dash pattern=on 2mm off 1mm, Purple, bend right = 15] (4,6.3) {};
\draw (5.5,2.7) edge [thick, dash pattern=on 2mm off 1mm, Blue, bend right = 15] (5.5,7.3) {};
\draw (7,1.7) edge [thick, dash pattern=on 2mm off 1mm, Green, bend right = 15] (7,8.3) {};
\draw [thick, Blue] (8,9) edge [] (5.5,7.33) node [] {};
\draw [thick, Blue] (5.5,7.33) edge [] (4,6.33) node [] {};
\draw [thick, Blue] (4,6.33) edge [] (2,5) node [] {};

\draw [thick, Blue] (7.95,9.1) edge (6.95,8.43) node [] {};
\draw [thick, Blue] (6.95,8.43) edge [] (5.45,7.43) node [] {};
\draw [thick, Blue] (5.45,7.43) edge [] (3.95,6.43) node [] {};
\draw [thick, Blue] (3.95,6.43) edge [] (1.8,5) node [] {};

\draw [thick, Purple] (2,5) edge (4,3.67) node [] {};
\draw [thick, Purple] (4,3.67) edge (5.5,2.67) node [] {};
\draw [thick, Green] (5.5,2.67) edge [] (7,1.67) node [] {};
\draw [thick, Green] (7,1.67) edge [] (8,1) node [] {};

\node () at (5,0.5) {\large (a)};
\node () at (6.4,2.4) {$\crossmark[Red]$};
\end{tikzpicture}\quad
\begin{tikzpicture}[baseline=11ex, scale =0.4]

\draw [thick, Blue] (1,5.1) edge (2,5) node [] {};
\draw [thick, Blue] (1,4.9) edge (2,5) node [] {};

\draw (4,3.7) edge [thick, dash pattern=on 2mm off 1mm, Blue, bend right = 15] (4,6.3) {};
\draw (5.5,2.7) edge [thick, dash pattern=on 2mm off 1mm, Blue, bend right = 15] (5.5,7.3) {};
\draw (7,1.7) edge [thick, dash pattern=on 2mm off 1mm, Blue, bend right = 15] (7,8.3) {};
\draw [thick, Blue] (8,9) edge [] (5.5,7.33) node [] {};
\draw [thick, Blue] (5.5,7.33) edge [] (4,6.33) node [] {};
\draw [thick, Blue] (4,6.33) edge [] (2,5) node [] {};

\draw [thick, Blue] (7.95,9.1) edge (6.95,8.43) node [] {};
\draw [thick, Blue] (6.95,8.43) edge [] (5.45,7.43) node [] {};
\draw [thick, Blue] (5.45,7.43) edge [] (3.95,6.43) node [] {};
\draw [thick, Blue] (3.95,6.43) edge [] (1.8,5) node [] {};

\draw [thick, Purple] (2,5) edge (4,3.67) node [] {};
\draw [thick, Purple] (4,3.67) edge (5.5,2.67) node [] {};
\draw [thick, Blue] (5.5,2.67) edge [] (7,1.67) node [] {};
\draw [thick, Green] (7,1.67) edge [] (8,1) node [] {};

\node () at (5,0.5) {\large (b)};
\node () at (6.4,2.4) {$\crossmark[Red]$};
\end{tikzpicture}\quad
\begin{tikzpicture}[baseline=11ex, scale =0.4]
\draw [thick, Blue] (1,5.1) edge (2,5) node [] {};
\draw [thick, Blue] (1,4.9) edge (2,5) node [] {};
\draw [thick, Blue] (1.8,5) edge (7.95,9.1) node [] {};
\draw [thick, Blue] (2,5) edge (8,9) node [] {};
\draw [thick, Purple] (2,5) edge (4,3.7) node [] {};
\draw [thick, Blue] (4,3.7) edge (8,1) node [] {};
\draw (4,6.3) edge [thick, Blue, dash pattern=on 2mm off 1mm, bend right = 10] (5.2,5) {};
\draw (5.2,5) edge [thick, Blue, dash pattern=on 2mm off 1mm, bend left = 10] (7,1.7) {};
\draw (7,8.3) edge [thick, Blue, dash pattern=on 2mm off 1mm, bend left = 10] (5.2,5) {};
\draw (5.2,5) edge [thick, Blue, dash pattern=on 2mm off 1mm, bend right = 10] (4,3.7) {};
\node [draw, circle, minimum size=2pt, color=Blue, fill=Blue, inner sep=0pt, outer sep=0pt] () at (5.2,5) {};
\node () at (5,0.5) {\large (c)};
\node () at (6.4,2.4) {$\crossmark[Red]$};
\end{tikzpicture}\ .
\end{aligned}
\label{eq:mass_typeIIA_nonregion_configurations_example}
\end{equation}
To see this, we consider the graphs $\overline{\gamma}_H^{}$ associated with the configurations above:
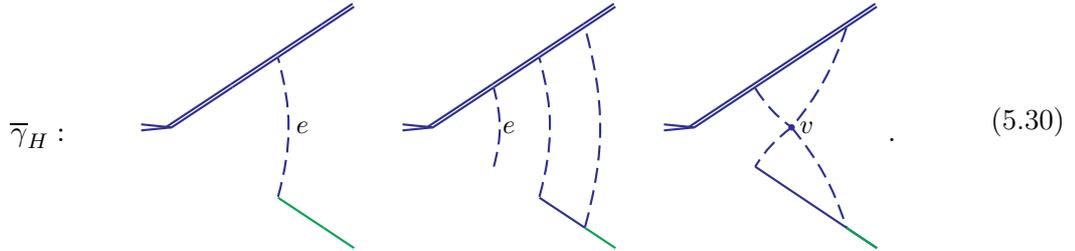
\begin{equation}
\begin{aligned}
\overline{\gamma}_H^{}:\qquad
\begin{tikzpicture}[baseline=11ex, scale =0.4]
\draw [thick, Blue] (1,5.1) edge (2,5) node [] {};
\draw [thick, Blue] (1,4.9) edge (2,5) node [] {};
\draw (5.5,2.7) edge [thick, dash pattern=on 2mm off 1mm, Blue, bend right = 15] (5.5,7.3) {};
\draw [thick, Blue] (8,9) edge [] (5.5,7.33) node [] {};
\draw [thick, Blue] (5.5,7.33) edge [] (4,6.33) node [] {};
\draw [thick, Blue] (4,6.33) edge [] (2,5) node [] {};
\draw [thick, Blue] (7.95,9.1) edge (6.95,8.43) node [] {};
\draw [thick, Blue] (6.95,8.43) edge [] (5.45,7.43) node [] {};
\draw [thick, Blue] (5.45,7.43) edge [] (3.95,6.43) node [] {};
\draw [thick, Blue] (3.95,6.43) edge [] (1.8,5) node [] {};
\draw [thick, Green] (5.5,2.67) edge [] (7,1.67) node [] {};
\draw [thick, Green] (7,1.67) edge [] (8,1) node [] {};
\node () at (6.3,5) {\small $e$};
\end{tikzpicture}\quad
\begin{tikzpicture}[baseline=11ex, scale =0.4]
\draw [thick, Blue] (1,5.1) edge (2,5) node [] {};
\draw [thick, Blue] (1,4.9) edge (2,5) node [] {};
\draw (4,3.7) edge [thick, dash pattern=on 2mm off 1mm, Blue, bend right = 15] (4,6.3) {};
\draw (5.5,2.7) edge [thick, dash pattern=on 2mm off 1mm, Blue, bend right = 15] (5.5,7.3) {};
\draw (7,1.7) edge [thick, dash pattern=on 2mm off 1mm, Blue, bend right = 15] (7,8.3) {};
\draw [thick, Blue] (8,9) edge [] (5.5,7.33) node [] {};
\draw [thick, Blue] (5.5,7.33) edge [] (4,6.33) node [] {};
\draw [thick, Blue] (4,6.33) edge [] (2,5) node [] {};
\draw [thick, Blue] (7.95,9.1) edge (6.95,8.43) node [] {};
\draw [thick, Blue] (6.95,8.43) edge [] (5.45,7.43) node [] {};
\draw [thick, Blue] (5.45,7.43) edge [] (3.95,6.43) node [] {};
\draw [thick, Blue] (3.95,6.43) edge [] (1.8,5) node [] {};
\draw [thick, Blue] (5.5,2.67) edge [] (7,1.67) node [] {};
\draw [thick, Green] (7,1.67) edge [] (8,1) node [] {};
\node () at (4.5,5) {\small $e$};
\end{tikzpicture}\quad
\begin{tikzpicture}[baseline=11ex, scale =0.4]
\draw [thick, Blue] (1,5.1) edge (2,5) node [] {};
\draw [thick, Blue] (1,4.9) edge (2,5) node [] {};
\draw [thick, Blue] (1.8,5) edge (7.95,9.1) node [] {};
\draw [thick, Blue] (2,5) edge (8,9) node [] {};
\draw [thick, Blue] (4,3.7) edge (8,1) node [] {};
\draw (4,6.3) edge [thick, Blue, dash pattern=on 2mm off 1mm, bend right = 10] (5.2,5) {};
\draw (5.2,5) edge [thick, Blue, dash pattern=on 2mm off 1mm, bend left = 10] (7,1.7) {};
\draw (7,8.3) edge [thick, Blue, dash pattern=on 2mm off 1mm, bend left = 10] (5.2,5) {};
\draw (5.2,5) edge [thick, Blue, dash pattern=on 2mm off 1mm, bend right = 10] (4,3.7) {};
\draw [thick, Green] (7,1.67) edge [] (8,1) node [] {};
\node [draw, circle, minimum size=2pt, color=Blue, fill=Blue, inner sep=0pt, outer sep=0pt] () at (5.2,5) {};
\node () at (5.7,5) {\small $v$};
\end{tikzpicture}\ .
\end{aligned}
\end{equation}
In each $\overline{\gamma}_H^{}$ above, there is an edge $e\in \gamma_H^{}$ or a vertex $v\in \gamma_H^{}$ as specified in the figure, such that $\overline{\gamma}_H^{} \setminus e$ or $\overline{\gamma}_H^{} \setminus v$ has two components, one of which is not solely attached by $P^\mu$ or $Q^\mu$. Thus, Requirement 4 is not satisfied, and consequently, the configurations in (\ref{eq:mass_typeIIA_nonregion_configurations_example}) are not type-IIA regions in the mass expansion. However, one can obtain valid type-IIA regions by modifying them into the following:
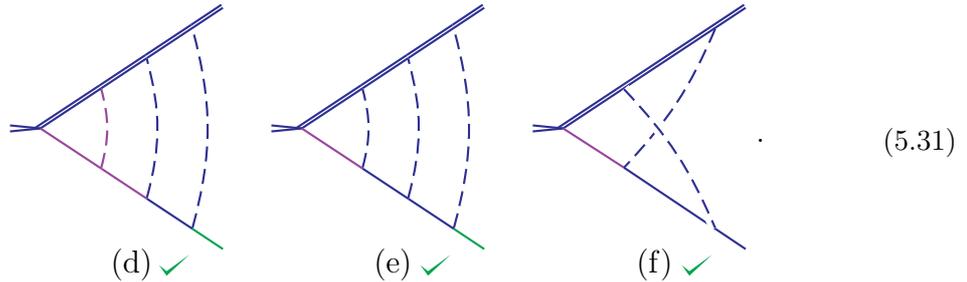
\begin{equation}
\begin{aligned}
\begin{tikzpicture}[baseline=11ex, scale =0.4]

\draw [thick, Blue] (1,5.1) edge (2,5) node [] {};
\draw [thick, Blue] (1,4.9) edge (2,5) node [] {};

\draw (4,3.7) edge [thick, dash pattern=on 2mm off 1mm, Purple, bend right = 15] (4,6.3) {};
\draw (5.5,2.7) edge [thick, dash pattern=on 2mm off 1mm, Blue, bend right = 15] (5.5,7.3) {};
\draw (7,1.7) edge [thick, dash pattern=on 2mm off 1mm, Blue, bend right = 15] (7,8.3) {};
\draw [thick, Blue] (8,9) edge [] (5.5,7.33) node [] {};
\draw [thick, Blue] (5.5,7.33) edge [] (4,6.33) node [] {};
\draw [thick, Blue] (4,6.33) edge [] (2,5) node [] {};

\draw [thick, Blue] (7.95,9.1) edge (6.95,8.43) node [] {};
\draw [thick, Blue] (6.95,8.43) edge [] (5.45,7.43) node [] {};
\draw [thick, Blue] (5.45,7.43) edge [] (3.95,6.43) node [] {};
\draw [thick, Blue] (3.95,6.43) edge [] (1.8,5) node [] {};

\draw [thick, Purple] (2,5) edge (4,3.67) node [] {};
\draw [thick, Purple] (4,3.67) edge (5.5,2.67) node [] {};
\draw [thick, Blue] (5.5,2.67) edge [] (7,1.67) node [] {};
\draw [thick, Green] (7,1.67) edge [] (8,1) node [] {};

\node () at (5,0.5) {\large (d)};
\node () at (6.4,2.4) {$\greencheckmark[Green]$};
\end{tikzpicture}\quad
\begin{tikzpicture}[baseline=11ex, scale =0.4]

\draw [thick, Blue] (1,5.1) edge (2,5) node [] {};
\draw [thick, Blue] (1,4.9) edge (2,5) node [] {};

\draw (4,3.7) edge [thick, dash pattern=on 2mm off 1mm, Blue, bend right = 15] (4,6.3) {};
\draw (5.5,2.7) edge [thick, dash pattern=on 2mm off 1mm, Blue, bend right = 15] (5.5,7.3) {};
\draw (7,1.7) edge [thick, dash pattern=on 2mm off 1mm, Blue, bend right = 15] (7,8.3) {};
\draw [thick, Blue] (8,9) edge [] (5.5,7.33) node [] {};
\draw [thick, Blue] (5.5,7.33) edge [] (4,6.33) node [] {};
\draw [thick, Blue] (4,6.33) edge [] (2,5) node [] {};

\draw [thick, Blue] (7.95,9.1) edge (6.95,8.43) node [] {};
\draw [thick, Blue] (6.95,8.43) edge [] (5.45,7.43) node [] {};
\draw [thick, Blue] (5.45,7.43) edge [] (3.95,6.43) node [] {};
\draw [thick, Blue] (3.95,6.43) edge [] (1.8,5) node [] {};

\draw [thick, Purple] (2,5) edge (4,3.67) node [] {};
\draw [thick, Blue] (4,3.67) edge (5.5,2.67) node [] {};
\draw [thick, Blue] (5.5,2.67) edge [] (7,1.67) node [] {};
\draw [thick, Green] (7,1.67) edge [] (8,1) node [] {};

\node () at (5,0.5) {\large (e)};
\node () at (6.4,2.4) {$\greencheckmark[Green]$};
\end{tikzpicture}\quad
\begin{tikzpicture}[baseline=11ex, scale =0.4]
\draw [thick, Blue] (1,5.1) edge (2,5) node [] {};
\draw [thick, Blue] (1,4.9) edge (2,5) node [] {};
\draw [thick, Blue] (1.8,5) edge (7.95,9.1) node [] {};
\draw [thick, Blue] (2,5) edge (8,9) node [] {};
\draw [thick, Purple] (2,5) edge (4,3.7) node [] {};
\draw [thick, Blue] (4,3.7) edge (8,1) node [] {};
\draw (7,8.3) edge [thick, Blue, dash pattern=on 2mm off 1mm, bend left = 10] (4,3.7) {};
\draw (4,6.3) edge [thick, draw=white, double=white, double distance=3pt, bend left = 10] (7,1.7) {}; \draw (4,6.3) edge [thick, Blue, dash pattern=on 2mm off 1mm, bend left = 10] (7,1.7) {};
\node () at (5,0.5) {\large (f)};
\node () at (6.4,2.4) {$\greencheckmark[Green]$};
\end{tikzpicture}\ .
\end{aligned}
\label{eq:mass_typeIIA_regions_example}
\end{equation}

We now propose that Requirements 1-4 are necessary and sufficient for any configuration of figure~\ref{figure-mass_typeIIA_general_configuration} to be a valid type-I region in the mass expansion.
\begin{proposition}
\label{proposition-mass_expansion_typeIIA}
    Every type-IIA region can be depicted by figure~\ref{figure-mass_typeIIA_general_configuration}, and Requirements 1-4 must be satisfied. Conversely, every configuration of figure~\ref{figure-mass_typeIIA_general_configuration} with Requirements 1-4 satisfied is a type-IIA region.
\end{proposition}

\subsubsection{Type-IIB regions}
\label{section-typeIIB_regions}

\begin{figure}[t]
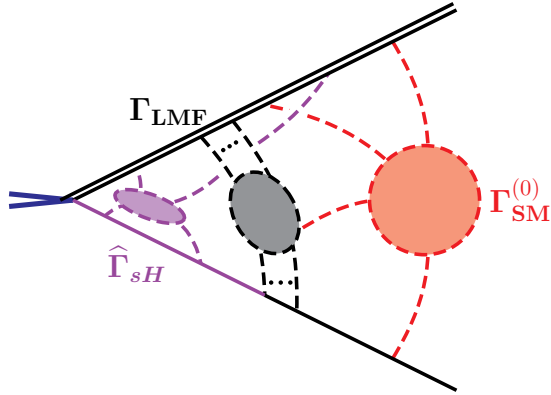

\centering
\include{figs/mass_typeIIB_general_configuration}
\vspace{-3em}\caption{The general configuration of type-IIB regions.}
\label{figure-mass_typeIIB_general_configuration}
\end{figure}

Finally, we explore the general prescription of the type-IIB regions, which can be depicted by figure~\ref{figure-mass_typeIIB_general_configuration}.

To start with, let us recall some key concepts developed at the beginning of section~\ref{section-characterizing_regions}: the graph $\Gamma_\text{LMF}^{}$, marked in black, represents the large momentum flow within $G$; the purple graph $\widehat{\Gamma}_{sH}$ is adjacent to both the $P$ branch and the $p$ branch of $\Gamma_\text{LMF}^{}$, whose edges are all in the semihard mode; the remaining part of the graph, $\Gamma_\text{SM}^{(0)} \equiv G\setminus \widehat{\Gamma}_{sH}\setminus \Gamma_\text{LMF}^{}$ which is in red, contains massless edges only. The subscript ``SM'' represents ``small momentum'' because for each edge $e\in \Gamma_\text{SM}^{(0)}$, all the components of its line momentum are small from this construction.

For convenience, let us use $\Gamma_\text{LMF}^{(0)}$ to denote the union of the massless edges in $\Gamma_\text{LMF}^{}$. Identifying the type-IIB regions of $G$ then amounts to specifying the modes in $\Gamma_\text{SM}^{(0)}$ and $\Gamma_\text{LMF}^{(0)}$. As we summarize from various examples, the modes in these subgraphs are dependent on each other as follows.
\begin{itemize}
    \item [1.] If all the edges of $\Gamma_\text{LMF}^{(0)}$ are in the same mode, then this mode can only be one of the following two possibilities:
    \begin{itemize}
        \item hard;
        \item semicollinear.
    \end{itemize}
    For both possibilities, the modes of each edge in $\Gamma_\text{SM}^{(0)}$ must be in the following form: semihard$\cdot$collinear$^{n_1}\cdot$soft$^{n_2}$ ($sHC^{n_1}S^{n_2}$), with $n_1\in \{0,1\}$ and $n_2\in \mathbb{N}$. The corresponding momentum scaling is:
    \begin{eqnarray}
    \label{eq:mass_expansion_semihard_collinear_power_n1_soft_power_n2_mode}
        l_{sHC^{n_1} S^{n_2}}^\mu\sim Q(\lambda^{n_2+1/2}, \lambda^{n_1+n_2+1/2}, \lambda^{n_1/2+n_2+1/2}).
    \end{eqnarray}
    For example, $(n_1,n_2)=(0,0)$ corresponds to the semihard mode, $(n_1,n_2)=(1,0)$ corresponds to the semihard$\cdot$collinear mode, $(n_1,n_2)=(0,1)$ corresponds to the semihard$\cdot$soft mode, etc. Below we show a few examples of this kind of type-IIB regions.
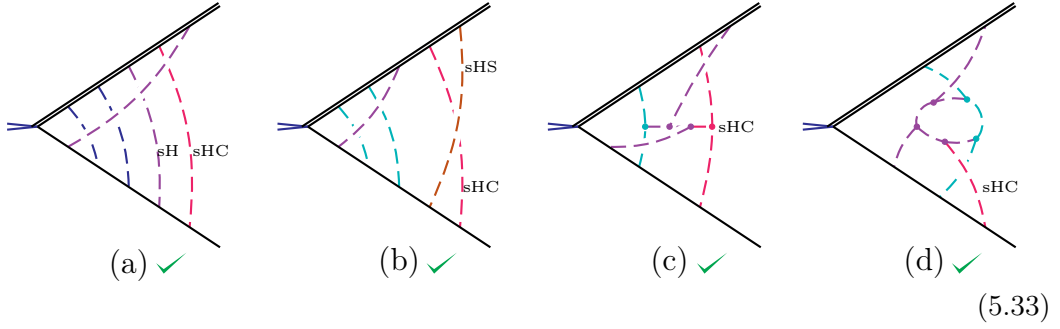
\begin{equation}
\begin{aligned}
\begin{tikzpicture}[baseline=11ex, scale =0.4]
\draw (3,5.67) edge [thick, Blue, dash pattern=on 2mm off 1mm, bend left = 15] (4,3.67) {};
\draw (4,6.33) edge [thick, Blue, dash pattern=on 2mm off 1mm, bend left = 15] (5,3) {};
\draw (5,7) edge [thick, Purple, dash pattern=on 2mm off 1mm, bend left = 15] (6,2.33) {};
\draw (6,7.67) edge [thick, WildStrawberry, dash pattern=on 2mm off 1mm, bend left = 15] (7,1.67) {};
\draw (7,8.33) edge [thick, draw=white, double=white, double distance=3pt, bend left = 15] (3,4.33) {}; \draw (7,8.33) edge [thick, Purple, dash pattern=on 2mm off 1mm, bend left = 15] (3,4.33) {};
\draw [thick, Blue] (1,5.1) edge (2,5) node [] {};
\draw [thick, Blue] (1,4.9) edge (2,5) node [] {};
\draw [thick] (1.8,5) edge (7.95,9.1) node [] {};
\draw [thick] (2,5) edge (8,9) node [] {};
\draw [thick] (2,5) edge (8,1) node [] {};
\node () at (6.3,4.2) {\tiny sH};
\node () at (7.7,4.2) {\tiny sHC};
\node () at (5,1-0.6) {\large (a)};
\node () at (6.4,2.4) {$\greencheckmark[Green]$};
\end{tikzpicture}\quad
\begin{tikzpicture}[baseline=11ex, scale =0.4]
\draw (3,5.67) edge [thick, BlueGreen, dash pattern=on 2mm off 1mm, bend left = 15] (4,3.67) {};
\draw (4,6.33) edge [thick, BlueGreen, dash pattern=on 2mm off 1mm, bend left = 15] (5,3) {};
\draw (5,7) edge [thick, draw=white, double=white, double distance=3pt, bend left = 15] (3,4.33) {}; \draw (5,7) edge [thick, Purple, dash pattern=on 2mm off 1mm, bend left = 15] (3,4.33) {};
\draw (6,7.67) edge [thick, WildStrawberry, dash pattern=on 2mm off 1mm, bend left = 15] (7,1.67) {};
\draw (7,8.33) edge [thick, draw=white, double=white, double distance=3pt, bend left = 15] (6,2.33) {}; \draw (7,8.33) edge [thick, Bittersweet, dash pattern=on 2mm off 1mm, bend left = 15] (6,2.33) {};
\draw [thick, Blue] (1,5.1) edge (2,5) node [] {};
\draw [thick, Blue] (1,4.9) edge (2,5) node [] {};
\draw [thick] (1.8,5) edge (7.95,9.1) node [] {};
\draw [thick] (2,5) edge (8,9) node [] {};
\draw [thick] (2,5) edge (8,1) node [] {};
\node () at (7.7,7) {\tiny sHS};
\node () at (7.7,3) {\tiny sHC};
\node () at (5,0.5) {\large (b)};
\node () at (6.4,2.4) {$\greencheckmark[Green]$};
\end{tikzpicture}\quad
\begin{tikzpicture}[baseline=11ex, scale =0.4]
\draw (4,6.33) edge [thick, BlueGreen, dash pattern=on 2mm off 1mm, bend left = 15] (4,3.67) {};
\draw (6,7.67) edge [thick, WildStrawberry, dash pattern=on 2mm off 1mm, bend left = 15] (6,2.33) {};
\draw (4.2,5) edge [thick, Purple, dash pattern=on 2mm off 1mm] (5.7,5) {};
\draw (5.7,5) edge [thick, WildStrawberry, dash pattern=on 2mm off 1mm] (6.4,5) {};
\draw (5.7,5) edge [thick, draw=white, double=white, double distance=3pt, bend left = 15] (3,4.33) {}; \draw (5.7,5) edge [thick, Purple, dash pattern=on 2mm off 1mm, bend left = 15] (3,4.33) {};
\draw (7,8.33) edge [thick, draw=white, double=white, double distance=3pt, bend right = 5] (5,5) {}; \draw (7,8.33) edge [thick, Purple, dash pattern=on 2mm off 1mm, bend right = 5] (5,5) {};

\node [draw, BlueGreen, circle, minimum size=2pt, fill, inner sep=0pt, outer sep=0pt] () at (4.2,5) {};
\node [draw, Purple, circle, minimum size=2pt, fill, inner sep=0pt, outer sep=0pt] () at (5,5) {};
\node [draw, Purple, circle, minimum size=2pt, fill, inner sep=0pt, outer sep=0pt] () at (5.7,5) {};
\node [draw, WildStrawberry, circle, minimum size=2pt, fill, inner sep=0pt, outer sep=0pt] () at (6.4,5) {};

\draw [thick, Blue] (1,5.1) edge (2,5) node [] {};
\draw [thick, Blue] (1,4.9) edge (2,5) node [] {};
\draw [thick] (1.8,5) edge (7.95,9.1) node [] {};
\draw [thick] (2,5) edge (8,9) node [] {};
\draw [thick] (2,5) edge (8,1) node [] {};
\node () at (7.2,5) {\tiny sHC};
\node () at (5,0.5) {\large (c)};
\node () at (6.4,2.4) {$\greencheckmark[Green]$};
\end{tikzpicture}\quad
\begin{tikzpicture}[baseline=11ex, scale =0.4]
\draw (6.4,5.9) edge [thick, BlueGreen, dash pattern=on 2mm off 1mm, bend right = 15] (5,7) {};
\draw (5.3,5.8) edge [thick, draw=white, double=white, double distance=3pt, bend right = 15] (7,8.33) {}; \draw (5.3,5.8) edge [thick, Purple, dash pattern=on 2mm off 1mm, bend right = 15] (7,8.33) {};
\draw (4.75,5) edge [thick, Purple, dash pattern=on 2mm off 1mm, bend right = 15] (4,3.67) {};
\draw (6.7,4.6) edge [BlueGreen, thick, dash pattern=on 2mm off 1mm, bend left = 15] (5.5,2.67) {};
\draw (5.67,4.5) edge [thick, draw=white, double=white, double distance=3pt, bend left = 15] (7,1.67) {}; \draw (5.67,4.5) edge [thick, WildStrawberry, dash pattern=on 2mm off 1mm, bend left = 15] (7,1.67) {};

\draw [thick, Blue] (1,5.1) edge (2,5) node [] {};
\draw [thick, Blue] (1,4.9) edge (2,5) node [] {};
\draw [thick] (1.8,5) edge (7.95,9.1) node [] {};
\draw [thick] (2,5) edge (8,9) node [] {};
\draw [thick] (2,5) edge (8,1) node [] {};

\draw (6.4,5.9) edge [Purple, thick, dash pattern=on 2mm off 1mm, bend right = 15] (5.3,5.8) {};
\draw (5.3,5.8) edge [Purple, thick, dash pattern=on 2mm off 1mm, bend right = 50] (4.75,5) {};
\draw (4.75,5) edge [Purple, thick, dash pattern=on 2mm off 1mm, bend right = 15] (5.67,4.5) {};
\draw (5.67,4.5) edge [Purple, thick, dash pattern=on 2mm off 1mm, bend right = 15] (6.7,4.6) {};
\draw (6.7,4.6) edge [BlueGreen, thick, dash pattern=on 2mm off 1mm, bend right = 60] (6.4,5.9) {};

\node [draw, BlueGreen, circle, minimum size=2pt, fill, inner sep=0pt, outer sep=0pt] () at (6.4,5.9) {};
\node [draw, Purple, circle, minimum size=2pt, fill, inner sep=0pt, outer sep=0pt] () at (5.3,5.8) {};
\node [draw, Purple, circle, minimum size=2pt, fill, inner sep=0pt, outer sep=0pt] () at (4.75,5) {};
\node [draw, Purple, circle, minimum size=2pt, fill, inner sep=0pt, outer sep=0pt] () at (5.67,4.5) {};
\node [draw, BlueGreen, circle, minimum size=2pt, fill, inner sep=0pt, outer sep=0pt] () at (6.7,4.6) {};
\node () at (7.5,3) {\tiny sHC};
\node () at (5,0.5) {\large (d)};
\node () at (6.4,2.4) {$\greencheckmark[Green]$};
\end{tikzpicture}
\end{aligned}
\label{eq:mass_typeIIB_single_mode_examples}
\end{equation}
    The region vectors corresponding to (a)-(d) are \textsf{(-}$\frac{\textsf{1}}{\textsf{2}}$\textsf{,-}$\frac{\textsf{1}}{\textsf{2}}$\textsf{,-}$\frac{\textsf{1}}{\textsf{2}}$\textsf{,0,0,-1,-1,0,-}$\frac{\textsf{1}}{\textsf{2}}$\textsf{,-}$\frac{\textsf{3}}{\textsf{2}}$\textsf{,-1,0,0,-1,-2;1)}, \textsf{(-}$\frac{\textsf{3}}{\textsf{2}}$\textsf{,-}$\frac{\textsf{1}}{\textsf{2}}$\textsf{,-}$\frac{\textsf{1}}{\textsf{2}}$\textsf{,0,0,-1,-1,-}$\frac{\textsf{1}}{\textsf{2}}$\textsf{,-}$\frac{\textsf{3}}{\textsf{2}}$\textsf{,-}$\frac{\textsf{3}}{\textsf{2}}$\textsf{,-1,-}$\frac{\textsf{1}}{\textsf{2}}$\textsf{,-}$\frac{\textsf{1}}{\textsf{2}}$\textsf{,-3,-2;1)}, \textsf{(-}$\frac{\textsf{1}}{\textsf{2}}$\textsf{,-}$\frac{\textsf{1}}{\textsf{2}}$\textsf{,0,-1,-1,-}$\frac{\textsf{3}}{\textsf{2}}$\textsf{,-}$\frac{\textsf{1}}{\textsf{2}}$\textsf{,-}$\frac{\textsf{1}}{\textsf{2}}$\textsf{,-1,-1,-1,-1,-2,-2,-2;1)}, and \textsf{(-}$\frac{\textsf{1}}{\textsf{2}}$\textsf{,0,-1,-1,-}$\frac{\textsf{3}}{\textsf{2}}$\textsf{,-1,-1,-1,-}$\frac{\textsf{1}}{\textsf{2}}$\textsf{,-1,-1,-2,-1,-}$\frac{\textsf{1}}{\textsf{2}}$\textsf{,-}$\frac{\textsf{1}}{\textsf{2}}$\textsf{;1)} respectively, as one can check in appendix~\ref{appendix-mass_expansion_examples}. In each of these regions, we have specified the modes in $\Gamma_{\text{SM}}^{(0)}$, which are all in the form of $sHC^{n_1}S^{n_2}$, as indicated above.
    
    \item [2.] If $\Gamma_\text{LMF}^{(0)}$ contain edges in distinct modes, then there are three following possibilities:
    \begin{itemize}
        \item semicollinear \& collinear;
        \item hard \& collinear;
        \item hard \& semicollinear.
    \end{itemize}
    For all these possibilities, the modes of each edge in $\Gamma_\text{SM}^{(0)}$ must be in the following form: semicollinear$^{n_1}\cdot$semihard$^{n_2}$ ($sC^{n_1}sH^{n_2}$), with $n_1\in \{0,1\}$ and $n_2\in \mathbb{N}_+$. The corresponding momentum scaling is:
    \begin{eqnarray}
    \label{eq:mass_expansion_semihard_power_n0_semicollinear_power_n1_soft_power_n2_mode}
        l_{sC^{n_1} sH^{n_2}}^\mu\sim Q(\lambda^{n_2/2}, \lambda^{n_1/2+n_2/2}, \lambda^{n_1/4+n_2/2}).
    \end{eqnarray}
    For example, $(n_1,n_2)=(0,1)$ corresponds to the semihard mode, $(n_1,n_2)=(1,1)$ corresponds to the semihard$\cdot$semicollinear mode, $(n_1,n_2)=(0,2)$ corresponds to the soft mode, etc. Below we show a few examples of this kind of type-IIB regions.
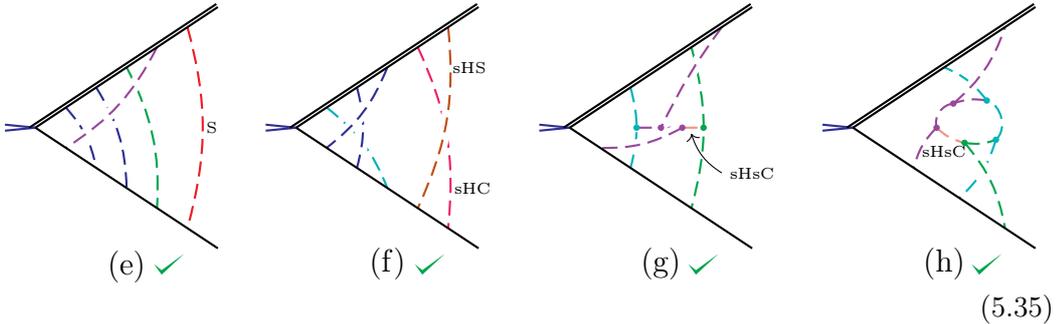
\begin{equation}
\begin{aligned}
\begin{tikzpicture}[baseline=11ex, scale =0.4]
\draw (3,5.67) edge [thick, Blue, dash pattern=on 2mm off 1mm, bend left = 15] (4,3.67) {};
\draw (4,6.33) edge [thick, Blue, dash pattern=on 2mm off 1mm, bend left = 15] (5,3) {};
\draw (5,7) edge [thick, Green, dash pattern=on 2mm off 1mm, bend left = 15] (6,2.33) {};
\draw (6,7.67) edge [thick, draw=white, double=white, double distance=3pt, bend left = 15] (3,4.33) {}; \draw (6,7.67) edge [thick, Purple, dash pattern=on 2mm off 1mm, bend left = 15] (3,4.33) {};
\draw (7,8.33) edge [thick, Red, dash pattern=on 2mm off 1mm, bend left = 15] (7,1.67) {};

\draw [thick, Blue] (1,5.1) edge (2,5) node [] {};
\draw [thick, Blue] (1,4.9) edge (2,5) node [] {};
\draw [thick] (1.8,5) edge (7.95,9.1) node [] {};
\draw [thick] (2,5) edge (8,9) node [] {};
\draw [thick] (2,5) edge (8,1) node [] {};
\node () at (7.8,5) {\tiny S};
\node () at (5,1-0.6) {\large (e)};
\node () at (6.4,2.4) {$\greencheckmark[Green]$};
\end{tikzpicture}\quad
\begin{tikzpicture}[baseline=11ex, scale =0.4]
\draw (3,5.67) edge [thick, BlueGreen, dash pattern=on 2mm off 1mm, bend left = 15] (5,3) {};
\draw (4,6.33) edge [thick, draw=white, double=white, double distance=3pt, bend left = 15] (4,3.67) {}; \draw (4,6.33) edge [thick, Blue, dash pattern=on 2mm off 1mm, bend left = 15] (4,3.67) {};
\draw (5,7) edge [thick, draw=white, double=white, double distance=3pt, bend left = 15] (3,4.33) {}; \draw (5,7) edge [thick, Blue, dash pattern=on 2mm off 1mm, bend left = 15] (3,4.33) {};
\draw (6,7.67) edge [thick, WildStrawberry, dash pattern=on 2mm off 1mm, bend left = 15] (7,1.67) {};
\draw (7,8.33) edge [thick, draw=white, double=white, double distance=3pt, bend left = 15] (6,2.33) {}; \draw (7,8.33) edge [thick, Bittersweet, dash pattern=on 2mm off 1mm, bend left = 15] (6,2.33) {};
\draw [thick, Blue] (1,5.1) edge (2,5) node [] {};
\draw [thick, Blue] (1,4.9) edge (2,5) node [] {};
\draw [thick] (1.8,5) edge (7.95,9.1) node [] {};
\draw [thick] (2,5) edge (8,9) node [] {};
\draw [thick] (2,5) edge (8,1) node [] {};
\node () at (7.7,7) {\tiny sHS};
\node () at (7.8,3) {\tiny sHC};
\node () at (5,0.5) {\large (f)};
\node () at (6.4,2.4) {$\greencheckmark[Green]$};
\end{tikzpicture}\quad
\begin{tikzpicture}[baseline=11ex, scale =0.4]
\draw (4,6.33) edge [thick, BlueGreen, dash pattern=on 2mm off 1mm, bend left = 15] (4,3.67) {};
\draw (6,7.67) edge [thick, Green, dash pattern=on 2mm off 1mm, bend left = 15] (6,2.33) {};
\draw (4.2,5) edge [thick, Purple, dash pattern=on 2mm off 1mm] (5.7,5) {};
\draw (5.7,5) edge [thick, Salmon, dash pattern=on 2mm off 1mm] (6.4,5) {};
\draw (5.7,5) edge [thick, draw=white, double=white, double distance=3pt, bend left = 15] (3,4.33) {}; \draw (5.7,5) edge [thick, Purple, dash pattern=on 2mm off 1mm, bend left = 15] (3,4.33) {};
\draw (7,8.33) edge [thick, draw=white, double=white, double distance=3pt, bend right = 5] (5,5) {}; \draw (7,8.33) edge [thick, Purple, dash pattern=on 2mm off 1mm, bend right = 5] (5,5) {};
\node [draw, BlueGreen, circle, minimum size=2pt, fill, inner sep=0pt, outer sep=0pt] () at (4.2,5) {};
\node [draw, Purple, circle, minimum size=2pt, fill, inner sep=0pt, outer sep=0pt] () at (5,5) {};
\node [draw, Purple, circle, minimum size=2pt, fill, inner sep=0pt, outer sep=0pt] () at (5.7,5) {};
\node [draw, Green, circle, minimum size=2pt, fill, inner sep=0pt, outer sep=0pt] () at (6.4,5) {};
\draw [thick, Blue] (1,5.1) edge (2,5) node [] {};
\draw [thick, Blue] (1,4.9) edge (2,5) node [] {};
\draw [thick] (1.8,5) edge (7.95,9.1) node [] {};
\draw [thick] (2,5) edge (8,9) node [] {};
\draw [thick] (2,5) edge (8,1) node [] {};
\draw (7,3.5) edge [bend left = 20, ->] (6,4.8) {};
\node () at (8,3.5) {\tiny sHsC};
\node () at (5,0.5) {\large (g)};
\node () at (6.4,2.4) {$\greencheckmark[Green]$};
\end{tikzpicture}\quad
\begin{tikzpicture}[baseline=11ex, scale =0.4]
\draw (6.4,5.9) edge [thick, BlueGreen, dash pattern=on 2mm off 1mm, bend right = 15] (5,7) {};
\draw (5.3,5.8) edge [thick, draw=white, double=white, double distance=3pt, bend right = 15] (7,8.33) {}; \draw (5.3,5.8) edge [thick, Purple, dash pattern=on 2mm off 1mm, bend right = 15] (7,8.33) {};
\draw (4.75,5) edge [thick, Purple, dash pattern=on 2mm off 1mm, bend right = 15] (4,3.67) {};
\draw (6.7,4.6) edge [BlueGreen, thick, dash pattern=on 2mm off 1mm, bend left = 15] (5.5,2.67) {};
\draw (5.67,4.5) edge [thick, draw=white, double=white, double distance=3pt, bend left = 15] (7,1.67) {}; \draw (5.67,4.5) edge [thick, Green, dash pattern=on 2mm off 1mm, bend left = 15] (7,1.67) {};
\draw [thick, Blue] (1,5.1) edge (2,5) node [] {};
\draw [thick, Blue] (1,4.9) edge (2,5) node [] {};
\draw [thick] (1.8,5) edge (7.95,9.1) node [] {};
\draw [thick] (2,5) edge (8,9) node [] {};
\draw [thick] (2,5) edge (8,1) node [] {};
\draw (6.4,5.9) edge [Purple, thick, dash pattern=on 2mm off 1mm, bend right = 15] (5.3,5.8) {};
\draw (5.3,5.8) edge [Purple, thick, dash pattern=on 2mm off 1mm, bend right = 50] (4.75,5) {};
\draw (4.75,5) edge [Salmon, thick, dash pattern=on 2mm off 1mm, bend right = 15] (5.67,4.5) {};
\draw (5.67,4.5) edge [Green, thick, dash pattern=on 2mm off 1mm, bend right = 15] (6.7,4.6) {};
\draw (6.7,4.6) edge [BlueGreen, thick, dash pattern=on 2mm off 1mm, bend right = 60] (6.4,5.9) {};
\node [draw, BlueGreen, circle, minimum size=2pt, fill, inner sep=0pt, outer sep=0pt] () at (6.4,5.9) {};
\node [draw, Purple, circle, minimum size=2pt, fill, inner sep=0pt, outer sep=0pt] () at (5.3,5.8) {};
\node [draw, Purple, circle, minimum size=2pt, fill, inner sep=0pt, outer sep=0pt] () at (4.75,5) {};
\node [draw, Green, circle, minimum size=2pt, fill, inner sep=0pt, outer sep=0pt] () at (5.67,4.5) {};
\node [draw, BlueGreen, circle, minimum size=2pt, fill, inner sep=0pt, outer sep=0pt] () at (6.7,4.6) {};
\node () at (5,4.2) {\tiny sHsC};
\node () at (5,0.5) {\large (h)};
\node () at (6.4,2.4) {$\greencheckmark[Green]$};
\end{tikzpicture}
\end{aligned}
\label{eq:mass_typeIIB_multiple_mode_examples}
\end{equation}
    The region vectors corresponding to (e)-(h) are \textsf{(-1,-}$\frac{\textsf{1}}{\textsf{2}}$\textsf{,0,0,0,-1,-1,0,-1,-1,-1,0,0,-1,-2;1)}, \textsf{(-}$\frac{\textsf{3}}{\textsf{2}}$\textsf{,-}$\frac{\textsf{1}}{\textsf{2}}$\textsf{,0,-}$\frac{\textsf{1}}{\textsf{2}}$\textsf{,0,-1,0,-1,-}$\frac{\textsf{3}}{\textsf{2}}$\textsf{,-}$\frac{\textsf{3}}{\textsf{2}}$\textsf{,0,0,-}$\frac{\textsf{1}}{\textsf{2}}$\textsf{,-3,-2;1)}, \textsf{(-}$\frac{\textsf{1}}{\textsf{2}}$\textsf{,0,0,-1,-1,-1,-}$\frac{\textsf{1}}{\textsf{2}}$\textsf{,-}$\frac{\textsf{1}}{\textsf{2}}$\textsf{,-1,-1,-1,-1,-}$\frac{\textsf{3}}{\textsf{2}}$\textsf{,-1,-1;1)}, and \textsf{(-}$\frac{\textsf{1}}{\textsf{2}}$\textsf{,0,-1,-1,-1,-1,-1,-1,-}$\frac{\textsf{1}}{\textsf{2}}$\textsf{,-1,-}$\frac{\textsf{3}}{\textsf{2}}$\textsf{,-1,-1,-}$\frac{\textsf{1}}{\textsf{2}}$\textsf{,-}$\frac{\textsf{1}}{\textsf{2}}$\textsf{;1)}, respectively, as one can check in appendix~\ref{appendix-mass_expansion_examples}. In each of these regions, we have specified the modes in $\Gamma_{\text{SM}}^{(0)}$, which are all in the form of $sC^{n_1}sH^{n_2}$, as indicated above.
\end{itemize}

Summarizing the examples provided, we reiterate two previously made statements. Firstly, the association of type-IIB regions with the nonplanar nature of $G$ is evident. This connection arises due to the adjacency of $\widehat{\Gamma}_{sH}$ to both the $P$ branch and the $p$ branch of $\Gamma_\text{LMF}$, as depicted in figure~\ref{figure-mass_typeIIB_general_configuration}. Secondly, the modes within $\Gamma_\text{SM}^{(0)}$ depend on the modes within $\Gamma_\text{LMF}^{(0)}$. This dependency manifests in the various possibilities we enumerate above, and, as validated by the examples in appendix~\ref{appendix-mass_expansion_examples}, it consistently holds for all regions examined.

While the examples shed light on the general configuration of type-IIB regions, additional constraints, like those we have proposed for type-I and IIA regions in sections~\ref{section-typeI_regions} and \ref{section-typeIIA_regions}, are necessary to understand the intricate structure comprehensively. Due to the complexity involved, we will leave a detailed exploration of these constraints to future works.

\subsection{A formalism to visualize the regions for planar graphs}
\label{section-formalism_describing_regions_planar_graphs}

As type-IIB regions arise exclusively from nonplanar Feynman graphs, only type-I and type-IIA regions are pertinent when we focus solely on planar graphs. In this subsection, we present a formalism in which each region of a planar graph can be naturally visualized as a 3-dimensional object—-a ``terrace''.

For any given loop of a planar graph $G$, we refer to it as a \emph{fundamental loop}, if the area enclosed by this loop cannot be further divided by any other edges of $G$. In momentum space, each region can be characterized by the modes of the fundamental loop momenta of $G$. According to the properties of type-I and type-IIA regions, these modes can be either semihard or collinear$^{n_1}\cdot$soft$^{n_2}$ ($n_1\in \{0,1\}$ and $n_2\in \mathbb{N}$).

To visualize the regions of $G$, a key operation is to place the area enclosed by each fundamental loop at a certain altitude $-i$, where $i$ is equal to the OSD of the $X$-mode massless edges. Namely, $i=1$ for $X=sH$ and $i=n_1+2n_2$ for $X=C^{n_1}S^{n_2}$. After this operation, the graph $G$ transforms into a 3-dimensional graph, referred to as the \emph{terrace} corresponding to the region $R$ and denoted as $\mathcal{T}_R$. The surface of $\mathcal{T}_R$ originating from the area enclosed by the $X$-mode loops is termed the \emph{$X$-surface} and denoted as $\mathcal{S}_X$. Among all the surfaces of $\mathcal{T}_R$, the ``highest one'', i.e., the one with the largest altitude, is either an $H$-surface (with altitude being $0$) or a $C$-surface (with altitude being $-1$). The altitude of any other surface of $\mathcal{T}_R$ is a negative integer.

\begin{figure}[t]
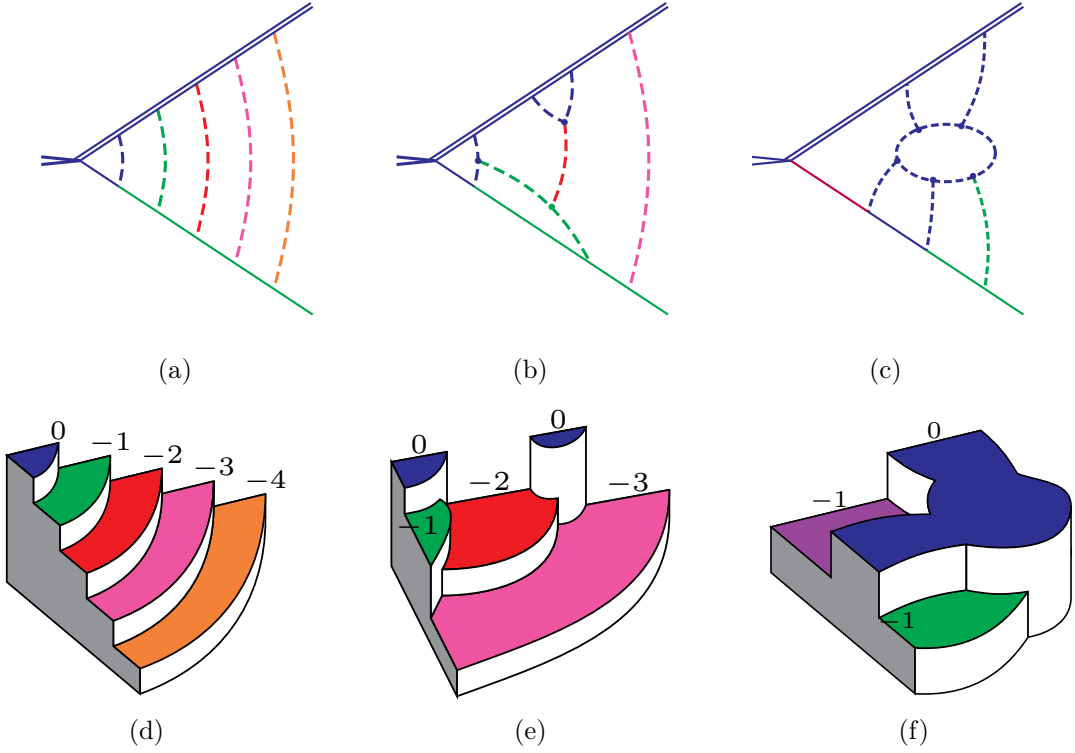

\centering
\begin{subfigure}[b]{0.25\textwidth}
\centering
\include{figs/mass_expansion_example_5loop_graph1}
\vspace{-2em}\caption{}
\label{mass_expansion_example_5loop_graph1}
\end{subfigure}
\qquad
\begin{subfigure}[b]{0.25\textwidth}
\centering
\include{figs/mass_expansion_example_5loop_graph2}
\vspace{-2em}\caption{}
\label{mass_expansion_example_5loop_graph2}
\end{subfigure}
\qquad
\begin{subfigure}[b]{0.25\textwidth}
\centering
\include{figs/mass_expansion_example_5loop_graph3}
\vspace{-2em}\caption{}
\label{mass_expansion_example_5loop_graph3}
\end{subfigure}
\\
\begin{subfigure}[b]{0.3\textwidth}
\include{figs/mass_expansion_example_5loop_graph1_terrace}
\vspace{-1em}\caption{}
\label{figure-mass_expansion_example_5loop_graph1_terrace}
\end{subfigure}
\quad
\begin{subfigure}[b]{0.3\textwidth}
\include{figs/mass_expansion_example_5loop_graph2_terrace}\vspace{10pt}
\vspace{-2em}\caption{}
\label{figure-mass_expansion_example_5loop_graph2_terrace}
\end{subfigure}
\quad
\begin{subfigure}[b]{0.3\textwidth}
\include{figs/mass_expansion_example_5loop_graph3_terrace}\vspace{10pt}
\vspace{-2em}\caption{}
\label{figure-mass_expansion_example_5loop_graph3_terrace}
\end{subfigure}
\caption{An illustration of the terrace formalism. Figures (a), (b),  and (c) depict three examples of regions in the mass expansion of five-loop graphs, where we have colored each hard-mode edge in \textbf{\color{Blue}blue}, each collinear-mode edge in \textbf{\color{Green}green}, each soft-mode edge in \textbf{\color{Red}red}, each collinear$\cdot$soft-mode edge in \textbf{\color{Rhodamine}rhodamine}, each soft${}^2$-mode edge in \textbf{\color{Orange}orange}, and each semihard-mode edge in \textbf{\color{Purple}purple}. Figures (d), (e), and (f) show their corresponding terraces, with surfaces colored according to the associated modes of the loop momenta.}
\label{figure-mass_expansion_example_5loop_graphs}
\end{figure}

Let us illustrate these concepts through the following examples. The region vectors \textsf{(-2,-1,-1,0,0,0,-1,-1,-2,-2,0,-1,-2,-3,-4;1)} (associated with figure~\ref{mass_expansion_examples_planar_1} in appendix~\ref{appendix-mass_expansion_examples}), \textsf{(-1,0,-1,0,0,-1,-2,0,0,-1,0,0,-2,-1,-3;1)} (associated with figure~\ref{mass_expansion_examples_planar_4}), and \textsf{(0,0,-1,0,-1,0,0,0,0,0,0,0,0,-1,0;1)} (associated with figure~\ref{mass_expansion_examples_planar_7}), correspond to the terraces shown in figures~\ref{mass_expansion_example_5loop_graph1}, \ref{mass_expansion_example_5loop_graph2}, and \ref{mass_expansion_example_5loop_graph3}, respectively. The surfaces of each terrace is marked in distinct colors, with the corresponding altitudes labelled next to them. Note that some $X$-surfaces can be disconnected. For example, the $H$-surface in figure~\ref{figure-mass_expansion_example_5loop_graph2_terrace} has two connected components.

This formalism enables us to visualize each region of a planar graph as a corresponding terrace. For later convenience, given a set of modes $X_1,X_2,\dots$, we define $\Gamma_{M,m}^\text{proj}[X_1,X_2,\dots]$ as the 2-dimensional graph obtained from projecting the $M$ boundary, the $m$ boundary, and the surfaces $\mathcal{S}_{X_1},\mathcal{S}_{X_2},\dots$ to the same altitude. With this construction, one can check that the graph $\Gamma_{M,m}^\text{proj}[H]$ is simply connected in figures~\ref{figure-mass_expansion_example_5loop_graph1_terrace} and \ref{figure-mass_expansion_example_5loop_graph2_terrace}, while not simply connected in figure~\ref{figure-mass_expansion_example_5loop_graph3_terrace}. This motivates us to construct terraces corresponding to the regions of $G$, as we will presented later.

As another key advantage, the requirements of generic type-I and type-IIA regions, as summarized in sections~\ref{section-typeI_regions} and \ref{section-typeIIA_regions}, can be translated into the requirements of $\mathcal{T}_R$, leading to a graph-finding algorithm to construct the regions. As we will see below, the complete set of terraces, each corresponding to a region of $G$, can be obtained from the following \emph{terrace-finding algorithm} consisting of the following steps.
\begin{itemize}
    \item \emph{Step 1}. Constructing the $H$-surface $\mathcal{S}_H$: we choose some fundamental loops of $G$, place their enclosed area at the altitude $0$, and define the result as $\mathcal{S}_H$. We further require that each component of $\mathcal{S}_H$ is adjacent to the $M$ boundary.
    \item \emph{Step 2}. Constructing the $sH$-surface $\mathcal{S}_{sH}$: if $\Gamma_{M,m}^\text{proj}[H]$ is simply connected, then $\mathcal{S}_{sH}=\varnothing$; otherwise, we place the area enclosed by $\Gamma_{M,m}^\text{proj}[H]$ at the altitude $-1$ and define the result as $\mathcal{S}_{sH}$. We further require that each component of $\mathcal{S}_{sH}$ is adjacent to the $m$ boundary.
    \item \emph{Step 3}. Constructing the $C$-surface $\mathcal{S}_C$: among the remaining fundamental loops of $G$, we choose some of them, place their enclosed area at the altitude $-1$, and define the result as $\mathcal{S}_H$. We further require that $\mathcal{S}_C$ is nonempty, and each its component is adjacent to the $m$ boundary.
    \item \emph{Step 4}. Construct the $C^{n_1}S^{n_2}$-surfaces $\mathcal{S}_{C^{n_1}S^{n_2}}$ for $n_1\in\{0,1\}$ and $n_2\in \mathbb{N}_+$: among the remaining fundamental loops of $G$, we choose some of them, place their enclosed area at the altitude $-n_1-2n_2$, and define the result as $\mathcal{S}_{C^{n_1}S^{n_2}}$. We further require the following:
    \begin{itemize}
        \item each of them is connected;
        \item for each $n\in\mathbb{N}_+$, $\mathcal{S}_{S^n}$ is adjacent to both $\mathcal{S}_{CS^{n-1}}$ and the $M$ boundary, and $\mathcal{S}_{CS^n}$ is adjacent to both $\mathcal{S}_{S^n}$ and the $m$ boundary;
        \item for each given $n_1$ and $n_2$, the graph $\Gamma_{M,m}^\text{proj}[H,sH,C,S,\dots,C^{n_1}S^{n_2}]$ is simply connected.
    \end{itemize}
    \item \emph{Step 5}. Checking the requirements of $R$: for the terrace $\mathcal{T}_R$ obtained from above, we require the corresponding region $R$ satisfies all Requirements 1-4, as stated in sections~\ref{section-typeI_regions} and \ref{section-typeIIA_regions}. The region would be ruled out if any of them are not satisfied.
\end{itemize}

We now demonstrate the equivalence between this algorithm and the construction of terraces from the regions. For convenience, we will denote the union of the $X$-mode fundamental loops of $G$ by $\Gamma_X$.
\begin{theorem}
\label{theorem-mass_expansion_equivalence_subgraphs_terraces}
    The terrace-finding algorithm above outputs precisely the set of terraces corresponding to the regions of a planar graph $G$.
\end{theorem}
\begin{proof}
    From Step 5 of the terrace-finding algorithm, it is obvious that its outputs automatically satisfy Requirements 1-4 in sections~\ref{section-typeI_regions} and \ref{section-typeIIA_regions}. We still need to prove that every region of $G$ can be obtained by the algorithm. It then suffices to check that all the requirements on $\mathcal{T}_R$ are automatically satisfied for the region $R$.

    First, in the construction of the $H$-surface $\mathcal{S}_H$, the only requirement is that each of its components is adjacent to the $M$ boundary. This is automatically satisfied in figure~\ref{figure-mass_typeI_general_configuration} or \ref{figure-mass_typeIIA_general_configuration}.

    If $R$ is a type-I region, then $\Gamma_{sH}=\varnothing$, and from figure~\ref{figure-mass_typeI_general_configuration}, the graph $\Gamma_{M,m}^\text{proj}[H]$ is always simply connected. If $R$ is a type-IIA region, then $\Gamma_{sH}\neq \varnothing$, each of whose components is adjacent to the $m$ boundary. Furthermore, from figure~\ref{figure-mass_typeIIA_general_configuration}, the graph $\Gamma_{M,m}^\text{proj}[H]$ is not simply connected, while $\Gamma_{M,m}^\text{proj}[H,sH]$ is. It is clear that either case conforms with the construction of $\mathcal{S}_{sH}$ in Step 2.
    
    Next, in the construction of the $C$-surface $\mathcal{S}_C$, the only requirement is that each of its components is adjacent to the $m$ boundary. This is, again, automatically satisfied in figure~\ref{figure-mass_typeI_general_configuration} or \ref{figure-mass_typeIIA_general_configuration}.

    Then, we study the properties of the graph $\Gamma_S$. A key observation is that $\Gamma_S$ contains the soft-sensitive subgraph, namely, $\gamma_S^{}\subset \Gamma_S$. To see this, note that by definition, the momentum of each edge $e\in \gamma_S^{}$ depends on some soft-mode momentum, thus $e\in \Gamma_S$. Then on the one hand, Requirement 1 implies that $\gamma_S^{}$ is adjacent to the $M$ boundary. On the other hand, Requirement 3 states that each component of $\gamma_{S}^{(0)}$ is adjacent to $\gamma_S^{(M)}$. As $G$ is planar, to satisfy both these conditions, $\Gamma_S$ must be connected and adjacent to the $M$ boundary. Furthermore, from Requirement 2, which ensures $G_2/G_0 = (\gamma_C^{}\cup \gamma_S^{})/\gamma_H^{}$ to be 1VI, the graphs $\Gamma_C$ and $\Gamma_S$ must have at least one edge in common. Additionally, note that the graph $\Gamma_{M,m}^\text{proj}[H,sH,C,S]$ must be simply connected, otherwise Requirement 3 would not be satisfied for some mode $X\in \{CS, S^2,\dots\}$.
    
    These properties of $\Gamma_S$, as summarized above, are precisely the requirements of $\mathcal{S}_S$ in Step 4. In other words, the graph $\Gamma_S$ can be obtained from the construction of the $S$-surface $\mathcal{S}_S$. By applying the same analysis, we verify that the graphs $\Gamma_{CS}$, $\Gamma_{S^2}$, etc., can also be obtained from constructing the corresponding surfaces.
    
    As a result, every region of $G$ can be obtained from the terrace-finding algorithm. The proof is completed.
\end{proof}

Based on theorem~\ref{theorem-mass_expansion_equivalence_subgraphs_terraces}, one can obtain the complete set of regions for a planar graph by finding all its corresponding terraces. This provides a graph-finding approach to the heavy-to-light regions without the need of constructing Newton polytopes. It is worth noting that for other types of mass expansion, a modified terrace formalism may be applicable to identify all the relevant regions. For instance, a comprehensive region analysis for the two-loop QED massive form factor has been conducted in a recent study~\cite{tHvLnMrnsVnzWang23}, where the internal edges are either massless, or of the mass $m\ll Q$. We believe that by modifying the algorithm above, one can also reproduce the complete set of regions in such cases. This will be left for future work.

Let us end this section by obtaining the 11 regions of figure~\ref{mass_expansion_example_graph2} using this approach. In the outputs of the terrace-finding algorithm, there is exactly one terrace with a surface whose altitude is $-3$:
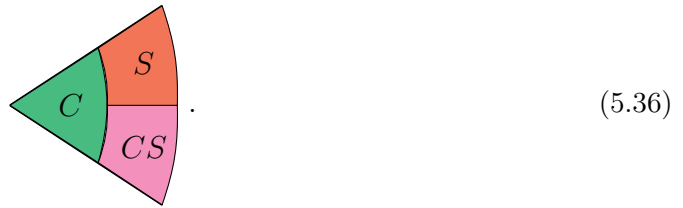
\begin{equation}
\label{eq:terrace_example1}
\begin{aligned}
\begin{tikzpicture}[baseline=11ex, scale =0.4]
\draw [thick] (2,5) edge (7,8.3) node [] {};
\draw [thick] (2,5) edge (7,1.7) node [] {};
\draw [fill=Green!67] (2,5) coordinate [] () -- (4.9,3.1) coordinate [] () to[bend right = 15] (4.9,6.9) coordinate [] () -- cycle;
\draw [fill=Rhodamine!67] (4.9,3.1) coordinate [] () -- (7,1.7) coordinate [] () to[bend right = 10] (7.5,5) coordinate [] () -- (5.2,5) coordinate [] () to[bend left = 10] cycle;
\draw [fill=Red!67] (4.9,6.9) coordinate [] () -- (7,8.3) coordinate [] () to[bend left = 10] (7.5,5) coordinate [] () -- (5.2,5) coordinate [] () to[bend right = 10] cycle;
\node () at (4,5) {\large $C$};
\node () at (6.4,6.4) {\large $S$};
\node () at (6.4,3.6) {\large $CS$};
\end{tikzpicture}
\end{aligned}\ .
\end{equation}
For simplicity, we have only shown the vertical view of the terrace. This terrace corresponds to the region where the momenta of the three fundamental loops are in the collinear mode, the soft mode, and the collinear$\cdot$soft mode respectively. Equivalently, the associated region vector is \textsf{(-1,0,-1,-2,-1,-1,-2,-3,-2;1)}.

There are four terraces such that the lowest altitude of their surfaces is $-2$:
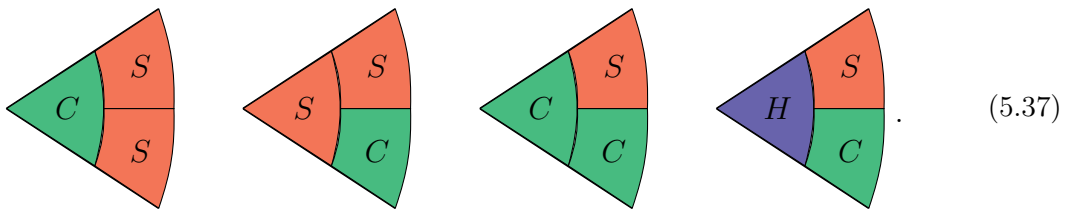
\begin{equation}
\label{eq:terrace_example2}
\begin{aligned}
\begin{tikzpicture}[baseline=11ex, scale =0.4]
\draw [thick] (2,5) edge (7,8.3) node [] {};
\draw [thick] (2,5) edge (7,1.7) node [] {};
\draw [fill=Green!67] (2,5) coordinate [] () -- (4.9,3.1) coordinate [] () to[bend right = 15] (4.9,6.9) coordinate [] () -- cycle;
\draw [fill=Red!67] (4.9,3.1) coordinate [] () -- (7,1.7) coordinate [] () to[bend right = 10] (7.5,5) coordinate [] () -- (5.2,5) coordinate [] () to[bend left = 10] cycle;
\draw [fill=Red!67] (4.9,6.9) coordinate [] () -- (7,8.3) coordinate [] () to[bend left = 10] (7.5,5) coordinate [] () -- (5.2,5) coordinate [] () to[bend right = 10] cycle;
\node () at (4,5) {\large $C$};
\node () at (6.4,6.4) {\large $S$};
\node () at (6.4,3.6) {\large $S$};
\end{tikzpicture}
\qquad
\begin{tikzpicture}[baseline=11ex, scale =0.4]
\draw [thick] (2,5) edge (7,8.3) node [] {};
\draw [thick] (2,5) edge (7,1.7) node [] {};
\draw [fill=Red!67] (2,5) coordinate [] () -- (4.9,3.1) coordinate [] () to[bend right = 15] (4.9,6.9) coordinate [] () -- cycle;
\draw [fill=Green!67] (4.9,3.1) coordinate [] () -- (7,1.7) coordinate [] () to[bend right = 10] (7.5,5) coordinate [] () -- (5.2,5) coordinate [] () to[bend left = 10] cycle;
\draw [fill=Red!67] (4.9,6.9) coordinate [] () -- (7,8.3) coordinate [] () to[bend left = 10] (7.5,5) coordinate [] () -- (5.2,5) coordinate [] () to[bend right = 10] cycle;
\node () at (4,5) {\large $S$};
\node () at (6.4,6.4) {\large $S$};
\node () at (6.4,3.6) {\large $C$};
\end{tikzpicture}
\qquad
\begin{tikzpicture}[baseline=11ex, scale =0.4]
\draw [thick] (2,5) edge (7,8.3) node [] {};
\draw [thick] (2,5) edge (7,1.7) node [] {};
\draw [fill=Green!67] (2,5) coordinate [] () -- (4.9,3.1) coordinate [] () to[bend right = 15] (4.9,6.9) coordinate [] () -- cycle;
\draw [fill=Green!67] (4.9,3.1) coordinate [] () -- (7,1.7) coordinate [] () to[bend right = 10] (7.5,5) coordinate [] () -- (5.2,5) coordinate [] () to[bend left = 10] cycle;
\draw [fill=Red!67] (4.9,6.9) coordinate [] () -- (7,8.3) coordinate [] () to[bend left = 10] (7.5,5) coordinate [] () -- (5.2,5) coordinate [] () to[bend right = 10] cycle;
\node () at (4,5) {\large $C$};
\node () at (6.4,6.4) {\large $S$};
\node () at (6.4,3.6) {\large $C$};
\end{tikzpicture}
\qquad
\begin{tikzpicture}[baseline=11ex, scale =0.4]
\draw [thick] (2,5) edge (7,8.3) node [] {};
\draw [thick] (2,5) edge (7,1.7) node [] {};
\draw [fill=Blue!67] (2,5) coordinate [] () -- (4.9,3.1) coordinate [] () to[bend right = 15] (4.9,6.9) coordinate [] () -- cycle;
\draw [fill=Green!67] (4.9,3.1) coordinate [] () -- (7,1.7) coordinate [] () to[bend right = 10] (7.5,5) coordinate [] () -- (5.2,5) coordinate [] () to[bend left = 10] cycle;
\draw [fill=Red!67] (4.9,6.9) coordinate [] () -- (7,8.3) coordinate [] () to[bend left = 10] (7.5,5) coordinate [] () -- (5.2,5) coordinate [] () to[bend right = 10] cycle;
\node () at (4,5) {\large $H$};
\node () at (6.4,6.4) {\large $S$};
\node () at (6.4,3.6) {\large $C$};
\end{tikzpicture}\ .
\end{aligned}
\end{equation}

There are four terraces corresponding to type-I regions, such that for each of them, the altitudes of its surfaces are $-1$ and $0$:
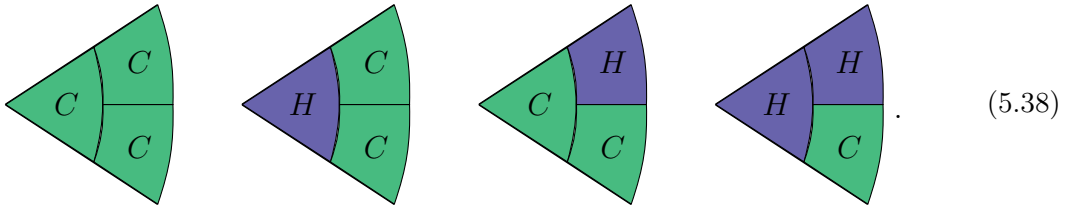
\begin{equation}
\label{eq:terrace_example3}
\begin{aligned}
\begin{tikzpicture}[baseline=11ex, scale =0.4]
\draw [thick] (2,5) edge (7,8.3) node [] {};
\draw [thick] (2,5) edge (7,1.7) node [] {};
\draw [fill=Green!67] (2,5) coordinate [] () -- (4.9,3.1) coordinate [] () to[bend right = 15] (4.9,6.9) coordinate [] () -- cycle;
\draw [fill=Green!67] (4.9,3.1) coordinate [] () -- (7,1.7) coordinate [] () to[bend right = 10] (7.5,5) coordinate [] () -- (5.2,5) coordinate [] () to[bend left = 10] cycle;
\draw [fill=Green!67] (4.9,6.9) coordinate [] () -- (7,8.3) coordinate [] () to[bend left = 10] (7.5,5) coordinate [] () -- (5.2,5) coordinate [] () to[bend right = 10] cycle;
\node () at (4,5) {\large $C$};
\node () at (6.4,6.4) {\large $C$};
\node () at (6.4,3.6) {\large $C$};
\end{tikzpicture}
\qquad
\begin{tikzpicture}[baseline=11ex, scale =0.4]
\draw [thick] (2,5) edge (7,8.3) node [] {};
\draw [thick] (2,5) edge (7,1.7) node [] {};
\draw [fill=Blue!67] (2,5) coordinate [] () -- (4.9,3.1) coordinate [] () to[bend right = 15] (4.9,6.9) coordinate [] () -- cycle;
\draw [fill=Green!67] (4.9,3.1) coordinate [] () -- (7,1.7) coordinate [] () to[bend right = 10] (7.5,5) coordinate [] () -- (5.2,5) coordinate [] () to[bend left = 10] cycle;
\draw [fill=Green!67] (4.9,6.9) coordinate [] () -- (7,8.3) coordinate [] () to[bend left = 10] (7.5,5) coordinate [] () -- (5.2,5) coordinate [] () to[bend right = 10] cycle;
\node () at (4,5) {\large $H$};
\node () at (6.4,6.4) {\large $C$};
\node () at (6.4,3.6) {\large $C$};
\end{tikzpicture}
\qquad
\begin{tikzpicture}[baseline=11ex, scale =0.4]
\draw [thick] (2,5) edge (7,8.3) node [] {};
\draw [thick] (2,5) edge (7,1.7) node [] {};
\draw [fill=Green!67] (2,5) coordinate [] () -- (4.9,3.1) coordinate [] () to[bend right = 15] (4.9,6.9) coordinate [] () -- cycle;
\draw [fill=Green!67] (4.9,3.1) coordinate [] () -- (7,1.7) coordinate [] () to[bend right = 10] (7.5,5) coordinate [] () -- (5.2,5) coordinate [] () to[bend left = 10] cycle;
\draw [fill=Blue!67] (4.9,6.9) coordinate [] () -- (7,8.3) coordinate [] () to[bend left = 10] (7.5,5) coordinate [] () -- (5.2,5) coordinate [] () to[bend right = 10] cycle;
\node () at (4,5) {\large $C$};
\node () at (6.4,6.4) {\large $H$};
\node () at (6.4,3.6) {\large $C$};
\end{tikzpicture}
\qquad
\begin{tikzpicture}[baseline=11ex, scale =0.4]
\draw [thick] (2,5) edge (7,8.3) node [] {};
\draw [thick] (2,5) edge (7,1.7) node [] {};
\draw [fill=Blue!67] (2,5) coordinate [] () -- (4.9,3.1) coordinate [] () to[bend right = 15] (4.9,6.9) coordinate [] () -- cycle;
\draw [fill=Green!67] (4.9,3.1) coordinate [] () -- (7,1.7) coordinate [] () to[bend right = 10] (7.5,5) coordinate [] () -- (5.2,5) coordinate [] () to[bend left = 10] cycle;
\draw [fill=Blue!67] (4.9,6.9) coordinate [] () -- (7,8.3) coordinate [] () to[bend left = 10] (7.5,5) coordinate [] () -- (5.2,5) coordinate [] () to[bend right = 10] cycle;
\node () at (4,5) {\large $H$};
\node () at (6.4,6.4) {\large $H$};
\node () at (6.4,3.6) {\large $C$};
\end{tikzpicture}\ .
\end{aligned}
\end{equation}

There is exactly one type-IIA region:
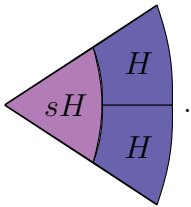
\begin{equation}
\label{eq:terrace_example4}
\begin{aligned}
\begin{tikzpicture}[baseline=11ex, scale =0.4]
\draw [thick] (2,5) edge (7,8.3) node [] {};
\draw [thick] (2,5) edge (7,1.7) node [] {};
\draw [fill=Purple!67] (2,5) coordinate [] () -- (4.9,3.1) coordinate [] () to[bend right = 15] (4.9,6.9) coordinate [] () -- cycle;
\draw [fill=Blue!67] (4.9,3.1) coordinate [] () -- (7,1.7) coordinate [] () to[bend right = 10] (7.5,5) coordinate [] () -- (5.2,5) coordinate [] () to[bend left = 10] cycle;
\draw [fill=Blue!67] (4.9,6.9) coordinate [] () -- (7,8.3) coordinate [] () to[bend left = 10] (7.5,5) coordinate [] () -- (5.2,5) coordinate [] () to[bend right = 10] cycle;
\node () at (4,5) {\large $sH$};
\node () at (6.4,6.4) {\large $H$};
\node () at (6.4,3.6) {\large $H$};
\end{tikzpicture}
\end{aligned}\ .
\end{equation}

Finally, there is one hard region:
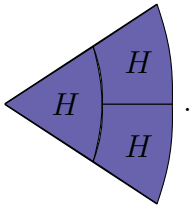
\begin{equation}
\label{eq:terrace_example5}
\begin{aligned}
\begin{tikzpicture}[baseline=11ex, scale =0.4]
\draw [thick] (2,5) edge (7,8.3) node [] {};
\draw [thick] (2,5) edge (7,1.7) node [] {};
\draw [fill=Blue!67] (2,5) coordinate [] () -- (4.9,3.1) coordinate [] () to[bend right = 15] (4.9,6.9) coordinate [] () -- cycle;
\draw [fill=Blue!67] (4.9,3.1) coordinate [] () -- (7,1.7) coordinate [] () to[bend right = 10] (7.5,5) coordinate [] () -- (5.2,5) coordinate [] () to[bend left = 10] cycle;
\draw [fill=Blue!67] (4.9,6.9) coordinate [] () -- (7,8.3) coordinate [] () to[bend left = 10] (7.5,5) coordinate [] () -- (5.2,5) coordinate [] () to[bend right = 10] cycle;
\node () at (4,5) {\large $H$};
\node () at (6.4,6.4) {\large $H$};
\node () at (6.4,3.6) {\large $H$};
\end{tikzpicture}
\end{aligned}\ .
\end{equation}

It is straightforward to check that, the 11 terraces in (\ref{eq:terrace_example1})-(\ref{eq:terrace_example5}) correspond to precisely the 11 region vectors in the second column of table~\ref{table-mass_expansion_example_graphs_regions}.

\section{Conclusions and outlook}
\label{section-conclusions_outlook}

In this work, we have mainly studied the regions for the following three expansions: the on-shell expansion for generic wide-angle scattering, the soft expansion for generic wide-angle scattering, and the mass expansion for heavy-to-light decay processes. For each expansion, we provide, either through a rigorous proof or a proposition, the complete picture of the regions for graphs at any loop order. Each region can be described by the scaling of both loop momenta and Lee-Pomeransky parameters.

The foundation of our entire analysis lies in a geometric approach to the MoR: each region $R$ contributing to the expansion is identified as $f_R$, a lower facet of the Newton polytope $\Delta(\mathcal{P})$ in the Lee-Pomeransky parameter space. Specifically, the Lee-Pomeransky parameter scaling associated with $R$ is reflected in the entries of the vector normal to~$f_R$. Identifying the list of regions can then be translated into recognizing the list of lower facets of $\Delta(\mathcal{P})$. To this end, we associate weights to both Lee-Pomeransky parameters and kinematic factors, and a pivotal observation is that each term of $\mathcal{P}(\x;\s)$ corresponds to a spanning (2-)tree, while each term of $\mathcal{P}^{(R)}(\x;\s)$ corresponds to a \emph{minimum} spanning (2-)tree. Leveraging this insight, we employ graph theory to pinpoint the lower facets of $\Delta(\mathcal{P})$ and subsequently determine the regions. Two essential criteria, namely the facet criterion and the minimum-weight criterion, as we demonstrated in section~\ref{section-two_fundamental_criteria}, are crucial components of our graph-theoretical approach.

In the study of the on-shell expansion for wide-angle scattering in section~\ref{section-regions_onshell}, we provided a rigorous proof establishing the exclusive relevance of hard, collinear, and soft modes within a given region. Furthermore, their corresponding subgraphs $H$, $J$, and $S$ are described by figure~\ref{figure-onshell_region_H_J_S_precise}, aligning with the generic pinch surfaces of $G|_{p^2=0}$. The proof, detailed in section~\ref{section-generic_form_region_rigorous_proof}, is extensive and technical, and for the first time, it brings forth a crucial clarification: no other modes play a role in the on-shell expansion at any order. Supplementing this, additional requirements of the subgraphs, which eliminate those configurations conforming with figure~\ref{figure-onshell_region_H_J_S_precise} but leading to scaleless integrals, have been formulated in ref.~\cite{GrdHzgJnsMaSchlk22}. These insights collectively culminate into the \emph{on-shell expansion region theorem}, as summarized in section~\ref{section-summary_onshell_regions}.

In section~\ref{section-regions_soft_expansion}, we delved into the soft expansion for wide-angle scattering. Based on our comprehensive testing of various examples, we postulate that only the hard, collinear, and soft modes are relevant in any given region, mirroring the scenario in the on-shell expansion. Rather than presenting an extensive proof, we encapsulate this observation in proposition~\ref{proposition-threshold_expansion_vector_generic_form}, which asserts that the subgraphs $H$, $J$, and $S$ adhere to the configuration in figure~\ref{figure-threshold_region_H_J_S_precise}. Additional requirements of $H$, $J$, and $S$ are then investigated in detail, and we showed that these requirements are necessary and sufficient for any configuration of figure~\ref{figure-threshold_region_H_J_S_precise} to qualify as a region in the soft expansion. These findings are summarized into the \emph{soft expansion region theorem} in section~\ref{section-summary_threshold}. We also demonstrated the interrelation between the on-shell expansion and the soft expansion, noting that the former can be viewed as a special case of the latter. Moreover, our soft expansion analysis can be extended to, for example, the collinear expansion.

Finally, in section~\ref{section-regions_mass_expansion}, we studied the mass expansion for the heavy-to-light decay processes. Through the examination of various examples, we validated the intricate mode structure within the regions. Notably, the hard, collinear, and soft modes persist, and additional modes such as collinear$\cdot$soft, semihard, semicollinear, and others may potentially emerge. To comprehend the relevance of these modes in a given region, we classified the regions into type I and type II, with the latter further subdivided into IIA and IIB. For each type, we characterized its general configuration, which can be verified from the examples provided in appendix~\ref{appendix-mass_expansion_examples}. In particular, for the special case of planar graphs, we introduced a terrace formalism in section~\ref{section-formalism_describing_regions_planar_graphs}, which aims to predict the precise list of regions by directly analyzing a graph, bypassing the need to construct Newton polytopes.

Nevertheless, let us still emphasize one potential limitation of our analysis. Given any Lee-Pomeransky polynomial $\mathcal{P}(\x;\s)$, the regions corresponding to the lower facets of $\Delta(\mathcal{P})$ (equivalently, the endpoint singularities in parameter space) are referred to as endpoint-type regions, while those hidden inside $\Delta(\mathcal{P})$, which are due to cancellations among $\mathcal{P}$ terms, are referred to as pinch-type regions (see section~\ref{section-validity_Newton_polytope_approach} for more detail). The Newton polytope approach, as we center on in this paper, is guaranteed to capture all the regions if there exists an analytic continuation from a Euclidean domain, where all the $\mathcal{P}$ terms have the same sign, because in such cases pinch-type regions can be excluded by definition. This statement still holds in the context of the on-shell expansion and soft expansion for most wide-angle scattering graphs. However, there are very few exceptions starting from three loops, for which special cares are needed~\cite{GrdHzgJnsMa24}. The exploration of probable pinch-type regions for generic wide-angle scattering is an intriguing topic, one that we leave for future investigations.

Below we comment on some other interesting topics for future research.
\bigbreak
\noindent \emph{Other expansions and processes.} The analyses presented in this work can be extended to encompass various expansions and processes. As discussed in section~\ref{section-extending_results_additional_expansions}, the soft expansion analysis is directly applicable to more general expansions that combine on-shell and soft expansions, and it can be further extended to the collinear expansion for wide-angle scattering. In the mass expansion discussed in section~\ref{section-regions_mass_expansion}, particle masses exist at two distinct scales, $m\sim \lambda Q$ and $M\sim Q$. The identification of relevant modes and the classification of regions can naturally extend to processes whose Feynman graphs share the same topology as depicted in figure~\ref{figure-problem_to_study_mass}, but with all masses at the same scale $m\sim \lambda Q$.\footnote{This scenario is referred to as a ``Sudakov limit'' in ref.~\cite{Smn02book}. It has been verified that collinear$^{n_1}\cdot$soft$^{n_2}$ modes remain relevant in the corresponding regions. Some recent progress on the $\mathcal{O}(\alpha_S^3)$ correction includes refs.~\cite{HennSmnSmnStnhs17,LeeSmnSmnStnhs18vector,AblgBllMqdRanaSchnd18,LeeSmnSmnStnhs18vectorscalar,AblgBllMqdRanaSchnd19,BllMqdRanaSchnd19,FaelLngSchwdStnhs22massive,FaelLngSchwdStnhs22singlet,BllDFrtAblMqdRanaSchnd23}.} Characterizing these regions using graph-theoretical approaches can also enhance the optimization of existing region-finding algorithms.

It is particularly intriguing for processes in which regions arise not only from the lower facets of $\Delta(P)$ but also from the cancellation of distinct $\mathcal{P}(\x;\s)$ terms. In certain expansions of one-loop graphs, for example, the self-energy graph with two massive edges in the threshold limit, the potential region arises from the cancellation, which has been identified in linear form as $(x_i-x_j)x_kx_l\dots$~\cite{JtzSmnSmn12,SmnvSmnSmn19,AnthnrySkrRmn19}. At higher loops, the cancellation structure can become much more intricate~\cite{GrdHzgJnsMa24}, and a systematic approach to identifying all possible cancellations is highly desirable. Alternatively, one may perform analytic continuations of the Mandelstam variables to ensure the Lee-Pomeransky polynomial is positive definite~\cite{Mshm19}, a technique successfully applied to some cases of massive four-point scattering~\cite{DvsMshmSthsWlm19,DvsHrchJnsKrnMshmSthsWlm19,DvsMshmSchwdSthsZhang22,DvsMshmSchwdSths23}.

One ultimate goal along this line is to provide a rigorous proof of the MoR or specify the cases in which the MoR fails to hold. In comparison to the focus of this paper, which addresses the question of identifying regions, proving or disproving the MoR is a distinct and challenging matter. Recent progress has been made toward this goal. Especially, in ref.~\cite{Jtz11}, the general formalism of the MoR is proposed in momentum space, with detailed justification provided under certain conditions on the relevant regions. Utilizing the Lee-Pomeransky representation, as in this manuscript, a rigorous mathematical proof of the MoR has been established for the leading terms~\cite{SmnvSmnSmn19} or for the special case of one-dimensional Feynman integrals~\cite{Smnv19,Smnv20}. These insightful analyses have the potential for extension to more general cases, complementing the all-order identification of regions presented in this work.
\bigbreak
\noindent\emph{Symanzik polynomials and their geometries.} In the context of this paper, the Newton polytope $\Delta(\mathcal{P})$ is defined as the convex hull of some specific points, each representing the powers of the Lee-Pomeransky parameters of a $\mathcal{P}$ term. Our findings related to the MoR can be interpreted as providing insights into the properties of $\Delta(\mathcal{P})$. This understanding may be valuable in studying other geometric methods based on the Symanzik polynomials, which find applications in sector decomposition, matroid invariants, UV/IR divergences, and maximal cuts of Feynman graphs, among other areas~\cite{KnkUeda10,Schtk18,AnthnryDasSkr20,Pnz19,GongYuan22,MzrTln22,AkHmHlmMzr22,AnthnryBnkBeraDtt23,Brt23}. Notably, in ref.~\cite{Brsk20}, it has been demonstrated that some Newton polytopes constructed from Symanzik polynomials can be considered as special \emph{generalized permutahedra}, opening up avenues for study from the perspective of tropical geometry~\cite{BrskMchTld23,AknHmdFstSvtrPlmdTms23,Svtr24} and providing more efficient numerical evaluations of Feynman integrals.
\bigbreak
\noindent\emph{Infrared structure of perturbative QCD, etc.} Our work can be useful for understanding various aspects of the infrared structure of perturbative QCD, such as the (violation of) hard-collinear-soft factorization and its application~\cite{ClsStm81,BdwBskLpg81,Sen83,CtnGzn00infrared,FgeSwtz14,BncLnnMgnMvlVnzWht15,BncLnnMgnMvlVnzWht16,Zeng15,LiuMnkPnn19,Ma20,BnkBrgJkwczVnz20}, soft anomalous dimension~\cite{BchNbt09,GrdMgn09,AmlDuhrGrd16,CrHGrdRch18,FcnGrdMly19,BchNbt20,FcnGrdMhrMlyVnz22disentagling}, cuts and discontinuities of Feynman integrals~\cite{AbrBrtDuhrGrd17cuts,AbrBrtDuhrGrd17algebraic,BjlHnsdtMLdSwtzVrg21,Brt23,HnsdtMLdSwtzVrg23}, and so on. As one specific example, the soft-expansion region proposition in section~\ref{section-regions_soft_expansion} describes the constraints on the subgraphs of a valid region, which can be employed to the state-of-the-art computation of soft currents~\cite{LiZhu13,DuhrGrm13,AntsDuhrDltFrlGrmHzgMstbg14,DxnHrmYanZhu20,Zhu20,CtnCieriCfrCrdsch23,CzkEschmSchlbg23,CtnCfrTrn20,DDcDuhrHndLiu23,ChenLuoYangZhu23,HzgMaMstlbgSrsh23} contributing to the threshold approximation of the Higgs boson cross section. Numerically, computer packages that utilize the idea of the MoR, such as \texttt{AMFlow}~\cite{LiuMa23AMFlow}, \texttt{DiffExp}~\cite{Hdg21DiffExp}, \texttt{SeaSyde}~\cite{AmdlBcnDvtRanaVcn23SeaSyde}, etc., may also benefit from the general prescription of the regions in a given kinematic limit.

In recent years, there has been growing interest in the local subtraction of infrared singularities based on factorization, where each subtraction term is intimately related to the expansions in the MoR; see refs.~\cite{Ma20,AntsStm18,AntsHndStmYangZeng20,AntsStm23,StmVkt23,AntsKrlStmVkt24} for example. In these works, the explicit form of a local subtraction term is usually set up on a case-by-case basis, and our knowledge of the MoR may help pioneer into a systematic approach to establishing local subtractions for more general cases. Notably, another recent work, ref.~\cite{GbtKswNvchkTcrd23}, develops an algorithm for identifying finite Feynman integrals (categorized into the so-called locally finite, evanescent, and evanescently finite integrals in that work) based on Landau singularity analysis in parameter space. Combining these works with ours and integrating their ideas would be very interesting.

This work may also extend beyond direct QCD calculations. In particular, as the MoR provides useful information to formulate an effective field theory, this work clears some potential risks when applying the Soft-Collinear Effective Theory (SCET). For example, the on-shell-expansion region theorem ensures that, under the condition specified in Eq.(\ref{eq:wideangle_onshell_kinematics}), only collinear and soft modes are relevant in an SCET Lagrangian. In other words, SCET$_\text{I}$ and direct QCD are guaranteed to yield identical results in such cases. Recently, the MoR has also been successfully applied to some other areas, such as the $\mathcal{N}=4$ super Yang-Mills theory~\cite{BltskSmn23,BltskBorkPklnSmn23,BltskBorkSmn23}, the Soft-Collinear gravity~\cite{BnkHgrSzfr22gravitational,BnkHgrSzfr22softcollinear,BnkHgrSzfr22,BnkHgrSwbch23}, especially the computation of cosmological correlators~\cite{BnkHgrSflp23}, and so on. The identification of regions, as presented in this paper, could offer valuable insights in these diverse applications.

\acknowledgments
I would like to express my sincere gratitude to Balasubramanian Ananthanarayan, Charalampos Anastasiou, Martin Beneke, Michael Borinsky, Einan Gardi, Franz Herzog, Stephen Jones, Bernhard Mistlberger, Sebastian Mizera, Ratan Sarkar, Johannes Schlenk, Vladimir Smirnov, George Sterman, Robert Szafron, Xing Wang, and HuaXing Zhu. Their invaluable insights and encouragement have been previous to me.

I am supported by the Swiss National Science Foundation (SNSF) grant No. 10001706. I am also much thankful to the Higgs Centre for Theoretical Physics, where the majority of this work was conducted with support from the UKRI FLF grant “Forest Formulas for the LHC” (Mr/S03479x/1). Additionally, I appreciate the hospitality of the Galileo Galilei Institute for Theoretical Physics and the support from INFN during my stay on September 4-15, 2023, when I benefited from many useful discussions and made progress toward the completion of this work.

\appendix

\section{Prim's algorithm and its relation to lemma~\ref{lemma-weight_hierarchical_partition_tree_structure}}
\label{appendix-Prim_algorithm}

Prim's algorithm is a systematic way of obtaining a minimum spanning tree for a weighted graph $\Gamma$. It starts with an arbitrary vertex $v\in \Gamma$ and adds one edge (including its endpoints) for each step until a minimum spanning tree is obtained. The implementation of Prim's algorithm involves the following steps.
\begin{enumerate}
    \item Choose a vertex $v\in \Gamma$ at random and initiate the minimum spanning tree with $v$. Namely, the set of vertices $\mathcal{V}=\{v\}$, and the set of edges $\mathcal{E}=\varnothing$.
    \item Repeat the following operations until $\mathcal{V}$ includes all the vertices of $\Gamma$:
    \begin{enumerate}
        \item [(1)] for all the edges whose endpoints are in $\mathcal{V}$ and $\Gamma\setminus \mathcal{V}$ respectively, find one with the minimum weight, and add this edge to the set $\mathcal{E}$;
        \item [(2)] add the endpoint of this edge, which is in $\Gamma\setminus \mathcal{V}$, to $\mathcal{V}$.
    \end{enumerate}
    \item The output $(\mathcal{V}, \mathcal{E})$ is a minimum spanning tree of $\Gamma$.
\end{enumerate}
It has been proved that the output of Prim's algorithm is always a minimum spanning tree.

Let us illustrate Prim's algorithm from the following example:
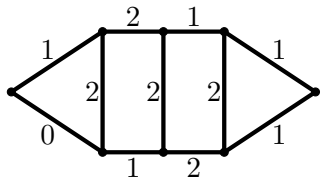
\begin{equation}
\begin{tikzpicture}[baseline=8ex, scale=0.4]
\draw [ultra thick] (0,5) -- (3,3);
\draw [ultra thick] (0,5) -- (3,7);
\draw [ultra thick] (3,3) -- (5,3);
\draw [ultra thick] (5,3) -- (7,3);
\draw [ultra thick] (3,7) -- (5,7);
\draw [ultra thick] (5,7) -- (7,7);
\draw [ultra thick] (7,3) -- (10,5);
\draw [ultra thick] (7,7) -- (10,5);
\draw [ultra thick] (3,7) -- (3,3);
\draw [ultra thick] (7,7) -- (7,3);
\draw [ultra thick] (5,7) -- (5,3);
\node [draw, circle, fill, minimum size=3pt, inner sep=0pt, outer sep=0pt] () at (0,5) {};
\node [draw, circle, fill, minimum size=3pt, inner sep=0pt, outer sep=0pt] () at (3,7) {};
\node [draw, circle, fill, minimum size=3pt, inner sep=0pt, outer sep=0pt] () at (3,3) {};
\node [draw, circle, fill, minimum size=3pt, inner sep=0pt, outer sep=0pt] () at (5,3) {};
\node [draw, circle, fill, minimum size=3pt, inner sep=0pt, outer sep=0pt] () at (5,7) {};
\node [draw, circle, fill, minimum size=3pt, inner sep=0pt, outer sep=0pt] () at (7,7) {};
\node [draw, circle, fill, minimum size=3pt, inner sep=0pt, outer sep=0pt] () at (7,3) {};
\node [draw, circle, fill, minimum size=3pt, inner sep=0pt, outer sep=0pt] () at (10,5) {};
\node () at (1.2,6.4) {$1$};
\node () at (1.2,3.6) {$0$};
\node () at (4,2.5) {$1$};
\node () at (6,2.5) {$2$};
\node () at (4,7.5) {$2$};
\node () at (6,7.5) {$1$};
\node () at (2.67,5) {$2$};
\node () at (4.67,5) {$2$};
\node () at (6.67,5) {$2$};
\node () at (8.8,6.4) {$1$};
\node () at (8.8,3.6) {$1$};
\end{tikzpicture}\nonumber
\end{equation}
where the weight of each edge is labelled in the figure. Let us start from the leftmost vertex and construct a minimum spanning tree using Prim's algorithm:
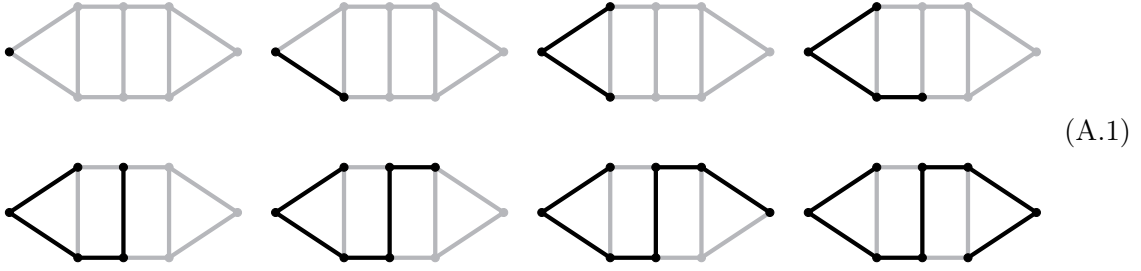
\begin{equation}
\begin{aligned}
&\begin{tikzpicture}[baseline=8ex, scale=0.3]
\draw [ultra thick, color=Black!33] (0,5) -- (3,3);
\draw [ultra thick, color=Black!33] (0,5) -- (3,7);
\draw [ultra thick, color=Black!33] (3,3) -- (5,3);
\draw [ultra thick, color=Black!33] (5,3) -- (7,3);
\draw [ultra thick, color=Black!33] (3,7) -- (5,7);
\draw [ultra thick, color=Black!33] (5,7) -- (7,7);
\draw [ultra thick, color=Black!33] (7,3) -- (10,5);
\draw [ultra thick, color=Black!33] (7,7) -- (10,5);
\draw [ultra thick, color=Black!33] (3,7) -- (3,3);
\draw [ultra thick, color=Black!33] (7,7) -- (7,3);
\draw [ultra thick, color=Black!33] (5,7) -- (5,3);
\node [draw, circle, fill, minimum size=3pt, inner sep=0pt, outer sep=0pt] () at (0,5) {};
\node [draw, circle, fill=Black!33, color=Black!33, minimum size=3pt, inner sep=0pt, outer sep=0pt] () at (3,7) {};
\node [draw, circle, fill=Black!33, color=Black!33, minimum size=3pt, inner sep=0pt, outer sep=0pt] () at (3,3) {};
\node [draw, circle, fill=Black!33, color=Black!33, minimum size=3pt, inner sep=0pt, outer sep=0pt] () at (5,3) {};
\node [draw, circle, fill=Black!33, color=Black!33, minimum size=3pt, inner sep=0pt, outer sep=0pt] () at (5,7) {};
\node [draw, circle, fill=Black!33, color=Black!33, minimum size=3pt, inner sep=0pt, outer sep=0pt] () at (7,7) {};
\node [draw, circle, fill=Black!33, color=Black!33, minimum size=3pt, inner sep=0pt, outer sep=0pt] () at (7,3) {};
\node [draw, circle, fill=Black!33, color=Black!33, minimum size=3pt, inner sep=0pt, outer sep=0pt] () at (10,5) {};
\end{tikzpicture}\quad
\begin{tikzpicture}[baseline=8ex, scale=0.3]
\draw [ultra thick] (0,5) -- (3,3);
\draw [ultra thick, color=Black!33] (0,5) -- (3,7);
\draw [ultra thick, color=Black!33] (3,3) -- (5,3);
\draw [ultra thick, color=Black!33] (5,3) -- (7,3);
\draw [ultra thick, color=Black!33] (3,7) -- (5,7);
\draw [ultra thick, color=Black!33] (5,7) -- (7,7);
\draw [ultra thick, color=Black!33] (7,3) -- (10,5);
\draw [ultra thick, color=Black!33] (7,7) -- (10,5);
\draw [ultra thick, color=Black!33] (3,7) -- (3,3);
\draw [ultra thick, color=Black!33] (7,7) -- (7,3);
\draw [ultra thick, color=Black!33] (5,7) -- (5,3);
\node [draw, circle, fill, minimum size=3pt, inner sep=0pt, outer sep=0pt] () at (0,5) {};
\node [draw, circle, fill=Black!33, color=Black!33, minimum size=3pt, inner sep=0pt, outer sep=0pt] () at (3,7) {};
\node [draw, circle, fill, minimum size=3pt, inner sep=0pt, outer sep=0pt] () at (3,3) {};
\node [draw, circle, fill=Black!33, color=Black!33, minimum size=3pt, inner sep=0pt, outer sep=0pt] () at (5,3) {};
\node [draw, circle, fill=Black!33, color=Black!33, minimum size=3pt, inner sep=0pt, outer sep=0pt] () at (5,7) {};
\node [draw, circle, fill=Black!33, color=Black!33, minimum size=3pt, inner sep=0pt, outer sep=0pt] () at (7,7) {};
\node [draw, circle, fill=Black!33, color=Black!33, minimum size=3pt, inner sep=0pt, outer sep=0pt] () at (7,3) {};
\node [draw, circle, fill=Black!33, color=Black!33, minimum size=3pt, inner sep=0pt, outer sep=0pt] () at (10,5) {};
\end{tikzpicture}\quad
\begin{tikzpicture}[baseline=8ex, scale=0.3]
\draw [ultra thick] (0,5) -- (3,3);
\draw [ultra thick] (0,5) -- (3,7);
\draw [ultra thick, color=Black!33] (3,3) -- (5,3);
\draw [ultra thick, color=Black!33] (5,3) -- (7,3);
\draw [ultra thick, color=Black!33] (3,7) -- (5,7);
\draw [ultra thick, color=Black!33] (5,7) -- (7,7);
\draw [ultra thick, color=Black!33] (7,3) -- (10,5);
\draw [ultra thick, color=Black!33] (7,7) -- (10,5);
\draw [ultra thick, color=Black!33] (3,7) -- (3,3);
\draw [ultra thick, color=Black!33] (7,7) -- (7,3);
\draw [ultra thick, color=Black!33] (5,7) -- (5,3);
\node [draw, circle, fill, minimum size=3pt, inner sep=0pt, outer sep=0pt] () at (0,5) {};
\node [draw, circle, fill, minimum size=3pt, inner sep=0pt, outer sep=0pt] () at (3,7) {};
\node [draw, circle, fill, minimum size=3pt, inner sep=0pt, outer sep=0pt] () at (3,3) {};
\node [draw, circle, fill=Black!33, color=Black!33, minimum size=3pt, inner sep=0pt, outer sep=0pt] () at (5,3) {};
\node [draw, circle, fill=Black!33, color=Black!33, minimum size=3pt, inner sep=0pt, outer sep=0pt] () at (5,7) {};
\node [draw, circle, fill=Black!33, color=Black!33, minimum size=3pt, inner sep=0pt, outer sep=0pt] () at (7,7) {};
\node [draw, circle, fill=Black!33, color=Black!33, minimum size=3pt, inner sep=0pt, outer sep=0pt] () at (7,3) {};
\node [draw, circle, fill=Black!33, color=Black!33, minimum size=3pt, inner sep=0pt, outer sep=0pt] () at (10,5) {};
\end{tikzpicture}\quad
\begin{tikzpicture}[baseline=8ex, scale=0.3]
\draw [ultra thick] (0,5) -- (3,3);
\draw [ultra thick] (0,5) -- (3,7);
\draw [ultra thick] (3,3) -- (5,3);
\draw [ultra thick, color=Black!33] (5,3) -- (7,3);
\draw [ultra thick, color=Black!33] (3,7) -- (5,7);
\draw [ultra thick, color=Black!33] (5,7) -- (7,7);
\draw [ultra thick, color=Black!33] (7,3) -- (10,5);
\draw [ultra thick, color=Black!33] (7,7) -- (10,5);
\draw [ultra thick, color=Black!33] (3,7) -- (3,3);
\draw [ultra thick, color=Black!33] (7,7) -- (7,3);
\draw [ultra thick, color=Black!33] (5,7) -- (5,3);
\node [draw, circle, fill, minimum size=3pt, inner sep=0pt, outer sep=0pt] () at (0,5) {};
\node [draw, circle, fill, minimum size=3pt, inner sep=0pt, outer sep=0pt] () at (3,7) {};
\node [draw, circle, fill, minimum size=3pt, inner sep=0pt, outer sep=0pt] () at (3,3) {};
\node [draw, circle, fill, minimum size=3pt, inner sep=0pt, outer sep=0pt] () at (5,3) {};
\node [draw, circle, fill=Black!33, color=Black!33, minimum size=3pt, inner sep=0pt, outer sep=0pt] () at (5,7) {};
\node [draw, circle, fill=Black!33, color=Black!33, minimum size=3pt, inner sep=0pt, outer sep=0pt] () at (7,7) {};
\node [draw, circle, fill=Black!33, color=Black!33, minimum size=3pt, inner sep=0pt, outer sep=0pt] () at (7,3) {};
\node [draw, circle, fill=Black!33, color=Black!33, minimum size=3pt, inner sep=0pt, outer sep=0pt] () at (10,5) {};
\end{tikzpicture}\\
&\\
&\begin{tikzpicture}[baseline=8ex, scale=0.3]
\draw [ultra thick] (0,5) -- (3,3);
\draw [ultra thick] (0,5) -- (3,7);
\draw [ultra thick] (3,3) -- (5,3);
\draw [ultra thick, color=Black!33] (5,3) -- (7,3);
\draw [ultra thick, color=Black!33] (3,7) -- (5,7);
\draw [ultra thick, color=Black!33] (5,7) -- (7,7);
\draw [ultra thick, color=Black!33] (7,3) -- (10,5);
\draw [ultra thick, color=Black!33] (7,7) -- (10,5);
\draw [ultra thick, color=Black!33] (3,7) -- (3,3);
\draw [ultra thick, color=Black!33] (7,7) -- (7,3);
\draw [ultra thick] (5,7) -- (5,3);
\node [draw, circle, fill, minimum size=3pt, inner sep=0pt, outer sep=0pt] () at (0,5) {};
\node [draw, circle, fill, minimum size=3pt, inner sep=0pt, outer sep=0pt] () at (3,7) {};
\node [draw, circle, fill, minimum size=3pt, inner sep=0pt, outer sep=0pt] () at (3,3) {};
\node [draw, circle, fill, minimum size=3pt, inner sep=0pt, outer sep=0pt] () at (5,3) {};
\node [draw, circle, fill, minimum size=3pt, inner sep=0pt, outer sep=0pt] () at (5,7) {};
\node [draw, circle, fill=Black!33, color=Black!33, minimum size=3pt, inner sep=0pt, outer sep=0pt] () at (7,7) {};
\node [draw, circle, fill=Black!33, color=Black!33, minimum size=3pt, inner sep=0pt, outer sep=0pt] () at (7,3) {};
\node [draw, circle, fill=Black!33, color=Black!33, minimum size=3pt, inner sep=0pt, outer sep=0pt] () at (10,5) {};
\end{tikzpicture}\quad
\begin{tikzpicture}[baseline=8ex, scale=0.3]
\draw [ultra thick] (0,5) -- (3,3);
\draw [ultra thick] (0,5) -- (3,7);
\draw [ultra thick] (3,3) -- (5,3);
\draw [ultra thick, color=Black!33] (5,3) -- (7,3);
\draw [ultra thick, color=Black!33] (3,7) -- (5,7);
\draw [ultra thick] (5,7) -- (7,7);
\draw [ultra thick, color=Black!33] (7,3) -- (10,5);
\draw [ultra thick, color=Black!33] (7,7) -- (10,5);
\draw [ultra thick, color=Black!33] (3,7) -- (3,3);
\draw [ultra thick, color=Black!33] (7,7) -- (7,3);
\draw [ultra thick] (5,7) -- (5,3);
\node [draw, circle, fill, minimum size=3pt, inner sep=0pt, outer sep=0pt] () at (0,5) {};
\node [draw, circle, fill, minimum size=3pt, inner sep=0pt, outer sep=0pt] () at (3,7) {};
\node [draw, circle, fill, minimum size=3pt, inner sep=0pt, outer sep=0pt] () at (3,3) {};
\node [draw, circle, fill, minimum size=3pt, inner sep=0pt, outer sep=0pt] () at (5,3) {};
\node [draw, circle, fill, minimum size=3pt, inner sep=0pt, outer sep=0pt] () at (5,7) {};
\node [draw, circle, fill, minimum size=3pt, inner sep=0pt, outer sep=0pt] () at (7,7) {};
\node [draw, circle, fill=Black!33, color=Black!33, minimum size=3pt, inner sep=0pt, outer sep=0pt] () at (7,3) {};
\node [draw, circle, fill=Black!33, color=Black!33, minimum size=3pt, inner sep=0pt, outer sep=0pt] () at (10,5) {};
\end{tikzpicture}\quad
\begin{tikzpicture}[baseline=8ex, scale=0.3]
\draw [ultra thick] (0,5) -- (3,3);
\draw [ultra thick] (0,5) -- (3,7);
\draw [ultra thick] (3,3) -- (5,3);
\draw [ultra thick, color=Black!33] (5,3) -- (7,3);
\draw [ultra thick, color=Black!33] (3,7) -- (5,7);
\draw [ultra thick] (5,7) -- (7,7);
\draw [ultra thick, color=Black!33] (7,3) -- (10,5);
\draw [ultra thick] (7,7) -- (10,5);
\draw [ultra thick, color=Black!33] (3,7) -- (3,3);
\draw [ultra thick, color=Black!33] (7,7) -- (7,3);
\draw [ultra thick] (5,7) -- (5,3);
\node [draw, circle, fill, minimum size=3pt, inner sep=0pt, outer sep=0pt] () at (0,5) {};
\node [draw, circle, fill, minimum size=3pt, inner sep=0pt, outer sep=0pt] () at (3,7) {};
\node [draw, circle, fill, minimum size=3pt, inner sep=0pt, outer sep=0pt] () at (3,3) {};
\node [draw, circle, fill, minimum size=3pt, inner sep=0pt, outer sep=0pt] () at (5,3) {};
\node [draw, circle, fill, minimum size=3pt, inner sep=0pt, outer sep=0pt] () at (5,7) {};
\node [draw, circle, fill, minimum size=3pt, inner sep=0pt, outer sep=0pt] () at (7,7) {};
\node [draw, circle, fill=Black!33, color=Black!33, minimum size=3pt, inner sep=0pt, outer sep=0pt] () at (7,3) {};
\node [draw, circle, fill, minimum size=3pt, inner sep=0pt, outer sep=0pt] () at (10,5) {};
\end{tikzpicture}\quad
\begin{tikzpicture}[baseline=8ex, scale=0.3]
\draw [ultra thick] (0,5) -- (3,3);
\draw [ultra thick] (0,5) -- (3,7);
\draw [ultra thick] (3,3) -- (5,3);
\draw [ultra thick, color=Black!33] (5,3) -- (7,3);
\draw [ultra thick, color=Black!33] (3,7) -- (5,7);
\draw [ultra thick] (5,7) -- (7,7);
\draw [ultra thick] (7,3) -- (10,5);
\draw [ultra thick] (7,7) -- (10,5);
\draw [ultra thick, color=Black!33] (3,7) -- (3,3);
\draw [ultra thick, color=Black!33] (7,7) -- (7,3);
\draw [ultra thick] (5,7) -- (5,3);
\node [draw, circle, fill, minimum size=3pt, inner sep=0pt, outer sep=0pt] () at (0,5) {};
\node [draw, circle, fill, minimum size=3pt, inner sep=0pt, outer sep=0pt] () at (3,7) {};
\node [draw, circle, fill, minimum size=3pt, inner sep=0pt, outer sep=0pt] () at (3,3) {};
\node [draw, circle, fill, minimum size=3pt, inner sep=0pt, outer sep=0pt] () at (5,3) {};
\node [draw, circle, fill, minimum size=3pt, inner sep=0pt, outer sep=0pt] () at (5,7) {};
\node [draw, circle, fill, minimum size=3pt, inner sep=0pt, outer sep=0pt] () at (7,7) {};
\node [draw, circle, fill, minimum size=3pt, inner sep=0pt, outer sep=0pt] () at (7,3) {};
\node [draw, circle, fill, minimum size=3pt, inner sep=0pt, outer sep=0pt] () at (10,5) {};
\end{tikzpicture}
\end{aligned}
\end{equation}
Note that after each step, the sets $\mathcal{V}$ and $\mathcal{E}$ include the black vertices and edges respectively.

We have stated below lemma~\ref{lemma-weight_hierarchical_partition_tree_structure} that, the proof of
\begin{eqnarray}
\widetilde{\gamma}_{i,j}\cap T^1\text{ is a spanning tree of }\widetilde{\gamma}_{i,j}\quad \Rightarrow \quad T^1\text{ is a minimum spanning tree of }\Gamma \nonumber
\end{eqnarray}
is straightforward with the help of Prim's algorithm. In more detail, the left-hand side can be seen as an output of Prim's algorithm. To see this, let us choose any vertex $v\in \Gamma_M$ and denote $\Gamma'\equiv v$. We then do the following operations recursively until $\Gamma'$ includes all the vertices of $\Gamma$: among all the edges of $T^1$ whose endpoints are in $\Gamma'$ and $\Gamma\setminus \Gamma'$ respectively, we find one with the minimum weight and include it (as well as its endpoints) to $\Gamma'$. It is clear that after each step, $\Gamma'$ is a tree graph and all its edges are in $T^1$. So in the end we have $\Gamma'=T^1$.

According to the condition that the graphs $\Gamma_M$, $\Gamma_M\cup \Gamma_{M-1}$, ..., $\Gamma_M\cup\dots\cup \Gamma_1$ are all connected, it follows from above that after each of the first $V(\Gamma_M)-1$ operations, an edge of $T^1\cap \Gamma_M$ is added to $T_0^1$; after each of the next $V(\Gamma_{M-1})-1$ operations, an edge of $T^1\cap \Gamma_{M-1}$ is added to $T_0^1$, etc. It is then clear that this procedure coincides with Prim's algorithm. Thus the output, which is $T^1$, must be a minimum spanning tree.

\section{Some details in proving the on-shell-expansion region theorem}
\label{appendix-details_in_proof}

In this appendix, we present some details of the proof in section~\ref{section-generic_form_region_rigorous_proof}, which shows that every region vector in the on-shell expansion must be described by eq.~(\ref{generic_form_region_vector_onshell}).

\subsection{The paths contained in \texorpdfstring{$G\setminus \widehat{\mathcal{V}}_0$}{TEXT}}
\label{appendix-GminusV0_disconnected}

Here we justify the statement made below lemma~\ref{lemma-onshell_Fp2_vertices_universal_property}: any path $P\subset G\setminus \widehat{\mathcal{V}}_0$ that joins two on-shell external momentum $p_i^\mu$ and $p_j^\mu$ must contain some edges satisfying $w<-1$.

Let us first note that none of the edges of $P$ can satisfy $w>-1$, otherwise the endpoints of this edge would belong to $\widehat{\mathcal{V}}_0$. So we only need to consider (and rule out) the possibility that every edge of $P$ satisfies $w(e)=-1$. For this possibility, a key observation is that $\mathcal{F}^{(p_i^2,R)}\neq 0$ (and similarly, $\mathcal{F}^{(p_j^2,R)}\neq 0$), because otherwise for any $\mathcal{U}^{(R)}$ term $\x^{\r}$, $p_i^\mu$ would attach to a vertex of the $q$ trunk of $T^1(\r)$, which must belong to $\widehat{\mathcal{V}}_0$ by definition. We then consider any canonical $\mathcal{F}^{(p_i^2,R)}$ term $\x^{\r'}$. Since $P$ joins $p_i^\mu$ and $p_j^\mu$, there must be an edge $e_0\in P$ whose endpoints are in $t(\r';p_i)$ and $t(\r';\widehat{p}_i)$ respectively. The graph $T^2(\r)\cup e_0$ then corresponds to a $\mathcal{U}^{(R)}$ term, and moreover, $e_0$ is incident with a vertex in its $q$ trunk. This implies that one vertex of $P$ is in $\widehat{\mathcal{V}}_0$, and contradicts the construction that $P\subset G\setminus \widehat{\mathcal{V}}_0$.

Therefore, $P$ must contain some edges satisfying $w<-1$.

\subsection{Details of proving lemma~\ref{lemma-onshell_Si_empty}: the existence of \texorpdfstring{$C_iD_i$}{TEXT}}
\label{appendix-details_lemma9_proof}

In proving lemma~\ref{lemma-onshell_Si_empty}, we have mentioned that for each path $A_iB_i$, there exists a path $C_iD_i \subset S\cap T^2(\r_*)$, such that the endpoints $C_i,D_i\in H\cup J$, and $A_iB_i\subseteq C_iD_i$. Here we demonstrate more detail of this statement.

\begin{figure}[t]
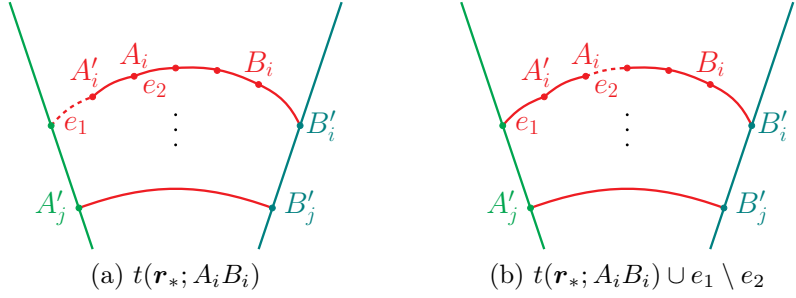

\centering
\begin{subfigure}[b]{0.3\textwidth}
\centering
\include{figs/onshell_lemma9_extension_within_S}
\vspace{-3em}\caption{$t(\r_*;A_iB_i)$}
\label{onshell_lemma9_extension_within_S}
\end{subfigure}
\hspace{3em}
\begin{subfigure}[b]{0.3\textwidth}
\centering
\include{figs/onshell_lemma9_extension_within_S_comparison}
\vspace{-3em}\caption{$t(\r_*;A_iB_i) \cup e_1\setminus e_2$}
\label{onshell_lemma9_extension_within_S_comparison}
\end{subfigure}
\caption{The comparison of $T^2(\r_*)$ and $T^2(\r_*)\cup e_1\setminus e_2$. Note that in (a) we have only shown the component of $T^2(\r_*)$ that contains the path $A_iB_i$, which we denote by $t(\r_*;A_iB_i)$, while in (b) we have only shown the component $t(\r_*;A_iB_i) \cup e_1\setminus e_2$.}
\label{figure-onshell_lemma9_proof}
\end{figure}

Let us consider the contrary: the path $A_iB_i$ can be extended to ``not longer than'' $A'_iB'_i$, where either $A'_i$ or $B'_i$ is in $S\setminus S_1$ rather than $H\cup J$. That is, there does not exist another path in $T^2(\r_*)$ which contains $A'_iB'_i$. An example is shown in figure~\ref{onshell_lemma9_extension_within_S}, where $A'_i\in S\setminus S$. According to lemma~\ref{lemma-onshell_sij_configuration_constraints}, the component of $S\setminus S_1$ that contains $A'_i$ is adjacent to $H\cup J$, so it follows that there must be an edge $e_1 \in (S\setminus S_1)\setminus T^2(\r_*)$, such that the loop in the graph $T^2(\r_*)\cup e_1$ contains an edge $e_2\in A_iB_i$.\footnote{Note that in this figure we have assumed that the endpoints of $e_1$ are in the same component of $T^2(\r_*)$ as $A_iB_i$. It is also possible that the endpoints of $e_1$ are in the distinct components of $T^2(\r_*)$, in which case the same analysis here applies.} The graph $T^2(\r_*)\cup e_1\setminus e_2$ then, as shown in figure~\ref{onshell_lemma9_extension_within_S_comparison}, is another spanning 2-tree corresponding to an $\mathcal{F}^{(q^2)}$ term, whose weight $w'$ can be related to $w(T^2(\r_*))$ as
\begin{eqnarray}
w'= w(T^2(\r_*)) +w(e_2) -w(e_1) <w(T^2(\r_*)),
\end{eqnarray}
where we have used that $w(e_2) =w_1< w(e_1)$. (Note that by definition, $e_1\in S\setminus S_1$ and $e_2\in S_1$.) The minimum-weight criterion is then violated. As a result, we have verified that for each $i\in \{1,\dots,k\}$ there exists a path $C_iD_i \subset S\cap T^2(\r_*)$, such that the endpoints $C_i,D_i\in H\cup J$, and $A_iB_i\subseteq C_iD_i$.
\section{Some examples in the mass expansion}
\label{appendix-mass_expansion_examples}

\begin{figure}[t]
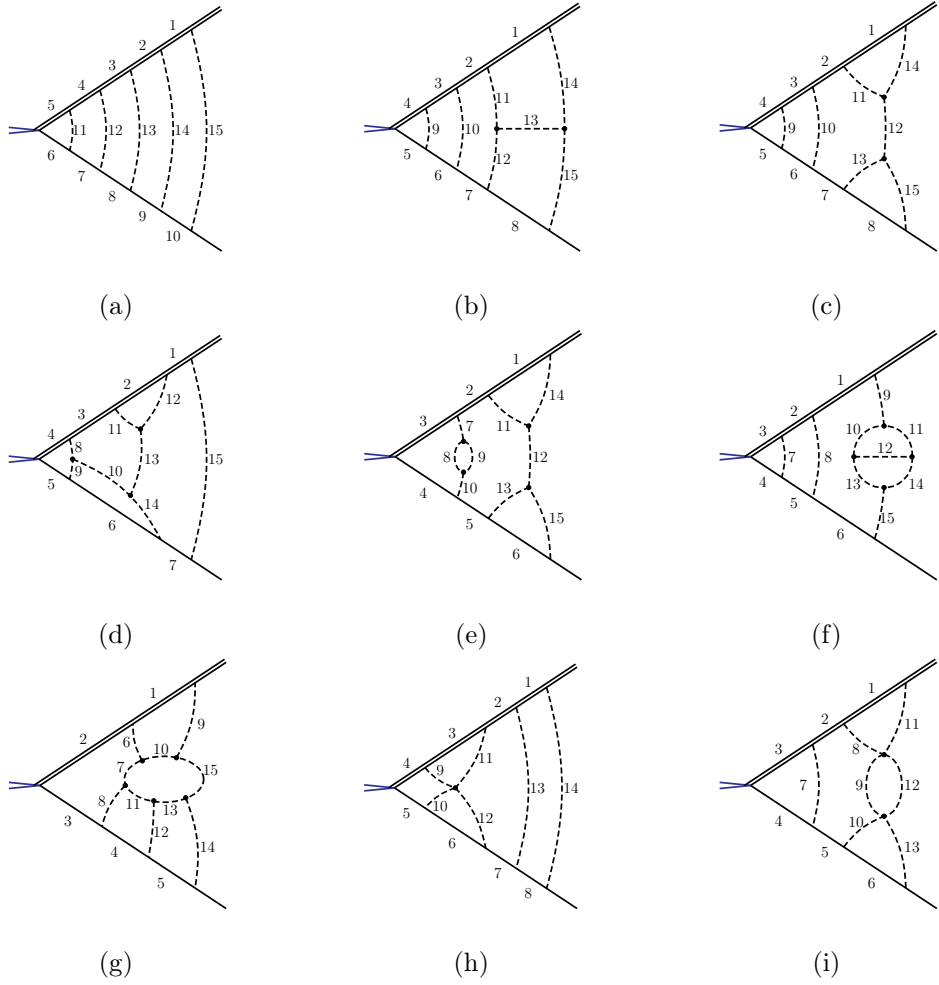

\centering
\begin{subfigure}[b]{0.25\textwidth}
\centering
\include{figs/mass_expansion_examples_planar_1}
\vspace{-2em}
\caption{}
\label{mass_expansion_examples_planar_1}
\end{subfigure}
\qquad
\begin{subfigure}[b]{0.25\textwidth}
\centering
\include{figs/mass_expansion_examples_planar_2}
\vspace{-2em}
\caption{}
\label{mass_expansion_examples_planar_2}
\end{subfigure}
\qquad
\begin{subfigure}[b]{0.25\textwidth}
\centering
\include{figs/mass_expansion_examples_planar_3}
\vspace{-2em}
\caption{}
\label{mass_expansion_examples_planar_3}
\end{subfigure}
\\
\begin{subfigure}[b]{0.25\textwidth}
\centering
\include{figs/mass_expansion_examples_planar_4}
\vspace{-2em}
\caption{}
\label{mass_expansion_examples_planar_4}
\end{subfigure}
\qquad
\begin{subfigure}[b]{0.25\textwidth}
\centering
\include{figs/mass_expansion_examples_planar_5}
\vspace{-2em}
\caption{}
\label{mass_expansion_examples_planar_5}
\end{subfigure}
\qquad
\begin{subfigure}[b]{0.25\textwidth}
\centering
\include{figs/mass_expansion_examples_planar_6}
\vspace{-2em}
\caption{}
\label{mass_expansion_examples_planar_6}
\end{subfigure}
\\
\begin{subfigure}[b]{0.25\textwidth}
\centering
\include{figs/mass_expansion_examples_planar_7}
\vspace{-2em}
\caption{}
\label{mass_expansion_examples_planar_7}
\end{subfigure}
\qquad
\begin{subfigure}[b]{0.25\textwidth}
\centering
\include{figs/mass_expansion_examples_planar_8}
\vspace{-2em}
\caption{}
\label{mass_expansion_examples_planar_8}
\end{subfigure}
\qquad
\begin{subfigure}[b]{0.25\textwidth}
\centering
\include{figs/mass_expansion_examples_planar_9}
\vspace{-2em}
\caption{}
\label{mass_expansion_examples_planar_9}
\end{subfigure}
\caption{Examples of planar graphs with the parameterization of their edges manifested. In the mass expansion, these graphs feature type-I and type-IIA regions but no type-IIB regions.}
\label{figure-mass_expansion_examples_planar}
\end{figure}

\begin{figure}[t]
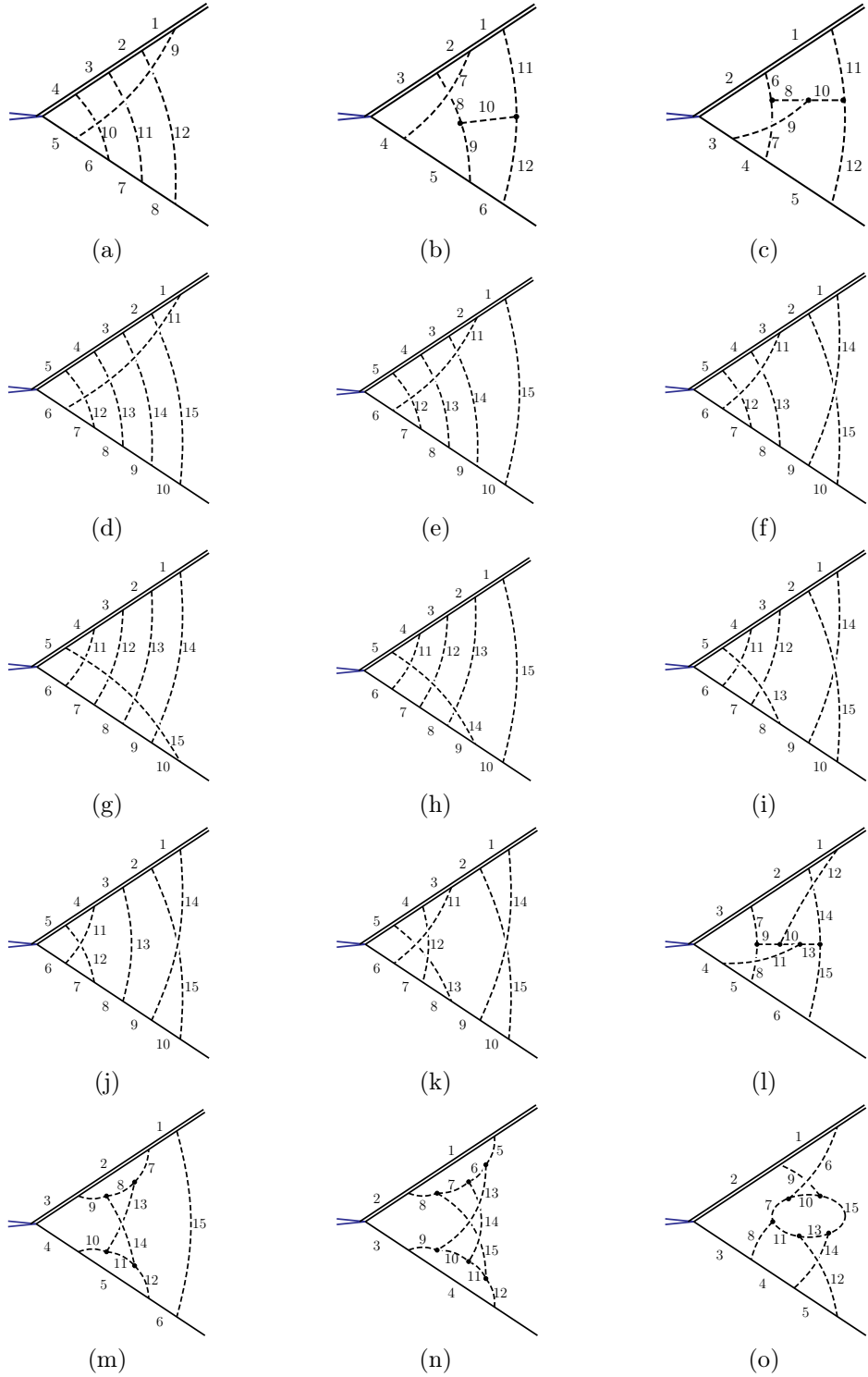

\centering
\begin{subfigure}[b]{0.25\textwidth}
\centering
\include{figs/mass_expansion_examples_nonplanar_1}
\vspace{-3em}
\caption{}
\label{mass_expansion_examples_nonplanar_1}
\end{subfigure}
\qquad
\begin{subfigure}[b]{0.25\textwidth}
\centering
\include{figs/mass_expansion_examples_nonplanar_2}
\vspace{-3em}
\caption{}
\label{mass_expansion_examples_nonplanar_2}
\end{subfigure}
\qquad
\begin{subfigure}[b]{0.25\textwidth}
\centering
\include{figs/mass_expansion_examples_nonplanar_3}
\vspace{-3em}
\caption{}
\label{mass_expansion_examples_nonplanar_3}
\end{subfigure}
\\
\begin{subfigure}[b]{0.25\textwidth}
\centering
\include{figs/mass_expansion_examples_nonplanar_4}
\vspace{-3em}
\caption{}
\label{mass_expansion_examples_nonplanar_4}
\end{subfigure}
\qquad
\begin{subfigure}[b]{0.25\textwidth}
\centering
\include{figs/mass_expansion_examples_nonplanar_5}
\vspace{-3em}
\caption{}
\label{mass_expansion_examples_nonplanar_5}
\end{subfigure}
\qquad
\begin{subfigure}[b]{0.25\textwidth}
\centering
\include{figs/mass_expansion_examples_nonplanar_6}
\vspace{-3em}
\caption{}
\label{mass_expansion_examples_nonplanar_6}
\end{subfigure}
\\
\begin{subfigure}[b]{0.25\textwidth}
\centering
\include{figs/mass_expansion_examples_nonplanar_7}
\vspace{-3em}
\caption{}
\label{mass_expansion_examples_nonplanar_7}
\end{subfigure}
\qquad
\begin{subfigure}[b]{0.25\textwidth}
\centering
\include{figs/mass_expansion_examples_nonplanar_8}
\vspace{-3em}
\caption{}
\label{mass_expansion_examples_nonplanar_8}
\end{subfigure}
\qquad
\begin{subfigure}[b]{0.25\textwidth}
\centering
\include{figs/mass_expansion_examples_nonplanar_9}
\vspace{-3em}
\caption{}
\label{mass_expansion_examples_nonplanar_9}
\end{subfigure}
\\
\begin{subfigure}[b]{0.25\textwidth}
\centering
\include{figs/mass_expansion_examples_nonplanar_10}
\vspace{-3em}
\caption{}
\label{mass_expansion_examples_nonplanar_10}
\end{subfigure}
\qquad
\begin{subfigure}[b]{0.25\textwidth}
\centering
\include{figs/mass_expansion_examples_nonplanar_11}
\vspace{-3em}
\caption{}
\label{mass_expansion_examples_nonplanar_11}
\end{subfigure}
\qquad
\begin{subfigure}[b]{0.25\textwidth}
\centering
\include{figs/mass_expansion_examples_nonplanar_12}
\vspace{-3em}
\caption{}
\label{mass_expansion_examples_nonplanar_12}
\end{subfigure}
\\
\begin{subfigure}[b]{0.25\textwidth}
\centering
\include{figs/mass_expansion_examples_nonplanar_13}
\vspace{-3em}
\caption{}
\label{mass_expansion_examples_nonplanar_13}
\end{subfigure}
\qquad
\begin{subfigure}[b]{0.25\textwidth}
\centering
\include{figs/mass_expansion_examples_nonplanar_14}
\vspace{-3em}
\caption{}
\label{mass_expansion_examples_nonplanar_14}
\end{subfigure}
\qquad
\begin{subfigure}[b]{0.25\textwidth}
\centering
\include{figs/mass_expansion_examples_nonplanar_15}
\vspace{-3em}
\caption{}
\label{mass_expansion_examples_nonplanar_15}
\end{subfigure}
\caption{Examples of nonplanar graphs with the parameterization of their edges manifested. In the mass expansion, in addition to type-I and type-IIA regions, type-IIB regions are also relevant here.}
\label{figure-mass_expansion_examples_nonplanar}
\end{figure}

In this appendix, we show some four-loop and five-loop graphs as examples in the mass expansion. For each of them, we list all its corresponding region vectors. The graphs are shown in figure~\ref{figure-mass_expansion_examples_planar} (planar) and figure~\ref{figure-mass_expansion_examples_nonplanar} (nonplanar), and each region vector is specified in one of the following two-column lists in the form
\begin{eqnarray}
    (v_1,\dots,v_N;1)_{X},
\end{eqnarray}
with $X\in \{\text{I},\text{IIA},\text{IIB}\}$ indicating the type of the corresponding region, as defined in section~\ref{section-characterizing_regions}.
Let us start with the region vectors associated with the examples of the planar graphs. There are 64 region vectors for figure~\ref{mass_expansion_examples_planar_1}, which are:
\begin{longtable}{ll}
    \textsf{(-2,-2,-1,-1,0,-1,-1,-2,-2,-3,-1,-2,-3,-4,-5;1)}$_\text{I}$,\  &\  \textsf{(-2,-2,-1,-1,0,-1,-1,-2,-2,-2,-1,-2,-3,-4,-4;1)}$_\text{I}$,\\
    \textsf{(-2,-1,-1,-1,0,-1,-1,-2,-2,-2,-1,-2,-3,-3,-4;1)}$_\text{I}$,\  &\  \textsf{(-2,-1,-1,-1,0,-1,-1,-1,-2,-2,-1,-2,-2,-3,-4;1)}$_\text{I}$,\\
    \textsf{(-2,-1,-1,0,0,-1,-1,-1,-2,-2,-1,-1,-2,-3,-4;1)}$_\text{I}$,\  &\  \textsf{(-2,-1,-1,0,0,0,-1,-1,-2,-2,0,-1,-2,-3,-4;1)}$_\text{I}$,\\
    \textsf{(-1,-1,-1,-1,0,-1,-1,-2,-2,-2,-1,-2,-3,-3,-3;1)}$_\text{I}$,\  &\  \textsf{(-1,-1,-1,-1,0,-1,-1,-1,-2,-2,-1,-2,-2,-3,-3;1)}$_\text{I}$,\\
    \textsf{(-1,-1,-1,-1,0,-1,-1,-1,-1,-2,-1,-2,-2,-2,-3;1)}$_\text{I}$,\  &\  \textsf{(-1,-1,-1,-1,0,-1,-1,-1,-1,-1,-1,-2,-2,-2,-2;1)}$_\text{I}$,\\
    \textsf{(-1,-1,-1,0,0,-1,-1,-1,-2,-2,-1,-1,-2,-3,-3;1)}$_\text{I}$,\  &\  \textsf{(-1,-1,-1,0,0,-1,-1,-1,-1,-2,-1,-1,-2,-2,-3;1)}$_\text{I}$,\\
    \textsf{(-1,-1,-1,0,0,-1,-1,-1,-1,-1,-1,-1,-2,-2,-2;1)}$_\text{I}$,\  &\  \textsf{(-1,-1,-1,0,0,0,-1,-1,-2,-2,0,-1,-2,-3,-3;1)}$_\text{I}$,\\
    \textsf{(-1,-1,-1,0,0,0,-1,-1,-1,-2,0,-1,-2,-2,-3;1)}$_\text{I}$,\  &\  \textsf{(-1,-1,-1,0,0,0,-1,-1,-1,-1,0,-1,-2,-2,-2;1)}$_\text{I}$,\\
\end{longtable}
There are 73 region vectors for figure~\ref{mass_expansion_examples_planar_2}, which are:
\begin{longtable}{ll}
    \textsf{(-2,-1,-1,0,-1,-1,-2,-3,-1,-2,-3,-3,-4,-4,-5;1)}$_\text{I}$,\  &\  \textsf{(-2,-1,-1,0,-1,-1,-2,-2,-1,-2,-3,-3,-4,-4,-4;1)}$_\text{I}$,\\
    \textsf{(-2,-1,-1,0,-1,-1,-2,-2,-1,-2,-3,-3,-3,-4,-3;1)}$_\text{I}$,\  &\  \textsf{(-2,-1,-1,0,-1,-1,-2,-1,-1,-2,-3,-1,-1,-4,-1;1)}$_\text{I}$,\\
    \textsf{(-2,-1,-1,0,-1,-1,-1,-2,-1,-2,-2,-2,-3,-4,-3;1)}$_\text{I}$,\  &\  \textsf{(-2,-1,0,0,-1,-1,-1,-2,-1,-1,-2,-2,-3,-4,-3;1)}$_\text{I}$,\\
    \textsf{(-2,-1,0,0,0,-1,-1,-2,0,-1,-2,-2,-3,-4,-3;1)}$_\text{I}$,\  &\  \textsf{(-1,-1,-1,-1,-1,-1,-1,-1,-2,-2,-2,-1,-1,-2,-1;1)}$_\text{I}$,\\
    \textsf{(-1,-1,-1,0,-1,-1,-2,-2,-1,-2,-3,-3,-3,-3,-3;1)}$_\text{I}$,\  &\  \textsf{(-1,-1,-1,0,-1,-1,-2,-2,-1,-2,-2,-3,-2,-2,-3;1)}$_\text{I}$,\\
    \textsf{(-1,-1,-1,0,-1,-1,-2,-1,-1,-2,-3,-1,-1,-3,-1;1)}$_\text{I}$,\  &\  \textsf{(-1,-1,-1,0,-1,-1,-1,-2,-1,-2,-2,-2,-3,-3,-3;1)}$_\text{I}$,\\
    \textsf{(-1,-1,-1,0,-1,-1,-1,-2,-1,-2,-2,-2,-2,-2,-3;1)}$_\text{I}$,\  &\  \textsf{(-1,-1,-1,0,-1,-1,-1,-1,-1,-2,-2,-2,-2,-2,-2;1)}$_\text{I}$,\\
    \textsf{(-1,-1,-1,0,-1,-1,-1,-1,-1,-2,-2,-1,-1,-2,-1;1)}$_\text{I}$,\  &\  \textsf{(-1,-1,-1,0,0,-1,-1,-1,0,-2,-2,-1,-1,-2,-1;1)}$_\text{I}$,\\
\end{longtable}
There are 54 region vectors for figure~\ref{mass_expansion_examples_planar_3}, which are:
\begin{longtable}{ll}
    \textsf{(-2,-1,-1,0,-1,-1,-2,-2,-1,-2,-3,-3,-3,-4,-3;1)}$_\text{I}$,\  &\  \textsf{(-2,-1,-1,0,-1,-1,-2,-1,-1,-2,-3,-3,-1,-4,-1;1)}$_\text{I}$,\\
    \textsf{(-1,-1,-1,-1,-1,-1,-1,-1,-2,-2,-2,-2,-1,-2,-1;1)}$_\text{I}$,\  &\  \textsf{(-1,-1,-1,0,-1,-1,-2,-2,-1,-2,-3,-3,-3,-3,-3;1)}$_\text{I}$,\\
    \textsf{(-1,-1,-1,0,-1,-1,-2,-2,-1,-2,-2,-3,-3,-2,-3;1)}$_\text{I}$,\  &\  \textsf{(-1,-1,-1,0,-1,-1,-2,-1,-1,-2,-3,-3,-1,-3,-1;1)}$_\text{I}$,\\
    \textsf{(-1,-1,-1,0,-1,-1,-2,-1,-1,-2,-2,-3,-1,-2,-1;1)}$_\text{I}$,\  &\  \textsf{(-1,-1,-1,0,-1,-1,-1,-2,-1,-2,-2,-2,-2,-2,-3;1)}$_\text{I}$,\\
    \textsf{(-1,-1,-1,0,-1,-1,-1,-1,-1,-2,-2,-2,-2,-2,-2;1)}$_\text{I}$,\  &\  \textsf{(-1,-1,-1,0,-1,-1,-1,-1,-1,-2,-2,-2,-1,-2,-1;1)}$_\text{I}$,\\
    \textsf{(-1,-1,-1,0,0,-1,-1,-1,0,-2,-2,-2,-1,-2,-1;1)}$_\text{I}$,\  &\  \textsf{(-1,-1,0,0,-1,-1,-1,-2,-1,-1,-2,-2,-2,-2,-3;1)}$_\text{I}$,\\
    \textsf{(-1,-1,0,0,-1,-1,-1,-1,-1,-1,-2,-2,-2,-2,-2;1)}$_\text{I}$,\  &\  \textsf{(-1,-1,0,0,-1,-1,-1,-1,-1,-1,-2,-2,-1,-2,-1;1)}$_\text{I}$,\\
    \textsf{(-1,-1,0,0,-1,0,-1,-1,0,0,-2,-2,-1,-2,-1;1)}$_\text{I}$,\  &\  \textsf{(-1,-1,0,0,0,-1,-1,-2,0,-1,-2,-2,-2,-2,-3;1)}$_\text{I}$,\\
\end{longtable}
There are 47 region vectors for figure~\ref{mass_expansion_examples_planar_4}, which are:
\begin{longtable}{ll}
    \textsf{(-2,-2,-1,0,-1,-2,-2,-1,-1,-2,-2,-4,-2,-3,-4;1)}$_\text{I}$,\  &\  \textsf{(-2,-1,-1,0,-1,-2,-2,-1,-1,-2,-2,-2,-2,-3,-4;1)}$_\text{I}$,\\
    \textsf{(-2,0,-1,0,-1,-2,-2,-1,-1,-2,0,0,-2,-3,-4;1)}$_\text{I}$,\ &\  \textsf{(-1,-1,-1,-1,-1,-1,-2,-2,-1,-1,-2,-2,-2,-1,-3;1)}$_\text{I}$,\\
    \textsf{(-1,-1,-1,-1,-1,-1,-1,-2,-1,-1,-2,-2,-2,-1,-2;1)}$_\text{I}$,\  &\  \textsf{(-1,-1,-1,0,-1,-2,-2,-1,-1,-2,-2,-2,-2,-3,-3;1)}$_\text{I}$,\\
    \textsf{(-1,-1,-1,0,-1,-1,-2,-1,-1,-2,-2,-2,-2,-2,-3;1)}$_\text{I}$,\  &\  \textsf{(-1,-1,-1,0,-1,-1,-2,-1,-1,-1,-2,-2,-2,-1,-3;1)}$_\text{I}$,\\
    \textsf{(-1,-1,-1,0,-1,-1,-1,-1,-1,-2,-2,-2,-2,-2,-2;1)}$_\text{I}$,\  &\  \textsf{(-1,-1,-1,0,-1,-1,-1,-1,-1,-1,-2,-2,-2,-1,-2;1)}$_\text{I}$,\\
    \textsf{(-1,-1,-1,0,0,-1,-2,0,0,-1,-2,-2,-2,-1,-3;1)}$_\text{I}$,\  &\  \textsf{(-1,-1,-1,0,0,-1,-1,0,0,-1,-2,-2,-2,-1,-2;1)}$_\text{I}$,\\
    \textsf{(-1,-1,0,0,-1,-2,-2,-1,-1,-1,-1,-2,-1,-3,-3;1)}$_\text{I}$,\  &\  \textsf{(-1,-1,0,0,-1,-1,-2,-1,-1,-1,-1,-2,-1,-1,-3;1)}$_\text{I}$,\\
    \textsf{(-1,-1,0,0,-1,-1,-1,-1,-1,-1,-1,-2,-1,-1,-2;1)}$_\text{I}$,\  &\  \textsf{(-1,-1,0,0,-1,-1,-1,0,-1,0,0,-2,0,-1,-2;1)}$_\text{I}$,\\
\end{longtable}
There are 20 region vectors for figure~\ref{mass_expansion_examples_planar_5}, which are:
\begin{longtable}{ll}
    \textsf{(-1,-1,-1,-1,-1,-1,-2,-2,-2,-2,-2,-2,-1,-2,-1;1)}$_\text{I}$,\  &\  \textsf{(-1,-1,0,-1,-1,-2,-1,-1,-1,-1,-2,-2,-2,-2,-3;1)}$_\text{I}$,\\
    \textsf{(-1,-1,0,-1,-1,-1,-1,-1,-1,-1,-2,-2,-2,-2,-2;1)}$_\text{I}$,\  &\  \textsf{(-1,-1,0,-1,-1,-1,-1,-1,-1,-1,-2,-2,-1,-2,-1;1)}$_\text{I}$,\\
    \textsf{(-1,-1,0,0,-1,-1,0,0,0,0,-2,-2,-1,-2,-1;1)}$_\text{I}$,\  &\  \textsf{(-1,0,0,-1,-1,-1,-1,-1,-1,-1,-1,-1,-1,-2,-1;1)}$_\text{I}$,\\
    \textsf{(-1,0,0,0,-1,-1,0,0,0,0,-1,-1,-1,-2,-1;1)}$_\text{I}$,\  &\  \textsf{(0,-1,-1,-1,-1,-1,-2,-2,-2,-2,0,-2,-1,0,-1;1)}$_\text{I}$,\\
    \textsf{(0,-1,0,-1,-1,-2,-1,-1,-1,-1,0,-2,-2,0,-3;1)}$_\text{I}$,\  &\  \textsf{(0,-1,0,-1,-1,-1,-1,-1,-1,-1,0,-2,-2,0,-2;1)}$_\text{I}$,\\
    \textsf{(0,-1,0,-1,-1,-1,-1,-1,-1,-1,0,-2,-1,0,-1;1)}$_\text{I}$,\  &\  \textsf{(0,-1,0,0,-1,-1,0,0,0,0,0,-2,-1,0,-1;1)}$_\text{I}$,\\
    \textsf{(0,0,0,-1,-1,-1,-1,-1,-1,-1,-1,-1,-1,-1,-1;1)}$_\text{I}$,\  &\  \textsf{(0,0,0,-1,-1,-1,-1,-1,-1,-1,0,-1,-1,0,-1;1)}$_\text{I}$,\\
    \textsf{(0,0,0,-1,0,-1,0,0,0,0,0,0,0,0,-1;1)}$_\text{IIA}$,\  &\  \textsf{(0,0,0,-1,0,0,0,0,0,0,0,0,0,0,0;1)}$_\text{IIA}$,\\
    \textsf{(0,0,0,0,-1,-1,0,0,0,0,-1,-1,-1,-1,-1;1)}$_\text{I}$,\  &\  \textsf{(0,0,0,0,-1,-1,0,0,0,0,0,-1,-1,0,-1;1)}$_\text{I}$,\\
    \textsf{(0,0,0,0,0,-1,0,0,0,0,0,0,0,0,-1;1)}$_\text{I}$,\  &\  \textsf{(0,0,0,0,0,0,0,0,0,0,0,0,0,0,0;1)}$_\text{I}$.
\end{longtable}
There are 12 region vectors for figure~\ref{mass_expansion_examples_planar_6}, which are:
\begin{longtable}{ll}
    \textsf{(-1,-1,0,-1,-1,-2,-1,-2,-3,-3,-3,-3,-3,-3,-3;1)}$_\text{I}$,\  &\  \textsf{(-1,-1,0,-1,-1,-1,-1,-2,-2,-2,-2,-2,-2,-2,-2;1)}$_\text{I}$,\\
    \textsf{(-1,0,0,-1,-1,-1,-1,-1,-2,-2,-2,-2,-2,-2,-2;1)}$_\text{I}$,\  &\  \textsf{(-1,0,0,0,-1,-1,0,-1,-2,-2,-2,-2,-2,-2,-2;1)}$_\text{I}$,\\
    \textsf{(0,0,0,-1,-1,-1,-1,-1,-1,-1,-1,-1,-1,-1,-1;1)}$_\text{I}$,\  &\  \textsf{(0,0,0,-1,-1,0,-1,0,0,0,0,0,0,0,0;1)}$_\text{IIA}$,\\
    \textsf{(0,0,0,-1,0,-1,0,0,-1,-1,-1,-1,-1,-1,-1;1)}$_\text{IIA}$,\  &\  \textsf{(0,0,0,-1,0,0,0,0,0,0,0,0,0,0,0;1)}$_\text{IIA}$,\\
    \textsf{(0,0,0,0,-1,-1,0,-1,-1,-1,-1,-1,-1,-1,-1;1)}$_\text{I}$,\  &\  \textsf{(0,0,0,0,-1,0,0,0,0,0,0,0,0,0,0;1)}$_\text{IIA}$,\\
    \textsf{(0,0,0,0,0,-1,0,0,-1,-1,-1,-1,-1,-1,-1;1)}$_\text{I}$,\  &\  \textsf{(0,0,0,0,0,0,0,0,0,0,0,0,0,0,0;1)}$_\text{I}$.
\end{longtable}
There are 43 region vectors for figure~\ref{mass_expansion_examples_planar_7}, which are:
\begin{longtable}{ll}
    \textsf{(-2,-1,-1,-1,-2,-2,-2,-1,-4,-3,-1,-1,-3,-3,-3;1)}$_\text{I}$,\  &\  \textsf{(-1,-1,-1,-1,-2,-2,-2,-1,-3,-3,-1,-1,-3,-3,-3;1)}$_\text{I}$,\\
    \textsf{(-1,-1,-1,-1,-2,-2,-2,-1,-2,-2,-1,-1,-3,-3,-3;1)}$_\text{I}$,\  &\  \textsf{(-1,-1,-1,-1,-2,-2,-2,-1,-2,-2,-1,-1,-2,-3,-2;1)}$_\text{I}$,\\
    \textsf{(-1,-1,-1,-1,-1,-2,-2,-2,-2,-2,-2,-1,-1,-1,-2;1)}$_\text{I}$,\  &\  \textsf{(-1,-1,-1,-1,-1,-2,-2,-1,-2,-2,-1,-1,-2,-2,-2;1)}$_\text{I}$,\\
    \textsf{(-1,-1,-1,-1,-1,-2,-2,-1,-2,-2,-1,-1,-1,-1,-2;1)}$_\text{I}$,\  &\  \textsf{(-1,-1,-1,-1,-1,-2,-1,-1,-2,-1,-1,-1,-1,-1,-1;1)}$_\text{I}$,\\
    \textsf{(-1,0,-1,-2,-2,-1,-1,-1,-2,-2,-3,-3,-3,-3,-3;1)}$_\text{I}$,\  &\  \textsf{(-1,0,-1,-2,-2,-1,-1,-1,-2,-2,-2,-3,-2,-3,-2;1)}$_\text{I}$,\\
    \textsf{(-1,0,-1,-2,-1,-1,-1,-1,-2,-2,-3,-1,-1,-1,-3;1)}$_\text{I}$,\  &\  \textsf{(-1,0,-1,-1,-2,-1,-1,-1,-2,-2,-2,-2,-2,-3,-2;1)}$_\text{I}$,\\
    \textsf{(-1,0,-1,-1,-2,-1,-1,-1,-2,-2,-1,-1,-3,-3,-3;1)}$_\text{I}$,\  &\  \textsf{(-1,0,-1,-1,-2,-1,-1,-1,-2,-2,-1,-1,-2,-3,-2;1)}$_\text{I}$,\\
    \textsf{(-1,0,-1,-1,-1,-1,-1,-1,-2,-2,-2,-2,-2,-2,-2;1)}$_\text{I}$,\  &\  \textsf{(-1,0,-1,-1,-1,-1,-1,-1,-2,-2,-1,-1,-2,-2,-2;1)}$_\text{I}$,\\
\end{longtable}
There are 38 region vectors for figure~\ref{mass_expansion_examples_planar_8}, which are:
\begin{longtable}{ll}
    \textsf{(-2,-1,-1,-1,-1,-1,-2,-2,-2,-1,-2,-1,-3,-4;1)}$_\text{I}$,\  &\  \textsf{(-2,-1,-1,0,-1,-2,-2,-2,-1,-1,-2,-3,-3,-4;1)}$_\text{I}$,\\
    \textsf{(-2,-1,-1,0,-1,-1,-2,-2,-1,-1,-2,-1,-3,-4;1)}$_\text{I}$,\  &\  \textsf{(-1,-1,-1,-1,-1,-1,-2,-2,-2,-1,-2,-1,-3,-3;1)}$_\text{I}$,\\
    \textsf{(-1,-1,-1,-1,-1,-1,-1,-2,-2,-1,-2,-1,-2,-3;1)}$_\text{I}$,\  &\  \textsf{(-1,-1,-1,-1,-1,-1,-1,-1,-2,-1,-2,-1,-2,-2;1)}$_\text{I}$,\\
    \textsf{(-1,-1,-1,0,-1,-2,-2,-2,-1,-1,-2,-3,-3,-3;1)}$_\text{I}$,\  &\  \textsf{(-1,-1,-1,0,-1,-1,-2,-2,-1,-1,-2,-1,-3,-3;1)}$_\text{I}$,\\
    \textsf{(-1,-1,-1,0,-1,-1,-1,-2,-1,-1,-2,-1,-2,-3;1)}$_\text{I}$,\  &\  \textsf{(-1,-1,-1,0,-1,-1,-1,-1,-1,-1,-2,-1,-2,-2;1)}$_\text{I}$,\\
    \textsf{(-1,-1,-1,0,0,-1,-1,-2,0,0,-2,-1,-2,-3;1)}$_\text{I}$,\  &\  \textsf{(-1,-1,-1,0,0,-1,-1,-1,0,0,-2,-1,-2,-2;1)}$_\text{I}$,\\
    \textsf{(-1,-1,0,0,-1,-1,-1,-2,-1,-1,-1,-1,-2,-3;1)}$_\text{I}$,\  &\  \textsf{(-1,-1,0,0,-1,-1,-1,-2,0,-1,0,-1,-2,-3;1)}$_\text{I}$,\\
    \textsf{(-1,-1,0,0,-1,-1,-1,-1,-1,-1,-1,-1,-2,-2;1)}$_\text{I}$,\  &\  \textsf{(-1,-1,0,0,-1,-1,-1,-1,0,-1,0,-1,-2,-2;1)}$_\text{I}$,\\
\end{longtable}
There are 20 region vectors for figure~\ref{mass_expansion_examples_planar_9}, which are:
\begin{longtable}{ll}
    \textsf{(-1,-1,-1,-1,-1,-1,-2,-2,-2,-1,-2,-2,-1;1)}$_\text{I}$,\  &\  \textsf{(-1,-1,0,-1,-1,-2,-1,-2,-2,-2,-2,-2,-3;1)}$_\text{I}$,\\
    \textsf{(-1,-1,0,-1,-1,-1,-1,-2,-2,-2,-2,-2,-2;1)}$_\text{I}$,\  &\  \textsf{(-1,-1,0,-1,-1,-1,-1,-2,-2,-1,-2,-2,-1;1)}$_\text{I}$,\\
    \textsf{(-1,-1,0,0,-1,-1,0,-2,-2,-1,-2,-2,-1;1)}$_\text{I}$,\  &\  \textsf{(-1,0,0,-1,-1,-1,-1,-1,-1,-1,-2,-1,-1;1)}$_\text{I}$,\\
    \textsf{(-1,0,0,0,-1,-1,0,-1,-1,-1,-2,-1,-1;1)}$_\text{I}$,\  &\  \textsf{(0,-1,-1,-1,-1,-1,-2,0,-2,-1,0,-2,-1;1)}$_\text{I}$,\\
    \textsf{(0,-1,0,-1,-1,-2,-1,0,-2,-2,0,-2,-3;1)}$_\text{I}$,\  &\  \textsf{(0,-1,0,-1,-1,-1,-1,0,-2,-2,0,-2,-2;1)}$_\text{I}$,\\
    \textsf{(0,-1,0,-1,-1,-1,-1,0,-2,-1,0,-2,-1;1)}$_\text{I}$,\  &\  \textsf{(0,-1,0,0,-1,-1,0,0,-2,-1,0,-2,-1;1)}$_\text{I}$,\\
    \textsf{(0,0,0,-1,-1,-1,-1,-1,-1,-1,-1,-1,-1;1)}$_\text{I}$,\  &\  \textsf{(0,0,0,-1,-1,-1,-1,0,-1,-1,0,-1,-1;1)}$_\text{I}$,\\
    \textsf{(0,0,0,-1,0,-1,0,0,0,0,0,0,-1;1)}$_\text{IIA}$,\  &\  \textsf{(0,0,0,-1,0,0,0,0,0,0,0,0,0;1)}$_\text{IIA}$,\\
    \textsf{(0,0,0,0,-1,-1,0,-1,-1,-1,-1,-1,-1;1)}$_\text{I}$,\  &\  \textsf{(0,0,0,0,-1,-1,0,0,-1,-1,0,-1,-1;1)}$_\text{I}$,\\
    \textsf{(0,0,0,0,0,-1,0,0,0,0,0,0,-1;1)}$_\text{I}$,\  &\  \textsf{(0,0,0,0,0,0,0,0,0,0,0,0,0;1)}$_\text{I}$.
\end{longtable}

\newpage
We then show the region vectors associated with the examples of nonplanar graphs. There are 40 region vectors for figure~\ref{mass_expansion_examples_nonplanar_1}, which are:
\begin{longtable}{ll}
    \textsf{(-}$\frac{\textsf{1}}{\textsf{2}}$\textsf{,-}$\frac{\textsf{1}}{\textsf{2}}$\textsf{,-}$\frac{\textsf{1}}{\textsf{2}}$\textsf{,0,-1,-1,-}$\frac{\textsf{3}}{\textsf{2}}$\textsf{,-}$\frac{\textsf{3}}{\textsf{2}}$\textsf{,-1,-}$\frac{\textsf{1}}{\textsf{2}}$\textsf{,-2,-2;1)}$_\text{IIB}$,\  &\  \textsf{(-}$\frac{\textsf{1}}{\textsf{2}}$\textsf{,-}$\frac{\textsf{1}}{\textsf{2}}$\textsf{,-}$\frac{\textsf{1}}{\textsf{2}}$\textsf{,0,-1,-1,-}$\frac{\textsf{1}}{\textsf{2}}$\textsf{,-}$\frac{\textsf{3}}{\textsf{2}}$\textsf{,-1,-}$\frac{\textsf{1}}{\textsf{2}}$\textsf{,-1,-2;1)}$_\text{IIB}$,\\
    \textsf{(-}$\frac{\textsf{1}}{\textsf{2}}$\textsf{,-}$\frac{\textsf{1}}{\textsf{2}}$\textsf{,-}$\frac{\textsf{1}}{\textsf{2}}$\textsf{,0,-1,-1,-}$\frac{\textsf{1}}{\textsf{2}}$\textsf{,-}$\frac{\textsf{1}}{\textsf{2}}$\textsf{,-1,-}$\frac{\textsf{1}}{\textsf{2}}$\textsf{,-1,-1;1)}$_\text{IIB}$,\  &\  \textsf{(-1,-1,-1,0,-1,-1,-2,-2,-2,-1,-3,-3;1)}$_\text{I}$,\\
    \textsf{(-1,-1,-1,0,-1,-1,-1,-2,-2,-1,-2,-3;1)}$_\text{I}$,\  &\  \textsf{(-1,-1,-1,0,-1,-1,-1,-1,-2,-1,-2,-2;1)}$_\text{I}$,\\
    \textsf{(-}$\frac{\textsf{1}}{\textsf{2}}$\textsf{,-}$\frac{\textsf{1}}{\textsf{2}}$\textsf{,0,0,-1,-1,-1,-}$\frac{\textsf{3}}{\textsf{2}}$\textsf{,-1,-1,-}$\frac{\textsf{1}}{\textsf{2}}$\textsf{,-2;1)}$_\text{IIB}$,\  &\  \textsf{(-}$\frac{\textsf{1}}{\textsf{2}}$\textsf{,-}$\frac{\textsf{1}}{\textsf{2}}$\textsf{,0,0,-1,-1,-1,-}$\frac{\textsf{3}}{\textsf{2}}$\textsf{,-1,-}$\frac{\textsf{1}}{\textsf{2}}$\textsf{,-1,-2;1)}$_\text{IIB}$,\\
    \textsf{(-}$\frac{\textsf{1}}{\textsf{2}}$\textsf{,-}$\frac{\textsf{1}}{\textsf{2}}$\textsf{,0,0,-1,-1,-1,-1,-1,-}$\frac{\textsf{1}}{\textsf{2}}$\textsf{,-1,-}$\frac{\textsf{3}}{\textsf{2}}$\textsf{;1)}$_\text{IIB}$,\  &\  \textsf{(-}$\frac{\textsf{1}}{\textsf{2}}$\textsf{,-}$\frac{\textsf{1}}{\textsf{2}}$\textsf{,0,0,-1,-1,-1,-}$\frac{\textsf{1}}{\textsf{2}}$\textsf{,-1,-1,-}$\frac{\textsf{1}}{\textsf{2}}$\textsf{,-1;1)}$_\text{IIB}$,\\
\end{longtable}
There are 36 region vectors for figure~\ref{mass_expansion_examples_nonplanar_2}, which are:
\begin{longtable}{ll}
    \textsf{(-1,-}$\frac{\textsf{1}}{\textsf{2}}$\textsf{,0,-1,-1,-1,-1,-}$\frac{\textsf{1}}{\textsf{2}}$\textsf{,-}$\frac{\textsf{1}}{\textsf{2}}$\textsf{,-1,-2,-1;1)}$_\text{IIB}$,\  &\  \textsf{(-1,-1,-1,-1,-1,-1,-2,-2,-1,-1,-2,-1;1)}$_\text{I}$,\\
    \textsf{(-}$\frac{\textsf{1}}{\textsf{2}}$\textsf{,-}$\frac{\textsf{1}}{\textsf{2}}$\textsf{,0,-1,-1,-}$\frac{\textsf{3}}{\textsf{2}}$\textsf{,-1,-}$\frac{\textsf{1}}{\textsf{2}}$\textsf{,-}$\frac{\textsf{1}}{\textsf{2}}$\textsf{,-2,-2,-2;1)}$_\text{IIB}$,\  &\  \textsf{(-}$\frac{\textsf{1}}{\textsf{2}}$\textsf{,-}$\frac{\textsf{1}}{\textsf{2}}$\textsf{,0,-1,-1,-}$\frac{\textsf{3}}{\textsf{2}}$\textsf{,-1,-}$\frac{\textsf{1}}{\textsf{2}}$\textsf{,-}$\frac{\textsf{1}}{\textsf{2}}$\textsf{,-1,-1,-2;1)}$_\text{IIB}$,\\
    \textsf{(-}$\frac{\textsf{1}}{\textsf{2}}$\textsf{,-}$\frac{\textsf{1}}{\textsf{2}}$\textsf{,0,-1,-1,-1,-1,-}$\frac{\textsf{1}}{\textsf{2}}$\textsf{,-1,-}$\frac{\textsf{1}}{\textsf{2}}$\textsf{,-1,-}$\frac{\textsf{1}}{\textsf{2}}$\textsf{;1)}$_\text{IIB}$,\  &\  \textsf{(-}$\frac{\textsf{1}}{\textsf{2}}$\textsf{,-}$\frac{\textsf{1}}{\textsf{2}}$\textsf{,0,-1,-1,-1,-1,-}$\frac{\textsf{1}}{\textsf{2}}$\textsf{,-}$\frac{\textsf{1}}{\textsf{2}}$\textsf{,-1,-}$\frac{\textsf{3}}{\textsf{2}}$\textsf{,-1;1)}$_\text{IIB}$,\\
    \textsf{(-}$\frac{\textsf{1}}{\textsf{2}}$\textsf{,-}$\frac{\textsf{1}}{\textsf{2}}$\textsf{,0,-1,-1,-1,-1,-}$\frac{\textsf{1}}{\textsf{2}}$\textsf{,-}$\frac{\textsf{1}}{\textsf{2}}$\textsf{,-}$\frac{\textsf{1}}{\textsf{2}}$\textsf{,-1,-1;1)}$_\text{IIB}$,\  &\  \textsf{(-}$\frac{\textsf{1}}{\textsf{2}}$\textsf{,-}$\frac{\textsf{1}}{\textsf{2}}$\textsf{,0,-1,-1,-}$\frac{\textsf{1}}{\textsf{2}}$\textsf{,-1,-}$\frac{\textsf{1}}{\textsf{2}}$\textsf{,-}$\frac{\textsf{1}}{\textsf{2}}$\textsf{,-1,-1,-1;1)}$_\text{IIB}$,\\
\end{longtable}
There are 32 region vectors for figure~\ref{mass_expansion_examples_nonplanar_3}, which are:
\begin{longtable}{ll}
    \textsf{(-}$\frac{\textsf{1}}{\textsf{2}}$\textsf{,-}$\frac{\textsf{1}}{\textsf{2}}$\textsf{,-}$\frac{\textsf{1}}{\textsf{2}}$\textsf{,-1,-1,-1,-1,-1,-}$\frac{\textsf{1}}{\textsf{2}}$\textsf{,-}$\frac{\textsf{1}}{\textsf{2}}$\textsf{,-1,-}$\frac{\textsf{1}}{\textsf{2}}$\textsf{;1)}$_\text{IIB}$,\  &\  \textsf{(-1,-1,-1,-1,-2,-2,-1,-1,-1,-3,-3,-3;1)}$_\text{I}$,\\
    \textsf{(-1,-1,-1,-1,-2,-2,-1,-1,-1,-2,-2,-3;1)}$_\text{I}$,\  &\  \textsf{(-1,-1,-1,-1,-1,-2,-2,-2,-1,-1,-2,-1;1)}$_\text{I}$,\\
    \textsf{(-1,-1,-1,-1,-1,-2,-1,-1,-1,-2,-2,-2;1)}$_\text{I}$,\  &\  \textsf{(-1,-1,-1,-1,-1,-2,-1,-1,-1,-1,-2,-1;1)}$_\text{I}$,\\
    \textsf{(-}$\frac{\textsf{1}}{\textsf{2}}$\textsf{,0,-1,-1,-}$\frac{\textsf{3}}{\textsf{2}}$\textsf{,-}$\frac{\textsf{1}}{\textsf{2}}$\textsf{,-}$\frac{\textsf{1}}{\textsf{2}}$\textsf{,-1,-1,-1,-1,-2;1)}$_\text{IIB}$,\  &\  \textsf{(-}$\frac{\textsf{1}}{\textsf{2}}$\textsf{,0,-1,-1,-}$\frac{\textsf{1}}{\textsf{2}}$\textsf{,-}$\frac{\textsf{1}}{\textsf{2}}$\textsf{,-}$\frac{\textsf{1}}{\textsf{2}}$\textsf{,-1,-1,-1,-1,-1;1)}$_\text{IIB}$,\\
    \textsf{(-1,0,-1,-2,-2,-1,-3,-1,-1,-2,-2,-3;1)}$_\text{I}$,\  &\  \textsf{(-1,0,-1,-1,-2,-1,-1,-2,-2,-2,-2,-3;1)}$_\text{I}$,\\
    \textsf{(-1,0,-1,-1,-2,-1,-1,-1,-1,-2,-2,-3;1)}$_\text{I}$,\  &\  \textsf{(-1,0,-1,-1,-1,-1,-1,-2,-2,-2,-2,-2;1)}$_\text{I}$,\\
    \textsf{(-1,0,-1,-1,-1,-1,-1,-1,-1,-2,-2,-2;1)}$_\text{I}$,\  &\  \textsf{(-1,0,-1,-1,-1,-1,-1,-1,-1,-1,-2,-1;1)}$_\text{I}$,\\
    \textsf{(-1,0,0,-1,-1,0,-1,0,0,-2,-2,-2;1)}$_\text{I}$,\  &\  \textsf{(-1,0,0,-1,-1,0,-1,0,0,-1,-2,-1;1)}$_\text{I}$,\\
\end{longtable}
There are 94 region vectors for figure~\ref{mass_expansion_examples_nonplanar_4}, which are:
\begin{longtable}{ll}
    \textsf{(-2,-2,-1,-1,0,-1,-1,-1,-2,-2,-4,-1,-2,-3,-4;1)}$_\text{I}$,\  &\  \textsf{(-}$\frac{\textsf{1}}{\textsf{2}}$\textsf{,-}$\frac{\textsf{1}}{\textsf{2}}$\textsf{,-}$\frac{\textsf{1}}{\textsf{2}}$\textsf{,-}$\frac{\textsf{1}}{\textsf{2}}$\textsf{,0,-1,-1,-}$\frac{\textsf{3}}{\textsf{2}}$\textsf{,-}$\frac{\textsf{3}}{\textsf{2}}$\textsf{,-}$\frac{\textsf{3}}{\textsf{2}}$\textsf{,-1,-}$\frac{\textsf{1}}{\textsf{2}}$\textsf{,-2,-2,-2;1)}$_\text{IIB}$,\\
    \textsf{(-}$\frac{\textsf{1}}{\textsf{2}}$\textsf{,-}$\frac{\textsf{1}}{\textsf{2}}$\textsf{,-}$\frac{\textsf{1}}{\textsf{2}}$\textsf{,-}$\frac{\textsf{1}}{\textsf{2}}$\textsf{,0,-1,-1,-}$\frac{\textsf{1}}{\textsf{2}}$\textsf{,-}$\frac{\textsf{3}}{\textsf{2}}$\textsf{,-}$\frac{\textsf{3}}{\textsf{2}}$\textsf{,-1,-}$\frac{\textsf{1}}{\textsf{2}}$\textsf{,-1,-2,-2;1)}$_\text{IIB}$,\  &\  \textsf{(-}$\frac{\textsf{1}}{\textsf{2}}$\textsf{,-}$\frac{\textsf{1}}{\textsf{2}}$\textsf{,-}$\frac{\textsf{1}}{\textsf{2}}$\textsf{,-}$\frac{\textsf{1}}{\textsf{2}}$\textsf{,0,-1,-1,-}$\frac{\textsf{1}}{\textsf{2}}$\textsf{,-}$\frac{\textsf{1}}{\textsf{2}}$\textsf{,-}$\frac{\textsf{3}}{\textsf{2}}$\textsf{,-1,-}$\frac{\textsf{1}}{\textsf{2}}$\textsf{,-1,-1,-2;1)}$_\text{IIB}$,\\
    \textsf{(-}$\frac{\textsf{1}}{\textsf{2}}$\textsf{,-}$\frac{\textsf{1}}{\textsf{2}}$\textsf{,-}$\frac{\textsf{1}}{\textsf{2}}$\textsf{,-}$\frac{\textsf{1}}{\textsf{2}}$\textsf{,0,-1,-1,-}$\frac{\textsf{1}}{\textsf{2}}$\textsf{,-}$\frac{\textsf{1}}{\textsf{2}}$\textsf{,-}$\frac{\textsf{1}}{\textsf{2}}$\textsf{,-1,-}$\frac{\textsf{1}}{\textsf{2}}$\textsf{,-1,-1,-1;1)}$_\text{IIB}$,\  &\  \textsf{(-1,-1,-1,-1,0,-1,-1,-2,-2,-2,-2,-1,-3,-3,-3;1)}$_\text{I}$,\\
    \textsf{(-1,-1,-1,-1,0,-1,-1,-1,-2,-2,-2,-1,-2,-3,-3;1)}$_\text{I}$,\  &\  \textsf{(-1,-1,-1,-1,0,-1,-1,-1,-1,-2,-2,-1,-2,-2,-3;1)}$_\text{I}$,\\
\end{longtable}
There are 104 region vectors for figure~\ref{mass_expansion_examples_nonplanar_5}, which are:
\begin{longtable}{ll}
    \textsf{(-}$\frac{\textsf{3}}{\textsf{2}}$\textsf{,-}$\frac{\textsf{1}}{\textsf{2}}$\textsf{,-}$\frac{\textsf{1}}{\textsf{2}}$\textsf{,-}$\frac{\textsf{1}}{\textsf{2}}$\textsf{,0,-1,-1,-}$\frac{\textsf{3}}{\textsf{2}}$\textsf{,-}$\frac{\textsf{3}}{\textsf{2}}$\textsf{,-}$\frac{\textsf{3}}{\textsf{2}}$\textsf{,-1,-}$\frac{\textsf{1}}{\textsf{2}}$\textsf{,-2,-2,-3;1)}$_\text{IIB}$,\  &\  \textsf{(-}$\frac{\textsf{3}}{\textsf{2}}$\textsf{,-}$\frac{\textsf{1}}{\textsf{2}}$\textsf{,-}$\frac{\textsf{1}}{\textsf{2}}$\textsf{,-}$\frac{\textsf{1}}{\textsf{2}}$\textsf{,0,-1,-1,-}$\frac{\textsf{1}}{\textsf{2}}$\textsf{,-}$\frac{\textsf{3}}{\textsf{2}}$\textsf{,-}$\frac{\textsf{3}}{\textsf{2}}$\textsf{,-1,-}$\frac{\textsf{1}}{\textsf{2}}$\textsf{,-1,-2,-3;1)}$_\text{IIB}$,\\
    \textsf{(-}$\frac{\textsf{3}}{\textsf{2}}$\textsf{,-}$\frac{\textsf{1}}{\textsf{2}}$\textsf{,-}$\frac{\textsf{1}}{\textsf{2}}$\textsf{,0,0,-1,-1,-1,-}$\frac{\textsf{3}}{\textsf{2}}$\textsf{,-}$\frac{\textsf{3}}{\textsf{2}}$\textsf{,-1,-1,-}$\frac{\textsf{1}}{\textsf{2}}$\textsf{,-2,-3;1)}$_\text{IIB}$,\  &\  \textsf{(-}$\frac{\textsf{3}}{\textsf{2}}$\textsf{,-}$\frac{\textsf{1}}{\textsf{2}}$\textsf{,-}$\frac{\textsf{1}}{\textsf{2}}$\textsf{,0,0,-1,-1,-1,-}$\frac{\textsf{3}}{\textsf{2}}$\textsf{,-}$\frac{\textsf{3}}{\textsf{2}}$\textsf{,-1,-}$\frac{\textsf{1}}{\textsf{2}}$\textsf{,-1,-2,-3;1)}$_\text{IIB}$,\\
\end{longtable}
There are 98 region vectors for figure~\ref{mass_expansion_examples_nonplanar_6}, which are:
\begin{longtable}{ll}
    \textsf{(-}$\frac{\textsf{3}}{\textsf{2}}$\textsf{,-}$\frac{\textsf{3}}{\textsf{2}}$\textsf{,-}$\frac{\textsf{1}}{\textsf{2}}$\textsf{,-}$\frac{\textsf{1}}{\textsf{2}}$\textsf{,0,-1,-1,-}$\frac{\textsf{3}}{\textsf{2}}$\textsf{,-}$\frac{\textsf{3}}{\textsf{2}}$\textsf{,-}$\frac{\textsf{5}}{\textsf{2}}$\textsf{,-1,-}$\frac{\textsf{1}}{\textsf{2}}$\textsf{,-2,-3,-4;1)}$_\text{IIB}$,\  &\  \textsf{(-}$\frac{\textsf{3}}{\textsf{2}}$\textsf{,-}$\frac{\textsf{3}}{\textsf{2}}$\textsf{,-}$\frac{\textsf{1}}{\textsf{2}}$\textsf{,-}$\frac{\textsf{1}}{\textsf{2}}$\textsf{,0,-1,-1,-}$\frac{\textsf{3}}{\textsf{2}}$\textsf{,-}$\frac{\textsf{3}}{\textsf{2}}$\textsf{,-}$\frac{\textsf{3}}{\textsf{2}}$\textsf{,-1,-}$\frac{\textsf{1}}{\textsf{2}}$\textsf{,-2,-3,-3;1)}$_\text{IIB}$,\\
    \textsf{(-}$\frac{\textsf{3}}{\textsf{2}}$\textsf{,-}$\frac{\textsf{1}}{\textsf{2}}$\textsf{,-}$\frac{\textsf{1}}{\textsf{2}}$\textsf{,-}$\frac{\textsf{1}}{\textsf{2}}$\textsf{,0,-1,-1,-}$\frac{\textsf{3}}{\textsf{2}}$\textsf{,-}$\frac{\textsf{3}}{\textsf{2}}$\textsf{,-}$\frac{\textsf{3}}{\textsf{2}}$\textsf{,-1,-}$\frac{\textsf{1}}{\textsf{2}}$\textsf{,-2,-3,-2;1)}$_\text{IIB}$,\  &\  \textsf{(-}$\frac{\textsf{3}}{\textsf{2}}$\textsf{,-}$\frac{\textsf{1}}{\textsf{2}}$\textsf{,-}$\frac{\textsf{1}}{\textsf{2}}$\textsf{,-}$\frac{\textsf{1}}{\textsf{2}}$\textsf{,0,-1,-1,-}$\frac{\textsf{1}}{\textsf{2}}$\textsf{,-}$\frac{\textsf{3}}{\textsf{2}}$\textsf{,-}$\frac{\textsf{3}}{\textsf{2}}$\textsf{,-1,-}$\frac{\textsf{1}}{\textsf{2}}$\textsf{,-1,-3,-2;1)}$_\text{IIB}$,\\
\end{longtable}
There are 67 region vectors in figure~\ref{mass_expansion_examples_nonplanar_7}, which are:
\begin{longtable}{ll}
    \textsf{(-2,-1,-1,0,0,-1,-1,-2,-2,-2,-1,-2,-3,-4,-3;1)}$_\text{I}$,\  &\  \textsf{(-}$\frac{\textsf{1}}{\textsf{2}}$\textsf{,-}$\frac{\textsf{1}}{\textsf{2}}$\textsf{,-}$\frac{\textsf{1}}{\textsf{2}}$\textsf{,-}$\frac{\textsf{1}}{\textsf{2}}$\textsf{,0,-1,-1,-1,-1,-1,-1,-1,-1,-1,-}$\frac{\textsf{1}}{\textsf{2}}$\textsf{;1)}$_\text{IIB}$,\\
    \textsf{(-1,-1,-1,-1,0,-1,-1,-1,-1,-1,-2,-2,-2,-2,-1;1)}$_\text{I}$,\  &\  \textsf{(-1,-1,-1,0,0,-1,-1,-2,-2,-2,-1,-2,-3,-3,-3;1)}$_\text{I}$,\\
    \textsf{(-1,-1,-1,0,0,-1,-1,-1,-2,-2,-1,-2,-2,-3,-3;1)}$_\text{I}$,\  &\  \textsf{(-1,-1,-1,0,0,-1,-1,-1,-1,-1,-1,-2,-2,-2,-1;1)}$_\text{I}$,\\
    \textsf{(-}$\frac{\textsf{1}}{\textsf{2}}$\textsf{,-}$\frac{\textsf{1}}{\textsf{2}}$\textsf{,-}$\frac{\textsf{1}}{\textsf{2}}$\textsf{,0,0,0,-1,-1,-1,-1,0,-1,-1,-1,-}$\frac{\textsf{1}}{\textsf{2}}$\textsf{;1)}$_\text{IIB}$,\  &\  \textsf{(-1,-1,-1,0,0,0,-1,-1,-1,-1,0,-2,-2,-2,-1;1)}$_\text{I}$,\\
    \textsf{(-}$\frac{\textsf{1}}{\textsf{2}}$\textsf{,-}$\frac{\textsf{1}}{\textsf{2}}$\textsf{,0,-}$\frac{\textsf{1}}{\textsf{2}}$\textsf{,0,-1,0,-1,-1,-1,0,0,-1,-1,-}$\frac{\textsf{1}}{\textsf{2}}$\textsf{;1)}$_\text{IIB}$,\  &\  \textsf{(-1,-1,0,0,0,-1,-1,-1,-2,-2,-1,-1,-2,-3,-3;1)}$_\text{I}$,\\
    \textsf{(-1,-1,0,0,0,-1,-1,-1,-1,-1,-1,-1,-2,-2,-1;1)}$_\text{I}$,\  &\  \textsf{(-1,-1,0,0,0,-1,0,-1,-1,-1,0,0,-2,-2,-1;1)}$_\text{IIA}$,\\
    \textsf{(-1,-1,0,0,0,0,-1,-1,-2,-2,0,-1,-2,-3,-3;1)}$_\text{I}$,\  &\  \textsf{(-1,-1,0,0,0,0,-1,-1,-1,-1,0,-1,-2,-2,-1;1)}$_\text{I}$,\\
    \textsf{(-}$\frac{\textsf{1}}{\textsf{2}}$\textsf{,-}$\frac{\textsf{1}}{\textsf{2}}$\textsf{,0,0,0,0,0,-1,-1,-1,0,0,-1,-1,-}$\frac{\textsf{1}}{\textsf{2}}$\textsf{;1)}$_\text{IIB}$,\  &\  \textsf{(-1,-1,0,0,0,0,0,-1,-1,-1,0,0,-2,-2,-1;1)}$_\text{I}$,\\
\end{longtable}
There are 73 region vectors in figure~\ref{mass_expansion_examples_nonplanar_8}, which are:
\begin{longtable}{ll}
    \textsf{(-2,-1,-1,0,0,-1,-1,-2,-2,-2,-1,-2,-3,-3,-4;1)}$_\text{I}$,\  &\  \textsf{(-}$\frac{\textsf{1}}{\textsf{2}}$\textsf{,-}$\frac{\textsf{1}}{\textsf{2}}$\textsf{,-}$\frac{\textsf{1}}{\textsf{2}}$\textsf{,-}$\frac{\textsf{1}}{\textsf{2}}$\textsf{,0,-1,-1,-1,-1,-}$\frac{\textsf{3}}{\textsf{2}}$\textsf{,-1,-1,-1,-}$\frac{\textsf{1}}{\textsf{2}}$\textsf{,-2;1)}$_\text{IIB}$,\\
    \textsf{(-}$\frac{\textsf{1}}{\textsf{2}}$\textsf{,-}$\frac{\textsf{1}}{\textsf{2}}$\textsf{,-}$\frac{\textsf{1}}{\textsf{2}}$\textsf{,-}$\frac{\textsf{1}}{\textsf{2}}$\textsf{,0,-1,-1,-1,-1,-}$\frac{\textsf{1}}{\textsf{2}}$\textsf{,-1,-1,-1,-}$\frac{\textsf{1}}{\textsf{2}}$\textsf{,-1;1)}$_\text{IIB}$,\  &\  \textsf{(-1,-1,-1,-1,0,-1,-1,-1,-1,-2,-2,-2,-2,-1,-3;1)}$_\text{I}$,\\
    \textsf{(-1,-1,-1,-1,0,-1,-1,-1,-1,-1,-2,-2,-2,-1,-2;1)}$_\text{I}$,\  &\  \textsf{(-1,-1,-1,0,0,-1,-1,-2,-2,-2,-1,-2,-3,-3,-3;1)}$_\text{I}$,\\
    \textsf{(-1,-1,-1,0,0,-1,-1,-1,-2,-2,-1,-2,-2,-3,-3;1)}$_\text{I}$,\  &\  \textsf{(-1,-1,-1,0,0,-1,-1,-1,-1,-2,-1,-2,-2,-1,-3;1)}$_\text{I}$,\\
    \textsf{(-1,-1,-1,0,0,-1,-1,-1,-1,-1,-1,-2,-2,-1,-2;1)}$_\text{I}$,\  &\  \textsf{(-}$\frac{\textsf{1}}{\textsf{2}}$\textsf{,-}$\frac{\textsf{1}}{\textsf{2}}$\textsf{,-}$\frac{\textsf{1}}{\textsf{2}}$\textsf{,0,0,0,-1,-1,-1,-}$\frac{\textsf{3}}{\textsf{2}}$\textsf{,0,-1,-1,-}$\frac{\textsf{1}}{\textsf{2}}$\textsf{,-2;1)}$_\text{IIB}$,\\
    \textsf{(-}$\frac{\textsf{1}}{\textsf{2}}$\textsf{,-}$\frac{\textsf{1}}{\textsf{2}}$\textsf{,-}$\frac{\textsf{1}}{\textsf{2}}$\textsf{,0,0,0,-1,-1,-1,-}$\frac{\textsf{1}}{\textsf{2}}$\textsf{,0,-1,-1,-}$\frac{\textsf{1}}{\textsf{2}}$\textsf{,-1;1)}$_\text{IIB}$,\  &\  \textsf{(-1,-1,-1,0,0,0,-1,-1,-1,-2,0,-2,-2,-1,-3;1)}$_\text{I}$,\\
\end{longtable}
There are 82 region vectors in figure~\ref{mass_expansion_examples_nonplanar_9}, which are:
\begin{longtable}{ll}
    \textsf{(-}$\frac{\textsf{3}}{\textsf{2}}$\textsf{,-}$\frac{\textsf{1}}{\textsf{2}}$\textsf{,-}$\frac{\textsf{1}}{\textsf{2}}$\textsf{,-}$\frac{\textsf{1}}{\textsf{2}}$\textsf{,0,-1,-1,-1,-}$\frac{\textsf{3}}{\textsf{2}}$\textsf{,-}$\frac{\textsf{3}}{\textsf{2}}$\textsf{,-1,-1,-}$\frac{\textsf{1}}{\textsf{2}}$\textsf{,-3,-2;1)}$_\text{IIB}$,\  &\  \textsf{(-}$\frac{\textsf{3}}{\textsf{2}}$\textsf{,-}$\frac{\textsf{1}}{\textsf{2}}$\textsf{,-}$\frac{\textsf{1}}{\textsf{2}}$\textsf{,0,0,0,-1,-1,-}$\frac{\textsf{3}}{\textsf{2}}$\textsf{,-}$\frac{\textsf{3}}{\textsf{2}}$\textsf{,0,-1,-}$\frac{\textsf{1}}{\textsf{2}}$\textsf{,-3,-2;1)}$_\text{IIB}$,\\
    \textsf{(-}$\frac{\textsf{3}}{\textsf{2}}$\textsf{,-}$\frac{\textsf{1}}{\textsf{2}}$\textsf{,0,-}$\frac{\textsf{1}}{\textsf{2}}$\textsf{,0,-1,0,-1,-}$\frac{\textsf{3}}{\textsf{2}}$\textsf{,-}$\frac{\textsf{3}}{\textsf{2}}$\textsf{,0,0,-}$\frac{\textsf{1}}{\textsf{2}}$\textsf{,-3,-2;1)}$_\text{IIB}$,\  &\  \textsf{(-2,-1,-1,-1,0,-1,-1,-1,-2,-2,-2,-2,-1,-4,-3;1)}$_\text{I}$,\\
    \textsf{(-2,-1,-1,0,0,-1,-1,-2,-2,-2,-1,-2,-3,-4,-3;1)}$_\text{I}$,\  &\  \textsf{(-2,-1,-1,0,0,-1,-1,-1,-2,-2,-1,-2,-1,-4,-3;1)}$_\text{I}$,\\
    \textsf{(-2,-1,-1,0,0,0,-1,-1,-2,-2,0,-2,-1,-4,-3;1)}$_\text{I}$,\  &\  \textsf{(-}$\frac{\textsf{1}}{\textsf{2}}$\textsf{,-}$\frac{\textsf{1}}{\textsf{2}}$\textsf{,-}$\frac{\textsf{1}}{\textsf{2}}$\textsf{,-}$\frac{\textsf{1}}{\textsf{2}}$\textsf{,0,-1,-1,-1,-}$\frac{\textsf{3}}{\textsf{2}}$\textsf{,-}$\frac{\textsf{3}}{\textsf{2}}$\textsf{,-1,-1,-}$\frac{\textsf{1}}{\textsf{2}}$\textsf{,-2,-2;1)}$_\text{IIB}$,\\
\end{longtable}
There are 80 region vectors in figure~\ref{mass_expansion_examples_nonplanar_10}, which are:
\begin{longtable}{ll}
    \textsf{(-}$\frac{\textsf{3}}{\textsf{2}}$\textsf{,-}$\frac{\textsf{3}}{\textsf{2}}$\textsf{,-}$\frac{\textsf{1}}{\textsf{2}}$\textsf{,-}$\frac{\textsf{1}}{\textsf{2}}$\textsf{,0,-1,-1,-}$\frac{\textsf{3}}{\textsf{2}}$\textsf{,-}$\frac{\textsf{3}}{\textsf{2}}$\textsf{,-}$\frac{\textsf{5}}{\textsf{2}}$\textsf{,-1,-}$\frac{\textsf{1}}{\textsf{2}}$\textsf{,-2,-3,-4;1)}$_\text{IIB}$,\  &\  \textsf{(-}$\frac{\textsf{3}}{\textsf{2}}$\textsf{,-}$\frac{\textsf{3}}{\textsf{2}}$\textsf{,-}$\frac{\textsf{1}}{\textsf{2}}$\textsf{,-}$\frac{\textsf{1}}{\textsf{2}}$\textsf{,0,-1,-1,-}$\frac{\textsf{3}}{\textsf{2}}$\textsf{,-}$\frac{\textsf{3}}{\textsf{2}}$\textsf{,-}$\frac{\textsf{3}}{\textsf{2}}$\textsf{,-1,-}$\frac{\textsf{1}}{\textsf{2}}$\textsf{,-2,-3,-3;1)}$_\text{IIB}$,\\
    \textsf{(-}$\frac{\textsf{3}}{\textsf{2}}$\textsf{,-}$\frac{\textsf{1}}{\textsf{2}}$\textsf{,-}$\frac{\textsf{1}}{\textsf{2}}$\textsf{,-}$\frac{\textsf{1}}{\textsf{2}}$\textsf{,0,-1,-1,-}$\frac{\textsf{3}}{\textsf{2}}$\textsf{,-}$\frac{\textsf{3}}{\textsf{2}}$\textsf{,-}$\frac{\textsf{3}}{\textsf{2}}$\textsf{,-1,-}$\frac{\textsf{1}}{\textsf{2}}$\textsf{,-2,-3,-2;1)}$_\text{IIB}$,\  &\  \textsf{(-}$\frac{\textsf{3}}{\textsf{2}}$\textsf{,-}$\frac{\textsf{1}}{\textsf{2}}$\textsf{,-}$\frac{\textsf{1}}{\textsf{2}}$\textsf{,-}$\frac{\textsf{1}}{\textsf{2}}$\textsf{,0,-1,-1,-}$\frac{\textsf{1}}{\textsf{2}}$\textsf{,-}$\frac{\textsf{3}}{\textsf{2}}$\textsf{,-}$\frac{\textsf{3}}{\textsf{2}}$\textsf{,-1,-}$\frac{\textsf{1}}{\textsf{2}}$\textsf{,-1,-3,-2;1)}$_\text{IIB}$,\\
    \textsf{(-2,-2,-1,-1,0,-1,-1,-2,-2,-3,-2,-1,-3,-4,-5;1)}$_\text{I}$,\  &\  \textsf{(-2,-2,-1,-1,0,-1,-1,-2,-2,-2,-2,-1,-3,-4,-4;1)}$_\text{I}$,\\
    \textsf{(-2,-1,-1,-1,0,-1,-1,-2,-2,-2,-2,-1,-3,-4,-3;1)}$_\text{I}$,\  &\  \textsf{(-2,-1,-1,-1,0,-1,-1,-1,-2,-2,-2,-1,-2,-4,-3;1)}$_\text{I}$,\\
\end{longtable}
There are 94 region vectors in figure~\ref{mass_expansion_examples_nonplanar_11}, which are:
\begin{longtable}{ll}
    \textsf{(-}$\frac{\textsf{3}}{\textsf{2}}$\textsf{,-}$\frac{\textsf{1}}{\textsf{2}}$\textsf{,-}$\frac{\textsf{1}}{\textsf{2}}$\textsf{,-}$\frac{\textsf{1}}{\textsf{2}}$\textsf{,0,-1,-1,-1,-}$\frac{\textsf{3}}{\textsf{2}}$\textsf{,-}$\frac{\textsf{3}}{\textsf{2}}$\textsf{,-1,-1,-}$\frac{\textsf{1}}{\textsf{2}}$\textsf{,-3,-2;1)}$_\text{IIB}$,\  &\  \textsf{(-}$\frac{\textsf{3}}{\textsf{2}}$\textsf{,-}$\frac{\textsf{1}}{\textsf{2}}$\textsf{,-}$\frac{\textsf{1}}{\textsf{2}}$\textsf{,0,0,-1,-1,-}$\frac{\textsf{3}}{\textsf{2}}$\textsf{,-}$\frac{\textsf{3}}{\textsf{2}}$\textsf{,-}$\frac{\textsf{3}}{\textsf{2}}$\textsf{,-1,-}$\frac{\textsf{1}}{\textsf{2}}$\textsf{,-2,-3,-2;1)}$_\text{IIB}$,\\
    \textsf{(-}$\frac{\textsf{3}}{\textsf{2}}$\textsf{,-}$\frac{\textsf{1}}{\textsf{2}}$\textsf{,-}$\frac{\textsf{1}}{\textsf{2}}$\textsf{,0,0,-1,-1,-1,-}$\frac{\textsf{3}}{\textsf{2}}$\textsf{,-}$\frac{\textsf{3}}{\textsf{2}}$\textsf{,-1,-}$\frac{\textsf{1}}{\textsf{2}}$\textsf{,-1,-3,-2;1)}$_\text{IIB}$,\  &\  \textsf{(-}$\frac{\textsf{3}}{\textsf{2}}$\textsf{,-}$\frac{\textsf{1}}{\textsf{2}}$\textsf{,-}$\frac{\textsf{1}}{\textsf{2}}$\textsf{,0,0,-1,-1,-}$\frac{\textsf{1}}{\textsf{2}}$\textsf{,-}$\frac{\textsf{3}}{\textsf{2}}$\textsf{,-}$\frac{\textsf{3}}{\textsf{2}}$\textsf{,-1,-}$\frac{\textsf{1}}{\textsf{2}}$\textsf{,-}$\frac{\textsf{1}}{\textsf{2}}$\textsf{,-3,-2;1)}$_\text{IIB}$,\\
    \textsf{(-}$\frac{\textsf{3}}{\textsf{2}}$\textsf{,-}$\frac{\textsf{1}}{\textsf{2}}$\textsf{,-}$\frac{\textsf{1}}{\textsf{2}}$\textsf{,0,0,-1,-1,0,-}$\frac{\textsf{3}}{\textsf{2}}$\textsf{,-}$\frac{\textsf{3}}{\textsf{2}}$\textsf{,-1,0,0,-3,-2;1)}$_\text{IIB}$,\  &\  \textsf{(-}$\frac{\textsf{3}}{\textsf{2}}$\textsf{,-}$\frac{\textsf{1}}{\textsf{2}}$\textsf{,0,-}$\frac{\textsf{1}}{\textsf{2}}$\textsf{,0,-1,0,-1,-}$\frac{\textsf{3}}{\textsf{2}}$\textsf{,-}$\frac{\textsf{3}}{\textsf{2}}$\textsf{,0,0,-}$\frac{\textsf{1}}{\textsf{2}}$\textsf{,-3,-2;1)}$_\text{IIB}$,\\
\end{longtable}
There are 71 region vectors in figure~\ref{mass_expansion_examples_nonplanar_12}, which are:
\begin{longtable}{ll}
    \textsf{(-}$\frac{\textsf{1}}{\textsf{2}}$\textsf{,-}$\frac{\textsf{1}}{\textsf{2}}$\textsf{,-}$\frac{\textsf{1}}{\textsf{2}}$\textsf{,-}$\frac{\textsf{1}}{\textsf{2}}$\textsf{,-1,-1,-1,-1,-1,-1,-}$\frac{\textsf{1}}{\textsf{2}}$\textsf{,-1,-}$\frac{\textsf{1}}{\textsf{2}}$\textsf{,-1,-}$\frac{\textsf{1}}{\textsf{2}}$\textsf{;1)}$_\text{IIB}$,\  &\  \textsf{(-1,-1,-1,-1,-1,-2,-2,-1,-1,-1,-1,-2,-3,-3,-3;1)}$_\text{I}$,\\
    \textsf{(-1,-1,-1,-1,-1,-2,-2,-1,-1,-1,-1,-2,-2,-2,-3;1)}$_\text{I}$,\  &\  \textsf{(-1,-1,-1,-1,-1,-1,-2,-2,-2,-2,-1,-2,-1,-2,-1;1)}$_\text{I}$,\\
    \textsf{(-1,-1,-1,-1,-1,-1,-2,-1,-1,-1,-1,-2,-2,-2,-2;1)}$_\text{I}$,\  &\  \textsf{(-1,-1,-1,-1,-1,-1,-2,-1,-1,-1,-1,-2,-1,-2,-1;1)}$_\text{I}$,\\
    \textsf{(-}$\frac{\textsf{1}}{\textsf{2}}$\textsf{,-}$\frac{\textsf{1}}{\textsf{2}}$\textsf{,0,-1,-1,-}$\frac{\textsf{3}}{\textsf{2}}$\textsf{,-}$\frac{\textsf{1}}{\textsf{2}}$\textsf{,-}$\frac{\textsf{1}}{\textsf{2}}$\textsf{,-1,-1,-1,-1,-2,-2,-2;1)}$_\text{IIB}$,\  &\  \textsf{(-}$\frac{\textsf{1}}{\textsf{2}}$\textsf{,-}$\frac{\textsf{1}}{\textsf{2}}$\textsf{,0,-1,-1,-}$\frac{\textsf{3}}{\textsf{2}}$\textsf{,-}$\frac{\textsf{1}}{\textsf{2}}$\textsf{,-}$\frac{\textsf{1}}{\textsf{2}}$\textsf{,-1,-1,-1,-1,-1,-1,-2;1)}$_\text{IIB}$,\\
    \textsf{(-}$\frac{\textsf{1}}{\textsf{2}}$\textsf{,-}$\frac{\textsf{1}}{\textsf{2}}$\textsf{,0,-1,-1,-}$\frac{\textsf{1}}{\textsf{2}}$\textsf{,-}$\frac{\textsf{1}}{\textsf{2}}$\textsf{,-}$\frac{\textsf{1}}{\textsf{2}}$\textsf{,-1,-1,-1,-1,-1,-1,-1;1)}$_\text{IIB}$,\  &\  \textsf{(-1,-1,0,-1,-2,-2,-1,-3,-1,-1,-1,-2,-3,-3,-3;1)}$_\text{I}$,\\
    \textsf{(-1,-1,0,-1,-2,-2,-1,-3,-1,-1,-1,-2,-2,-2,-3;1)}$_\text{I}$,\  &\  \textsf{(-1,-1,0,-1,-1,-2,-1,-1,-2,-3,-3,-2,-3,-3,-3;1)}$_\text{I}$,\\
    \textsf{(-1,-1,0,-1,-1,-2,-1,-1,-2,-2,-2,-2,-3,-3,-3;1)}$_\text{I}$,\  &\  \textsf{(-1,-1,0,-1,-1,-2,-1,-1,-2,-2,-2,-2,-2,-2,-3;1)}$_\text{I}$,\\
\end{longtable}
There are 40 region vectors for figure~\ref{mass_expansion_examples_nonplanar_13}, which are:
\begin{longtable}{ll}
    \textsf{(-1,-1,-1,-1,-1,-2,-2,-2,-2,-1,-1,-1,-2,-2,-3;1)}$_\text{I}$,\  &\  \textsf{(-1,-1,-1,-1,-1,-2,-2,-1,-2,-1,-1,-1,-1,-1,-3;1)}$_\text{I}$,\\
    \textsf{(-1,-1,-1,-1,-1,-1,-2,-2,-2,-1,-1,-1,-2,-2,-2;1)}$_\text{I}$,\  &\  \textsf{(-1,-1,-1,-1,-1,-1,-2,-1,-2,-1,-1,-1,-1,-1,-2;1)}$_\text{I}$,\\
    \textsf{(-}$\frac{\textsf{1}}{\textsf{2}}$\textsf{,-}$\frac{\textsf{1}}{\textsf{2}}$\textsf{,0,-1,-1,-}$\frac{\textsf{3}}{\textsf{2}}$\textsf{,-1,-1,-}$\frac{\textsf{1}}{\textsf{2}}$\textsf{,-1,-1,-}$\frac{\textsf{1}}{\textsf{2}}$\textsf{,-1,-}$\frac{\textsf{1}}{\textsf{2}}$\textsf{,-2;1)}$_\text{IIB}$,\  &\  \textsf{(-}$\frac{\textsf{1}}{\textsf{2}}$\textsf{,-}$\frac{\textsf{1}}{\textsf{2}}$\textsf{,0,-1,-1,-}$\frac{\textsf{1}}{\textsf{2}}$\textsf{,-1,-1,-}$\frac{\textsf{1}}{\textsf{2}}$\textsf{,-1,-1,-}$\frac{\textsf{1}}{\textsf{2}}$\textsf{,-1,-}$\frac{\textsf{1}}{\textsf{2}}$\textsf{,-1;1)}$_\text{IIB}$,\\
    \textsf{(-1,-1,0,-1,-2,-2,-2,-2,-1,-1,-1,-3,-2,-1,-3;1)}$_\text{I}$,\  &\  \textsf{(-1,-1,0,-1,-2,-2,-2,-1,-1,-1,-3,-3,-1,-3,-3;1)}$_\text{I}$,\\
    \textsf{(-1,-1,0,-1,-2,-2,-2,-1,-1,-1,-1,-3,-1,-1,-3;1)}$_\text{I}$,\  &\  \textsf{(-1,-1,0,-1,-1,-2,-2,-2,-1,-2,-2,-1,-2,-1,-3;1)}$_\text{I}$,\\
    \textsf{(-1,-1,0,-1,-1,-2,-2,-2,-1,-1,-1,-1,-2,-1,-3;1)}$_\text{I}$,\  &\  \textsf{(-1,-1,0,-1,-1,-2,-2,-1,-1,-1,-1,-1,-1,-1,-3;1)}$_\text{I}$,\\
    \textsf{(-1,-1,0,-1,-1,-1,-2,-2,-1,-2,-2,-1,-2,-1,-2;1)}$_\text{I}$,\  &\  \textsf{(-1,-1,0,-1,-1,-1,-2,-2,-1,-1,-1,-1,-2,-1,-2;1)}$_\text{I}$,\\
    \textsf{(-1,-1,0,-1,-1,-1,-2,-1,-1,-1,-1,-1,-1,-1,-2;1)}$_\text{I}$,\  &\  \textsf{(-1,-1,0,0,-1,-1,-2,-2,0,0,0,-1,-2,0,-2;1)}$_\text{I}$,\\
\end{longtable}
There are 24 region vectors for figure~\ref{mass_expansion_examples_nonplanar_14}, which are:
\begin{longtable}{ll}
    \textsf{(-1,-1,-1,-1,-2,-2,-2,-2,-1,-1,-1,-1,-2,-2,-2;1)}$_\text{I}$,\  &\  \textsf{(-1,-1,-1,-1,-2,-2,-1,-2,-1,-1,-1,-1,-2,-1,-1;1)}$_\text{I}$,\\
    \textsf{(-1,-1,-1,-1,-2,-1,-2,-2,-1,-1,-1,-1,-1,-1,-2;1)}$_\text{I}$,\  &\  \textsf{(-1,-1,-1,-1,-2,-1,-1,-2,-1,-1,-1,-1,-1,-1,-1;1)}$_\text{I}$,\\
    \textsf{(-}$\frac{\textsf{1}}{\textsf{2}}$\textsf{,0,-1,-1,-1,-1,-1,-}$\frac{\textsf{1}}{\textsf{2}}$\textsf{,-1,-1,-1,-}$\frac{\textsf{1}}{\textsf{2}}$\textsf{,-1,-1,-}$\frac{\textsf{1}}{\textsf{2}}$\textsf{;1)}$_\text{IIB}$,\  &\  \textsf{(-}$\frac{\textsf{1}}{\textsf{2}}$\textsf{,0,-1,-1,-1,-1,-}$\frac{\textsf{1}}{\textsf{2}}$\textsf{,-}$\frac{\textsf{1}}{\textsf{2}}$\textsf{,-1,-1,-}$\frac{\textsf{1}}{\textsf{2}}$\textsf{,-}$\frac{\textsf{1}}{\textsf{2}}$\textsf{,-1,-}$\frac{\textsf{1}}{\textsf{2}}$\textsf{,-}$\frac{\textsf{1}}{\textsf{2}}$\textsf{;1)}$_\text{IIB}$,\\
    \textsf{(-1,0,-1,-1,-2,-2,-2,-1,-2,-2,-2,-1,-2,-2,-1;1)}$_\text{I}$,\  &\  \textsf{(-1,0,-1,-1,-2,-2,-2,-1,-1,-1,-1,-1,-2,-2,-1;1)}$_\text{I}$,\\
    \textsf{(-1,0,-1,-1,-2,-2,-1,-1,-2,-2,-1,-1,-2,-1,-1;1)}$_\text{I}$,\  &\  \textsf{(-1,0,-1,-1,-2,-2,-1,-1,-1,-1,-1,-1,-2,-1,-1;1)}$_\text{I}$,\\
    \textsf{(-1,0,-1,-1,-2,-1,-1,-1,-1,-1,-1,-1,-1,-1,-1;1)}$_\text{I}$,\  &\  \textsf{(0,-1,-1,-1,0,0,0,0,-1,-1,-1,-1,-2,-2,-2;1)}$_\text{I}$,\\
    \textsf{(0,0,-1,-1,-1,-1,-1,-1,-1,-1,-1,-1,-1,-1,-1;1)}$_\text{I}$,\  &\  \textsf{(0,0,-1,-1,0,0,0,0,-1,-1,-1,-1,-1,-1,-1;1)}$_\text{I}$,\\
    \textsf{(0,0,-1,-1,0,0,0,0,-1,-1,0,-1,-1,0,0;1)}$_\text{I}$,\  &\  \textsf{(0,0,-1,-1,0,0,0,0,-1,0,-1,-1,0,0,-1;1)}$_\text{I}$,\\
\end{longtable}
There are 42 region vectors for figure~\ref{mass_expansion_examples_nonplanar_15}, which are:
\begin{longtable}{ll}
    \textsf{(-}$\frac{\textsf{1}}{\textsf{2}}$\textsf{,-}$\frac{\textsf{1}}{\textsf{2}}$\textsf{,-}$\frac{\textsf{1}}{\textsf{2}}$\textsf{,-1,-1,-1,-1,-}$\frac{\textsf{1}}{\textsf{2}}$\textsf{,-1,-1,-}$\frac{\textsf{1}}{\textsf{2}}$\textsf{,-}$\frac{\textsf{1}}{\textsf{2}}$\textsf{,-1,-1,-1;1)}$_\text{IIB}$,\  &\  \textsf{(-1,-1,-1,-1,-1,-2,-2,-2,-2,-2,-2,-1,-1,-1,-2;1)}$_\text{I}$,\\
    \textsf{(-1,-1,-1,-1,-1,-2,-2,-1,-2,-2,-1,-1,-2,-2,-2;1)}$_\text{I}$,\  &\  \textsf{(-1,-1,-1,-1,-1,-2,-2,-1,-2,-2,-1,-1,-1,-1,-2;1)}$_\text{I}$,\\
    \textsf{(-1,-1,-1,-1,-1,-2,-1,-1,-2,-1,-1,-1,-1,-1,-1;1)}$_\text{I}$,\  &\  \textsf{(-}$\frac{\textsf{1}}{\textsf{2}}$\textsf{,0,-1,-1,-}$\frac{\textsf{3}}{\textsf{2}}$\textsf{,-1,-1,-1,-}$\frac{\textsf{1}}{\textsf{2}}$\textsf{,-1,-2,-2,-2,-}$\frac{\textsf{1}}{\textsf{2}}$\textsf{,-}$\frac{\textsf{1}}{\textsf{2}}$\textsf{;1)}$_\text{IIB}$,\\
    \textsf{(-}$\frac{\textsf{1}}{\textsf{2}}$\textsf{,0,-1,-1,-}$\frac{\textsf{3}}{\textsf{2}}$\textsf{,-1,-1,-1,-}$\frac{\textsf{1}}{\textsf{2}}$\textsf{,-1,-1,-2,-1,-}$\frac{\textsf{1}}{\textsf{2}}$\textsf{,-}$\frac{\textsf{1}}{\textsf{2}}$\textsf{;1)}$_\text{IIB}$,\  &\  \textsf{(-}$\frac{\textsf{1}}{\textsf{2}}$\textsf{,0,-1,-1,-1,-1,-1,-1,-}$\frac{\textsf{1}}{\textsf{2}}$\textsf{,-1,-}$\frac{\textsf{3}}{\textsf{2}}$\textsf{,-1,-1,-}$\frac{\textsf{1}}{\textsf{2}}$\textsf{,-}$\frac{\textsf{1}}{\textsf{2}}$\textsf{;1)}$_\text{IIB}$,\\
    \textsf{(-}$\frac{\textsf{1}}{\textsf{2}}$\textsf{,0,-1,-1,-1,-1,-1,-1,-}$\frac{\textsf{1}}{\textsf{2}}$\textsf{,-1,-1,-1,-}$\frac{\textsf{1}}{\textsf{2}}$\textsf{,-}$\frac{\textsf{1}}{\textsf{2}}$\textsf{,-}$\frac{\textsf{1}}{\textsf{2}}$\textsf{;1)}$_\text{IIB}$,\  &\  \textsf{(-}$\frac{\textsf{1}}{\textsf{2}}$\textsf{,0,-1,-1,-1,-1,-1,-1,-}$\frac{\textsf{1}}{\textsf{2}}$\textsf{,-1,-1,-}$\frac{\textsf{1}}{\textsf{2}}$\textsf{,-}$\frac{\textsf{1}}{\textsf{2}}$\textsf{,-1,-}$\frac{\textsf{1}}{\textsf{2}}$\textsf{;1)}$_\text{IIB}$,\\
\end{longtable}

\bibliographystyle{JHEP}
\bibliography{refs}

\end{document}